\documentclass[figuresleft,aps,nofootinbib,prd,preprintnumbers,sotoncolor]{uosthesis2}      
\graphicspath{{Figures/}}   
\usepackage{bibentry}          
\nobibliography*               
\usepackage{attrib}            
\hypersetup{colorlinks=true}   
\newcommand{\BibTeX}{{\rm B\kern-.05em{\sc i\kern-.025em b}\kern-.08em T\kern-.1667em\lower.7ex\hbox{E}\kern-.125emX}}

\newcounter{address}
\newcounter{alg}

\department  {School of Physics and Astronomy}
\DEPARTMENT  {\MakeUppercase{\deptname}}
\group       {[Group name]}
\GROUP       {\MakeUppercase{\groupname}}
\faculty     {Faculty of Engineering and Physical Sciences}
\FACULTY     {\MakeUppercase{\facname}}
\title      {Fermion mass hierarchies from vector-like families and possible explanations for the electron and muon anomalous magnetic moments}
\authors    {Huchan Lee} 
\addresses  {\groupname\\\deptname\\\univname}
\date       {\today}

\qualifications{The degree of M.S. in physics, Yonsei University}
\orcidid{0000-0002-7197-2937}
\doi{10.1002/0470841559.ch1}
\subject    {}

\newcommand{\func}{\operatorname}
\DeclareUnicodeCharacter{2212}{-}
\DeclareUnicodeCharacter{2032}{'}
\usepackage{verbatim,smartdiagram,float,tikz,amsmath,slashed,booktabs,cite,graphicx}
\usepackage{rotating}
\usepackage[utf8]{inputenc}
\usepackage[compat=1.1.0]{tikz-feynman}
\usepackage[export]{adjustbox} 
\usetikzlibrary{decorations.pathreplacing}
\newcommand{\tikzmark}[1]{\tikz[overlay,remember picture] \node[baseline] (#1) {};}
\newcommand\scalemath[2]{\scalebox{#1}{\mbox{\ensuremath{\displaystyle #2}}}}

\tikzset{My Node Style/.style={midway, right, xshift=3.0ex, align=left, font=\small, draw=none, thin, text=black}}

\newcommand\VerticalBrace[4][]{%
\begin{tikzpicture}[overlay,remember picture]
  \draw[decorate,decoration={brace, amplitude=1.5ex}, #1]
    ([yshift=1ex]#2.north east)  -- ([yshift=-1ex]#3.south east)
        node[My Node Style] {#4};
\end{tikzpicture}
}

\begin{document}
\pagenumbering{gobble} 
\copyrightDeclaration{} 
\raggedright                  
\frontmatter
\maketitle
\begin{abstract}
Many great efforts to find an answer on what are the most fundamental particles and forces in our nature have shaped the very important and beautiful theory known as the Standard Model (SM). The observables in the SM are consistent with their experimental bounds with high accuracies. However, many particle physicists agree that the SM is not an ultimate answer to our nature, since there are many observables which can not be addressed by the SM such as mass of neutrinos, a few of well-known anomalies in the SM, the puzzle of the CKM and PMNS mixing matrices, the Dark Matter (DM) and the Dark Energy (DE), etc. In order to bring these interesting topics to understanding of the human beings, it assumes that expansion of the SM is not avoidable and we call this expanded theory ``Beyond Standard Model (BSM)". Many possible BSM models have been suggested to connect with al least one of the listed observables and this idea motivates us to search for physics beyond the SM. Recapitulating the whole story, the SM itself is a great success, however it should be expanded by any means to explain the observables whose mechanisms are not confirmed.
\\~\\
We start from this consideration: how can we expand the SM without violating the gauge symmetry and the current experimental bounds for the SM observables. It is evident that the SM must be expanded for the observables which can not be addressed by the SM as discussed above. A possible answer to the question is a minimal extension to the SM and then to study the well-known anomalies and studying the anomalies was a main target over my two works~\cite{CarcamoHernandez:2019ydc,Hernandez:2021tii}. The other way is to study the FCNC observables within a minimally extended SM, as they are very sensitive to new physics and this approach is a main target of my third project. We made use of the model-dependent approach since there are new operators, which can not be fully replaced by the effective operators appearing in the model-independent approach. As we take the model-dependent approach, it is necessary to extend at least one of the following sectors: SM fermion, scalar and gauge symmetry. An important difference between our first and second (as well as third) work is whether we considered the hierarchical structure of the SM, as we regard the strong hierarchical structure of the SM as a very clear hint at new physics at higher energy scales.
\\~\\
A main motivation of our first work is to explain the muon and electron anomalous magnetic moment $g-2$ simultaneously. In order to achieve this goal, we extend the SM fermion sector by the fourth vector-like family and the scalar sector by a singlet flavon and the SM gauge symmetry by the local $U(1)^{\prime}$ symmetry. Under an assumption that our $Z^{\prime}$ neutral gauge boson only couples to the SM charged leptons, we defined the $Z^{\prime}$ coupling constants by using the mixing formalism in the mass basis. In order to make our analysis as simple as possible, we constrained the relevant mixing angles between $i$th chiral SM family and $4$th vector-like family to be $\theta_{12,14,24}$ for the charged leptons. In this analysis, the mixing angles $\theta_{12,14,24}$ are free parameters and they are constrained by experimental bounds of the anomalies, the branching ratio of $\mu \rightarrow e \gamma$, and neutrino trident production. Using the mass insertion approximation, we distinguished two mass sources, one of which is the chirality flip mass $M_{4}^{C}$ whereas the other is the vector-like mass $M_{4}^{L}$. What we found there is increasing the mixing angle $\theta_{12}$ slightly gives an unacceptably high prediction for the branching ratio of the charged lepton flavor violation (CLFV) $\mu \rightarrow e \gamma$ decay, and this becomes a good motivation to vanish the mixing angle through rest of the analysis. The dominant contribution to each anomaly arises from the $Z^{\prime}$ left-right interactions including an enhancement factor $M_{4}^{C}/m_{\mu}$, and the chirality flip mass $M_{4}^{C}$ can not increase as much as the vector-like mass $M_{4}^{L}$ does, as it is governed by the SM Higgs vev. For this reason, we constrained the chirality flip mass to be ranged from $0$ to $200\func{GeV}$, and then we found no any value between them can satisfy both anomalies, so leading to a conclusion this BSM model can not explain them simultaneously.
\\
Our second BSM model in our second work goes one step further from the first BSM model by implementing the hierarchical structure of the SM in a kinematic way. In order to achieve this goal, we need to assume the SM Lagrangian is the effective Yukawa interactions arising as a result of broken $U(1)^{\prime}$ symmetry and what this implement is the general Yukawa interactions can not take place due to the $U(1)^{\prime}$ charge. Under this consideration, our second model features that the SM fermions are augmented by two vector-like families and the scalar sector are enlarged by one more SM-like Higgs $H_d$ and a singlet flavon $\phi$ and lastly the SM gauge symmetry is extended by the global $U(1)^{\prime}$ symmetry (Notice that this $U(1)^{\prime}$ is global). One vector-like family can provide two effective seesaw operators, so this is why we introduce two vector-like families, and then all SM generations can acquire their masses. A clear difference between our first and second work is whether we built a mass matrix for each sector of the SM and the construction was done in our second work, so the mixing angles appearing in the second work become a ratio between the Yukawa and vector-like masses. We defined all required mixings, while diagonalizing the mass matrix for the charged lepton sector, and then discussed both anomalies mediated by the SM $W$ gauge boson and by the non-SM scalars at one-loop level. First of all, the $W$ contributions to both anomalies turn out to be too small to its experimental bound, so we conclude another approach is required to explain both anomalies simultaneously and come up with the non-SM scalar exchange at one-loop level and then finally confirm both anomalies can be explained by the non-SM scalar exchange simultaneously.
\\
A main motivation of our third work arises from studying the flavor changing neutral currents (FCNCs) to constrain masses of the vector-like family, while keeping the hierarchical structure of the SM implemented in the second work. What we considered especially important is to diagonalize a mass matrix for each fermion sector without any assumptions. For the correct diagonalization, we mainly focus on the second and third generation of the SM at cost of having massless particles in the first SM generation with only one vector-like family. In order to study the FCNC observables, we consider the SM $Z$ gauge boson, however it is evident that the SM $Z$ gauge boson can not generate the flavor violating interactions. The SM $Z$ gauge boson can cause the renormalizable flavor violating interactions by extending the SM fermions by the vector-like family and by operating $SU(2)$ violating mixings, and then it can have small non-zero off-diagonal $Z$ coupling constants in the mass basis. Using the defined $Z$ coupling constants in the mass basis, we analyze the charged lepton sector first via the CLFV $\tau \rightarrow \mu \gamma, \tau \rightarrow 3\mu$ and $Z \rightarrow \mu \tau$ decays, predicting the singlet or doublet vector-like charged lepton masses. Our numerical predictions are not significantly constrained by the experimental bounds for the CLFV decays, however it comes as the CMS experimental bound for the vector-like doublet charged leptons might be able to exclude our predictions to a significant extent. As for the quark sector, we use the rare $t \rightarrow c Z$ decay and the CKM mixing matrix and the CKM mixing matrix appears as a challenging observable to fit our predictions. After fitting our prediction to the CKM mixing matrix as much as possible, we confirm that no any point of our predictions is excluded by the experimental bound for the rare $t \rightarrow c Z$ decay, predicting mass range of vector-like quarks as in the charged leptons.
\\~\\
Based on the minimal extension of the SM in my three works, it has confirmed that physics beyond the SM can be explored in simple scenarios, leading to interesting scientific predictions related to the hypothetical particles such as vector-like particles, CP-even and -odd scalars, $Z^{\prime}$, etc. and these findings can be verified or ruled out in close future experiments.
\end{abstract}
\tableofcontents
\listoffigures
\listoftables
\authorshipdeclaration{\bibentry{Gunn:2001:pdflatex}\newline\bibentry{Lovell:2011:updated}\newline\bibentry{Gunn:2011:updated2}}
\acknowledgements{My PhD time at university of Southampton consists of my sincere appreciation to my supervisor prof. Steve, doctor Antonio, doctor Sam and my families. When I was in South Korea, my wish to learn the very beautiful structure of our mother nature in a mathematical way led me to come to university of Southampton. This was my first time to study abroad and it was never as easy as I had expected in many senses. Whenever I felt difficulties while studying here, the inherent power for me to endure the difficult times came from my supervisor Steve's comments and his teaching. From most of his words, this sentence ``You have worked very hard" has cheered me sincerely and motivated me forward again and again. What I must remember about his teaching is Steve can give me very certain as well as correct answers to my questions. Based on his insightful answers, I have been able to make my background to the phenomenology of the particle physics firm and established. No matter how much I have prepared for the meeting every week, Steve has made me surprise through his insightful and intuitive knowledge during the meeting and this fact has always convinced me taking prof. Steve was a right determination. Doctor Antonio has also taught me many important physical backgrounds and invaluable techniques. I am very grateful for his dedicated assistance for the three works we have worked together and especially the scalar sector of our second work could not be finalized without his help. On top of that, he has not mind to discuss with me more than $2$ hours sometimes and I am very happy to know Antonio and discuss many important features of our works. Doctor Sam is a person I should not forget in many ways. A quite memorable story for me is the time when I was a first year PhD student. At that time, I was not good at the tool Mathematica to carry out a numerical study in our first work and Sam really helped me teaching his code with a very kind remark to make me understand meaning of the code. Sam's teaching and discussion at that time have been important backgrounds to learn deeper knowledge from prof. Steve and Doctor Antonio. Plus, I have never ever forgotten my families's sincere and wholehearted support and help. If prof Steve, Doctor Antonio and Sam have helped me outside, my family have supported me inside and this has made me not broken no matter what difficulty I have met during my study time. I really and sincerely appreciate to the ones who have waited me until I am able to stand alone so far in many ways.}
\dedicatory{To my lovely family...
\\~\\
Without any of you, this beautiful study would never be achieved.}
\listofsymbols{ll}
{
2HDM & 2 Higgs Doublet Model \\
BSM & Beyond Standard Model \\
CKM & Cabbibo-Kobayashi-Maskawa \\
CLFV & Charged Lepton Flavor Violation \\
CM & Center-of-Momentum \\
DM & Dark Matter \\
DE & Dark Energy \\
FCNC & Flavor-Changing-Neutral-Current \\
GIM & Glashow-Lliopoulos-Maiani \\
GUT & Grand Unification Theory \\
LH & Left-Handed \\
LLP & Long-Lived Particle \\
LQ & LeptoQuark \\
NLO & Next Leading Order \\
PMNS & Pontecorvo-Maki-Nakagawa-Sakata \\
QFT & Quantum Field Theory \\
RH & Right-Handed \\
SM & Standard Model \\
SSB & Spontaneous Symmetry Breaking \\
ToE & Theory of Everything \\
VEV & Vacuum Expectation Value \\
VL & Vector-Like \\
WIMP & Weakly Interacting Massive Particle \\
}
\mainmatter
\chapter{Introduction} \label{Chapter:Introduction}

Many dedicated efforts to find an answer on what are the most fundamental particles and forces consisting of our nature have shaped the awesome and beautiful theory known as the Standard Model (SM). The discovery of the Higgs particle at CMS and ATLAS in 2012~\cite{ATLAS:2012yve,CMS:2012qbp}, especially, was one of the great successes in the history of particle physics as it is the only scalar particle in the SM at the moment and motivated us to search for physics beyond the Standard Model (SM). With the last puzzle of the Higgs particle, the SM looks complete and it has actually explained many observables with high accuracies and one of them is the correct measurement of the CKM mixing matrix.
\begin{figure}[H]
\centering
\includegraphics[keepaspectratio,width=0.7\textwidth]{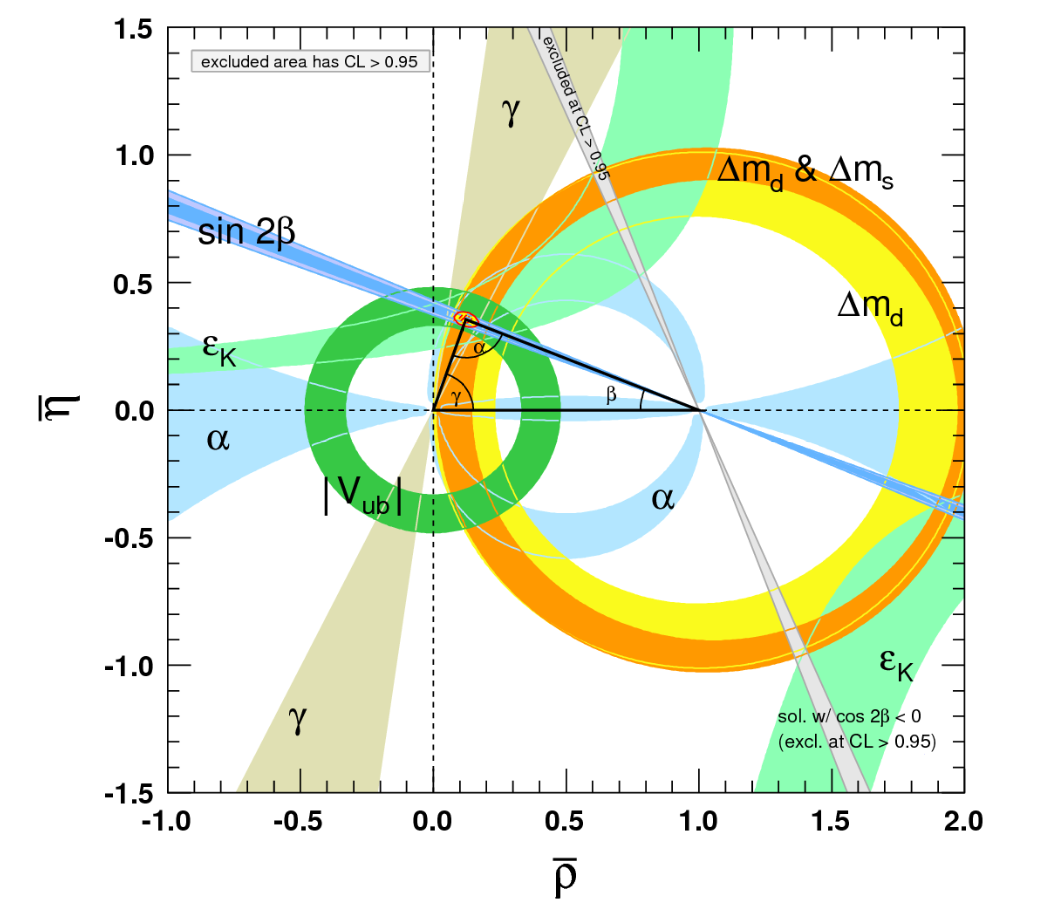}
\caption{The correct measurement of the CKM mixing matrix in terms of the parameters $\bar{\rho}$ and $\bar{\eta}$}
\label{fig:CKM_measured}
\end{figure}
What the stunning feature of Figure~\ref{fig:CKM_measured} tells is there exists a mixing between each generation of the SM fermions and the mixing mechanism is a principal rule when we extend the fermion sector of the SM. Interestingly, this mixing mechanism appears not only in the fermion sector but also in the electroweak gauge symmetry $SU(2)_L \times U(1)_Y$ sector.
\begin{enumerate}
\item The $SU(2)_L$ eigenstates $W_{1,2,3}$ and the $U(1)_Y$ eigenstate $B$
\item The mixing between the gauge particles in the flavor basis leads to the physical gauge particles such as $W^{\pm}, Z$ and $\gamma$.
\end{enumerate}
The mixing mechanism also tells that there must exist two eigenstates, one of which is the mass eigenstate (or equally the physical eigenstate) and the other is the flavor eigenstate (or equally the interaction eigenstate). The most sizeable mixing of the SM appears in the neutrino sector, confirmed in the neutrino oscillation experiment~\cite{Esteban:2020cvm}:
\begin{equation}
\theta_{23} \approx 49^{\circ}, \qquad \theta_{12} \approx 33^{\circ}, \qquad \theta_{13} \approx 9^{\circ}
\end{equation}
The large mixing angles of the neutrinos lead to the large off-diagonal elements in the PMNS mixing matrix and the different mixing patterns appearing in the CKM and PMNS mixing matrix are the well-known puzzle and will be discussed in detail in chapter \ref{Chapter:TheSM}. Even though the SM is greatly successful, many particle physicists agree that the observables such as masses of the SM neutrinos, a few of well-kwnon anomalies, etc. point out the SM itself is not enough to cover all aspects of our universe and the SM must be extended to the ``Beyond Standard Model" (BSM) in order to understand observables whose mechanisms are not confirmed. The way to new physics arises from both the theoretical and experimental approaches.
\\~\\
As for the theoretical approach, we start from this consideration: how can we expand the SM without violating the gauge symmetry and the current experimental bounds for the observables. It is evident that the SM must be expanded due to the limitations of the SM, which will be discussed in chapter~\ref{Chapter:TheSM}. A possible answer to the question is a minimal extension to the SM and then to study the well-known anomalies in the BSM model. This approach was done over our two works~\cite{CarcamoHernandez:2019ydc,Hernandez:2021tii}. Another answer is to study the FCNC observables in a minimally extended SM, as they are very sensitive to new physics, and this approach was done in our third work. Therefore, the minimal extension to the SM becomes quite important and this can be done by enlarging at least one of the following sectors: the SM fermion, the scalar and lastly the gauge symmetry. For the enlargement of the SM fermion sector, two hypothetical particles has been widely considered, which are the vector-like (VL) and leptoquark (LQ) particles. As we have made use of the vector-like particles over my three works, the vector-like fermions would be a main target in this thesis and a detailed description of the vector-like fermions will be studied in subsection~\ref{sec:pre_VL} as a prerequisite. As in the fermion sector, there have been many attempts to extend the SM scalar sector. The discovery of the SM Higgs particle in 2012 has led to a lot of questions, one of which is the small SM Higgs mass $125\func{GeV}$ is a result of the extremely fine tuned parameter following the QFT calculation. Based on this result, many particle physicists assume that more possible scenarios for the extension to the SM scalar sector will be lied ahead. With this motivation, we augmented the SM scalar sector by more scalar particles. In our first work, a singlet flavon is added for the enlargement of the scalar sector. For the second and third works, the SM scalar sector is extended by one more SM-like Higgs and the singlet flavon, so the BSM models feature the well-known 2HDM. Featuring the 2HDM is also motivated by the strong hierarchical structure of the SM fermions and this will be discussed in subsection~\ref{sec:pre_2HDM} as another prerequisite. Lastly, extension of the SM gauge symmetry has gotten intense attention for the need of the unification. This is based on the feature of the running coupling constants for the electromagnetic, weak and strong force and the feature tells they can be converged at a point at GUT scale. However, attempts to connect directly from the SM to either the GUT or ToE will be very likely to mislead our understanding to the known phenomenology, since there are many intermediate breaking patterns as well as hypothetical particles, and this connection is also opposite to the minimal extension. For this reason, we consider the simplest possibility: $U(1)^{\prime}$ symmetry. This feature will also be discussed in subsection~\ref{sec:pre_U1p} as the last prerequisite.
\\~\\
Now that we look at a few of aspects for the theoretical approach to arrive at the BSM, it is also important to figure out how we can find out new physics from experiments. There are three ways as follows:
\begin{itemize}
\item The energy frontier
\item The luminosity (intensity) frontier
\item The cosmic frontier
\end{itemize}
The energy frontier is simply to increase the CM collision energy in order to find a new particle in person and the direct discovery of the SM Higgs particle at CERN in 2012 is a great success of the energy frontier. After that discovery, no any new particle has been found by the experiments with $13\func{TeV}$ so far and this implies not just the energy frontier but also the luminosity frontier should be considered as of equal importance. The luminosity frontier is simply to increase the number of events to find out some anomaly within the events. This luminosity frontier is strongly preferred for the FCNC observables, since they are very suppressed in the SM by the Glashow-Iliopoulos-Maiani (GIM) mechanism. The FCNC observables, very sensitive to new physics, are a main target in our third work. Lastly, the cosmic frontier is to study the ultra-relativistic cosmic particles and this is deeply related to the cosmology and high energy neutrinos.
\\~\\
In this introduction, we simply look at how successful the SM is by discussing the mixing formalism appearing in both the fermion sector and the gauge symmetry sector. Even though the SM is quite successful in many fields, there are some important limitations such as the masses of the SM neutrinos, a few of well-known anomalies, dark matter, dark energy, etc. and these limitation tells the SM must be expanded to the BSM. For the extension, at least one of the following sectors, the SM fermion, scalar and gauge symmetry, must be extended and we discuss the theoretical aspects of the BSM models and its implications through my two published works~\cite{CarcamoHernandez:2019ydc,Hernandez:2021tii} plus my third work~\cite{Hernandez:2021oyv}.
\\~\\
This thesis is organized as follows. In chapter~\ref{Chapter:TheSM}, we discuss main features of the SM and its principal limitations. In chapter~\ref{Chapter:The1stBSMmodel}, the three common theoretical tools, which are the vector-like fermions, 2HDM and $U(1)^{\prime}$ symmetry, appearing in my three works are covered as prerequisites and the first BSM model as well as its mixing formalism in my first work are discussed. In chapter~\ref{Chapter:The1stpaper}, we try to explain both the muon and electron anomalous magnetic moments in a $Z^{\prime}$ model and discuss the experimental $Z^{\prime}$ mass bound and lastly conclude main features of our first work. In chapter~\ref{Chapter:The2ndBSMmodel}, we introduce our second BSM model and discuss how this model is different compared to our first BSM model. Plus, it will be emphasized that this model is strongly motivated by the hierarchical structure of the SM. In chapter~\ref{Chapter:The2ndpaper}, we discuss the first non-SM $W$ gauge boson contributions to both anomalies as an attempt to explain them simultaneously. In chapter~\ref{Chapter:The2ndpaper2}, we discuss the second non-SM scalar contributions to both anomalies, while investigating the scalar potential, and then conclude main features of the second work. 
In chapter~\ref{Chapter:The3rdBSMmodel}, we discuss our third BSM model with one vector-like family and the SM $Z$ gauge coupling constants in the mass basis. In chapter~\ref{Chapter:The3rdpaper}, we discuss diverse FCNC observables in both quark and lepton sectors as well as the CKM mixing matrix without unitarity to constrain our BSM model predictions and conclude our third paper.
Finally, we conclude main features of this thesis in chapter~\ref{Chapter:Conclusions}.
\chapter{The Standard Model and its limitations} \label{Chapter:TheSM}

\begin{figure}[H]
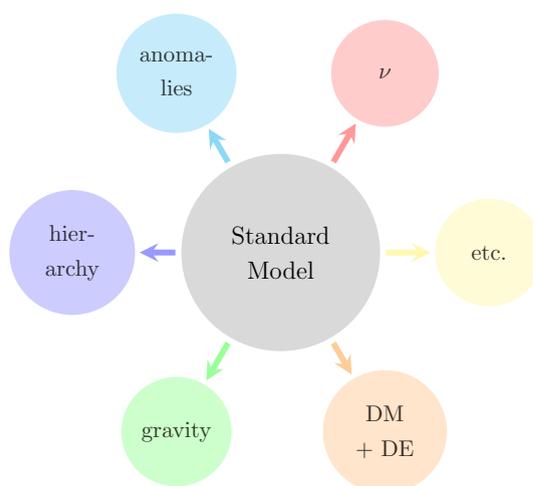

\centering
 \resizebox{.5\linewidth}{!}{%
\smartdiagram[constellation diagram]{Standard Model,
  $\nu$, anomalies, hierarchy, gravity, DM + DE, etc.}
  }%
\caption{The SM and the unspecified observables}
\label{fig:SM_unspecified_pheno}
\end{figure}

After discovering the charm quark through $J/\psi$ meson in 1974, the bottom quark in 1977, the top quark in 1995 and lastly the neutral component of Higgs particle in 2012, the current form of the Standard Model (SM) was established. The SM has been tested for many experiments and explained them with very high accuracies. The firmly accepted SM becomes now the cornerstone of the particle physics to figure out how our beautiful nature works in this universe. At the same time, however, the particle physicists have understood that the SM is not an ultimate answer for our nature due to lots of unspecified observables such as masses of the SM neutrinos, a few of well-known anomalies, hierarchical structure of the SM, gravity, matter-anti matter asymmetry and dark matter (DM) plus dark energy (DE), etc. A thing is certain, though, that all unidentified observables must start from the SM at the electroweak scale. This relation between the SM and the observables which can not be addressed by the SM can be pictorialized in Figure \ref{fig:SM_unspecified_pheno}.
These interesting observables which can not be addressed by the SM require for the SM to be expanded in anyway and the expanded SM is called, especially, ``Beyond the Standard Model (BSM)". A lot of new BSM models have been suggested to explain the observables and I will cover our three attempts to explain some of the observables based on my papers across next chapters. Before we go into the details of the papers, we need to look into some important properties of the SM as well as the limitations of the SM in detail. The rest of this chapter is assigned for the explanations.

\section{The Standard Model} \label{sec:the_SM}
The beautiful and awesome SM can be first represented by its gauge symmetry
\begingroup
\begin{equation}
SU(3)_C \times SU(2)_L \times U(1)_Y,
\label{eqn:gauge_symmetry_SM}
\end{equation}
\endgroup
where $SU$ means the special unitary and the subscript $C,L,Y$ mean the color charge, the left-handed chirality and the hypercharge, respectively. This gauge symmetry is especially important since it determines which kind of interactions can take place and there are two interactions, the Yukawa interactions and gauge interactions. The fermions which appear in the SM consist of both quarks and leptons, charged under the gauge symmetry. The SM particle spectrum is given in Figure \ref{fig:SM_particles}.
\begingroup
\begin{figure}[H]
\centering
\includegraphics[keepaspectratio,width=0.7\textwidth]{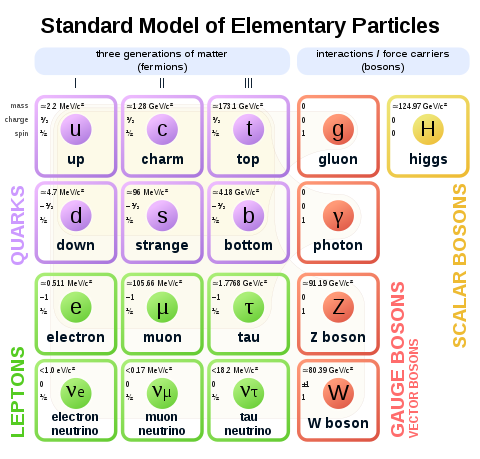}
\caption{The SM particle spectrum}
\label{fig:SM_particles} 
\end{figure}
\endgroup
The SM particles can be categorized by some principle standards. The first principal standard is whether they belong to either fermions or bosons. The fermions are the spin half-integer particles obeying the Pauli's exclusion principle, therefore they can not occupy the same quantum numbers and follow the Fermi-Dirac statistics, whereas the bosons are opposite to the fermions in a couple of senses that they do not respect the exclusion principle and they are the spin integer particles, so it is possible for them to take up the same quantum numbers and they follow the Bose-Einstein statistics. From the SM particles, they are separated as follows.
\begingroup
\begin{itemize}
\item Fermion : $u,c,t,d,s,b,\nu_e,\nu_\mu,\nu_\tau,e,\mu,\tau$
\item Boson : $h^0, \gamma, W^\pm, Z, g$
\end{itemize}
\endgroup
The SM particles again can be separated by mentioning their spins. In quantum field theory (QFT), the particles corresponding to spin $0$ are called scalars and there is only an unique scalar of ``Higgs" in the SM. The spin $1$ particles are called the vector particles and there are the photon $\gamma$, the $W$ bosons $W^\pm$ and lastly the gluons $g$ in the SM. The scalar and vector particles are inclusively grouped as the boson particles in that they have integer number of spin. All fermions of the SM have the spin $1/2$ as mentioned in the above context. These can be categorized as follows.
\begingroup
\begin{itemize}
\item spin $0$ particle : $h^0$ \hspace{1ex} \tikzmark{start}
\item spin $1$ particles : $\gamma, W^\pm, Z, g$ \hspace{1ex} \tikzmark{end}
\item spin $1/2$ particles : $u,c,t,d,s,b,\nu_e,e,\nu_\mu,\mu,\nu_\tau,\tau$ - Fermion
\end{itemize}
\endgroup
\VerticalBrace[thick]{start -| end}{end}{Boson}
The fermion particles can be separated again depending on whether or not they experience the strong interaction $SU(3)$ and the ones charged under the strong interaction are called ``quarks", while the ones not charged are called ``leptons".
\begingroup
\begin{itemize}
\item spin $0$ particle : $h^0$ \hspace{1ex} \tikzmark{top1}
\item spin $1$ particles : $\gamma, W^\pm, Z, g$ \hspace{1ex} \tikzmark{bot1}
\item spin $1/2$ particles (quarks) : $u,d,c,s,t,b$ \hspace{1ex} \tikzmark{top2}
\item spin $1/2$ particles (leptons) : $\nu_e,e,\nu_\mu,\mu,\nu_\tau,\tau$ \hspace{1ex} \tikzmark{bot2}
\end{itemize}
\endgroup
\VerticalBrace[thick]{top1 -| bot1}{bot1}{Boson}
\VerticalBrace[thick,]{top2 -| bot2}{bot2}{Fermion}
One of the most important differences between quarks and leptons is the lepton particles can be isolated, so it is possible to appear as an independent particle, whereas the quark particles can not be isolated due to the confinement, therefore they are observed as the shape of baryon or meson. The baryon is an composite particle consisting of three quarks (or three anti-quarks) and the meson is a composite particle consisting of one quark and one anti-quark. This interesting feature can be understood by the running coupling constant of each interaction and this will be detailed further in the next subsection of unification.
The fermion sector can be further divided by generation and how to divide the fermions by generation is associated with how stable the fermions are. For example, the $u$ quark which belongs to the first generation of the SM is most stable among the three generations, whereas the $t$ quark of third generation is most unstable therefore it decays into $b$ quark soon by the exchange of the $W$ gauge boson.
\begingroup
\begin{itemize}
\item spin $0$ particle : $h^0$ \hspace{1ex} \tikzmark{top3}
\item spin $1$ particles : $\gamma, W^\pm, Z, g$ \hspace{1ex} \tikzmark{bot3}
\item spin $1/2$ particles (quarks) : $\underbrace{u,d}_{1st},\underbrace{c,s}_{2nd},\underbrace{t,b}_{3rd}$ \hspace{1ex} \tikzmark{top4}
\item spin $1/2$ particles (leptons) : $\underbrace{\nu_e,e}_{1st},\underbrace{\nu_\mu,\mu}_{2nd},\underbrace{\nu_\tau,\tau}_{3rd}$ \hspace{1ex} \tikzmark{bot4}
\end{itemize}
\endgroup
\VerticalBrace[thick]{top3 -| bot3}{bot3}{Boson}
\VerticalBrace[thick,]{top4 -| bot4}{bot4}{Fermion}
It is interesting that the generation is related to the flavor, which describes each species of the SM fermions, thus there are six species in either quark or lepton sector. Since the particles in each generation share the similar patterns about their stability, relative lightness, etc., there have been many attempts to figure out quark and lepton sector based on their relative similarities and these kind of studies have been known as the flavor physics. My research fields are also deeply related to the flavor physics and I am going to talk about my three works based on features of the flavor physics. 
\\
The SM is also known as the chiral theory, which means the left-handed (LH) particles behave differently when compared to the right-handed (RH) particles. Taking the different chirality of each quark and lepton into account, they can be written symbolically in terms of the quantum numbers under the SM gauge symmetry ($i=1,2,3$).
\begingroup
\begin{itemize}
\item LH quark sector $Q_{iL} = \left( u_{iL}, d_{iL} \right)^T = \left( \mathbf{3}, \mathbf{2}, \frac{1}{6} \right)$
\item RH up-type quark sector $u_{iR} = \left( \mathbf{3}, \mathbf{1}, \frac{2}{3} \right)$
\item RH down-type quark sector $d_{iR} = \left( \mathbf{3}, \mathbf{1}, -\frac{1}{3} \right)$
\item LH lepton sector $L_{iL} = \left( \nu_{iL}, e_{iL} \right)^T = \left( \mathbf{1}, \mathbf{2}, -\frac{1}{2} \right)$
\item RH down-type lepton sector $e_{iR} = \left( \mathbf{1}, \mathbf{1}, -1 \right)$
\end{itemize}
\endgroup
The bold numbers $\mathbf{3}$ or $\mathbf{1}$ in parentheses mean the fields behave as a triplet or singlet under the color charge respectively, and the bold numbers $\mathbf{2}$ or $\mathbf{1}$ at the middle of the parentheses mean they transformed as a doublet or singlet under the left-handed chirality and finally the fractions or integer numbers are the hypercharge of each fermion field. The capital letters $Q,L$ are conventionally used to implement they are the left-handed doublets, whereas the small letters stand for the right-handed singlets. The SM Higgs field can be represented by the gauge symmetry notation.
\begingroup
\begin{itemize}
\item The Higgs field $H= \left( h^+, h^0 \right)^T = \left( \mathbf{1}, \mathbf{2}, \frac{1}{2} \right)$
\end{itemize}
\endgroup
The Yukawa interactions between two fermions and one scalar can arise by making the interactions be the gauge singlet. In other words, the interaction which is not the gauge singlet is not allowed to be written in the Lagrangian. Taking this property into account, a few of the allowed Yukawa interactions can be written as follows $(i,j=1,2,3)$
\begingroup
\begin{equation}
\mathcal{L} = (y_u)_{ij} \overline{Q}_{iL} \widetilde{H} u_{jR} + (y_d)_{ij} \overline{Q}_{iL} H d_{jR} + (y_l)_{ij} \overline{L}_{iL} H e_{jR} + \func{h.c.},
\label{eqn:Yukawa_interactions_SM}
\end{equation}
\endgroup
where $y_u,y_d,y_l$ are the Yukawa coefficients which determine the strength of the interactions for up, down quarks and charged leptons  respectively and $\widetilde{H}$ is defied as $i\sigma^2 H^*$. These terms are a few of the well-known Yukawa interactions in the SM and are going to be discussed in detail when introducing the mass insertion process in the next subsection. 
\\~\\
The last aspect of the SM particles is whether they are either Dirac or Majorana particles. The Dirac particles are the ones whose anti-particles is completely different to those, so they are clearly distinguishable, whereas the Majorana particles are the ones whose anti-particles are exactly same to the original particles. From this feature, it is possible to derive the Dirac and Majorana mass. The Dirac mass consists of two completely different particles, which means they behave differently under the SM gauge group therefore it is impossible to write their mass by hand since it violates the gauge conservation. As for the Majorana mass, it is possible to write their mass by hand as long as they are the trivial gauge singlets under the SM (or equally sterile), however there is no way to constrain the Majorana mass by experiments. Besides that, the Majorana mass requires one more condition, which is that the particles consisting of the Majorana mass must be neutral otherwise it violates the charge conservation unlike to the Dirac mass. The only candidate to meet this condition could be the neutrinos in the SM and all other fermions are Dirac particles. Even though the neutrinos are assumed to be the Majorana particles in the SM, their Majorana mass still violates the gauge singlet conservation since they are charged under the SM gauge group as $\left( \mathbf{1}, \mathbf{2}, -\frac{1}{2} \right)$ which is not the trivial gauge singlet. This problem is covered in the section ``first limitation of the SM - neutrinos" in detail.
\\~\\
Now that the particles consisting of the SM are introduced in some brief and compact way, it is necessary to look into a couple of main mechanisms like the Yukawa interactions, spontaneous symmetry breaking (SSB), and the broken gauge symmetry, step-by-step through in the next subsections.

\subsection{The Yukawa interactions and the spontaneous symmetry breaking} \label{sec:Yukawa_ssb}
In order to grab a sense of how the mass term can be written in the Lagrangian, consider the next term written by hand as just an example.
\begingroup
\begin{equation}
\mathcal{L} = m^2 \overline{e}e
\end{equation}
\endgroup
This term looks apparently right at first sight, however a problem starts to emerge when it takes the polarized basis using the left- and right-handed projector $P_{L,R}$. Expanding the term by inserting the projects, the result is
\begingroup
\begin{equation}
\begin{split}
\mathcal{L} &= m^2 \overline{e} \left( P_L + P_R \right) \left( P_L + P_R \right) e \\
&= m^2 \overline{e}_R e_L + \overline{e}_L e_R.
\label{eqn:toy_mass_ee}
\end{split}
\end{equation}
\endgroup
A clear problem of the result of Equation \ref{eqn:toy_mass_ee} is that each term is not the gauge singlet under the SM gauge group, therefore these kind of mass terms are not allowed to be written. Thus the mass terms for the Dirac fermions in any Lagrangian must arise from the Yukawa interactions discussed in Equation~\ref{eqn:Yukawa_interactions_SM}, which respect the gauge singlet condition. Consider the third term of Equation~\ref{eqn:Yukawa_interactions_SM} for simplicity.
\begingroup
\begin{equation}
\begin{split}
\mathcal{L} &= (y_l)_{ij} \overline{L}_{iL} H e_{jR} 
= (y_l)_{ij} 
\begin{pmatrix}
\overline{\nu}_{iL} & \overline{e}_{iL}
\end{pmatrix}
\begin{pmatrix}
h^+ \\ h^0
\end{pmatrix}
e_{jR}
\\
&= (y_l)_{ij} \left( \overline{\nu}_{iL} h^+ e_{jR} - \overline{e}_{iL} h^0 e_{jR} \right) 
\\
&\overset{?}{=} (y_l)_{ij} v \overline{e}_{iL} e_{jR} + \cdots = (m_e)_{ij} \overline{e}_{iL} e_{jR} + \cdots
\label{eqn:process_Diracmass}
\end{split}
\end{equation}
\endgroup
If the field $h^0$ can develop its vacuum expectation value (vev) $v$ as seen in Equation \ref{eqn:process_Diracmass}, the mass can be written by the form of the Yukawa constant $y_l$ multiplied by the vev $v$ in the Lagrangian. One thing to care about is whether the initial Yukawa and final mass terms conserve the gauge singlet condition. The initial Yukawa interaction $(y_l)_{ij} \overline{L}_{iL} H e_{jR}$ definitely maintains the gauge singlet condition, whereas the mass term $(m_e)_{ij} \overline{e}_{iL} e_{jR}$ does not. What this means is all mass terms in the Lagrangian appear as a result of broken symmetries. The symmetry breaking process takes place when the Higgs field $h^0$ develop its vev $v$ and this process is known as the spontaneous symmetry breaking (ssb), which occurs in the Higgs field potential. The spontaneous symmetry breaking process is especially important as it can allow masses for the SM Dirac fermions, written by the Yukawa constants multiplied by the non-zero vev, without violating the underlying gauge symmetry and this feature will be discussed in detail in the next subsection.
\subsection{The Higgs potential and the spontaneous symmetry breaking} \label{sec:Higgs_potential}
Consider first the Higgs potential which consists of a single real Higgs field $h$
\begingroup
\begin{equation}
V(h) = \frac{1}{2}m^2 h^2 + \frac{1}{4!}\lambda h^4,
\label{eqn:Higgs_potential_realfield}
\end{equation}
\endgroup
where $m^2$ is a positive mass parameter which has mass dimension of $2$ and $\lambda$ is a positive dimensionless quartic coupling constant. In the Higgs potential of Equation \ref{eqn:Higgs_potential_realfield}, no something special takes place; the minimum of the potential is appeared as the Higgs field $h$ approaches to $0$ and the minimum is simply $0$. To make this potential more interesting, we rewrite the Higgs potential with the mass parameter $m^2$ replaced by $-\mu^2$ where $\mu^2$ is a positive mass parameter.
\begingroup
\begin{equation}
V(h) = -\frac{1}{2}\mu^2 h^2 + \frac{1}{4!}\lambda h^4
\label{eqn:Higgs_potential_rewriten}
\end{equation}
\endgroup
The rewritten Higgs potential of Equation \ref{eqn:Higgs_potential_rewriten} gives rise to nonzero minimum of the Higgs potential $V(h)$ at nonzero value of the Higgs field $h$. The comparison between the potential of Equation \ref{eqn:Higgs_potential_realfield} and Equation \ref{eqn:Higgs_potential_rewriten} can be seen clearly by the below graph in Figure \ref{fig:comparison_two_HiggsPotentials}.
\begingroup
\begin{figure}[H]
\centering
\includegraphics[keepaspectratio,width=0.8\textwidth]{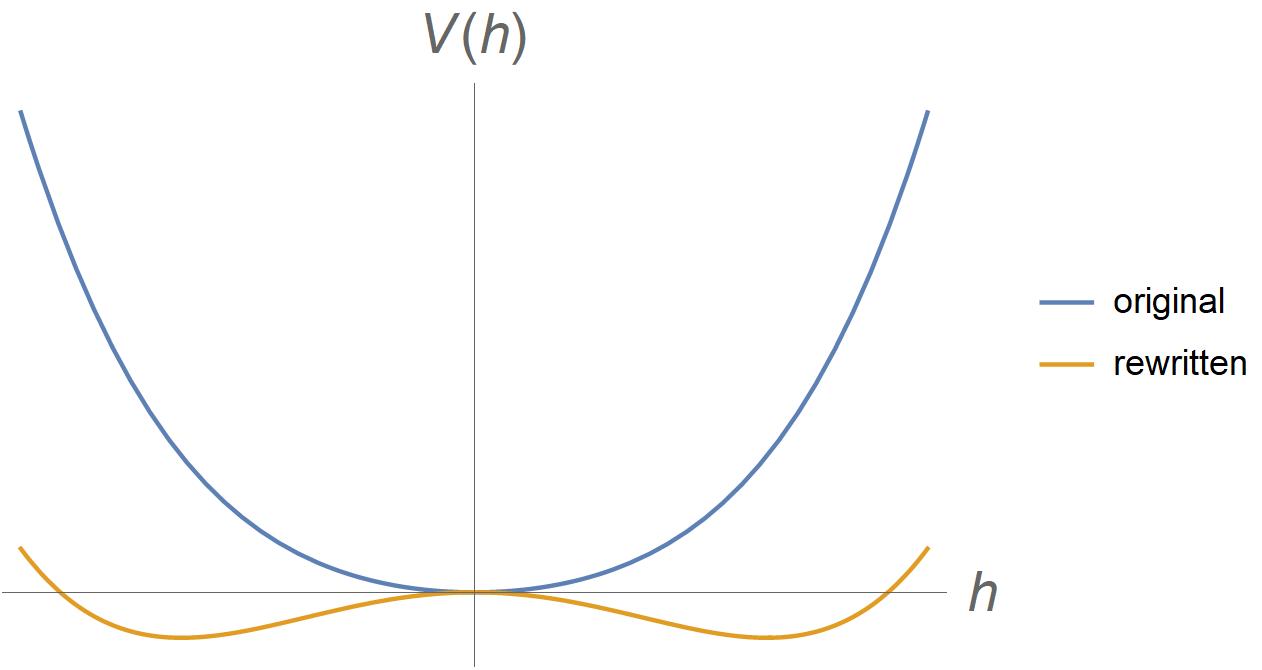}
\caption{The ``original" means the Higgs potential of Equation \ref{eqn:Higgs_potential_realfield} and the ``rewritten" means that of Equation \ref{eqn:Higgs_potential_rewriten}}
\label{fig:comparison_two_HiggsPotentials}
\end{figure}
\endgroup
A difference between the ``original" and ``rewritten" Higgs potential arises from the energy (or equally temperature) difference. In other words, the ``original" Higgs potential takes place at high energy (or equally at high temperature) and the potential turns into the ``rewritten" Higgs potential as the high energy goes down to low energy (or equally at low temperature). The low energy scale corresponds to the electroweak scale whose mass order is about a hundreds of $\func{GeV}$ scale and the interesting ssb process can take place at this energy scale. Imagine a ball is put at the center of the Higgs field in the ``rewritten" Higgs potential. As time goes on, the ball is likely to fall down either to left or right direction in the potential since the center position is unstable. Besides that, the direction which the ball will take can not be predicted and this is the reason why this symmetry breaking process takes place spontaneously. After the ball is put in the stable (or minimum) position, it can have nonzero value of the Higgs potential at the nonzero value of the Higgs field $h$ and the nonzero value of the Higgs potential is reinterpreted as the vacuum expectation value $v$. This process can also be understood mathematically by the Equation \ref{eqn:Higgs_potential_rewriten}. Differentiating the Higgs potential of Equation \ref{eqn:Higgs_potential_rewriten} with respect to the Higgs field $h$, the minimization condition reads off (Suppose that the Higgs field $h$ developed its vev $v$):
\begin{equation}
\begin{split}
\frac{\partial V(v)}{\partial h} &= -\mu^2 v + \frac{\lambda}{6} v^3 = v \left( -\mu^2 + \frac{\lambda}{6} v^2 \right) = 0 \\
&\rightarrow v = \pm \sqrt{\frac{6\mu^2}{\lambda}},0.
\label{eqn:vev_derivation}
\end{split}
\end{equation}
Since I am only interested in the nonzero vev $v$, the value $0$ will be excluded. It confirms that the vev $v$ is given from both the mass parameter $\mu^2$ and the quartic coupling constant $\lambda$ and the experimentally known value of the vev is $246.22\func{GeV}$. The order of this vev is $100\func{GeV}$ and this corresponds to the electroweak scale. An important feature of the Higgs potential is the Higgs field appearing in the SM is a complex field, so the Higgs potential features 3-dimensional space as shown in Figure~\ref{fig:Higgs_Complex}.
\begingroup
\begin{figure}[H]
\includegraphics[keepaspectratio,width=\textwidth]{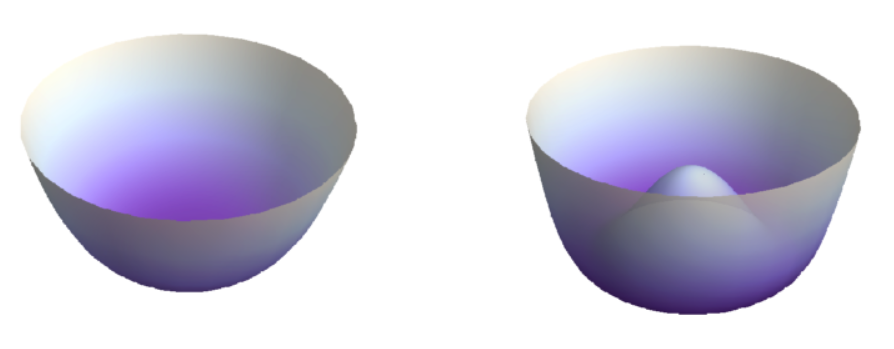}
\caption{The left is a complex Higgs potential at high temperature and the right is the complex Higgs potential at low temperature.}
\label{fig:Higgs_Complex}
\end{figure}
\endgroup
\subsection{The broken gauge symmetry of the SM and its interpretation} \label{sec:broken_gaugesymmetry}
The ssb process covered in the previous subsection applies exactly to the general SM Yukawa interactions with a couple of modifications. The first change is the single real Higgs field $h^0$ must be changed to the complex field in order to implement the $U(1)_Q$ electromagnetic symmetry. The second change is the single complex Higgs field $h$ needs to be doublet in order to make the SM Yukawa interactions be the gauge singlets, therefore one more charged Higgs $h^+$ is required. The generally accepted form of the SM Higgs field is given by
\begingroup
\begin{equation}
H = \begin{pmatrix}
h^+ \\ h^0
\end{pmatrix}.
\end{equation}
\endgroup
After the neutral component of the Higgs field $h^0$ develops its vev $v$ by the spontaneous symmetry breaking process, the Higgs doublet takes the below shape in unitary gauge.
\begingroup
\begin{equation}
H = \begin{pmatrix}
h^+ \\ h^0
\end{pmatrix} \rightarrow
\begin{pmatrix}
0 \\ \frac{1}{\sqrt{2}} \left( v + h \right)
\end{pmatrix}
\end{equation}
\endgroup
The neutral Higgs field $h$ after ssb is what the particle physicists found at CERN experiment in 2012. The SM Higgs field $H$ consists of one neutral Higgs and another charged Higgs, however what we found is the only neutral Higgs field since the charged Higgs is integrated out below the electroweak scale. If the energy scale goes up more than the electroweak scale, then we might be able to observe the charged Higgs $h^+$ in person and many studies have been suggested to find out the charged Higgs and thus to make the Higgs field $H$ complete. Taking a look at the value of $\frac{1}{\sqrt{2}}v$, it is about $174\func{GeV}$, which is quite close to the top quark mass (in other words, the top quark Yukawa coupling is nearly $1$) and this is the reason why the Higgs physics is sensitive to the correct measurement of the top quark mass. After the currently accepted form of the SM Higgs doublet field develops its vev, it confirms that how the SM gauge symmetry is broken to the smaller symmetry.
\begin{equation}
SU(3)_C \times SU(2)_L \times U(1)_Y \rightarrow SU(3)_C \times U(1)_Q \quad (\text{After ssb})
\label{eqn:broken_SMgaugesymmetry}
\end{equation}
After the spontaneous symmetry breaking, the electroweak gauge symmetry $SU(2)_L \times U(1)_Y$ is broken to the electromagnetic gauge symmetry $U(1)_Q$ and three of the massless $SU(2)$ gauge bosons $A^{1,2,3}$ plus $U(1)$ gauge boson $B$ in the SM get massive via the mixing mechanism after the Higgs develops its vev and one of the mixed gauge bosons remains massless. The three massive gauge bosons $W^+,W^-,Z$ and one massless gauge boson $\gamma$ appear as a linear combination of the four gauge bosons $A^1,A^2,A^3,B$ in the interaction basis as follows:
\begin{equation}
\begin{split}
\begin{pmatrix}
W^+ \\ W^-
\end{pmatrix}
&=
\frac{1}{\sqrt{2}}
\begin{pmatrix}
1 & -i \\
1 & +i 
\end{pmatrix}
\begin{pmatrix}
A^1 \\ A^2
\end{pmatrix}, \quad m_{W^\pm} = g\frac{v}{2} \\
\begin{pmatrix}
Z \\ A
\end{pmatrix}
&=
\frac{1}{\sqrt{g^2+g^{\prime 2}}}
\begin{pmatrix}
g & -g^\prime \\
g^\prime & g 
\end{pmatrix}
\begin{pmatrix}
A^3 \\ B
\end{pmatrix}, \quad m_{Z} = \sqrt{g^2+g^{\prime2}}\frac{v}{2}, \quad m_A = 0,
\label{eqn:WpWmZA_linearmixing}
\end{split}
\end{equation} 
where the coupling constant $g$ and $g^\prime$ are the ones in the covariant derivative of the Higgs field $h$
\begin{equation}
D_\mu h = \left( \partial_\mu - igA_\mu^a \tau^a - i\frac{1}{2}g^\prime B_\mu \right)h,
\label{eqn:covariant_derivative_h}
\end{equation}
where $\tau^a = \frac{\sigma^a}{2} \equiv T^a$ and the index $a$ runs from $1$ to $3$. The mixing matrix for the massless fields $A^3,B$ can be rewritten in terms of the Weinberg angle or weak mixing angle $\theta_W$.
\begin{equation}
\begin{pmatrix}
Z \\ A
\end{pmatrix}
=
\begin{pmatrix}
\cos\theta_W & -\sin\theta_W \\
\sin\theta_W & \cos\theta_W 
\end{pmatrix}
\begin{pmatrix}
A^3 \\ B
\end{pmatrix},
\label{eqn:ZA_Weinbergmixing}
\end{equation}
where $\cos\theta_W = \frac{g}{\sqrt{g^2+g^{\prime2}}}$ and $\sin\theta_W = \frac{g^\prime}{\sqrt{g^2+g^{\prime2}}}$. Then, the covariant derivative of $h$ can be rewritten in terms of the mass eigenstates $W^\pm,Z,A$.
\begin{equation}
\begin{split}
D_\mu h &= \left( \partial_\mu - ig A_\mu^1 T^1 - ig A_\mu^2 T^2 -ig A_\mu^3 T^3 - i\frac{1}{2}g^\prime B_\mu \right)h \\
&= \left( \partial_\mu - ig A_\mu^1 T^1 - ig A_\mu^2 T^2 -ig A_\mu^3 T^3 - iY g^\prime B_\mu \right)h \\
&= \bigg( \partial_\mu - ig \frac{1}{\sqrt{2}}\left( W_\mu^+ + W_\mu^- \right) T^1 - ig \frac{i}{\sqrt{2}}\left( W_\mu^+ - W_\mu^- \right) T^2 \\ 
&-ig \left( \cos\theta_W Z_\mu + \sin\theta_W A_\mu \right) T^3 - iY g^\prime \left( -\sin\theta_W Z_\mu + \cos\theta_W A_\mu \right) \bigg)h \\
&= \bigg( \partial_\mu - ig \frac{1}{\sqrt{2}}\left( W_\mu^+ T^+ + W_\mu^- T^- \right) -i\frac{g}{\cos\theta_W} \left( T^3 - \sin^2\theta_W Q \right) Z_\mu -ie Q A_\mu \bigg)h
\label{eqn:covariant_derivative_mass}
\end{split}
\end{equation}
While deriving the final result of Equation \ref{eqn:covariant_derivative_mass}, there are many important implications; from the first equality I rewrite the factor $\frac{1}{2}$ as the quantum number of $U(1)_Y$; from the second equality the covariant term is rewritten in terms of the mass eigenstates using the Equations \ref{eqn:WpWmZA_linearmixing} and \ref{eqn:ZA_Weinbergmixing}; from the last equality I rearrange terms with respect to $Z_\mu$ and $A_\mu$ using $T^\pm = T^1 \pm iT^2$ and defined the charge quantum number $Q$ by $T^3+Y$ and the electron charge $e$ by $gg^\prime/\sqrt{g^2+g^{\prime2}}$. With the redefined quantity $Q$, I also rearrange the interaction terms with $Z_\mu$. Summarizing the redefined quantities, they are given by:
\begin{equation}
e = \frac{gg^\prime}{\sqrt{g^2+g^{\prime2}}} = g\sin\theta_W = g^\prime\cos\theta_W, \quad Q=T^3+Y, \quad m_W = m_Z \cos\theta_W.
\label{eqn:eQmW}
\end{equation}
From the quantities of Equation \ref{eqn:eQmW}, the mass of weak gauge bosons are connected to each other by the weak mixing angle $\theta_W$ and the electric charge $e$ and quantum number $Q$ tells that the electromagnetic $U(1)_Q$ symmetry appear as a result of broken bigger gauge symmetry $SU(2)_L \times U(1)_Y$ by the SM Higgs vev $v$.

\subsection{The CKM mixing matrix} \label{sec:CKM}

One of the great successes in the SM is the discovery of the mixing in the fermion sector of the SM, as mentioned in the introduction, which is seen in person from the CKM mixing matrix as well as the PMNS mixing matrix. The CKM mixing matrix is appeared as a result of the mixing between the flavor and mass eigenstates of the fermions in the SM and it can be seen manifestly by the charged current $J_W^{\mu +}$. For the task, it needs to define the two bases connected by the unitary transformation as follows:
\begin{equation}
u_L^i = U_u^{ij} u_L^{\prime j}, \qquad d_L^i = U_d^{ij} d_L^{\prime j}
\end{equation}
where the indices $i,j$ run from $1$ to $3$ and $U_{u,d}$ are the unitary mixing matrices for the up- and down-quark sector in the SM. Then we are ready to write down the CKM mixing matrix in terms of the mixing matrices in the charged current $J_W^{\mu +}$ as follows:
\begin{equation}
\begin{split}
J_W^{\mu +} &= \frac{1}{\sqrt{2}} \overline{u}_L^i \gamma^{\mu} d_L^i = \frac{1}{\sqrt{2}} \overline{u}_L^{\prime i} \gamma^{\mu} \left( U_u^\dagger U_d \right)_{ij} d_L^{\prime j} \\
&= \frac{1}{\sqrt{2}} \overline{u}_L^{\prime i} \gamma^{\mu} V_{ij} d_L^{\prime j}
\end{split},
\end{equation}
where $V_{ij}$ is the unitary mixing matrix known as the Cabibbo-Kobayashi-Maskawa (CKM) mixing matrix. This unitary mixing matrix is first predicted by Cabbibo with two SM generations and is expanded by Kobayashi and Maskawa by assuming three generations of the SM. Therefore, the mixing angle in the unitary matrix consisting of only two generations is known as the Cabibbo mixing angle and its value is about $0.22$. The full CKM mixing matrix can be parameterized by two conventions; one of which is the Wolfenstein parameterization which manifests the hierarchical structure of the CKM mixing matrix in terms of the parameter $\lambda$, and the other is the Euler rotation which reveals the Cabibbo mixing angle well.
\begin{equation}
\begin{split}
V_{\func{CKM}} & \equiv V_{L}^{u \dagger} V_{L}^{d} 
=
\begin{pmatrix}
V_{ud} & V_{us} & V_{ub} \\
V_{cd} & V_{cs} & V_{cb} \\
V_{tb} & V_{ts} & V_{tb} 
\end{pmatrix}
	\\
&= 
\begin{pmatrix}
1 & 0 & 0 \\
0 & c_{23} & -s_{23} \\
0 & s_{23} & c_{23}
\end{pmatrix}
\begin{pmatrix}
c_{13} & 0 & s_{13} e^{-i \delta_{\func{CP}}} \\
0 & 1 & 0 \\
-s_{13} e^{i \delta_{\func{CP}}} & 0 & c_{13}
\end{pmatrix}
\begin{pmatrix}
c_{12} & s_{12} & 0 \\
-s_{12} & c_{12} & 0 \\
0 & 0 & 1
\end{pmatrix}
	\\
&=
\begin{pmatrix}
c_{12} c_{13} & s_{12} c_{13} & s_{13} e^{-i \delta_{\func{CP}}} \\
-s_{12} c_{23} - c_{12} s_{23} s_{13} e^{i \delta_{\func{CP}}} & c_{12} c_{23} - s_{12} s_{23} s_{13} e^{i \delta_{\func{CP}}} & s_{23} c_{13} \\
s_{12} s_{23} - c_{12} c_{23} s_{13} e^{i \delta_{\func{CP}}} & -c_{12} s_{23} - s_{12} c_{23} s_{13} e^{i \delta_{\func{CP}}} & c_{23} c_{13}
\end{pmatrix}
	\\
&=
\begin{pmatrix}
1 - \lambda^2/2 & \lambda & A \lambda^3 \left( \rho - i \eta \right) \\
-\lambda & 1 - \lambda^2/2 & A \lambda^2 \\
A \lambda^3 \left( 1 - \rho - i \eta \right) & -A \lambda^2 & 1
\end{pmatrix}
\end{split}
\end{equation}
where $c_{ij} = \cos \theta_{ij}$, $s_{ij} = \sin \theta_{ij}$, $\delta_{\func{CP}}$ is the CP-violating phase angle. It is possible to connect the mixing angles used in the Euler rotation and the parameters $A, \lambda, \rho, \eta$ used in the Wolfenstein parameterization using the experimentally known hierarchical structure among the mixing angles $1 \gg s_{12} \gg s_{23} \gg s_{13}$ as follows~\cite{Wolfenstein:1983yz,Buras:1994ec,Charles:2004jd}:
\begin{equation}
\begin{split}
s_{12} &\simeq \lambda \simeq \frac{\lvert V_{us} \rvert}{\lvert V_{ud} \rvert^2 + \lvert V_{us} \rvert^2}, 
	\\
s_{23} &\simeq A \lambda^2  = \lambda \lvert \frac{V_{cb}}{V_{us}} \rvert, 
	\\
s_{13} e^{i\delta_{\func{CP}}} &= V_{ub}^* = A \lambda^3 \left( \rho + i \eta \right) 
	\\
	&= \frac{A \lambda^3 \left( \bar{\rho} + i \bar{\eta} \right)}{1-A^2\lambda^4\left( \bar{\rho} + i \bar{\eta} \right)} \sqrt{\frac{1-A^2\lambda^4}{1-\lambda^2}}
\end{split}
\label{eqn:relation_s12s23s13_rhobaretabar}
\end{equation}
where the new parameters $\bar{\rho}, \bar{\eta}$ are introduced to keep unitarity of the CKM mixing matrix in terms of $A, \lambda, \bar{\rho}, \bar{\eta}$ to all orders in $\lambda$ and they are defined as follows:
\begin{equation}
\begin{split}
\bar{\rho} &= \rho \left( 1 - \lambda^2/2 + \cdots \right), \\
\bar{\eta} &= \eta \left( 1 - \lambda^2/2 + \cdots \right).
\end{split}
\end{equation}
Using the parameters $\bar{\rho}$ and $\bar{\eta}$, the Wofenstein parameterization can be rewritten by:
\begin{equation}
V_{\func{CKM}} = 
\begin{pmatrix}
1 - \lambda^2/2 & \lambda & A \lambda^3 \left( \bar{\rho} - i \bar{\eta} \right) \\
-\lambda & 1 - \lambda^2/2 & A \lambda^2 \\
A \lambda^3 \left( 1 - \bar{\rho} - i \bar{\eta} \right) & -A \lambda^2 & 1
\end{pmatrix},
\end{equation}
and then it can easily be confirmed that any unitary mixing matrix follows the well-known constraint.
\begin{equation}
\sum_{i=1,2,3} V_{ij} V_{ik}^* = \delta_{jk}, \quad \text{for $j,k=1,2,3$}
\end{equation}
From the constraint, we are able to write one unitary condition such as:
\begin{equation}
\begin{split}
V_{ud} V_{ub}^* + V_{cd} V_{cb}^* + V_{td} V_{tb}^* &= 0,
	\\
\frac{V_{ud} V_{ub}^*}{V_{cd} V_{cb}^*} + 1 + \frac{V_{td} V_{tb}^*}{V_{cd} V_{cb}^*} &= 0,
\end{split}
\end{equation}
and then to sketch the famous triangle in the complex plane in terms of the parameters $\bar{\rho}, \bar{\eta}$ using the relation defined in Equation~\ref{eqn:relation_s12s23s13_rhobaretabar}
\begin{figure}[H]
\centering
\begin{subfigure}{0.50\textwidth}
\includegraphics[keepaspectratio,width=\textwidth]{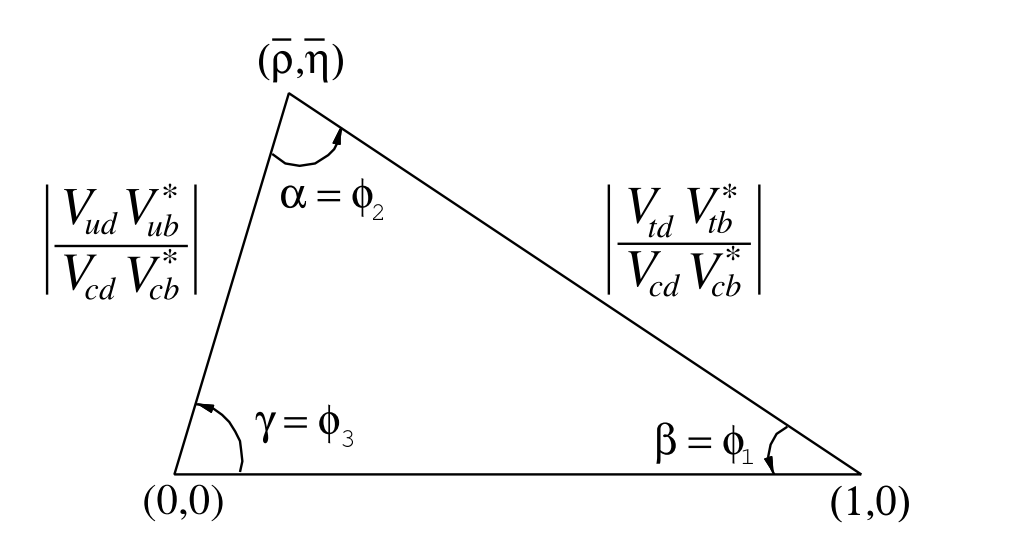}
\caption{The unitary triangle in terms of the parameters $\bar{\rho}$ and $\bar{\eta}$}
\end{subfigure}
\begin{subfigure}{0.80\textwidth}
\includegraphics[keepaspectratio,width=\textwidth]{CKM0.PNG}
\caption{The unitary triangle experimentally observed}
\end{subfigure}
\end{figure}
where the angles $\alpha, \beta$ and $\gamma$ are defined as follows:
\begin{equation}
\begin{split}
\beta = \phi_1 = \func{arg} \left( -\frac{V_{cd} V_{cb}^*}{V_{td} V_{tb}^*} \right), \\
\alpha = \phi_2 = \func{arg} \left( -\frac{V_{td} V_{tb}^*}{V_{ud} V_{ub}^*} \right), \\
\gamma = \phi_3 = \func{arg} \left( -\frac{V_{ud} V_{ub}^*}{V_{cd} V_{cb}^*} \right).
\end{split}
\end{equation}
The most fitted CKM mixing matrix with the unitarity of the SM is given in Particle Data Group (PDG) as well as the Wolfenstein parameters $(A=0.832 \pm 0.009, \ \lambda = 0.22465 \pm 0.00039, \ \bar{\rho} = 0.139 \pm 0.016, \ \bar{\eta} = 0.346 \pm 0.010)$:
\begin{equation}
V_{\func{CKM}} = 
\begin{pmatrix}
0.97446 \pm 0.00010 & 0.22452 \pm 0.00044 & 0.00365 \pm 0.00012 \\
0.22438 \pm 0.00044 & 0.97359_{-0.00011}^{+0.00010} & 0.04214 \pm 0.00076 \\
0.00896_{-0.00023}^{+0.00024} & 0.04133 \pm 0.00074 & 0.999105 \pm 0.000032
\end{pmatrix}.
\end{equation}
There is another interesting case, where the unitarity of the CKM mixing matrix is relaxed, as shown in Equation~\ref{eqn:CKM_wx_uni}~\cite{Branco:2021vhs,ParticleDataGroup:2020ssz}
\begingroup
\setlength\arraycolsep{5pt}
\begin{equation}
\lvert K_{\func{CKM}} \rvert
=
\begin{pmatrix}
0.97370 \pm 0.00014 & 0.22450 \pm 0.00080 & 0.00382 \pm 0.00024 \\
0.22100 \pm 0.00400 & 0.98700 \pm 0.01100 & 0.04100 \pm 0.00140 \\
0.00800 \pm 0.00030 & 0.03880 \pm 0.00110 & 1.01300 \pm 0.03000 
\end{pmatrix}.
\label{eqn:CKM_wx_uni}
\end{equation}
\endgroup
A possible deviation can arise from the first row of the CKM mixing matrix without unitarity and the deviation can be expressed by~\cite{Branco:2021vhs}
\begin{equation}
\Delta = 1 - \lvert V_{ud} \rvert^2 - \lvert V_{us} \rvert^2 - \lvert V_{ub} \rvert^2,
\end{equation}
and its experimental value is known as~\cite{Branco:2021vhs}
\begin{equation}
\sqrt{\Delta} \sim 0.04.
\label{eqn:dev_exp}
\end{equation}
This interesting case is discussed in my third work.
\section{First limitation of the SM - neutrinos} \label{sec:limitation_neutrinos}
The SM neutrinos $\nu_{e,\mu,\tau}$ (or equally $\nu_{1,2,3}$ in the interaction basis, respectively) are the most intuitive and instructive observables to hint at physics beyond the SM. It had been believed that the SM neutrinos are massless for a long time, since no any right-handed neutrinos had been observed, before the Super-Kamiokande experiment revealed the tiny mass differences between the SM neutrinos confirmed by the neutrino oscillation experiment. What the experiment revealed is there are tiny mass differences among the three neutrinos and the mass differences at $1\sigma$ can be confirmed by the NuFIT 5.0~\cite{Esteban:2020cvm}.
\begin{equation}
\begin{split}
\Delta m_{21}^2 &= m_2^2-m_1^2 = \left( 7.42_{-0.20}^{+0.21} \right) \times 10^{-5} \func{eV}^2, \\
\Delta m_{31}^2 &= m_3^2-m_2^2 = \left( 2.517_{0.028}^{+0.026} \right) \times 10^{-3} \func{eV}^2 \quad \text{for normal ordering}, \\
\Delta m_{32}^2 &= m_3^2-m_2^2 = -2.498_{-0.028}^{+0.028} \times 10^{-3} \func{eV}^2\quad \text{for inverted ordering},
\label{eqn:neutrino_mass_splitting}
\end{split}
\end{equation}
The neutrino mass splitting reported in Equation \ref{eqn:neutrino_mass_splitting} tells that the neutrino sector of the SM must be enlarged to cover the neutrino's tiny mass, which makes predict the right-handed neutrinos which has not yet been observed.
\subsection{The mass mechanism for neutrinos} \label{sec:mass_mechanism_neutrinos}
Now that the particle physicists know that the SM neutrinos are massive, they have considered the mass insertion mechanism for the neutrinos. In order to consider the mass insertion mechanism for the neutrinos, the first thing to do is to confirm that whether the neutrinos follow either the Dirac or Majorana nature. As mentioned in the earlier context, the SM neutrinos are the only candidates which could be the Majorana particles. Therefore, both the Dirac and Majorana mass insertion scenarios for the SM neutrinos must be considered until future experiments reveal its nature further. First of all, consider a case where the SM neutrinos are the Dirac particles. If the SM neutrinos have the Dirac mass, it necessarily requires to assume the right-handed neutrinos $\nu_{iR}$ ($i=1,2,3$). Then, the Lagrangian for the Dirac SM neutrinos can be written as follows ($i,j=1,2,3$):
\begin{equation}
\begin{split}
\mathcal{L}_{\nu} &= \left( y_\nu \right)_{ij} \overline{L}_{iL} \widetilde{H} \nu_{jR} + \func{h.c.}, \\
&= \left( y_\nu \right)_{ij}
\begin{pmatrix}
\overline{\nu}_{iL} \overline{l}_{iL}
\end{pmatrix}
\begin{pmatrix}
h^0 \\ -h^-
\end{pmatrix}
\nu_{jR} + \func{h.c.}, \quad \text{after ssb} \\
&= \left( y_\nu \right)_{ij} \left\langle h^0 \right\rangle \overline{\nu}_{iL} \nu_{jR} + \cdots + \func{h.c.}, \\
&= \left( m_\nu \right)_{ij} \overline{\nu}_{iL} \nu_{jR} + \cdots + \func{h.c.},
\label{eqn:Dirac_neutrino_mass}
\end{split}
\end{equation}
The derived Dirac mass for the light SM neutrinos from Equation \ref{eqn:Dirac_neutrino_mass} looks correct and no problem takes place as long as the right-handed neutrinos $\nu_{jR}$ are assumed. However, there is one thing which should be considered. Taking that the vev $\langle h \rangle$ is around order of hundreds $\func{GeV}$ into account, the Yukawa constant for the neutrinos must be suppressed by order of nearly $10^{-12}$, which looks quite ``unnatural" when compared to that of the other SM Dirac fermions. The extraordinarily suppressed neutrino Yukawa constants do not cause any physical problems, however it is less convincible to acknowledge the Dirac mass for the SM neutrinos due to the strong hierarchical structure.

Next it is possible to come up with the Majorana mass for the SM neutrinos since they can be the Majorana particles. The Majorana particle has its own anti-particle and this property can be seen by the following definition.
\begin{equation}
\psi = \psi_L + \psi_R = \psi_L + C\overline{\psi_L}^T = \psi_L + \psi_L^C,
\label{eqn:Majorana_definition}
\end{equation}
where $C$ is the charge conjugation operator and is defined by $C = i \gamma^2 \gamma^0$. Operating the charge conjugation operator $C$ to the $\psi$, the Majorana particle is exactly same as its own anti-particle.
\begin{equation}
\psi^C = \left( \psi_L + \psi_L^C \right)^C = \left( \psi_L^C + \psi_L \right) = \psi
\end{equation}
From Equation \ref{eqn:Majorana_definition}, it is clear that the right-handed component $\psi_R$ can be interpreted as $C\overline{\psi_L}^T$ and the right-handed component written in terms of the left-handed field can be applied to the Dirac mass in order to write down the Majorana mass.
\begin{equation}
\mathcal{L} = m \overline{\nu}_L \nu_R \rightarrow \mathcal{L} = \frac{1}{2}m \overline{\nu}_L C \overline{\nu_L}^T 
\label{eqn:Majorana_mass_neutrinos}
\end{equation}
The Majorana mass of Equation \ref{eqn:Majorana_mass_neutrinos} implements that the mass term can be written only in terms of the left-handed fields therefore it does not take a risk of assuming the right-handed neutrinos. However, the Majorana mass term violates the gauge singlet condition as well as the lepton number conservation which has not been observed yet. Therefore, both approaches have its own one advantage and one disadvantage. 

In order to solve the problems of either the Dirac or Majorana masses, the Seesaw mechanism was suggested, which can explain the tiny mass of the light SM neutrinos in a dynamical way. Consider a Lagrangian including the heavy right-handed Majorana neutrino $N$ after ssb (we only consider a neutrino species for simplicity).
\begin{equation}
\mathcal{L}_{\nu,M} = y_{D} v \overline{\nu} N + M_N \overline{N} N + \func{h.c.}
\end{equation}
Then the interactions can be written in the matrix basis as follows:
\begin{equation}
\begin{pmatrix}
0 & M_D \\
M_D & M_N
\end{pmatrix},
\label{eqn:seesaw_conventional}
\end{equation}
where $M_D$ is the Dirac mass defined by $M_D = y_D v$ and $M_N$ is the Majorana mass which can be as heavy as possible if the right-handed neutrino is trivial gauge singlet under the SM gauge group. Diagonalizing the mass matrix of Equation \ref{eqn:seesaw_conventional}, it gives
\begin{equation}
\begin{pmatrix}
-\frac{M_D^2}{M_N} & 0 \\
0 & M_N
\end{pmatrix},
\label{eqn:conventioanl_seesaw}
\end{equation}
which tells that the light mass of the SM neutrinos can be accompanied by mass of the heavy Majorana neutrinos. Taking a closer look at the light SM neutrino mass given from the seesaw mechanism, we can also read off the effective operator for the SM neutrinos.
\begin{equation}
\mathcal{L}_{\nu}^{\func{effective}} = \frac{1}{M_N} \left( L \widetilde{H} L^T \widetilde{H}^T \right)
\label{eqn:Weinberg_operator}
\end{equation}
The Lagrangian of Equation \ref{eqn:Weinberg_operator} is known as the Weinberg's five dimensional effective operator (or equally Type $1$ seesaw mechanism) for the neutrinos and the effective operator predicts that the right-handed Majorana neutrinos are of order $10^{14}\func{GeV}$ as long as the Dirac masses are of order $100\func{GeV}$. There are three well-known seesaw models named type $1,2,3$ seesaw mechanism in Figure~\ref{fig:type123seesawmechanism},
\begingroup
\begin{figure}[H]
\centering
\begin{subfigure}{0.30\textwidth}
	\begin{tikzpicture}[
  arrowlabel/.style={
    /tikzfeynman/momentum/.cd, 
    arrow shorten=#1,arrow distance=2.0mm
  },
  arrowlabel/.default=0.4
]
		\begin{feynman}
		
		\vertex (a1) {\(L\)};
		\vertex [above right = 1.2cm and 1.2cm of a1] (a2) ;
		\vertex [above left = 1cm and 1cm of a2] (a3) {\(H\)} ;
		\vertex [right = 1cm of a2] (a4) ;
		\vertex [above right = 1cm and 1cm of a4] (a5) {\(H\)} ;
		\vertex [below right = 1cm and 1cm of a4] (a6) {\(L\)} ;
	
		\vertex [below right = 0.1cm and 0.2cm of a2] (c3) {\(N_{R}\)};
		
		\diagram* {
		(a1) -- [fermion] (a2) -- [scalar] (a3),
		(a2) -- [majorana] (a4),
		(a5) -- [scalar] (a4) -- [anti fermion] (a6),		
		};
		\end{feynman} 
	\end{tikzpicture} 
\end{subfigure}
\hspace{0.1cm}
\begin{subfigure}{0.30\textwidth}
	\begin{tikzpicture}[
  arrowlabel/.style={
    /tikzfeynman/momentum/.cd, 
    arrow shorten=#1,arrow distance=2.0mm
  },
  arrowlabel/.default=0.4
]
		\begin{feynman}
		
		\vertex (a1) {\(L\)};
		\vertex [above right = 0.7cm and 1.7cm of a1] (a2) ;
		\vertex [below right = 0.5cm and 1.5cm of a2] (a6) {\(L\)} ;			\vertex [above = 1cm of a2] (a4) ;	
		\vertex [above left = 0.5cm and 1.5cm of a4] (a3) {\(H\)} ;
		\vertex [above right = 0.5cm and 1.5cm of a4] (a5) {\(H\)} ;
	
		\vertex [above right = 0.2cm and 0.1cm of a2] (c3) {\(\Delta\)};
		
		\diagram* {
		(a1) -- [fermion] (a2) -- [anti fermion] (a6),
		(a2) -- [scalar] (a4),
		(a3) -- [scalar] (a4) -- [scalar] (a5),		
		};
		\end{feynman} 
	\end{tikzpicture} 
\end{subfigure}
\hspace{0.1cm}
\begin{subfigure}{0.30\textwidth}
	\begin{tikzpicture}[
  arrowlabel/.style={
    /tikzfeynman/momentum/.cd, 
    arrow shorten=#1,arrow distance=2.0mm
  },
  arrowlabel/.default=0.4
]
		\begin{feynman}
		
		\vertex (a1) {\(L\)};
		\vertex [above right = 1.2cm and 1.2cm of a1] (a2) ;
		\vertex [above left = 1cm and 1cm of a2] (a3) {\(H\)} ;
		\vertex [right = 1cm of a2] (a4) ;
		\vertex [above right = 1cm and 1cm of a4] (a5) {\(H\)} ;
		\vertex [below right = 1cm and 1cm of a4] (a6) {\(L\)} ;
	
		\vertex [below right = 0.1cm and 0.2cm of a2] (c3) {\(\Sigma_{R}\)};
		
		\diagram* {
		(a1) -- [fermion] (a2) -- [scalar] (a3),
		(a2) -- [majorana] (a4),
		(a5) -- [scalar] (a4) -- [anti fermion] (a6),		
		};
		\end{feynman} 
	\end{tikzpicture} 
\end{subfigure}
\caption{Type $1,2,3$ seesaw mechanism from left to right, respectively}
\label{fig:type123seesawmechanism}
\end{figure}
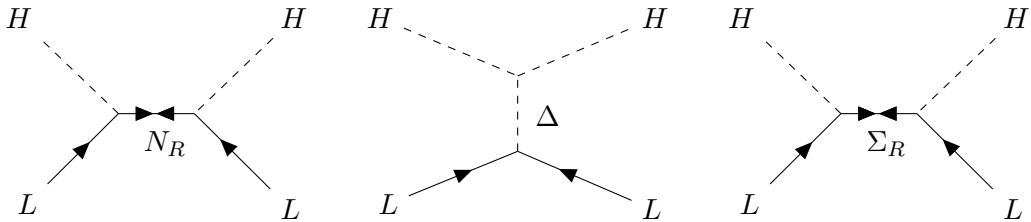
\endgroup
where the type $1,2,3$ seesaw mechanism exchanges right-handed singlet neutrinos $N_{R}$, scalar triplets $\Delta$ and fermion triplets $\Sigma_{R}$, respectively. On top of the well-known type $1,2,3$ seesaw mechanisms, there are diverse variations from the standard seesaw mechanisms such as type $1$b seesaw mechanism and inverse seesaw mechanism for the purpose of lowering mass order of the right-handed neutrinos and of taking moderate Yukawa couplings.
We will make use of the Weinberg-like operator known as type $1$b seesaw mechanism~\cite{Hernandez-Garcia:2019uof} from our second work (we call the Weinberg operator ``type 1a seesaw mechanism" for comparison). Our attempt to explain mass and mixing of the SM neutrinos in our second work~\cite{Hernandez:2021tii} are based on the following assumptions.
\begin{enumerate}
\item We assume the SM neutrinos are Majorana particles.
\item The SM neutrinos are extended by the vector-like neutrinos with the type $1$b seesaw mechanism.
\end{enumerate}

\subsection{mass ordering for the neutrinos and lepton flavor mixing} \label{sec:mass_ordering_lfm}
\begin{figure}[H]
\centering
\includegraphics[keepaspectratio,width=0.9\textwidth]{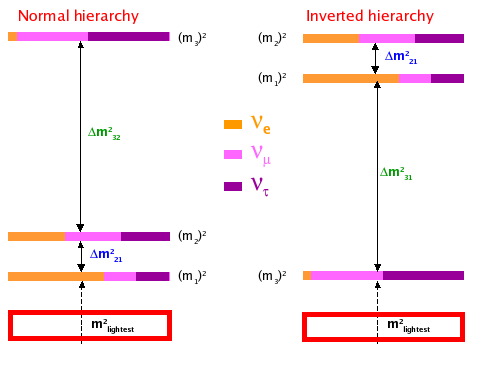}
\caption{The left is the normal hierarchy where $m_1^2$ is the lightest and the right is the inverted hierarchy where $m_3^2$ is the lightest.}
\label{fig:NH_and_IH}
\end{figure}

The SM neutrinos are strictly massless in the SM since there are no right-handed neutrinos. What this feature implies is there is no mixing among the three SM neutrinos, which is exactly opposite to the observed PMNS mixing matrix, taking into account that the sizeable off-diagonal mixings of the PMNS mixing matrix come from the neutrino sector and the sizeable neutrino mixings have been confirmed by the neutrino oscillation experiment for the first time. Thus, the PMNS mixing matrix itself is a great hint at new physics and it explicitly tells why the SM neutrino sector requires physics beyond the SM. Since the neutrino oscillation experiments are only sensitive to the SM neutrinos' mass differences and their mixings, the absolute mass scale of the SM neutrinos have been known yet, and this feature allows two possible mass orderings, which are known as the normal and inverted hierarchy and they are given in Figure \ref{fig:NH_and_IH}. The neutrino oscillation experiments have revealed that the SM neutrinos $\nu_{e,\mu,\tau}$ change their flavors while propagating some distance, and this characteristic can be also identified in the Figure \ref{fig:NH_and_IH} in that the SM neutrinos $\nu_{e,\mu,\tau}$ in the mass basis consist of the flavor eigenstates of the neutrinos $\nu_{1,2,3}$.
\begin{equation}
\begin{pmatrix}
\nu_e \\ \nu_\mu \\ \nu_\tau
\end{pmatrix}
=
\begin{pmatrix}
U_{e1} & U_{e2} & U_{e3} \\
U_{\mu1} & U_{\mu2} & U_{\mu3} \\
U_{\tau1} & U_{\tau2} & U_{\tau3}
\end{pmatrix}
\begin{pmatrix}
\nu_1 \\ \nu_2 \\ \nu_3
\end{pmatrix}
=
U_{\func{PMNS}}
\begin{pmatrix}
\nu_1 \\ \nu_2 \\ \nu_3
\end{pmatrix}
\label{eqn:neutrino_mixing_matrix}
\end{equation}
The PMNS unitary mixing matrix of Equation \ref{eqn:neutrino_mixing_matrix} is especially important since it provides some clues on why the quark mixing matrix (or equally the CKM matrix) is quite different to the lepton mixing matrix (or equally the PMNS matrix).
\begin{figure}[H]
\centering
\includegraphics[keepaspectratio,width=0.9\textwidth]{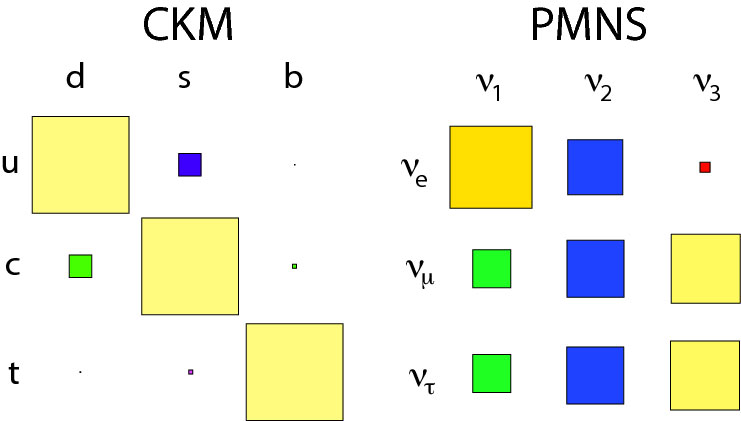}
\caption{The left is the quark mixing matrix and the right is the lepton mixing matrix.}
\label{fig:CKM_PMNS}
\end{figure}
Before going into the details of the CKM and PMNS mixing matrix, it is necessary to understand how the mixing takes place at the interaction basis, thus consider an weak current as an example.
\begin{equation}
J_W^{\mu+} = \frac{1}{\sqrt{2}} \overline{u}_L^i \gamma^\mu d_L^i,
\label{eqn:weak_current}
\end{equation}
where each basis is defined as follows: $u_L^i = \left( u_{L}^{1}, u_{L}^{2}, u_{L}^{3} \right)$, $d_L^i = \left( d_{L}^{1}, d_{L}^{2}, d_{L}^{3} \right)$. The defined bases $u_L^i$ and $d_L^i$ in the interaction basis can be transformed to the physical basis $u_L^{i\prime} = \left( u_{L}, c_{L}, t_{L} \right)$ and $d_L^{i\prime} = \left( d_{L}, s_{L}, b_{L} \right)$ via the unitary mixing matrices $U_{u,d}$.
\begin{equation}
u_L^{i} = U_{u}^{ij} u_L^{j\prime}, \quad d_L^{i} = U_{d}^{ij} d_L^{j\prime}
\label{eqn:physical_bases_ud}
\end{equation}
Substituting the Equation \ref{eqn:physical_bases_ud} back into the Equation \ref{eqn:weak_current}, the weak current generates an additional mixing.
\begin{equation}
J_W^{\mu+} = \frac{1}{\sqrt{2}} \overline{u}_L^{j\prime} (U_u^{ij})^\dagger \gamma^\mu U_d^{ik} d_L^{k\prime} = \frac{1}{\sqrt{2}} \overline{u}_L^{j\prime
} \gamma^\mu ( U_u^\dagger U_d )_{jk} d_L^{k\prime} = \frac{1}{\sqrt{2}} \overline{u}_L^{j\prime
} \gamma^\mu ( U_{\func{CKM}} )_{jk} d_L^{k\prime}
\label{eqn:weak_CKM_current}
\end{equation}
The Equation \ref{eqn:weak_CKM_current} implements that the quark (lepton) mixing matrix arises as a result of mixing between up-quark sector (neutrinos) and down-type quark sector (charged leptons). Focussing on the magnitude of each component of the CKM and PMNS mixing matrix, their behaviours are quite different; the diagonal components are more dominant that the off-diagonal components in the CKM matrix, whereas most of the off-diagonal components are compatible to the diagonal components in the PMNS matrix. It is possible to parameterize the PMNS mixing matrix in terms of the lepton mixing angles $\theta_{12},\theta_{13}$ and $\theta_{23}$ and it is given by Equation \ref{eqn:parameterization_PMNS}
\begin{equation}
\begin{split}
U_{\func{PMNS}} &= 
\begin{pmatrix}
U_{e1} & U_{e2} & U_{e3} \\
U_{\mu1} & U_{\mu2} & U_{\mu3} \\
U_{\tau1} & U_{\tau2} & U_{\tau3}
\end{pmatrix} \\
&= 
\begin{pmatrix}
1 & 0 & 0 \\
0 & c_{23} & s_{23} \\
0 & -s_{23} & c_{23} \\
\end{pmatrix}
\begin{pmatrix}
c_{13} & 0 & s_{13} e^{-i\delta_{\func{CP}}} \\
0 & 1 & 0 \\
-s_{13} e^{i\delta_{\func{CP}}} & 0 & c_{13}
\end{pmatrix}
\begin{pmatrix}
c_{12} & s_{12} & 0 \\
-s_{12} & c_{12} & 0 \\
0 & 0 & 1
\end{pmatrix} \\
&= \begin{pmatrix}
c_{12} c_{13} & s_{12} c_{13} & s_{13} e^{-i\delta_{\func{CP}}} \\
-s_{12} c_{23} - c_{12} s_{23} s_{13} e^{i\delta_{\func{CP}}} & c_{12} c_{23} - s_{12} s_{23} s_{13} e^{i\delta_{\func{CP}}} & s_{23} c_{13} \\
s_{12} s_{23} - c_{12} c_{23} s_{13} e^{i\delta_{\func{CP}}} & -c_{12} s_{23} - s_{12} c_{23} s_{13} e^{i\delta_{\func{CP}}} & c_{23} c_{13}
\end{pmatrix},
\end{split}
\label{eqn:parameterization_PMNS}
\end{equation}
where $c,s$ are the shortened notations for $\cos\theta,\sin\theta$, respectively; two Majorana phases $\alpha,\alpha^\prime$ are required for completeness in the PMNS mixing matrix, however I consider them just $0$ for simplicity. The experimentally fitted mixing angles from NuFIT 5.0~\cite{Esteban:2020cvm} are given as follows:
\begin{equation}
\sin^2\theta_{12} = 0.304_{-0.012}^{+0.012}, \quad \sin^2\theta_{23} = 0.573_{-0.020}^{+0.016}, \quad \sin^2\theta_{13} = 0.02219_{-0.00063}^{+0.00062}, \quad \delta_{\func{CP}} = 195{^\circ} _{-24^\circ}^{+27^\circ}.
\label{eqn:neutrino_mixing_angles}
\end{equation}
The current mixing angles $\theta_{13,23}$ exclude the possibility of the ``tri-bimaximal" or ``bimaximal" mixing, which are assumed when $\theta_{13} = 0$ and $\theta_{23} = 45^\circ$ some years ago. However, the similar second and third rows in the PMNS mixing matrix might be able to implement some hidden symmetry and many flavor models with some discrete or continuous symmetry have been considered. In this section, we simply reviewed what is the current position of the SM neutrinos and how it can connect from the SM to new physics. The neutrinos are definitely one of the clear hints at physics beyond the SM and are required to be searched carefully and passionately.
\section{Second limitation of the SM - the muon and electron anomalous magnetic moments $g-2$} \label{sec:muon_e_g2}
\begin{figure}[H]
\centering
\includegraphics[keepaspectratio,width=0.9\textwidth]{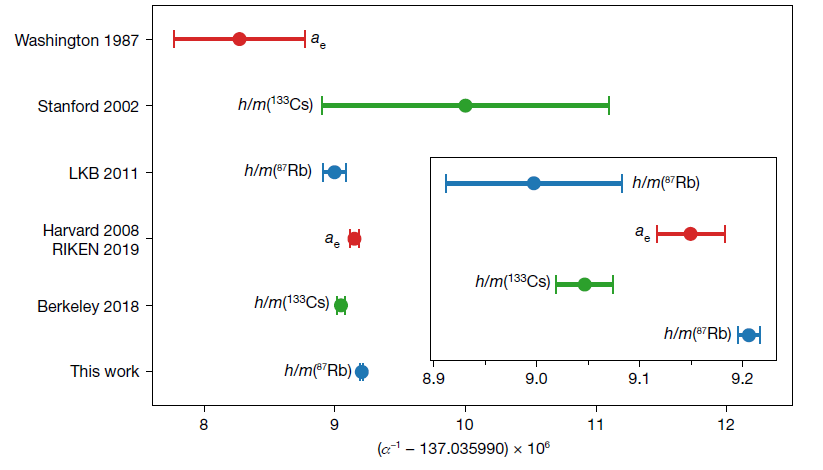}
\caption{Many experiments to improve the precision of the fine structure constant. The red points are contributed from $g_e - 2$ experiments and QED impacts. The green points and blue points are given from caesium and rubidium recoil experiments, respectively.}
\label{fig:correct_measurements_alpha}
\end{figure}

There are a few of well-known anomalies which can not be addressed by the SM. One of the famous anomalies is known as the long-established muon anomalous magnetic moment $a_\mu = \left( g-2 \right)_\mu$ and the other is the less-established electron anomalous magnetic moment $a_e = \left( g-2 \right)_e$. It is important to understand the origin of this anomalies before I go into the details of both anomalies. The magnetic dipole moment at the classical level can arise from the circulating current of the charged particle with the electric charge $e$ and mass $m$ in the natural units $(\hbar = 1, c = 1)$
\begin{equation}
\boldsymbol{\mu}_L = \frac{e}{2} \boldsymbol{r} \times \boldsymbol{v} = \frac{e}{2m} \boldsymbol{L},
\end{equation}
where the orbital angular momentum $\boldsymbol{L}$ is given by $m \boldsymbol{r} \times \boldsymbol{v}$. At the quantum level, the intrinsic quantity ``spin" $\boldsymbol{S}$ corresponds to the classical orbital angular momentum $\boldsymbol{L}$ and the magnetic moment can be rewritten in terms of the spin
\begin{equation}
\boldsymbol{\mu}_l = g_l \frac{e}{2m_l} \boldsymbol{S},
\end{equation}
where $l = e,\mu,\tau$ and the magnitude of the spin $\boldsymbol{S}$ is $1/2$. The $g_l$ was first suggested to be $2$ by Paul Dirac in 1928 and the experimentally observed values for the magnetic moment revealed that it is slightly shifted from the value $2$ and the difference is called the ``anomalous magnetic moment". The anomalous magnetic moment is defined in terms of the gyromagnetic ratio by
\begin{equation}
a_l = \mu_l/\mu_B - 1 = \frac{1}{2}\left( g_l - 2 \right),
\end{equation}
where $\mu_B$ is the Bohr magneton $\mu_B = \frac{e}{2m_e}$ in natural units. The anomalous magnetic moment $a_l$ has revealed that the higher order contributions take actually place and those higher contributions to the magnetic moment is the main reason that the $g_l$ is not exactly 2 but with some small deviation from the central value $2$. Julian Schwinger calculated the one-loop correction to the magnetic moment which consists of the SM particles like electron and photon and he found the first loop effect yields $\alpha/2\pi$, 
\begin{figure}[H]
\centering
\begin{subfigure}{0.48\textwidth}
	\scalebox{0.8}{
	\begin{tikzpicture}
		\begin{feynman}
		\vertex (a1) {\(e\)};
		\vertex [right = 2cm of a1] (a2);
		\vertex [right = 4cm of a2] (a3);	
		\vertex [right = 2cm of a3] (a4) {\(e\)};
		\vertex [above = 2cm of a2] (b2) ;
		\vertex [above right = 2cm and 2cm of a2] (b4);
		\vertex [above right = 2cm and 2cm of b4] (e1) {\(\gamma\)};
		\vertex [above = 2cm of a3] (b3) ;
		\diagram* {
		(a1) -- [fermion] (a2) -- [edge label'=\(\gamma\),boson] (a3) -- [fermion] (a4),
		(a2) -- [quarter left,fermion,edge label=\({e}\)] (b4) -- [quarter left,fermion,edge label=\({e}\)] (a3),
		(b4) -- [photon] (e1),
		};
		\end{feynman}
	\end{tikzpicture}}
\end{subfigure}$=\alpha/2\pi$
\caption{The one-loop correction to the electron magnetic moment calculated by Julian Schwinger and the value is $\alpha/2\pi$}.
\end{figure}
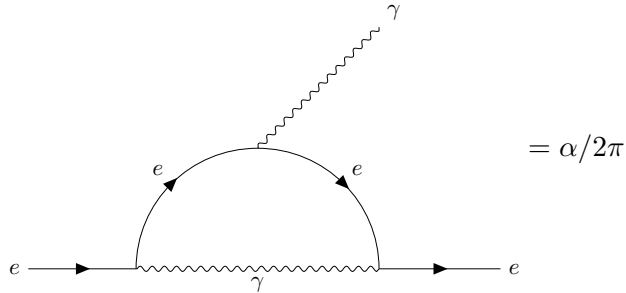
where the $\alpha$ is the fine structure constant. The higher SM loop corrections to the magnetic moment for the lepton $l$ were calculated up to five loops corrections and the BNL experiment reported the muon anomalous magnetic moment with the $1\sigma$ of error bar in 2018.
\begin{equation}
\Delta a_\mu = a_\mu^{\func{Exp}} - a_\mu^{\func{SM}} = \left( 26.1 \pm 8.0 \right) \times 10^{-10}
\label{eqn:muong2_1sigma}
\end{equation}
The muon anomalous magnetic moment of Equation \ref{eqn:muong2_1sigma} implements $3.2\sigma$ deviation from its central value and the deviation clearly exceeds the SM prediction. The muon anomaly deviation above slightly $3\sigma$ has been established for a long time and regarded as a signal for new physics. The electron anomalous magnetic moment $a_e$ is somewhat less interesting since it is not fully established as compared to the muon anomaly. The fundamental reason causing the difference in the electron anomaly arises from the correct measurement of the fine structure constant going on globally and this feature is given in Figure \ref{fig:correct_measurements_alpha} \cite{Morel:2020dww}.
This precision measurement of the fine structure constant also affects my second work; we made use of the fine structure constant of the Berkely 2018 experiment $\alpha_{\func{Berkeley}}$ at the beginning of the work, which gives rise to $\Delta a_e = a_e^{\func{Exp}} - a_e^{\func{SM}}[\alpha_{\func{Berkeley}}] = \left( -8.8 \pm 3.6 \right) \times 10^{-13}$ and $-2.4\sigma$, and then the new result of the fine structure constant of the LKB 2020 experiment $\alpha_{\func{LKB2020}}$ was released at the last of the work, which gives rise to $\Delta a_e = a_e^{\func{Exp}} - a_e^{\func{SM}}[\alpha_{\func{LKB2020}}] = \left( 4.8 \pm 3.0 \right) \times 10^{-13}$ and $+1.6\sigma$, and the new result to the electron anomaly looks like it is under the prediction of the SM. As nobody knew which experiment is exact at the moment, we stuck to the result of the Berkeley 2018 experiment. The correct measurement of the fine structure constant is quite important for all fields of the modern physics and it is especially important to the particle physics in that the electron anomaly we have regarded as a new physics signal is likely to be less interesting if the new result of the LKB 2020 experiment is correct. The muon anomaly is also likely to get affected by the change of the fine structure constant and the FNAL reports a long-awaited new result for the muon anomalous magnetic $g-2$ with $4.2\sigma$ of SM deviation~\cite{Muong-2:2021ojo}.
\\~\\
As an interesting case, there have been many attempts to explain the muon and electron anomalous magnetic moments in a unified way, since they share the structural similarity as a leading order contribution. If we consider the one-loop correction to both anomalies as the dominant next leading order (NLO) contribution, explaining the different sign of the muon and electron anomalies under the results of BNL 2018 and Berkeley 2018 experiment has been challenging and this was our good motivation governed in both my first and second work. The motivation will be covered in detail in my first and second works.
\section{Third limitation of the SM - the hierarchy} \label{sec:limitation_hierarchy}
\begin{figure}[H]
\centering
\begin{subfigure}{0.52\textwidth}
\includegraphics[keepaspectratio,width=\textwidth]{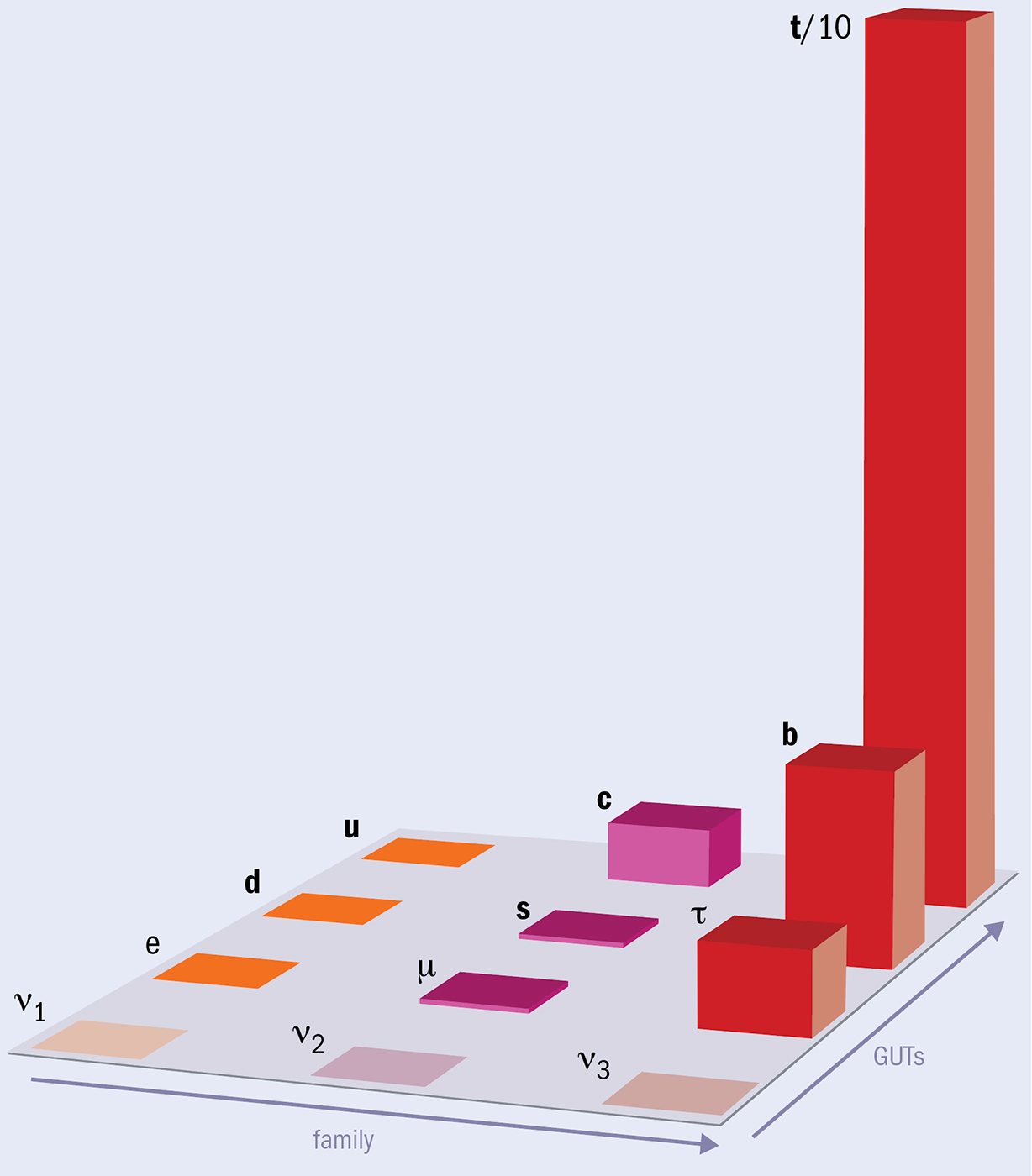} 
\end{subfigure} 
\begin{subfigure}{0.44\textwidth}
\includegraphics[keepaspectratio,width=\textwidth]{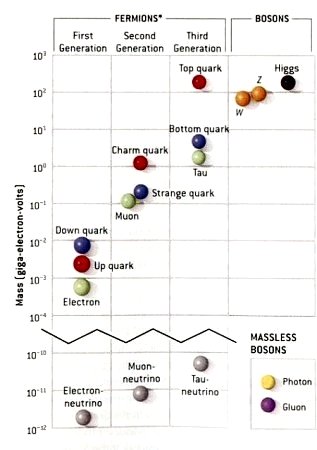} 
\end{subfigure}
\caption{The left is the relative magnitude of each particle's Yukawa constant and the right is the Yukawa constant is expressed in logarithmic scale.}
\label{fig:hierarchy_SM}
\end{figure}

The hierarchical structure of the SM might be able to look less convincible when compared to other limitations. The main reason for this is that all particles under the SM can be explained by the Yukawa interactions and spontaneous symmetry breaking with the relative Yukawa constants. The hierarchical structure of the SM can be seen in Figure \ref{fig:hierarchy_SM}.
However, this kind of view has not come to an agreement for the mass insertion mechanism of each particle, especially neutrino, by many particle physicists since it looks ``unnatural". This hierarchical problem of the SM can be further clarified by mentioning the order of the Yukawa constant for the SM neutrino, electron and lastly top quark with the Higgs vev $246/\sqrt{2}\simeq 174\func{GeV}$, respectively.
\begin{equation}
\mathcal{O}(y_\nu) \simeq 10^{-12}, \quad
\quad \mathcal{O}(y_e) \simeq 10^{-6}, \quad
\mathcal{O}(y_t) \simeq 1 
\end{equation}
The big gap between the order of SM neutrino and that of top quark has made many particle physicists consider whether there is another more convincible and reasonable mass insertion mechanism to cover the tiny mass of neutrinos more naturally and the consideration has taken the shape of the seesaw mechanism. Starting from the conventional seesaw mechanism covered in Equation \ref{eqn:conventioanl_seesaw}, there have been many variants for the seesaw mechanism like type 1b seesaw, inverse seesaw mechanism, etc. for the purpose of lowering the mass scale of Majorana neutrinos assumed to exist at high energy scale. Through these diverse seesaw mechanisms, it is possible to lower the mass scale of the sterile neutrinos up to $\func{TeV}$ scale which is an accessible energy scale in close future experiments.
\\~\\
Considering the hierarchical structure of the SM seriously leads to the seesaw mechanism to explain the very tiny mass of neutrinos in a more natural and dynamical way and the seesaw mechanism might be able to reveal the sterile neutrinos which resides in the $\func{TeV}$ scale. As long as the hierarchy of the SM might be able to reveal some new physics, I believe it should be considered as serious as other limitations.

\section{Fourth limitation of the SM - unification} \label{sec:limitation_unification}
\begin{figure}[H]
\centering
\includegraphics[keepaspectratio,width=0.8\textwidth]{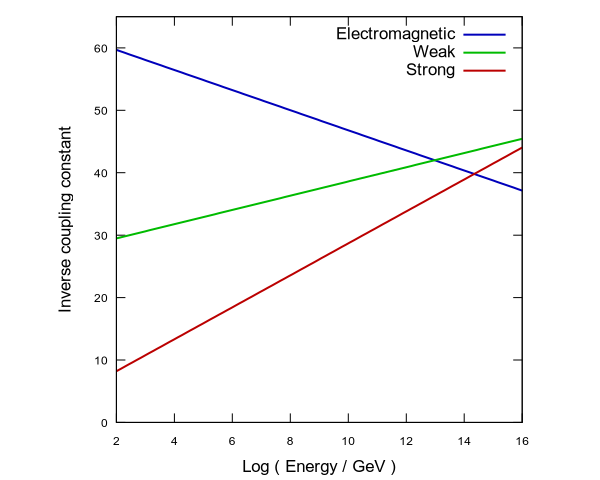}
\caption{Running coupling constants of the SM interactions}
\label{fig:running_coupling_constants}
\end{figure}

One of the greatest successes of the SM is the forces which had been regarded as the separate forces actually could be united as a more fundamental force like the electroweak interaction at the electroweak energy scale. A very beautiful aspect of this unified interaction can be clarified by the gauge symmetry $SU(2)_L \times U(1)_Y$ in a mathematical way, however it requires to look into the interaction in detail to determine whether the combined gauge interaction means really two separate forces are united. This investigation can be done through the connection of the coupling constants $g$ and $g^\prime$ which appear in the covariant derivative of the Higgs field of Equation \ref{eqn:covariant_derivative_h} as seen in the Equation \ref{eqn:eQmW}.
\begin{equation}
e = \frac{gg^\prime}{\sqrt{g^2+g^{\prime2}}} = g\sin\theta_W = g^\prime \cos\theta_W, \quad \frac{g^\prime}{g} = \tan\theta_W
\end{equation}
Suppose that there is a larger gauge symmetry $G$ involving the electroweak gauge symmetry $SU(2)_L \times U(1)_Y$ which consists of two independent interactions.
\begin{equation}
G \supset SU(2)_L \times U(1)_Y
\end{equation}
Then, all observables can be described by a new coupling constant $g_{\func{new}}$ under the larger gauge group $G$. Including the strong interaction $SU(3)$, I can define the larger gauge group at the scale of grand unified theory (GUT).
\begin{equation}
G_{\func{GUT}} \supset SU(3)_C \times SU(2)_L \times U(1)_L
\end{equation}
At the GUT scale, the observables can be described by the new coupling constant $g_{GUT}$ where each coupling constant $g_{EM}, g_{W}$ and $g_{S}$ meet simultaneously. Particularly, the Callan-Symanzik equation revealed that the coupling constants $g_{\func{EM}}, g_{W}$ and $g_{S}$ have the dependence on the momentum scale and this property is shown by the running coupling constants in Figure \ref{fig:running_coupling_constants}
In Figure \ref{fig:running_coupling_constants}, the weak and strong coupling constants tells an interesting feature known as asymptotic behaviour, in which the coupling constant $g_{W}$ and $g_{S}$ get stronger as the distance increases, whereas the characteristic is exactly opposite to the electromagnetic coupling constant $g_{\func{EM}}$. The reason that the weak and strong interactions behave differently arises from their self-interactions which cause negative values of the beta function.
\\~\\
As shown in the unification for the electromagnetic, weak and strong force, the particle physicists have dreamed of unifying the last known force ``gravity" with the SM forces. However, it has remained unsuccessful for a long time since the gravity can not be quantized. The diverse efforts to find the quantized gravity has evolved as the string theory which looks relatively successful rather than other theories, however this field still has some critical issues like it is difficult to observe or experiment the results derived from the string theory since the energy scale is too high to experiment with the current machine power. Despite all these difficulties, unifying all known forces has been very attractive since this unification itself is a great motivation for new physics.
\\~\\
My BSM models~\cite{CarcamoHernandez:2019ydc,Hernandez:2021tii,Hernandez:2021oyv} did not touch this unification in person since it requires to investigate how the SM gauge symmetry can arise as a result of more fundamental symmetries spontaneously broken and inevitably makes the investigation much more challenging due to lots of breaking patterns and assumed particles (as well as lots of assumptions). A research to investigate more fundamental symmetries is carried out within an extension of the Pati-Salam model~\cite{King:2021jeo} and this research can suggest one possible direction to constructing more fundamental symmetries despite many difficulties such as lots of assumptions. This unification is definitely an important subject, related to the origin of the universe, and this research will be carried out with more fundamental symmetries in my future works.

\section{Fifth limitation of the SM - Dark matter and Dark energy} \label{sec:limitation_DM_DE}
\begin{figure}[H]
\centering
\begin{subfigure}{0.549\textwidth}
\includegraphics[keepaspectratio,width=\textwidth]{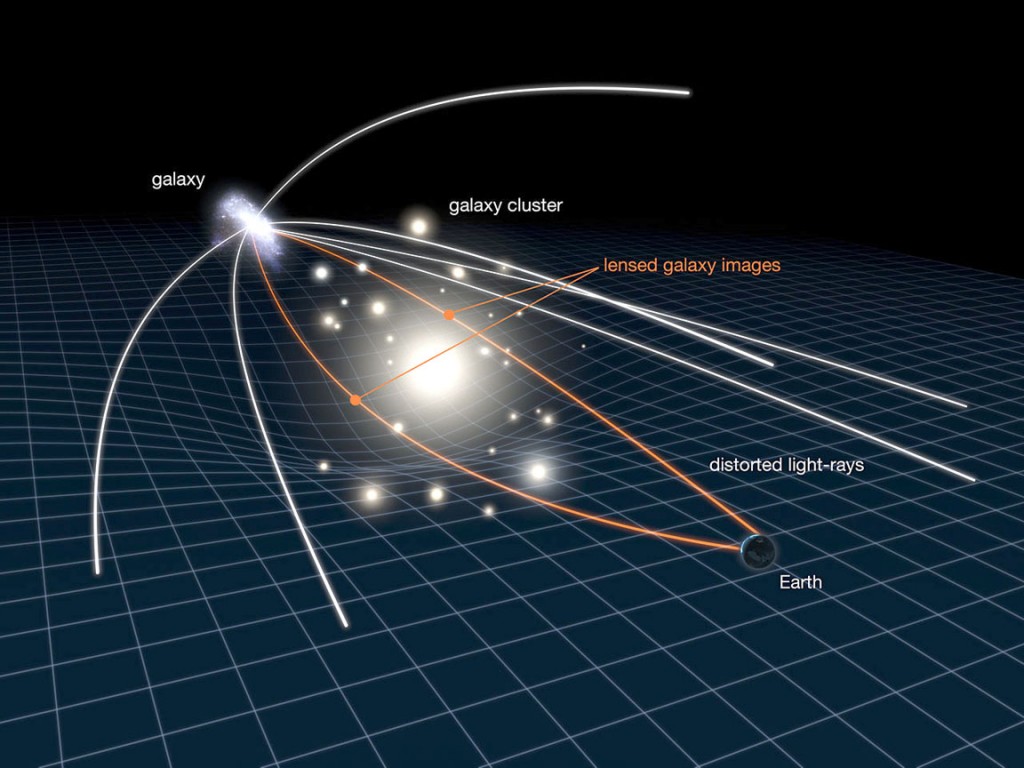} 
\end{subfigure} 
\begin{subfigure}{0.411\textwidth}
\includegraphics[keepaspectratio,width=\textwidth]{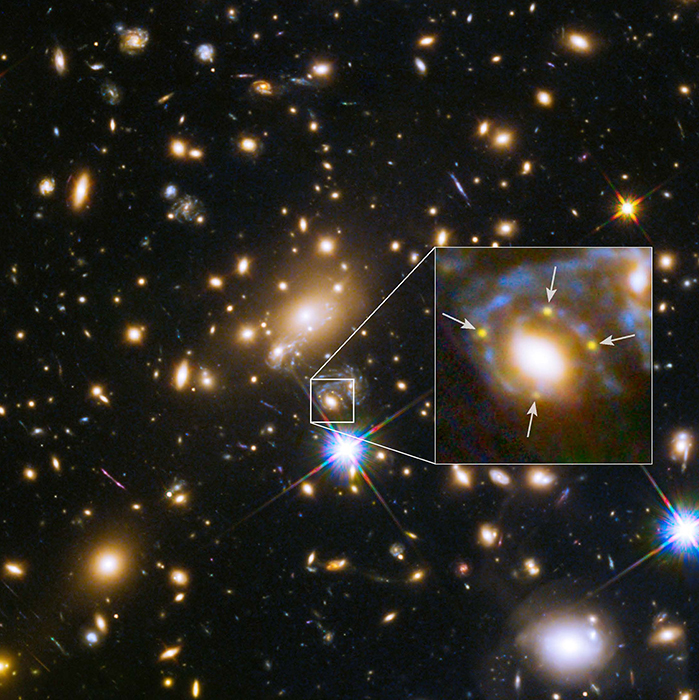} 
\end{subfigure}
\caption{The left is the principle of the gravity lensing effect and the right is the actual lensing effect observed. }
\label{fig:gravity_lensing}
\end{figure}

\begin{figure}[H]
\centering
\begin{subfigure}{0.465\textwidth}
\includegraphics[keepaspectratio,width=\textwidth]{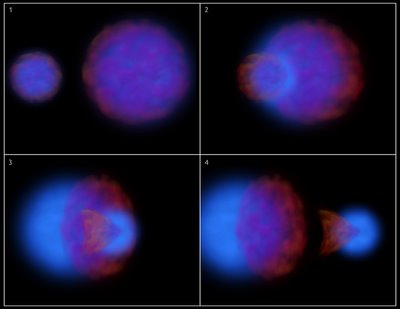} 
\end{subfigure} 
\begin{subfigure}{0.495\textwidth}
\includegraphics[keepaspectratio,width=\textwidth]{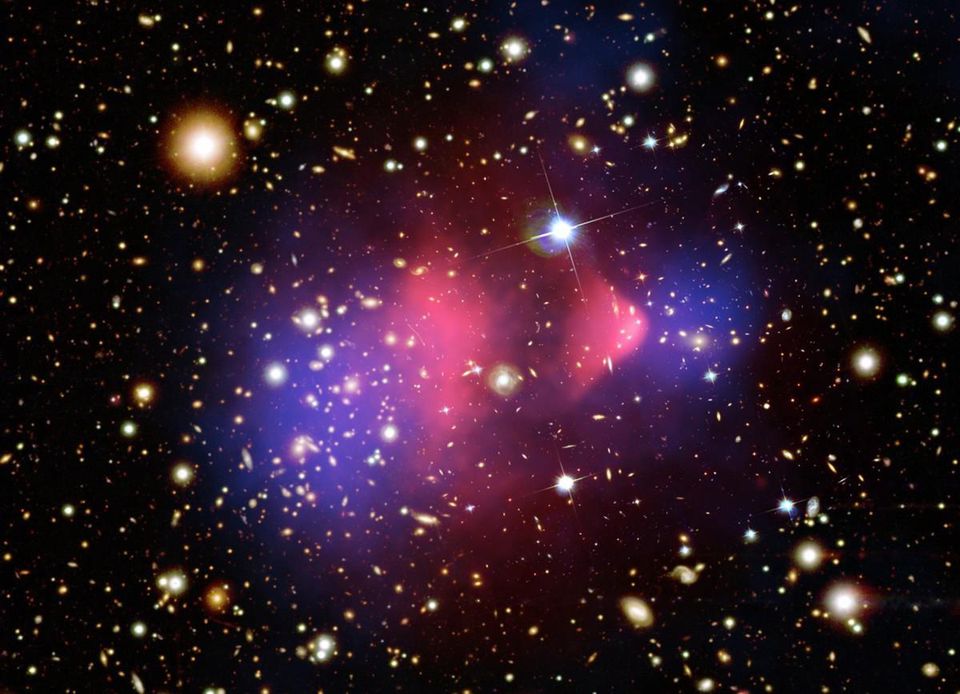} 
\end{subfigure}
\caption{The left is how the bullet clusters takes place and the right is the visualized image for the bullet clusters. }
\label{fig:bullet_clusters}
\end{figure}

\begin{figure}[H]
\centering
\includegraphics[keepaspectratio,width=0.8\textwidth]{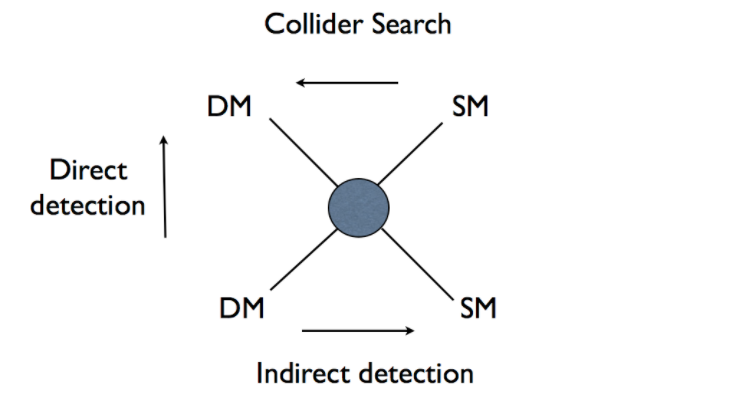}
\caption{Three DM search methods}
\label{fig:DM_search}
\end{figure}

\begin{figure}[H]
\centering
\includegraphics[keepaspectratio,width=0.8\textwidth]{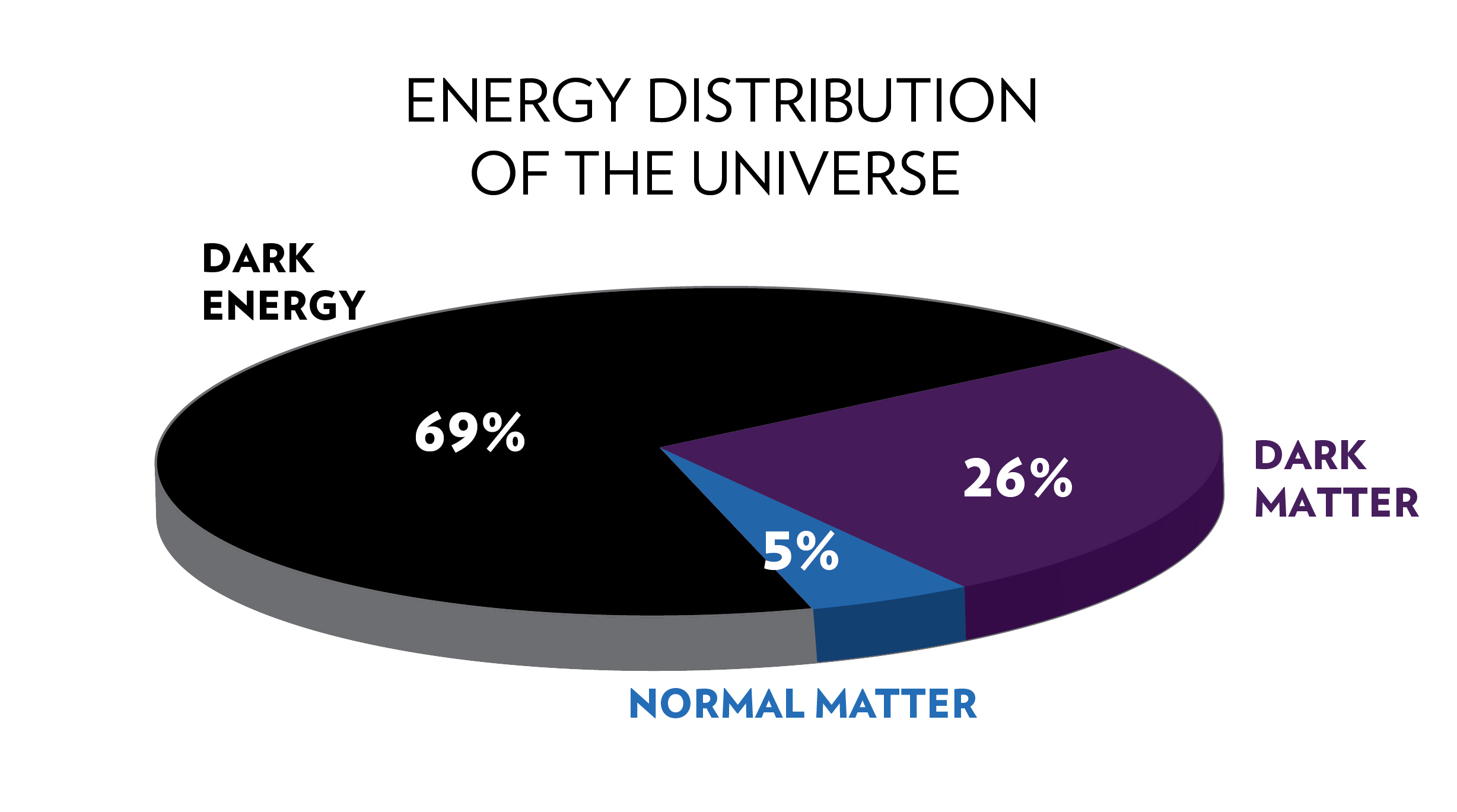}
\caption{The energy distribution in our universe}
\label{fig:energy_distribution}
\end{figure}

The all limitations of the SM treated so far are based on the known SM particles and forces. However, I talk the rarely known objects Dark Matter (DM) and Dark Energy (DE) in this section. The main motivation for the DM started from the rotation curve of the spiral galaxy.
\begin{figure}[H]
\centering
\includegraphics[keepaspectratio,width=0.8\textwidth]{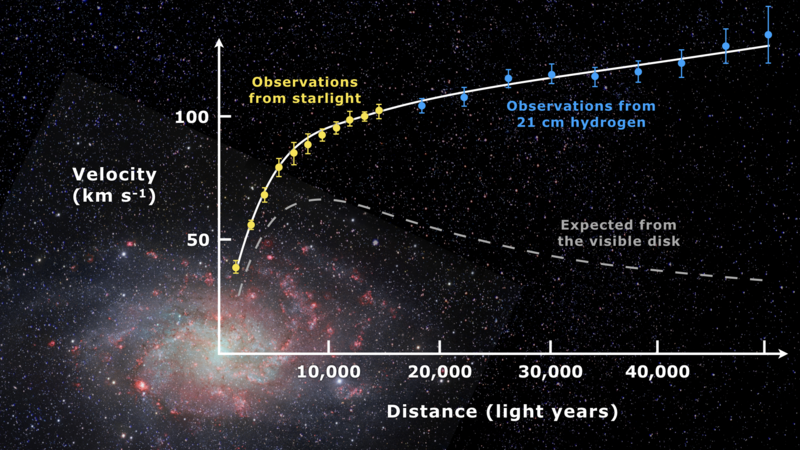}
\caption{Rotation curve of the spiral galaxy}
\label{fig:rotation_curve}
\end{figure}
As seen in Figure \ref{fig:rotation_curve}, many physicists expected the speed of the spiral galaxy would be slower as the distance between the galaxy and the center of the galaxy increases. However, the actual observation of the speed showed that it continues to increase as the distance increases. In order to explain the observed result of the rotation curve of the spiral galaxy, two theories known as the modified gravity and the Dark Matter have been discussed. As time goes on, many evidences supporting the DM started to emerge like the gravity lensing effect, the bullet clusters, etc. in Figure \ref{fig:gravity_lensing} and in Figure \ref{fig:bullet_clusters}.
The gravity lensing effect in Figure \ref{fig:gravity_lensing} explains how the unseen objects can be seen by distorting the path of the light and is actually observed in the right of Figure \ref{fig:gravity_lensing}.
When two clusters collide with each other, they just pass by one another as if no collisions take place and the right of Figure~\ref{fig:bullet_clusters} is the visualized image for the bullet clusters. Except for these observations such as the gravity lensing effect and the bullet clusters, there are many other observations supporting the DM theory, whereas the modified gravity theory has some limitations to explain the observed effects. Therefore, many particle physicists have leaned on the DM model and this field has been one of the most open and interesting phenomenology in the particle physics. Despite all these interesting and attractive properties, the actual difficulty for the research of the DM arises from that few things have been known so far; the particle physicists have believed its existence, however there is no any clue or hint about whether they are either boson or fermion, their spin, the candidate particle for the DM, etc. Assuming stability of the DM and it does not have the electromagnetic interaction with the charged particles, many sterile candidates which do not respect the SM gauge symmetry like the weakly interacting massive particle (WIMP) have been suggested, however these are remained as one of possibilities and it requires some physical observables on it to confirm whether the candidate is really consistent with the DM. Three search methods for the DM were suggested as seen in Figure \ref{fig:DM_search}.
Each method has its own advantage and disadvantage and I will not address them here since DM is not directly related to my researches I have done so far. In general, the DM is deeply related to the cosmology, which studies the origin and evolution of the universe. The cosmology has revealed the energy distribution in our universe in Figure \ref{fig:energy_distribution}.
Following the energy distribution of Figure \ref{fig:energy_distribution}, the universe consists of $5\%$ of the SM particles, $26\%$ of the DM and $69\%$ of the Dark Energy which is calculated in the Einstein's field equation to explain the current accelerating universe. It is worth mentioning that the vector-like fermions, frequently used to extend the SM fermion sector in my works~\cite{CarcamoHernandez:2019ydc,Hernandez:2021tii,Hernandez:2021oyv}, does not belong to the $5\%$ of matter particles. Based on the features, the presence of DM and DE has been believed they must exist. However, many particle physicists agree that their candidates are not in the SM; the neutrinos had been considered as the only possible candidate for the DM in the SM, however their too light mass had some issues with the stability of the DM so the possibility was excluded. The DM and DE which have been firmly established phenomenologically should be explained by some BSM models and these are one of the clear reasons why the SM should be extended.
\\~\\
As in the unification covered in the subsection~\ref{sec:limitation_unification}, I did not touch the DM in my works~\cite{CarcamoHernandez:2019ydc,Hernandez:2021tii,Hernandez:2021oyv} in person since we focused on more flavor observables such as a few of anomalies as well as the FCNCs. However, my BSM models have big room for the DM since the SM fermion and scalar sector all are extended by the vector-like fermions and one more SM-like Higgs (as well as one singlet flavon), respectively. The hypothetical vector-like fermions and non-SM scalars are generally assumed to be much heavier than the top quark, thus they are suitable to be candidates for the DM and there was an attempt~\cite{Fu:2021uoo} to explain DM by the vector-like neutrinos via the type $1$b seesaw mechanism, used in our second work~\cite{Hernandez:2021tii}. In my future researches, the DM will be studied within an extension of the SM. 
\chapter{The first BSM model - the fermiophobic $Z^\prime$ model} \label{Chapter:The1stBSMmodel}

We saw that the SM itself is not enough to cover every detail of our universe via the discussions of chapter~\ref{Chapter:TheSM}, therefore we need to come up with how to extend the SM and we start from this consideration: how can we extend the SM without violating the gauge symmetry and experimental bounds for the SM observables. A possible answer to the question is a minimal extension to the SM, which is a main principle over my three works. As emphasized in the introduction, at least one of the following sectors, the SM fermion, scalar and gauge symmetry, must be expanded for the minimal extension. Before we go to the prerequisites to cover each hypothetical tool required for our BSM models, it is good to look at main features in my three works. 
\begin{itemize}
\item First work (Phys. Rev. \textbf{D}, 115016)~\cite{CarcamoHernandez:2019ydc}
	\begin{itemize}
	\item Main target : the muon and electron $g-2$, the $\mu \rightarrow e \gamma$ decay and the neutrino trident production, the $Z^{\prime}$ gauge boson
	\item extension of the fermion sector : fourth vector-like family
	\item extension of the scalar sector : a singlet flavon
	\item extension of the SM gauge symmetry : $U(1)^{\prime}$ local symmetry (gauge symmetry)
	\end{itemize}
\item Second work (Phys. Rev. \textbf{D}, 115024)~\cite{Hernandez:2021tii}
	\begin{itemize}
	\item Main target : the muon and electron $g-2$, the $\mu \rightarrow e \gamma$ decay and the hierarchical structure of the SM, the SM $W$ gauge bosons, the non-SM scalars
	\item extension of the fermion sector : fourth and fifth vector-like families
	\item extension of the scalar sector : a more SM-like Higgs and the singlet flavon
	\item extension of the SM gauge symmetry : $U(1)^{\prime}$ global symmetry (family symmetry)
	\end{itemize}
\item Third work (In arXiv)~\cite{Hernandez:2021oyv}
	\begin{itemize}
	\item Main target : the diverse FCNC observables such as $\tau \rightarrow \mu \gamma, \tau \rightarrow 3\mu, Z \rightarrow \mu \tau$ and the rare $t \rightarrow c Z$ decay and the CKM mixing matrix and the hierarchical structure of the SM, the SM $Z$ gauge boson
	\item extension of the fermion sector : fourth vector-like family
	\item extension of the scalar sector : the SM-like Higgs and the singlet flavon
	\item extension of the SM gauge symmetry : $U(1)^{\prime}$ global symmetry (family symmetry)
	\end{itemize}
\end{itemize}
The common features appearing in all three works are the vector-like family, the SM-like Higgs plus the singlet flavon and lastly the $U(1)^{\prime}$ symmetry. Therefore, it is quite important to understand them and they are discussed in the following prerequisites.
\section{Prerequisite : vector-like (VL) fermions}
\label{sec:pre_VL}
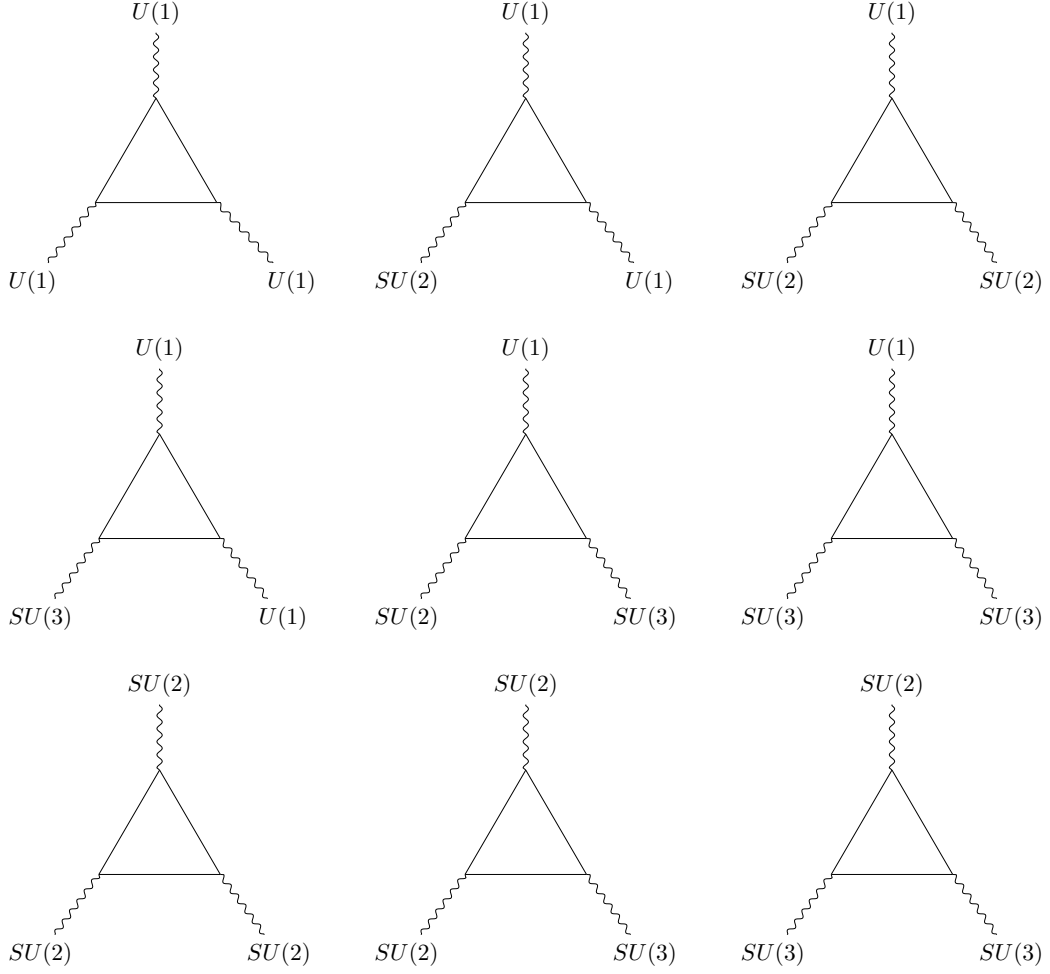
\begin{figure}[h]
\centering
\begin{subfigure}{0.32\textwidth}
\scalebox{0.8}{
	\begin{tikzpicture}
		\begin{feynman}
		\vertex (a1) {\(U(1)\)};
		\vertex [below = 1.414cm of a1] (a2);
		\vertex [below left = 1.732cm and 1cm of a2] (a3);	
		\vertex [below right = 1.732cm and 1cm of a2] (a4);
		\vertex [below left = 1cm and 0.5cm of a3] (b1) {\(U(1)\)};
		\vertex [below right = 1cm and 0.7cm of a4] (b2) {\(U(1)\)};
		\diagram* {
		(a1) -- [boson] (a2),
		(a2) -- [] (a3) -- [] (a4) -- [] (a2),
		(b1) -- [boson] (a3),
		(b2) -- [boson] (a4),
		};
		\end{feynman}
	\end{tikzpicture}}
\end{subfigure}
\begin{subfigure}{0.32\textwidth}
\scalebox{0.8}{
	\begin{tikzpicture}
		\begin{feynman}
		\vertex (a1) {\(U(1)\)};
		\vertex [below = 1.414cm of a1] (a2);
		\vertex [below left = 1.732cm and 1cm of a2] (a3);	
		\vertex [below right = 1.732cm and 1cm of a2] (a4);
		\vertex [below left = 1cm and 0.3cm of a3] (b1) {\(SU(2)\)};
		\vertex [below right = 1cm and 0.5cm of a4] (b2) {\(U(1)\)};
		\diagram* {
		(a1) -- [boson] (a2),
		(a2) -- [] (a3) -- [] (a4) -- [] (a2),
		(b1) -- [boson] (a3),
		(b2) -- [boson] (a4),
		};
		\end{feynman}
	\end{tikzpicture}}
\end{subfigure}
\begin{subfigure}{0.32\textwidth}
\scalebox{0.8}{
	\begin{tikzpicture}
		\begin{feynman}
		\vertex (a1) {\(U(1)\)};
		\vertex [below = 1.414cm of a1] (a2);
		\vertex [below left = 1.732cm and 1cm of a2] (a3);	
		\vertex [below right = 1.732cm and 1cm of a2] (a4);
		\vertex [below left = 1cm and 0.3cm of a3] (b1) {\(SU(2)\)};
		\vertex [below right = 1cm and 0.3cm of a4] (b2) {\(SU(2)\)};
		\diagram* {
		(a1) -- [boson] (a2),
		(a2) -- [] (a3) -- [] (a4) -- [] (a2),
		(b1) -- [boson] (a3),
		(b2) -- [boson] (a4),
		};
		\end{feynman}
	\end{tikzpicture}}
\end{subfigure}
\\[2ex]
\begin{subfigure}{0.32\textwidth}
\scalebox{0.8}{
	\begin{tikzpicture}
		\begin{feynman}
		\vertex (a1) {\(U(1)\)};
		\vertex [below = 1.414cm of a1] (a2);
		\vertex [below left = 1.732cm and 1cm of a2] (a3);	
		\vertex [below right = 1.732cm and 1cm of a2] (a4);
		\vertex [below left = 1cm and 0.3cm of a3] (b1) {\(SU(3)\)};
		\vertex [below right = 1cm and 0.5cm of a4] (b2) {\(U(1)\)};
		\diagram* {
		(a1) -- [boson] (a2),
		(a2) -- [] (a3) -- [] (a4) -- [] (a2),
		(b1) -- [boson] (a3),
		(b2) -- [boson] (a4),
		};
		\end{feynman}
	\end{tikzpicture}}
\end{subfigure}
\begin{subfigure}{0.32\textwidth}
\scalebox{0.8}{
	\begin{tikzpicture}
		\begin{feynman}
		\vertex (a1) {\(U(1)\)};
		\vertex [below = 1.414cm of a1] (a2);
		\vertex [below left = 1.732cm and 1cm of a2] (a3);	
		\vertex [below right = 1.732cm and 1cm of a2] (a4);
		\vertex [below left = 1cm and 0.3cm of a3] (b1) {\(SU(2)\)};
		\vertex [below right = 1cm and 0.3cm of a4] (b2) {\(SU(3)\)};
		\diagram* {
		(a1) -- [boson] (a2),
		(a2) -- [] (a3) -- [] (a4) -- [] (a2),
		(b1) -- [boson] (a3),
		(b2) -- [boson] (a4),
		};
		\end{feynman}
	\end{tikzpicture}}
\end{subfigure}
\begin{subfigure}{0.32\textwidth}
\scalebox{0.8}{
	\begin{tikzpicture}
		\begin{feynman}
		\vertex (a1) {\(U(1)\)};
		\vertex [below = 1.414cm of a1] (a2);
		\vertex [below left = 1.732cm and 1cm of a2] (a3);	
		\vertex [below right = 1.732cm and 1cm of a2] (a4);
		\vertex [below left = 1cm and 0.3cm of a3] (b1) {\(SU(3)\)};
		\vertex [below right = 1cm and 0.3cm of a4] (b2) {\(SU(3)\)};
		\diagram* {
		(a1) -- [boson] (a2),
		(a2) -- [] (a3) -- [] (a4) -- [] (a2),
		(b1) -- [boson] (a3),
		(b2) -- [boson] (a4),
		};
		\end{feynman}
	\end{tikzpicture}}
\end{subfigure}
\\[2ex]
\begin{subfigure}{0.32\textwidth}
\scalebox{0.8}{
	\begin{tikzpicture}
		\begin{feynman}
		\vertex (a1) {\(SU(2)\)};
		\vertex [below = 1.414cm of a1] (a2);
		\vertex [below left = 1.732cm and 1cm of a2] (a3);	
		\vertex [below right = 1.732cm and 1cm of a2] (a4);
		\vertex [below left = 1cm and 0.3cm of a3] (b1) {\(SU(2)\)};
		\vertex [below right = 1cm and 0.3cm of a4] (b2) {\(SU(2)\)};
		\diagram* {
		(a1) -- [boson] (a2),
		(a2) -- [] (a3) -- [] (a4) -- [] (a2),
		(b1) -- [boson] (a3),
		(b2) -- [boson] (a4),
		};
		\end{feynman}
	\end{tikzpicture}}
\end{subfigure}
\begin{subfigure}{0.32\textwidth}
\scalebox{0.8}{
	\begin{tikzpicture}
		\begin{feynman}
		\vertex (a1) {\(SU(2)\)};
		\vertex [below = 1.414cm of a1] (a2);
		\vertex [below left = 1.732cm and 1cm of a2] (a3);	
		\vertex [below right = 1.732cm and 1cm of a2] (a4);
		\vertex [below left = 1cm and 0.3cm of a3] (b1) {\(SU(2)\)};
		\vertex [below right = 1cm and 0.3cm of a4] (b2) {\(SU(3)\)};
		\diagram* {
		(a1) -- [boson] (a2),
		(a2) -- [] (a3) -- [] (a4) -- [] (a2),
		(b1) -- [boson] (a3),
		(b2) -- [boson] (a4),
		};
		\end{feynman}
	\end{tikzpicture}}
\end{subfigure}
\begin{subfigure}{0.32\textwidth}
\scalebox{0.8}{
	\begin{tikzpicture}
		\begin{feynman}
		\vertex (a1) {\(SU(2)\)};
		\vertex [below = 1.414cm of a1] (a2);
		\vertex [below left = 1.732cm and 1cm of a2] (a3);	
		\vertex [below right = 1.732cm and 1cm of a2] (a4);
		\vertex [below left = 1cm and 0.3cm of a3] (b1) {\(SU(3)\)};
		\vertex [below right = 1cm and 0.3cm of a4] (b2) {\(SU(3)\)};
		\diagram* {
		(a1) -- [boson] (a2),
		(a2) -- [] (a3) -- [] (a4) -- [] (a2),
		(b1) -- [boson] (a3),
		(b2) -- [boson] (a4),
		};
		\end{feynman}
	\end{tikzpicture}}
\end{subfigure}
\caption{Gauge anomalies which can take place at $SU(2) \times U(1)$ interaction theory}
\label{fig:gauge_anomalies_SU2U1}
\end{figure}

In order to extend the SM fermion sector, many hypothetical ingredients have been studied and considered such as the vector-like (VL) fermion, leptoquark (LQ) and long-lived particle (LLP). Each of those is based on the fundamental property of the SM and we focus mainly on the vector-like fermions since the SM fermion sector of all my works are extended by only the vector-like fermions.
\\~\\
The vector-like fermions are a well-known candidate as they have the exact same quantum numbers as in the SM fermions and one vector-like family consists of two partner particles which share the same quantum number, so they cancel out the possible gauge anomalies in the $SU(2) \times U(1)$ interaction theory as given in Figure~\ref{fig:gauge_anomalies_SU2U1}. 
As the name ``vector-like" tells, one vector-like family generally consists of its left-handed (LH) and right-handed (RH) particles where both share the same quantum numbers, so they can have two source of mass; one is the chirality flip mass and the other is the vector-like mass. This interesting mass sources with the vector-like particles can be seen clearly by referring to the particle content used in our first paper.
\begin{table}[H]
	\setlength\heavyrulewidth{0.25ex}
	\centering\renewcommand{\arraystretch}{1.3} 
	\resizebox{\textwidth}{!}{%
	\begin{tabular}{ccccccccccccccccccccc}
		\toprule
		\toprule
		Field & $Q_{iL}$ & $u_{iR}$ & $d_{iR}$ & $L_{iL}$ & $e_{iR}$ & $\nu_{iR}$ & 
		$H$ & $Q_{4L}$ & $\widetilde{Q}_{4R}$ & $\widetilde{u}_{4L}$ & $u_{4R}$ & $%
		\widetilde{d}_{4L}$ & $d_{4R}$ & $L_{4L}$ & $\widetilde{L}_{4R}$ & $%
		\widetilde{E}_{4L}$ & $E_{4R}$ & $\nu _{4R}$ & $\widetilde{\nu }_{4L}$ & $%
		\phi _{f}$ \\ \midrule
		$SU(3)_{C}$ & $\mathbf{3}$ & $\mathbf{3}$ & $\mathbf{3}$ & $\mathbf{1}$ & $%
		\mathbf{1}$ & $\mathbf{1}$ & $\mathbf{1}$ & $\mathbf{3}$ & $\mathbf{3}$ & $%
		\mathbf{3}$ & $\mathbf{3}$ & $\mathbf{3}$ & $\mathbf{3}$ & $\mathbf{1}$ & $%
		\mathbf{1}$ & $\mathbf{1}$ & $\mathbf{1}$ & $\mathbf{1}$ & $\mathbf{1}$ & $%
		\mathbf{1}$ \\ 
		$SU(2)_{L}$ & $\mathbf{2}$ & $\mathbf{1}$ & $\mathbf{1}$ & $\mathbf{2}$ & $%
		\mathbf{1}$ & $\mathbf{1}$ & $\mathbf{2}$ & $\mathbf{2}$ & $\mathbf{2}$ & $%
		\mathbf{1}$ & $\mathbf{1}$ & $\mathbf{1}$ & $\mathbf{1}$ & $\mathbf{2}$ & $%
		\mathbf{2}$ & $\mathbf{1}$ & $\mathbf{1}$ & $\mathbf{1}$ & $\mathbf{1}$ & $%
		\mathbf{1}$ \\ 
		$U(1)_{Y}$ & $\frac{1}{6}$ & $\frac{2}{3}$ & $-\frac{1}{3}$ & $-\frac{1}{2}$
		& $-1$ & $0$ & $\frac{1}{2}$ & $\frac{1}{6}$ & $\frac{1}{6}$ & $\frac{2}{3}$
		& $\frac{2}{3}$ & $-\frac{1}{3}$ & $-\frac{1}{3}$ & $-\frac{1}{2}$ & $-\frac{%
			1}{2}$ & $-1$ & $-1$ & $0$ & $0$ & $0$ \\ 
		$U(1)'$ & $0$ & $0$ & $0$ & $0$ & $0$ & $0$ & $0$ & $q_{Q_{4}}$ & $%
		q_{Q_{4}} $ & $q_{u_{4}}$ & $q_{u_{4}}$ & $q_{d_{4}}$ & $q_{d_{4}}$ & $%
		q_{L_{4}}$ & $q_{L_{4}}$ & $q_{e_{4}}$ & $q_{e_{4}}$ & $q_{\nu _{4}}$ & $%
		q_{\nu _{4}}$ & $-q_{f_4}$ \\ 
		\bottomrule
		\bottomrule
	\end{tabular}}
	\caption{Particle content to explain two mass sources consisting of the vector-like particles}
	\label{tab:first_model}
\end{table}
As shown in Table~\ref{tab:first_model}, the subscript $4$ means fourth vector-like family (It means fourth after the three SM generations). For simplicity, we focus on the fourth vector-like lepton doublets ($L_{4L}, \widetilde{L}_{4R}$). The first interesting feature of them is they have the exact same quantum numbers, so it allows for them to have the vector-like mass as follows:
\begin{equation}
\mathcal{L}_{\func{mass}} = M_{4}^{L} \overline{L}_{4L} \widetilde{L}_{4R} + \func{h.c.},
\end{equation}
and the vector-like mass $M_{4}^{L}$ is not constrained by any symmetry, so it can be as heavy as possible and allows more freedom to our numerical scans. Along the vector-like mass, the vector-like particles can have one more mass source called the chirality flip mass as follows:
\begin{equation}
\mathcal{L}_{\func{mass}} = x_L \overline{L}_{4L} \widetilde{H} E_{4R} + \func{h.c.} = M_{4}^{C} \overline{E}_{4L} E_{4R} + \func{h.c.},
\end{equation}
Unlike the vector-like mass $M_{4}^{L}$, the chirality flip mass $M_{4}^{C}$ is governed by the vev of the neutral component of the SM Higgs, therefore it can not be very heavy. Before going into a further detailed analysis of the vector-like and chirality flip mass, it is good to remind of their symbolic notation appearing in many Feynman diagrams.
\vspace{10pt}
\begin{figure}[H]
\centering
	\begin{tikzpicture}
		\begin{feynman}
		\vertex (a1) ;
		\vertex [right = 3cm of a1] (a2) ;
		\vertex [right = 3cm of a2] (a3) ;	
		\vertex [below right = 1cm and 3cm of a2] (b3) ;
		\diagram* {
		(a1) -- [insertion=0.5] (a3),
		};
		\end{feynman} 
	\end{tikzpicture} 
\caption{A symbolic notation for either the vector-like or chirality flip mass}
\label{fig:symbol_vl_or_cf}
\end{figure}
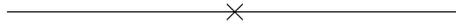
As Figure~\ref{fig:symbol_vl_or_cf} tells, the symbolic notation ``cross" means either the vector-like or chirality flip mass, so it should be careful when we consider interactions involving either the vector-like or chirality flip mass. This property of the vector-like and chirality flip mass is reflected on our analysis of the muon and electron $g-2$ in our first work, by considering the masses separately in the one-loop diagrams. Strictly speaking, this is not a very correct way of dealing with the masses in our analytic analysis, since we did not carry out diagonalization of the mass matrix, however it is a good approximation under the assumption $M_{4}^{L} \gg M_{4}^{C}$. As you are aware of from the diagonalization, it is possible to build the mass matrix involving both the vector-like and chirality flip mass and let me construct it as follows (I do not consider three generations of the SM at the moment, but only focus on the fourth vector-like leptons for simplicity):
\begin{equation}
M^{L }
=
\left( 
\begin{array}{c|cc}
& \widetilde{E} _{4R} & E _{4R} \\ \hline
E_{4L} & M_{4}^{L} & y_{4}^{(e)} v \\
\widetilde{E }_{4L} & 0 & M_{4}^{E} \\
\end{array}%
\right)
=
\left( 
\begin{array}{c|cc}
& \widetilde{E} _{4R} & E _{4R} \\ \hline
E_{4L} & M_{4}^{L} & M_{4}^{C} \\
\widetilde{E }_{4L} & 0 & M_{4}^{E} \\
\end{array}%
\right).
\label{eqn:an_example_massmatrix}
\end{equation}
We can find the physical vector-like masses by diagonalizing the mass matrix given in Equation~\ref{eqn:an_example_massmatrix}. As mentioned previously, we did not carry out the diagonalization of our whole mass matrix in our first work since we had not constructed the whole mass matrix, however, it was a good approximation to the diagonalization under the assumption $M_{4}^{L,E} \gg M_{4}^{C}$. 
\\~\\
It is a common feature that the vector-like family was used over my three works, however there is a big difference between the first and second plus third BSM model. In the first model, the general SM Yukawa interactions are allowed as the SM Higgs $H$ is neutral under the $U(1)^{\prime}$ charge, whereas the general SM Yukawa interactions are not allowed in both the second and third BSM model as the SM Higgs is charged under the $U(1)^{\prime}$ symmetry in order to take the SM as an effective theory and then to bring the hierarchical structure of the SM under our understanding in a kinematic way. On top of that, we used the mass insertion approximation in the analysis of the muon and electron $g-2$ in our first work, which means both the chirality and vector-like masses appear in the analytic form of the anomalies. In the second work, two vector-like families were used in order to make all SM generations massive (One vector-like family can provide two effective seesaw operators) and the mass matrices of second work were diagonalized with an assumption all mixing angles appearing there are quite small. In our third work, we made use of just one vector-like family for the purpose of diagonalizing the mass matrices without any assumptions. The first SM generation can not be massive as a result, however it is a good approximation taking into account the first SM generation is quite light.
\\~\\
There is an important feature in my three works~\cite{CarcamoHernandez:2019ydc,Hernandez:2021tii,Hernandez:2021oyv}, that both the vector-like doublet and singlet fermions were considered at the same time. Taking into account that the most simplest extension of the SM fermion sector is to consider only the vector-like singlet fermions, it looks like our BSM models are less minimal (as well as less economical) and the vector-like doublet fermions are not necessary to extend the SM fermions. However we prefer to consider both for two reasons.
\begin{enumerate}
\item The first reason is to have a realistic fermion mass spectrum. Referring to the numerical plots for the charged lepton as well as quark sector in our third work~\cite{Hernandez:2021oyv}, the doublet vector-like fermion's masses are relatively lighter compared to their singlet vector-like masses, which means the doublet vector-like fermions will be likely to be observed first if there exist the vector-like fermions.
\item As the BSM models under consideration in my three works were based on UV completion theory, which means the BSM models must be effective theories to be able to explain the SM without violating gauge symmetry and the current experimental bounds if the energy scale goes down to the electroweak scale. In this view point considering both provides many natural analytical explanations for the observables. In our third work~\cite{Hernandez:2021oyv} we were able to find out how deviation of the first row of the CKM mixing matrix can arise in an analytic way by considering both the doublet and singlet vector-like fermions.
\end{enumerate}
For these reasons, we firmly believe that considering a complete vector-like family is much more realistic to extend the SM fermions.

Summarizing this subsection, the vector-like particles have been an important ingredient to extend the SM fermion sector in that they can implement richer phenomenology such as the vector-like or chirality flip mass and purely new interactions at tree-level, so we can explore its new possibility to new physics. We will take a closer look at how the vector-like particles can contribute to new physics throughout the rest of this thesis.
\section{Prerequisite : 2 Higgs Doublet Model (2HDM)}
\label{sec:pre_2HDM}
The observed SM Higgs reports a quite light mass $125\func{GeV}$ and this observation has been a quite interesting open question: why is the SM Higgs so light? Following one of the convincing explanations for the Higgs mass, the SM Higgs is an extremely fine tuned parameter by subtracting from the SM Higgs's bare mass to its loop correction as follows:
\begingroup
\begin{equation}
m_{\func{phys}}^2 = m_{\func{bare}}^2 - m_{\func{loop}}^2,
\end{equation}
\endgroup
where order of $m_{\func{bare,loop}}$ is about $10^{19}$, which is quite close to the Planck scale. This extreme suppression between $m_{\func{bare}}$ and $m_{\func{loop}}$ might imply new possibilities and one of which is there must be some other extra symmetry in the SM scalar sector if we approximate the SM Higgs mass to be zero. Plus, the SM Higgs has vev of $246\func{GeV}$ and what this means is the SM can not predict a heavier mass like $1\func{TeV}$. However, many particle physicists agree that there must be at least a new physics between the electroweak scale to the Planck scale and this naturally leads to needs to extend the SM scalar sector in order to explain the heavy particles assumed in new physics. As long as we consider only the Yukawa interactions and the spontaneous symmetry breaking to assign each fermion a mass, it is necessary to include more scalars, whose vevs are generally assumed to be heavier than the SM Higgs vev, which plays a crucial role in my three works. And the SM Higgs vev is known as $246\func{GeV}$ and we can come up with the possibility known as the 2 Higgs Doublet Model (2HDM) if we consider the up- and down-type quark sectors (as well as the charged lepton sector) are governed by the different Higgs vevs. This is a quite nice idea when we try to explain the strong hierarchical structure of the SM and it has been one of our main motivations over my second and third works. Our second and third BSM models feature the 2HDM, in which the up-type Higgs $H_u$ couples to the up-type quarks, whereas the down-type Higgs $H_d$ interacts with the down-type quarks as well as the charged leptons (the SM neutrinos need to be treated separately and this will be discussed when we explore our second work). Therefore, the up- and down-type Higgs vevs hold for this relation:
\begingroup
\begin{equation}
v_{u}^2 + v_{d}^{2} = (246\func{GeV})^2
\end{equation}
\endgroup
Therefore, the SM Higgs is enlarged by the two SM-like Higgses $H_{u,d}$ and this has a couple of advantages; one of which is the hierarchical structure of the SM can be explained dynamically and the other is we can expect richer phenomenology of the CP-even and -odd Higgses, which appear as a result of the mixing between $H_{u}$ and $H_{d}$, and this feature is covered in our second work.
\\~\\
Summarizing this subsection, the light SM Higgs mass $125\func{GeV}$ can imply the other extra symmetry in the SM scalar sector and its vev $246\func{GeV}$ can not predict heavier particles whose masses are order of $1\func{TeV}$ or above than that. The assumed heavy particles are frequently considered in some new physics model and this means some other scalars, whose vevs are generally assumed to be heavier than that of the SM Higgs, should be introduced. An extended 2HDM by the singlet flavon has become a main BSM model for my second and third work in order to explain the hierarchical structure of the SM and this feature will be discussed in detail when we explore the second and third work.
\section{Prerequisite : $U(1)^{\prime}$ gauge symmetry}
\label{sec:pre_U1p}
As in the enlargement of the fermion and scalar sectors, extension of the SM gauge symmetry has been considered important and all the extensions are based on the unification covered in chapter~\ref{Chapter:TheSM}. The generally accepted standard theory to explain the origin of the universe is known as the Big Bang theory as given in Figure~\ref{fig:BigBang},
\begingroup
\begin{figure}[H]
\includegraphics[keepaspectratio,width=\textwidth]{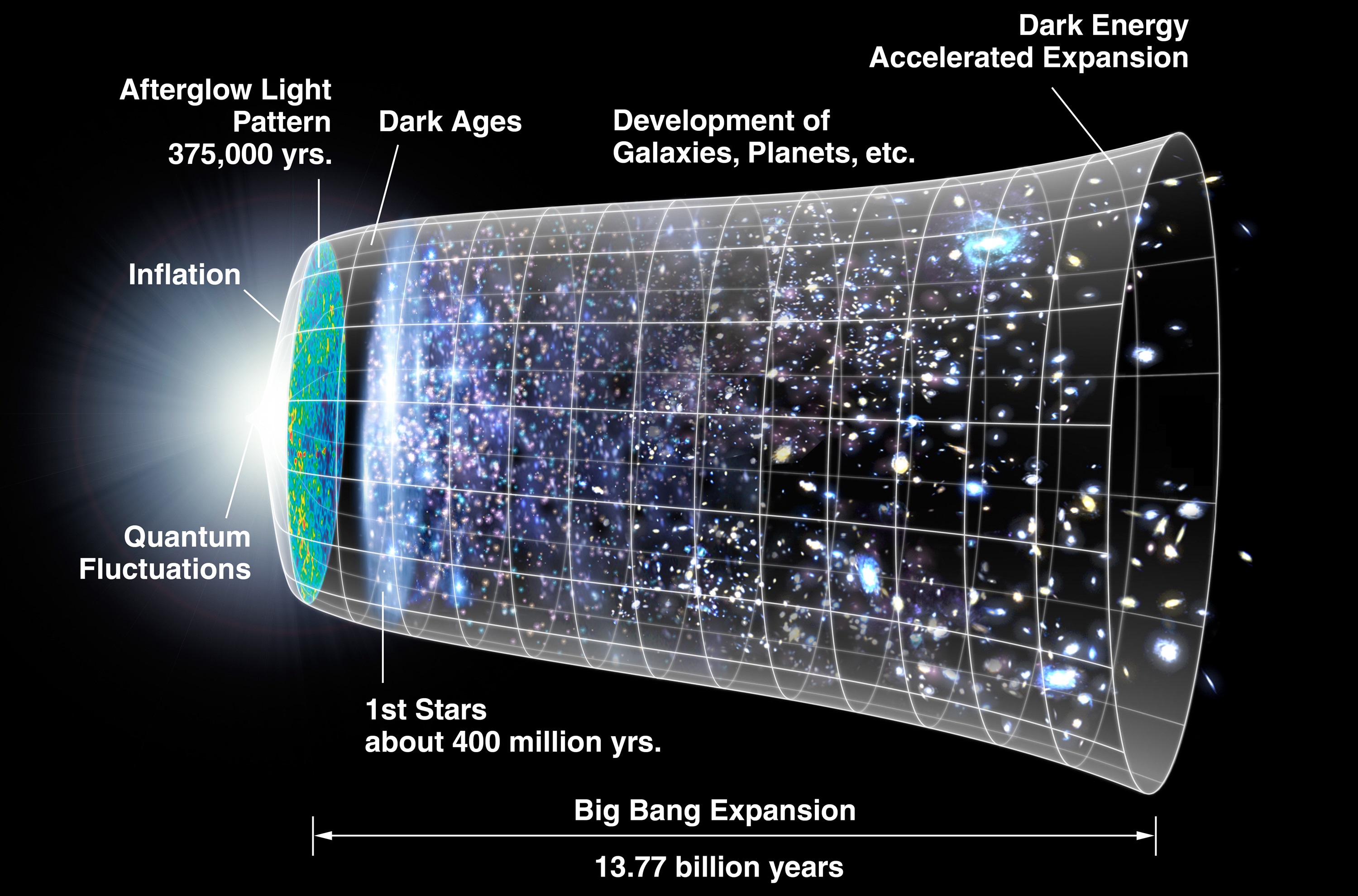}
\caption{The Big Bang theory which has been accepted as a standard theory for the origin of the universe}
\label{fig:BigBang}
\end{figure}
\endgroup
and it is believed there must be a most fundamental symmetry corresponding for the start of the Big Bang theory. The four forces, which are electromagnetic, weak, strong and gravity, have been known so far and it is proved that the electromagnetic and weak forces can be unified at the electroweak scale. These unifications can be seen in Figure~\ref{fig:uni}.
\begingroup
\begin{figure}[H]
\includegraphics[keepaspectratio,width=\textwidth]{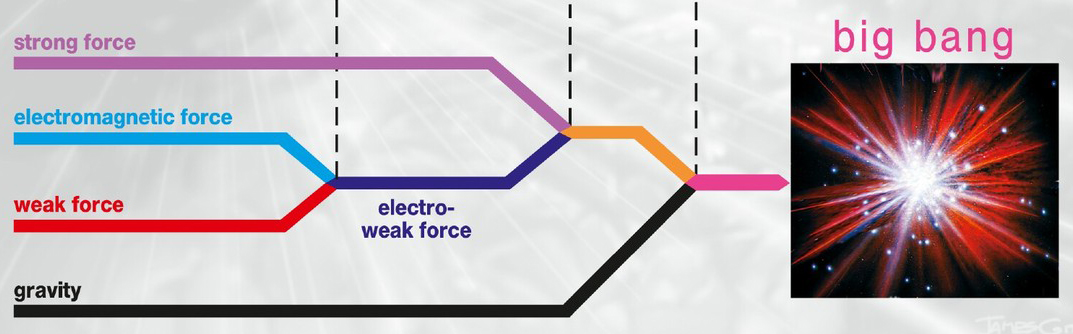} 
\caption{Unification of all the fundamental forces at higher energy scales}
\label{fig:uni}
\end{figure}
\endgroup
The unification between the electroweak and strong forces is known as the Grand Unified Theory (GUT), whose energy scale is corresponding for $10^{15}\func{GeV}$ and that between the GUT and gravity is known as the Theory of Everything (ToE), whose energy scale is corresponding for the Planck scale $10^{19}\func{GeV}$. There have been many attempts to explain the GUT or ToE by the larger symmetries such as $E_{6}, E_{8}$ or $SO(10)$. However, a critical problem of direct connection from the SM gauge symmetry to the higher energy scale is very likely to mislead our understanding to the phenomenology of the particle physics and it is not possible to confirm the derived new particles and intermediate symmetries experimentally, as their energy scale is too high. For this reason, we focus on the minimal extension for the gauge symmetry and a simplest possibility is $U(1)^{\prime}$ symmetry. The $U(1)^{\prime}$ symmetry can be further separated depending on whether it is local or global. If the $U(1)^{\prime}$ symmetry is local, the neutral $Z^{\prime}$ boson appears and the $Z^{\prime}$ boson can lead to some new physics and this idea was carried out in my first work. If the $U(1)^{\prime}$ symmetry is global, the $Z^{\prime}$ boson does not appear and a main role of the symmetry is to constrain some Yukawa interactions, which was carried out in my second and third works.
\\~\\
Summarizing this subsection, the extension of the SM gauge symmetry is based on the feature of the unification as well as the fermion and scalar sectors. Since the direct connection from the SM gauge symmetry to either GUT or ToE is very likely to mislead our understanding to the phenomenology, so we take a simplest possibility, the $U(1)^{\prime}$ symmetry, in order to avoid the misleading. The $U(1)^{\prime}$ symmetry can be further separated depending on whether it is a local or global. The local $U(1)^{\prime}$ symmetry features the neutral $Z^{\prime}$ gauge boson and this will be discussed in our first work. Our second and third work features the $U(1)^{\prime}$ global symmetry and the $U(1)^{\prime}$ symmetry plays a role of constraining some Yukawa interactions.
\\~\\
We have discussed common features shared by my three works and we start to discuss our first BSM model in our first work from the next section.
\section{Introduction and motivation} \label{sec:first_paper_moti_int}
The Standard Model (SM) provides an excellent explanation of all experimental data, apart from neutrino mass and lepton mixing.
Yet there are a few possible anomalies in the flavour sector that may indicate new physics beyond the SM.
For example, recently, there have been hints of universality violation in the charged lepton sector from $B\rightarrow K^{(\ast )}l^{+}l^{-}$ decays by the LHCb collaboration\cite{Descotes-Genon:2013wba,Altmannshofer:2013foa,Ghosh:2014awa}. Specifically, the $R_{K}$ \cite{Aaij:2019wad} and $R_{K^{\ast }}$ \cite{Bifani} ratios of $\mu ^{+}\mu ^{-}$ to $e^{+}e^{-}$ final states in the $B\rightarrow K^{(\ast)}l^{+}l^{-}$ decays are observed to be about $70\%$ of their expected values with a roughly $2.5\sigma$ deviation from the Standard Model (SM) in each channel. Following the recent measurement of $R_{K^{\ast }}$ \cite{Bifani}, a number of phenomenological analyses have been presented\cite{Hiller:2017bzc,Ciuchini:2017mik,Geng:2017svp,Capdevila:2017bsm,Ghosh:2017ber,Ciuchini:2019usw,Bardhan:2017xcc} that favour a new effective field theory (EFT) physics operator of the $C_{9\mu }^{NP}=-C_{10\mu }^{NP}$ form\cite{Glashow:2014iga,DAmico:2017mtc,Aebischer:2019mlg}. The most recent global fit of this operator combination yields ${C_9 = (34.0\text{TeV}) ^{-2}}$\cite{Aebischer:2019mlg}, though other well-motivated solutions are also possible\cite{Descotes-Genon:2015uva}.

In previous works\cite{Glashow:2014iga}, it has been suggested that such observations of charged lepton universality violation (CLUV) must be accompanied by charged lepton flavour violation (CLFV) such as $\mu\rightarrow e \gamma$ in the same sector, however, such a link cannot be established in a model-independent way because the low-energy effective operators for each class of processes are different. Nevertheless, in concrete models the connection is often manifest. This motivates studies of specific models. For example, studies of CLFV in B-decays using generic $Z^{\prime }$ models (published before the $R_{K^{\ast }}$ measurement but compatible with it) are provided in Ref. \cite{Crivellin:2015era}. A concise review of BSM scenarios that aim to explain CLUV and possible connections to dark matter is provided in Ref. \cite{Vicente:2018xbv}. Other theoretical explanations for universality violation in the lepton sector are discussed in Refs. \cite{Glashow:2014iga,Crivellin:2015era,Crivellin:2015lwa,Bonilla:2017lsq,Assad:2017iib, King:2017anf,Romao:2017qnu,Antusch:2017tud,Ko:2017quv,King:2018fcg,Falkowski:2018dsl,Barman:2018jhz,CarcamoHernandez:2018aon,deMedeirosVarzielas:2018bcy,Rocha-Moran:2018jzu,Hu:2018veh,Carena:2018cow,Babu:2018vrl,Allanach:2018lvl,Fornal:2018dqn,Aydemir:2018cbb,CarcamoHernandez:2019cbd,CarcamoHernandez:2019xkb,Aydemir:2019ynb}.

Independently of these anomalies, for some time now, it has been known that the experimentally measured anomalous magnetic moments 
g-2 of both the muon and electron each observe a discrepancy of a few standard deviations with respect to the Standard Model predictions. The longstanding non-compliance of the muon g-2 with the SM was first observed by the Brookhaven E821 experiment at BNL~\cite{Bennett:2006fi}. The electron g-2 has more recently revealed a discrepancy with the SM, following an accurate measurement 
of the fine structure constant~\cite{Parker:2018vye}.
However the different magnitude and opposite signs of the electron and muon g-2 deviations makes it difficult to explain both of these anomalies in any model, which also satisfies the constraints of CLFV, with all existing simultaneous explanations involving 
new scalars~\cite{Crivellin:2018qmi,Giudice:2012ms,Davoudiasl:2018fbb,Liu:2018xkx,Bauer:2019gfk,Han:2018znu,Dutta:2018fge,Badziak:2019gaf,Endo:2019bcj}, or conformal extended technicolour~\cite{Appelquist:2004mn}. We know of no study which discusses both anomalies in a $Z'$ model.
One possible reason is that the CLFV process 
$\mu\rightarrow e \gamma$, which would be concrete of BSM physics in the charged fermion sector,
is very constraining. Neutrino phenomena do give rise to CLFV but in the most minimal extensions this would occur at a very low rate in the charged sector, making it practically unobservable. Given the considerable resources committed to looking for CLFV, it is crucial to study relevant, well-motivated BSM scenarios which allow for CLFV at potentially observable rates. For example, such decays can be enhanced by several orders of magnitude if one considers extensions of the SM with an extra $U(1)'$ gauge symmetry spontaneously broken at the TeV scale. To summarise, although such extensions are able to successfully accommodate the experimental value of the muon magnetic moment\cite{Falkowski:2018dsl,Allanach:2015gkd,Raby:2017igl,CarcamoHernandez:2019xkb,Kawamura:2019rth}, we know of no study of a $Z'$ model which discusses both the electron and muon magnetic moments, including the constraints from $\mu\rightarrow e \gamma$.

In this work, we ask the question: is it possible to explain the anomalous muon and electron $g-2$ in a $Z^{\prime }$ model?
It is difficult to answer this question in general, since there are many possible $Z'$ models.
However it is possible to consider a model in which the $Z'$ only has couplings to the electron and muon and their associated 
neutrinos, arising from mixing with a vector-like fourth family of leptons, thereby 
eliminating the quark couplings and allowing us to focus on 
the connection between CLUV, CLFV and the electron and muon g-2 anomalies.
Such a renormalisable and gauge invariant 
model is possible within a $U(1)'$ gauge extension of the SM augmented by a fourth, vector-like family of fermions and right-handed neutrinos as proposed in \cite{King:2017anf}. In the fermiophobic version of this model~\cite{King:2017anf}, only the fourth family carry $U(1)'$ charges, with the three chiral families not coupling to the $Z'$ in the absence of mixing. Then one can switch on mixing between the first 
and second family of charged leptons and the fourth family, allowing controlled couplings of the $Z'$ to only the electron and muon
(and fourth family leptons) of the kind
we desire. Such a model allows charged lepton universality violation (CLUV) at tree-level with CLFV and contributions to 
the electron and muon magnetic moments at loop level. 
Within such a model we attempt to explain 
the anomalous magnetic moments of both the muon and electron within the relevant parameter space of the model,
while satisfying the constraints of $BR(\mu\rightarrow e\gamma)$ and neutrino trident production.
Using both analytic and numerical arguments, 
we find that it is not possible to simultaneously explain the electron and muon g-2 results consistent with these constraints.

The remainder of this article is organised as follows; in Section \ref{sec:model} we outline the renormalisable and gauge invariant 
fermiophobic model in which the $Z'$ couples only to a vector-like fourth family. 
In Section \ref{sec:LFV_in_this_model}, we show how it is possible to switch on the couplings of 
the $Z'$ to the electron and muon and their associated 
neutrinos, thereby eliminating all unnecessary couplings and allowing us to focus on 
the connection between CLUV, CLFV and the electron and muon g-2 anomalies. A simplified analytical analysis of the CLFV and the electron and muon g-2 anomalies in the fermiophobic $Z'$ Model is presented in Section \ref{sec:analytics}. In Section \ref{sec:statistical_analysis} we analyse the parameter space numerically, presenting detailed predictions for each of the examined leptonic phenomena. In Section~\ref{sec:Zp_exp_bound} we discuss the experimental and theoretical $Z^{\prime}$ mass bound. Section~\ref{sec:conclusion} concludes the paper.
 
\section{The Fermiophobic $Z'$ Model}
\label{sec:model}
Consider an extension of the SM with a $U(1)'$ gauge symmetry, where fermion content is expanded by right-handed neutrinos and a fourth, vector-like family. The scalar sector is augmented by gauge singlet fields with non-trivial charge assignments under the new symmetry. The basic framework for such a theory was defined in  \cite{King:2017anf}.
Henceforth we consider the case where the 
SM fermions in our model are uncharged under the additional symmetry, whereas the vector-like fermions are charged under this symmetry, corresponding to so called ``fermiophobic $Z'$'' model considered in \cite{King:2017anf}. 
The field content and charge assignments are given in Table \ref{tab:model_content}.
Note that such a theory is anomaly free; left- and right-handed fields of the vector-like fermion family have identical charges under $U(1)'$, and hence chiral anomalies necessarily cancel. 
\begingroup
\begin{table}[H]
	\setlength\heavyrulewidth{0.25ex}
	\centering\renewcommand{\arraystretch}{1.3} 
	\resizebox{\textwidth}{!}{%
	\begin{tabular}{ccccccccccccccccccccc}
		\toprule
		\toprule
		Field & $Q_{iL}$ & $u_{iR}$ & $d_{iR}$ & $L_{iL}$ & $e_{iR}$ & $\nu_{iR}$ & 
		$H$ & $Q_{4L}$ & $\widetilde{Q}_{4R}$ & $\widetilde{u}_{4L}$ & $u_{4R}$ & $%
		\widetilde{d}_{4L}$ & $d_{4R}$ & $L_{4L}$ & $\widetilde{L}_{4R}$ & $%
		\widetilde{E}_{4L}$ & $E_{4R}$ & $\nu _{4R}$ & $\widetilde{\nu }_{4L}$ & $%
		\phi _{f}$ \\ \midrule
		$SU(3)_{c}$ & $\mathbf{3}$ & $\mathbf{3}$ & $\mathbf{3}$ & $\mathbf{1}$ & $%
		\mathbf{1}$ & $\mathbf{1}$ & $\mathbf{1}$ & $\mathbf{3}$ & $\mathbf{3}$ & $%
		\mathbf{3}$ & $\mathbf{3}$ & $\mathbf{3}$ & $\mathbf{3}$ & $\mathbf{1}$ & $%
		\mathbf{1}$ & $\mathbf{1}$ & $\mathbf{1}$ & $\mathbf{1}$ & $\mathbf{1}$ & $%
		\mathbf{1}$ \\ 
		$SU(2)_{L}$ & $\mathbf{2}$ & $\mathbf{1}$ & $\mathbf{1}$ & $\mathbf{2}$ & $%
		\mathbf{1}$ & $\mathbf{1}$ & $\mathbf{2}$ & $\mathbf{2}$ & $\mathbf{2}$ & $%
		\mathbf{1}$ & $\mathbf{1}$ & $\mathbf{1}$ & $\mathbf{1}$ & $\mathbf{2}$ & $%
		\mathbf{2}$ & $\mathbf{1}$ & $\mathbf{1}$ & $\mathbf{1}$ & $\mathbf{1}$ & $%
		\mathbf{1}$ \\ 
		$U(1)_{Y}$ & $\frac{1}{6}$ & $\frac{2}{3}$ & $-\frac{1}{3}$ & $-\frac{1}{2}$
		& $-1$ & $0$ & $\frac{1}{2}$ & $\frac{1}{6}$ & $\frac{1}{6}$ & $\frac{2}{3}$
		& $\frac{2}{3}$ & $-\frac{1}{3}$ & $-\frac{1}{3}$ & $-\frac{1}{2}$ & $-\frac{%
			1}{2}$ & $-1$ & $-1$ & $0$ & $0$ & $0$ \\ 
		$U(1)'$ & $0$ & $0$ & $0$ & $0$ & $0$ & $0$ & $0$ & $q_{Q_{4}}$ & $%
		q_{Q_{4}} $ & $q_{u_{4}}$ & $q_{u_{4}}$ & $q_{d_{4}}$ & $q_{d_{4}}$ & $%
		q_{L_{4}}$ & $q_{L_{4}}$ & $q_{e_{4}}$ & $q_{e_{4}}$ & $q_{\nu _{4}}$ & $%
		q_{\nu _{4}}$ & $-q_{f_4}$ \\ 
		\bottomrule
		\bottomrule
	\end{tabular}}
	\caption{Particle assignments under $SU(3)_{C}\times SU(2)_{L}\times U(1)_{Y}\times U(1)'$ gauge symmetry. $i=1,2,3$. The SM singlet scalars $\protect\phi_f $ ($f=Q,u,d,L,e$) have $U(1)'$ charges $-q_{f_4}=-q_{Q_{4,}u_{4},d_{4},L_{4},e_{4}}$.}
	\label{tab:model_content}
\end{table}
\endgroup
Although the $Z'$ couples only to the vector-like fourth family to start with, due to the mixing between SM fermions and those of the fourth vector-like family (arising from the Lagrangian below) the $Z'$ will get induced couplings to chiral
SM fermions.
After mixing, the model can allow for a viable dark matter candidate and operators crucial for explaining the $R_{K}$ and $R_{K^{\ast }} $ flavour anomalies\cite{Falkowski:2018dsl}. 
As we shall see, this setup can also generate CLFV signatures such as $\mu \rightarrow e\gamma $ and accommodate the experimental value of the anomalous muon and electron
magnetic dipole moments.

With the particle content, symmetries and charge assignments in Table \ref{tab:model_content}, the following renormalisable Lagrangian terms are available: 
\begingroup
\begin{align}
	\begin{split}
		\mathcal{L}_{Y} &=\sum_{i=1}^{3}\sum_{j=1}^{3}y_{ij}^{(u) }%
		\overline{Q}_{iL}\widetilde{H}u_{jR}+\sum_{i=1}^{3}\sum_{j=1}^{3}y_{ij}^{%
		(d) }\overline{Q}_{iL}Hd_{jR}
		\\
		&+\sum_{i=1}^{3}\sum_{j=1}^{3}%
		y_{ij}^{(e)}\overline{L}_{iL}He_{jR}+%
		\sum_{i=1}^{3}\sum_{j=1}^{3}y_{ij}^{(\nu) }\overline{L}_{iL}%
		\widetilde{H}\nu _{jR} \\
		&+y_{4}^{(u)}\overline{Q}_{4L}\widetilde{H}u_{4R}+y_{4}^{		(d)}\overline{Q}_{4L}Hd_{4R}+y_{4}^{(e)}\overline{L}_{4L}HE_{4R}+y_{4}^{(\nu) }\overline{L}_{4L}\widetilde{H}\nu
		_{4R}\\
		&+\sum_{i=1}^{3}x_{i}^{(Q)}\phi _{Q}\overline{Q}_{Li}\tilde{Q}%
		_{4R}+\sum_{i=1}^{3}x_{i}^{(u)}\phi _{u}\overline{\tilde{u}}%
		_{4L}u_{Ri}+\sum_{i=1}^{3}x_{i}^{(d)}\phi _{d}\overline{\tilde{d%
		}}_{4L}d_{Ri}%
		\\&+\sum_{i=1}^{3}x_{i}^{(L)}\phi _{L}\overline{L}%
		_{Li}\tilde{L}_{4R}+\sum_{i=1}^{3}x_{i}^{(e)}\phi _{e}\overline{%
			\widetilde{E}}_{4L}e_{Ri} 
		+M_{4}^{Q}\overline{Q}_{4L}\tilde{Q}_{4R}+M_{4}^{u}\overline{\tilde{u}}_{4L}u_{4R}%
		\\
		&+M_{4}^{d}\overline{\tilde{d}}_{4L}d_{4R}+M_{4}^{L}\overline{L}_{4L}\tilde{L}%
		_{4R}+M_{4}^{E}\overline{\widetilde{E}}_{4L}E_{4R}+M_{4}^{\nu}\overline{\tilde{\nu}}_{4L}\nu _{4R}+H.c.
		\label{eqn:Yukawa_Lagrangian}
	\end{split}
\end{align}
\endgroup
where the requirement of $U(1)'$ invariance of the Yukawa interactions involving the fourth family yields the following constraints on the $U(1)'$ charges of fourth fermion families:
\begingroup
\begin{equation}
q_{Q_4}=q_{u_4}=q_{d_4}\hspace{1cm}q_{L_4}=q_{e_4}=q_{\nu_4}  
\end{equation}
\endgroup 
It is clear from Equation \eqref{eqn:Yukawa_Lagrangian} that fields in the 4th, vector-like family obtain masses from two sources; firstly, Yukawa terms involving the SM Higgs field such as $y_{4}^{(e)}\overline{L}_{4L}He_{4R}$ which get promoted to 
chirality flipping fourth family 
mass terms $M_4^C$ once the Higgs acquires a vev, and secondly from vector-like mass terms like $M_{4}^{L}\overline{L}_{4L}\tilde{L}_{4R}$ (these terms show up in lines 2 and 4 of Equation \eqref{eqn:Yukawa_Lagrangian} respectively). 
For the purposes of clarity, we shall treat $M_4^C$ and $M_{4}^{L}\overline{L}_{4L}\tilde{L}_{4R}$ as independent mass terms
in the analysis of the physical quantities of interest, rather than constructing the full fourth family mass matrix and 
diagonalising it, since such quantities rely on a chirality flip and are sensitive to $M_4^C$ rather than 
the vector-like masses $M_{4}^{L}\overline{L}_{4L}\tilde{L}_{4R}$. Spontaneous breaking of $U(1)'$ by the scalars $\phi_i$ spontaneously acquiring vevs gives rise to a massive $Z^{\prime }$ boson featuring couplings with the chiral and vector-like fermion fields. In the interaction basis such terms will be diagonal and of the following form:
\begingroup
\begin{equation}
\begin{split}
	\mathcal{L}_{Z^{\prime }}^{gauge}&=g^{\prime }Z_{\mu }^{\prime }(
	\overline{Q}_{L}D_{Q}\gamma ^{\mu }{Q}_{L}+\overline{u}_{R}D_{u}\gamma ^{\mu
	}u_{R}+\overline{d}_{R}D_{d}\gamma ^{\mu }d_{R}
	\\
	&+\overline{L}_{L}D_{L}\gamma
	^{\mu }L_{L}+\overline{e}_{R}D_{e}\gamma ^{\mu }e_{R} + \overline{\nu}_{R}D_{\nu}\gamma ^{\mu }\nu_{R})
	\label{eqn:Zprime_couplings_interaction_basis}
\end{split}
\end{equation}
\endgroup
Here, $g'$ is the `pure' gauge coupling of $U(1)'$ and each of the $D$s are 4x4 matrices. However, only the fourth family has non-vanishing $U(1)'$ charges as per Table \ref{tab:model_content} and hence these matrices are given by:
\begingroup
\begin{equation}
	\begin{gathered}
		D_{Q}=\mathrm{diag}(0,0,0,q_{Q_4}),\quad
		D_{u}=\mathrm{diag}(0,0,0,q_{u_4}),\quad D_{d}=\mathrm{diag}(0,0,0,q_{d_4}),\\
		D_{L}=\mathrm{diag}(0,0,0,q_{L_4}),\quad D_{e}=\mathrm{diag}(0,0,0,q_{e_4}),\quad
		D_{\nu}=\mathrm{diag}(0,0,0,q_{\nu_4})
		\label{eqn:Zprime_coupling_matrices_interaction_basis}
	\end{gathered}
\end{equation}
\endgroup
At this stage, the SM quarks and leptons do not couple to the $Z'$. However, the Yukawa couplings detailed in Equation \eqref{eqn:Yukawa_Lagrangian} have no requirement to be diagonal. Before we can determine the full masses of the propagating vector-like states and SM fermions, we need to transform the field content of the model such that the Yukawa couplings become diagonal. Therefore, fermions in the mass basis (denoted by primed fields) are related to particles in the interaction basis by the following unitary transformations:
\begingroup 
\begin{equation}
\begin{gathered}
	Q_{L}^{\prime }=V_{Q_{L}}{Q}_{L},\qquad u_{R}^{\prime }=V_{u_{R}}{u}%
	_{R},\qquad d_{R}^{\prime }=V_{d_{R}}{d}_{R},
	\\
	L_{L}^{\prime }=V_{L_{L}}%
	{L}_{L},\qquad e_{R}^{\prime }=V_{e_{R}}{e}_{R}
	,\qquad \nu_{R}^{\prime }=V_{\nu_{R}}{\nu}_{R}
	\label{eqn:field_transformations_to_mass_basis}
\end{gathered}
\end{equation}
\endgroup
This mixing induces couplings of SM mass eigenstate fermions to the massive $Z^{\prime }$ which can be expressed as follows
\begingroup
\begin{equation}
	\begin{split}
		D^{\prime }_Q=V_{Q_L}D_QV^{\dagger}_{Q_L}, \qquad 
		D^{\prime }_{u}=V_{u_R}D_uV^{\dagger}_{u_R}, \qquad 
		D^{\prime }_d=V_{d_R}D_dV^{\dagger}_{d_R},\\ 
		D^{\prime }_L=V_{L_L}D_LV^{\dagger}_{L_L}, \qquad
		D^{\prime }_e=V_{e_R}D_eV^{\dagger}_{e_R}, \qquad  
		D^{\prime }_\nu=V_{\nu_R}D_\nu V^{\dagger}_{\nu_R}
	\end{split}
	\label{eqn:coupling_conversion_to_mass_basis}
\end{equation}
\endgroup
Thus far all discussion of interactions and couplings has been general. In Sections \ref{sec:LFV_in_this_model} and \ref{sec:statistical_analysis}, we will prohibit mixing in some sectors to simplify our phenomenological analysis. In particular, we shall only consider induced $Z'$ couplings to the electron and muon.
\chapter{Is it possible to explain the muon and electron $g-2$ in a $Z^\prime$ model?} \label{Chapter:The1stpaper}

As an attempt to explain the connection between CLUV, CLFV and both anomalies with the $Z^\prime$ gauge boson, we set up the fermiophobic $Z^\prime$ model covered in chapter~\ref{Chapter:The1stBSMmodel}. In the model we showed the chiral SM particles have no interaction with the fourth vector-like family in the interaction basis due to the diagonal charge matrix, however the mixing arise in the physical basis since the diagonal charge matrix get to induce the off-diagonal mixing through the unitary transformation. From this stage, I expand our analysis for both anomalies as well as the two constraints, which are the CLFV $\mu \rightarrow e \gamma$ decay and neutrino trident production. 

\section{$Z'$ couplings to the electron and muon}
\label{sec:LFV_in_this_model}
In this paper we are particularly interested in the electron and muon g-2. We therefore take a minimal scenario and consider mixing only between first and second families of charged leptons, and ignore all quark and neutrino mixing, leading to a leptophillic $Z'$ model, in which the $Z'$ couples only to the electron, muon and their associated neutrinos. Therefore, only $V_{L_L}$ and $V_{e_R}$ will be non-diagonal, and LHC results will not constrain the $Z'$ mass as there is no direct coupling between SM quarks and the new vector boson, nor mixing between SM and vector-like quarks, because SM quarks are uncharged under $U(1)'$ as seen in Table \ref{tab:model_content}. Among the CLFV processes, we will focus on studying the $\mu\to e\gamma$ decay, which put tighter constrains than the $\tau\to\mu\gamma$ and $\tau\to e\gamma$ decays. For this reason, to simplify the parameter space, we also forbid the third family fermions from mixing with any other fermionic content. As such, all mixing at low energies can be expressed as per Equation \eqref{eqn:leptonic_mixing_matrices}.
\begingroup
\begin{equation}
\resizebox{0.9\hsize}{!}{$
	V_{L_L,e_R} =
	\begin{pmatrix}
	\cos\theta_{12}^{L,R} & \sin\theta_{12}^{L,R} & 0 & 0 \\ 
	-\sin\theta_{12}^{L,R} & \cos\theta_{12}^{L,R} & 0 & 0 \\ 
	0 & 0 & 1 & 0 \\ 
	0 & 0 & 0 & 1
	\end{pmatrix}
	\begin{pmatrix}
	\cos\theta_{14}^{L,R} & 0 & 0 & \sin\theta_{14}^{L,R} \\ 
	0 & 1 & 0 & 0 \\ 
	0 & 0 & 1 & 0 \\ 
	-\sin\theta_{14}^{L,R} & 0 & 0 & \cos\theta_{14}^{L,R}
	\end{pmatrix}
	\begin{pmatrix}
	1 & 0 & 0 & 0 \\ 
	0 & \cos\theta_{24}^{L,R} & 0 & \sin\theta_{24}^{L,R} \\ 
	0 & 0 & 1 & 0 \\ 
	0 & -\sin\theta_{24}^{L,R} & 0 & \cos\theta_{24}^{L,R}
	\end{pmatrix}$}
	\label{eqn:leptonic_mixing_matrices}
\end{equation}
\endgroup
The mixing angle $\theta_{14,24}$ can be expressed in terms of mass parameters written in the Lagrangian of Equation~\ref{eqn:Yukawa_Lagrangian} as follows.
\begingroup
\begin{equation}
	\tan\theta_{14}^L = \frac{x_1^{(L)}\langle\phi_L\rangle}{M_4^L}\,\,,\qquad \tan\theta_{24}^L = \frac{x_2^{(L)}\langle\phi_L\rangle}{\sqrt{\big(x_1^{(L)}\langle\phi_L\rangle\big)^2+\big(M_4^L\big)^2}}
	\label{eqn:angles_from_lag_params}
\end{equation}
\endgroup
The angles defined here take the theory from the interaction basis in Equation \eqref{eqn:Yukawa_Lagrangian} to the mass eigenbasis of primed fields introduced with Equation \eqref{eqn:field_transformations_to_mass_basis}. They directly parameterise the mixing between the 4th, vector-like family and the usual three chiral families of SM fermions. Such mixing parameters will cause the $D'$ matrices from Equation \eqref{eqn:coupling_conversion_to_mass_basis} to become off-diagonal. This incites couplings between the massive $Z'$ vector boson and the SM leptons, suppressed by the mixing angles. These mixing angles can be expressed in terms of parameters from the Lagrangian (Equation \eqref{eqn:Yukawa_Lagrangian}), as per Equation \eqref{eqn:angles_from_lag_params} \cite{King:2017anf}.
\begingroup
\begin{equation}
	\tan\theta_{14}^L = \frac{x_1^{(L)}\langle\phi_L\rangle}{M_4^L}\,\,,\qquad \tan\theta_{24}^L = \frac{x_2^{(L)}\langle\phi_L\rangle}{\sqrt{\big(x_1^{(L)}\langle\phi_L\rangle\big)^2+\big(M_4^L\big)^2}}
	\label{eqn:angles_from_lag_params}
\end{equation}
\endgroup
With the restrictions defined in Equation \eqref{eqn:leptonic_mixing_matrices} and above, all of the relevant couplings between the massive $Z^{\prime}$ and fermions in the mass basis of propagating fields can be determined as the following:
\begin{equation}
\mathcal{L}_{Z^{\prime }}^{gauge}=Z_{\mu }^{\prime } \overline{l}_{L,R}(g_{L,R})_{ll'}\gamma ^{\mu }l'_{L,R}
\end{equation}
where $l,l'=e, \mu, E$, the mass eigenstate leptons electron, muon and vector-like lepton respectively with the following couplings to the massive $Z'$ boson:
\begingroup
\begin{align}
	(g_{L,R})_{\mu\mu} &= g'q_{L_4,e_4}\Big(\cos\theta_{12}^{L,R}\sin\theta_{24}^{L,R}-\cos\theta_{24}^{L,R}\sin\theta_{12}^{L,R}\sin\theta_{14}^{L,R}\Big)^2\label{eqn:mu_mu_zprime_coupling}\\[7pt]
	(g_{L,R})_{ee} &= g'q_{L_4,e_4}\Big(\sin\theta_{12}^{L,R}\sin\theta_{24}^{L,R}+\cos\theta_{12}^{L,R}\cos\theta_{24}^{L,R}\sin\theta_{14}^{L,R}\Big)^2\label{eqn:e_e_zprime_coupling}\\[7pt]
	(g_{L,R})_{EE} &= g'q_{L_4,e_4}\Big(\cos\theta_{14}^{L,R}\Big)^2\Big(\cos\theta_{24}^{L,R}\Big)^2\label{eqn:E_E_zprime_coupling}\\[7pt]
	(g_{L,R})_{eE} &= g'q_{L_4,e_4}\cos\theta_{14}^{L,R}\cos\theta^{L,R}_{24}\Big(\sin\theta_{12}^{L,R}\sin\theta_{24}^{L,R}+\cos\theta_{12}^{L,R}\cos\theta_{24}^{L,R}\sin\theta_{14}^{L,R}\Big)\label{eqn:e_E_zprime_coupling}\\[7pt]
	(g_{L,R})_{\mu E} &= g'q_{L_4,e_4}\cos\theta_{14}^{L,R}\cos\theta_{24}^{L,R}\Big(\cos\theta_{12}^{L,R}\sin\theta_{24}^{L,R}-\cos\theta_{24}^{L,R}\sin\theta_{12}^{L,R}\sin\theta_{14}^{L,R}\Big)\label{eqn:mu_E_zprime_coupling}\\[7pt]
	(g_{L,R})_{\mu e} &= g'q_{L_4,e_4}\Big(\sin\theta_{12}^{L,R}\sin\theta_{24}^{L,R}+\cos\theta_{12}^{L,R}\cos\theta_{24}^{L,R}\sin\theta^{L,R}_{14}\Big) \label{eqn:mu_e_zprime_coupling} \\
	& \times \Big(\cos\theta_{12}^{L,R}\sin\theta_{24}^{L,R}-\cos\theta_{24}^{L,R}\sin\theta_{12}^{L,R}\sin\theta_{14}^{L,R}\Big) 
	 \nonumber
\end{align}
\endgroup
It is important to note that only the first and second family of SM leptons $e,\mu$ couple to the massive $Z^{\prime}$, 
with their non-universal and flavour changing couplings controlled by the mixing angles $\theta^{L,R}_{14}, \theta^{L,R}_{24}$
with the vector-like family. Throughout the remainder of this work, we assume that $g'q_{L4,e4}=1$ for simplicity.

\subsection{Muon decay to electron plus photon}
In this subsection we study charged lepton flavor violating process $\mu\to e\gamma$ in the context of our BSM scenario. It is worth mentioning that a future observation of the $\mu\to e\gamma$ decay will be indisputable evidence of physics beyond the SM . The SM does predict non-zero branching ratios for the processes $\mu\to e\gamma$, $\tau\to\mu\gamma$ and $\tau\to e\gamma$, but such predictions are several orders of magnitude below projected experimental sensitivities \cite{Lindner:2016bgg,Calibbi:2017uvl}. The $\mu\rightarrow e\gamma$ decay rate is enhanced with respect to the SM by additional contributions due to virtual $Z^\prime$ and charged exotic lepton exchange at the one-loop level. General $l_{i}\rightarrow l_{j}\gamma$ decay can be described by the following effective operator \cite{Lindner:2016bgg}: 
\begingroup
\begin{equation}
	\mathcal{L}_{EFT}=\frac{\mu _{ij}^{M}}{2}\overline{l}_{i}\sigma ^{\mu \nu}l_{j}F_{\mu \nu }+\frac{\mu_{ij}^{E}}{2}i\overline{l}_{i}\gamma ^{5}\sigma^{\mu \nu }l_{j}F_{\mu \nu }
	\label{eqn:anomalous_moments_effective_lagrangian}
\end{equation}
\endgroup
where $F_{\mu\nu}$ denotes the electromagnetic field strength tensor, $\mu _{ij}^{E}$ and $\mu _{ij}^{M}$ are the transition electric and magnetic moments, respectively and $i,j=1,2,3$ denote family indices. Diagonal elements in the transition magnetic moment $\mu _{ij}^{M}$ give rise to the anomalous dipole moments $\Delta a_{l}=\frac{1}{2}(g_{l}-2)$ of leptons, whilst off-diagonal elements in the transition moments contribute to the $l_{i}\rightarrow l_{j}\gamma $ decay amplitude. Based on the effective Lagrangian in Equation \eqref{eqn:anomalous_moments_effective_lagrangian}, one has that the amplitude for a generic lepton decay $f_{1}\rightarrow f_{2}\gamma $ has the form \cite{Lavoura:2003xp}: 
\begingroup
\begin{equation}
	\mathcal{A} =\, e\varepsilon _{\mu }^{\ast }(q)\overline{v }_{2}(p_{2})\left[i\sigma^{\mu\nu}q_{\nu}(\sigma_{L}P_{L}+\sigma_{R}P_{R})\right]u_{1}(p_{1})	\label{eqn:LFV_decay_amplitude}
\end{equation}
\endgroup
where $\sigma _{L}$ and $\sigma _{R}$ are numerical quantities with dimension of inverse mass that can be expressed in terms of loop integrals \cite{Lavoura:2003xp}. $u_1$ and $v_2$ are spinors, furthermore, we have the following relations: 
\begingroup
\begin{equation}
	\begin{gathered}
		\sigma ^{\mu \nu } =\frac{i}{2}\left[\gamma^{\mu},\gamma^{\nu}\right],\hspace{1.5cm}P_{L,R}=\frac{1}{2}(1\mp\gamma_{5}),\hspace{ 1.5cm}q=p_{1}-p_{2} 
	\end{gathered}  
	\label{eqn:operators_and_projectors}
\end{equation}
\endgroup
In such a general case, the decay rate expression for the $\mu\to e\gamma$ process is the following \cite{Lavoura:2003xp,Chiang:2011cv,Lindner:2016bgg,Raby:2017igl}:
\begingroup
\begin{equation}
	\Gamma(\mu\rightarrow e\gamma)=\frac{\alpha_{em}}{1024\pi^{4}}\frac{m_{\mu }^{5}}{M_{Z^{\prime }}^{4}}(\left\vert \widetilde{\sigma }_{L}\right\vert ^{2}+\left\vert \widetilde{\sigma }_{R}\right\vert^{2})
	\label{eqn:mu_e_gamma_decay_rate_prediction}
\end{equation}
\endgroup
where $\widetilde{\sigma }_{L}$ and $\widetilde{\sigma }_{R}$ are given by: 
\begingroup
\begin{align}
	\begin{split}
	\widetilde{\sigma }_{L}& =\sum_{a=e,\mu,E}\left[(g_{L})_{ea}(g_{L})_{a\mu}F(x_{a})+\frac{m_{a}}{m_{\mu }}(g_{L})_{ea}(g_{R})_{a\mu}G(x_{a})\right] , \\
	\widetilde{\sigma }_{R}& =\sum_{a=e,\mu,E}\left[ (g_{R})_{ea}(g_{R})_{a\mu}F(x_{a})+\frac{m_{a}}{m_{\mu }}(g_{R})_{ea}(g_{L})_{a\mu}G(x_{a})\right],\hspace{1.5cm}x_{a}=\frac{m_{a}^{2}}{M_{Z^{\prime }}^{2}}
	\label{eqn:contributions_to_muegamma_(sigmas)}
	\end{split}
\end{align}
\endgroup
$F(x)$ and $G(x)$ are loop functions related to the Feynman diagrams for $\mu\rightarrow e\gamma$ as per Figure \ref{fig:muegamma_feynman_diagrams}, and have the functional form given in Equation \eqref{eqn:loop_functions}. $g_{L,R}$ are couplings in the fermion mass basis, as detailed in Equations \eqref{eqn:mu_mu_zprime_coupling} through \eqref{eqn:mu_e_zprime_coupling}. $m_a$ here corresponds to the full propagating mass of the vector-like partners. In the approximation where the vector like mass $M_4^L$ is always much greater than the chirality-flipping mass $M_4^C$ ($M_4^L\gg M_4^C$) that we will adopt here, this full propagating mass is almost equivalent to the vector-like mass. Therefore when $a=E$, we approximate $m_E\simeq M_4^L$. The loop functions are given by \cite{Raby:2017igl}: 
\begingroup
\begin{equation}
	\begin{gathered} 
		F(x)=\frac{5x^{4}-14x^{3}+39x^{2}-38x-18x^{2}\ln x+8}{ 12(1-x)^{4}}, \\ G(x)=\frac{x^{3}+3x-6x\ln x-4}{2(1-x)^{3}} 
	\end{gathered}  
\label{eqn:loop_functions}
\end{equation}
\endgroup
Equation \eqref{eqn:mu_e_gamma_decay_rate_prediction} has some generic features; the loop function $F(x)$ varies between 0.51 and 0.67 when $x$ is varied in the range $10^{-3}\le x \le 2$, whilst in the same region, $G(x)$ varies between -1.98 and -0.84. Consequently, in the case of charged fermions running in loops, contributions proportional to $G(x)$ will likely dominate over those proportional to $F(x)$. The dominant contributions involve left-right and right-left $Z^{\prime }$ couplings, whereas the subleading ones include either left-left or right-right couplings. Dividing Equation \eqref{eqn:mu_e_gamma_decay_rate_prediction} by the known decay rate of the muon yields a prediction for the $\mu\rightarrow e\gamma$ branching fraction \cite{Lavoura:2003xp,Chiang:2011cv,Lindner:2016bgg,Raby:2017igl}: 
\begingroup
\addtolength{\jot}{5pt}
\begin{align}
	\begin{split}
		\operatorname{BR}(\mu \rightarrow e \gamma) &= \frac{\alpha}{1024\pi^4} \frac{m_{\mu}^5}{M_{Z^{\prime}}^4 \Gamma_{\mu}} \Bigg[ \Big\vert(g_L)_{\mu\mu}(g_L)_{\mu e} F(x_\mu) + (g_L)_{\mu E} (g_L)_{eE} F(x_{E})+ (g_L)_{\mu e} (g_L)_{e e} F(x_e) \\
		&+ \frac{m_{\mu}}{m_{\mu}}(g_L)_{\mu e} (g_R)_{\mu \mu} G(x_\mu) + \frac{M_4^C}{m_{\mu}}(g_L)_{eE} (g_R)_{\mu E} G(x_{E}) + \frac{m_e}{m_{\mu}}(g_L)_{e e} (g_R)_{\mu e} G(x_e) \Big\vert^2 \\
		&+ \Big\vert (g_R)_{\mu \mu} (g_R)_{\mu e} F(x_\mu) + (g_R)_{\mu E} (g_R)_{eE} F(x_{E}) + (g_R)_{\mu e} (g_R)_{e e} F(x_e)\\
		&+\frac{m_{\mu}}{m_{\mu}}(g_R)_{\mu e} (g_L)_{\mu \mu} G(x_\mu)+ \frac{M_4^C}{m_{\mu}}(g_R)_{eE} (g_L)_{\mu E} G(x_{E}) + \frac{m_e}{m_{\mu}}(g_R)_{e e} (g_L)_{\mu e} G(x_e) \Big\vert^2 \Bigg]
		\label{eqn:mu_e_gamma}
	\end{split}
\end{align}
\endgroup
\begingroup
\begin{figure}[tbp]
	\centering
	\includegraphics[width=\textwidth]{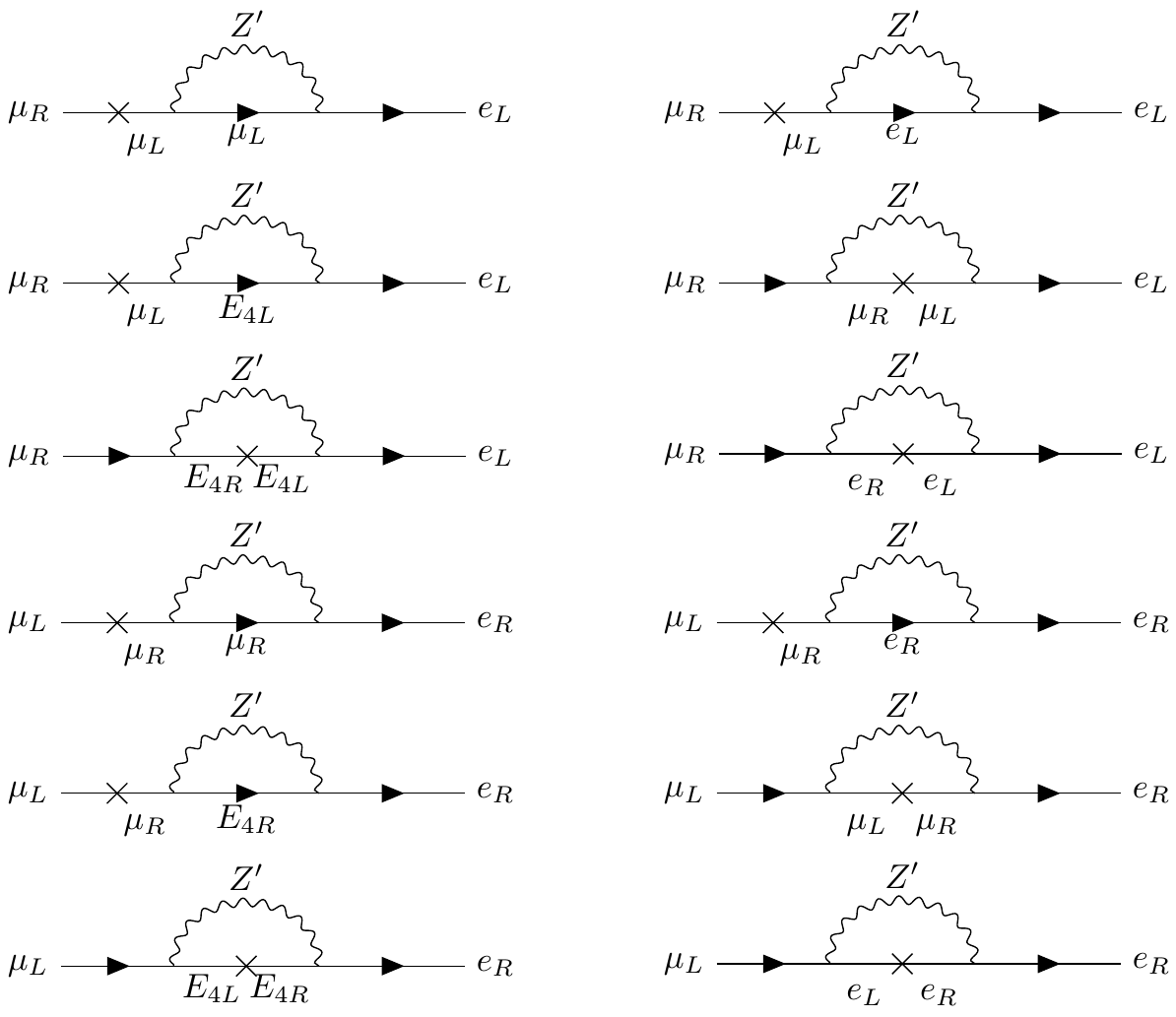}
	\caption{Feynamn diagrams contributing to the $\mu \rightarrow e\gamma $ decay. Note that 
	these diagrams all rely on a chirality flipping mass (LR). Where the chirality flip involves the fourth family, 
	the relevant mass is $M_4^C$.}
	\label{fig:muegamma_feynman_diagrams}
\end{figure}
\endgroup
where the total muon decay width is $\Gamma _{\mu }=\frac{G_{F}^{2}m_{\mu}^{5}}{192\pi ^{3}}=3\times 10^{-19}\text{GeV}$. The mass $M_4^C$ that appears in the Feynman diagrams  with a chirality flip on the 4th family fermions $E_4$ (Figure \ref{fig:muegamma_feynman_diagrams}, 5th and 11th diagrams) is \emph{not} the vector-like mass, but instead arises from the Yukawa-like couplings from Equation \eqref{eqn:Yukawa_Lagrangian}, $M_4^C=y^{(e)}_{44}v_{\phi}$, where $v_{\phi}$ is the vacuum expectation value of the SM Higgs field, which acquires a vev and spontaneously breaks electroweak symmetry in the established manner. Under the assumption that $M_4^C>m_\mu$, such terms proportional to the chirality flipping mass in Equation \eqref{eqn:mu_e_gamma} give by far the largest contributions to $\mu\rightarrow e\gamma$. The experimental limit on $\operatorname{BR}(\mu\rightarrow e\gamma)$ is determined from non-observation at the MEG experiment at a 90\% confidence level\cite{TheMEG:2016wtm,Tanabashi:2018oca}:
\begingroup
\begin{equation}
\operatorname{BR}(\mu\rightarrow e\gamma) < 4.2\times10^{-13}
\label{eqn:muegamma_limit}
\end{equation}
\endgroup

\subsection{Anomalous magnetic moment of the muon $\Delta a_\mu$}
In this subsection we study the muon anomalous magnetic moment in the context of our BSM scenario. In a model such as this, the Feynman diagrams for $\mu\rightarrow e\gamma$ are easily modified to give contributions to the anomalous magnetic moment of the muon as per Figure \ref{fig:muong2_feynman_diagrams}. The prediction for such an observable in our model therefore takes the form \cite{Raby:2017igl}:
\begingroup
\addtolength{\jot}{5pt}
\begin{align}
	\begin{split}
		\Delta a_{\mu}^{Z^{\prime}} &= -\frac{m_{\mu}^2}{8\pi^2 M_{Z^{\prime}}^2} \bigg[ \big( \vert (g_L)_{\mu\mu} \vert^2 + \vert (g_R)_{\mu\mu} \vert^2 \big) F(x_{\mu}) 
		+ \big( \vert (g_L)_{\mu E} \vert^2 + \vert (g_R)_{\mu E} \vert^2 \big) F(x_{E}) \\
		&+ \big( \vert (g_L)_{\mu e} \vert^2 + \vert (g_R)_{\mu e} \vert^2 \big) F(x_{e}) 
		+ \operatorname{Re}\big( (g_L)_{\mu \mu} (g_R^*)_{\mu \mu} \big) G(x_{\mu}) \\
		&+ \operatorname{Re}\big( (g_L)_{\mu E} (g_R^*)_{\mu E} \big) \frac{M_4^C}{m_{\mu}} G(x_{E}) 
		+ \operatorname{Re}\big( (g_L)_{\mu e} (g_R^*)_{\mu e} \big) \frac{m_{e}}{m_{\mu}} G(x_{e}) \bigg]
	\label{eqn:muon_g-2_contributions}
	\end{split}
\end{align}
\endgroup
\begingroup
\begin{figure}[tbp]
	\centering
	\includegraphics[width=\textwidth]{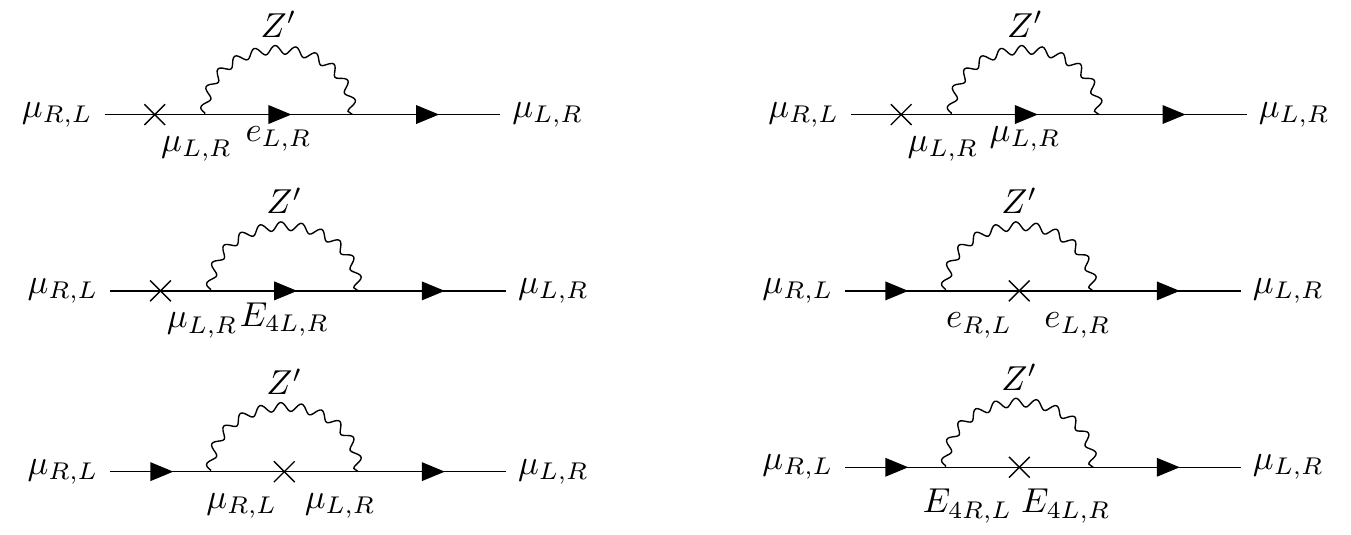}
	\caption{Feyman diagrams contributing to the muon $(g-2)_{\mu}$}
	\label{fig:muong2_feynman_diagrams}
\end{figure}
\endgroup
Once more, the dominant terms will be those proportional to the enhancement factor of $\frac{M_4^C}{m_\mu}$, corresponding to the final diagram in Figure \ref{fig:muong2_feynman_diagrams}, provided $M_4^C>m_\mu$. Recent experimental evidence has shown that the muon magnetic moment as measured by the E821 experiment at BNL is at around a 3.5$\sigma$ deviation from the SM prediction\cite{Hagiwara:2011af,Nomura:2018lsx,Nomura:2018vfz,Bennett:2006fi,Davier:2010nc,Davier:2017zfy,Davier:2019can}: 
\begingroup
\begin{equation}
	(\Delta a_{\mu })_{\exp}=(26.1\pm 8)\times10^{-10}
	\label{eqn:experimental_muon_g-2_data}
\end{equation}
\endgroup
\subsection{Anomalous magnetic moment of the electron $\Delta a_{e}$}
Analogously to the muon, there is also an amendment to the electron $(g-2)_e$ in this scenario, from Feynman diagrams given in Figure \ref{fig:electrong-2_feynman_diagrams}. The analytic expression for $\Delta a_e$ is the following \cite{Raby:2017igl}:
\begingroup
\addtolength{\jot}{5pt}
\begin{align}
	\begin{split}
		\Delta a_{e}^{Z^{\prime}} &= -\frac{m_{e}^2}{8\pi^2 M_{Z^{\prime}}^2} \bigg[ \big( \vert (g_L)_{ee} \vert^2 + \vert (g_R)_{ee} \vert^2 ) F(x_{e}) 
		+ \big( \vert (g_L)_{e \mu} \vert^2 + \vert (g_R)_{e \mu} \vert^2 \big) F(x_{\mu}) \\
		&+ \big( \vert (g_L)_{eE} \vert^2 + \vert (g_R)_{eE} \vert^2 \big) F(x_{E}) 
		+ \operatorname{Re}\big( (g_L)_{ee} (g_R^*)_{ee} \big) \frac{m_{e}}{m_{e}} G(x_{e}) \\
		&+ \operatorname{Re}\big( (g_L)_{e \mu} (g_R^*)_{e \mu} \big) \frac{m_{\mu}}{m_{e}} G(x_{\mu}) 
		+ \operatorname{Re}\big( (g_L)_{eE} (g_R^*)_{eE} \big) \frac{M_4^C}{m_{e}} G(x_{E}) \bigg]
	\end{split}
	\label{eqn:electron_g-2_contributions}
\end{align}
\endgroup
\begingroup
\begin{figure}[tbp]
	\centering
	\includegraphics[width=\textwidth]{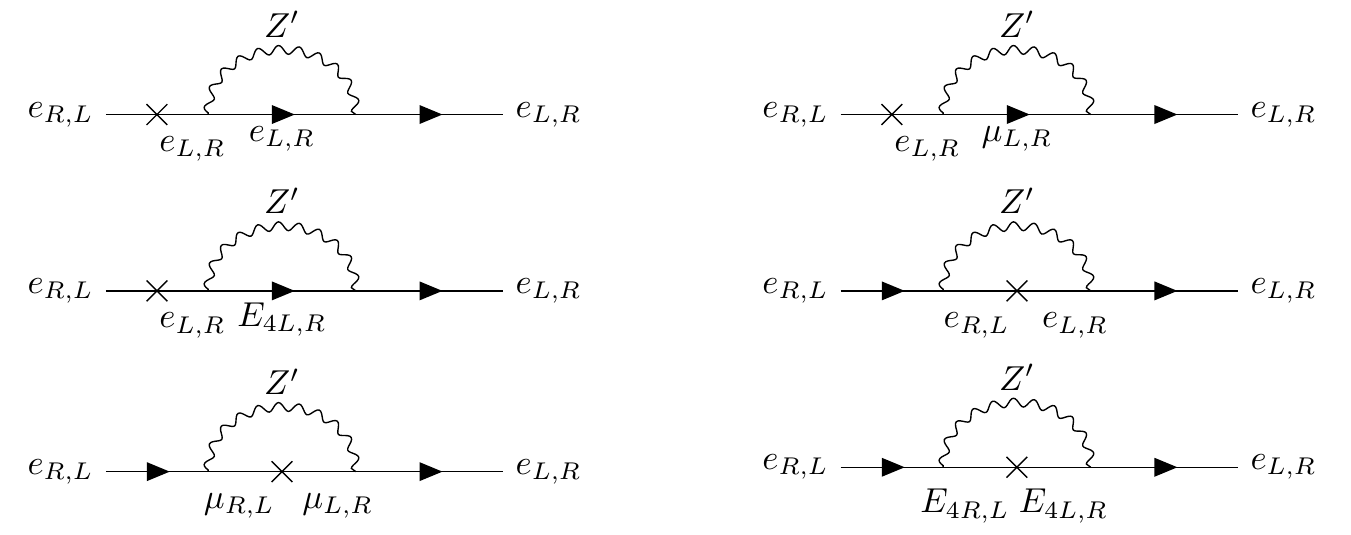}
	\caption{Feynamn diagrams contributing to the electron $g-2$}
	\label{fig:electrong-2_feynman_diagrams}
\end{figure}
\endgroup
As per the muon moment, if $M_4^C>m_\mu$ the largest contribution to the electron moment will be the final term in Equation \eqref{eqn:electron_g-2_contributions}, corresponding to the last diagram in Figure \ref{fig:electrong-2_feynman_diagrams}. The most recent experimental result of the $(g-2)_e$, obtained from measurement of the fine structure constant of QED, shows a $2.5\sigma$ deviation from the SM, similarly to the muon magnetic moment\cite{Parker:2018vye}: 
\begingroup
\begin{equation}
	(\Delta a_e)_{\text{exp}} = (-0.88\pm0.36)\times 10^{-12}
	\label{eqn:experimental_electron_g-2_data}
\end{equation}
\endgroup
Notice especially that Equations \eqref{eqn:experimental_muon_g-2_data} and \eqref{eqn:experimental_electron_g-2_data} have deviations from the SM in opposite directions, therefore explaining both phenomena simultaneously can be difficult for a given model to achieve.

\subsection{Neutrino trident production}
So-called trident production of neutrinos by process $\nu_{\mu}\gamma^{\ast}\rightarrow\nu_{\mu}\mu^{+}\mu^{-}$ through nuclear scattering is also relevant. The Feynamn diagram contributing to neutrino trident production in our model is shown in Figure \ref{fig:trident_feynman_diagrams}. This process constrains the following effective four lepton interaction, which in this scenario arises from leptonic $Z^{\prime}$ interactions \cite{Geiregat:1990gz,Mishra:1991bv,Altmannshofer:2014pba}: 
\begingroup
\begin{equation}
	\Delta\mathcal{L}_{eff}\supset-\frac{(g_L)_{\mu\mu}^{2}}{2M_{Z^{\prime}}^{2}}(\overline{\mu}_{L}\gamma^{\lambda}\mu_{L})(\overline{\nu}_{\mu L}\,\gamma_{\lambda}\,\nu_{\mu L}) -\frac{(g_R)_{\mu\mu}(g_L)_{\mu\mu}}{2M_{Z^{\prime}}^{2}}(\overline{\mu}_{R}\gamma^{\lambda}\mu_{R})(\overline{\nu}_{\mu L}\,\gamma_{\lambda}\,\nu_{\mu L})
	\label{eqn:trident_effective_Lagrangian}
\end{equation}
\endgroup
Said coupling is constrained as in the $SU(2)_L$ symmetric SM, left-handed muons and left-handed muon neutrinos couple identically to the $Z'$ vector boson. Experimental data on neutrino trident production $\nu_{\mu}\gamma^{\ast}\rightarrow\nu_{\mu}\mu^{+}\mu^{-}$ yields the following constraint at $95\%$ CL \cite{Falkowski:2017pss}: 
\begingroup
\begin{figure}[tbp]
	\centering
	\includegraphics[width=\textwidth]{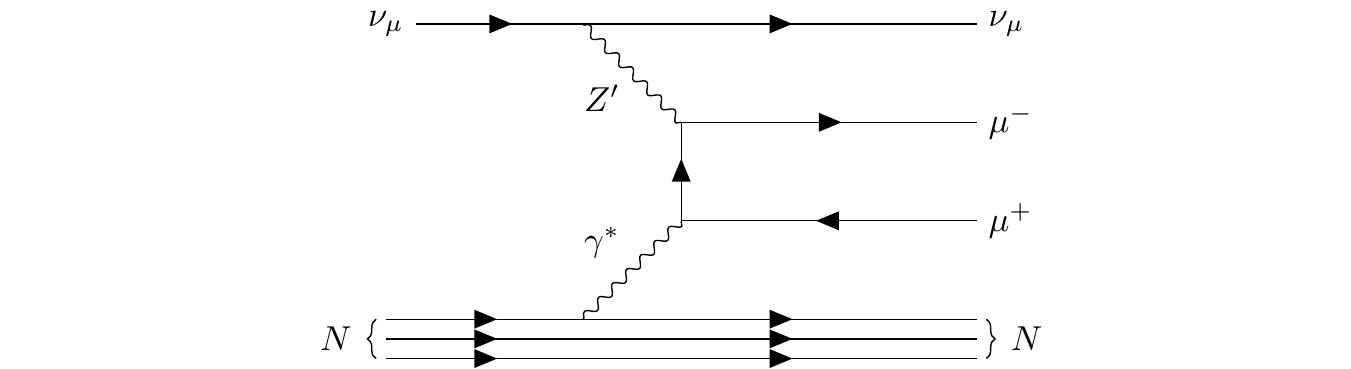}
	\caption{Feynamn diagram contributing to neutrino trident production, $N$ denotes a nucleus.}
	\label{fig:trident_feynman_diagrams}
\end{figure}
\endgroup
\begingroup
\begin{equation}
	-\frac{1}{(390\text{GeV})^{2}}\,\lesssim\frac{(g_L)_{\mu\mu}^{2}+(g_L)_{\mu\mu}(g_R)_{\mu\mu}}{M_{Z^{\prime}}^{2}}\lesssim\frac{1}{(370\text{GeV})^{2}}
	\label{eqn:trident_constraint_on_effective_coupling}
\end{equation}
\endgroup
This limit can be applied to the model's parameter space in a similar manner to other CLFV constraints discussed previously.

\section{Analytic arguments for $(g-2)_\mu$, $(g-2)_e$ and $\operatorname{BR}(\mu\rightarrow e\gamma)$}
\label{sec:analytics}
In order to gain an analytic understanding of the interplay between $(g-2)_\mu$, $(g-2)_e$ and $\operatorname{BR}(\mu\rightarrow e\gamma)$, in this section we shall make some simplifying assumptions about the parameters appearing in Equations \eqref{eqn:muon_g-2_contributions}, \eqref{eqn:electron_g-2_contributions} and \eqref{eqn:mu_e_gamma}.
If we assume large fourth family chirality flipping masses
$M_4^C\gg m_\mu$, then the expressions for these phenomena reduce to a minimal number of terms, all proportional to $M_4^C$. Furthermore, we assume that left- and right- handed couplings are related by some real, positive constants $k_1$ and $k_2$ defined thus:
\begingroup
\begin{align}
\begin{split}
(g_L)_{\mu E} = g_{\mu E}, \quad&\quad (g_R)_{\mu E} = k_1g_{\mu E},\\
(g_L)_{eE} = g_{eE}, \quad&\quad (g_R)_{eE} = -k_2g_{eE}
\label{eqn:analytic_coupling_assumptions}
\end{split}
\end{align}
\endgroup
The final coupling in Equation \eqref{eqn:analytic_coupling_assumptions} is defined with a sign convention such that, seeing as it is known numerically that the $G$ loop function is always negative, we automatically recover the correct signs for all of our observables. We also define the following prefactor constants to further simplify our expressions:
\begingroup
\begin{gather}
C_1 = \frac{\alpha}{1024\pi^2}\frac{m_\mu^5}{M_{Z'}^4\Gamma_\mu}\; ,\quad C_2 = \frac{m_\mu^2}{8\pi^2M_{Z'}^2}\; ,\quad C_3 = \frac{m_e^2}{8\pi^2M_{Z'}^2}
\label{eqn:simplified_prefactor_constants}
\end{gather}
\endgroup
Under such assumptions, Equations \eqref{eqn:muon_g-2_contributions}, \eqref{eqn:electron_g-2_contributions} and \eqref{eqn:mu_e_gamma} reduce to the following:
\begingroup
\begin{gather}
\operatorname{BR}(\mu\rightarrow e\gamma) = C_1\Bigg(\Big\lvert\frac{M_4^C}{m_\mu}k_1g_{eE}g_{\mu E}G(x_E)\Big\rvert^2+\Big\lvert\frac{M_4^C}{m_\mu}k_2g_{eE}g_{\mu E}G(x_E)\Big\rvert^2\Bigg)
\label{eqn:simplified_mu_e_gamma_expression}
\\[2ex]
\operatorname{|\Delta a_\mu|} = C_2k_1g_{\mu E}^2\frac{M_4^C}{m_\mu}|G(x_E)|
\label{eqn:simplified_muong2_expression}
\\[2ex]
\operatorname{|\Delta a_e|} = C_3k_2g_{eE}^2\frac{M_4^C}{m_e}|G(x_E)|
\label{eqn:simplified_electrong2_expression}
\end{gather}
\endgroup
We can then invert Equations \eqref{eqn:simplified_muong2_expression} and \eqref{eqn:simplified_electrong2_expression} to obtain expressions for the couplings in terms of the observables as per Equation \eqref{eqn:couplings_from_observables}.
\begingroup
\begin{gather}
g_{\mu E} = \sqrt{\frac{|\Delta a_\mu|}{C_2k_1}\frac{1}{|G(x_E)|}\frac{m_\mu}{M_4^C}}\;,\quad\quad g_{eE} = \sqrt{\frac{|\Delta a_e|}{C_3k_2}\frac{1}{|G(x_E)|}\frac{m_e}{M_4^C}}
\label{eqn:couplings_from_observables}
\end{gather}
\endgroup
Substituting into the flavour violating muon decay in Equation \eqref{eqn:simplified_mu_e_gamma_expression} and expanding the constants defined earlier yields:
\begingroup
\begin{align}
	\operatorname{BR}(\mu\rightarrow e\gamma) = \frac{\alpha\pi^2}{16}\frac{(k_1^2+k_2^2)}{k_1k_2}|\Delta a_\mu||\Delta a_e|\frac{m_\mu^2}{\Gamma_\mu m_e}
	\label{eqn:really_simplified_mu_e_gamma_expression}
\end{align}
\endgroup
independently of $M_{Z'}$ and $M_4^C$ which cancel.
Rearranging Equation \ref{eqn:really_simplified_mu_e_gamma_expression} and setting the physical quantities 
$|\Delta a_\mu|$, $|\Delta a_e|$ equal to their desired 
central values, yields a simple condition on $r=k_1/k_2$ in order to satisfy the bound on 
$\operatorname{BR}(\mu\rightarrow e\gamma)$:
\begingroup
\begin{align}
\| r+\frac{1}{r} \|< 5.57\times10^{-10}
\end{align}
\endgroup
Since the left hand side is minimised for $r=1$, 
the bound on $\operatorname{BR}(\mu\rightarrow e\gamma)$
can never be satisfied while accounting for $(g-2)_\mu$, $(g-2)_e$
(although clearly it is possible to satisfy it with either $(g-2)_\mu$ or $(g-2)_e$ but not both).
However this conclusion is based on the assumption that the physical quantities are dominated by the diagrams involving the 
chirality flipping fourth family masses $M_4^C\gg m_\mu$. In order to relax this assumption, a more complete 
analysis of the parameter space is required, one that considers all relevant terms in our expressions for observables in a numerical exploration of the parameter space. Such investigations are detailed in Section \ref{sec:statistical_analysis}.

\section{Numerical Analysis of the Fermiophobic $Z'$ Model}
\label{sec:statistical_analysis}
Given the expressions for observables that we have outlined above, we use these phenomena to constrain the parameter space of the model. As mentioned, a minimal parameter space is considered here, limiting mixing to the lepton sector and omitting the third chiral family from any mixing. From coupling expressions in Section \ref{sec:LFV_in_this_model}, the angular mixing parameters such as $\theta_{24L}$ and particle masses form a minimal parameter space for this model. We set direct mixing between the electron and muon ($\theta_{12L,R}$) to be vanishing for all tests, as even small direct mixing can easily violate the strict MEG constraint on $\operatorname{BR}(\mu\rightarrow e\gamma)$.

\subsection{Anomalous muon magnetic moment}
Initially, we focus on the longest-standing anomaly, that of $(g-2)_\mu$. We first utilise a simple parameter space, as we require only mixing between the muon and vector-like lepton fields. To keep the analysis in a region potentially testable by upcoming future experiments, we take a vector-like fourth family lepton mass of $M_4^L=1\text{TeV}$ and a chirality-flipping fourth family mass of $M_4^C=200\text{GeV}$
(as discussed earlier we make a distinction between these two sources of mass). The smaller value of $M_4^C$
is well motivated by the need for perturbativity in Yukawa couplings, as the SM Higgs vev is 176GeV, since $M_4^C$ is proportional to the Higgs vev. For this investigation, the parameter space under test is detailed in Table  \ref{tab:parameters_for_muong-2_only_test}.
\begingroup
\setlength{\tabcolsep}{30pt}
\begin{table}[H]
	\centering
	\renewcommand{\arraystretch}{1.2}
	\begin{tabular}{l c}
		\toprule
		\toprule
		\textbf{Parameter} & \textbf{Value/Scanned Region}\\
		\midrule
		$M_{Z'}$ & $50\rightarrow1000$ GeV \\
		$M_4^C$ & $200$ GeV \\
		$M_4^L$ & $1000$ GeV \\
		$\sin^2\theta_{12L,R}$ & $0.0$ \\
		$\sin^2\theta_{14L}$ & $0.0$ \\
		$\sin^2\theta_{14R}$ & $0.0$ \\
		$\sin^2\theta_{24L,R}$ & $0.0\rightarrow 1.0$ \\
		\bottomrule
		\bottomrule
	\end{tabular}
	\caption{Explored parameter space for muon $g-2$ test.}
	\label{tab:parameters_for_muong-2_only_test}
\end{table}
\endgroup
Within the stated parameter space, expressions for the observables under test are simplified considerably, and with fixed $M_4^C$ and $M_4^L$ we constrain the space in terms of the three variables $\sin^2\theta_{24L}$, $\sin^2\theta_{24R}$ and $M_{Z'}$, as shown in Figure \ref{fig:muong-2_and_trident}. Note that, as $\theta_{12L,R}$ and $\theta_{14L,R}$ are set vanishing, contributions to $(g-2)_e$ and $\operatorname{BR}(\mu\rightarrow e\gamma)$ are necessarily vanishing, as can be readily seen from Equations \eqref{eqn:electron_g-2_contributions} and \eqref{eqn:mu_e_gamma}. The dominant contribution to $(g-2)_\mu$ under these assumptions is shown in the final Feynman diagram in Figure \ref{fig:muong2_feynman_diagrams}, that with the enhancement factor of $M_4^C/m_\mu$.
\begingroup
\begin{figure}[H]
	\centering
	\begin{subfigure}{0.48\textwidth}
		\includegraphics[width=1.0\textwidth]{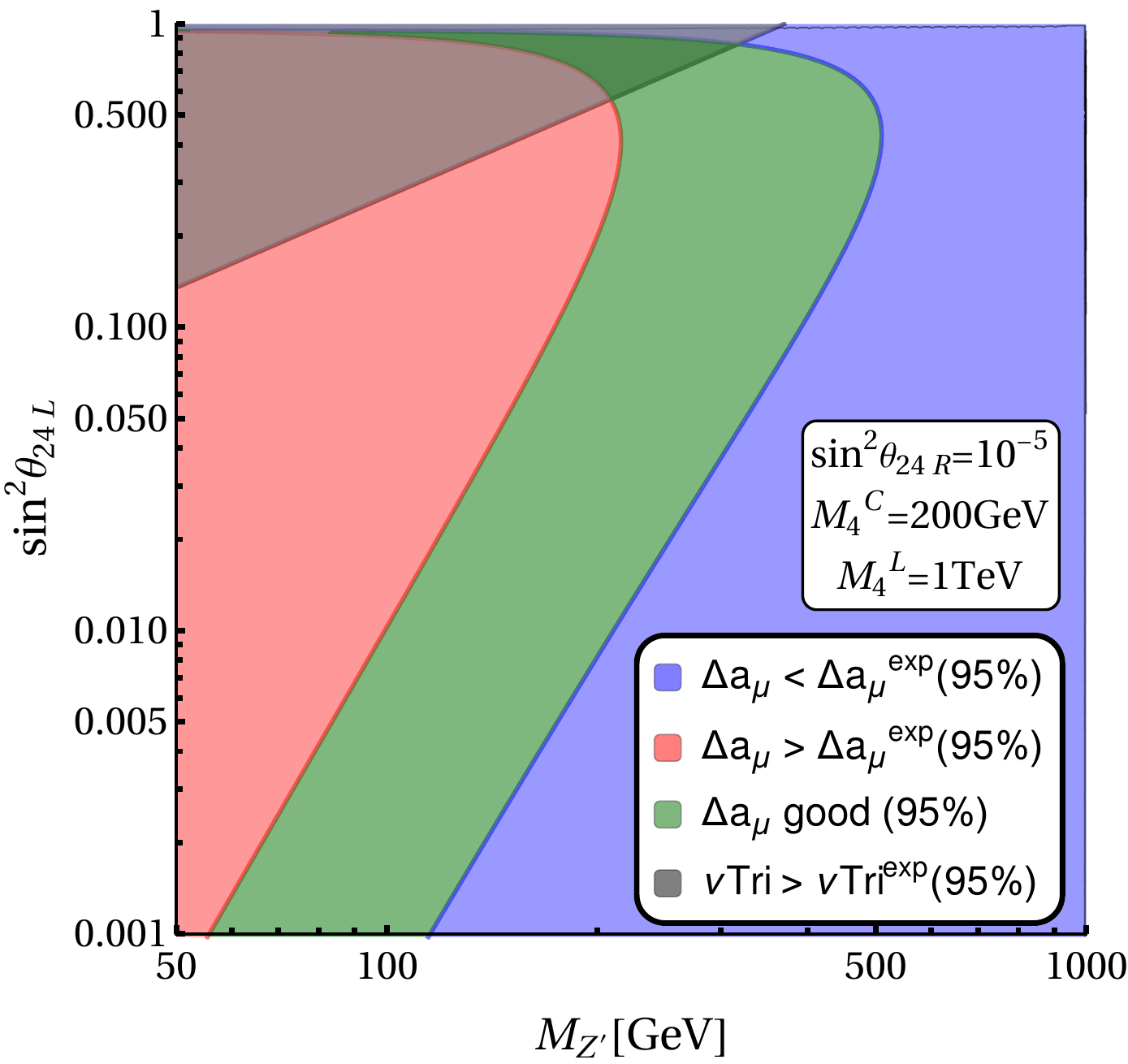}
		\captionsetup{width=0.8\linewidth}
		\caption{Trident exclusion and regions of $\Delta a_\mu$, with a fixed $\sin^2\theta_{24R}$.}
		\label{fig:muong-2_and_trident1}
	\end{subfigure}
	\begin{subfigure}{0.48\textwidth}
		\includegraphics[width=0.96\textwidth]{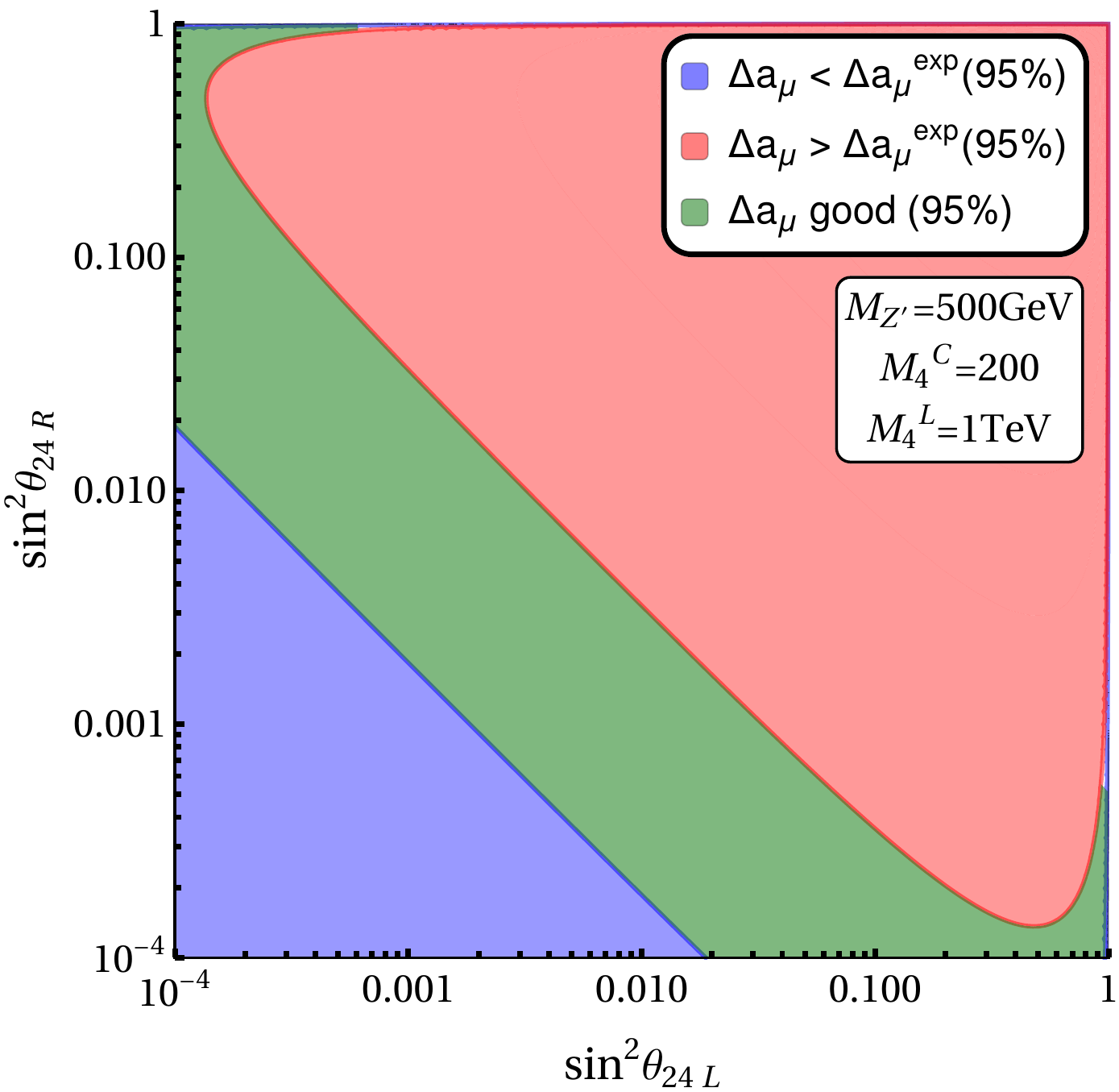}
		\captionsetup{width=0.8\linewidth}
		\caption{$\Delta a_\mu$ in angular parameter space with fixed $Z'$ mass}
		\label{fig:muong-2_and_trident2}
	\end{subfigure}
	\captionsetup{width=0.8\linewidth}
	\caption{Constraints in the $M_{Z'}$, $\sin^2\theta_{24L}$ and $\sin^2\theta_{24R}$ parameter space, mixing between the electron and vector-like lepton switched off. Note we will discuss the $Z^{\prime}$ experimental bound in a next subsection.}
	\label{fig:muong-2_and_trident}
\end{figure}
\endgroup
The legend in Figure \ref{fig:muong-2_and_trident} shows the constraint from neutrino trident production as `$\nu\text{Tri}$' for brevity. Using only mixing between the muon and the vector-like lepton, it is not possible to predict a value for the electron $g-2$ consistent with the observed value as the electron-$Z'$ coupling does not exist. In order to recover this, we must consider mixing of the vector-like lepton with the electron, detailed in the following subsection.

\subsection{Anomalous electron magnetic moment}
Here we concentrate on the $(g-2)_e$. In order to test this observable alone, we investigate only mixing between the electron and vector-like lepton, and ignore any muon contributions. The region of parameter space under test is given in Table \ref{tab:parameters_for_electrong-2_only_test}, note also that mixing with the right-handed electron field is not required to obtain a good prediction.
\begingroup
\setlength{\tabcolsep}{30pt}
\begin{table}[H]
	\centering
	\renewcommand{\arraystretch}{1.2}
	\begin{tabular}{lc}
		\toprule
		\toprule
		\textbf{Parameter} & \textbf{Value/Scanned Region}\\
		\midrule
		$M_{Z'}$ & $50\rightarrow1000$ GeV \\
		$M_4^C$ & $200$ GeV \\
		$M_4^L$ & $1000$ GeV \\
		$\sin^2\theta_{12L,R}$ & $0.0$ \\
		$\sin^2\theta_{14L}$ & $0.0\rightarrow1.0$ \\
		$\sin^2\theta_{14R}$ & $0.0$ \\
		$\sin^2\theta_{24L,R}$ & $0.0$ \\
		\bottomrule
		\bottomrule
	\end{tabular}
	\caption{Explored parameter space for electron $g-2$ test.}
	\label{tab:parameters_for_electrong-2_only_test}
\end{table}
\endgroup
In Figure \ref{fig:electron_g-2_only}, we colour the electron $g-2$ being greater than the observed value (i.e. `less negative' than the experimental data) as the blue region, as such values are more SM-like. Blue regions therefore ameliorate the SM's tension with the experimental data but do not fully resolve it.
\begingroup
\begin{figure}[H]
	\centering
	\includegraphics[width=0.8\textwidth]{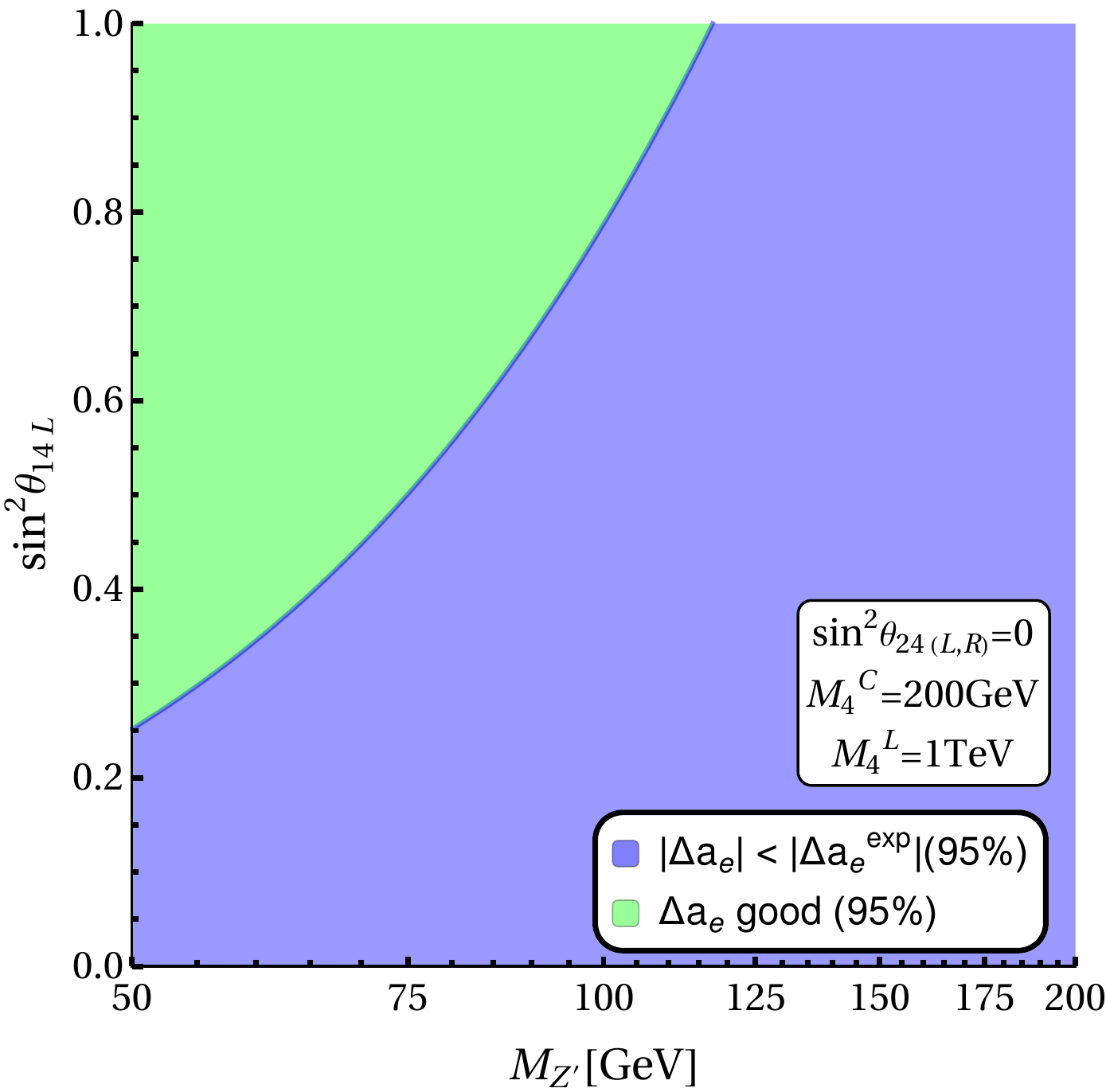}
	\captionsetup{width=0.8\linewidth}
	\caption{$\Delta a_e$ impact on $\sin^2\theta_{14L}$, $M_{Z'}$ parameter space, mixing between the muon and vector-like lepton switched off. Note we will discuss the $Z^{\prime}$ experimental bound in a next subsection.}
	\label{fig:electron_g-2_only}
\end{figure}
\endgroup
Similarly to the preceeding section, because there are no couplings between the electron and the muon (even at the loop level), there are no contributions to the CLFV decay $\mu\rightarrow e\gamma$. Similarly, there are no amendments to the SM expressions for the muon $g-2$ or neutrino trident decay. From this analysis one can conclude that only through using mixing between both muons and electrons with the vector-like leptons is it possible to simultaneously predict observed values of both the anomalous magnetic moments.

\subsection{Attempt to explain both anomalous moments}
In an attempt satisfy all constraints simultaneously, we set specific values for $M_{Z'}$, $M^C_4$ and $\sin^2\theta_{14}^L$ that inhabit allowed regions of parameter space in Figures \ref{fig:muong-2_and_trident1}, \ref{fig:muong-2_and_trident2} and \ref{fig:electron_g-2_only}, then scan through angular mixing parameters as before. The investigated region is summarised in Table \ref{tab:parameters_for_first_attempt_scan}. The choice of $Z'$ mass here is motivated by studying the regions of Figures \ref{fig:muong-2_and_trident} and \ref{fig:electron_g-2_only} that admit muon and electron $(g-2)$s respectively.
\begingroup
\setlength{\tabcolsep}{30pt}
\begin{table}[H]
	\centering
	\renewcommand{\arraystretch}{1.2}
	\begin{tabular}{lc}
		\toprule
		\toprule
		\textbf{Parameter/Observable} & \textbf{Value/Scanned Region}\\
		\midrule
		$M_{Z'}$ & $75$ GeV \\
		$M_4^C$ & $200$ GeV \\
		$M_4^L$ & $1000$ GeV \\
		$\sin^2\theta_{12L,R}$ & $0.0$ \\
		$\sin^2\theta_{14L}$ & $0.75$ \\
		$\sin^2\theta_{14R}$ & $0.0$ \\
		$\sin^2\theta_{24L,R}$ & $10^{-7}\rightarrow 1.0$ \\
		$\operatorname{BR}(\mu\rightarrow e\gamma)$ & $10^{-3}\rightarrow 1.0$\\
		\bottomrule
		\bottomrule
	\end{tabular}
	\captionsetup{width=0.8\linewidth}
	\caption{Parameter space and $\operatorname{BR}(\mu\rightarrow e\gamma)$ in a parameter space where the electron and muon both mix with the vector-like lepton. Initial attempt to satisfy both anomalous moments.}
	\label{tab:parameters_for_first_attempt_scan}
\end{table}
\endgroup
This story concludes quite quickly with all points being excluded. The enchancement factor of $M^C_4/m_\mu$ in Equation \eqref{eqn:muon_g-2_contributions} is largely responsible for $(g-2)_\mu$ in this scenario, however such a term also gives an unacceptably large contribution to $\operatorname{BR}(\mu\rightarrow e\gamma)$ as per Equation \eqref{eqn:mu_e_gamma}, resulting in a branching fraction far above the experimental limit; the minimum $\operatorname{BR}(\mu\rightarrow e\gamma)$ for any parameter points in this scenario is around $10^{-3}$, as shown in Table \ref{tab:parameters_for_first_attempt_scan}. Such a situation persists even if $\sin^2\theta_{14}^L$ is scanned through it's entire range, and furthermore is unchanged by the choice of $M_4^L$, and is insensitive to the $Z'$ mass in the case of large $M_{4}^{C}$. We conclude therefore, that with a large chirality-flipping mass circa 200 GeV, it is not possible to simultaneously satisfy constraints and make predictions consistent with current data. This conclusion is consistent with the analytic arguments of the previous section, where the large contributions coming from large chirality flipping fourth family masses $M_4^C$ were assumed to dominate. We now go beyond this approximation, considering henceforth very small $M_4^C$.

If one sets $M_4^C$ vanishing, terms proportional to the aforementioned enhancement factor also vanish, eliminating the largest contribution to $\mu\rightarrow e\gamma$, as follows from Equation \eqref{eqn:mu_e_gamma}. Motivated by this reduction in the most restrictive decay the above analysis is repeated, but with the chirality-flipping mass removed.

\subsubsection{Vanishing $M^C_4$}
If we choose to turn off the chirality-flipping mass of the vector-like leptons, their mass becomes composed entirely of $M_4^L$. Terms proportional to the enhancement factor $M_4^C/m_\mu$ in Equation \eqref{eqn:muon_g-2_contributions} are sacrificed, which makes achieving a muon $g-2$ that is consistent with the experimental result more challenging. Larger mixing between the muon and vector-like leptons is required, but more freedom exists with respect to $BR(\mu\rightarrow e\gamma)$. We investigated a region of parameter space defined as per Table \ref{tab:parameters_for_large_scan_no_M4C}, to test its viability.
\begingroup
\setlength{\tabcolsep}{30pt}
\begin{table}[H]
	\centering
	\renewcommand{\arraystretch}{1.2}
	\begin{tabular}{lc}
		\toprule
		\toprule
		\textbf{Parameter} & \textbf{Value/Scanned Region}\\
		\midrule
		$M_{Z'}$ & $50\rightarrow 100$ GeV \\
		$M_4^C$ & $0$ GeV \\
		$M_4^L$ & $1000$ GeV \\
		$\sin^2\theta_{12L,R}$ & $0.0$ \\
		$\sin^2\theta_{14L}$ & $0.5\rightarrow 1.0$ \\
		$\sin^2\theta_{14R}$ & $0.0$ \\
		$\sin^2\theta_{24L,R}$ & $0.0\rightarrow 1.0$ \\
		\bottomrule
		\bottomrule
	\end{tabular}
	\caption{Parameters for scan without chirality-flipping mass.}
	\label{tab:parameters_for_large_scan_no_M4C}
\end{table}
\endgroup
For the results of this scan we consider the impact of each constraint separately, then check for overlap of allowed regions. Note that in Figure \ref{fig:scan_without_M4C_results}, angular parameters \emph{and} the heavy vector $Z^\prime$ mass are varied simultaneously, hence here we randomly select points and evaluate relevant phenomena, rather than excluding regions in the space. This also explains the spread of parameter points as compared to the previous exclusions.  Note that the range of $\sin^2\theta_{14}^L$ has been restricted in Tables \ref{tab:parameters_for_large_scan_no_M4C} and \ref{tab:parameters_for_large_scan_small_M4C} due to the fact that no points that satisfy $\operatorname{BR}(\mu\rightarrow e\gamma)$ could be found with $\sin^2\theta_{14}^L<0.5$, omitting this region increases the efficiency of our parameter scan. We also limit the ranges of $M_{Z'}$ in Tables \ref{tab:parameters_for_large_scan_no_M4C} and \ref{tab:parameters_for_large_scan_small_M4C} as $Z'$ masses much higher than this were found to be incompatible with $(g-2)_\mu$, and masses much below saturated the bound from $\mu\rightarrow e\gamma$.
\begingroup
\begin{figure}[H]
	\centering
	\begin{subfigure}{0.48\textwidth}
		\includegraphics[width=1.0\textwidth]{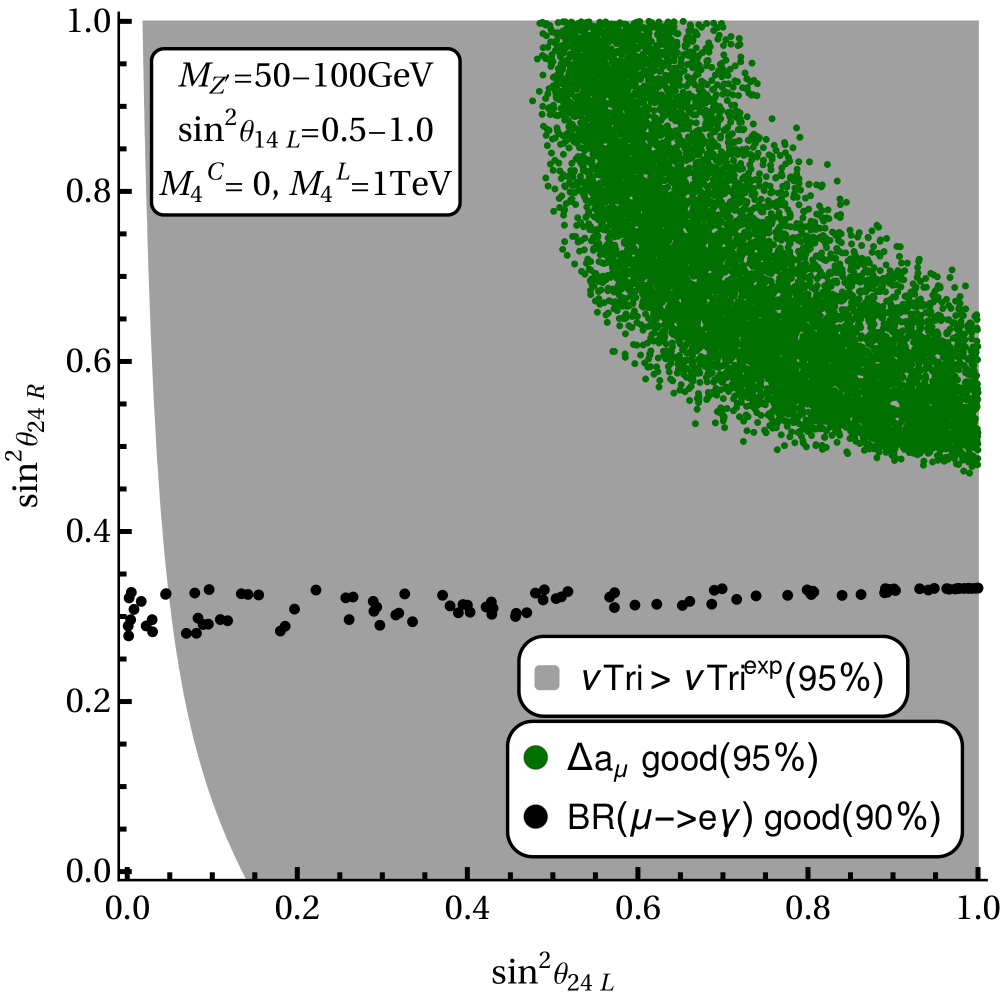}
		\captionsetup{width=0.8\textwidth}
		\caption{Parameter points that resolve $\Delta a_\mu$ and separately, points allowed under the $\mu\rightarrow e\gamma$ constraint. Fixed parameters given in legend. Chirality-flipping mass is set vanishing. All good $\Delta a_\mu$ points are excluded
		by trident and $\mu\rightarrow e\gamma$.}
		\label{fig:scan_without_M4C_g-2mu_and_muegamma}
	\end{subfigure}
	\begin{subfigure}{0.48\textwidth}
		\includegraphics[width=1.0\textwidth]{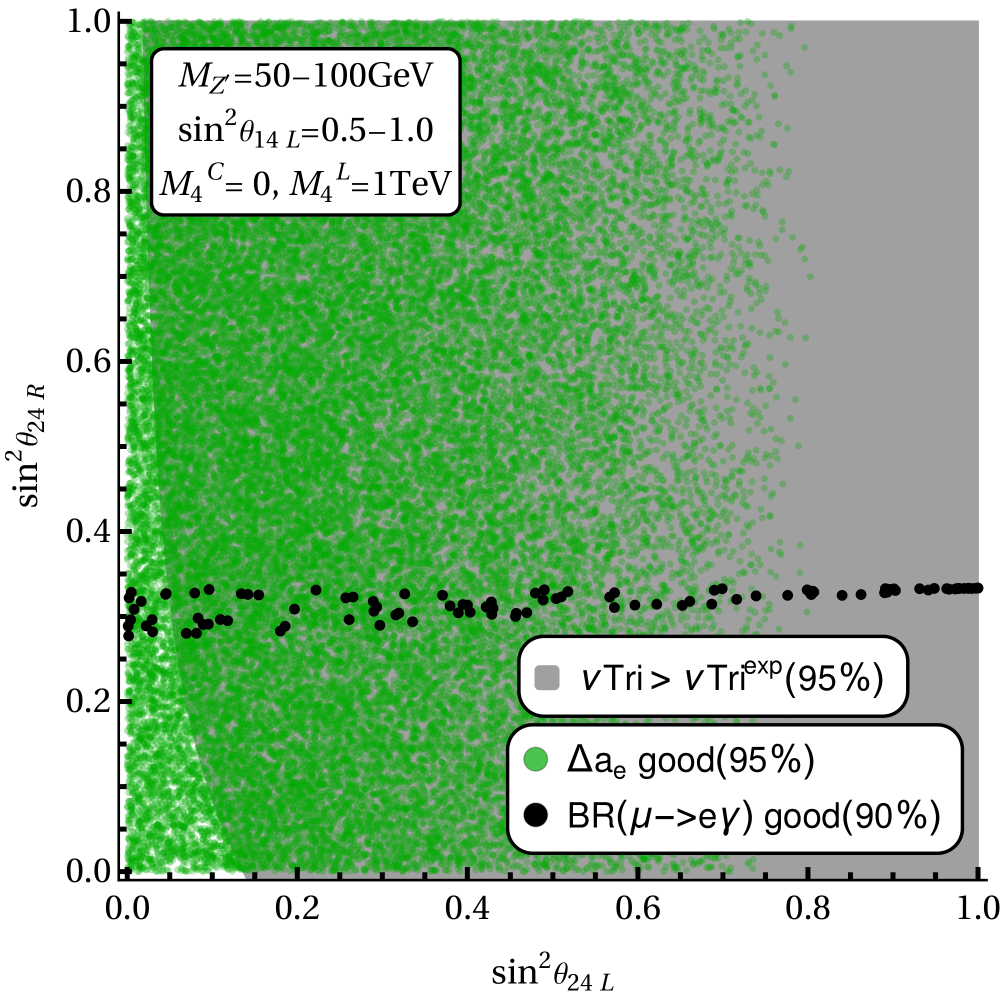}
		\captionsetup{width=0.8\textwidth}
		\caption{Parameter points that resolve $\Delta a_e$ and separately, points allowed under $\mu\rightarrow e\gamma$. Fixed parameters given in legend. Chirality-flipping mass is set vanishing.
		Some good $\Delta a_e$ points are allowed 
		by trident and $\mu\rightarrow e\gamma$.}
		\label{fig:scan_without_M4C_g-2e_and_muegamma}
	\end{subfigure}
	\captionsetup{width=0.8\textwidth}
	\caption{Parameter scan results for $M_4^C=0$. Note we will discuss the $Z^{\prime}$ experimental bound in a next subsection.}
	\label{fig:scan_without_M4C_results}
\end{figure}
\endgroup
In Figure \ref{fig:scan_without_M4C_g-2mu_and_muegamma}, one can see that, as suspected, larger $\sin^2\theta_{24L,R}$ mixings are required to obtain a muon $(g-2)_\mu$ consistent with current data. 
However, there is no overlapped region in Figure \ref{fig:scan_without_M4C_g-2mu_and_muegamma}, and $(g-2)_\mu$ cannot be solved without violating the muon decay constraint for a vanishing chirality-flipping mass, or the shown exclusion for neutrino trident production. On the other hand, Figure~\ref{fig:scan_without_M4C_g-2e_and_muegamma} shows that there are points that resolve the SM's tension with $(g-2)_e$, and are allowed by the strict $\operatorname{BR}(\mu\rightarrow e\gamma)$ limit and neutrino trident production. The lack of terms with the enhancement factor of $M^C_4/m_\mu$ in Equation \eqref{eqn:mu_e_gamma} means that points have been found with an acceptable branching fraction of $\mu\rightarrow e\gamma$ that was not possible with a large $M_4^C$.
 
Note that in both panels of Figure \ref{fig:scan_without_M4C_results} the most conservative neutrino trident limit is shown, where we assume that $M_{Z'}$ is fixed at 50GeV. We have also found that there is also no obvious correlation between $M_{Z'}$ and $\sin^2\theta_{14L}$ for $\mu\rightarrow e\gamma$, and points appear to be randomly distributed in this space. Since we have seen that neither large nor vanishing $M_4^C$ are viable, in the next subsection we switch on a small but non-zero $M_4^C$, to investigate if it may be possible to increase $(g-2)_\mu$ to an acceptable level, without giving an overlarge contribution to the CLFV muon decay.

\subsubsection{Small $M^C_4$ $\mathcal{O}(m_\mu)$}
Here we perform analogous tests to those above but with a small chirality flipping mass, motivated by $(g-2)_\mu$ with the requirement that $\operatorname{BR}(\mu\rightarrow e\gamma)$ remains below the experimental limit. Ranges of parameters scanned in this investiagtion are given in Table \ref{tab:parameters_for_large_scan_small_M4C}.
\begingroup
\setlength{\tabcolsep}{30pt}
\begin{table}[H]
	\centering
	\renewcommand{\arraystretch}{1.2}
	\begin{tabular}{lc}
		\toprule
		\toprule
		\textbf{Parameter} & \textbf{Value/Scanned Region}\\
		\midrule
		$M_{Z'}$ & $50\rightarrow 100$ GeV \\
		$M_4^C$ & $5m_\mu$ \\
		$\sin^2\theta_{14L}$ & $0.5\rightarrow 1.0$ \\
		$\sin^2\theta_{14R}$ & $0.0$ \\
		$\sin^2\theta_{24L,R}$ & $0.0\rightarrow 1.0$ \\
		$\sin^2\theta_{12L,R}$ & $0.0$ \\
		\bottomrule
		\bottomrule
	\end{tabular}
	\caption{Parameters for larger scan with a small chirality-flipping mass.}
	\label{tab:parameters_for_large_scan_small_M4C}
\end{table}
\endgroup
Figure \ref{fig:small_M4C_results} shows points allowed under each separate observable in an analogous parameter space to Figure \ref{fig:scan_without_M4C_results}, but with $M_4^C=5m_\mu$. Once more neutrino trident production excludes a large region of the parameter space in this scenario. From initial study of the parameter space it seems that there is overlap between the allowed regions of $(g-2)_\mu$, $(g-2)_e$ and $\operatorname{BR}(\mu\rightarrow e\gamma)$, however, upon closer inspection of the parameter points allowed by $\mu\rightarrow e\gamma$, those points always yield negative (wrong sign) 
$(g-2)_\mu$ that is far away from the experimental value, and hence all points are excluded. 
\begingroup
\begin{figure}[H]
	\centering
	\begin{subfigure}{0.48\textwidth}
		\includegraphics[width=1.0\textwidth]{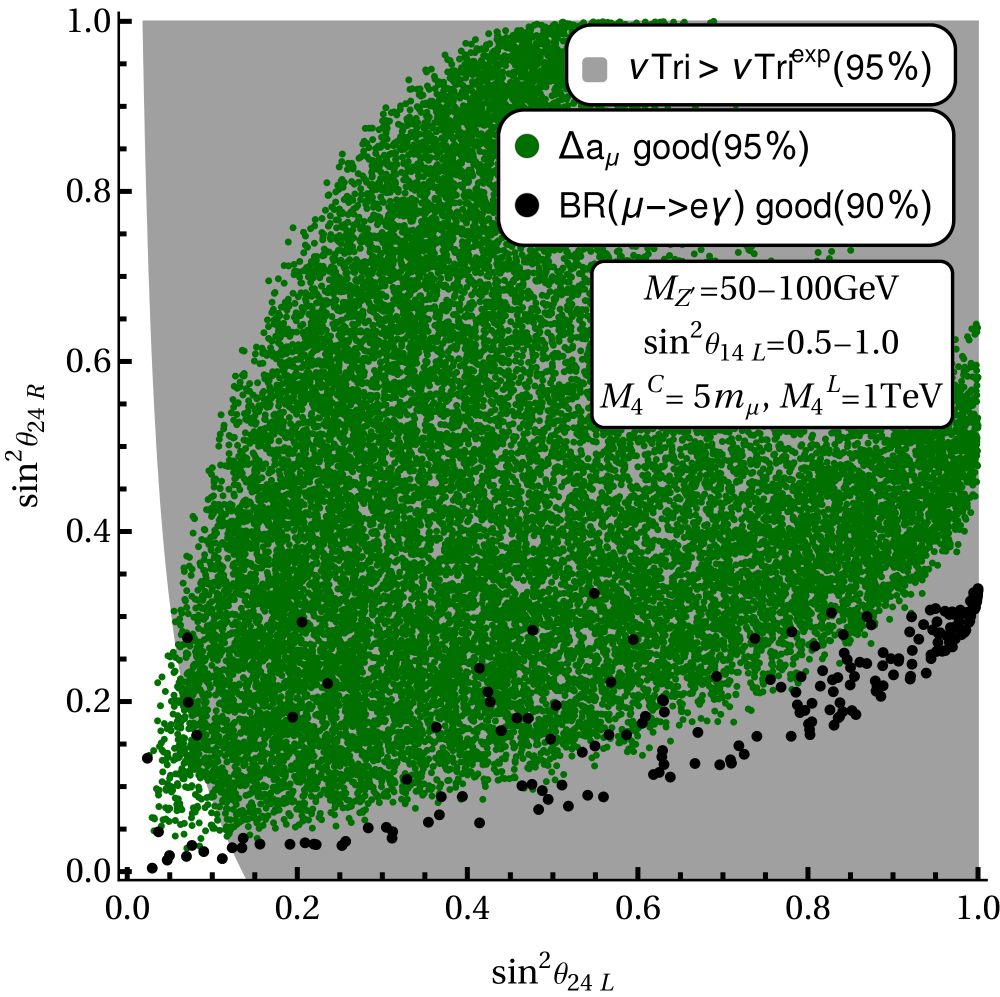}
		\caption{Parameter points that resolve $\Delta a_\mu$ and separately, points allowed under the $\mu\rightarrow e\gamma$ constraint. Fixed parameters given in legend, small chirality flipping mass. Unfortunately none of the points shown which have viable
		$\mu\rightarrow e\gamma$ and satisfy trident also have good $\Delta a_\mu$ (see text).
		}
		\label{fig:small_M4C_g-2mu_and_muegamma}
	\end{subfigure}
	\begin{subfigure}{0.48\textwidth}
		\includegraphics[width=1.0\textwidth]{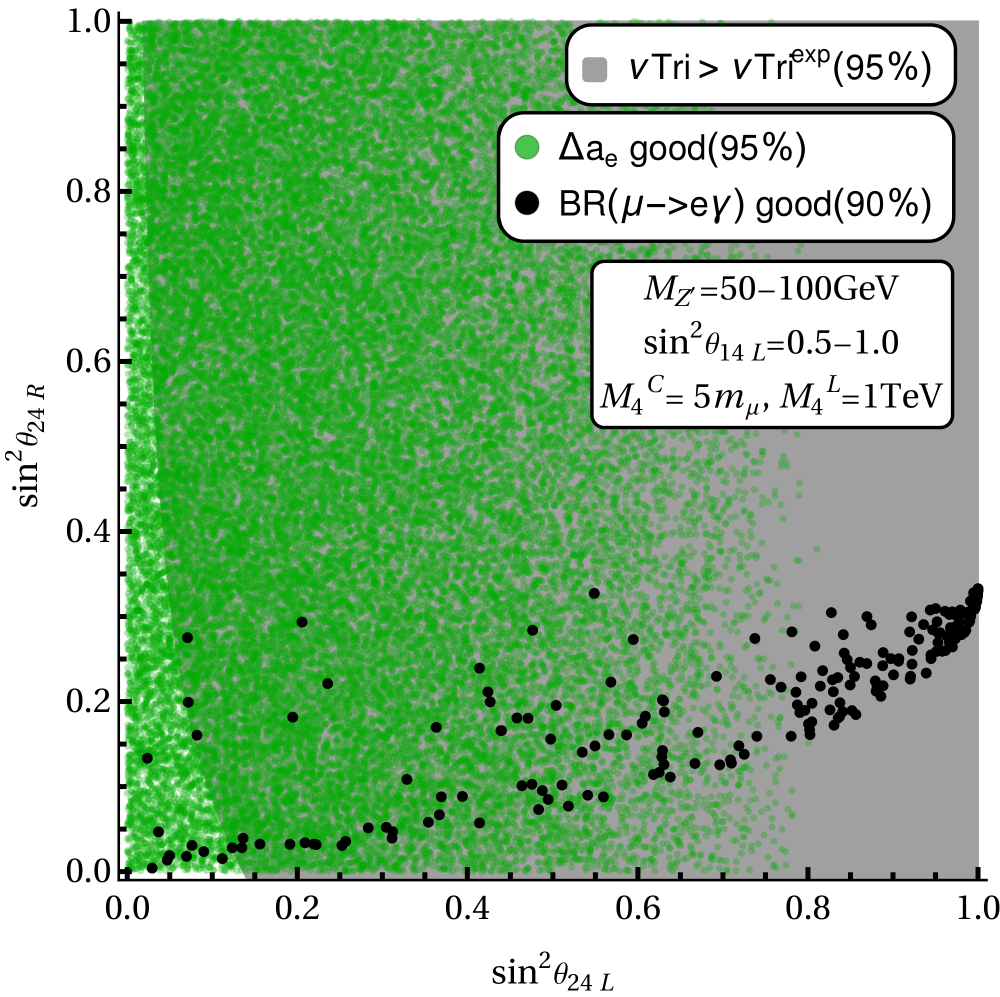}
		\captionsetup{width=0.8\textwidth}
		\caption{Parameter points that resolve $\Delta a_e$ and separately, points allowed under the $\mu\rightarrow e\gamma$ constraint. Fixed parameters given in legend, small chirality flipping mass.}
		\label{fig:small_M4C_g-2e_and_muegamma}
	\end{subfigure}
	\captionsetup{width=0.8\textwidth}
	\caption{Parameter scan results for small $M_4^C=5m_\mu$. Note we will discuss the $Z^{\prime}$ experimental bound in a next subsection.}
	\label{fig:small_M4C_results}
\end{figure}
\endgroup
In Table \ref{tab:view_allowed_points}, we examine more closely the points that are allowed under the most stringent constraint of  
$\mu\rightarrow e\gamma$. As 4th family mixing with the muons exists in this space, neutrino trident production is also a consideration, and the constraint of this observable in our space is given in Figure \ref{fig:small_M4C_results}. All points valid when considering $\operatorname{BR}(\mu\rightarrow e\gamma)$ exist with a small $\sin^2\theta_{24R}$ mixing angle, but can have a wide range of $Z'$ masses and $\sin^2\theta_{14L}$.
\begingroup
\begin{table}[H]
	\centering
	\renewcommand{\arraystretch}{1}
	\resizebox{\textwidth}{!}{$
	\begin{tabular}{ccccccc}
		\toprule
		\toprule
		\multicolumn{4}{c}{\textbf{Parameter}} & \multicolumn{3}{c}{\textbf{Observable}} \\
		\cmidrule(lr{\tabcolsep}){1-4} \cmidrule(l{\tabcolsep}r){5-7}
		$M_{Z'}/\text{GeV}$ & $\sin^2\theta_{14L}$ & $\sin^2\theta_{24L}$ & $\sin^2\theta_{24R}$ & $\operatorname{BR}(\mu\rightarrow e\gamma)$ & $\Delta a_e$ & $\Delta a_\mu$ \\
		\midrule
		$69.5$ & $0.61$ & $0.11$ & $0.02$ & $3.25\times10^{-13}$ & $-2.15\times10^{-13}$ &  $-1.80\times10^{-10}$ \\
		$68.5$ & $0.80$ & $0.05$ & $0.01$ & $1.69\times10^{-13}$ & $-3.32\times10^{-13}$ & $-1.63\times10^{-10}$ \\
		$91.0$ & $0.99$ & $0.08$ & $0.16$ & $3.34\times10^{-13}$ & $-2.41\times10^{-13}$ & $-1.19\times10^{-9}$ \\
		$63.0$ & $0.99$ & $0.02$ & $0.13$ & $1.38\times10^{-13}$ & $-5.390\times10^{-13}$ & $-2.03\times10^{-9}$ \\
		$65.5$ & $0.78$ & $0.07$ & $0.02$ & $4.94\times10^{-14}$ & $-3.43\times10^{-13}$ & $-2.36\times10^{-10}$ \\
		$64.8$ & $0.78$ & $0.09$ & $0.02$ & $3.61\times10^{-13}$ & $-3.46\times10^{-13}$ & $-3.19\times10^{-10}$ \\
		$77.9$ & $0.85$ & $0.005$ & $0.02$ & $6.13\times10^{-14}$ & $-2.77\times10^{-13}$ & $-1.77\times10^{-10}$ \\
		$91.4$ & $0.81$ & $0.14$ & $0.04$ & $5.80\times10^{-14}$ & $-1.73\times10^{-13}$ & $-2.71\times10^{-10}$ \\
		$97.2$ & $0.86$ & $0.08$ & $0.03$ & $1.07\times10^{-13}$ & $-1.73\times10^{-13}$ & $-2.71\times10^{-10}$ \\
		$76.0$ & $0.63$ & $0.03$ & $0.004$ & $1.72\times10^{-13}$ & $-2.01\times10^{-13}$ & $-3.97\times10^{-11}$ \\
		$56.8$ & $0.96$ & $0.04$ & $0.05$ & $3.77\times10^{-14}$ & $-6.22\times10^{-13}$ & $-8.36\times10^{-10}$ \\
		$78.1$ & $0.99$ & $0.07$ & $0.20$ & $1.84\times10^{-14}$ & $-3.32\times10^{-13}$ & $-2.04\times10^{-9}$ \\
		$89.4$ & $1.0$ & $0.07$ & $0.28$ & $2.95\times10^{-13}$ & $-2.56\times10^{-13}$ & $-2.25\times10^{-9}$ \\
		\bottomrule
		\bottomrule
	\end{tabular} $}
	\captionsetup{width=\textwidth}
	\caption{Parameter points that are below the upper bound on $\operatorname{BR}(\mu\rightarrow e\gamma)$ for $M_4^C=5m_\mu$. The points in this table correspond to the 13 black points in Figure \ref{fig:small_M4C_results} that are also below the grey neutrino trident exclusion. These points do not satisfy the experimental value of $(\Delta a_{\mu })_{\exp}=(26.1\pm 8)\times10^{-10}$.}
	\label{tab:view_allowed_points}
\end{table}
\endgroup
We see that for the points in Table \ref{tab:view_allowed_points}, electron $g-2$ prefers regions of the space with small $\sin^2\theta_{24L}$, similarly to the preferred points under the neutrino trident constraint, given in the same plot as an excluded region derived in the same way as previous results for $M_4^C=0$. Many of these points are simultaneously consistent with the $\mu\rightarrow e \gamma$ limit, and also provide a $(g-2)_e$ consistent with experimental data (denoted in green), whilst a subset of these points do not violate the neutrino trident production limit. From these results, we can conclude that the best points lie in the region of small $\sin^2\theta_{24L}$ and $\sin^2\theta_{24R}$, and that such points simultaneously comply with $\operatorname{BR}(\mu\rightarrow e\gamma)$,  $(g-2)_e$ and neutrino trident. Such candidate points however do not allow for resolution of $\Delta a_\mu$, as they all have negative values for $\Delta a_\mu$, as opposed to the experimental value which is positive. 

A number of other chirality flipping masses were examined in this work, in the region $5m_\mu < M_4^C < 200\text{GeV}$, including a parameter scan whereby $M_4^C$ was randomly selected between these limits, and these tests yielded similar results to those shown in the last three sections, whereby it was not possible to obtain predictions that were simultaneously consistent with $(g-2)_e$, $(g-2)_\mu$ and $\operatorname{BR}(\mu\rightarrow e\gamma)$.

\section{The experimental and theoretical bound for the neutral $Z^{\prime}$ gauge boson} \label{sec:Zp_exp_bound}
The neutral $Z^{\prime}$ gauge boson can be constrained by both the effective four fermion effective interactions and the theoretical oblique corrections $S,T,U$. However, the unknown coupling constant and mass of $Z^{\prime}$ gauge boson leave the predictions not fully determined. Both will be explored in order.
\subsection{Four fermion effective interactions}
In order to make our analysis as simple as possible, we assumed the neutral $Z^{\prime}$ gauge boson are generated via the leptonic collision process $e^{+} e^{-} \rightarrow Z^{\prime} \rightarrow e^{+} e^{-}$ and the experimental bound for the $Z^{\prime}$ boson is given in PDG~\cite{CELLO:1984rge,Adeva:1984fy,PLUTO:1984grm,MARK-J:1984fmz,Derrick:1985xj,VENUS:1989wgn}.
\begingroup
\begin{equation}
M_{Z^{\prime}}^{\func{EXP}} = 48 \func{GeV}
\label{eqn:MZp_old}
\end{equation}
\endgroup
However, a critical problem is the result of Equation~\ref{eqn:MZp_old} is too old to trust. Plus, it looks like the fact that the CM energy of the $e^{+}e^{-}$ collision process of the LEP experiment has reached  up to $209\func{GeV}$ makes our numerical prediction for the $Z^{\prime}$ mass, $75\func{GeV}$, excluded completely. However, we came to two agreements on the fact through our discussions as follows:
\begin{enumerate}
\item The experimental $Z^{\prime}$ bound of Equation~\ref{eqn:MZp_old} is a somewhat weak bound.
\item The CM energy $209\func{GeV}$ is not the ultimate experimental bound for the $Z^{\prime}$ neutral gauge boson at the moment.
\end{enumerate}
Based on the agreements, we find a suitable relation to constrain the $Z^{\prime}$ mass and the effective four fermion leptonic Lagrangian for the constraint is given by~\cite{CarcamoHernandez:2019xkb}:
\begingroup
\begin{equation}
\begin{split}
\mathcal{L}_{\func{eff}} &= -\frac{g_X^2}{M_{Z^{\prime}}^2} \sum_{j=1}^{3} [ x_{l_{1L}} x_{l_{jL}} ( \overline{l}_{1} \gamma^{\mu} P_L l_1 )( \overline{l}_{jL} \gamma_{\mu} l_{jL} ) + x_{l_{1L}} x_{l_{jR}} ( \overline{l}_{1} \gamma^{\mu} P_L l_{1} )( \overline{l}_{jR} \gamma_{\mu} l_{jR} )] \\
&-\frac{g_X^2}{M_{Z^{\prime}}^2} \sum_{j=1}^{3} [ x_{l_{1R}} x_{l_{jL}} ( \overline{l}_{1} \gamma^{\mu} P_R l_1 )( \overline{l}_{jL} \gamma_{\mu} l_{jL} ) + x_{l_{1R}} x_{l_{jR}} ( \overline{l}_{1} \gamma^{\mu} P_R l_{1} )( \overline{l}_{jR} \gamma_{\mu} l_{jR} )]
\label{eqn:eff_MZp}
\end{split}
\end{equation}
\endgroup
The resulting limit for the leptonic collision process $e^+ e^- \rightarrow \mu^+ \mu^-$ from Equation~\ref{eqn:eff_MZp} has the form of~\cite{CarcamoHernandez:2019xkb,ALEPH:2013dgf}:
\begingroup
\begin{equation}
\frac{2M_{Z^{\prime}}}{g_X \sqrt{x_{l_{1L}} x_{l_{2L}}+x_{l_{1R}} x_{l_{2R}}+x_{l_{1R}} x_{l_{2L}}+x_{l_{1L}} x_{l_{2R}}}} > 4.6 \func{TeV},
\label{eqn:MZp_limit}
\end{equation}
\endgroup 
\subsection{Oblique corrections $S,T$ and $U$}
It is well-known that any BSM model can be significantly constrained by the oblique corrections $S,T$ and $U$. The oblique corrections $S,T$ and $U$ were first suggested by Peskin and Takeuchi in 1991 and one of the great success of the oblique corrections was to find top quark's mass. Therefore, we can expect that masses of the hypothetical particles such as $Z^{\prime}$ gauge boson and non-SM scalars can be constrained by the oblique corrections. The oblique corrections $S$ and $T$ come from dimension six operators whereas $U$ comes from dimension eight operator, so the corrections $S$ and $T$ play a more important role in constraining the hypothetical particles. The definition for the corrections $S$ and $T$ are given by~\cite{CarcamoHernandez:2015smi}
\begin{equation}
\begin{split}
S &= \frac{2\sin 2\theta_W}{\alpha_{\func{EM}}\left( M_{Z} \right)} \left. \frac{d\Pi_{30}\left( q^2 \right)}{dq^2} \right|_{q^2=0},
\\
T &= \left. \frac{\Pi_{33}\left( q^2 \right) - \Pi_{11}\left( q^2 \right)}{\alpha_{\func{EM}}\left( M_{Z} \right) M_{W}^2} \right|_{q^2=0},
\end{split}
\end{equation}
where $\Pi$s are the vacuum polarization amplitudes with external gauge bosons $W_{1,2,3}$ and $B$. The oblique corrections $S$ and $T$ consist of its SM part and new physics effect
\begin{equation}
\begin{split}
S &= S_{\func{SM}} + \Delta S, \\
T &= T_{\func{SM}} + \Delta T,
\end{split}
\end{equation}
and the $S_{\func{SM}}$ and $T_{\func{SM}}$ are calculated by~\cite{CarcamoHernandez:2017pei}
\begin{equation}
\begin{split}
S_{\func{SM}} &= \frac{1}{12\pi} \ln \left( \frac{m_{h}^2}{m_{W}^2} \right) + \frac{1}{2\pi} \left[ 3 - \frac{1}{3} \ln \left( \frac{m_{t}^2}{m_{b}^2} \right) \right] \simeq 0.106
\\
T_{\func{SM}} &= -\frac{3}{16\pi \cos^2\theta_{W}} \ln \left( \frac{m_{h}^2}{m_{W}^2} \right) + \frac{3m_{t}^2}{32\pi^2 \alpha_{\func{em}}\left( m_{Z} \right) v^2} \simeq 0.537
\end{split}
\end{equation}
where $\alpha\left(M_{Z}^2\right)^{-1} = 128.944$. The current best-fit results of oblique parameters $S,T$ and $U$ are given by~\cite{ParticleDataGroup:2020ssz,Jueid:2021avn}
\begin{equation}
S = -0.01 \pm 0.10, \quad T = 0.03 \pm 0.12, \quad U = 0.02 \pm 0.11.
\end{equation}
The neutral $Z^{\prime}$ gauge boson can appear in the vacuum polarized amplitudes via mixing with the external gauge bosons $W_{1,2,3}$ and $B$.
\begin{figure}[H]
\centering
	\begin{tikzpicture}
		\begin{feynman}
		\vertex (a1) {\(W_{1,2,3},B\)};
		\vertex [right = 2cm of a1] (a2);
		\vertex [right = 4cm of a2] (a3);
		\vertex [right = 2cm of a3] (a4) {\(W_{1,2,3},B\)};
		\diagram* {
		(a1) -- [boson,momentum=\(q\)] (a2) -- [fermion,half left,out=90,in=90,edge label=\(k+q\)] (a3) -- [boson,momentum=\(q\)] (a4),
		(a3) -- [fermion,half left,out=90,in=90,edge label=\(k\)] (a2),
		};
		\end{feynman}
	\end{tikzpicture}
\caption{Diagrams contributing to the gauge boson 2-point function}
\end{figure}
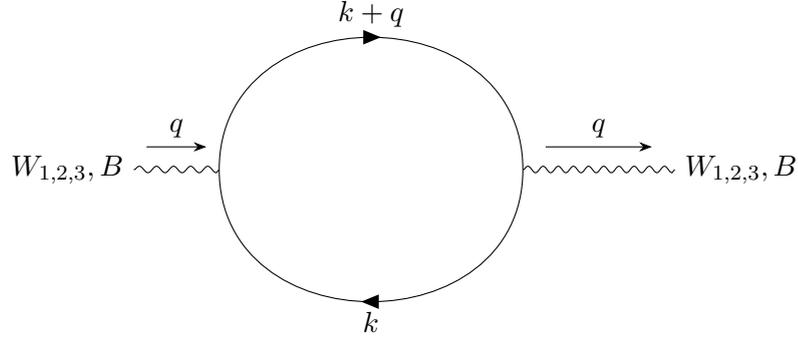 
However, a main difficulty of determining the $Z^{\prime}$ theoretical bound comes from its unknown coupling constants and masses. Plus, I could not find any convincing papers related to constraining the light $Z^{\prime}$ gauge boson with the oblique parameters (There are a few papers, however all of them depend on a specific scenario and their constraints are of order $\func{TeV}$, which is quite difficult to generalize and apply to our BSM model). In my fourth project, I will try to constrain the $M_{Z^\prime}$ with the oblique parameters after finding preferred order of $Z^{\prime}$ coupling constants by diverse flavor observables such as muon $g-2$, FCNC observables, etc.
\section{Concluding Remarks}
\label{sec:conclusion}
In this paper, we have addressed the question: is it possible to explain the anomalous muon and electron $g-2$ in a $Z^{\prime }$ model? Although it is difficult to answer this question in general, since there are many possible $Z'$ models, we have seen that it is possible to consider a simple renormalisable and gauge invariant model in which the $Z'$ only has couplings to the electron and muon and their associated neutrinos, arising from mixing with a vector-like fourth family of leptons. This is achieved by assuming that only the vector-like leptons have non vanishing $U(1)'$ charges and are assumed to only mix with the first and second family of SM charged leptons. In this scenario, the heavy $Z^{\prime }$ gauge boson couples with the first and second family of SM charged leptons only through mixing with the vector-like generation. 

A feature of our analysis is to distinguish the two sources of mass for the 4th, vector-like family: 
the chirality flipping fourth family mass terms $M_4^C$ arising from the Higgs Yukawa couplings and are proportional to the Higgs vev
and the vector-like masses $M_{4}^{L}$ which are not proportional to the Higgs vev.
For the purposes of clarity we have
treated $M_4^C$ and $M_{4}^{L}$ as independent mass terms
in the analysis of the physical quantities of interest, rather than constructing the full fourth family mass matrix and 
diagonalising it, since such quantities rely on a chirality flip and are sensitive to $M_4^C$ rather than $M_{4}^{L}$. 

We began by assuming large fourth family chirality flipping masses
$M_4^C\gg m_\mu$, and showed that the expressions for $(g-2)_\mu$, $(g-2)_e$ and $\operatorname{BR}(\mu\rightarrow e\gamma)$
reduced to a minimal number of terms, all proportional to $M_4^C$. We were then able to construct an analytic argument which shows that 
it is not possible to explain the anomalous muon and electron $g-2$ in the $Z^{\prime }$ model, while respecting the 
bound on $\operatorname{BR}(\mu\rightarrow e\gamma)$.

We then performed a detailed numerical analysis of the parameter space of the above model, beginning with large 
$M_4^C =200$ GeV, where we showed that it is possible to account for $(g-2)_\mu$ in a region of parameter space
where the electron couplings were zero. Similarly, for $M_4^C =200$ GeV, we showed that it is possible to account for $(g-2)_e$ in a region of parameter space where the muon couplings were zero. In both cases $\operatorname{BR}(\mu\rightarrow e\gamma)$
was identically zero. 

Keeping $M_4^C =200$ GeV, we then attempted to explain both anomalous magnetic moments by switching on 
the couplings to the electron and muon simultaneously, but saw that it was not possible to do this while satisfying
$\operatorname{BR}(\mu\rightarrow e\gamma)$, as expected from the analytic arguments.

We then went beyond the regime of the analytic arguments by considering very small values of $M_4^C$.
With $M_4^C=0$, we saw that it is not possible to account for $(g-2)_\mu$ without violating the bounds from 
$\operatorname{BR}(\mu\rightarrow e\gamma)$ and trident, however it is possible to account for $(g-2)_e$
while respecting all constraints. With small but non-zero $M_4^C$ we reached similar conclusions, although the analysis
was more complicated, and it was necessary to examine specific benchmark points to reach this conclusion.

We stress that the fermiophobic
$Z'$ model is a good candidate to explain either $(g-2)_\mu$ or $(g-2)_e$, consistently with 
$\operatorname{BR}(\mu\rightarrow e\gamma)$ and trident, with the choice determined by the specific mixing scenario. 
However to explain the $(g-2)_\mu$ always requires a significant non-vanishing 
chirality flipping mass involving the 4th vector-like family of leptons.

We would like to comment on the generality of our conclusion that, for the $Z^\prime$ framework considered in this paper, we cannot simultaneously explain the electron and muon g-2 results within the relevant parameter space of the model,
while satisfying the constraints of $BR(\mu\rightarrow e\gamma)$ and neutrino trident production.
Does this conclusion apply to all $Z^\prime$ models? While it is impossible to answer this question absolutely, 
there are reasons why our results here might be considered very general and indicative of a large class
of $Z^\prime$ models. The main reason for this is that, in the considered 
framework, the $Z^\prime$ is only allowed to couple to the electron and muon and their associated 
neutrinos, arising from mixing with a vector-like fourth family of leptons, thereby 
eliminating the quark couplings and allowing us to focus on 
the connection between CLUV, CLFV and the electron and muon $g-2$ anomalies only, independently of other constraints.
Moreover, the allowed $Z^\prime$ couplings are free parameters in our approach and so may represent the couplings in a large class
of $Z^\prime$ models. Furthermore, we have presented a general analytic argument that provides some insight into our numerical
results. For example, we do not require the $Z^\prime$ to couple identically to left- and right-handed leptons, and the masses for intermediate particles in the one-loop diagrams cancel in the final expression for $\operatorname{BR}(\mu\rightarrow e\gamma)$ in Equation \ref{eqn:really_simplified_mu_e_gamma_expression}, which lends this result some generality. 
We also note that this paper represents the first paper to attempt to explain both electron and muon $g-2$ anomalies simultaneously
within a $Z'$ model. Thus, although the problem of the CLFV constraint in preventing an explanation of electron and muon $g-2$ anomalies is well known in general, it had not been studied within the framework of $Z^\prime$ models before the present paper. Indeed this is the first work we know of that attempts to explain the muon and electron anomalous magnetic moments simultaneously using a simple $Z^\prime$ model.
On top of that, we also discuss the $Z^{\prime}$ mass limit. The current $Z^{\prime}$ experimental bound is known as $48\func{GeV}$ at PDG, however we agree that this is a somewhat weak bound and need to determined the correct $Z^{\prime}$ mass bound, based on the fact that the LEP experimental has reached up to $209\func{GeV}$ for the CM energy of the leptonic collision process $e^+ e^- \rightarrow e^+ e^-$. Using the experimental limit suggested by LEP experiment~\cite{CarcamoHernandez:2019xkb,ALEPH:2013dgf}, we derived the numerical mass bound for the $Z^{\prime}$ gauge boson, which is $M_{Z^{\prime}} > 287.5\func{GeV}$. Since this numerical result depends on lots of assumptions though, we conclude the correct $Z^{\prime}$ mass bound is not yet completely determined.

Finally we comment that since 
there are models in the literature which account for all these observables based on having scalars, it might be interesting to extend
the scalar sector of a $Z'$ model. 
The lepton flavour violating processes could then be used to set constraints on the masses for the CP even and CP odd heavy neutral scalars, as in \cite{CarcamoHernandez:2019xkb}. 
However, such a study 
is beyond the scope of the present paper. 

In conclusion, within a model where the $Z'$ only has tunable couplings to the electron and muon and their associated 
neutrinos, arising from mixing with a vector-like fourth family of leptons,
it is not possible to simultaneously satisfy the experimentally observed values of 
$(g-2)_\mu$ and $(g-2)_e$, while respecting the $\operatorname{BR}(\mu\rightarrow e\gamma)$
and trident constraints, within any of the exhaustively explored parameter space
(only one or other of $(g-2)_\mu$ or $(g-2)_e$ can be explained). 
Since the model allows complete freedom in the choice of couplings, and the diagrams involving 
fourth family lepton exchange can be chosen to contribute or not,
this model may be regarded as indicative of any $Z'$ model with gauge coupling and charges of order one.
\chapter{The second BSM model - SM fermion mass hierarchies from VL families with an extended 2HDM} \label{Chapter:The2ndBSMmodel}

In this chapter, we start discussing our second BSM model with two vector-like families, mainly motivated by the hierarchical structure of the SM, taking the SM as an effective theory. In the BSM model, we discuss the hierarchical structure of the SM with the mixing formalism and how the BSM model can give rise to the effective SM interactions.
\section{Introduction and motivation}
\label{sec:Introduction}

The Standard Model (SM) has made many successful predictions for the phenomenology 
of both quark and lepton sectors with very high accuracy.
However there are long-established anomalies which are not addressed by 
the SM such as muon and electron anomalous magnetic moments
$a_{\mu }=\left( g-2 \right)_\mu/2, a_e=\left( g-2 \right)_e/2$. 
The muon
anomalous magnetic moment reported by the Brookhaven E821 experiment at BNL%
\cite{Bennett:2006fi} and the electron anomaly have confirmed $+3.5 \sigma$
and $-2.5 \sigma$ deviations from the SM, respectively. Detailed data analysis of the Standard Model predictions for the muon anomalous magnetic moment are provided in \cite{Hagiwara:2011af,Davier:2017zfy,Blum:2018mom,Keshavarzi:2018mgv,Aoyama:2020ynm}. The experimentally
observed values for the muon and electron anomalies at $1\sigma$ of
experimental error bars, respectively, read \footnote{It is worth mentioning that the experimental value of the anomalous magnetic moment of the electron is sensitive to the measurement of the fine-structure constant $\alpha$. The experimental value of $\Delta a_e=a_{e,exp}-a_e(\alpha_{\textit{Berkeley}})$ used in this work and given in Equation \ref{eqn:deltaamu_deltaae_at_1sigma} is obtained 
using $\alpha_{\textit{Berkeley}}$ from caesium recoil measurements by the Berkeley 2018 experiment \cite{Parker:2018vye}.
As this paper was being completed a different experiment \cite{Morel:2020dww} reported a result that implies 
$\Delta a_e = a_e^{\func{Exp}} - a_e^{\func{SM}} = \left( 0.48 \pm
0.30 \right) \times 10^{-12}$ which differs from the SM by $+1.6\sigma$.
The two experiments appear to be inconsistent with each other, and our results here are based on the earlier result in 
Equation \ref{eqn:deltaamu_deltaae_at_1sigma}.}:
\begingroup
\begin{equation}
\begin{split}
\Delta a_\mu &= a_\mu^{\func{Exp}} - a_\mu^{\func{SM}} = \left( 26.1 \pm
8.0 \right) \times 10^{-10} \\
\Delta a_e &= a_e^{\func{Exp}} - a_e^{\func{SM}} = \left( -0.88 \pm
0.36 \right) \times 10^{-12}.  \label{eqn:deltaamu_deltaae_at_1sigma}
\end{split}%
\end{equation}
\endgroup
When trying to explain both anomalies to within $1\sigma $, a main
difficulty arises 
from the sign of each anomaly: the muon anomaly requires positive definite non-standard contributions, whereas the electron anomaly requires such contributions to contribute with a negative sign~ \cite{Morel:2020dww}.
Without loss of generality, the Feynman diagrams corresponding to the
contributions for the muon and electron anomalies take the same internal
structure at one-loop except from the fact that the external particles are
different. 
The similar structure 
of the one-loop level contributions to 
the muon and electron anomalous magnetic moments might be able to be
explained by the same new physics, but accounting for the relative negative sign is challenging. 
For example, considering the one-loop exchange of $W$ or $%
Z^{\prime }$ gauge bosons results 
in theoretical predictions for the muon and electron anomalies 
having the same sign.

In this paper we take the view that both anomalies should be
explained to $1\sigma $ using the same internal structure at the 
one-loop level by some new physics which is capable of accounting for the correct signs of the anomalies.
To explain the muon and electron anomalies, we focus on a well motivated model which is also capable 
of accounting for origin of Yukawa
couplings and hierarchies in the SM. 
The model we consider will account for the Yukawa coupling
constant for the top quark being nearly $1$ while that for the electron is around $10^{-6}$, as well as all the other fermion hierarchies
in between, as well as the neutrino masses and mixing.
In order to achieve this we shall 
introduce vector-like particles, which are charged under a 
global $U(1)^{\prime }$ symmetry. In a related previous work~\cite{King:2018fcg}, with a gauged $U(1)^{\prime }$ symmetry, the first
family of quarks and leptons remained massless when only one vector-like family is included.
Here we shall modify the model to include two vector-like families charged under a global $U(1)^{\prime }$ to allow 
also the first family to be
massive and avoid $Z'$ constraints. Then we shall apply the resulting model to the problem of muon and electron anomalous magnetic moments.
The considered model is based on 
a 2 Higgs doublet model (2HDM) extension of the SM, supplemented 
by a global $U(1)^\prime$ symmetry, where the particle spectrum is enlarged by
the inclusion of 
two 
vector-like fermion families, as well as one singlet Higgs to break the $U(1)^\prime$ symmetry \footnote{An example of a multiHiggs doublet model that uses a flavor dependent global $U(1)^{\prime }$ symmetry to explain the SM charged fermion mass hierarchy by hierarchies of the vacuum expectation values of the Higgs doublets is provided in \cite{CentellesChulia:2020bnf}}. The SM Yukawa interactions are forbidden, but the Yukawa interactions with vector-like families charged under the $U(1)^\prime$ symmetry are allowed. Once the flavon develops a vev and the heavy vector-like fermions are integrated out, 
the effective SM Yukawa interactions are generated, as indicated in 
Figure \ref%
{fig:mass_insertion_diagrams}. Furthermore, this model also highlights the
shape of the 2HDM model type II, since in our proposed model, one Higgs
doublet (which in the alignment limit corresponds to the SM Higgs doublet)
couples with the up type quarks whereas the other one features Yukawa
interactions with down type quarks and SM charged leptons. 
Regarding the neutrino sector, since we consider 
the SM neutrinos as Majorana particles, we have that this sector 
requires another approach relying on the inclusion of 
a new five dimensional Weinberg-like operator, which is allowed in this model
and which requires 
both SM Higgs doublets to be present, namely the so called Type Ib seesaw model~\cite{Hernandez-Garcia:2019uof}.
\\~\\
We shall show that the heavy vector-like leptons are useful and necessary to explain the
anomalous electron and muon magnetic moment deviations from the SM, of magnitude and opposite
signs given in Equation \ref{eqn:deltaamu_deltaae_at_1sigma}. A study of such $g-2$ anomalies in terms
of New Physics and a possible UV complete explanation via vector-like
leptons was performed in \cite{Crivellin:2018qmi}, although the model presented here is quite different, since our model is 
motivated by the requirement of accounting also for the fermion mass hierarchies. Other theories with extended
symmetries and particle spectrum have also been proposed to find an explanation
for the muon and electron anomalous magnetic moments \cite%
{Appelquist:2004mn,Giudice:2012ms,Freitas:2014pua,Falkowski:2018dsl,Crivellin:2018qmi,Allanach:2015gkd,Chen:2016dip,Raby:2017igl,Chiang:2017tai,Chen:2017hir,Davoudiasl:2018fbb,Liu:2018xkx,CarcamoHernandez:2019xkb,Nomura:2019btk,Kawamura:2019rth,Bauer:2019gfk,Han:2018znu,Dutta:2018fge,Badziak:2019gaf,Endo:2019bcj,Hiller:2019mou,CarcamoHernandez:2019ydc,CarcamoHernandez:2019lhv,Kawamura:2019hxp,Cornella:2019uxs,CarcamoHernandez:2020pxw,Arbelaez:2020rbq,Hiller:2020fbu,Jana:2020pxx,deJesus:2020ngn,deJesus:2020upp,Hati:2020fzp,Botella:2020xzf,Dorsner:2020aaz,Calibbi:2020emz,Dinh:2020pqn,Jana:2020joi,Chun:2020uzw,Chua:2020dya,Daikoku:2020nhr,Banerjee:2020zvi,Chen:2020jvl,Bigaran:2020jil,Kawamura:2020qxo,Endo:2020mev,Chakrabarty:2020jro,Li:2020dbg}%
. 
In the following we provide a brief comparison of our model to other works, starting
with the model proposed in \cite{Chun:2020uzw}
where vector-like leptons are also present. The model of \cite{Chun:2020uzw} corresponds to an extended type X lepton specific 2HDM model of \cite{Chun:2020uzw} having a $Z_2$ discrete symmetry under which one of the scalar doublets and the leptonic fields are charged. In such model the vector-like leptons induces a one-loop level contribution to the electron anomalous magnetic moment whereas the muon anomalous magnetic moment is generated at two-loop via the exchange of a light pseudoscalar. On the other hand, in our proposed model a spontaneously broken global $U(1)^\prime$ symmetry is considered instead of the $Z_2$ symmetry and the vector-like leptons generate one-loop level contributions to the muon and electron anomalous magnetic moments and at the same type produce the SM charged lepton masses, thus providing a connection of the charged lepton mass generation mechanism and the $g-2$ anomalies,
which is not given in the model of \cite{Chun:2020uzw}. It is also worth emphasising that our model is  
very different from other models proposed in the literature based on the Universal Seesaw mechanism
\cite{Davidson:1987mh,Davidson:1987mi,Berezhiani:1991ds,Sogami:1991yq,Gu:2010zv,Alvarado:2012xi,Hernandez:2013mcf,Kawasaki:2013apa,Mohapatra:2014qva,Dev:2015vjd,Borah:2017inr,Patra:2017gak,Babu:2018vrl,deMedeirosVarzielas:2018bcy,CarcamoHernandez:2018aon,CarcamoHernandez:2019vih,CarcamoHernandez:2019pmy,Hernandez:2021uxx}.
Universal Seesaw models are typically based on the left-right symmetric model with electroweak singlet fermions only,
while our vector-like fermions involves complete families, including electroweak doublets which are typically the lightest ones.
Some examples of theories relying on the Universal Seesaw mechanism to explain the SM charged fermion mass hierarchy are provided in 
~\cite{Davidson:1987mh,Davidson:1987mi,Berezhiani:1991ds,Sogami:1991yq,Gu:2010zv,Alvarado:2012xi,Hernandez:2013mcf,Kawasaki:2013apa,Mohapatra:2014qva,Dev:2015vjd,Borah:2017inr,Patra:2017gak,Babu:2018vrl,deMedeirosVarzielas:2018bcy,CarcamoHernandez:2018aon,CarcamoHernandez:2019vih,CarcamoHernandez:2019pmy,Hernandez:2021uxx}.\newline

In the approach followed in this paper the large third family quark and lepton Yukawa couplings are effectively generated via mixing with a vector-like fourth family of electroweak doublet fermions, which are assumed to be relatively light, with masses around the TeV scale. The smallness of the second family quark and lepton Yukawa couplings is due to their coupling to heavier vector-like fourth family electroweak singlet fermions. Similar considerations apply to the lightest first family quarks and leptons which couple to heavy fifth family vector-like fermions. 
It may seem that the problem of the hierarchies of SM fermions is not solved but simply reparameterised in terms of unknown vector-like fermion masses. However, there are four advantages to this approach. Firstly, the approach is dynamical, since the vector-like masses are new physical quantities which could in principle be determined by a future theory. Secondly, it has experimental consequences, since the new vector-like fermions can be discovered either directly, or (as in this paper) indirectly via their loop contributions. Thirdly, this approach can also account for small quark mixing angles~\cite{King:2018fcg}, as well as large lepton mixing angles via the type Ib seesaw mechanism~\cite{Hernandez-Garcia:2019uof}. Fourthly, the effective Yukawa couplings are proportional to a product of two other dimensionless couplings, so a small hierarchy in those couplings can give a quadratically larger hierarchy in the effective couplings. For all these reasons, the approach we follow in this paper is both well motivated and interesting.

Returning to 
our proposed model framework, we first consider the contribution of $W$ boson exchange with neutrinos to the
electron and muon anomalous magnetic moments at the one-loop level. Since this model involves the vector-like
neutrinos, the sensitivity of the branching ratio of $\mu \rightarrow e\gamma $
decay can be enhanced with respect to the observable level and the muon and
electron anomalous magnetic moments are studied while keeping the $\mu
\rightarrow e\gamma $ constraint. As a result, we find that the impact of
our predictions with $W$ exchange at one-loop level is 
negligible when compared to their experimental bound.
We then consider the contributions from the 2HDM scalar exchange.
To study the implications of the one-loop level 
scalar exchange in the muon and electron anomalous magnetic moments, we
first construct a scalar potential and derive the mass squared matrix for
CP-even, CP-odd and charged Higgses assuming there is no mixing between the
SM Higgs $h$ and two non-SM physical scalars $H_{1,2}$. A 
diagonal Yukawa matrix for charged leptons implies 
the absence of mixing between charged leptons, resulting in vanishing 
branching ratio for the 
$\mu \rightarrow e\gamma $ decay, which in turn leads to a fulfillment of
the charged lepton flavor violating constraints 
in this scenario. In such a framework we show that both anomalies can successfully explain both 
anomalies, including their opposite signs, 
at the $1\sigma $ level.
We present some benchmark points for both the muon and the
electron anomalies, together with some numerical scans around these points,
which indicate the 
mass regions of the Higgs scalars of the 2HDM in this scenario.
We also provide some analytic arguments to augment the numerical results.

The layout of the remainder of the paper is as follows.
In Section~\ref{II} we discuss the origin of Yukawa couplings from a fourth and fifth vector-like family, within a mass insertion formalism.
In Section~\ref{III} we construct the effective Yukawa matrices using a more detailed mixing formalism which goes beyond the 
mass insertion formalism.
In Section~\ref{IV} we consider $W$ exchange contributions to 
$\left( g-2 \right)_\protect\mu, \left( g-2 \right)_e$ and $\func{BR}\left(\protect\mu \rightarrow e \protect\gamma \right)$
based on the type Ib seesaw mechanism within our model and show that the contributions are too small.
In Section~\ref{sec:Analytic_arguments_muon_electron_g2_scalars} we turn to Higgs scalar exchange contributions to
$\left( g-2 \right)_\protect\mu, \left( g-2
\right)_e$ and $\func{BR}\left(\protect\mu \rightarrow e \protect\gamma \right)$, focussing on analytical formulae.
Then in Section~\ref{sec:Numerical_analysis_of_scalars} we give a full numerical analysis of such contributions, showing that they can 
successfully explain the anomalies, presenting some benchmark points for both the muon and the
electron anomalies, together with some numerical scans around these points,
which indicate the 
mass regions of the Higgs scalars of the 2HDM in this scenario. Section~\ref{sec:non_SM_scalars_exp_bound} discusses the experimental and theoretical non-SM scalars' mass bound. Section~\ref{sec:Conclusion} concludes the main body of the paper.
Appendix~\ref{A} provides a discussion of the quark mass matrices in two bases.
Appendix~\ref{B} includes a brief discussion of heavy scalar production at a proton-proton collider.

\section{The origin of Yukawa couplings from a fourth and fifth vector-like family}
\label{II}

\label{sec:The_origin_of_Yukawa_coupling}

We start by asking a question: what is the origin of the SM Yukawa
couplings? 
In addressing such question, we assume that 
the SM Yukawa Lagrangian is the low energy limit of an 
extended theory with enlarged symmetry and particle spectrum, and arises
after the spontanous breaking of an $U(1)^{\prime }$ global symmetry at an
energy scale 
as low as $\func{TeV}$. Therefore, understanding the origin of the Yukawa
interaction naturally leads to the presence of another Higgses whose masses
are higher than 
the mass of the SM Higgs. Furthermore, the SM Yukawa interactions are
forbidden by the global $U(1)^{\prime }$ symmetry, however the Yukawa
interaction with the vector-like particles are allowed. With these
considerations in place, the possible diagrams generating 
the Yukawa interactions can be drawn as indicated in 
Figure \ref{fig:mass_insertion_diagrams}.
\begingroup
\begin{figure}[H]
\begin{subfigure}{0.48\textwidth}
	\includegraphics[width=1.0\textwidth]{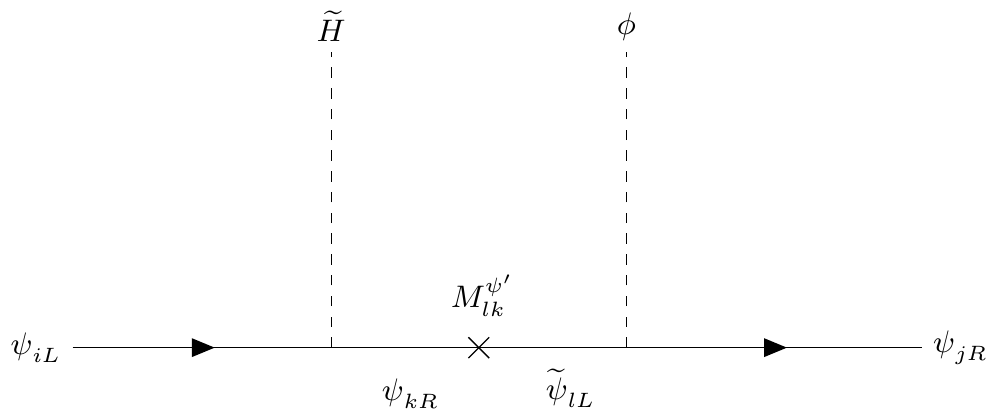}
\end{subfigure} \hspace{0.1cm} 
\begin{subfigure}{0.48\textwidth}
	\includegraphics[width=1.0\textwidth]{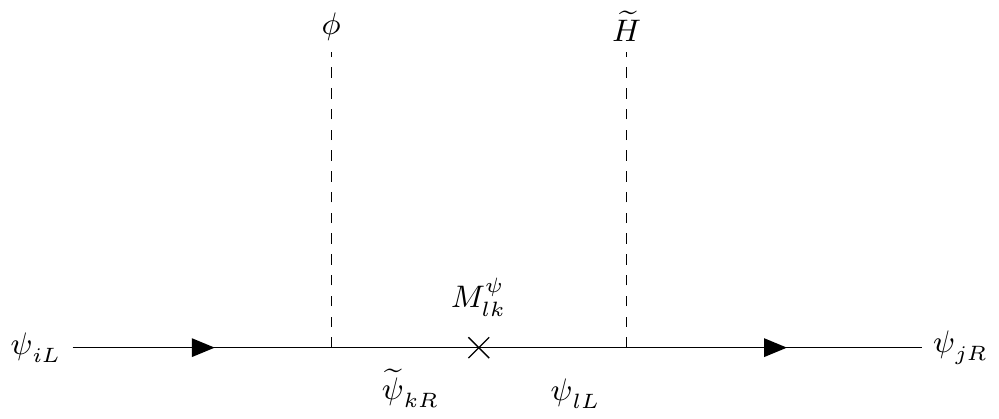}
\end{subfigure}
\caption{Diagrams in this model which lead to the effective Yukawa
interactions, where $\protect\psi,\protect\psi^\prime = Q,u,d,L,e$(neutrinos
will be treated separately) $i,j=1,2,3$, $k,l=4,5$, $M_{lk}$ is vector-like
mass and $\widetilde{H} = i\protect\sigma_2 H^*, H = H_{u,d}$}
\label{fig:mass_insertion_diagrams}
\end{figure}
\endgroup
There are two key features in Figure \ref{fig:mass_insertion_diagrams}, one
of which is the presence of the assumed flavon $\phi$ and the other one is
the vector-like mass $M$. Once the flavon $\phi$ develops its vev, the
effective Yukawa interactions $\overline{\psi}_{iL} \widetilde{H} \psi_{jR}$ are generated with a
coupling constant proportional to 
$\left\langle \phi \right\rangle/M$, assumed to be less than unity, which appears in front of the usual
Yukawa constant. The proportional factor $\left\langle \phi \right\rangle/M$
plays a crucial role 
in explaining why one Yukawa constant can be relatively smaller or bigger
than the other ones 
since the magnitude of each Yukawa constant is accompanied by the mass of
the vector-like particles. The effective Lagrangian in this diagram reads in the mass insertion formalism:

\begin{equation}
\mathcal{L}_{\func{eff}}^{\func{Yukawa}} =y_{ik}^\psi (M_{\psi^\prime}^{-1})_{kl} {x_{lj}^{\psi^\prime}
\left\langle \phi \right\rangle} \overline{%
\psi}_{iL} \widetilde{H} \psi_{jR} + {x_{ik}^{\psi} \left\langle \phi
\right\rangle}(M_{\psi}^{-1} )_{kl}  y_{lj}^\psi  \overline{\psi}_{iL} \widetilde{H}
\psi_{jR} + \func{h.c.}  \label{eqn:the_effective_Yukawa_Lagrangian}
\end{equation}

where $\psi,\psi^\prime = Q,u,d,L,e$ (neutrinos will be treated separately)
and $x$ is a Yukawa constant in the interaction with $\phi$ and $y$ is in
the interaction with $\widetilde{H}$ as per Figure \ref%
{fig:mass_insertion_diagrams}. Throughout this work, we take a view that the
Yukawa constant $y$ can be ideally of order unity while the $x$ is
small compared to the $y$. We shall also use a mixing formalism rather than the mass insertion formalism.

\subsection{The model with $U(1)^{\prime }$ global symmetry}

For an analysis of the phenomenology described above, we extend the SM fermion sector by adding two vector-like fermions, the SM gauge symmetry by including the global $U(1)^\prime$ symmetry and the scalar sector of the 2HDM model is enlarged by considering a gauge scalar singlet, whose VEV triggers the spontaneous breaking of the $U(1)^\prime$ symmetry. 
The scalar sector of the model is composed of 
by two $SU(2)$ doublet scalars 
$H_{u,d}$ and one flavon $\phi$. 
 Our extended 2HDM with enlarged particle spectrum and symmetries has the interesting feature that 
the SM Yukawa interactions are forbidden due to the global $U(1)^\prime$ symmetry
whereas the Yukawa interactions of SM fermions with vector-like families are allowed. Furthermore, such vector-like families have mass terms which are allowed by the symmetry. Thus, the SM charged fermions masses are generated from a Universal Seesaw mechanism mediated by heavy vector-like fermions.
 Unlike the $U(1)^\prime $ model proposed in \cite%
{King:2017anf}, we assume that the $U(1)^\prime $ symmetry is global instead
of local. This allows us more flexibility in the allowed range for the scale where the $U(1)^\prime $ symmetry is broken.
On top of that, the up-type
quarks feature Yukawa interaction with the up-type Higgs 
whereas the down-type ones interact with down-type Higgs. In this BSM
model, the SM particles are neutral under the $U(1)^\prime$ symmetry, while the vector-like particles and all other scalars are charged under the symmetry. The particle content and symmetries of the model are 
shown in Table \ref{tab:model_content}.
\begingroup
\begin{table}[H]
\centering\renewcommand{\arraystretch}{0.5} 
\resizebox{0.6\textwidth}{!}{
\begin{tabular}{ccccc}
\toprule
\toprule
Field & $SU(3)_C$ & $SU(2)_L$ & $U(1)_Y$ & $U(1)^{\prime}$ \\
\midrule
$Q_{iL}$ & $\mathbf{3}$ & $\mathbf{2}$ & $\frac{1}{6}$ & $0$ \\[5pt]
$u_{iR}$ & $\mathbf{3}$ & $\mathbf{1}$ & $\frac{2}{3}$ & $0$ \\[5pt]
$d_{iR}$ & $\mathbf{3}$ & $\mathbf{1}$ & $-\frac{1}{3}$ & $0$ \\[5pt]
$L_{iL}$ & $\mathbf{1}$ & $\mathbf{2}$ & $-\frac{1}{2}$ & $0$ \\[5pt]
$e_{iR}$ & $\mathbf{1}$ & $\mathbf{1}$ & $1$ & $0$ \\
\midrule
$Q_{kL}$ & $\mathbf{3}$ & $\mathbf{2}$ & $\frac{1}{6}$ & $1$ \\[5pt]
$u_{kR}$ & $\mathbf{3}$ & $\mathbf{1}$ & $\frac{2}{3}$ & $-1$ \\[5pt]
$d_{kR}$ & $\mathbf{3}$ & $\mathbf{1}$ & $-\frac{1}{3}$ & $-1$ \\[5pt]
$L_{kL}$ & $\mathbf{1}$ & $\mathbf{2}$ & $-\frac{1}{2}$ & $1$ \\[5pt]
$e_{kR}$ & $\mathbf{1}$ & $\mathbf{1}$ & $-1$ & $-1$ \\[5pt]
$\nu_{kR}$ & $\mathbf{1}$ & $\mathbf{1}$ & $0$ & $-1$ \\
\midrule
$\widetilde{Q}_{kR}$ & $\mathbf{3}$ & $\mathbf{2}$ & $\frac{1}{6}$ & $1$ \\[5pt]
$\widetilde{u}_{kL}$ & $\mathbf{3}$ & $\mathbf{1}$ & $\frac{2}{3}$ & $-1$ \\[5pt]
$\widetilde{d}_{kL}$ & $\mathbf{3}$ & $\mathbf{1}$ & $-\frac{1}{3}$ & $-1$ \\[5pt]
$\widetilde{L}_{kR}$ & $\mathbf{1}$ & $\mathbf{2}$ & $-\frac{1}{2}$ & $1$ \\[5pt]
$\widetilde{e}_{kL}$ & $\mathbf{1}$ & $\mathbf{1}$ & $-1$ & $-1$ \\[5pt]
$\widetilde{\nu}_{kL}$ & $\mathbf{1}$ & $\mathbf{1}$ & $0$ & $-1$ \\
\midrule
$\phi$ & $\mathbf{1}$ & $\mathbf{1}$ & $0$ & $1$ \\[5pt]
$H_u$ & $\mathbf{1}$ & $\mathbf{2}$ & $\frac{1}{2}$ & $-1$ \\[5pt]
$H_d$ & $\mathbf{1}$ & $\mathbf{2}$ & $-\frac{1}{2}$ & $-1$ \\[5pt]
\bottomrule
\bottomrule
\end{tabular}}%
\caption{This model is an extended 2HDM by the global $U(1)^\prime$ symmetry
with two vector-like families plus one flavon and reflects the property that the SM Yukawa interactions are forbidden. All SM particles $\protect\psi_i(i=1,2,3)$ are neutral under the $U(1)^\prime$ symmetry and the right neutrinos $\protect\nu_{iR}$ are not considered. Notice that this model involves two right-handed vector-like neutrinos $\protect\nu_{kR}, \widetilde{\protect\nu}_{kR}$. The SM particles are extended by two vector-like families where $k=4,5$ and two SM Higgses $H_{u,d}$ are charged negatively under $U(1)^\prime $ to forbid the renormalizable SM Yukawa interactions. The flavon field $\protect\phi$ plays a role of braking the $U(1)^\prime$ symmetry at $\func{TeV}$ scale.}
\label{tab:model_content}
\end{table}
\endgroup
The right-handed neutrinos $\nu_{iR}$ are absent in this model since we treat
the left-handed neutrinos in the lepton doublet as Majorana particles and
they are only extended by vector-like neutrinos. The vector-like particles
and their partners have exact opposite charge to each other under the
extended gauge symmetry to cancel out chiral anomaly. Lastly, the SM Higgses 
$H_{u,d}$ are negatively charged under the $U(1)^\prime$ symmetry to forbid
the renormalizable SM Yukawa interactions. 

\subsection{mass insertion formalism}

The renormalizable Yukawa interactions and mass terms 
for both up and down quark sectors read: 
\begin{equation}
\begin{split}
\mathcal{L}_{q}^{\func{Yukawa+Mass}} &= y_{ik}^{u} \overline{Q}_{iL} 
\widetilde{H}_u u_{kR} + x_{ki}^{u} \phi \overline{\widetilde{u}}_{kL}
u_{iR} + x_{ik}^Q \phi \overline{Q}_{iL} \widetilde{Q}_{kR} + y_{ki}^u 
\overline{Q}_{kL} \widetilde{H}_u u_{iR} \\
&+ y_{ik}^{d} \overline{Q}_{iL} \widetilde{H}_d d_{kR} + x_{ki}^{d} \phi 
\overline{\widetilde{d}}_{kL} d_{iR} + y_{ki}^d \overline{Q}_{kL} \widetilde{%
H}_d d_{iR} \\
&+ M_{kl}^{u} \overline{\widetilde{u}}_{lL} u_{kR} + M_{kl}^{d} \overline{%
\widetilde{d}}_{lL} d_{kR} + M_{kl}^Q \overline{Q}_{kL} \widetilde{Q}_{lR} + 
\func{h.c.}  \label{eqn:general_Quark_Yukawa__Mass_Lagrangian}
\end{split}%
\end{equation}

where $i,j=1,2,3$, $k,l=4,5$ and $\widetilde{H}=i\sigma _{2}H^{\ast }$.
The possible diagrams contributing to the low energy quark Yukawa interaction
are given in Figure \ref{fig:diagrams_quark_mass_insertion}:
\begingroup
\begin{figure}[H]
\centering
\begin{subfigure}{0.48\textwidth}
	\includegraphics[width=1.0\textwidth]{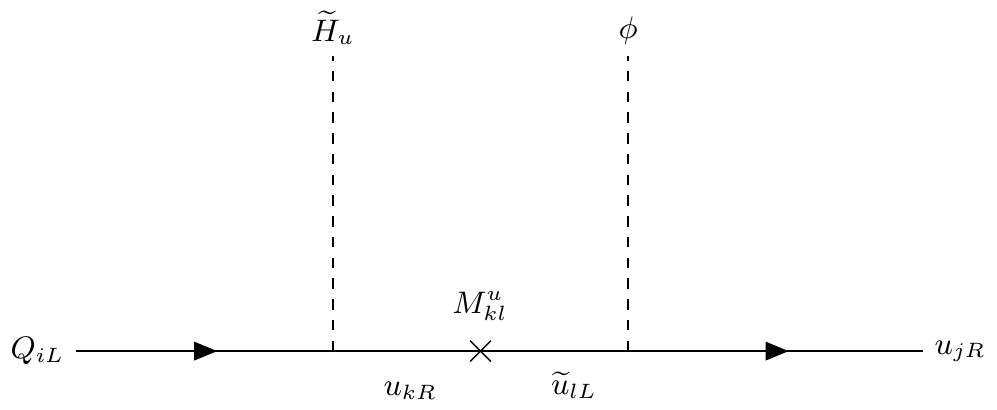}
\end{subfigure} \hspace{0.1cm} 
\begin{subfigure}{0.48\textwidth}
	\includegraphics[width=1.0\textwidth]{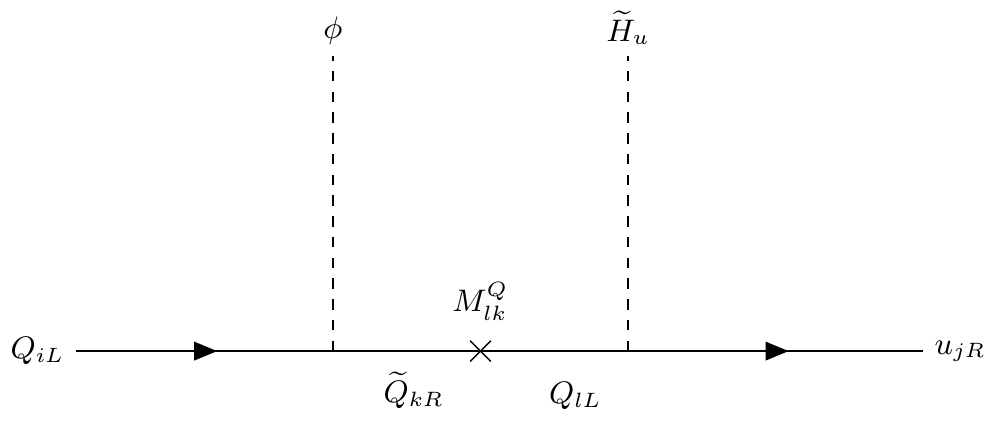}
\end{subfigure}
\begin{subfigure}{0.48\textwidth}
	\includegraphics[width=1.0\textwidth]{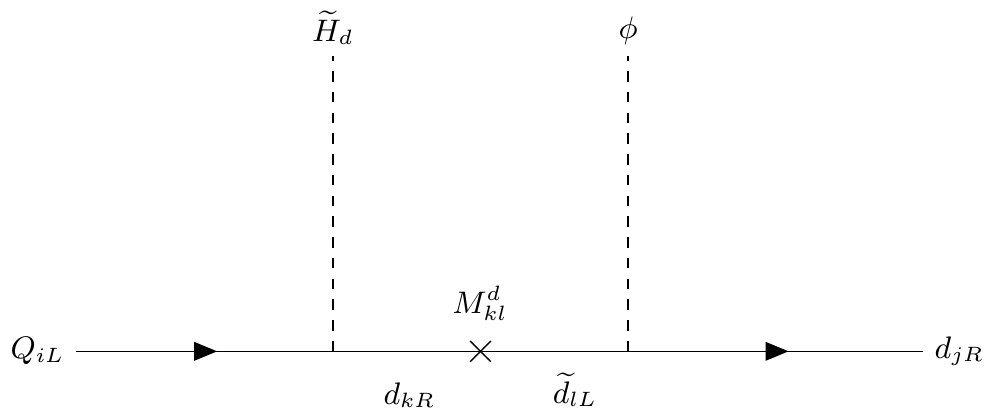}
\end{subfigure} \hspace{0.1cm} 
\begin{subfigure}{0.48\textwidth}
	\includegraphics[width=1.0\textwidth]{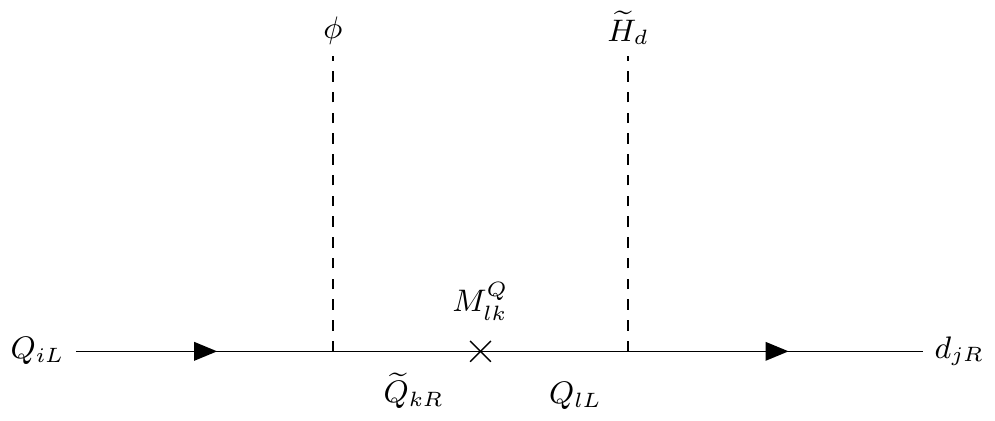}
\end{subfigure}
\caption{Diagrams in this model which lead to the effective Yukawa
interactions for the up quark sector(two above diagrams) and the down quark
sector(two below diagrams) in mass insertion formalism, where $i,j=1,2,3$
and $k,l=4,5$ and $M_{lk}$ is vector-like mass.}
\label{fig:diagrams_quark_mass_insertion}
\end{figure}
\endgroup
The above two diagrams correspond to the up-type quark sector 
whereas the below two diagrams correespond 
to the down-type quark sector. The model under consideration is an extended 2HDM where the 
up-type Higgs $H_u$ is relevant for the up-type quark sector whereas the%
down-type Higgs $H_d$ is suitable for the down-type quark and charged lepton sectors. Like in the quark sector, the Yukawa interactions
and mass terms 
for charged leptons can be written in a similar way:
\begingroup
\begin{equation}
\begin{split}
\mathcal{L}_{e}^{\func{Yukawa+Mass}} &= y_{ik}^{e} \overline{L}_{iL} 
\widetilde{H}_{d} e_{kR} + x_{ki}^{e} \phi \overline{\widetilde{e}}_{kL}
e_{iR} + x_{ik}^L \phi \overline{L}_{iL} \widetilde{L}_{kR} + y_{ki}^e 
\overline{L}_{kL} \widetilde{H}_{d} e_{iR} \\
&+ M_{kl}^{e} \overline{\widetilde{%
e}}_{lL} e_{kR} + M_{kl}^L \overline{L}_{kL} \widetilde{L}_{lR} + \func{h.c.}
\end{split}
\label{eqn:general_charged_lepton_Yukawa_Mass_Lagrangian}
\end{equation}
\endgroup
Then, the possible diagrams giving rise to the 
charged lepton Yukawa interactions are shown in Figure \ref%
{fig:diagrams_charged_leptons_mass_insertion}: 
\begingroup
\begin{figure}[tbph]
\centering
\begin{subfigure}{0.48\textwidth}
	\includegraphics[width=1.0\textwidth]{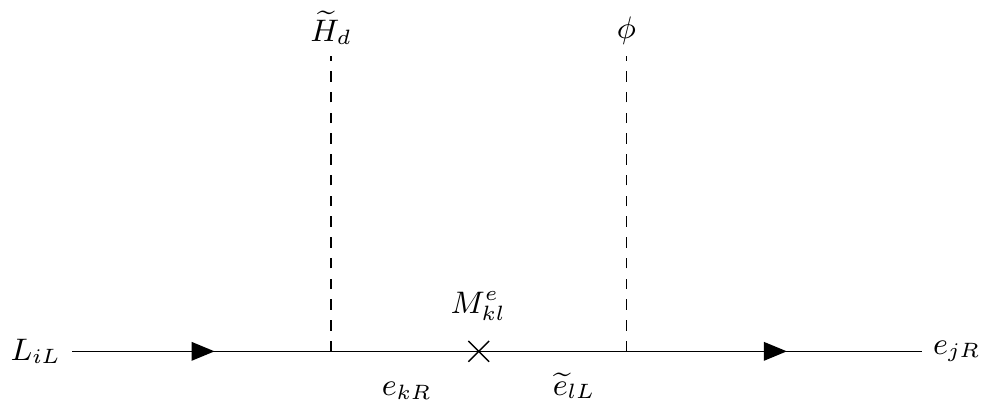}
\end{subfigure} \hspace{0.1cm} 
\begin{subfigure}{0.48\textwidth}
	\includegraphics[width=1.0\textwidth]{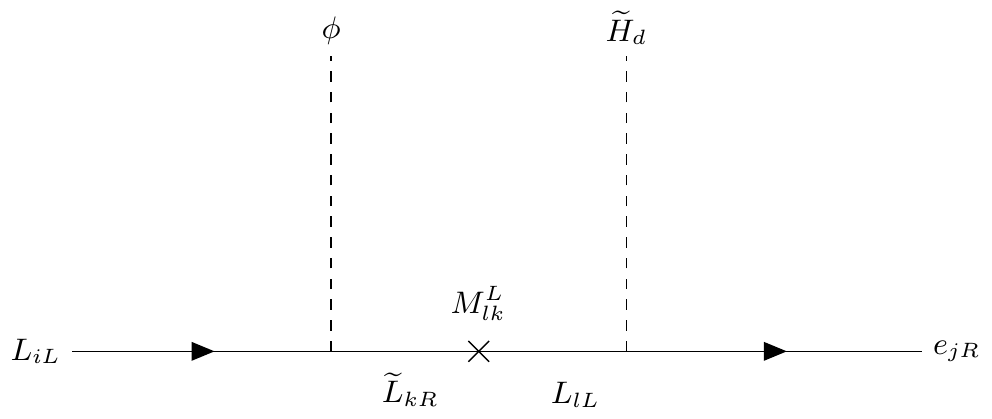}
\end{subfigure}
\caption{Diagrams in this model which lead to the effective Yukawa
interactions for the charged lepton sector in mass insertion formalism,
where $i,j=1,2,3$ and $k,l=4,5$ and $M_{lk}$ is vector-like mass.}
\label{fig:diagrams_charged_leptons_mass_insertion}
\end{figure}
\endgroup
As for the neutrinos, its behaviour is different as compared to the quarks or
charged leptons since there exists only Majorana neutrinos in this model so
initial and final neutrinos in mass insertion formalism diagrams must be same. The
Yukawa interactions and mass terms for the neutrino sector 
are given by:
\begingroup
\begin{equation}
\begin{split}
\mathcal{L}_{\nu}^{\func{Yukawa+Mass}} = y_{ik}^{\nu} \overline{L}_{iL} 
\widetilde{H}_u \nu_{kR} + x_{ik}^L \overline{L}_{iL} H_d \overline{%
\widetilde{\nu}}_{kR} + M_{kl}^{M} \overline{\widetilde{\nu}}_{lR} \nu_{kR}
+ \func{h.c.}
\end{split}
\label{eqn:general_neutrinos_Yukawa_Mass_Lagrangian}
\end{equation}
\endgroup
Here, one important feature in Equation \ref%
{eqn:general_neutrinos_Yukawa_Mass_Lagrangian} is the presence of the 
vector-like mass $M$. From the two Yukawa interactions in Equation \ref%
{eqn:general_neutrinos_Yukawa_Mass_Lagrangian}, it follows that both $\nu_R$
and $\widetilde{\nu}_R $ have a lepton number $+1$ and they are different
particles. And then taking a look at the vector-like mass term in Equation %
\ref{eqn:general_neutrinos_Yukawa_Mass_Lagrangian}, it can be confirmed that
the vector-like mass is not a strict Majorana mass because $\nu_R$ and $%
\widetilde{\nu}_R$ are different particles but plays a role of Majorana mass
since the mass term violates the lepton number conservation. The
corresponding diagram for the 
neutrino sector in the mass insertion formalism is given in Figure \ref{fig:diagrams_neutrinos_mass_insertion}.
However for our calculations we use a mixing formalism (see next section).
\begingroup
\begin{figure}[tbph]
\centering
\begin{subfigure}{0.48\textwidth}
	\includegraphics[width=1.0\textwidth]{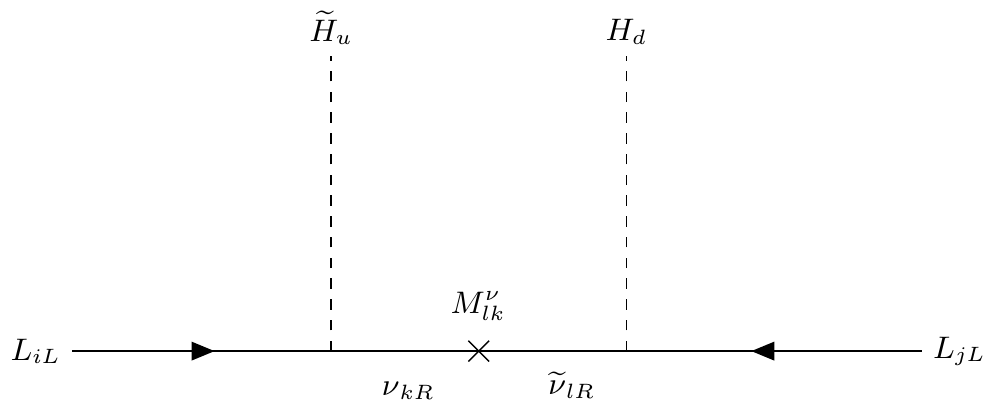}
\end{subfigure}
\caption{Type Ib seesaw diagram~\cite{Hernandez-Garcia:2019uof} which leads to the effective Yukawa
interactions for the Majorana neutrinos in mass insertion formalism,
where $i,j=1,2,3$ and $k,l=4,5$ and $M_{lk}$ is vector-like mass.}
\label{fig:diagrams_neutrinos_mass_insertion}
\end{figure}
\endgroup
The operator $ \overline{L}_i  \overline{L}_j \widetilde{H}_u H_d$ resulting from Figure \ref%
{fig:diagrams_neutrinos_mass_insertion} gives rise to the so 
called type Ib seesaw mechanism~\cite{Hernandez-Garcia:2019uof}
which differs from the usual 
type Ia seesaw mechanism corresponding to the Weinberg operator 
$ \overline{L}_i  \overline{L}_j \widetilde{H}_u\widetilde{H}_u$ and will be discussed later in detail.

\section{Effective Yukawa matrices using a mixing formalism}
\label{III}

\label{sec:higgs_and_flavour_mixing}

As seen from Equation \ref{eqn:the_effective_Yukawa_Lagrangian}, we need to
mix Higgses with the 
flavon to generate the 
effective Yukawa Lagrangian required to produce the SM fermion mass
hierarchy. Since there is no an extra symmetry or constraint to keep the
mixing between Higgses and flavon from taking place, it is natural to assume
their mixing. 

\subsection{The $7\times7$ matrix}

\label{subsec:the_77_matrix}

Consider the $7\times 7$ mass matrix for Dirac fermions:
\begingroup
\begin{equation}
M^{\psi }=\left( 
\begin{array}{c|ccccccc}
& \psi _{1R} & \psi _{2R} & \psi _{3R} & \psi _{4R} & \psi _{5R} & 
\widetilde{\psi }_{4R} & \widetilde{\psi }_{5R} \\ \hline
\overline{\psi }_{1L} & 0 & 0 & 0 & y_{14}^{\psi }\langle \widetilde{H}%
^{0}\rangle & y_{15}^{\psi }\langle \widetilde{H}^{0}\rangle
& x_{14}^{\psi }\langle \phi \rangle & x_{15}^{\psi }\langle
\phi \rangle \\ 
\overline{\psi }_{2L} & 0 & 0 & 0 & y_{24}^{\psi }\langle \widetilde{H}%
^{0}\rangle & y_{25}^{\psi }\langle \widetilde{H}^{0}\rangle
& x_{24}^{\psi }\langle \phi \rangle & x_{25}^{\psi }\langle
\phi \rangle \\ 
\overline{\psi }_{3L} & 0 & 0 & 0 & y_{34}^{\psi }\langle \widetilde{H}%
^{0}\rangle & y_{35}^{\psi }\langle \widetilde{H}^{0}\rangle
& x_{34}^{\psi }\langle \phi \rangle & x_{35}^{\psi }\langle
\phi \rangle \\ 
\overline{\psi }_{4L} & y_{41}^{\psi }\langle \widetilde{H}%
^{0}\rangle & y_{42}^{\psi }\langle \widetilde{H}^{0}\rangle
& y_{43}^{\psi }\langle \widetilde{H}^{0}\rangle & 0 & 0 & 
M_{44}^{\psi } & M_{45}^{\psi } \\ 
\overline{\psi }_{5L} & y_{51}^{\psi }\langle \widetilde{H}%
^{0}\rangle & y_{52}^{\psi }\langle \widetilde{H}^{0}\rangle
& y_{53}^{\psi }\langle \widetilde{H}^{0}\rangle & 0 & 0 & 
M_{54}^{\psi } & M_{55}^{\psi } \\ 
\overline{\widetilde{\psi }}_{4L} & x_{41}^{\psi ^{\prime }}\langle
\phi \rangle & x_{42}^{\psi ^{\prime }}\langle \phi \rangle
& x_{43}^{\psi ^{\prime }}\langle \phi \rangle & M_{44}^{\psi
^{\prime }} & M_{45}^{\psi ^{\prime }} & 0 & 0 \\ 
\overline{\widetilde{\psi }}_{5L} & x_{51}^{\psi ^{\prime }}\langle
\phi \rangle & x_{52}^{\psi ^{\prime }}\langle \phi \rangle
& x_{53}^{\psi ^{\prime }}\langle \phi \rangle & M_{54}^{\psi
^{\prime }} & M_{55}^{\psi ^{\prime }} & 0 & 0
\end{array}%
\right) ,  \label{eqn:general_77_mass_matrix}
\end{equation}
\endgroup
with the coefficients $y$ and $x$ being 
Yukawa constants where the former is
expected to be of order unity whereas 
the latter is 
smaller than 
$y$. Furthermore, the $125$ GeV SM like Higgs boson $H$ will corresponds to the lightest of the CP even neutral scalar states arising from $H_u$, $H_d$ and $\phi$, whereas
$M$ is the
vector-like mass. The column vector located at the lower left block in
Equation \ref{eqn:general_77_mass_matrix} consists of left-handed particles
while the row vector at the upper right block are made up of right-handed
particles. The zeros in the $3\times3$ upper block in Equation \ref%
{eqn:general_77_mass_matrix} mean that no SM Yukawa interactions take place
due to 
 charge conservation as well as zeros in two $2\times2$ blocks. Since
we are interested in explaining the muon and electron anomalous magnetic
moments in this model, we first focus on the lepton sector 
in the next subsection and the method used for obtaining the low energy SM Yukawa matrices in the lepton sector can be
applied to the quark sector in the same way with a slight change so that the
quark sector will be discussed in Appendix~\ref{A}.

\subsection{A convenient basis for charged leptons}

\label{subsec:a_convenient_basis_for_charged_leptons}

From Equation \ref{eqn:general_77_mass_matrix}, we can take a specified
basis by rotating some fields as below:
\begingroup
\begin{equation}
M^{e}=\left( 
\begin{array}{c|ccccccc}
& e_{1R} & e_{2R} & e_{3R} & e_{4R} & e_{5R} & \widetilde{L}_{4R} & 
\widetilde{L}_{5R} \\ \hline
\overline{L}_{1L} & 0 & 0 & 0 & 0 & y_{15}^{e}v_{d} & 0 & x_{15}^{L}v_{\phi }
\\ 
\overline{L}_{2L} & 0 & 0 & 0 & y_{24}^{e}v_{d} & y_{25}^{e}v_{d} & 0 & 
x_{25}^{L}v_{\phi } \\ 
\overline{L}_{3L} & 0 & 0 & 0 & y_{34}^{e}v_{d} & y_{35}^{e}v_{d} & 
x_{34}^{L}v_{\phi } & x_{35}^{L}v_{\phi } \\ 
\overline{L}_{4L} & 0 & 0 & y_{43}^{e}v_{d} & 0 & 0 & M_{44}^{L} & M_{45}^{L}
\\ 
\overline{L}_{5L} & y_{51}^{e}v_{d} & y_{52}^{e}v_{d} & y_{53}^{e}v_{d} & 0
& 0 & 0 & M_{55}^{L} \\ 
\overline{\widetilde{e}}_{4L} & 0 & x_{42}^{e}v_{\phi } & x_{43}^{e}v_{\phi }
& M_{44}^{e} & 0 & 0 & 0 \\ 
\overline{\widetilde{e}}_{5L} & x_{51}^{e}v_{\phi } & x_{52}^{e}v_{\phi } & 
x_{53}^{e}v_{\phi } & M_{54}^{e} & M_{55}^{e} & 0 & 0
\end{array}%
\right) ,  \label{eqn:a_specified_basis_charged_leptons_first}
\end{equation}
\endgroup
where $v_{d}=\left\langle H_{d}^{0}\right\rangle $ and $\nu _{\phi }=$ $%
\left\langle \phi \right\rangle $. We start by pointing out the reason why
we take this specific basis for the charged leptons. The reason is that the
strong 
hierarchical structure of the SM fermion Yukawa couplings can be implemented
by the rotations with a simple assumption in this model to be specified
below. 
In order to arrive from Equation \ref{eqn:general_77_mass_matrix} to
Equation \ref{eqn:a_specified_basis_charged_leptons_first}, we rotate the leptonic fields $%
L_{4L} $ and $L_{5L}$ to turn off $M_{54}^{L}$ and rotate $e_{4R}$ and $%
e_{5R}$ to turn off $M_{45}^{e}$. Then, we can rotate $L_{1L}$ and $L_{3L}$
to set $x_{14}^{L}v_{\phi }$ to zero and then rotate $L_{2L}$ and $L_{3L}$
to set $x_{24}^{L}v_{\phi }$ to zero. The same rotation can be applied to $%
e_{1R,2R,3R}$ to set $y_{41,42}^{e}v_{d}$ to zero. Finally, we can further
rotate $L_{1L}$ and $L_{2L}$ to switch off $y_{14}^{e}v_{d}$ and this
rotation also goes for $e_{1R,2R}$ to switch off $x_{41}^{e}v_{\phi }$. 
The above given mass matrix includes three distinct mass scales which are the vev $v_{d}$
of the neutral component of the Higgs doublet $H_{d}$, the vev $v_{\phi }$
of the flavon $\phi $ and the vector-like masses $M$, whose orders of magnitude can be in principle be different.  
Therefore, the mass matrix will be diagonalized by the seesaw
mechanism step-by-step instead of diagonalising it at once. This mechanism
is also known as Universal Seesaw, and was proposed for the first time, in
the context of a left-right symmetric model in \cite{Davidson:1987mh}.

\subsection{A basis for decoupling heavy fourth and fifth vector-like family}

\label{subsec:a_decoupling_basis}

As mentioned in the previous section \ref%
{subsec:a_convenient_basis_for_charged_leptons}, the mass matrix in Equation \ref%
{eqn:a_specified_basis_charged_leptons_first}
involves
three distinct mass scales $v_{d}$, $v_{\phi }$ and $M$ so it is possible to
split 
this whole mass matrix by partial blocks to group mass terms with vev of $H_{d}$ as
in Equation \ref{eqn:a_specified_basis_charged_leptons_third}
\begingroup
\begin{equation}
M^{e}=\left( 
\begin{array}{c|ccccc|cc}
& e_{1R} & e_{2R} & e_{3R} & e_{4R} & e_{5R} & \widetilde{L}_{4R} & 
\widetilde{L}_{5R} \\ \hline
\overline{L}_{1L} & 0 & 0 & 0 & 0 & y_{15}^{e}v_{d} & 0 & x_{15}^{L}v_{\phi }
\\ 
\overline{L}_{2L} & 0 & 0 & 0 & y_{24}^{e}v_{d} & y_{25}^{e}v_{d} & 0 & 
x_{25}^{L}v_{\phi } \\ 
\overline{L}_{3L} & 0 & 0 & 0 & y_{34}^{e}v_{d} & y_{35}^{e}v_{d} & 
x_{34}^{L}v_{\phi } & x_{35}^{L}v_{\phi } \\ 
\overline{L}_{4L} & 0 & 0 & y_{43}^{e}v_{d} & 0 & 0 & M_{44}^{L} & M_{45}^{L}
\\ 
\overline{L}_{5L} & y_{51}^{e}v_{d} & y_{52}^{e}v_{d} & y_{53}^{e}v_{d} & 0
& 0 & 0 & M_{55}^{L} \\ \hline
\overline{\widetilde{e}}_{4L} & 0 & x_{42}^{e}v_{\phi } & x_{43}^{e}v_{\phi }
& M_{44}^{e} & 0 & 0 & 0 \\ 
\overline{\widetilde{e}}_{5L} & x_{51}^{e}v_{\phi } & x_{52}^{e}v_{\phi } & 
x_{53}^{e}v_{\phi } & M_{54}^{e} & M_{55}^{e} & 0 & 0
\end{array}%
\right) ,  \label{eqn:a_specified_basis_charged_leptons_third}
\end{equation}
\endgroup
and then elements of the blocks involving 
$\phi$ can be rotated away to make
those zeros by unitary mixing matrices of Equation \ref%
{eqn:unitary_mixing_matrix_charged_leptons} as per Equation \ref%
{eqn:basis_decoupling_heavy_VL_fermions}:
\begingroup
\begin{equation}
M^{e }=\left( 
\begin{array}{c|ccccccc}
& e _{1R} & e _{2R} & e _{3R} & e _{4R} & e _{5R} & 
\widetilde{L }_{4R} & \widetilde{L }_{5R} \\ \hline
\overline{L }_{1L} &  &  &  &  &  & 0 & 0 \\ 
\overline{L }_{2L} &  &  &  &  &  & 0 & 0 \\ 
\overline{L }_{3L} &  &  & \widetilde{y}_{\alpha \beta }^{\prime e}v_{d}
&  &  & 0 & 0 \\ 
\overline{L }_{4L} &  &  &  &  &  & \widetilde{M}_{44}^{L} & 
M_{45}^{\prime L} \\ 
\overline{L }_{5L} &  &  &  &  &  & 0 & \widetilde{M}_{55}^{L} \\ 
\overline{\widetilde{e }}_{4L} & 0 & 0 & 0 & \widetilde{M}_{44}^{e} & 0 & 
0 & 0 \\ 
\overline{\widetilde{e }}_{5L} & 0 & 0 & 0 & M_{54}^{\prime \prime e} & \widetilde{M}_{55}^{e} & 0 & 0%
\end{array}%
\right) ,  \label{eqn:basis_decoupling_heavy_VL_fermions}
\end{equation}
\endgroup
where the indices $\alpha,\beta$ run from 1 to 5, and tilde, primes repeated
in the mass matrix mean that the parameters are rotated. The unitary $%
5\times5$ matrices are defined to be
\begingroup
\begin{equation}
V_L = V_{45}^L V_{35}^L V_{25}^L V_{15}^L V_{34}^L V_{24}^L V_{14}^L, 
\hspace{0.5cm} V_{e} = V_{45}^{e} V_{35}^{e} V_{25}^{e} V_{15}^{e}
V_{34}^{e} V_{24}^{e} V_{14}^{e},
\label{eqn:unitary_mixing_matrix_charged_leptons}
\end{equation}
\endgroup
where each of the unitary matrices $V_{i4,5}$ are parameterized by a single
angle $\theta_{i4,5}$ describing the mixing between the $i$th chiral family
and the $4,5$th vector-like family. The $5 \times 5$ Yukawa constant matrix
in a mass basis (primed) can be diagonalized by the unitary rotation
matrices as below:
\begingroup
\begin{equation}
\widetilde{y}_{\alpha\beta}^{e} = V_L \widetilde{y}_{\alpha\beta}^{\prime e}
V_{e}^{\dagger}  \label{eqn:diagonalization_Yukawa_matrices}
\end{equation}
\endgroup
From Equation \ref{eqn:a_specified_basis_charged_leptons_first}, we can read
off the $5 \times 5$ upper block and confirm that the $(3,4), (1,5), (2,5),
(3,5) $ mixings in the $L$ sector and $(2,4), (3,4), (1,5), (2,5), (3,5)$
mixings in the $e$ sector are required to go to the decoupling basis. The
unitary matrices of Equation \ref{eqn:unitary_mixing_matrix_charged_leptons} and
mixing angles appearing in the unitary matrices are parameterized by
\begingroup
\begin{equation}
\begin{split}
V_L &= V_{35}^{L} V_{25}^{L} V_{15}^{L} V_{34}^{L} \\
&= 
\begin{pmatrix}
1 & 0 & 0 & 0 & 0 \\ 
0 & 1 & 0 & 0 & 0 \\ 
0 & 0 & c_{35}^L & 0 & s_{35}^L \\ 
0 & 0 & 0 & 1 & 0 \\ 
0 & 0 & -s_{35}^L & 0 & c_{35}^L%
\end{pmatrix}
\begin{pmatrix}
1 & 0 & 0 & 0 & 0 \\ 
0 & c_{25}^L & 0 & 0 & s_{25}^L \\ 
0 & 0 & 1 & 0 & 0 \\ 
0 & 0 & 0 & 1 & 0 \\ 
0 & -s_{25}^L & 0 & 0 & c_{25}^L%
\end{pmatrix}
\begin{pmatrix}
c_{15}^L & 0 & 0 & 0 & s_{15}^L \\ 
0 & 1 & 0 & 0 & 0 \\ 
0 & 0 & 1 & 0 & 0 \\ 
0 & 0 & 0 & 1 & 0 \\ 
-s_{15}^L & 0 & 0 & 0 & c_{15}^L%
\end{pmatrix}
\\
&\times
\begin{pmatrix}
1 & 0 & 0 & 0 & 0 \\ 
0 & 1 & 0 & 0 & 0 \\ 
0 & 0 & c_{34}^L & s_{34}^L & 0 \\ 
0 & 0 & -s_{34}^L & c_{34}^L & 0 \\ 
0 & 0 & 0 & 0 & 1%
\end{pmatrix}
\approx 
\begin{pmatrix}
1 & 0 & 0 & 0 & s_{15}^L \\ 
0 & 1 & 0 & 0 & s_{25}^L \\ 
0 & 0 & 1 & s_{34}^L & s_{35}^L \\ 
0 & 0 & -s_{34}^L & 1 & 0 \\ 
-s_{15}^L & -s_{25}^L & -s_{35}^L & 0 & 1%
\end{pmatrix}%
, \\
&s_{34}^L = \frac{x_{34}^L \left\langle \phi \right\rangle}{\sqrt{\left(
x_{34}^L \left\langle \phi \right\rangle \right)^2 + \left( M_{44}^L
\right)^2}}, \hspace{0.5cm} s_{15}^L = \frac{x_{15}^L \left\langle \phi
\right\rangle}{\sqrt{\left( x_{15}^L \left\langle \phi \right\rangle
\right)^2 + \left( M_{55}^L \right)^2}}, \\
&s_{25}^L = \frac{x_{25}^L \left\langle \phi \right\rangle}{\sqrt{\left(
x_{25}^L \left\langle \phi \right\rangle \right)^2 + \left( M_{55}^{\prime
L} \right)^2}}, \hspace{0.5cm} s_{35}^L = \frac{x_{35}^{\prime L}
\left\langle \phi \right\rangle}{\sqrt{\left( x_{35}^{\prime L} \left\langle
\phi \right\rangle \right)^2 + \left( M_{55}^{\prime\prime L} \right)^2}}, \\
& x_{35}^{\prime L} \left\langle \phi \right\rangle = c_{34}^{L} x_{35}^{L}
\left\langle \phi \right\rangle + s_{34}^{L} M_{45}^{L}, \hspace{0.5cm}
M_{45}^{\prime L} = -s_{34}^{L} x_{35}^{L} \left\langle \phi \right\rangle +
c_{34}^{L} M_{45}^{L} \\
&\widetilde{M}_{44}^{L} = \sqrt{\left( x_{34}^L \left\langle \phi
\right\rangle \right)^2 + \left( M_{44}^L \right)^2}, \\
&M_{55}^{\prime L} = \sqrt{\left( x_{15}^L \left\langle \phi \right\rangle
\right)^2 + \left( M_{55}^L \right)^2}, \hspace{0.1cm} M_{55}^{\prime\prime
L} = \sqrt{\left( x_{25}^L \left\langle \phi \right\rangle \right)^2 +
\left( M_{55}^{\prime L} \right)^2}, \hspace{0.1cm} 
\\
&\widetilde{M}_{55}^{L} = 
\sqrt{\left( x_{35}^{\prime L} \left\langle \phi \right\rangle \right)^2 +
\left( M_{55}^{\prime\prime L} \right)^2}
\end{split}
\label{eqn:unitary_rotation_VQ}
\end{equation}
\endgroup
\begingroup
\begin{equation}
\begin{split}
V_{e} &= V_{35}^{e} V_{25}^{e} V_{15}^{e} V_{34}^{e} V_{24}^{e} \\
&= 
\begin{pmatrix}
1 & 0 & 0 & 0 & 0 \\ 
0 & 1 & 0 & 0 & 0 \\ 
0 & 0 & c_{35}^{e} & 0 & s_{35}^{e} \\ 
0 & 0 & 0 & 1 & 0 \\ 
0 & 0 & -s_{35}^{e} & 0 & c_{35}^{e}%
\end{pmatrix}
\begin{pmatrix}
1 & 0 & 0 & 0 & 0 \\ 
0 & c_{25}^{e} & 0 & 0 & s_{25}^{e} \\ 
0 & 0 & 1 & 0 & 0 \\ 
0 & 0 & 0 & 1 & 0 \\ 
0 & -s_{25}^{e} & 0 & 0 & c_{25}^{e}%
\end{pmatrix}
\begin{pmatrix}
c_{15}^{e} & 0 & 0 & 0 & s_{15}^{e} \\ 
0 & 1 & 0 & 0 & 0 \\ 
0 & 0 & 1 & 0 & 0 \\ 
0 & 0 & 0 & 1 & 0 \\ 
-s_{15}^{e} & 0 & 0 & 0 & c_{15}^{e}%
\end{pmatrix}
\\
& \times 
\begin{pmatrix}
1 & 0 & 0 & 0 & 0 \\ 
0 & 1 & 0 & 0 & 0 \\ 
0 & 0 & c_{34}^{e} & s_{34}^{e} & 0 \\ 
0 & 0 & -s_{34}^{e} & c_{34}^{e} & 0 \\ 
0 & 0 & 0 & 0 & 1%
\end{pmatrix}
\begin{pmatrix}
1 & 0 & 0 & 0 & 0 \\ 
0 & c_{24}^{e} & 0 & s_{24}^{e} & 0 \\ 
0 & 0 & 1 & 0 & 0 \\ 
0 & -s_{24}^{e} & 0 & c_{24}^{e} & 0 \\ 
0 & 0 & 0 & 0 & 1%
\end{pmatrix}
\approx 
\begin{pmatrix}
1 & 0 & 0 & 0 & \theta_{15}^{e} \\ 
0 & 1 & 0 & \theta_{24}^{e} & \theta_{25}^{e} \\ 
0 & 0 & 1 & \theta_{34}^{e} & \theta_{35}^{e} \\ 
0 & -\theta_{24}^{e} & -\theta_{34}^{e} & 1 & 0 \\ 
-\theta_{15}^{e} & -\theta_{25}^{e} & -\theta_{35}^{e} & 0 & 1%
\end{pmatrix}%
, \\
&s_{24}^{e} \approx \frac{x_{42}^{e} \left\langle \phi \right\rangle}{%
M_{44}^{e}}, \hspace{0.5cm} s_{34}^{e} \approx \frac{x_{43}^{e} \left\langle
\phi \right\rangle}{M_{44}^{\prime e}}, \hspace{0.5cm} s_{15}^{e} \approx \frac{x_{51}^{e}
\left\langle \phi \right\rangle}{M_{55}^{e}}, \hspace{0.5cm} s_{25}^{e}
\approx \frac{x_{52}^{\prime e} \left\langle \phi \right\rangle}{%
M_{55}^{\prime e}}, \hspace{0.5cm} s_{35}^{e} \approx \frac{x_{53}^{e}
\left\langle \phi \right\rangle}{M_{55}^{\prime\prime e}}, \\
& x_{52}^{\prime e} \left\langle \phi \right\rangle = c_{24}^{e} x_{52}^{e}
\left\langle \phi \right\rangle + s_{24}^{e} M_{54}^{e}, \hspace{0.5cm}
M_{54}^{\prime e} = -s_{24}^{e} x_{52}^{e} \left\langle \phi \right\rangle +
c_{24}^{e} M_{54}^{e}, \\
& x_{53}^{\prime e} \left\langle \phi \right\rangle = c_{34}^{e} x_{53}^{e}
\left\langle \phi \right\rangle + s_{34}^{e} M_{54}^{\prime e}, \hspace{0.5cm%
} M_{54}^{\prime\prime e} = -s_{34}^{e} x_{53}^{e} \left\langle \phi
\right\rangle + c_{34}^{e} M_{54}^{\prime e}, \\
&M_{44}^{\prime e} = \sqrt{\left( x_{42}^{e} \left\langle \phi \right\rangle
\right)^2 + \left( M_{44}^{e} \right)^2} \hspace{0.5cm}, \widetilde{M}%
_{44}^{e} = \sqrt{\left( x_{43}^{e} \left\langle \phi \right\rangle
\right)^2 + \left( M_{44}^{e} \right)^2}, \\
&M_{55}^{\prime e} = \sqrt{\left( x_{51}^{e} \left\langle \phi \right\rangle
\right)^2 + \left( M_{55}^{e} \right)^2}, \hspace{0.1cm} M_{55}^{\prime%
\prime e} = \sqrt{\left( x_{52}^{\prime e} \left\langle \phi \right\rangle
\right)^2 + \left( M_{55}^{\prime e} \right)^2}, \hspace{0.1cm} 
\\
&\widetilde{M}%
_{55}^{e} = \sqrt{\left( x_{53}^{\prime e} \left\langle \phi \right\rangle
\right)^2 + \left( M_{55}^{\prime\prime e} \right)^2}.
\end{split}
\label{eqn:unitary_rotation_Vec}
\end{equation}
\endgroup
Given the above unitary rotations, the $5\times5$ Yukawa matrices are
computed in terms of the mixing angles and the upper $3\times3$ block would
be the effective SM Yukawa matrix. Assuming all $\cos\theta$ to be 1 and
neglecting order of $\theta$ square or more than that, we have a simple $3
\times 3$ Yukawa matrix of Equation \ref%
{eqn:effective_Yukawa_constant_decoupling_basis_charged_leptons}.
\begingroup
\begin{equation}
\begin{split}
y_{ij}^e &= 
\begin{pmatrix}
s_{15}^L y_{51}^e + y_{15}^e \theta_{15}^{e} & s_{15}^L y_{52}^e + y_{15}^e
\theta_{25}^{e} & s_{15}^L y_{53}^e + y_{15}^e \theta_{35}^{e} \\ 
s_{25}^L y_{51}^e + y_{25}^e \theta_{15}^{e} & s_{25}^L y_{52}^e + y_{24}^e
\theta_{24}^{e} + y_{25}^e \theta_{25}^{e} & s_{25}^L y_{53}^e + y_{24}^e
\theta_{34}^{e} + y_{25}^e \theta_{35}^{e} \\ 
s_{35}^L y_{51}^e + y_{35}^e \theta_{15}^{e} & s_{35}^L y_{52}^e + y_{34}^e
\theta_{24}^{e} + y_{35}^e \theta_{25}^{e} & s_{34}^L y_{43}^e + s_{35}^L
y_{53}^e + y_{34}^e \theta_{34}^{e} + y_{35}^e \theta_{35}^{e}%
\end{pmatrix}%
\end{split}
\label{eqn:effective_Yukawa_constant_decoupling_basis_charged_leptons}
\end{equation}
\endgroup

\subsection{A convenient basis for neutrinos}

\label{subsec:a_convenient_basis_neutrinos}

The relevant Yukawa and mass terms of the neutrino sector give rise to the
following neutrino mass matrix:
\begingroup 
\begin{equation}
M^{\nu }=\left( 
\begin{array}{c|ccc|cccc}
& L_{1L} & L_{2L} & L_{3L} & \overline{\nu }_{4R} & \overline{\nu }_{5R} & 
\widetilde{\nu }_{4R} & \widetilde{\nu }_{5R} \\ \hline
L_{1L} & 0 & 0 & 0 & y_{14}^{\nu }v_{u} & y_{15}^{\nu }v_{u} & 
x_{14}^{L}v_{d} & x_{15}^{L}v_{d} \\ 
L_{2L} & 0 & 0 & 0 & y_{24}^{\nu }v_{u} & y_{25}^{\nu }v_{u} & 
x_{24}^{L}v_{d} & x_{25}^{L}v_{d} \\ 
L_{3L} & 0 & 0 & 0 & y_{34}^{\nu }v_{u} & y_{35}^{\nu }v_{u} & 
x_{34}^{L}v_{d} & x_{35}^{L}v_{d} \\ \hline
\overline{\nu }_{4R} & y_{14}^{\nu }v_{u} & y_{24}^{\nu }v_{u} & y_{34}^{\nu
}v_{u} & 0 & 0 & M_{44}^{\nu } & M_{54}^{\nu } \\ 
\overline{\nu }_{5R} & y_{15}^{\nu }v_{u} & y_{25}^{\nu }v_{u} & y_{35}^{\nu
}v_{u} & 0 & 0 & M_{45}^{\nu } & M_{55}^{\nu } \\ 
\widetilde{\nu }_{4R} & x_{14}^{L}v_{d} & x_{24}^{L}v_{d} & x_{34}^{L}v_{d}
& M_{44}^{\nu } & M_{45}^{\nu } & 0 & 0 \\ 
\widetilde{\nu }_{5R} & x_{15}^{L}v_{d} & x_{25}^{L}v_{d} & x_{35}^{L}v_{d}
& M_{54}^{\nu } & M_{55}^{\nu } & 0 & 0
\end{array}%
\right)  \label{eqn:mass_matrix_neutrinos}
\end{equation}
\endgroup
Here, the zeros in the upper $3 \times 3$ block of Equation \ref%
{eqn:mass_matrix_neutrinos} mean that neutrinos remain massless in the SM.
Therefore, the SM neutrinos can be massive via the inclusion 
of two
vector-like families. In order to make this mass matrix as simple as
possible, the only choice left is to rotate $\nu_{4R}$ and $\nu_{5R}$ to
turn off $M_{45}^{\nu}$ since rotations between $L_{1L,2L,3L}$ are already
used in the charged lepton sector. 
\chapter{Non-standard contributions to the muon and electron anomalous magnetic moments : $W$ gauge boson exchange} \label{Chapter:The2ndpaper}
In the second BSM model covered in chapter~\ref{Chapter:The2ndBSMmodel}, we discuss the first non-SM contributions to the muon and electron anomalous magnetic moments $g-2$ with the SM $W$ gauge boson at one-loop level via the type $1$b seesaw mechanism.
\section{$W$ boson exchange contributions to the $\left( g-2 \right)_\protect\mu, \left( g-2
\right)_e$ and $\func{BR}\left(\protect\mu \rightarrow e \protect\gamma %
\right)$}
\label{IV}

\label{sec:Analytic_arguments_muon_electron_g2_W} 
Within the framework of our proposed model, we start by investigating the
muon and electron anomalous magnetic moments with $W$ boson exchange first. 
Given that such $W$ boson exchange contribution also involves virtual
neutrinos in the internal lines of the loop, we revisit the mass matrix for
neutrinos. In this mass matrix, we remove fifth vector-like neutrinos $%
\nu_{5R}$ and $\widetilde{\nu}_{5R}$ since they are too heavy to contribute
to the phenomenology under study. As mentioned in the previous section, we
stick to a condition where the coefficient $y$ is expected to be of order
unity, 
whereas the coupling $x$ is expected to be smaller than $y$. 
Such condition can be easily seen by substituting the coefficients $%
y_{i4}^\nu$ by $y_{i}^\nu$ and the coefficients $x_{i4}^L$ by $\epsilon
y_{i}^{\nu \prime}$ where $\epsilon$ is a suppression factor. Putting all
these considerations together, the mass matrix for neutrinos in Equation \ref%
{eqn:mass_matrix_neutrinos} after electroweak symmetry breaking takes the
form:
\begingroup
\begin{equation}
M^\nu \approx \left( 
\begin{array}{c|ccc|cc}
& \nu_{1L} & \nu_{2L} & \nu_{3L} & \overline{\nu}_{4R} & \widetilde{\nu}_{4R}
\\ \hline
\nu_{1L} & 0 & 0 & 0 & y_{1}^\nu v_u & \epsilon y_{1}^{\nu \prime} v_d \\ 
\nu_{2L} & 0 & 0 & 0 & y_{2}^\nu v_u & \epsilon y_{2}^{\nu \prime} v_d \\ 
\nu_{3L} & 0 & 0 & 0 & y_{3}^\nu v_u & \epsilon y_{3}^{\nu \prime} v_d \\ 
\hline
\overline{\nu}_{4R} & y_{1}^\nu v_u & y_{2}^\nu v_u & y_{3}^\nu v_u & 0 & 
M_{44}^{\nu} \\ 
\widetilde{\nu}_{4R} & \epsilon y_{1}^{\nu \prime} v_d & \epsilon y_{2}^{\nu
\prime} v_d & \epsilon y_{3}^{\nu \prime} v_d & M_{44}^{\nu} & 0%
\end{array}
\right) \equiv \left( 
\begin{array}{cc}
0 & m_D \\ 
m_D^T & M_N%
\end{array}
\right) , \label{eqn:mass_matrix_neutrinos_reduced}
\end{equation}
\endgroup
where $v_u(v_d)$ is the vev of $\widetilde{H}_u(H_d)$, $v_u$ runs from $246/%
\sqrt{2}\func{GeV} \simeq 174 \func{GeV}$ to $246 \func{GeV}$ and $%
v_u^2+v_d^2 = \left( 246 \func{GeV} \right)^2$.

\subsection{Type 1b seesaw mechanism}

\label{subsec:type_1b_seesaw_mechanism}

Now that we constructed the neutrino mass matrix for this task, the next
step is to read off the operator which gives rise to the neutrino mass from
the mass matrix. Generally, the well-known operator for neutrino mass is the
Weinberg operator(type 1a seesaw mechanism) $\frac{1}{\Lambda}L_i L_j H H$.
A main feature of the Weinberg operator is the same SM Higgs should be repeated
in the operator, however that 
property is not present in our model 
since the Higgs doublets $H_{u,d}$ are negatively charged under the $%
U(1)^{\prime}$ symmetry, which implies 
the corresponding Weinberg operator having such fields will not be invariant
under the $U(1)^{\prime}$ unless an insertion of a quadratic power of the
gauge singlet scalar $\phi$ is considered. However we do not consider the
operators $\frac{1}{\Lambda^3}(\bar{L}_i\tilde{H}_u)(\tilde{H}%
_uL^C_j)(\phi^*)^2$ and $\frac{1}{\Lambda^3}(\bar{L}_iH_d)(H_dL^C_j)\phi^2$
in the neutrino sector, since they are very subleading and thus will give a
tiny contribution to the light active neutrino masses. 
Instead of relying on a seven dimensional Weinberg to generate the tiny
masses for the light active neutrinos, we take another approach named type
1b seesaw mechanism (we call the Weinberg operator ``type 1a seesaw
mechanism" to differentiate with) 
where the mixing of different $SU(2)$ Higgs doublets can appear satisfying
charge conservation. Diagrams for the operators are given in Figure \ref%
{fig:diagrams_type1ab_operators} for comparison:
\begingroup
\begin{figure}[]
\centering
\begin{subfigure}{0.48\textwidth}
	\includegraphics[width=1.0\textwidth]{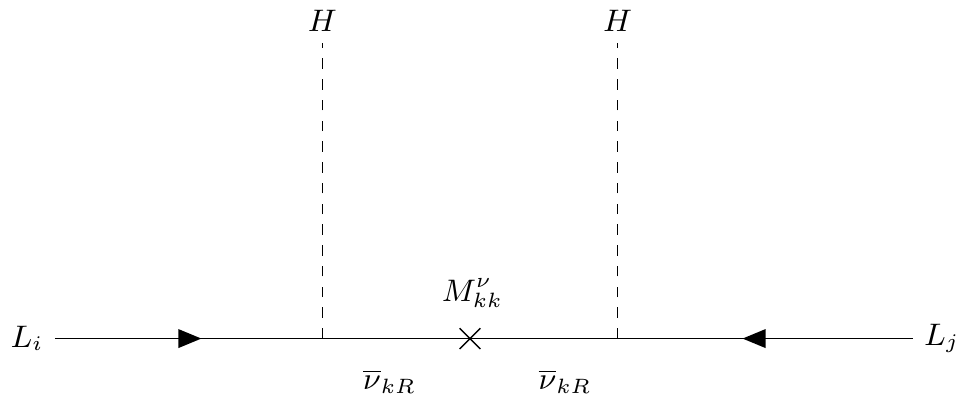}
\end{subfigure} \hspace{0.1cm} 
\begin{subfigure}{0.48\textwidth}
	\includegraphics[width=1.0\textwidth]{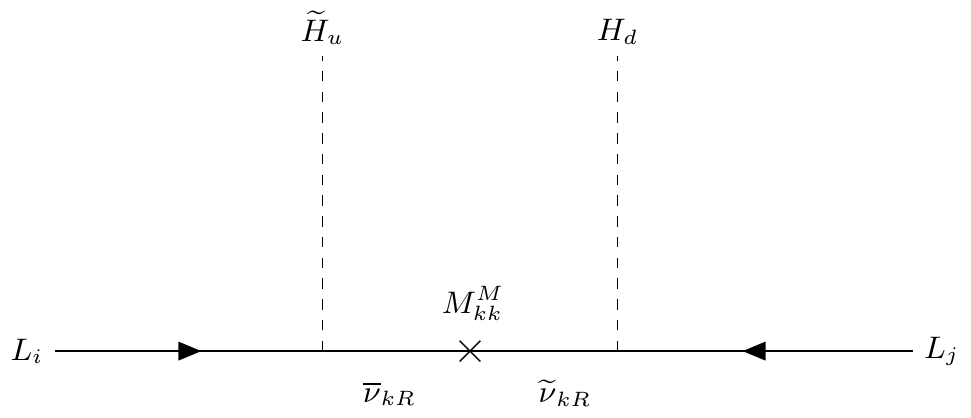}
\end{subfigure}
\caption{Diagrams which lead to effective Weinberg operators for the
Majorana and vector-like mass in the mass insertion formalism, where $%
i,j=1,2,3$ and $k=4$, respectively. The left is the Weinberg operator(or
type 1a seesaw mechanism) in which mass $M$ is Majorana mass and the right
is Weinberg-like operator(or type 1b seesaw mechanism) in which mass $M$ is
vector-like mass.}
\label{fig:diagrams_type1ab_operators}
\end{figure}
\endgroup
The diagrams in Figure \ref{fig:diagrams_type1ab_operators} clearly tell the
difference between Majorana mass and vector-like mass. They share a common
property that they violate the lepton number conservation, 
whereas the particles appearing in 
a Majorana mass term are same but those ones involved in vector-like mass
terms are different. As the type 1b seesaw mechanism only works in this
model, we make use of this seesaw mechanism for the analysis of neutrinos.
With the operator, the renormalizable Lagrangian for neutrinos can be
written as:
\begingroup
\begin{equation}
\begin{split}
\mathcal{L}_{\nu}^{\func{Yukawa+Mass}} = y_{i}^{\nu} \overline{L}_{iL} 
\widetilde{H}_u \nu_{kR} + \epsilon y_i^{\nu\prime} \overline{L}_{iL} H_d 
\overline{\widetilde{\nu}}_{kR} + M_{kk}^{M} \overline{\widetilde{\nu}}_{kR}
\nu_{kR} + \func{h.c.} , \label{eqn:Lagrangian_neutrinos}
\end{split}%
\end{equation}
\endgroup
where $i = 1,2,3$ and $k=4$. The renormalizable Lagrangian of Equation %
\ref{eqn:Lagrangian_neutrinos} above the electroweak scale 
generates an effective Lagrangian after decoupling the heavy vector-like
neutrinos, which is suitable for study of low energy neutrino phenomenology.
The effective Lagrangian for neutrino at electroweak scale is given by\cite%
{Hernandez-Garcia:2019uof}
\begingroup
\begin{equation}
\mathcal{L}^{d=5} = c_{ij}^{d=5} \left( \left( L_i^T \widetilde{H}_u \right)
\left( H_d^T L_j \right) + \left( L_i^T H_d \right) \left( \widetilde{H}_u^T
L_j \right) \right),  \label{eqn:effective_operators_neutrinos}
\end{equation}
\endgroup
where the coefficient $c_{ij}^{d=5}$ is suppressed by a factor of the
vector-like mass $M$. The neutrino mass matrix of Equation \ref%
{eqn:mass_matrix_neutrinos_reduced} can be diagonalized by the unitary
matrix $U$ as below:
\begingroup
\begin{equation}
U^T 
\begin{pmatrix}
0 & m_D^T \\ 
m_D & M_N%
\end{pmatrix}
U = 
\begin{pmatrix}
m_\nu^{\func{diag}} & 0 \\ 
0 & M_N^{\func{diag}}%
\end{pmatrix},
\label{eqn:diagonalization_neutrinos}
\end{equation}
\endgroup
where $m_\nu^{\func{diag}}$ is a diagonal matrix for the light left-handed
neutrinos $\nu_{iL}$ and $M_N^{\func{diag}}$ is that for the heavy
vector-like neutrinos $\nu_{4R}, \widetilde{\nu}_{4R}$. Here, the unitary
mixing matrix $U$ is defined by multiplication of two unitary matrices which we call $U_A$ and $U_B$, respectively\cite{Blennow:2011vn}:
\begingroup
\begin{equation}
\begin{split}
U &= U_A \cdot U_B \\
U_A &= \func{exp}%
\begin{pmatrix}
0 & \Theta \\ 
-\Theta^\dagger & 0%
\end{pmatrix}
\simeq 
\begin{pmatrix}
I-\frac{\Theta \Theta^\dagger}{2} & \Theta \\ 
-\Theta^\dagger & I-\frac{\Theta \Theta^\dagger}{2}%
\end{pmatrix}
\quad \text{at leading order in $\Theta$} \\
U_B &= 
\begin{pmatrix}
U_{\func{PMNS}} & 0 \\ 
0 & I%
\end{pmatrix}%
\end{split}
\label{eqn:unitary_matrix_neutrinos}
\end{equation}
\endgroup
The unitary matrix $U_{\func{PMNS}}$ in $U_B$ is the well-known
Pontecorvo-Maki-Nakagawa-Sakata matrix and is parameterized by \cite%
{Hernandez-Garcia:2019uof,Chau:1984fp}
\begingroup
\begin{equation}
\begin{split}
U_{\func{PMNS}} 
&= 
\begin{pmatrix}
1 & 0 & 0 \\ 
0 & \cos\theta_{23} & \sin\theta_{23} \\ 
0 & -\sin\theta_{23} & \cos\theta_{23}%
\end{pmatrix}
\begin{pmatrix}
\cos\theta_{13} & 0 & \sin\theta_{13} e^{-i\delta_{\func{CP}}} \\ 
0 & 1 & 0 \\ 
-\sin\theta_{13} e^{i\delta_{\func{CP}}} & 0 & \cos\theta_{13}%
\end{pmatrix}
\\
&\times
\begin{pmatrix}
\cos\theta_{12} & \sin\theta_{12} & 0 \\ 
-\sin\theta_{12} & \cos\theta_{12} & 0 \\ 
0 & 0 & 1%
\end{pmatrix}
\begin{pmatrix}
e^{-i\alpha^\prime/2} & 0 & 0 \\ 
0 & e^{-i\alpha/2} & 0 \\ 
0 & 0 & 1%
\end{pmatrix}%
,  \label{eqn:PMNS_mixing_matrix}
\end{split}
\end{equation}
\endgroup
where the Majorana phase $\alpha^\prime$ is set to zero in this model. The mixing matrices $%
U_{A,B}$ are unitary, however the $3\times 3$ upper block of the unitary
matrix $U$ is not unitary due to the 
factor $\left( I - \Theta \Theta^\dagger/2 \right)$ for the light neutrinos. An
interesting feature of the unitary matrix $U$ is it is unitary globally, but
non-unitary locally and this non-unitarity contributes to explain muon and
electron anomalous magnetic moments. Replacing the unitary matrices in
Equation \ref{eqn:unitary_matrix_neutrinos} back to Equation \ref%
{eqn:diagonalization_neutrinos}, the result is simplified with the
assumption $M_N \gg m_D$ to the conventional seesaw mechanism:
\begingroup
\begin{equation}
\begin{split}
\Theta &\simeq m_D^\dagger M_N^{-1} \\
U_{\func{PMNS}}^* m_{\nu}^{\func{diag}} U_{\func{PMNS}}^\dagger &\simeq
-m_D^T M_N^{-1} m_D \equiv -m \\
M_N^{\func{diag}} &\simeq M_N,  \label{eqn:parameterisation_neutrinos}
\end{split}%
\end{equation}
\endgroup
where $m$ is the effective mass matrix resulted from Equation \ref%
{eqn:mass_matrix_neutrinos_reduced}.
\begingroup
\begin{equation}
m_{ij} = \frac{\epsilon v_u v_d}{M_{44}^{\nu}} \left( y_i^\nu
y_j^{\nu\prime} + y_i^{\nu\prime} y_j^\nu \right)
\label{eqn:effective_neutrino_mass_matrix}
\end{equation}
\endgroup
Therefore, smallness of the light neutrino masses can be understood not only
from mass of vector-like mass $M_{44}^{\nu}$ but also from the suppression
factor $\epsilon$ and the presence of $\epsilon$ allows more flexibility in the allowed mass values of the vector-like neutrinos. Revisiting non-unitarity part for the light neutrinos from
the unitary matrix $U$~\cite{Blennow:2011vn,Fernandez-Martinez:2007iaa}, it
reads: 
\begingroup
\begin{equation}
\begin{split}
&\left( I - \frac{\Theta_i \Theta_j^\dagger}{2} \right) U_{\func{PMNS}} =
\left( I - \eta_{ij} \right) U_{\func{PMNS}}
\label{eqn:non_unitarity_light_neutrinos}
\end{split}%
\end{equation}
\endgroup
The non-unitarity $\eta$ is associated with the presence of the heavy vector-like
neutrinos and can be derived from a coefficient of the effective Lagrangian
at dimension 6\cite{Broncano:2002rw}:
\begingroup
\begin{equation}
\mathcal{L}^{d=6} = c_{ij}^{d=6} \left( \left( L_i^\dagger \widetilde{H}_u
\right) i\slashed{\partial} \left( \widetilde{H}_u^\dagger L_j \right) +
\left( L_i^\dagger H_d \right) i\slashed{\partial} \left( H_d^\dagger L_j
\right) \right)  \label{eqn:effective_operators_dimension6}
\end{equation}
\endgroup
Once the SM Higgs doublets in Equation \ref%
{eqn:effective_operators_dimension6} develop its vev, the Lagrangian at
dimension 6 causes non-diagonal kinetic terms for the light neutrinos and it
gives rise to deviations of unitarity when it is diagonalized. The
deviations of unitarity can be expressed in terms of the coefficient at
dimension 6 $\eta_{ij} \equiv v^2 c_{ij}^{d=6}/2$.
\begingroup
\begin{equation}
\eta_{ij} = \frac{\Theta_i \Theta_j^\dagger}{2} = \frac{1}{2} \frac{%
m_D^\dagger m_D}{M_N^2} = \frac{1}{2M_{44}^{\nu 2}} \left( v_u^2 y_i^{\nu*}
y_j^\nu + \epsilon^2 v_d^2 y_i^{\nu\prime*} y_j^{\nu\prime} \right) \simeq 
\frac{v_u^2}{2M_{44}^{\nu 2}} y_i^{\nu *} y_j^{\nu}
\label{eqn:deviation_of_unitarity}
\end{equation}
\endgroup
From the fourth term in Equation \ref{eqn:deviation_of_unitarity}, the term
with $\epsilon^2$ can be safely ignored due to both relative
smallness of $v_d$ and the suppression factor $\epsilon$. Thus, the
deviation of unitarity $\eta$ consists of the vector-like mass $M_{44}^{\nu}$
and the Yukawa couplings $y_{i,j}^\nu$. As an interesting example, it is
possible that the Yukawa couplings $y_{i,j}^\nu$ can be obtained from the
observables such as the PMNS mixing matrix and two mass squared splitting, $%
\Delta m_{sol}^2$ and $\Delta m_{atm}^2$, in the neutrino oscillation
experiments. Since the hierarchy between the light neutrinos is not yet
determined, there are two possible scenarios, normal hierarchy(NH) and
inverted hierarchy(IH), and the lightest neutrino remains massless, whereas two
other neutrinos get massive. The Yukawa couplings $y_{i}^{\nu,\nu\prime}$
for the NH($m_1 = 0$) are determined by
\begingroup
\begin{equation}
\begin{split}
y_i^\nu &= \frac{y}{\sqrt{2}} \left( \sqrt{1+\rho} \left( U_{\func{PMNS}}^*
\right)_{i3} + \sqrt{1-\rho} \left( U_{\func{PMNS}}^* \right)_{i2} \right) \\
y_i^{\nu\prime} &= \frac{y^\prime}{\sqrt{2}} \left( \sqrt{1+\rho} \left( U_{%
\func{PMNS}}^* \right)_{i3} - \sqrt{1-\rho} \left( U_{\func{PMNS}}^*
\right)_{i2} \right),
\end{split}%
\end{equation}
\endgroup
where $y$ and $y^\prime$ are real numbers and $\rho=(1-\sqrt{r})/(1+\sqrt{r}%
) $ with $r \equiv \lvert \Delta m_{sol}^2 \rvert/\lvert \Delta m_{atm}^2
\rvert = \Delta m_{21}^2/\Delta m_{31}^2$, whereas the Yukawa couplings $y_i^{\nu,\nu\prime}$ for the IN($m_3=0$) are
\begingroup
\begin{equation}
\begin{split}
y_i^\nu &= \frac{y}{\sqrt{2}} \left( \sqrt{1+\rho} \left( U_{\func{PMNS}}^*
\right)_{i2} + \sqrt{1-\rho} \left( U_{\func{PMNS}}^* \right)_{i1} \right) \\
y_i^{\nu\prime} &= \frac{y^\prime}{\sqrt{2}} \left( \sqrt{1+\rho} \left( U_{%
\func{PMNS}}^* \right)_{i2} - \sqrt{1-\rho} \left( U_{\func{PMNS}}^*
\right)_{i1} \right),
\end{split}%
\end{equation}
\endgroup
where $\rho=(1-\sqrt{1+r})/(1+\sqrt{1+r})$ with $r \equiv \lvert \Delta
m_{sol}^2 \rvert/\lvert \Delta m_{atm}^2 \rvert = \Delta m_{21}^2/\Delta
m_{32}^2$.

\subsection{The charged lepton flavour violation(CLFV) $\protect\mu %
\rightarrow e \protect\gamma$ decay}

\label{subsec:CLFV_muegamma_decay}

\begin{figure}[]
\centering
\begin{subfigure}{0.48\textwidth}
	\includegraphics[width=1.0\textwidth]{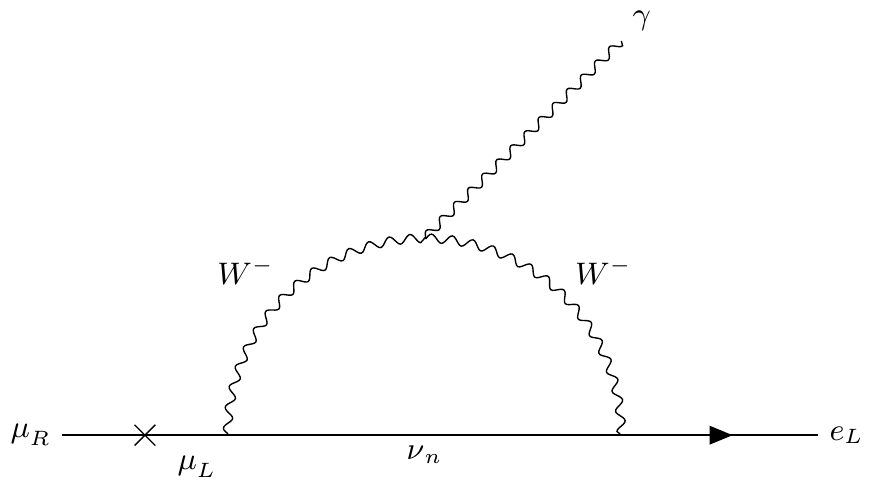}
\end{subfigure} \hspace{0.1cm}
\begin{subfigure}{0.48\textwidth}
	\includegraphics[width=1.0\textwidth]{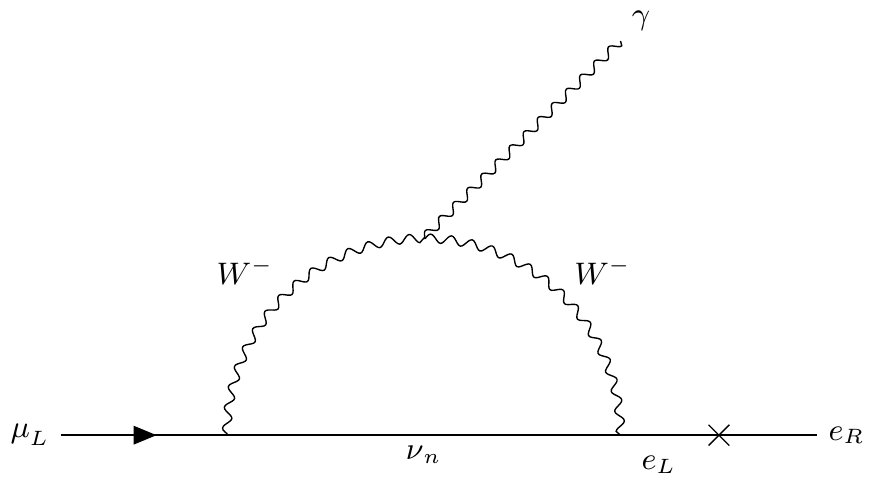}
\end{subfigure}
\caption{Diagrams for CLFV $\protect\mu \rightarrow e \protect\gamma$ decay
with all neutrinos. Here $n=1,2,3,4,5$.}
\label{fig:diagrams_muegamma_with_allneutrinos}
\end{figure}

Consider the three light neutrinos in the SM for the CLFV $\mu \rightarrow e \gamma$
decay first. In this case, the unitary mixing matrix becomes just the PMNS
mixing matrix and the GIM mechanism which suppresses flavour-changing
process works, therefore it leads quite suppressed sensitivity for $\func{BR}%
\left(\mu \rightarrow e \gamma\right)$ about $10^{-55}$\cite{Calibbi:2017uvl}%
, which is impossible to observe with the current sensitivity of $\mu
\rightarrow e \gamma$ decay. This impractical sensitivity can be enhanced to
the observable level by introducing the heavy vector-like neutrinos which give
rise to deviation of unitarity. With the presence of heavy vector-like
neutrinos, the GIM mechanism is gone and the factor suppressed by GIM
mechanism can survive with a factor of deviation of unitarity, which plays a
crucial role to increase significantly order of theoretical prediction for $%
\mu \rightarrow e \gamma$ decay\cite{Glashow:1970gm}. Therefore, the
strongest constraint for deviation of unitarity in the modified PMNS mixing
matrix comes from CLFV $\mu \rightarrow e \gamma$ decay. The possible one-loop
diagrams for the CLFV $\mu \rightarrow e \gamma$ with all neutrinos in this
model are given in Figure \ref{fig:diagrams_muegamma_with_allneutrinos}.

The amplitude from above diagrams in Figure \ref%
{fig:diagrams_muegamma_with_allneutrinos} reads\cite{Calibbi:2017uvl}:
\begingroup
\begin{equation}
\begin{split}
\mathcal{M}\left( \mu \rightarrow e \gamma \right) &= \overline{u}_e i
\sigma_{\mu\nu} q^{\nu} \left( F_1 + F_2 \gamma^5 \right) u_\mu
\epsilon^{*\mu} \\
&= \overline{u}_e i \sigma_{\mu\nu} q^{\nu} \left( A_R P_R + A_L P_L \right)
u_\mu \epsilon^{* \mu},  \label{eqn:amplitude_muegamma_W}
\end{split}%
\end{equation}
\endgroup
where $u$ is Dirac spinor for the muon and electron, $q$ is
four momentum of an outgoing photon, $F_{1,2}$ are form factors, $%
A_{L,R}$ are left- and right-handed amplitude defined to be $A_{L,R} = F_1
\pm F_2$ and lastly $P_{L,R}$ are projection operators. From the amplitude,
the helicity flip between initial particle and final particle should arise
and this makes the helicity flip process takes place on one of external legs
since the $W$ gauge boson couples only to left-handed fields. Comparing the
left diagram with the right, the left is proportional to the muon mass,
while the right is proportional to the electron mass, which means that
impact of the right is ignorable. The unpolarized squared amplitude $\lvert 
\mathcal{M} \rvert^2$ takes the form:
\begingroup
\begin{equation}
\lvert \mathcal{M} \rvert^2 = m_\mu^4 \left( A_R + A_L \right)^2 \simeq
m_\mu^4 \left( A_R \right)^2  \label{eqn:unpolarised_squared_amplitude}
\end{equation}
\endgroup
Then, the decay rate is given by
\begingroup
\begin{equation}
\Gamma\left( \mu \rightarrow e \gamma \right) = \frac{\lvert \mathcal{M}
\rvert^2}{16\pi m_\mu} = \frac{m_\mu^3}{16\pi} \lvert A_R \rvert^2
\label{eqn:decay_rate_muegamma}
\end{equation}
\endgroup
where $A_R$ is expressed by~\cite{Cheng:1984vwu,Calibbi:2017uvl}\footnote{%
Since the PMNS mixing matrix is multiplied by a factor of deviation of
unitarity, it is not unitary any more. Therefore, the first term of sum over
neutrino eigenstates in Equation (28) of \cite{Calibbi:2017uvl} does not
vanish and come in our prediction with a loop function $F(x_n)$.}
\begingroup
\begin{equation}
A_R = \frac{g^2 e}{128\pi^2} \frac{m_\mu}{M_W^2} \sum_{n=1,2,3,4,5} U_{2 n}
U_{1 n}^* F(x_n) \left[ 1 - \frac{1}{3} \frac{\ln \xi}{\xi-1} + \frac{1}{%
\xi-1} \left( \frac{\xi \ln \xi}{\xi-1} - 1 \right) \right]
\label{eqn:left_handed_amplitdue_AR}
\end{equation}
\endgroup
Taking the unitary gauge into account, $\xi \rightarrow \infty$, the
additional $\xi$-dependent terms in $A_R$ all are cancelled by contribution
of Goldstone bosons so $A_R$ is gauge invariant. Substituting the gauge
invariant $A_R$ back into the decay rate of Equation \ref%
{eqn:decay_rate_muegamma} and dividing the expanded decay rate by the total
muon decay rate $\Gamma\left(\mu \rightarrow e \nu \overline{\nu} \right) =
G_F^2 m_\mu^5/192\pi^3$, we have the prediction for $\mu \rightarrow e
\gamma $ decay\cite{Hernandez-Garcia:2019uof,Calibbi:2017uvl}:
\begingroup
\begin{equation}
\func{BR}\left(\mu \rightarrow e \gamma \right) = \frac{\Gamma \left( \mu
\rightarrow e \gamma \right)}{\Gamma \left( \mu \rightarrow e \nu_\mu 
\overline{\nu}_e \right)} = \frac{3\alpha}{32\pi} \frac{\lvert \sum_{n=1}^5
U_{2n} U_{n1}^{\dagger} F(x_n) \rvert^2}{\left( U U^{\dagger} \right)_{11}
\left( U U^{\dagger} \right)_{22}} , \label{eqn:prediction_muegamma_neutrinos}
\end{equation}
\endgroup
where $x_n = M_n^2/M_W^2$ and the loop function $F(x_n)$ is
\begingroup
\begin{equation}
F(x_n) = \frac{10 - 43 x_n + 78 x_n^2 - (49 - 18 \log x_n) x_n^3 + 4 x_n^4}{%
3(x_n-1)^4}.  \label{eqn:loop_function_F}
\end{equation}
\endgroup
Numerator in Equation \ref{eqn:prediction_muegamma_neutrinos} can be
simplified by separating the light neutrinos and heavy vector-like neutrinos as
below(Contribution of the fifth neutrino $\widetilde{\nu}_{4R}$ is safely ignored both by the
suppression factor $\epsilon$ and by relative smallness of $v_d$ compared to 
$v_u$):
\begingroup
\begin{equation}
\begin{split}
\lvert \sum_{n=1}^5 U_{2n} U_{n1}^{\dagger} F(x_n) \rvert^2 &\simeq \lvert
U_{2i} U_{i1}^\dagger F(0) + U_{24} U_{41}^\dagger F(x_4) \rvert^2 \\
U_{2i} U_{i1}^{\dagger} &= -\eta_{12}^* - \eta_{21} = -2\eta_{21} \\
U_{24} U_{41}^{\dagger} &= \Theta_{24} \Theta_{14}^* = 2\eta_{21} \\
\lvert \sum_{n=1}^5 U_{2n} U_{n1}^{\dagger} F(x_n) \rvert^2 &\simeq \lvert
4\eta_{21} \rvert^2 \left( F(x_4) - F(x_0) \right)^2
\label{eqn:numerator_simplified}
\end{split}%
\end{equation}
\endgroup
The final form for the CLFV $\mu \rightarrow e \gamma$ decay in this model reads:
\begingroup
\begin{equation}
\func{BR}\left(\mu \rightarrow e \gamma \right) = \frac{3\alpha_{\func{em}}}{%
8\pi} \lvert \eta_{21} \rvert^2 \left( F(x_4) - F(0) \right)^2,
\label{eqn:prediction_muegamma_final}
\end{equation}
\endgroup
where $\alpha_{\func{em}}$ is the fine structure constant. We find that our
theoretical prediction for the $\mu \rightarrow e \gamma$ decay can be expressed
in terms of the deviation of unitarity $\eta_{21}$.

\subsection{The anomalous muon magnetic moment $g-2$}

\label{subsec:muon_g2_Wboson}

We derive our prediction for the muon anomalous magnetic moment in this
section and confirm the derived expression can be consistent with an
expression of the theoretical prediction for $\mu \rightarrow e \gamma$ in
references\cite{Calibbi:2017uvl,Lindner:2016bgg}. Consider two possible
diagrams for muon anomalous magnetic moment at one-loop level in Figure \ref%
{fig:diagrams_muong2_with_allneutrinos}.
\begingroup
\begin{figure}[]
\centering
\begin{subfigure}{0.48\textwidth}
	\includegraphics[width=1.0\textwidth]{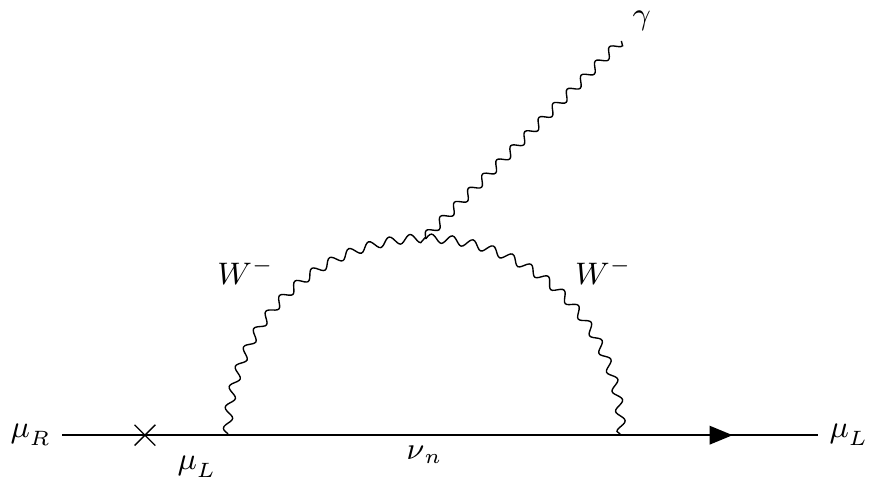}
\end{subfigure} \hspace{0.1cm}
\begin{subfigure}{0.48\textwidth}
	\includegraphics[width=1.0\textwidth]{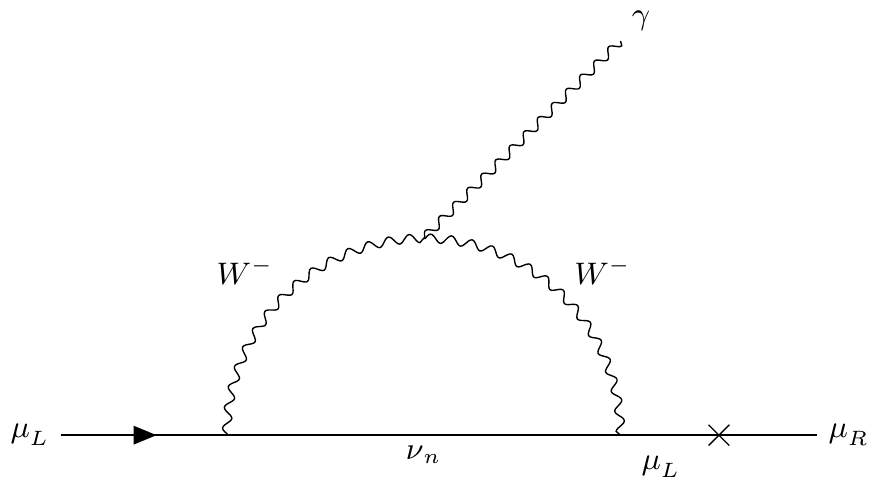}
\end{subfigure}
\caption{Diagrams for muon anomalous magnetic moment with all neutrinos. Here $n=1,2,3,4,5$.}
\label{fig:diagrams_muong2_with_allneutrinos}
\end{figure}
\endgroup
The amplitude for the muon anomalous magnetic moment at one-loop level is:
\begingroup
\begin{equation}
\begin{split}
\mathcal{M}\left( \Delta a_\mu \right) &= \overline{u}_\mu i \sigma_{\mu\nu}
q^{\nu} \left( F_1 + F_2 \gamma^5 \right) u_\mu \epsilon^{*\mu} \\
&= \overline{u}_\mu i \sigma_{\mu\nu} q^{\nu} \left( A_R P_R + A_L P_L
\right) u_\mu \epsilon^{* \mu}  \label{eqn:amplitude_muong2_W}
\end{split}%
\end{equation}
\endgroup
Unlike the CLFV $\mu \rightarrow e \gamma$ decay, muon anomaly diagrams have
the same structure for helicity flip process. So we conclude $A_R$ is equal
to $A_L$ and can make use of other expression of this amplitude to derive
our own expression for $\Delta a_\mu$\cite{Lindner:2016bgg}.
\begingroup
\begin{equation}
\begin{split}
V &= \overline{u}_\mu i \sigma_{\alpha\beta} q^\beta e m_\mu \left(
A_{\mu\mu}^M + \gamma_5 A_{\mu\mu}^E \right) u_\mu \epsilon^{*\alpha} \\
&= \overline{u}_\mu i \sigma_{\alpha \beta} q^{\beta} e m_\mu \left( \left(
A_{\mu\mu}^M + A_{\mu\mu}^E \right) P_R + \left( A_{\mu\mu}^M - A_{\mu\mu}^E
\right) P_L \right) u_\mu \epsilon^{*\alpha}
\end{split}
\label{eqn:amplitude_muong2_W_LiviewonLFV}
\end{equation}
\endgroup
Comparing Equation \ref{eqn:amplitude_muong2_W} with Equation \ref%
{eqn:amplitude_muong2_W_LiviewonLFV}, we confirm that
\begingroup
\begin{equation}
\begin{split}
A_R &= e m_\mu \left( A_{\mu\mu}^M + A_{\mu\mu}^E \right) \\
A_L &= e m_\mu \left( A_{\mu\mu}^M - A_{\mu\mu}^E \right)
\label{eqn:equality_ALR_AME}
\end{split}%
\end{equation}
\endgroup
Here, we can use the condition that $A_R = A_L$ identified in Figure \ref%
{fig:diagrams_muong2_with_allneutrinos} and can rearrange $A_{L,R}$ in terms
of $A_{\mu\mu}^{M,E}$, which are essential to derive our theoretical muon
anomaly prediction. Then, we find our desirable form $A_{\mu\mu}^{M,E}$ for
the muon anomalous magnetic moment.
\begingroup
\begin{equation}
\begin{split}
A_{\mu\mu}^M &= \frac{1}{e m_\mu} A_R = \frac{g^2}{128\pi^2} \frac{1}{M_W^2}
\sum_{n=1,2,3,4,5} U_{2n} U_{2n}^* F(x_n) \\
A_{\mu\mu}^E &= 0  \label{eqn:redefinition_AME}
\end{split}%
\end{equation}
\endgroup
Using the definition for both the muon anomalous magnetic moment and branching
ratio of $\mu \rightarrow e \gamma$ decay in \cite{Lindner:2016bgg}, we can
check our analytic argument for the observable and constraint are correct.
\begingroup
\begin{equation}
\begin{split}
\Delta a_\mu &= A_{\mu\mu}^M m_\mu^2 \\
\func{BR}\left( \mu \rightarrow e \gamma \right) &= \frac{3(4\pi)^3 \alpha_{%
\func{em}}}{4G_F^2} \left( \lvert A_{\mu e}^M \rvert^2 + \lvert A_{\mu e}^E
\rvert^2 \right)  \label{eqn:definition_muong2_muegamma_reivewonLFV}
\end{split}%
\end{equation}
\endgroup
One difference between $A_{\mu\mu}^{M,E}$ and $A_{L,R}$ is $%
that A_{\mu\mu}^{M,E} $ is only determined by the internal structure of the loop in
Figure \ref{fig:diagrams_muong2_with_allneutrinos}, whereas $A_{L,R}$ is the
extended factor by multiplying $A_{\mu\mu}^{M,E}$ by the helicity flip mass
in one of the external legs. Therefore, it is natural to think $A_{\mu\mu}^{M,E}$
is the same as $A_{\mu e}^{M,E}$ since their internal structure of loop are
exactly same\footnote{%
One can concern the coefficient at the vertex with electron. However, this
change is already reflected on the loop integration $A_R$ of Equation \ref%
{eqn:left_handed_amplitdue_AR} by $U_{1n}$. For the muon anomaly, the
coefficient is simply replaced by $U_{2n}$, therefore, modification of the
coefficient at the vertex does not harm our argument.}. The muon anomalous
magnetic moment and the branching ratio of $\mu \rightarrow e \gamma$ take
the form:
\begingroup
\begin{equation}
\begin{split}
\Delta a_\mu &= \frac{\alpha_W}{32\pi} \frac{m_\mu^2}{M_W^2}
\sum_{n=1,2,3,4,5} U_{2n} U_{2n}^* F(x_n) \\
\func{BR}\left(\mu \rightarrow e \gamma \right) &= \frac{3\alpha_{\func{em}}%
}{32\pi} \lvert \sum_{n=1,2,3,4,5} U_{2n} U_{1n}^* F(x_n) \rvert^2
\label{eqn:muong2_muegamma_W_ourprediction}
\end{split}%
\end{equation}
\endgroup
where the $\alpha_W$ is the weak coupling constant. As for the branching
ratio of $\mu \rightarrow e \gamma$ in Equation \ref%
{eqn:muong2_muegamma_W_ourprediction}, we showed that substituting $A_{\mu
e} $ back into the branching ratio in Equation \ref%
{eqn:definition_muong2_muegamma_reivewonLFV} is exactly consistent with the
one in Equation \ref{eqn:prediction_muegamma_neutrinos}. Expanding the unitary
mixing matrices in the muon anomaly prediction in Equation \ref%
{eqn:muong2_muegamma_W_ourprediction}, yields the following relation:
\begingroup
\begin{equation}
\Delta a_\mu = \frac{\alpha_W}{32\pi} \frac{m_\mu^2}{M_W^2} \left( (1 -
2\eta_{22})F(0) + 2\eta_{22}F(x_4) \right).
\label{eqn:expansion_muong2_ourprediction}
\end{equation}
\endgroup
Looking at Equation \ref%
{eqn:expansion_muong2_ourprediction}, it is clear that the SM part which is without $\eta$
and the BSM having $\eta$ are entangled together. We arrive at the right
prediction for the muon anomaly at one-loop by removing the SM part from
Equation \ref{eqn:expansion_muong2_ourprediction}
\begingroup
\begin{equation}
\Delta a_\mu = \frac{\alpha_W}{16\pi} \frac{m_\mu^2}{M_W^2} \eta_{22} \left(
F(x_4) - F(0) \right).  \label{eqn:expansion_muong2_correct_prediction}
\end{equation}
\endgroup
Similarly to the branching ratio of $\mu \rightarrow e \gamma$ decay, it can
be confirmed that the prediction for the muon anomaly also consists of the
factor of deviation of unitarity $\eta$.

\subsection{The anomalous electron magnetic moment $g-2$}

\label{subsec:electron_g2_Wboson}

As in the muon anomalous magnetic moment, the same diagrams with external
particles replaced by electrons can be generated in Figure \ref{fig:diagrams_electrong2_with_allneutrinos}.
\begingroup
\begin{figure}[]
\centering
\begin{subfigure}{0.48\textwidth}
	\includegraphics[width=1.0\textwidth]{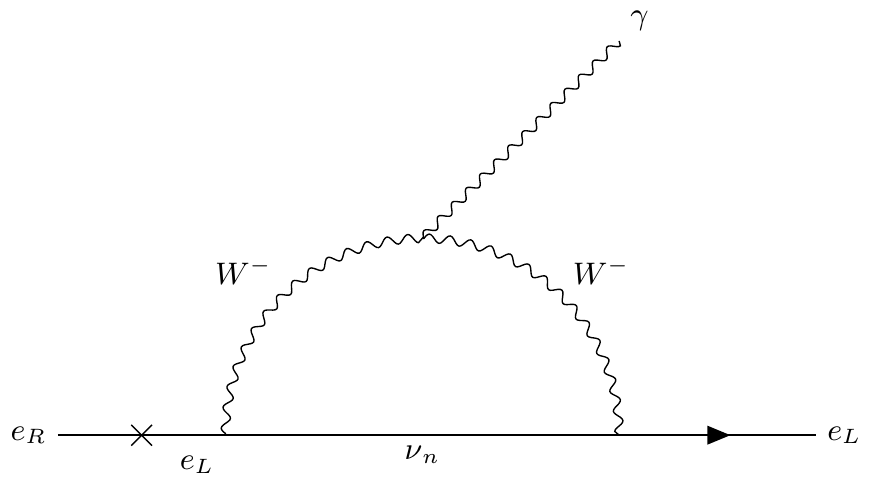}
\end{subfigure} \hspace{0.1cm}
\begin{subfigure}{0.48\textwidth}
	\includegraphics[width=1.0\textwidth]{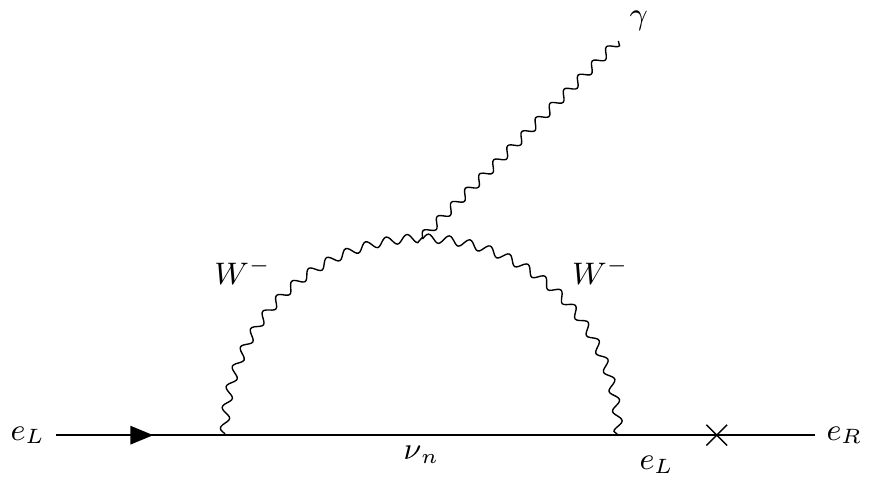}
\end{subfigure}
\caption{Diagrams for electron anomalous magnetic moment with all neutrinos. Here $n=1,2,3,4,5$.}
\label{fig:diagrams_electrong2_with_allneutrinos}
\end{figure}
\endgroup
Using the complete form of the muon anomaly prediction in Equation \ref%
{eqn:expansion_muong2_correct_prediction}, we can derive the right
prediction for the electron anomalous magnetic moment with slight modifications $%
m_\mu \rightarrow m_e, \eta_{22} \rightarrow \eta_{11}$.
\begingroup
\begin{equation}
\Delta a_e = \frac{\alpha_W}{16\pi} \frac{m_e^2}{M_W^2} \eta_{11} \left(
F(x_4) - F(0) \right).  \label{eqn:expansion_electrong2_correct_prediction}
\end{equation}
\endgroup
\subsection{Numerical analysis of $W$ exchange contributions}


The presence of heavy vector-like neutrinos leads to the deviation of
unitarity and the observables $\Delta a_{\mu,e}$ and constraint $\func{BR}%
\left(\mu \rightarrow e \gamma\right)$ can be written in terms of the factor
of non-unitarity $\eta$.
\begingroup
\begin{equation}
\begin{split}
\func{BR}\left(\mu \rightarrow e \gamma \right) &= \frac{3\alpha_{\func{em}}%
}{8\pi} \lvert \eta_{21} \rvert^2 \left( F(x_4) - F(0) \right)^2 \\
\Delta a_\mu &= \frac{\alpha_W}{16\pi} \frac{m_\mu^2}{M_W^2} \eta_{22}
\left( F(x_4) - F(0) \right) \\
\Delta a_e &= \frac{\alpha_W}{16\pi} \frac{m_e^2}{M_W^2} \eta_{11} \left(
F(x_4) - F(0) \right).  \label{eqn:prediction_muegamma_muong2_electrong2}
\end{split}%
\end{equation}
\endgroup
\subsubsection{The branching ratio of $\protect\mu \rightarrow e \protect\gamma$
decay}

We consider the branching ratio of $\mu \rightarrow e \gamma$ decay first.
Since we assume that mass of heavy vector-like neutrinos are heavier than $1%
\func{TeV}$, the value of $F(0)$ for the light neutrinos converges to approximately $%
3.3$, while that of $F(x_4)$ for the heavy vector-like neutrino converges to $%
1.3 $. Therefore, the branching ratio of $\mu \rightarrow e \gamma$ decay
can be reduced to\cite{Hernandez-Garcia:2019uof}
\begingroup
\begin{equation}
\func{BR}\left(\mu \rightarrow e \gamma \right) = \frac{3\alpha_{\func{em}}}{%
8\pi} \lvert \eta_{21} \rvert^2 \left( F(x_4) - F(0) \right)^2 \leq \frac{%
3\alpha_{\func{em}}}{2\pi} \lvert \eta_{21} \rvert^2.
\label{eqn:muegamma_simplified}
\end{equation}
\endgroup
The non-unitarity $\eta$ of Equation \ref{eqn:deviation_of_unitarity}
consists of four free parameters: mass of heavy vector-like neutrinos $%
M_{44}^{\nu}$, a real number $y$, a CP violation phase $\delta$, and a
Majorana phase $\alpha$. The experimental branching ratio of $\mu
\rightarrow e \gamma$ decay constrains the minimal parameter space in terms
of $M_{44}^{\nu}$ and $y$, while setting up two phases $\delta,\alpha$ which
maximize or minimize the branching ratio of $\mu \rightarrow e \gamma$\cite%
{Hernandez-Garcia:2019uof}, and the minimal parameter space is shown in Figure \ref{fig:parameter_space_muegamma_M44_vu_tanbeta}.
\begingroup
\begin{figure}[]
\centering
\begin{subfigure}{0.48\textwidth}
	\includegraphics[keepaspectratio,width=\textwidth]{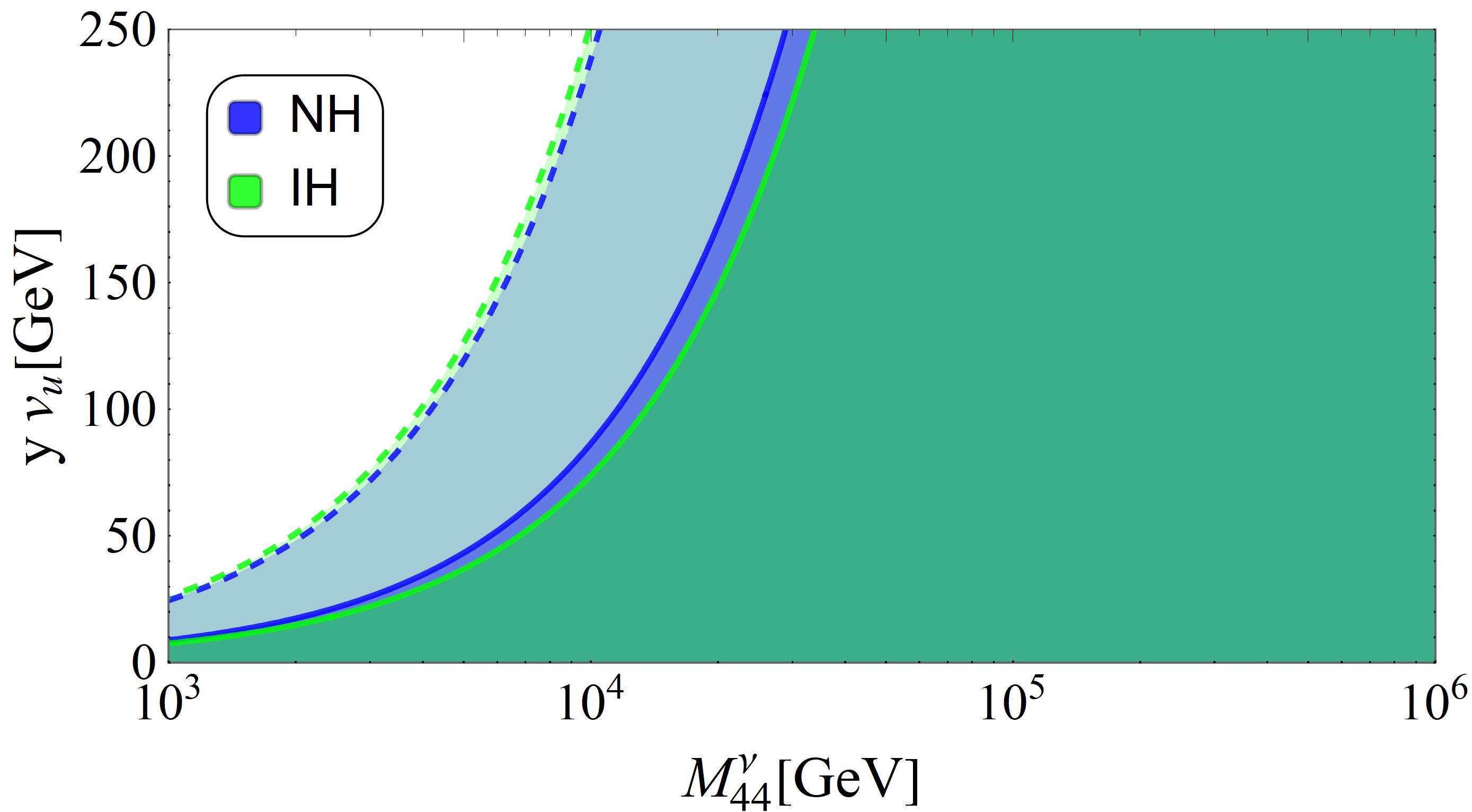}
\end{subfigure} \hspace{0.1cm}
\begin{subfigure}{0.48\textwidth}
	\includegraphics[keepaspectratio,width=\textwidth]{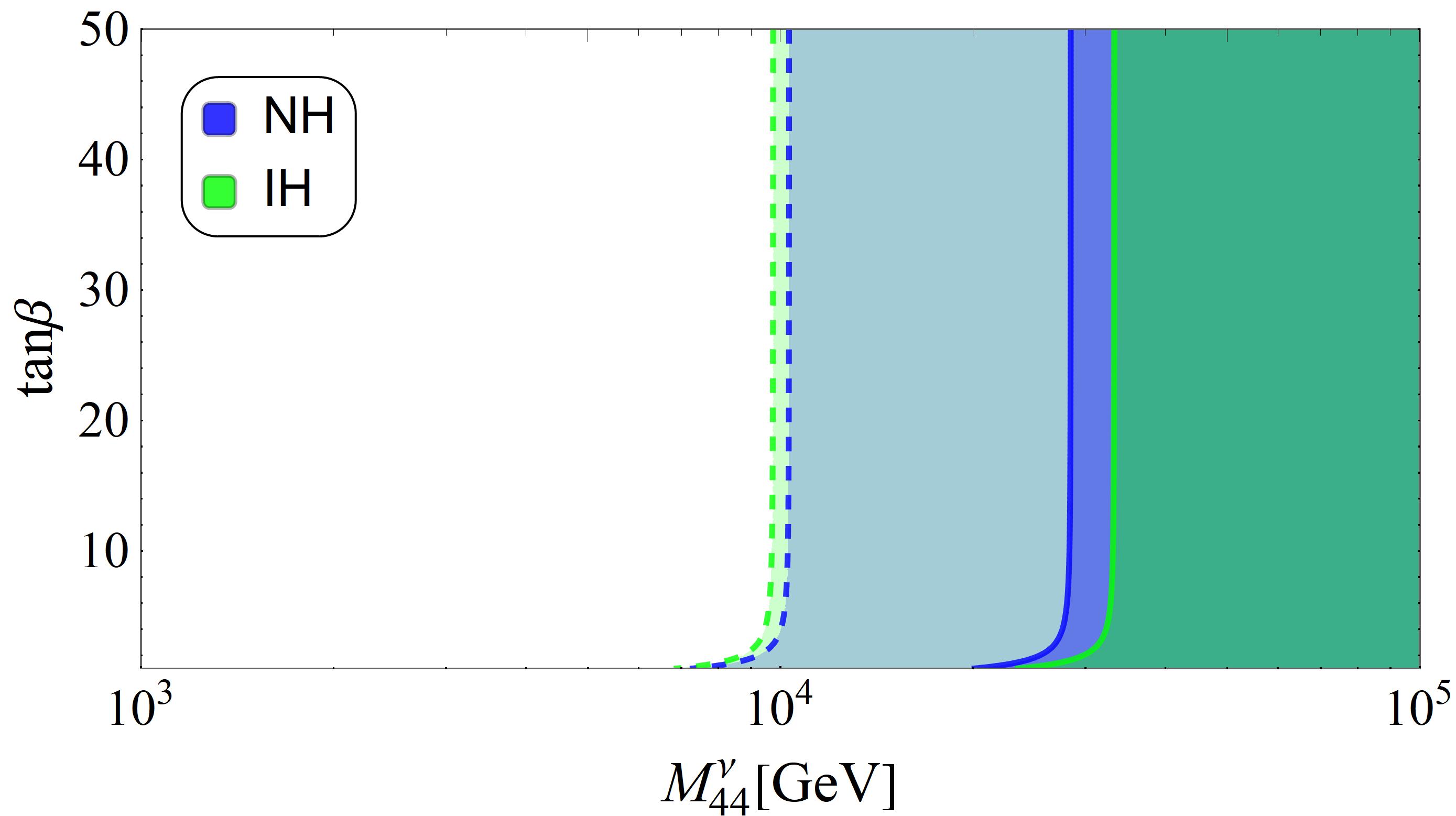}
\end{subfigure}
\caption{The left plot is an available parameter space for two free
parameters: mass of vector-like neutrino $M_{44}^{\protect\nu}$ and SM
up-type Higgs vev $v_u$. Here, the free parameter $y$ is set to $1$. The
right plot is the case where vev of the up-type Higgs is constrained from $%
246/\protect\sqrt{2} \simeq 174$ to $246\func{GeV}$ or from $\tan\protect%
\beta = 1$ to $50$ in a same way}
\label{fig:parameter_space_muegamma_M44_vu_tanbeta}
\end{figure}
\endgroup
The left plot in Figure \ref{fig:parameter_space_muegamma_M44_vu_tanbeta} is
an available parameter space for mass of the vector-like neutrino versus the free
parameter $y$ times SM up-type Higgs vev $v_u$. The blue bold line
corresponds to bound of the branching ratio of $\mu \rightarrow e \gamma$
decay at the normal hierarchy with CP violation phase $\delta = 0$ and
Majorana phase $\alpha = 0$ and this line can be relaxed up to the blue
dotted line where $\delta = 0$, $\alpha = 2\pi$. The green bold(dotted) line
corresponds to the inverted hierarchy with $\delta = \pi/2(0)$ and $\alpha = 
\frac{9\pi}{10}(0)$. Since we are especially interested in the range of SM
up-type Higgs vev $v_u$ from $174$ to $246\func{GeV}$, the right plot
consistent with the interested range is extracted from the left after
replacing $v_u$ by $\tan\beta=v_u/v_d$ using the relation $v_u^2 + v_d^2 =
(246 \func{GeV})^2$. \newline
~\newline
As for the constraint of deviation of unitarity $\eta$ with the CLFV $\mu
\rightarrow e \gamma$ decay at $1 \sigma$, it is given by\cite%
{Fernandez-Martinez:2016lgt,Tanabashi:2018oca}
\begingroup
\begin{equation}
\lvert \eta_{21} \rvert \leq 8.4 \times 10^{-6}.
\label{eqn:constraint_nonunitarity_eta21}
\end{equation}
\endgroup
\subsubsection{The muon and electron anomalous magnetic moments $\Delta a_{%
\protect\mu,e}$}

\label{subsec:muon_and_electrong2_W}

As in the constraint for $\eta_{21}$ in Equation \ref%
{eqn:constraint_nonunitarity_eta21}, the other non-unitarities $\eta_{11,22}$
for the electron and muon anomalous mangetic moment are given by\cite%
{Hernandez-Garcia:2019uof,Fernandez-Martinez:2016lgt}
\begingroup
\begin{equation}
\begin{split}
\eta_{11} &< 4.2 \times 10^{-4} \text{ (for NH) }, \quad < 4.8 \times
10^{-4} \text{ (for IH) } \\
\eta_{22} &< 2.9 \times 10^{-7} \text{ (for NH) }, \quad < 2.4 \times
10^{-7} \text{ (for IH) }  \label{eqn:non_unitarity_eta11_eta_22}
\end{split}%
\end{equation}
\endgroup
With the constraints $\eta_{11,22}$ in Equation \ref%
{eqn:non_unitarity_eta11_eta_22}, we can calculate impact of the muon and
electron anomalous magnetic moments at NH(IH) using Equation \ref%
{eqn:prediction_muegamma_muong2_electrong2}.
\begingroup
\begin{equation}
\begin{split}
\Delta a_\mu &= \frac{\alpha_W}{16\pi} \frac{m_\mu^2}{M_W^2} \eta_{22}
\left( F(x_4) - F(0) \right) \simeq -6.6(-5.5) \times 10^{-16} \\
\Delta a_e &= \frac{\alpha_W}{16\pi} \frac{m_e^2}{M_W^2} \eta_{11} \left(
F(x_4) - F(0) \right) \simeq -2.2(-2.6) \times 10^{-17}
\label{eqn:magnitude_muon_electrong2_with_constraints}
\end{split}%
\end{equation}
\endgroup
There are two interesting features in the above prediction for the muon and
electron anomalous magnetic moments. One feature is sign of each prediction.
As mentioned in the introduction, this prediction with the $W$ exchange can not flip the sign of
each anomaly. In order to explain both anomalies at $1 \sigma$, the
prediction for both anomalies with $W$ exchange requires additional
contributions such as $Z^\prime$ or scalar exchange. Another feature is
magnitude of each prediction. For the muon anomaly, the experimental order
of magnitude at $1 \sigma$ is about $10^{-9}$, however our prediction is
much smaller than that of the experimental bound as well as the electron
anomaly, which means the non-unitarity derived from the presence of heavy
vector-like neutrino can not bring the anomalies to the observable level.
This inadequate prediction with $W$ exchange has been a good motivation to
search for another possibility such as scalar exchange.
\chapter{Non-standard contributions to the muon and electron anomalous magnetic moments : non-SM scalar exchange} \label{Chapter:The2ndpaper2}
In the second BSM model covered in chapter~\ref{Chapter:The2ndBSMmodel}, we discuss the second non-SM contributions to the muon and electron anomalous magnetic moments $g-2$ with the non-SM scalars at one-loop level via the scalar potential. We also discuss how the non-SM physical scalars appear in the scalar potential and carry out numerical scans to find relevant mass parameters for the vector-like charged leptons as well as the non-SM scalars. And then we conclude the non-SM scalar exchange can actually accommodate both anomalies at $1\sigma$ error bar of the anomalies.
\section{Higgs exchange to contributions to $\left( g-2 \right)_\protect\mu, \left( g-2
\right)_e$ and $\func{BR}\left(\protect\mu \rightarrow e \protect\gamma %
\right)$}

\label{sec:Analytic_arguments_muon_electron_g2_scalars}

The relevant sector for the muon and electron anomalous magnetic moments with
scalar exchange is the charged lepton Yukawa matrix which can be expressed in the mass insertion formalism as,
\begingroup
\begin{equation}
\begin{split}
y_{ij}^e &= 
\begin{pmatrix}
0 & 0 & 0 \\ 
0 & y_{24}^e x_{42}^{e} & y_{24}^e x_{43}^{e} \\ 
0 & y_{34}^e x_{42}^{e} & y_{34}^e x_{43}^{e}
\end{pmatrix}
\frac{\left\langle \phi \right\rangle}{M_{44}^{e}} + 
\begin{pmatrix}
y_{15}^e x_{51}^{e} & y_{15}^e x_{52}^{e} & y_{15}^e x_{53}^{e} \\ 
y_{25}^e x_{51}^{e} & y_{25}^e x_{52}^{e} & y_{25}^e x_{53}^{e} \\ 
y_{35}^e x_{51}^{e} & y_{35}^e x_{52}^{e} & y_{35}^e x_{53}^{e}
\end{pmatrix}
\frac{\left\langle \phi \right\rangle}{M_{55}^{e}} \\
&+ 
\begin{pmatrix}
y_{51}^e x_{15}^{L} & y_{52}^e x_{15}^{L} & y_{53}^e x_{15}^{L} \\ 
y_{51}^e x_{25}^{L} & y_{52}^e x_{25}^{L} & y_{53}^e x_{25}^{L} \\ 
y_{51}^e x_{35}^{L} & y_{52}^e x_{35}^{L} & y_{53}^e x_{35}^{L}
\end{pmatrix}
\frac{\left\langle \phi \right\rangle}{M_{55}^{L}} + 
\begin{pmatrix}
0 & 0 & 0 \\ 
0 & 0 & 0 \\ 
0 & 0 & x_{34}^L y_{43}^e
\end{pmatrix}
\frac{\left\langle \phi \right\rangle}{M_{44}^{L}}
\end{split}
\label{eqn:effective_Yukawa_matrix_charged_leptons_assumption_second}
\end{equation}
\endgroup
The effective Yukawa matrix of Equation \ref%
{eqn:effective_Yukawa_matrix_charged_leptons_assumption_second} in the
mass basis is diagonalized by the universal seesaw mechanism due to involving a
few of different mass scales. Therefore, the only diagonal
components should alive in the mass matrix. In order to make the mass matrix
diagonal, we assume that $y_{34}^e = x_{43}^{e} = y_{15,25,35}^e =
x_{51,52,53}^{e} = x_{25,35}^L = y_{52,53}^e = 0$. Then, the mass matrix is
reduced to

\begin{equation}
\begin{split}
y_{ij}^e &= 
\begin{pmatrix}
0 & 0 & 0 \\ 
0 & y_{24}^e x_{42}^{e} & 0 \\ 
0 & 0 & 0
\end{pmatrix}
\frac{\left\langle \phi \right\rangle}{M_{44}^{e}} + 
\begin{pmatrix}
0 & 0 & 0 \\ 
0 & 0 & 0 \\ 
0 & 0 & 0
\end{pmatrix}
\frac{\left\langle \phi \right\rangle}{M_{55}^{e}} 
\\
&+ 
\begin{pmatrix}
y_{51}^e x_{15}^{L} & 0 & 0 \\ 
0 & 0 & 0 \\ 
0 & 0 & 0
\end{pmatrix}
\frac{\left\langle \phi \right\rangle}{M_{55}^{L}} + 
\begin{pmatrix}
0 & 0 & 0 \\ 
0 & 0 & 0 \\ 
0 & 0 & x_{34}^L y_{43}^e
\end{pmatrix}
\frac{\left\langle \phi \right\rangle}{M_{44}^{L}} \\
y_{ij}^e &= 
\begin{pmatrix}
y_{51}^e s_{15}^L & 0 & 0 \\ 
0 & y_{24}^e s_{24}^{e} & 0 \\ 
0 & 0 & y_{43}^e s_{34}^L%
\end{pmatrix},
\label{eqn:diagonalized_Yukawa_matrix_charged_leptons}
\end{split}%
\end{equation}

where $s_{15}^L \simeq x_{15}^L \left\langle \phi \right\rangle/M_{55}^{L}$, 
$s_{24}^{e} \simeq x_{42}^{e} \left\langle \phi \right\rangle/M_{44}^{e}$, $%
s_{34}^L \simeq x_{34}^L \left\langle \phi \right\rangle/M_{44}^L$ and the
diagonal elements from top-left to bottom-right should be responsible for
electron, muon and tau Yukawa constants, respectively. After removing all
irrelevant terms to both anomalies and applying the assumption, the $7
\times 7$ mass matrix in the interaction basis is also reduced to as below:
\begingroup
\begin{equation}
M^{e}=\left( 
\begin{array}{c|cccc}
& e_{1R} & e_{2R} & e_{4R} & \widetilde{L}_{5R} \\ \hline
\overline{L}_{1L} & 0 & 0 & 0 & x_{15}^{L}v_{\phi } \\ 
\overline{L}_{2L} & 0 & 0 & y_{24}^{e}v_{d} & 0 \\ 
\overline{L}_{5L} & y_{51}^{e}v_{d} & 0 & 0 & M_{55}^{L} \\ 
\overline{\widetilde{e}}_{4L} & 0 & x_{42}^{e}v_{\phi } & M_{44}^{e} & 0
\end{array}%
\right) \label{eqn:reduced_charged_lepton_mass_matrix}
\end{equation}
\endgroup
The reduced charged lepton mass matrix of Equation \ref%
{eqn:reduced_charged_lepton_mass_matrix} clearly tells
that no mixing between charged leptons arise so the branching ratio of $\mu
\rightarrow e \gamma$ is naturally satisfied under this scenario. The scalar
exchange for both anomalies can be realized by closing the Higgs sectors in
Figure \ref{fig:diagrams_charged_leptons_mass_insertion} as per Figure \ref%
{fig:muon_electrong2_scalar_exchange}.
\begingroup
\begin{figure}[H]
\centering
\begin{subfigure}{0.48\textwidth}
	\includegraphics[width=1.0\textwidth]{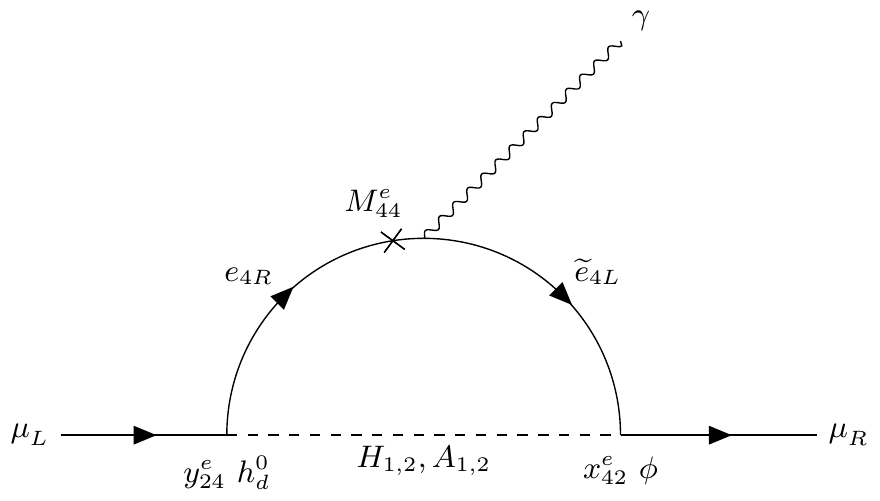}
\end{subfigure} \hspace{0.1cm}
\begin{subfigure}{0.48\textwidth}
	\includegraphics[width=1.0\textwidth]{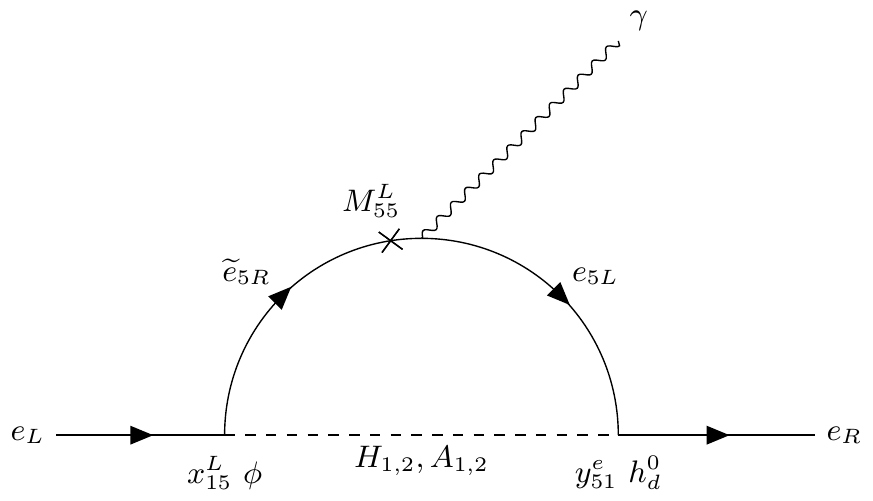}
\end{subfigure}
\caption{Diagrams contributing to the muon anomaly (left) and
the electron anomaly (right) where $H_{1,2}$ are CP-even non-SM
scalars and $A_{1,2}$ are CP-odd scalars in the physical basis}
\label{fig:muon_electrong2_scalar_exchange}
\end{figure}
\endgroup
In Figure \ref{fig:muon_electrong2_scalar_exchange}, the CP-even non-SM scalars $%
H_{1,2}$ and CP-odd scalars $A_{1,2}$ appear as a result of mixing between
Higgses $H_u,H_d$ and $\phi$ in the interaction basis. The Higgs sector in the interaction
basis is defined by
\begingroup
\begin{equation}
\begin{split}
H_u &= 
\begin{pmatrix}
H_u^+ \\ 
v_u + \frac{1}{\sqrt{2}}\left( \func{Re}H_{u}^0 + i\func{Im}H_{u}^0 \right)%
\end{pmatrix}%
,
\\
H_d &= 
\begin{pmatrix}
v_d + \frac{1}{\sqrt{2}}\left( \func{Re}H_{d}^0 + i\func{Im}H_{d}^0 \right)
\\ 
H_d^-%
\end{pmatrix}%
,
\\
\phi &= \frac{1}{\sqrt{2}} \left( v_\phi + \func{Re}\phi + i\func{Im}%
\phi \right).  \label{eqn:Higgs_fields_mass_basis}
\end{split}%
\end{equation}
\endgroup
For consistency, we equate $v_u, v_d$ and $v_\phi$ to $v_1, v_2$ and $v_3$,
respectively.

\subsection{The 2HDM scalar potential}

\label{subsec:the_scalar_potential} The scalar potential of the model under
consideation takes the form:%
\begingroup
\begin{equation}
\begin{split}
V &=\mu _{1}^{2}\left( H_{u}H_{u}^{\dagger }\right) +\mu _{2}^{2}\left(
H_{d}H_{d}^{\dagger }\right) +\mu _{3}^{2}\left( \phi \phi ^{\ast }\right)
+\mu _{\func{sb}}^{2}\left[ \phi ^{2}+\left( \phi ^{\ast }\right) ^{2}\right]
\\
&+\lambda _{1}\left( H_{u}H_{u}^{\dagger }\right) ^{2}+\lambda _{2}\left(
H_{d}H_{d}^{\dagger }\right) ^{2}+\lambda _{3}\left( H_{u}H_{u}^{\dagger }\right) \left(H_{d}H_{d}^{\dagger }\right) +\lambda _{4}\left( H_{u}H_{d}^{\dagger
}\right) \left( H_{d}H_{u}^{\dagger }\right) 
\\
&+\lambda _{5}\left( \varepsilon
_{ij}H_{u}^{i}H_{d}^{j}\phi ^{2}+\func{h.c}\right)+\lambda _{6}\left( \phi \phi ^{\ast }\right) ^{2}
\\
&+\lambda _{7}\left( \phi
\phi ^{\ast }\right) \left( H_{u}H_{u}^{\dagger }\right) +\lambda _{8}\left(
\phi \phi ^{\ast }\right) \left( H_{d}H_{d}^{\dagger }\right) ,
\end{split}
\end{equation}%
\endgroup
where the $\lambda _{i}$ ($i=1,2,\cdots ,8$) are dimensionless parameters
whereas the $\mu _{j}$ ($j=1,2,3$) are dimensionful parameters and $\mu _{%
\func{sb}}$ is a dimensionfull soft-breaking parameter. We consider the $%
U(1)^{\prime }$ symmetry as global in this model so our model does not
feature $Z^{\prime }$ boson and the scalar potential requires the inclusion
of the soft-breaking mass term $-\mu _{\func{sb}}^{2}\left[ \phi ^{2}+\left( \phi^{\ast }\right) ^{2}\right] $ in order to prevent the appearance of a massless scalar state arising from the imaginary part of $\phi $. The minimization conditions of the scalar potential yield the following
relations: 
\begingroup
\begin{equation}
\begin{split}
\mu _{1}^{2} &=-2\lambda _{1}v_{1}^{2}-\lambda _{3}v_{2}^{2}-\frac{1}{2}%
\lambda _{7}v_{3}^{2}+\frac{\lambda _{5}v_{2}v_{3}^{2}}{2v_{1}}, \\
\mu _{2}^{2} &=-2\lambda _{2}v_{2}^{2}-\lambda _{3}v_{1}^{2}-\frac{1}{2}%
\lambda _{8}v_{3}^{2}+\frac{\lambda _{5}v_{3}^{2}v_{1}}{
2v_{2}}, \\
\mu _{3}^{2} &=-\lambda _{8}v_{2}^{2}-\lambda _{6}v_{3}^{2}+v_{1}\left(
2\lambda _{5}v_{2}-\lambda _{7}v_{1}\right) -2\mu _{\func{sb}}^{2}.
\end{split}
\end{equation}
\endgroup
\subsection{Mass matrix for CP-even, CP-odd neutral and charged scalars}

\label{subsec:mass_mamtrix_for_CP_even_odd_charged} The squared mass matrix
for the CP-even scalars in the basis $\left( \func{Re}H_{u}^{0},\func{Re}%
H_{d}^{0},\func{Re}\phi \right) $ takes the form:%
\begingroup
\begin{equation}
\mathbf{M}_{\func{CP-even}}^{2}=\left( 
\begin{array}{ccc}
4\lambda _{1}v_{1}^{2}+\frac{\lambda _{5}v_{2}v_{3}^{2}}{2v_{1}} & -\frac{1}{2%
}\lambda _{5}v_{3}^{2}+2\lambda _{3}v_{1}v_{2} & \sqrt{2}v_{3}\left( -\lambda
_{5}v_{2}+\lambda _{7}v_{1}\right) \\ 
-\frac{1}{2}\lambda _{5}v_{3}^{2}+2\lambda _{3}v_{1}v_{2} & 4\lambda
_{2}v_{2}^{2}+\frac{\lambda _{5}v_{1}v_{3}^{2}}{2v_{2}} & \sqrt{2}%
v_{3}\left( -\lambda _{5}v_{1}+\lambda _{8}v_{2}\right) \\ 
\sqrt{2}v_{3}\left( -\lambda _{5}v_{2}+\lambda _{7}v_{1}\right) & \sqrt{2}%
v_{3}\left( -\lambda _{5}v_{1}+\lambda _{8}v_{2}\right) & 2\lambda
_{6}v_{3}^{2}
\end{array}%
\right).  \label{eqn:mass_matrix_CP_even}
\end{equation}
\endgroup
From the mass matrix given above, we find that the CP-even scalar spectrum
is composed of the $125$ GeV\ SM-like Higgs $h$ and two non-SM CP-even
Higgses $H_{1,2}$. Furthermore, we assume that no mixing between the SM
physical Higgs $h$ and the two non-SM CP-even Higgses $H_{1,2}$ arise and
this assumption constrains the $(1,2)$, $(1,3)$, $(2,1)$ and $(3,1)$ elements of CP-even mass matrix of Equation \ref{eqn:mass_matrix_CP_even}. The constraints are given by the following decoupling limit scenario
\begingroup
\begin{equation}
\begin{split}
\lambda_5 &= \frac{4 v_1 v_2}{v_3^2} \lambda_3 \\
\lambda_7 &= \frac{v_2}{v_1} \lambda_5 = \frac{4 v_2^2}{v_3^2} \lambda_3,
\label{eqn:constraints_in_mass_matrix_CP_even}
\end{split}%
\end{equation}
\endgroup
and then the CP-even mass matrix of Equation \ref{eqn:mass_matrix_CP_even}
with the constraints is simplified to
\begingroup
\begin{equation}
\mathbf{M}_{\func{CP-even}}^{2}=\left( 
\begin{array}{ccc}
4\lambda _{1}v_{1}^{2} + 2v_{2}^{2} \lambda_{3} & 0 & 0 \\ 
0 & 4\lambda _{2}v_{2}^{2} + 2v_{1}^{2} \lambda_{3} & \sqrt{2}v_{3}\left( -%
\frac{4 v_{1}^{2} v_{2}}{v_{3}^{2}} \lambda_{3} +\lambda _{8}v_{2}\right) \\ 
0 & \sqrt{2}v_{3}\left( -\frac{4 v_{1}^{2} v_{2}}{v_{3}^{2}} \lambda_{3}
+\lambda _{8}v_{2}\right) & 2\lambda _{6}v_{3}^{2}%
\end{array}%
\right).  \label{eqn:mass_matrix_CP_even_simplified}
\end{equation}
\endgroup
In the above given decoupling limit scenario, chosen in order to simplify our analysis, 
the CP-even neutral scalar states contained in the $SU(2)$
doublet $H_{u}$ will not mix with the CP-even neutral ones contained in $%
H_{d}$. In such limit, the neutral CP-even states of $H_{u}$ will not
feature mixing with the gauge singlet scalar $\phi $. Thus, the lightest $125\func{GeV}$ CP-even scalar of our model will have couplings to the SM
particles close to the SM expectation, which is consistent with the current
experimental data. Diagonalizing the simplified CP-even mass matrix, it reveals masses of the physical SM Higgs $h$ and non-SM CP-even scalars $H_{1,2}$ in the physical basis $\left( h, H_1, H_2 \right)$
\begingroup
\begin{equation}
R_{\func{CP-even}}^\dagger \mathbf{M}_{\func{CP-even}}^2 R_{\func{CP-even}}
= \func{diag} \left( m_h^2, m_{H_1}^2, m_{H_2}^2 \right).
\label{eqn:diagonalising_CP_even_mass_matrix}
\end{equation}
\endgroup
The SM Higgs $h$ is appeared as $\func{Re}H_{u}^{0}$ itself and the non-SM
CP-even scalars $H_{1,2}$ are the states which $\func{Re}H_{d}^{0}$ is mixed with $\func{Re}\phi $. Regarding the CP-odd scalar sector, we find that the squared mass matrix for the CP-odd scalars in the basis $\left( \func{Im} H_{u}^{0},\func{Im}H_{d}^{0},\func{Im}\phi \right) $ is given by: 
\begingroup
\begin{equation}
\mathbf{M}_{\func{CP-odd}}^{2}=\left( 
\begin{array}{ccc}
\frac{\lambda _{5}v_{2}v_{3}^{2}}{2v_{1}} & \frac{1}{2}\lambda
_{5}v_{3}^{2} & \sqrt{2}\lambda _{5}v_{2}v_{3} \\ 
\frac{1}{2}\lambda _{5}v_{3}^{2} & \frac{\lambda _{5}v_{1}v_{3}^{2}}{2v_{2}%
} & \sqrt{2}\lambda _{5}v_{1}v_{3} \\ 
\sqrt{2}\lambda _{5}v_{2}v_{3} & \sqrt{2}\lambda _{5}v_{1}v_{3} & 
4\lambda _{5}v_{1}v_{2}-4\mu _{\func{sb}}^{2}%
\end{array}%
\right).
\end{equation}
\endgroup
The squared CP-odd mass matrix is diagonalized in the same way as in the
CP-even mass matrix and the CP-odd physical basis is given by $\left(
G_{Z},A_{1},A_{2}\right) $ where $G_{Z}$ is the massless Goldstone bosons
associated with the longitudinal components of the $Z$ gauge boson, whereas $%
A_{1}$ and $A_{2}$ are massive non-SM CP-odd scalars
\begingroup
\begin{equation}
R_{\func{CP-odd}}^{\dagger }\mathbf{M}_{\func{CP-odd}}^{2}R_{\func{CP-odd}}=%
\func{diag}\left( 0,m_{A_{1}}^{2},m_{A_{2}}^{2}\right) .
\label{eqn:diagonalizing_CP_odd_mass_matrix}
\end{equation}
\endgroup
Furthermore, the squared mass matrix for the electrically charged scalars is
given by:
\begingroup 
\begin{equation}
\mathbf{M}_{\func{charged}}^{2}=\left( 
\begin{array}{cc}
\lambda _4 v_2^2+\frac{\lambda _5 v_2 v_3^2}{2 v_1} & \lambda _4 v_1 v_2+%
\frac{1}{2} \lambda _5 v_3^2 \\ 
\lambda _4 v_1 v_2+\frac{1}{2} \lambda _5 v_3^2 & \lambda _4 v_1^2+\frac{%
\lambda _5 v_1 v_3^2}{2 v_2}%
\end{array}
\right).
\end{equation}
\endgroup
The charged scalar mass matrix can be diagonalized in the basis $\left(
H_{1}^{\pm}, H_{2}^{\pm} \right)$ as in CP-even or -odd mass matrix:
\begingroup
\begin{equation}
R_{\func{charged}}^\dagger \mathbf{M}_{\func{charged}}^{2} R_{\func{charged}%
} = \func{diag} \left( 0, m_{H^\pm}^2 \right).
\label{eqn:diagonalizing_charged_mass_matrix}
\end{equation}
\endgroup
Then, the electrically charged scalar sector contains the massive scalars $%
H^{\pm }$ and the massless electrically charged scalars $G_{W}^{\pm }$ which
correspond to the Goldstone bosons associated with the longitudinal
components of the $W^{\pm }$ gauge bosons. In the following sections we will
analyze the phenomenological implications of our model in the Higgs diphoton
decay as well as in the muon and electron anomalous magnetic moments. 

\subsection{The Higgs diphoton signal strength}

\label{Diphpoton} The rate for the $h\rightarrow \gamma \gamma $ decay is
given by: 
\begingroup
\begin{equation}
\Gamma (h\rightarrow \gamma \gamma )=\dfrac{\alpha _{\func{em}}^{2}m_{h}^{3}%
}{256\pi ^{3}v^{2}}\left\vert \sum_{f}a_{hff}N_{C}Q_{f}^{2}F_{1/2}(\rho
_{f})+a_{hWW}F_{1}(\rho _{W})+\frac{C_{hH^{\pm }H^{\mp }}v}{2m_{H^{\pm }}^{2}%
}F_{0}(\rho _{H_{k}^{\pm }})\right\vert ^{2},
\end{equation}%
\endgroup
where $\rho _{i}$ are the mass ratios $\rho _{i}=\frac{m_{h}^{2}}{4M_{i}^{2}}
$ with $M_{i}=m_{f},M_{W}$; $\alpha _{\func{em}}$ is the fine structure
constant; $N_{C}$ is the color factor ($N_{C}=1$ for leptons and $N_{C}=3$
for quarks) and $Q_{f}$ is the electric charge of the fermion in the loop.
From the fermion-loop contributions we only consider the dominant top quark
term. Furthermore, $C_{hH^{\pm }H^{\mp }}$ is the trilinear coupling between
the SM-like Higgs and a pair of charged Higges, whereas $a_{htt}$ and $%
a_{hWW}$ are the deviation factors from the SM Higgs-top quark coupling and
the SM Higgs-W gauge boson coupling, respectively (in the SM these factors
are unity). Such deviation factors are close to unity in our model and they
are defined as below:
\begingroup
\begin{equation}
a_{htt} \simeq 1, \quad a_{hWW} = \frac{1}{\sqrt{v_1^2 + v_2^2}} \frac{%
\partial}{\partial h} \left( \sum_{i,j=1,2,3} v_i \left( R_{\func{CP-even}%
}^T \right)_{ij} \left( h, H_1, H_2 \right)_{j} \right) = \frac{v_1}{\sqrt{%
v_1^2 + v_2^2}}
\end{equation}
\endgroup
Furthermore, $F_{1/2}(z)$ and $F_{1}(z)$ are the dimensionless loop factors
for spin-$1/2$ and spin-$1$ particles running in the internal lines of the
loops. These loop factors take the form:
\begingroup
\begin{equation}
\begin{split}
F_{1/2}(z)& =2(z+(z-1)f(z))z^{-2}, \\
F_{1}(z)& =-2(2z^{2}+3z+3(2z-1)f(z))z^{-2}, \\
F_{0}(z)& =-(z-f(z))z^{-2},
\end{split}%
\end{equation}%
\endgroup
with 
\begingroup
\begin{equation}
f(z)=\left\{ 
\begin{array}{lcc}
\arcsin ^{2}\sqrt{2} & \text{for} & z\leq 1 \\ 
&  &  \\ 
-\frac{1}{4}\left( \ln \left( \frac{1+\sqrt{1-z^{-1}}}{1-\sqrt{1-z^{-1}}%
-i\pi }\right) ^{2}\right) & \text{for} & z>1
\end{array}%
\right.
\end{equation}
\endgroup
In order to study the implications of our model in the decay of the $125$
GeV Higgs into a photon pair, one introduces the Higgs diphoton signal
strength $R_{\gamma \gamma }$, which is defined as:
\begingroup 
\begin{equation}
R_{\gamma \gamma }=\frac{\sigma (pp\rightarrow h)\Gamma (h\rightarrow \gamma
\gamma )}{\sigma (pp\rightarrow h)_{\func{SM}}\Gamma (h\rightarrow \gamma
\gamma )_{\func{SM}}}\simeq a_{htt}^{2}\frac{\Gamma (h\rightarrow \gamma
\gamma )}{\Gamma (h\rightarrow \gamma \gamma )_{\func{SM}}}.  \label{eqn:hgg}
\end{equation}%
\endgroup
That Higgs diphoton signal strength, normalizes the $\gamma \gamma $ signal
predicted by our model in relation to the one given by the SM. Here we have
used the fact that in our model, single Higgs production is also dominated
by gluon fusion as in the Standard Model. The ratio $R_{\gamma \gamma }$ has been measured by CMS and ATLAS collaborations with the best fit signals \cite{Sirunyan:2018ouh,Aad:2019mbh}%
: 
\begingroup
\begin{equation}
R_{\gamma \gamma }^{\func{CMS}}=1.18_{-0.14}^{+0.17}\quad \text{and}\quad R_{\gamma
\gamma }^{\func{ATLAS}}=0.96\pm 0.14.  \label{eqn:rgg}
\end{equation}
\endgroup
As it will be shown in the next subsection, the constraints arising from the Higgs diphoton decay rate will be considered in our numerical analysis. 

\subsection{The muon and electron anomalous magnetic moments}

\label{gminus2} 

The Yukawa interactions relevant for the computation of the muon anomalous
magnetic moment are:
%
%
%
\begingroup
\begin{equation}
\mathcal{L}_{\Delta a_\mu} = y_{24}^e \mu \left( \func{Re} H_{d}^0 - i
\gamma^5 \func{Im} H_{d}^0 \right) \overline{e}_4 + x_{42}^e \widetilde{e}_4
\left( \func{Re} \phi - i \gamma^5 \func{Im} \phi \right) \overline{e}_2 +
M_{44}^e \widetilde{e}_4 \overline{e}_4 + \func{h.c.}
\label{eqn:Lagrangian_muon_anomaly_expansion}
\end{equation}
\endgroup
where the Yukawa coupling constants $y_{24}^e,x_{42}^e$ are assumed to be real, the scalar fieds have been expanded by their real and imaginary parts and the properties of the projection operators $P_{L,R}$ acting on the charged leptonic fields have been used. 
By expressing the scalar fields in the interaction basis in terms of the scalar fields in the physical basis, the charged lepton Yukawa interactions relevant for the computation of the $g-2$ anomalies take the form:
\begingroup
\begin{equation}
\begin{split}
\mathcal{L}_{\Delta a_{\mu }}& =y_{24}^{e}\mu \left(
(R_{e}^{T})_{22}H_{1}+(R_{e}^{T})_{23}H_{2}-i\gamma
^{5}(R_{o}^{T})_{22}A_{1}-i\gamma ^{5}(R_{o}^{T})_{23}A_{2}\right) \overline{%
e}_{4} \\
& +x_{42}^{e}\widetilde{e}_{4}\left(
(R_{e}^{T})_{32}H_{1}+(R_{e}^{T})_{33}H_{2}-i\gamma ^{5}( R_{o}^{T})
_{32}A_{1}-i\gamma ^{5}(R_{o}^{T})_{33}A_{2}\right) \overline{e}%
_{2}
\\
&+M_{44}^{e}\widetilde{e}_{4}\overline{e}_{4}+\func{h.c.}
\end{split}%
\end{equation}
\endgroup
where we are using the unitary gauge where the contributions arising from
unphysical Goldstone bosons to the muon anomaly are excluded and we shorten
the notations $R_{\func{CP}}$ by $R_{e(o)}$. Here $R_{e}$ and $R_{o}$ are the rotation matrices that diagonalize the squared mass matrices for the CP even and CP odd scalars, respectively. Then, it follows that the muon and electron anomalous magnetic moments in the scenario of diagonal SM charged lepton mass matrix take the form:
\begingroup
\begin{equation}
\begin{split}
\Delta a_{\mu }=y_{24}^{e}x_{42}^{e}\frac{m_{\mu }^{2}}{8\pi ^{2}}\Big[&
\left( R_{e}^{T}\right) _{22}\left( R_{e}^{T}\right) _{32}I_{S}^{(\mu
)}\left( m_{e_{4}},m_{H_{1}}\right) +\left( R_{e}^{T}\right) _{23}\left(
R_{e}^{T}\right) _{33}I_{S}^{(\mu )}\left( m_{e_{4}},m_{H_{2}}\right) \\
& -\left( R_{o}^{T}\right) _{22}\left( R_{o}^{T}\right) _{32}I_{P}^{(\mu
)}\left( m_{e_{4}},m_{A_{1}}\right) -\left( R_{o}^{T}\right) _{23}\left(
R_{o}^{T}\right) _{33}I_{P}^{\left( \mu \right) }\left(
m_{e_{4},}m_{A_{2}}\right) \Big] \\
\Delta a_{e}=y_{51}^{e}x_{15}^{L}\frac{m_{e}^{2}}{8\pi ^{2}}\Big[& \left(
R_{e}^{T}\right) _{22}\left( R_{e}^{T}\right) _{32}I_{S}^{(e)}\left(
m_{e_{5}},m_{H_{1}}\right) +\left( R_{e}^{T}\right) _{23}\left(
R_{e}^{T}\right) _{33}I_{S}^{(e)}\left( m_{e_{5}},m_{H_{2}}\right) \\
& -\left( R_{o}^{T}\right) _{22}\left( R_{o}^{T}\right)
_{32}I_{P}^{(e)}\left( m_{e_{5}},m_{A_{1}}\right) -\left( R_{o}^{T}\right)
_{23}\left( R_{e}^{T}\right) _{33}I_{P}^{\left( E\right) }\left(
m_{e_{5}},m_{A_{2}}\right) \Big],
\end{split}
\label{eqn:muon_electron_anomalous_magnetic_moments}
\end{equation}
\endgroup
where the loop integrals are given by \cite{Diaz:2002uk,Jegerlehner:2009ry,Kelso:2014qka,Lindner:2016bgg,Kowalska:2017iqv}: 
\begingroup
\begin{equation}
I_{S\left( P\right) }^{\left( e,\mu \right) }\left( m_{E_{4,5}},m_{S}\right)
=\int_{0}^{1}\frac{x^{2}\left( 1-x\pm \frac{m_{E_{4,5}}}{m_{e,\mu }}\right) }{%
m_{e,\mu }^{2}x^{2}+\left( m_{E_{4,5}}^{2}-m_{e,\mu }^{2}\right) x+m_{S,P}^{2}\left(
1-x\right) }dx  \label{eqn:loop_integrals}
\end{equation}
\endgroup
and $S(P)$ means scalar (pseudoscalar) and $E_{4,5}$ stands for the vector-like family. It is worth mentioning that $E_{4}$ and $E_{5}$ only contribute to the muon and electron anomalous magnetic moments, respectively.

%

\section{Numerical analysis of the Higgs exchange contributions}

\label{sec:Numerical_analysis_of_scalars}

For the sake of simplicity, we consider the scenario of absence of mixing between SM charged leptons, which automatically prevents charged lepton flavour violating decays.   
In our numerical analysis we have
found that the non-SM CP-even scalar mass can reach values around $200 \func{%
GeV}$. Despite the fact that the non SM CP-even scalar is quite light and can have
a sizeable decay mode into a bottom-anti bottom quark pair, its single LHC
production via gluon fusion mechanism is strongly suppressed since it is
dominated by the triangular bottom quark loop. Such non SM CP-even scalar $H$
can also be produced by vector boson fusion but such production is expected
to have a low total cross section due to small $HWW$ and $HZZ$ couplings,
which are proportional to $v_d$. In this section we will discuss the
implications of our model in the muon and electron anomalous magnetic
moments.

\subsection{The fitting function $\protect\chi^2$ and free parameter setup}

\label{subsec:fitting_function_chi_and_parameters}

For the first approach to both anomalies, we construct the fitting function $%
\chi^2$
\begingroup
\begin{equation}
\begin{split}
\chi^2 &= \frac{\left( m_h^{\func{Thy}}-m_h^{\func{Cen}} \right)^2}{\left(
\delta m_h^{\func{Dev}} \right)^2} + \frac{\left( a_{hWW}^{\func{Thy}%
}-a_{hWW}^{\func{Cen}} \right)^2}{\left( \delta a_{hWW}^{\func{Dev}}
\right)^2} + \frac{\left( R_{\gamma\gamma}^{\func{Thy}}-R_{\gamma\gamma}^{%
\func{Cen}} \right)^2}{\left( \delta R_{\gamma\gamma}^{\func{Dev}} \right)^2}
\\
&+ \frac{\left( \Delta a_{\mu}^{\func{Thy}}-\Delta a_{\mu}^{\func{Cen}}
\right)^2}{\left( \delta \Delta a_{\mu}^{\func{Dev}} \right)^2} + \frac{%
\left( \Delta a_{e}^{\func{Thy}}-\Delta a_{e}^{\func{Cen}} \right)^2}{\left(
\delta \Delta a_{e}^{\func{Dev}} \right)^2},
\end{split}%
\end{equation}
\endgroup
where the superscripts $\func{Thy}$, $\func{Cen}$ and $\func{Dev}$ mean
theoretical prediction, central value of experimental bound and deviation
from the central value at one of $1,2,3\sigma$, respectively. The parameters
used in this fitting function are defined as below(the integer number
multiplied in delta terms means $\sigma$):
\begingroup
\begin{equation}
\begin{split}
m_h^{\func{Cen}} = 125.38 \func{GeV}, &\quad \delta m_h^{\func{Dev}} = 3
\times 0.14 \func{GeV}, \\
a_{hWW}^{\func{Cen}} = 0.59, &\quad \delta a_{hWW}^{\func{Dev}} = 1 \times
0.35, \\
R_{\gamma \gamma}^{\func{Cen}} = \frac{1}{2} \left( R_{\gamma \gamma}^{\func{%
CMS}} + R_{\gamma \gamma}^{\func{ATLAS}} \right) = 1.07, &\quad \delta
R_{\gamma \gamma}^{\func{Dev}} = 1 \times 0.14, \\
\Delta a_\mu^{\func{Cen}} = 26.1 \times 10^{-10}, &\quad \delta\Delta a_\mu^{%
\func{Dev}} = 1 \times \left( 8.0 \times 10^{-10} \right) \\
\Delta a_e^{\func{Cen}} = -0.88 \times 10^{-12}, &\quad \delta\Delta a_e^{%
\func{Dev}} = 2 \times \left( 0.36 \times 10^{-12} \right)
\end{split}%
\end{equation}
\endgroup
For an initial scan, we set up the starting parameter region as below:
\begingroup
\begin{center}
{\renewcommand{\arraystretch}{1.5} 
\begin{tabular}{cc}
\toprule
\toprule
\textbf{Parameter} & \textbf{Value/Scanned Region($\func{GeV}$)} \\ 
\midrule
$v_u = v_1$ & $\frac{\tan\beta}{\sqrt{1+\tan\beta^2}} \times 246$ \\ 
$v_d = v_2$ & $\frac{1}{\sqrt{1+\tan\beta^2}} \times 246$ \\ 
$v_\phi = v_3$ & $\pm [0.01,1.00] \times 1000$ \\ 
\midrule
$\tan\beta = v_u/v_d$ & $\left[ 5, 50 \right]$ \\ 
\midrule
$\lambda_1$ & $\left( m_h^2 - \frac{v_2 v_3^2 \lambda_5}{2 v_1}
\right)/\left( 4 v_1^2 \right)$ \\ 
$\lambda_2$ & $\pm \left[ 0.50, 12.00 \right]$ \\ 
$\lambda_3$ & $\pm \left[ 0.50, 12.00 \right]$ \\ 
$\lambda_4$ & $\pm \left[ 0.50, 12.00 \right]$ \\ 
$\lambda_5$ & $4 v_1 v_2 \lambda_3/(v_3)^2$ \\ 
$\lambda_6$ & $\pm \left[ 0.50, 12.00 \right]$ \\ 
$\lambda_7$ & $v_2 \lambda_5/v_1$ \\ 
$\lambda_8$ & $\pm \left[ 0.50, 12.00 \right]$ \\ 
\midrule
$M_{44}^{e}$ & $\left[ 2 \times 10^2, 2 \times 10^3 \right]$ \\ 
$M_{55}^{L}$ & $\left[ 2 \times 10^2, 2 \times 10^3 \right]$ \\ 
$\mu_{\func{sb}}$ & $i^{[0,1]} \times \left[300, 500 \right]$ \\ 
\midrule
$y_{e}$ & $\sqrt{2} m_{e}/v_2$ \\ 
$y_{\mu}$ & $\sqrt{2} m_{\mu}/v_2$ \\ 
$y_{24}^{e} = y_2$ & $\pm \left[ 1.0, 3.5 \right]$ \\ 
$y_{51}^{e} = y_1$ & $\pm \left[ 1.0, 3.5 \right]$ \\ 
$x_{42}^{e} = x_2$ & $\lvert y_{\mu} M_{44}^{e} / \left( y_{24}^{e} v_3 \right)
\rvert$ \\ 
$x_{15}^{L} = x_1$ & $\lvert y_{e} M_{55}^{L} / \left( y_{51}^{e} v_3 \right)
\rvert$ \\ 
\bottomrule
\bottomrule
\end{tabular}
\captionof{table}{Initial parameter setup} \label%
{tab:parameter_region_initial_scan}}
\end{center}
\endgroup
\begingroup
\begin{enumerate}
\item For the Higgs vevs, we are interested in the range of $\tan\beta$ from 
$5$ to $50$ as in the $W$ boson exchange in Figure \ref%
{fig:parameter_space_muegamma_M44_vu_tanbeta}

\item For $\lambda_1$, we fixed mass of the SM physical Higgs $h$ to be $125%
\func{GeV}$ to save time and to make the calculation faster. For $%
\lambda_{5,7} $, the assumption that no mixing between the SM Higgs $h$ and
non-SM Higgses $H_{1,2}$ arise is reflected on these parameters. All quartic coupling constants $\lambda_{1,\cdots,8}$ are set up not to go over $%
4\pi$ for perturbativity.

\item For the vector-like masses $M_{44}^{e}$ and $M_{55}^{L}$, there is a
constraint that the lightest should be greater than $200\func{GeV}$ \cite%
{Xu:2018pnq}.

\item In our numerical analysis we consider solutions where the non SM
scalar masses are larger than about $200\func{GeV}$ as done in \cite%
{Hernandez-Sanchez:2020vax}.

\item The soft-breaking mass term $\mu_{\func{sb}}$ is a free parameter,
which does not generate any problem and appropiate values of this parameters
yields masses for scalars and vector-like fermions consistent with the
experimental constraints. 

\item The diagonal Yukawa constants appearing in Equation \ref%
{eqn:diagonalized_Yukawa_matrix_charged_leptons} should be the Yukawa
constant for electron, muon and tau, respectively. The Yukawa constants $%
y_{24,51}$ and $x_{42,15}$ interacting with vector-like families are defined
under this consideration. For perturbativity, the Yukawa constants $%
y_{24,51} $ are considered not to go over $\sqrt{4\pi}$.
\end{enumerate}
\endgroup
After saturating value of the $\chi^2$ function less than or nearly $2$
which we believe it is converged enough, we find a best peaked value for
each free parameter. For the given parameters, we rename them by adding an
index ``p" to the end of subscript of each parameter like $\func{\tan\beta_p}
$ and then the expansion factor $\kappa$ is multiplied to find a correlation
between the observables and the mass parameters. Then, the parameter region
is refreshed by both the specific value of each parameter and the expansion
factor $\kappa$ as per Table \ref{tab:parameter_region_subsequent_scan}.
\begingroup
\begin{center}
{\renewcommand{\arraystretch}{1.5} 
\begin{tabular}{cc}
\toprule
\toprule
\textbf{Parameter} & \textbf{Value/Scanned Region($\func{GeV}$)} \\ 
\midrule
$v_u = v_1$ & $\frac{\tan\beta_p}{\sqrt{1+\tan\beta_p^2}} \times 246$ \\ 
$v_d = v_2$ & $\frac{1}{\sqrt{1+\tan\beta_p^2}} \times 246$ \\ 
$v_\phi = v_3$ & $\left[ (1 - \kappa), (1 + \kappa) \right] \times v_{3p}$
\\ 
\midrule
$\tan\beta = v_u/v_d$ & $\left[ (1 - \kappa), (1 + \kappa) \right] \times 
\func{\tan\beta}_p$ \\ 
\midrule
$\lambda_1$ & $\left( m_h^2 - \frac{v_{2} v_{3}^2 \lambda_{5}}{2 v_{1}}
\right)/\left( 4 v_{1}^2 \right)$ \\ 
$\lambda_2$ & $\left[ (1 - \kappa), (1 + \kappa) \right] \times \lambda_{2p}$
\\ 
$\lambda_3$ & $\left[ (1 - \kappa), (1 + \kappa) \right] \times \lambda_{3p}$
\\ 
$\lambda_4$ & $\left[ (1 - \kappa), (1 + \kappa) \right] \times \lambda_{4p}$
\\ 
$\lambda_5$ & $4 v_{1} v_{2} \lambda_{3}/(v_{3})^2$ \\ 
$\lambda_6$ & $\left[ (1 - \kappa), (1 + \kappa) \right] \times \lambda_{6p}$
\\ 
$\lambda_7$ & $v_{2} \lambda_{5}/v_{1}$ \\ 
$\lambda_8$ & $\left[ (1 - \kappa), (1 + \kappa) \right] \times \lambda_{8p}$
\\ \hline
$M_{44}^{e}$ & $\left[ (1 - \kappa), (1 + \kappa) \right] \times M_{44p}^{e}$
\\ 
$M_{55}^{L}$ & $\left[ (1 - \kappa), (1 + \kappa) \right] \times M_{55p}^{L}$
\\ 
$\mu_{\func{sb}}$ & $\left[ (1 - \kappa), (1 + \kappa) \right] \times \mu_{%
\func{sb}p}$ \\ 
\midrule
$y_{e}$ & $\sqrt{2} m_{e}/v_{2}$ \\ 
$y_{\mu}$ & $\sqrt{2} m_{\mu}/v_{2}$ \\ 
$y_{24}^{e} = y_2$ & $\left[ (1 - \kappa), (1 + \kappa) \right] \times y_{24p}^{e}$
\\ 
$y_{51}^{e} = y_1$ & $\left[ (1 - \kappa), (1 + \kappa) \right] \times y_{51p}^{e}$
\\ 
$x_{42}^{e} = x_2$ & $y_{\mu} M_{44}^{e} / \left( y_{24}^{e} v_3 \right)$ \\ 
$x_{15}^{L} = x_1$ & $y_{e} M_{55}^{L} / \left( y_{51}^{e} v_3 \right)$ \\ \midrule
$\kappa$ & $0.1$ \\ 
\bottomrule
\bottomrule
\end{tabular}
\captionof{table}{Next parameter setup after the initial scan result} \label%
{tab:parameter_region_subsequent_scan}}
\end{center}
\endgroup
\subsection{A scanned result on the free parameters as well as observables
across over the first and second scan}

\label{subsec:scanned_result}

The best peaked value for each parameter is listed in Table \ref%
{tab:parameter_region_best_peaked} and energy scale is in unit of $\func{GeV}
$. Note that all cases are carried out independently and all points of plots
in each case are collected within $1\sigma$ constraint of each anomaly.
\begingroup
\begin{center}
{\renewcommand{\arraystretch}{1.5}
\resizebox{!}{0.6\textwidth}{ 
\begin{tabular}{|c|c|c|c|c|c|}
\hline
\textbf{Parameter} & case A & case B & case C & case D & case E \\ \hline
$v_u = v_1$ & $245.925$ & $245.936$ & $245.951$ & $245.917$ & $245.948$ \\ 
$v_d = v_2$ & $6.086$ & $5.595$ & $4.921$ & $6.387$ & $5.077$ \\ 
$v_\phi = v_3$ & $-57.761$ & $-36.470$ & $-57.919$ & $-30.746$ & $-17.146$ \\ \hline
$\tan\beta = v_u/v_d$ & $40.410$ & $43.957$ & $49.977$ & $38.503$ & $48.441$ \\ \hline
$\lambda_1$ & $0.063$ & $0.064$ & $0.066$ & $0.064$ & $0.065$ \\ 
$\lambda_2$ & $-7.978$ & $8.414$ & $-2.000$ & $2.948$ & $10.382$ \\ 
$\lambda_3$ & $-6.344$ & $-2.675$ & $6.242$ & $-1.724$ & $-0.706$ \\ 
$\lambda_4$ & $1.859$ & $2.158$ & $-3.633$ & $10.837$ & $-2.796$ \\ 
$\lambda_5$ & $-11.384$ & $-11.070$ & $9.009$ & $-11.460$ & $-12.000$ \\ 
$\lambda_6$ & $2.888$ & $1.228$ & $0.866$ & $1.351$ & $1.324$ \\ 
$\lambda_7$ & $-0.282$ & $-0.252$ & $0.180$ & $-0.298$ & $-0.248$ \\ 
$\lambda_8$ & $-1.363$ & $-1.346$ & $-10.845$ & $-11.510$ & $7.033$ \\ \hline
$M_{44}^{e}$ & $1475.010$ & $1355.470$ & $1495.770$ & $1134.340$ & $1681.760$ \\ 
$M_{55}^{L}$ & $279.386$ & $211.263$ & $204.706$ & $323.292$ & $331.462$ \\ 
$\mu_{\func{sb}}$ & $424.618i$ & $443.435i$ & $480.993$ & $480.062i$ & $491.533$  \\ 
\hline
$y_{e} \left[ 10^{-4} \right]$ & $1.135$ & $1.234$ & $1.403$ & $1.081$ & $1.360$ \\ 
$y_{\mu} \left[ 10^{-2} \right]$ & $2.391$ & $2.600$ & $2.956$ & $2.278$ & $2.865$ \\ 
$y_{24}^{e} = y_2$ & $-3.161$ & $-3.101$ & $-2.942$ & $-1.548$ & $1.662$ \\ 
$y_{51}^{e} = y_1$ & $2.315$ & $2.164$ & $2.050$ & $1.352$ & $3.377$ \\ 
$x_{42}^{e} = x_2$ & $0.193$ & $0.312$ & $0.260$ & $0.543$ & $1.691$ \\ 
$x_{15}^{L} = x_1 \left[ 10^{-4} \right]$ & $2.371$ & $3.304$ & $2.419$ & $8.408$ & $7.787$ \\ \hline
$m_{H_1}$ & $213.390$ & $222.924$ & $212.147$ & $238.523$ & $205.477$ \\ 
$m_{H_2}$ & $911.585$ & $614.516$ & $891.413$ & $518.147$ & $354.709$ \\ 
$m_{A_1}$ & $741.343$ & $537.111$ & $807.268$ & $435.887$ & $282.964$ \\ 
$m_{A_2}$ & $1003.790$ & $939.553$ & $1035.800$ & $1006.240$ & $1015.760$ \\ 
$m_{H^\pm}$ & $938.259$ & $674.054$ & $987.625$ & $929.786$ & $504.684$ \\ \hline
$\Delta a_\mu \left[ 10^{-9} \right]$ & $2.734$ & $2.688$ & $2.935$ & $2.891$ & $2.393$ \\ 
$\Delta a_e \left[ 10^{-13} \right]$ & $-5.073$ & $-8.310$ & $-5.543$ & $%
-6.365$ & $-9.232$ \\ 
$a_{hWW}$ & $1.000$ & $1.000$ & $1.000$ & $1.000$ & $1.000$ \\ 
$R_{\gamma\gamma}$ & $0.999$ & $0.999$ & $0.999$ & $0.999$ & $0.999$ \\ \hline
$\chi^2$ & $1.794$ & $1.516$ & $1.870$ & $1.740$ & $1.579$ \\ \hline
\end{tabular}}}
\captionof{table}{A best peaked value for each parameter at each case. All
energy scale is in $\func{GeV}$ units. Notice that in all cases $v_3$ is smaller than the vector like mass parameters $M_{44}^{e}$ and $M_{55}^{L}$, which is consistent with the assumption made in section I, regarding the fact that the corresponding expansion parameter $v_3/M_{\psi}$ is less than unity.} \label{tab:parameter_region_best_peaked}
\end{center}
\endgroup
Here, we put two constraints on the lightest vector-like mass and the
lightest non-SM scalar mass; the vector-like mass should be greater than $200%
\func{GeV}$ as well as the non-SM scalar mass\cite{Xu:2018pnq,Hernandez-Sanchez:2020vax}. After we
carry out second parameter scan based on the first scan result of Table \ref%
{tab:parameter_region_best_peaked}, range of the parameters are given in
Table \ref{tab:range_secondscan}.
\begingroup
\begin{sidewaystable}
\begin{center}
{\renewcommand{\arraystretch}{2.0}
\resizebox{\textwidth}{!}{
\begin{tabular}{|c|c|c|c|c|c|}
\hline
\textbf{Parameter} & case A & case B & case C & case D & case E \\ \hline
$v_u = v_1$ & $\left[ 245.907 \rightarrow 245.938 \right]$ & $\left[ 245.921
\rightarrow 245.947 \right]$ & $\left[ 245.939 \rightarrow 245.959 \right]$
& $\left[ 245.898 \rightarrow 245.931 \right]$ & $\left[ 245.935 \rightarrow 245.957 \right]$ \\ 
$v_d = v_2$ & $\left[ 5.533 \rightarrow 6.761 \right]$ & $\left[ 5.087
\rightarrow 6.216 \right]$ & $\left[ 4.474 \rightarrow 5.468 \right]$ & $%
\left[ 5.807 \rightarrow 7.096 \right]$ & $\left[ 4.616 \rightarrow 5.641 \right]$ \\ 
$v_\phi = v_3$ & $\left[ -63.525 \rightarrow -51.985 \right]$ & $\left[
-40.117 \rightarrow -32.823 \right]$ & $\left[ -63.706 \rightarrow -52.128 %
\right]$ & $\left[ -33.820 \rightarrow -27.671 \right]$ & $\left[ -18.860 \rightarrow -15.438 \right]$ \\ \hline
$\tan\beta = v_u/v_d$ & $\left[ 36.371 \rightarrow 44.451 \right]$ & $\left[
39.561 \rightarrow 48.353 \right]$ & $\left[ 44.980 \rightarrow 54.975 %
\right]$ & $\left[ 34.653 \rightarrow 42.354 \right]$ & $\left[ 43.597 \rightarrow 53.284 \right]$ \\ \hline
$m_{H_1}$ & $\left[ 200.000 \rightarrow 242.653 \right]$ & $\left[ 201.520
\rightarrow 246.046 \right]$ & $\left[ 200.000 \rightarrow 230.754 \right]$
& $\left[ 215.523 \rightarrow 261.920 \right]$ & $\left[ 200.000 \rightarrow 220.017 \right]$ \\ 
$m_{H_2}$ & $\left[ 752.061 \rightarrow 1088.130 \right]$ & $\left[ 516.289
\rightarrow 724.997 \right]$ & $\left[ 735.831 \rightarrow 1059.900 \right]$
& $\left[ 441.371 \rightarrow 604.981 \right]$ & $\left[ 338.724 \rightarrow 374.424 \right]$ \\ 
$m_{A_1}$ & $\left[ 638.813 \rightarrow 853.637 \right]$ & $\left[ 442.527
\rightarrow 640.705 \right]$ & $\left[ 670.550 \rightarrow 945.705 \right]$
& $\left[ 357.697 \rightarrow 516.760 \right]$ & $\left[ 266.086 \rightarrow 297.589 \right]$ \\ 
$m_{A_2}$ & $\left[ 892.847 \rightarrow 1141.780 \right]$ & $\left[ 847.825
\rightarrow 1032.140 \right]$ & $\left[ 927.768 \rightarrow 1154.640 \right]$
& $\left[ 907.576 \rightarrow 1105.770 \right]$ & $\left[ 918.667 \rightarrow 1114.960 \right]$ \\ 
$m_{H^\pm}$ & $\left[ 783.823 \rightarrow 1111.600 \right]$ & $\left[
580.316 \rightarrow 779.945 \right]$ & $\left[ 842.585 \rightarrow 1143.880 %
\right]$ & $\left[ 856.237 \rightarrow 1007.360 \right]$ & $\left[ 478.900 \rightarrow 529.178 \right]$ \\ \hline
$M_{44}^e$ & $\left[ 1327.510 \rightarrow 1622.510 \right]$ & $\left[
1219.930 \rightarrow 1491.020 \right]$ & $\left[ 1346.190 \rightarrow
1645.330 \right]$ & $\left[ 1029.900 \rightarrow 1247.770 \right]$ & $\left[ 1513.590 \rightarrow 1849.930 \right]$ \\ 
$M_{55}^L$ & $\left[ 251.447 \rightarrow 307.323 \right]$ & $\left[ 200.000
\rightarrow 232.389 \right]$ & $\left[ 200.000 \rightarrow 225.176 \right]$
& $\left[ 290.963 \rightarrow 355.621 \right]$ & $\left[ 298.317 \rightarrow 364.604 \right]$ \\ 
$\lvert \mu_{\func{sb}} \rvert$ & $\left[ 382.158 \rightarrow 467.079 \right]
$ & $\left[ 399.091 \rightarrow 487.777 \right]$ & $\left[ 432.895
\rightarrow 529.091 \right]$ & $\left[ 432.059 \rightarrow 528.067 \right]$ & $\left[ 442.381 \rightarrow 540.679 \right]$
\\ \hline
$\Delta a_\mu \left[ 10^{-9} \right]$ & $\left[ 1.811 \rightarrow 3.410 %
\right]$ & $\left[ 1.810 \rightarrow 3.410 \right] $ & $\left[ 1.810
\rightarrow 3.410 \right]$ & $\left[ 1.810 \rightarrow 3.410 \right]$ & $\left[ 1.810 \rightarrow 3.410 \right]$ \\ 
$\Delta a_e \left[ 10^{-13} \right]$ & $\left[ -6.730 \rightarrow -5.200 %
\right]$ & $\left[ -11.142 \rightarrow -5.985 \right]$ & $\left[ -7.207
\rightarrow -5.200 \right]$ & $\left[ -8.721 \rightarrow -5.200 \right]$ & $\left[ -12.393 \rightarrow -5.442 \right]$
\\ 
$a_{hWW}$ & $\left[ 1.000 \rightarrow 1.000 \right]$ & $\left[ 1.000
\rightarrow 1.000 \right]$ & $\left[ 0.999 \rightarrow 1.000 \right]$ & $%
\left[ 1.000 \rightarrow 1.000 \right]$ & $\left[ 1.000 \rightarrow 1.000 \right]$ \\ 
$R_{\gamma\gamma}$ & $\left[ 0.999 \rightarrow 0.999 \right]$ & $\left[
0.999 \rightarrow 0.999 \right]$ & $\left[ 0.999 \rightarrow 1.000 \right]$
& $\left[ 1.000 \rightarrow 1.000 \right]$ & $\left[ 0.999 \rightarrow 1.000 \right]$ \\ \hline
$\chi^2$ & $\left[ 1.604 \rightarrow 2.750 \right]$ & $\left[ 1.501
\rightarrow 2.635 \right]$ & $\left[ 1.580 \rightarrow 2.761 \right]$ & $%
\left[ 1.509 \rightarrow 2.749 \right]$ & $\left[ 1.501 \rightarrow 2.720 \right]$ \\ \hline
\end{tabular}
}
\captionof{table}{A scanned range of each parameter at case A, B, C, D and E.
$H_{1,2}$ mean non SM CP-even scalars and $A_{1,2}$ are non SM CP-odd scalars and $H^\pm$ stand for non SM charged scalars in this model. All data of $\Delta
a_{\mu,e}$ are collected within the $1\sigma$ constraint of each anomaly.} %
\label{tab:range_secondscan}}
\end{center}
\end{sidewaystable}
\endgroup
\subsection{The muon and electron anomalous magnetic moments}

\label{subsec:muon_electrong2_scalar}

In order to confirm that our theoretical prediction for both anomalies can accommodate their constraints at $1\sigma$ and to analyze correlations between both anomalies and mass parameters, we consider cases B and E in Table \ref{tab:parameter_region_best_peaked} since their benchmark point have relative lower values of the $\chi^2$ function when compared to other cases. The reason that the cases B and E have the lower values of the $\chi^2$ function arises from the obtained value of the electron anomaly, which is very close to the central experimental value. All cases reveal nearly central value of muon anomaly constraint at $1\sigma$, whereas the other cases except B and E reveal nearly edge value of electron anomaly constraint at $1\sigma$. Therefore, the reason why the cases B and E are more converged is related to whether our theoretical prediction for both anomalies can gain access to their central value of each anomaly constraint at $1\sigma$. More importantly, the case E is only one satisfying vacuum stability conditions and a detailed investigation for the vacuum stability of each case will be studied in a subsection. For these reasons, we take the case E in Table \ref{tab:range_secondscan} to study the correlations. The relevant parameter spaces are listed in Figure \ref{fig:relevant_parameters_delamue} and \ref{fig:relevant_parameters_vl_delamue}.
\begingroup
\begin{figure}[]
\centering
\begin{subfigure}{0.48\textwidth}
\includegraphics[keepaspectratio,scale=0.45]{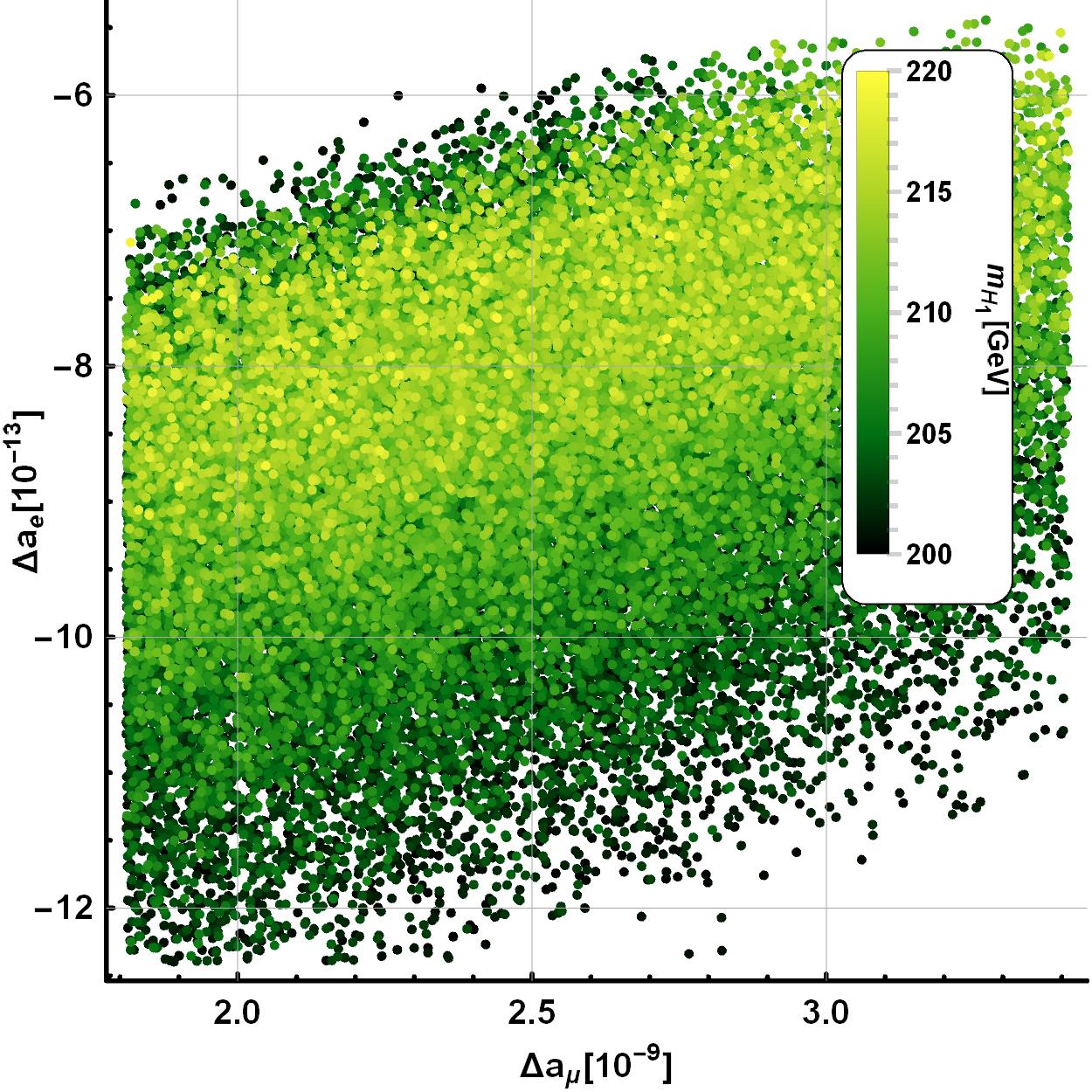} 
\end{subfigure}
\hspace{0.1cm}
\begin{subfigure}{0.48\textwidth}
\includegraphics[keepaspectratio,scale=0.45]{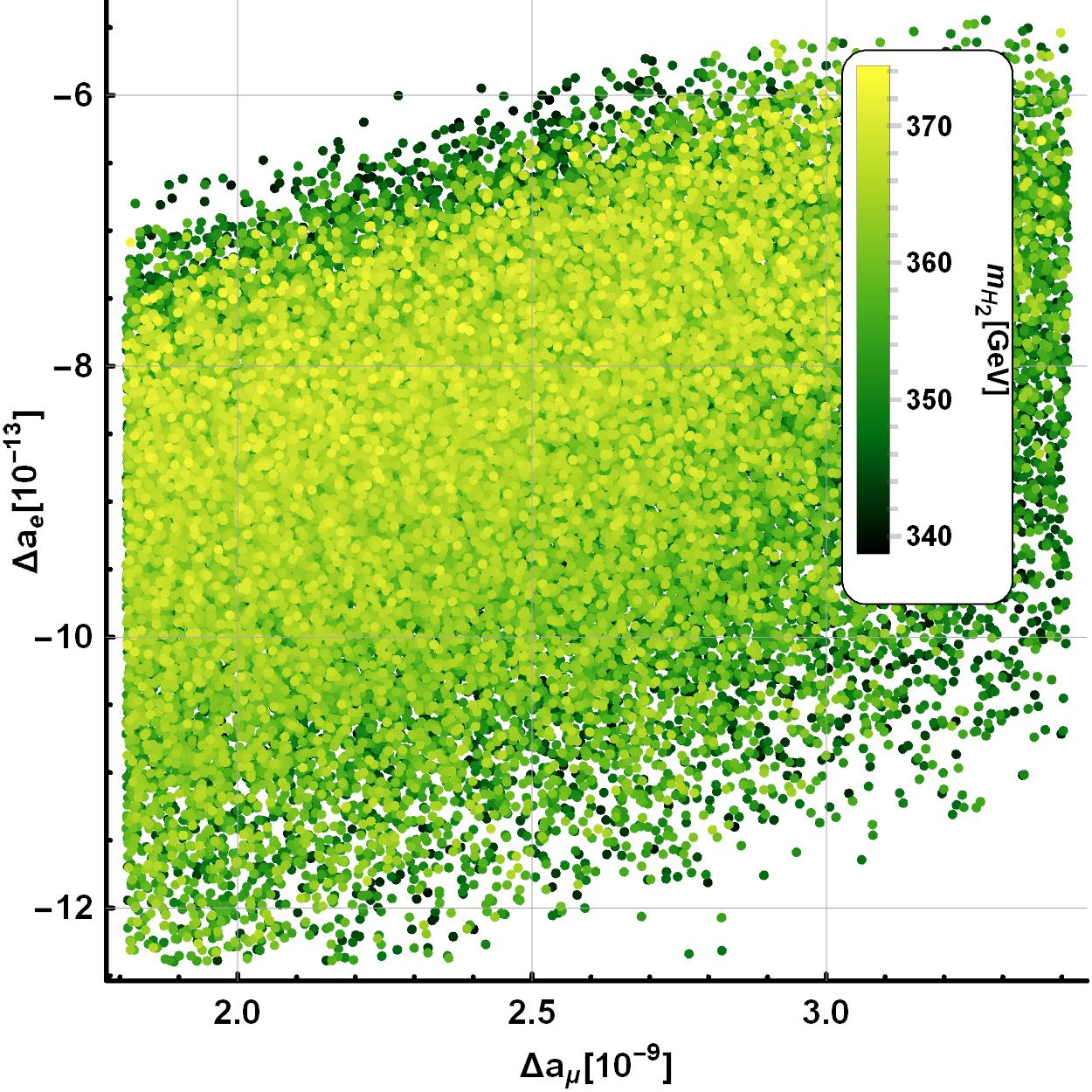} 
\end{subfigure} 
\par
\begin{subfigure}{0.48\textwidth}
\includegraphics[keepaspectratio,scale=0.45]{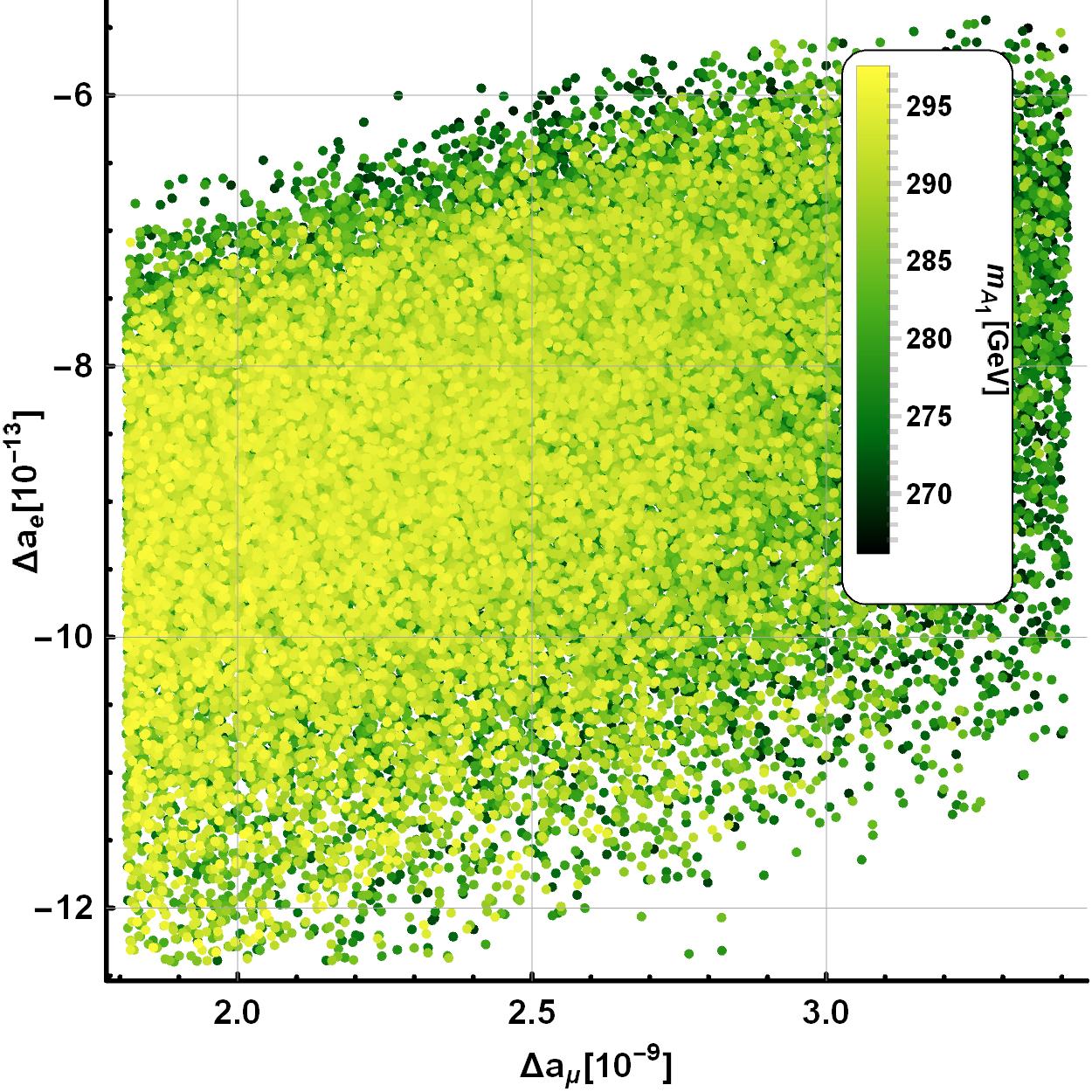} 
\end{subfigure}
\hspace{0.1cm}
\begin{subfigure}{0.48\textwidth}
\includegraphics[keepaspectratio,scale=0.45]{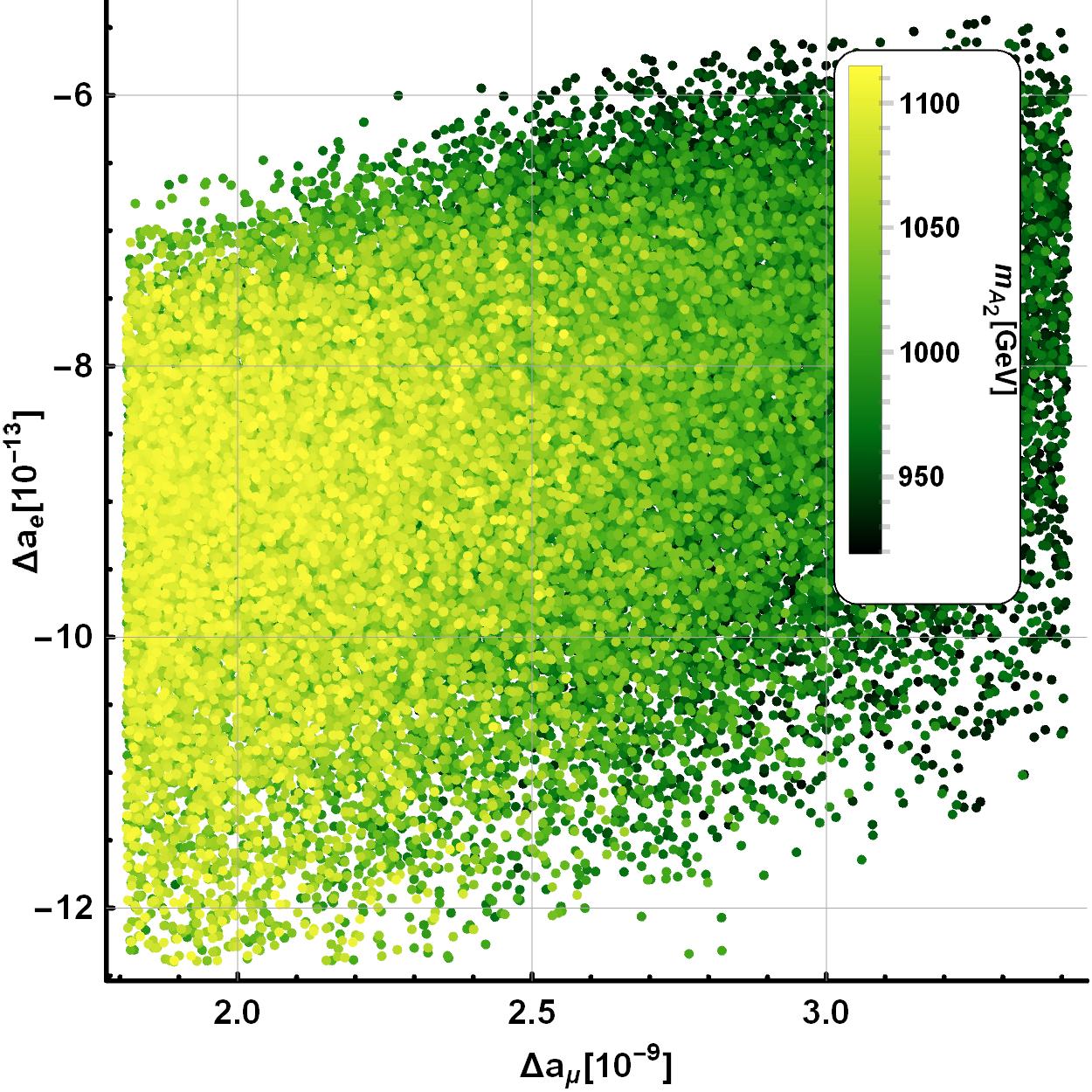} 
\end{subfigure}
\begin{subfigure}{0.48\textwidth}
\includegraphics[keepaspectratio,scale=0.45]{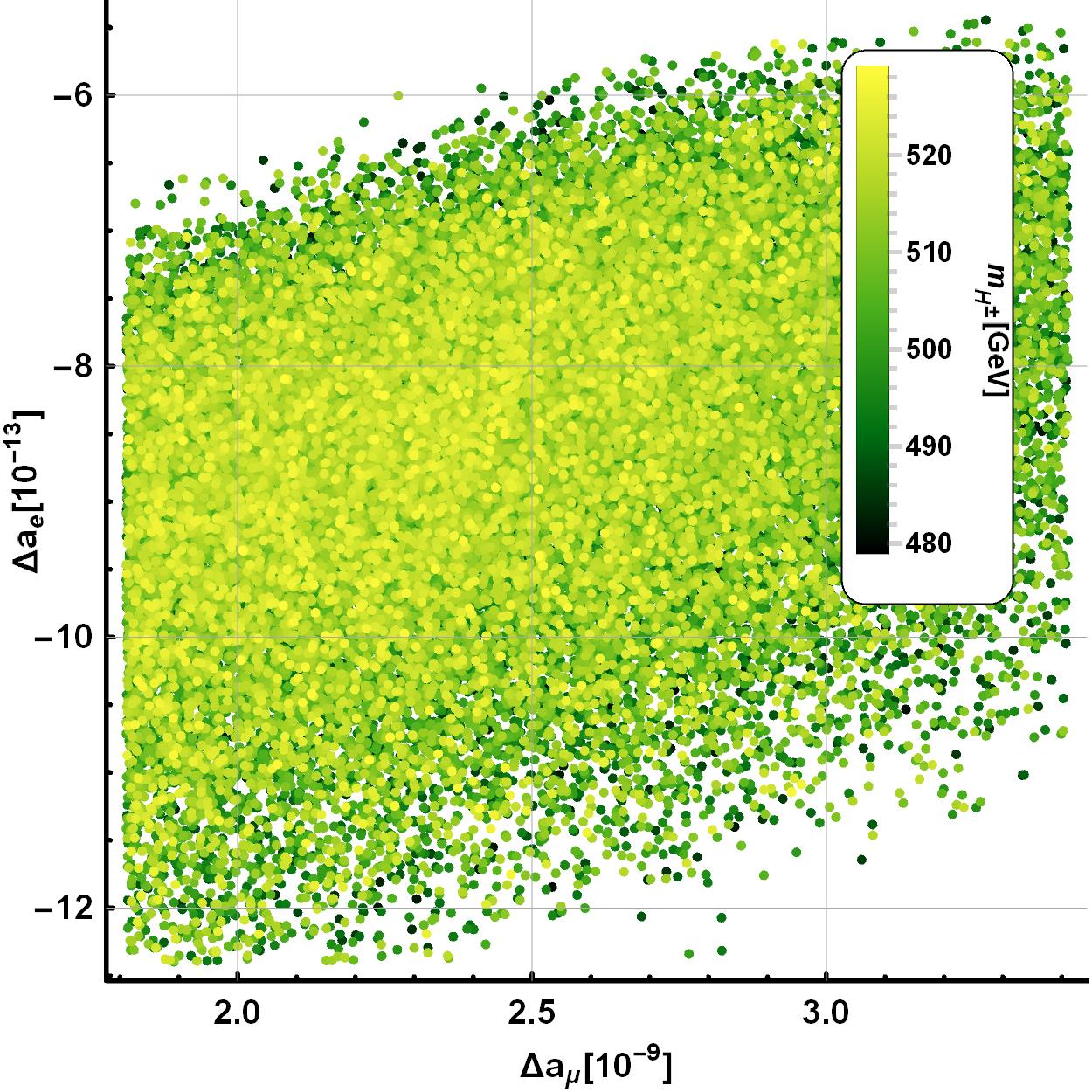} 
\end{subfigure}
\caption{Available parameter spaces for the muon anomaly versus electron
anomaly with a mass parameter which attends the both anomalies$%
(H_{1,2},A_{1,2})$ and does not$(H^\pm)$. $H_{1,2}$ are non-SM CP even scalars, $%
A_{1,2}$ are non-SM CP odd scalars and $H^{\pm}$ are non-SM charged scalars.
All points in each plot are collected within $1\protect\sigma$ constraint of
each anomaly.}
\label{fig:relevant_parameters_delamue}
\end{figure}
\endgroup
\begingroup
\begin{figure}[]
\centering
\begin{subfigure}{0.48\textwidth}
\includegraphics[keepaspectratio,scale=0.46]{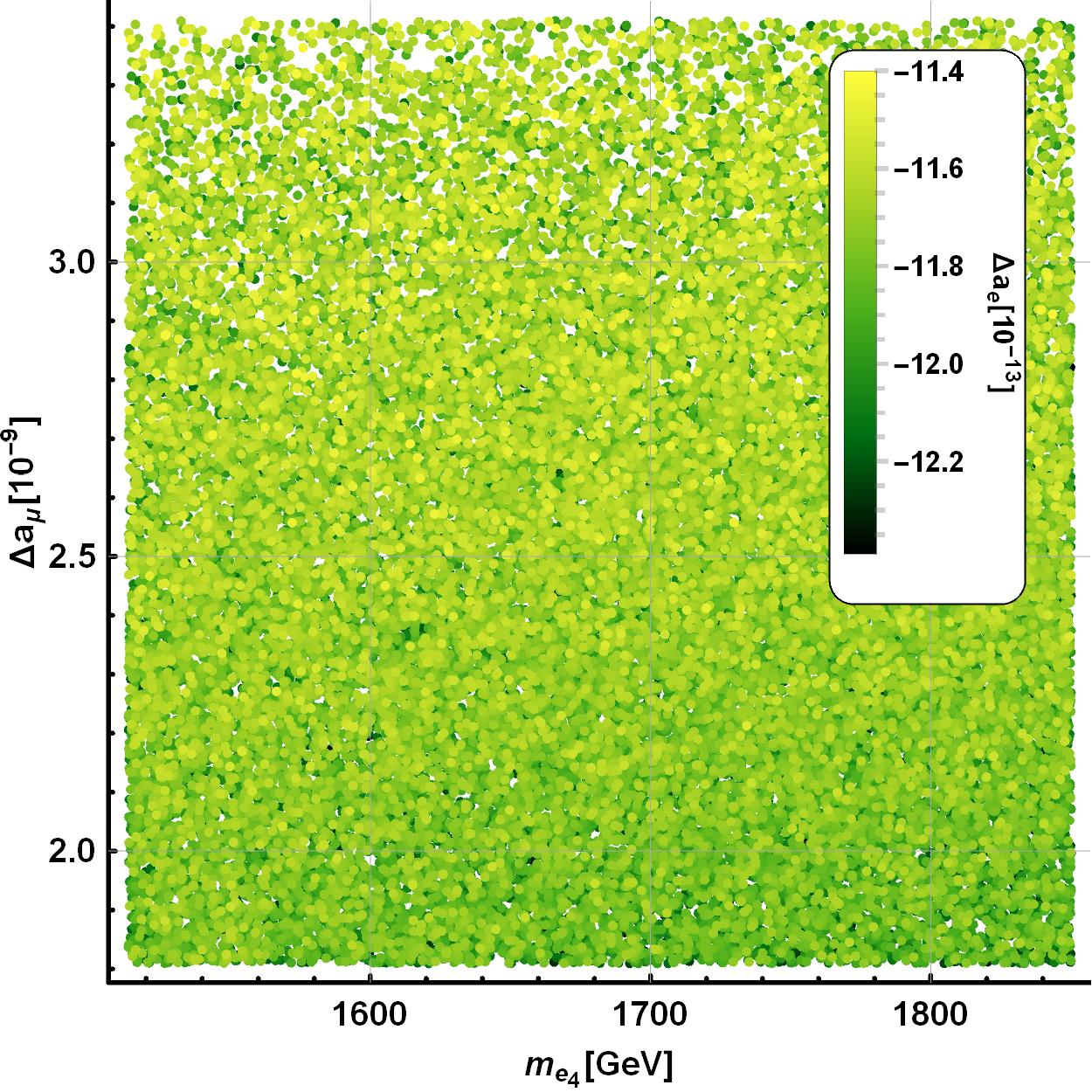} 
\end{subfigure}
\begin{subfigure}{0.48\textwidth}
\includegraphics[keepaspectratio,scale=0.46]{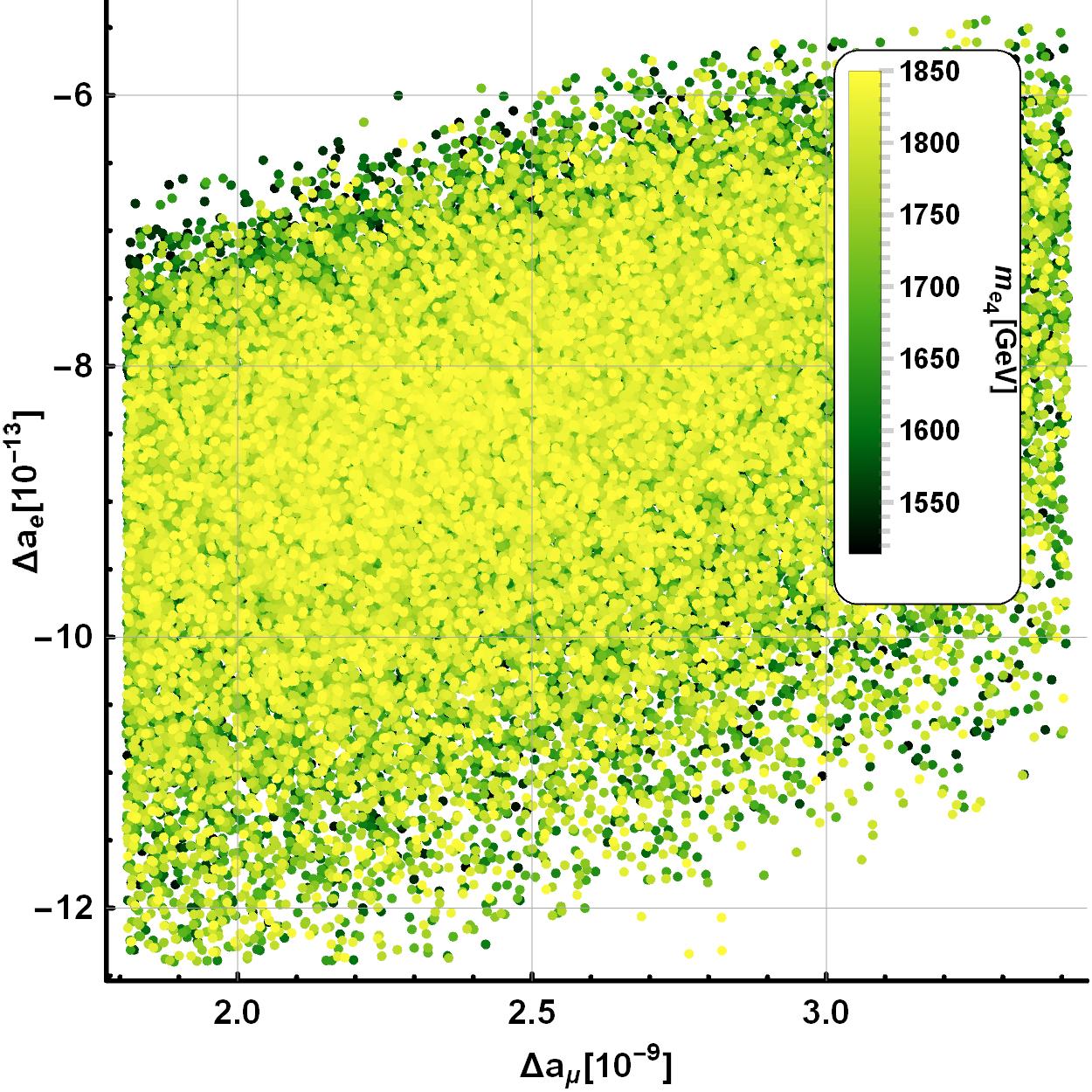} 
\end{subfigure} 
\par
\begin{subfigure}{0.48\textwidth}
\includegraphics[keepaspectratio,scale=0.46]{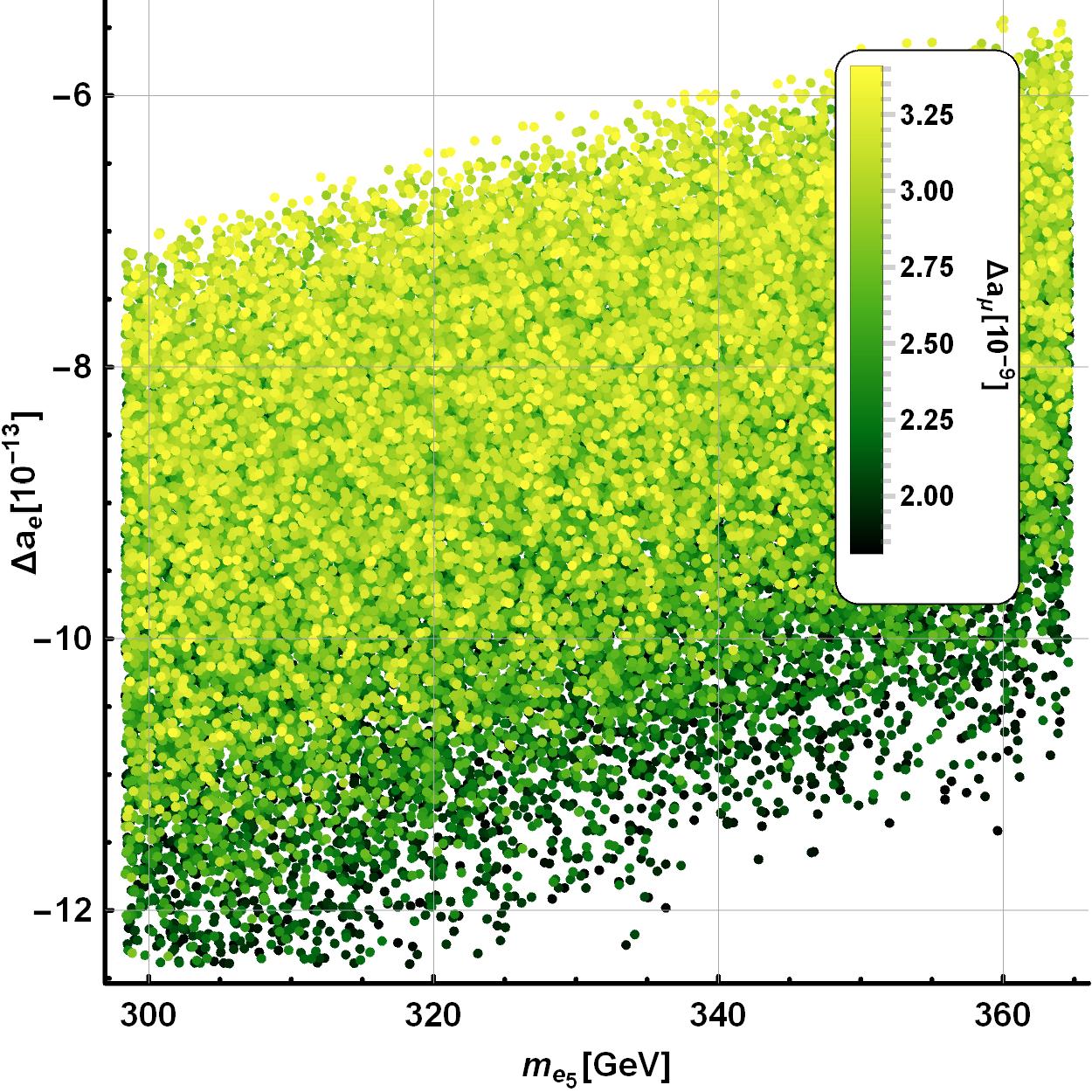} 
\end{subfigure}
\begin{subfigure}{0.48\textwidth}
\includegraphics[keepaspectratio,scale=0.46]{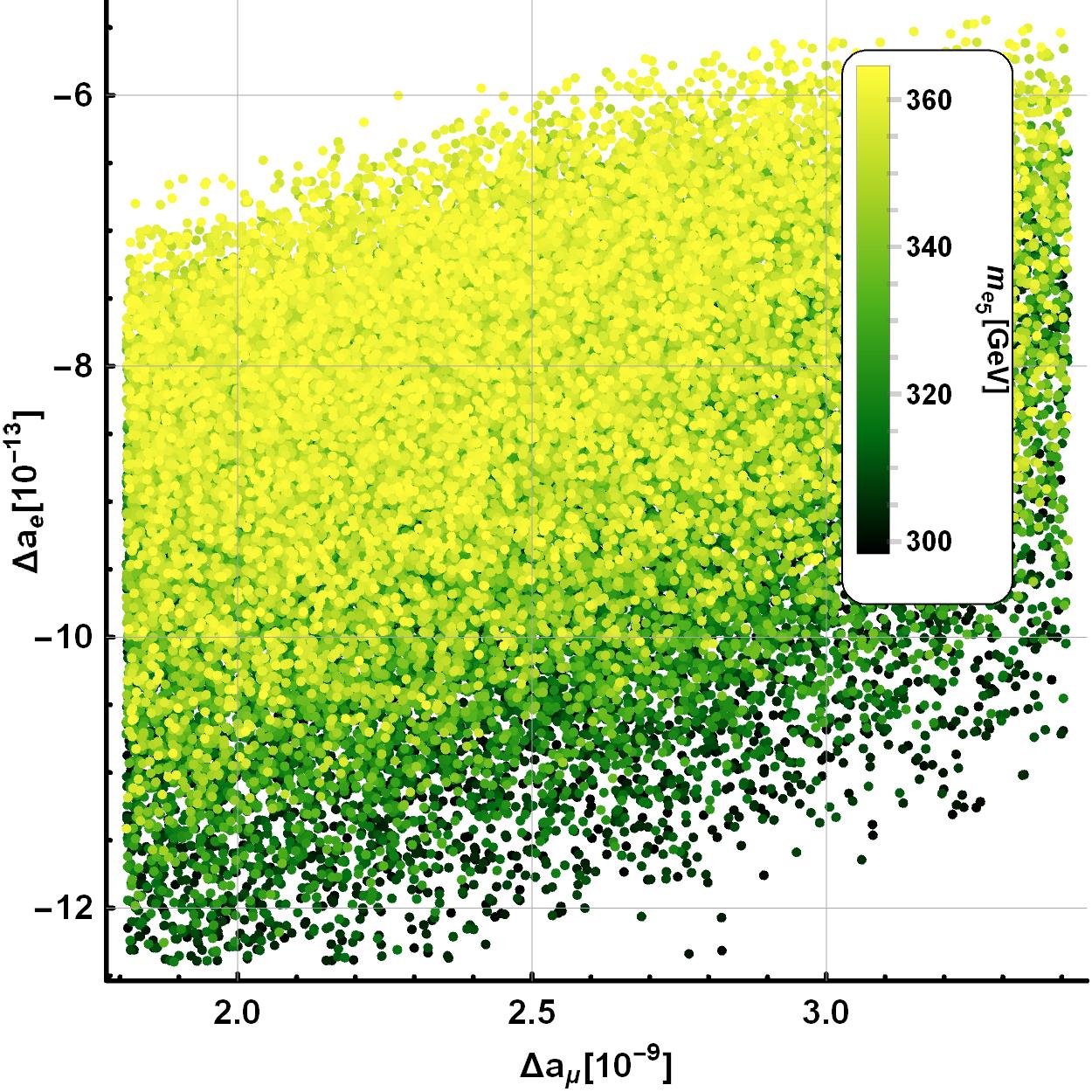} 
\end{subfigure}
\caption{Available parameter spaces for the muon anomaly(electron anomaly)
versus a relevant vector-like mass $m_{e_4}(m_{e_5})$ with another
anomaly(two left plots) in bar where $m_{e_4}(m_{e_5})$ is simplified
notation for $M_{44}^{e}(M_{55}^{L})$, while the two right plots for the
muon anomaly versus electron anomaly with a vector-like mass $%
m_{e_4}(m_{e_5})$}
\label{fig:relevant_parameters_vl_delamue}
\end{figure}
\endgroup
To begin with, we consider the parameter spaces for the muon anomaly versus electron anomaly with a mass parameter which attends both anomalies $(H_{1,2}, A_{1,2})$ and does not $(H^\pm)$ in Figure \ref{fig:relevant_parameters_delamue}. Even thought the non-SM charged scalar does not attend both anomalies, the
similar pattern which the other scalars implement in Figure \ref%
{fig:relevant_parameters_delamue} is also appeared. We confirmed that mass of $H_2$ is nearly proportional to that of $H^\pm$, which causes the correlation identified in plots of the other non-SM scalars in Figure \ref{fig:relevant_parameters_delamue} is still maintained for the non-SM charged scalar. Interestingly, the cases A, B and C in Table \ref{tab:range_secondscan} reported $m_{H_2}$ is nearly proportional to $m_{H^\pm}$ one-to-one ratio, whereas the cases D and E revealed a fat proportion between them and still maintained the correlation.
\newline
As mentioned at the beginning of this section, we take the case E for the plots in Figure \ref{fig:relevant_parameters_delamue} and \ref{fig:relevant_parameters_vl_delamue} and a main distinction between the case E and others arises from the value of electron anomaly. If we take other cases instead of the case E to investigate the parameter spaces, the parameter region appeared in top-left plot of Figure \ref{fig:relevant_parameters_delamue} will be shifted upward by locating at the value of $-5$ or $-6 \times 10^{-13}$ for the electron anomaly. In other words, the whole colored region in Figure \ref{fig:relevant_parameters_delamue} is shifted upwards to meet the scanned value of electron anomaly constraint at $1\sigma$, holding the correlations. Therefore, the white region appeared in Figure \ref{fig:relevant_parameters_delamue} is not strictly excluded region and affected by how well a benchmark point is converged and by a factor of $\kappa$. However, these plots still tells a correlation between both anomalies and a tendency that the lighter mass of $H_1$ is located at edge region of the parameter space. Mass of the lightest non-SM scalar $H_1$ implied in top-left plot of Figure \ref{fig:relevant_parameters_delamue} is ranged from $200$ to $220\func{GeV}$~\cite{Hernandez-Sanchez:2020vax} and the cross section for this light non-SM scalar will be compared to that for SM Higgs in appendix. As for mass range of the other non-SM scalars confirmed in rest of other plots in Figure \ref{fig:relevant_parameters_delamue}, they all implied heavier mass than that of $H_1$ which can be flexible depending on how the parameters are converged as seen in each case of Table \ref{tab:range_secondscan}.
\newline~\newline
We investigate a correlation for an anomaly versus a relevant mass parameter with another anomaly in bar in Figure \ref{fig:relevant_parameters_vl_delamue}. Note that the fourth vector-like mass is relevant only for the muon anomaly, whereas the fifth is only for the electron anomaly. Even though the fourth (fifth) is irrelevant to the electron (muon) anomaly, it is good to express them together since we rearrange the mass parameters and the anomalies in bar for comparison. The top-left plot in Figure \ref{fig:relevant_parameters_vl_delamue} just fills in whole parameter region, thus no any correlation between the fourth vector-like mass and the muon anomaly is identified. After we rearranged the order of $m_{e_4}$ and $\Delta a_{\mu,e}$ from the top-left plot, we can confirm the similar correlation identified in Figure \ref{fig:relevant_parameters_delamue} from the top-right plot in Figure \ref{fig:relevant_parameters_vl_delamue}. The bottom-left plot identifies some correlation between the fifth vector-like mass and the electron anomaly contrary to the top-left plot. For the fifth vector-like mass, we put the constraint that the lightest vector-like mass should be greater than $200\func{GeV}$ \cite%
{Xu:2018pnq} and the mass region below $200\func{GeV}$ is all excluded. After rearranging the order of $m_{e_5}$ and $\Delta a_{\mu}$ as in the above plot, we confirmed the similar correlation appears in the bottom-right plot. Interestingly, the top-right and the bottom-right plots check the similar correlation.
\newline~\newline
We confirmed that the muon and electron anomalous magnetic moments with vector-like particles can be explained to within $1\sigma$ constraint of each anomaly in a unified way, which is based on two attributes; the first one is the extended scalar sector and the second one is related with the contributions of the vector-like leptons. The first one which is reflected in our prediction for both anomalies, consists of four non-SM scalars and these contributions play a crucial role for determining the magnitude of each anomaly. The second one is seen by two vertices of both anomaly diagrams. The other Yukawa interactions can take place at each vertex since the vector-like leptons come in the loop, which is differentiated by the case where the normal SM particles enter in the loop. To be more specific, the helicity flip mass caused by the vector-like fermions in the CP-even and CP-odd basis couples the initial particle inside the loop to another particle of different chirality, thus allowing different interactions at each vertex.  This means that the different sign problem can be solved by only considering multiplication of the Yukawa constants of each vertex and this property will be covered in detail in next subsection.

\subsubsection{Vacuum stability}

An important feature of our extended 2HDM theory 
is that it predicts 
large values for the Yukawa coupling constants $y_{2,1}, x_{2,1}$ which can be ideally order of unity in our model. If the Yukawa coupling constants are much lower than unity, which means $y_{2,1},x_{2,1} \ll 1$, it will not cause any problem for stabilization of the scalar potential. 
However, large values of the leptonic Yukawa couplings are required in our model to successfully explain both $g-2$ anomalies within the $1\sigma$ experimentally allowed range 
and since they are somehow related with the electroweak sector parameters, it might be able to destabilize the Higgs potential. As 
previously mentioned, in our analysis of the scalar sector and $g-2$ anomalies we are restricting to the scenario of decoupling limit, 
which implies that the 
large values of the leptonic Yukawa couplings will have a very small impact in 
the stability of up-type Higgs $H_u$ potential, whereas the conditons for the stability of the down type Higgs $H_d$ potential need to be determined. 
To discuss the stability of the scalar potential, one has to analyze its
quartic terms because they will dominate the behaviour of the scalar
potential in the region of very large values of the field components. To
this end, the quartic terms of the scalar potential are written in terms of
the Hermitian bilinear combination of the scalar fields. To simplify our
analysis, we discuss the stability conditions of the resulting 2HDM scalar
potential arising after the gauge singlet scalar field $\phi $ acquire
vacuum expectation value. Such stability conditions have been analyzed in detail in the framework of 2HDM in \cite{Maniatis:2006fs,Bhattacharyya:2015nca}. In order to analyze the stability 
of the $H_d$ potential, what we need to check if 
the quartic scalar couplings in each case of Table~\ref{tab:parameter_region_best_peaked} fullfill the stability conditions to be determined below. Given that our Higgs potential corresponds to the one of an extended 2HDM with the flavon field $\phi$, in order to apply the stability conditions used in the reference~\cite{Bhattacharyya:2015nca} to our Higgs potential, we need to reduce the number of scalar degrees of freedom by considering the resulting 2HDM scalar potential arising after the gauge singlet scalar field $\phi$ is integrated out. 
From the scalar potential it follows that the relevant quartic coupling constant $\lambda_6$ must be positive, otherwise the vev $v_3$ would fall into negative infinity when the field $\phi$ value increases. For the same reason, the quartic coupling constants $\lambda_{1,2}$ must also be positive. 
From the aforementioned stability conditions 
we conclude that the cases A and C must be excluded since their corresponding quartic coupling constants $\lambda_2$ are negative. Assuming the flavon field $\phi$ develops its vev $v_3$, we can rewrite the Higgs potential in terms of $H_u$ and $H_d$ fields as follows:
\begingroup
\begin{equation}
\begin{split}
V &=\mu _{1}^{2}\left( H_{u}H_{u}^{\dagger }\right) +\mu _{2}^{2}\left(
H_{d}H_{d}^{\dagger }\right) +\mu _{3}^{2}\frac{v_3^{2}}{2}
+2\mu _{\func{sb}}^{2}\frac{v_{3}^{2}}{2}
+\lambda _{1}\left( H_{u}H_{u}^{\dagger }\right) ^{2}+\lambda _{2}\left(
H_{d}H_{d}^{\dagger }\right) ^{2} \\
&+\lambda _{3}\left( H_{u}H_{u}^{\dagger }\right) \left(
H_{d}H_{d}^{\dagger }\right) +\lambda _{4}\left( H_{u}H_{d}^{\dagger
}\right) \left( H_{d}H_{u}^{\dagger }\right) +\lambda _{5}\left( \varepsilon
_{ij}H_{u}^{i}H_{d}^{j}\frac{v_{3}^{2}}{2}+\func{h.c}\right)  \\
&+\lambda _{6}\frac{v_{3}^{4}}{4}+\lambda _{7}\frac{v_{3}^{2}}{2} \left( H_{u}H_{u}^{\dagger }\right) +\lambda _{8}\frac{v_{3}^{2}}{2} \left( H_{d}H_{d}^{\dagger }\right).
\end{split}
\end{equation}%
\endgroup
Dropping all numbers and combining same order terms, the Higgs potential becomes much simpler as follows:
\begingroup
\begin{equation}
\begin{split}
V &=\left( \mu _{1}^{2} + \lambda _{7}\frac{v_{3}^{2}}{2} \right)\left( H_{u}H_{u}^{\dagger }\right) +\left( \mu _{2}^{2} + \lambda _{8}\frac{v_{3}^{2}}{2} \right)\left(
H_{d}H_{d}^{\dagger }\right) 
\\
&+ \lambda _{1} \left( H_{u}H_{u}^{\dagger }\right) ^{2}+\lambda _{2}\left(
H_{d}H_{d}^{\dagger }\right) ^{2}
\\
&+\lambda _{3}\left( H_{u}H_{u}^{\dagger }\right) \left(
H_{d}H_{d}^{\dagger }\right) +\lambda _{5} \frac{v_{3}^{2}}{2} \left( \varepsilon
_{ij}H_{u}^{i}H_{d}^{j}+\func{h.c}\right)  \\
\end{split}
\end{equation}%
\endgroup
where it is worth mentioning that the 
$\lambda_4$ term can be safely removed in the Higgs potential since it does not play a role in the CP-even, odd but charged mass matrix (Now our focus is the neutral scalar sectors). Here, we can 
impose one 
extra condition for the stabilization check, which is that the redefined mass terms must be negative, otherwise we get zero vev as a global minimum.
\begingroup
\begin{equation}
\begin{split}
&\mu_1^2 + \lambda_7 \frac{v_3^2}{2} = -2\lambda_1 v_1^2 - \lambda_3 v_2^2 + 2\lambda_3 v_2^2 = -2\lambda_1 v_1^2 + \lambda_3 v_2^2 < 0 \\
&\mu_2^2 + \lambda_8 \frac{v_3^2}{2} = -2\lambda_2 v_2^2 + \lambda_3 v_1^2 - \frac{1}{2} \lambda_8 v_3^2 + \lambda_8 \frac{v_3^2}{2} = -2\lambda_2 v_2^2 + \lambda_3 v_1^2 < 0
\label{eqn:redefined_mass_parameters}
\end{split}
\end{equation}
\endgroup
We have used the decoupling limit of Equation~\ref{eqn:constraints_in_mass_matrix_CP_even} at the first equality of Equation~\ref{eqn:redefined_mass_parameters}. From this equation, it is possible to 
determine the appropriate sign for the quartic coupling constant $\lambda_3$. In our numerical analysis, the vev $v_1$ is much dominant than the vev $v_2$ so it leads to a negative sign for the quartic coupling constant $\lambda_3$, otherwise the below equation of Equation~\ref{eqn:redefined_mass_parameters} would become positive. The sign of the quartic coupling constant $\lambda_3$ also determines the one of $\lambda_{5,7}$ in the decoupling scenario, which means that $\lambda_{5,7}$ must also be negative. On top of that, the large Yukawa coupling constants $y,x$ can be understood in connection with the vev $v_3$. To this end, we consider the definition for the Yukawa coupling constants $x_1$ and $x_2$, which are given by: 
\begingroup
\begin{equation}
x_{2} = \left\lvert \frac{y_{\mu} M_{44}}{y_{2} v_3} \right\rvert, \quad x_{1} = \left\lvert \frac{y_{e} M_{55}}{y_{1} v_3} \right\rvert,
\end{equation}
\endgroup
where in order to successfully explain both $g-2$ anomalies within the $1\sigma$ experimentally allowed range, one has to rely on small values of $v_3$, which are $\mathcal{O}(10\func{GeV})$, and the small values of $v_3$ do not significantly spoil the down-type Higgs $H_d$ potential as seen in Equation~\ref{eqn:redefined_mass_parameters}. In other words, the mass parameters $\mu_{1,2}^2$ are much larger than the 
parameters $\lambda_{7,8}v_3^2/2$, 
thus allowing more freedom in the sign of $\lambda_8$. 
Then, we are now ready to match our simplified Higgs potential with the one given in the reference~\cite{Bhattacharyya:2015nca}. Taking into consideration that our Higgs alignment is different than the one of ~\cite{Bhattacharyya:2015nca}, our mass parameters can be redefined as follows:
\begingroup
\begin{gather}
m_{11}^2 = \mu_1^2+\lambda_7 \frac{v_3^2}{2}, \quad m_{22}^2 = \mu_2^2+\lambda_8 \frac{v_3^2}{2}, \quad m_{12}^2 = \lambda_5 \frac{v_3^2}{2} \\
\beta_1 = 2\lambda_1, \quad \beta_2 = 2\lambda_2, \quad \beta_3 = \lambda_3, \quad \beta_4 = 0, \quad \beta_5 = 0
\end{gather}
\endgroup
Then, following \cite{Maniatis:2006fs,Bhattacharyya:2015nca}, it is found that the scalar potential is stable, when the following relations are fullfilled:
\begingroup
\begin{eqnarray}
\beta _{1} &\geq &0,\hspace{1.5cm}\beta _{2}\geq 0,\hspace{1.5cm}\beta _{3}+\sqrt{\beta _{1}\beta _{2}}\geq 0
\label{eqn:stability_condition_A}
\end{eqnarray}
\endgroup
\begingroup
\begin{equation}
\beta_3 + \beta_4 + \sqrt{\beta_1 \beta_2} > \lvert \beta_5 \rvert \rightarrow \beta_3 + \sqrt{\beta_1 \beta_2} > 0,
\label{eqn:stability_condition_C}
\end{equation}
\endgroup
The last stability condition can be rewritten as shown on the right side since the $\beta_{4,5}$ are zero in our Higgs potential and the cases B and D must be excluded by this 
last condition shown in Equation~\ref{eqn:stability_condition_C}. The conditions given in Eqs. (\ref{eqn:stability_condition_A}) and (\ref{eqn:stability_condition_C}) are crucial to guarantee the stability of the electroweak vacuum. Furthermore, one has to require that the squared masses for the
physical scalars are positive. Besides that, according to \cite{Bhattacharyya:2015nca}, the minimum of the scalar potential is a global
minumum when the following condition is fulfilled:
\begingroup
\begin{equation}
m_{12}^2 \left( m_{11}^2 - m_{22}^2 \sqrt{\frac{\beta_1}{\beta_2}} \right) \left( \tan\beta - \sqrt[4]{\frac{\beta_1}{\beta_2}} \right) > 0 \rightarrow m_{12}^2 \left( m_{11}^2 - m_{22}^2 \sqrt{\frac{\beta_1}{\beta_2}} \right) > 0
\end{equation}
\endgroup
where the latter condition on the left hand side is always successfully fulfilled for all cases, so we can simply drop off the condition as shown on the right side. Then, it is enough to confirm whether each case satisfies the reduced global minimum condition and the case E successfully fulfills that requirement as shown below: 
\begingroup
\begin{equation}
\begin{gathered}
m_{12}^2 = -1763.9\func{GeV}^2, \quad m_{11}^2 = -7896.5\func{GeV}^2, \quad m_{22}^2 = -43258.8\func{GeV}^2,
\\
\sqrt{\frac{\beta_1}{\beta_2}} = 0.0791994
\\
m_{12}^2 \left( m_{11}^2 - m_{22}^2 \sqrt{\frac{\beta_1}{\beta_2}} \right) \approx 7.886 \times 10^6\func{GeV}^4 > 0
\label{eqn:vaccum_stability_check}
\end{gathered}
\end{equation}
\endgroup
Thus, we have numerically checked that the best fit point corresponding to the case E obtained in the numerical analysis of the scalar potential and $g-2$ muon and electron anomalies is consistent
with the above given stability conditions of the scalar potential and at the same time ensure positive values for the squared masses of the physical scalars, consistent with the current experimental data. Finally, to close this section, it is worth mentioning that  
the large Yukawa coupling constants $y,x$ involve 
the small vev $v_3$ in our model and 
this ensures that not only the $H_u$ potential is stable in the decoupling scenario but also the $H_d$ potential successfully fullfill the requirements of vacuum stability for 
both the small vev $v_3$ and appropriate values of the quartic scalar couplings.

\subsubsection{How is the scalar exchange possible to accommodate both
anomalies at $1\protect\sigma$ constraint analytically?}

In order to 
analyze how the scalar exchange is able 
to explain both anomalies within the $1\sigma$ range, 
we revisit the analytic expressions for both muon and electron anomalous
magnetic moments:
\begingroup
\begin{equation}
\begin{split}
\Delta a_{\mu }=y_{24}^{e}x_{42}^{e}\frac{m_{\mu }^{2}}{8\pi ^{2}}\Big[&
\left( R_{e}^{T}\right) _{22}\left( R_{e}^{T}\right) _{32}I_{S}^{(\mu
)}\left( m_{e_{4}},m_{H_{1}}\right) +\left( R_{e}^{T}\right) _{23}\left(
R_{e}^{T}\right) _{33}I_{S}^{(\mu )}\left( m_{e_{4}},m_{H_{2}}\right) \\
& -\left( R_{o}^{T}\right) _{22}\left( R_{o}^{T}\right) _{32}I_{P}^{(\mu
)}\left( m_{e_{4}},m_{A_{1}}\right) -\left( R_{o}^{T}\right) _{23}\left(
R_{o}^{T}\right) _{33}I_{P}^{\left( \mu \right) }\left(
m_{e_{4},}m_{A_{2}}\right) \Big], \\
\Delta a_{e}=y_{51}^{e}x_{15}^{L}\frac{m_{e}^{2}}{8\pi ^{2}}\Big[& \left(
R_{e}^{T}\right) _{22}\left( R_{e}^{T}\right) _{32}I_{S}^{(e)}\left(
m_{e_{5}},m_{H_{1}}\right) +\left( R_{e}^{T}\right) _{23}\left(
R_{e}^{T}\right) _{33}I_{S}^{(e)}\left( m_{e_{5}},m_{H_{2}}\right) \\
& -\left( R_{o}^{T}\right) _{22}\left( R_{o}^{T}\right)
_{32}I_{P}^{(e)}\left( m_{e_{5}},m_{A_{1}}\right) -\left( R_{o}^{T}\right)
_{23}\left( R_{e}^{T}\right) _{33}I_{P}^{\left( E\right) }\left(
m_{e_{5}},m_{A_{2}}\right) \Big],
\label{eqn:muon_electron_anomaly_prediction}
\end{split}%
\end{equation}
\endgroup
where
\begingroup
\begin{equation}
I_{S\left( P\right) }^{\left( e,\mu \right) }\left( m_{E_{4,5}},m_{S}\right)
=\int_{0}^{1}\frac{x^{2}\left( 1-x\pm \frac{m_{E_{4,5}}}{m_{e,\mu }}\right) }{%
m_{e,\mu }^{2}x^{2}+\left( m_{E_{4,5}}^{2}-m_{e,\mu }^{2}\right) x+m_{S,P}^{2}\left(
1-x\right) }dx  \label{eqn:loop_integrals}
\end{equation}
\endgroup
with $S(P)$ corresponding to scalar (pseudoscalar) and $E_{4,5}$ standing for the vector-like family. Furthermore, $E_{4}$ and $E_{5}$ only contribute to the muon and electron anomalous magnetic moments, respectively.

First of all, we focus on the sign 
of each anomaly. The different signs of each anomaly indicated by the $%
1\sigma$ experimentally allowed range 
can be understood at the level of Yukawa constants apart from the loop
structures. As seen in Table \ref{tab:parameter_region_initial_scan}, the
Yukawa coefficient $y$ can be either positive or negative, while $x$ only
remains positive since we take the absolute value to the $x$. 
We also considered the case where the coefficients $x,y$ are purely
positive, assuming $v_3$ is positive, without taking absolute value and the
multiplication of the Yukawa coefficients $x \times y$ cannot change the
sign of each anomaly since the denominator of $x$ includes $y$ and they are
cancel out. Then, the sign problem depends on summing over loop functions and we found that the order of the muon anomaly prediction is
suitable, 
whereas the corresponding to the electron anomaly is about $10^{-16}$ which
is too small to be accommodated within the 
$1\sigma$ experimentally allowed range. Therefore, we found that taking an
absolute value to one of the Yukawa coefficients is an appropriate strategy
for the sign and allows to reproduce the correct order of magnitude of each
anomaly 
allowed by the $1\sigma$ experimentally allowed range, for an appropiate
choice of the model parameters. This feature is a crucial difference compared with the $W$ or $Z^\prime$ gauge boson exchange\cite{CarcamoHernandez:2019ydc}. The $W$ gauge boson exchange covered in the main body of this work keeps the same coupling constant at each vertex, therefore it is completelly different from 
the scalar exchange with vector-like leptons. For the $Z^\prime$ exchange covered in \cite{CarcamoHernandez:2019ydc}, it has the common property that the coupling constant of each vertex is different to each other, whereas the coupling constants of the $Z^\prime$ are more constrained by the mixing angle between $i$th chiral family and fourth vector-like family, so it is impossible to explain both anomalies at the same time. As a result, allowing different Yukawa constants with appropiate signs enables both anomalies to be explained in a unified way. 
\newline
~\newline
Next we turn our attention to the order of magnitude of our predictions for both anomalies.
Considering that the sign problem is solved by having each Yukawa constant $%
y $ either positive or negative, it can be easily understood that inside
the structure in parentheses of Equation \ref%
{eqn:muon_electron_anomaly_prediction} should imply the same direction, which is 
is determined by the contribution of all loop functions in
parentheses. Since the mass difference among non-SM scalars and vector-like
particles is not so big, we have to consider their masses in the computation of muon and electron anomalous magnetic moments, as follows from Equation \ref%
{eqn:muon_electron_anomaly_prediction}. 
For an easy analysis, we take the case E reported in Table \ref%
{tab:parameter_region_best_peaked} and suppose that
\begingroup
\begin{equation}
\begin{split}
&\left( R_e^T \right)_{22} \left( R_e^T \right)_{32} = c_1, \quad \left(
R_e^T \right)_{23} \left( R_e^T \right)_{33} = -c_1,
\\
&\left( R_o^T \right)_{22} \left( R_o^T \right)_{32} = c_2, \quad \left( R_o^T \right)_{23} \left( R_o^T \right)_{33} = -c_2, 
\\
&I_S^{\mu} \left( m_{e_4}, m_{H_1} \right) = d_1, \quad I_S^{\mu} \left(
m_{e_4}, m_{H_2} \right) = d_2, 
\\
&I_P^{\mu} \left( m_{e_4}, m_{A_1}
\right) = -d_3, \quad I_P^{\mu} \left( m_{e_4}, m_{A_2} \right) = -d_4, 
\\
&I_S^{e} \left( m_{e_5}, m_{H_1} \right) = e_1, \quad I_S^{e} \left( m_{e_5},
m_{H_2} \right) = e_2,
\\
&I_P^{e} \left( m_{e_5}, m_{A_1} \right) =
-e_3, \quad I_P^{e} \left( m_{e_5}, m_{A_2} \right) = -e_4, 
\\
&d_1 > d_3 > d_2 > d_4, \quad e_1 > e_3 > e_2 > e_4
\end{split}%
\end{equation}
\endgroup
where $c_{1,2}$ are arbitrary constant between $0$ and $1$ either positive
or negative and mass ordering among $d(e)_i, (i=1,2,3,4)$ can be easily
understood by considering mass difference between non-SM scalars and
vector-like particles. The muon and electron anomaly prediction can be
rewritten in terms of these redefined constants:
\begingroup
\begin{equation}
\begin{split}
\Delta a_\mu &= y_2 x_2 \frac{m_\mu^2}{8\pi^2} \left[ c_1 d_1 - c_1 d_2 + c_2
d_3 - c_2 d_4 \right] 
\\
&= y_2 x_2 \frac{m_\mu^2}{8\pi^2} \left[ c_1 \left( d_1
- d_2 \right) + c_2 \left( d_3 - d_4 \right) \right] = y_2 x_2 \frac{m_\mu^2%
}{8\pi^2} \left[ c_1 d_{12} + c_2 d_{34} \right] \\
\Delta a_e &= y_1 x_1 \frac{m_e^2}{8\pi^2} \left[ c_1 e_1 - c_1 e_2 + c_2 e_3
- c_2 e_4 \right] 
\\
&= y_1 x_1 \frac{m_e^2}{8\pi^2} \left[ c_1 \left( e_1 - e_2
\right) + c_2 \left( e_3 - e_4 \right) \right] = y_1 x_1 \frac{m_e^2}{8\pi^2}
\left[ c_1 e_{12} + c_2 e_{34} \right]
\end{split}%
\end{equation}
\endgroup
where $y_2,x_2,y_1,x_1$ are simplified notation for $%
y_{24}^e,x_{42}^e,y_{51}^e,x_{15}^L$, respectively, and $d(e)_{ij} \equiv
d(e)_i - d(e)_j$ and $d(e)_{ij}$ are positive. Since the inside structure in
parentheses depends on relative magnitude of both $c_{1,2}$ and $d(e)_{ij}$
at this stage where no more analytic simplication is possible, it is good to
implement a specific value for them. Referring the values used to derive the
result of case E, they are
\begingroup
\begin{equation}
\begin{split}
y_2 x_2 \frac{m_\mu^2}{8 \pi^2} c_1 d_1 = -4.629 \times 10^{-7}, &\quad y_1
x_1 \frac{m_e^2}{8 \pi^2} c_1 e_1 = -8.532 \times 10^{-12} \\
-y_2 x_2 \frac{m_\mu^2}{8 \pi^2} c_1 d_2 = 4.520 \times 10^{-7}, &\quad
-y_1 x_1 \frac{m_e^2}{8 \pi^2} c_1 e_2 = 6.808 \times 10^{-12} \\
y_2 x_2 \frac{m_\mu^2}{8 \pi^2} c_2 d_3 = 7.984 \times 10^{-8}, &\quad y_1
x_1 \frac{m_e^2}{8 \pi^2} c_2 e_3 = 1.323 \times 10^{-12} \\
-y_2 x_2 \frac{m_\mu^2}{8 \pi^2} c_2 d_4 = -6.659 \times 10^{-8}, &\quad -y_1
x_1 \frac{m_e^2}{8 \pi^2} c_2 e_4 = -5.217 \times 10^{-13}
\label{eqn:middle_values_for_prediction}
\end{split}%
\end{equation}
\endgroup
and summing over all values in left or right column of Equation \ref%
{eqn:middle_values_for_prediction} yields the prediction for muon and
electron anomaly at $1\sigma$
\begingroup
\begin{equation}
\begin{split}
\Delta a_\mu &= y_2 x_2 \frac{m_\mu^2}{8\pi^2} \left[ c_1 d_1 - c_1 d_2 +
c_2 d_3 - c_2 d_4 \right] = 2.393 \times 10^{-9} \\
\Delta a_e &= y_1 x_1 \frac{m_e^2}{8\pi^2} \left[ c_1 e_1 - c_1 e_2 + c_2
e_3 - c_2 e_4 \right] = -9.232 \times 10^{-13}.
\end{split}%
\end{equation}
\endgroup

\section{The experimental and theoretical bound for the non-SM CP-even and -odd scalars} \label{sec:non_SM_scalars_exp_bound}

The scalar sector in this BSM model is an extended 2HDM with one singlet flavon. To be more specific, it is a type II 2HDM, which reproduces the SM Yukawa structure in the alignment limit ($\beta - \alpha = \pi/2$). The case E of Table~\ref{tab:range_secondscan}, regarded as valid in this work, tells that vev of the singlet flavon $\phi$ is order of $10\func{GeV}$, which is quite small so it does not significantly impact the parameter space considered in this work. Therefore, we can safely claim that our BSM model is a well-approximated type II 2HDM and mass interval of the non-SM scalars are investigated by the Gfitter Group~\cite{Haller:2018nnx}
\begin{equation}
\begin{gathered}
130\func{GeV} < M_{H}, M_{A} < 1000\func{GeV}, \\
100\func{GeV} < M_{H^\pm} < 1000\func{GeV}, \\
0 < \beta - \alpha < \pi, \\
0.001 < \tan\beta < 50, \\
-8 \times 10^{5}\func{GeV}^{2} < M_{12}^{2} < 8 \times 10^{5}\func{GeV}^{2},
\label{eqn:const_2HDM}
\end{gathered}
\end{equation}
where $M_{12}^2$ can be found from Equation~\ref{eqn:vaccum_stability_check} and it is worth mentioning that since the 2HDM model under consideration has generally large freedom in the parameter space, the given constraints~\ref{eqn:const_2HDM} suggest weak exclusion limits on the parameter space. The LEP experiments in search for the charged scalar in the type II 2HDM strengthen mass bound for the charged scalar $M_{H^\pm}$~\cite{ALEPH:2013htx}.
\begin{equation}
M_{H^\pm} \gtrsim 150\func{GeV}
\end{equation}
Comparing the case E and the given bounds, none of our predictions from the case E is excluded by the given bounds except for $m_{A_2}$, some of which exceeds the $1000\func{GeV}$, however it should notice that the $m_{A_2}$ depends on the free parameter $\mu_{\func{sb}}$ appearing in the scalar potential to prevent additional Goldstone bosons appearing in this work. Therefore, we can conclude our numerical predictions carried out in this work are well consistent with the current experimental bounds. As our BSM model features an extended 2HDM, it might cause dangerous scalar-mediated flavor-changing-neutral-currents (FCNCs) and CP-violation. For the dangerous scalar-mediated FCNCs, there are three safety devices in this BSM model, which are the alignment limit, decoupling limit, and lastly proper diagonalization. The BSM model under consideration can reproduce the SM Yukawa structure in the alignment limit as mentioned earlier and keep mixing between up-type SM-like Higgs $H_{u}$ and down-type SM-like Higgs $H_{d}$ arising by the decoupling limit. Plus, the proper diagonalization suppresses off-diagonal elements of mass matrices within this BSM model so that effects of the dangerous FCNCs become small enough to ignore. For the CP-violation, we have only considered real Yukawa coupling constants for the scalar and fermion sectors, and this feature naturally leads to the CP-conserving scenario. Therefore the BSM model under consideration are free from the dangerous FCNCs and CP-violation. 
\section{Conclusion}

\label{sec:Conclusion}
We have proposed a model to account for the hierarchical structure of the SM Yukawa couplings.
In our approach 
the SM is an effective theory arising from a theory with extended particle
spectrum and symmetries.
The considered model includes 
an extension of the 2HDM where the particle spectrum is enlarged by the
inclusion of two vector-like fermion families, right handed Majorana neutrinos and a
gauge singlet scalar field, together with 
the inclusion of a global $U(1)^\prime$ symmetry spontaneously broken at the 
$\func{TeV}$ scale. 
Since the $U(1)^\prime$ symmetry is global, this model does not feature 
a $Z^\prime$ boson and it is softly broken in the 2HDM potential to avoid a Goldstone boson. 
Its main effect is to forbid SM Yukawa interactions due
to the $U(1)^\prime$ charge conservation. 
Besides that, this model has the
property of the 2HDM type II where one Higgs doublet couples with the up-type fermions whereas the remaining one has Yukawa interactions with down-type fermions, where such couplings are allowed between chiral fermions and vector-like fermions due to the choice of 
$U(1)^\prime$ charges (chiral fermions having zero charges while vector-like fermions, Higgs and flavons have charges $\pm 1$).
Below the mass scale of the vector-like fermions, such couplings result in effective Yukawa couplings suppressed by a factor 
$\left\langle \phi \right\rangle/M$ where the numerator is the vev of the flavon
and the denominator is the vector-like mass. This factor naturally determines
the magnitude of SM interactions and the mass scale for the vector-like
fermions under a suitable choice of the flavon vev. 
We have developed a mixing formalism based on $7 \times 7$ mass matrices to describe the mixing 
of the three chiral families with the two vector-like
families. 
\newline~\newline
Within the above proposed model, 
we have focused on accommodating the long-established muon and less established electron anomalous
magnetic moments at one-loop level. A main difficulty arises from the sign of each anomalous deviation of the experimental value from its SM prediction. Generally, the Feynman diagrams for the muon
and electron anomalous magnetic moments have the same structure except from the fact that the external particles are different, which makes it difficult to flip the sign of each contribution. Specifically we have required that both deviations in 
Equation \ref{eqn:deltaamu_deltaae_at_1sigma}) 
at one-loop should be accommodated within the $1\sigma$ experimentally allowed range,
which is a challenging requirement. %
\newline
~\newline
We first considered in detail the $W$ boson exchange contributions to the muon and electron 
anomalous magnetic moments at one-loop.
The relevant sector for the $W$ boson exchange is that of the neutrino and we analyzed a novel
operator that generates the masses of the light active neutrinos in this model. 
The well-known five dimensional Weinberg
operator which we refer as type Ia seesaw mechanism does not work in this model 
since it is forbiden by the $U(1)$ symmetry due to the fact that both $SU(2)$ scalar doublets are negatively charged under this symmetry. 
For this reason, we made use of the Weinberg-like operator known as
type Ib seesaw mechanism allowed in this model. With the type Ib seesaw
mechanism, we built the neutrino mass matrix with two vector-like neutrinos and ignored fifth vector-like neutrinos since they are too heavy to
contribute to the phenomenology. The deviation of unitarity $\eta$ derived
from the heavy vector-like neutrinos plays a crucial role for enhancing the
sensitivity of the CLFV $\mu \rightarrow e \gamma$ decay to the observable
level. Furthermore, the Yukawa constants of Dirac neutrino mass matrix can be connected to the observables measured in neutrino oscillation experiments. 
One of the neutrino Yukawa constants is defined with a
suppression factor $\epsilon$. Therefore, the effective $3 \times 3$
neutrino mass matrix tells that the tiny masses of the light active neutrinos depend on the mass scale of vector-like neutrinos as well as on the suppression factor $\epsilon$. %
This implies that mass scale of vector-like neutrinos is not required to be of the order 
of $10^{14}$ GeV, as in 
the conventional type Ia seesaw mechanism. 
In our proposed model, the vector-like neutrinos can have masses at the $\func{TeV}$ scale, thus allowing to test our model at colliders. Those vector-like neutrinos can be pair produced at the LHC via Drell-Yan annihilation mediated by a virtual $Z$ gauge boson. They can also be produced in association with a SM charged lepton via Drell-Yan annihilation mediated by a $W$ gauge boson. These heavy vector like sterile neutrinos can decay into a SM charged lepton and light active neutrinos. Thus, the heavy neutrino pair production at a proton-proton collider will give rise to an opposite sign dilepton final state, which implies that the observation of an excess of events in this final state over the SM background can be a smoking gun signature
of this model, whose observation will be crucial to assess its viability. 
It is confirmed that the branching ratio of $\mu \rightarrow e \gamma$ decay
can be expressed in terms of the deviation of unitarity $\eta$ as shown in 
\cite{Hernandez-Garcia:2019uof,Calibbi:2017uvl} and our prediction
for the muon and electron anomalous magnetic moments can also be written in
terms of non-unitarity. We derived the analytic expression for the anomalies
and found that the order of magnitude of these predictions is too small to accommodate the
experimental bound within the $1\sigma$ range and the sign of each prediction also points
out in the same direction. Therefore, we concluded that the $W$ boson exchange at
one-loop is not enough to explain both anomalies at $1\sigma$ and this
conclusion has been a good motivation to search for another possibility such
as scalar exchange, which is one of the main purposes of this work.  
\newline
~\newline
We then turned our attention to the 2HDM contributions (inclusion also of the singlet scalar $\phi$) to the muon and electron 
anomalous magnetic moments, assuming by a choice of parameters a diagonal charged lepton mass matrix to 
suppress the branching ratio of $\mu \rightarrow e \gamma$. In our analysis
we considered in detail the scalar sector of our model, which is composed of 
two $SU(2)$ scalar doublets $H_u$ and $H_d$ and one electrically neutral complex scalar $\phi$ by studying the corresponding scalar potential, deriving the squared mass matrices for the CP-even, CP-odd neutral and electrically charged scalars and determining the resulting scalar mass spectrum.
We have restricted to the scenario corresponding to the decoupling limit where 
no mixing between the physical SM Higgs $h$ and the physical non-SM scalars $H_{1,2}$
arise and within this scenario we have imposed the restrictions arising from 
the Higgs diphoton decay rate, the $hWW$ coupling, the $125$ GeV mass of the SM-like Higgs and the experimental lower bounds on non SM scalar masses, to determine the allowed parameter space consistent with the muon and electron anomalous magnetic moments. To this end, we have constructed a $\chi^2$ fitting function, which measures the deviation of the values of the physical observables obtained in the model, i.e., $(g-2)_{e,\mu}$, the $125$ GeV SM-like Higgs mass, the Higgs diphoton signal strength, the $hWW$ coupling, with respect to their experimental values. Its minimization allows to determine the values of the model parameters consistent with the measured experimental values of these observables. 
After saturating the $\chi^2$ value less than or nearly
2, we obtained five independent benchmark points and carried out second scan with
the benchmark points to find a correlation between observables and mass
parameters. For the plots, we took an appropriate case which is more converged when compared to other
ones and satisfying the vacuum stability conditions. We found that our prediction for both anomalies can be explained
within the $1\sigma$ constraint of each anomaly and a correlation 
proportional for muon versus electron anomaly is appeared in Figure \ref%
{fig:relevant_parameters_delamue} and \ref%
{fig:relevant_parameters_vl_delamue}. Here, we put two constraints on mass
of the lightest non-SM scalar and of the lightest vector-like family; $%
m_{H_1}, m_{e_5} > 200\func{GeV}$ based on references. The second scan result tells that the available parameter space is not significantly constrained by current experimental
results on non-SM scalar mass and vector-like mass, while keeping
perturbativity for quartic couplings and Yukawa constants. An important feature of our BSM model is it predicts the large Yukawa coupling constants $y,x$, which might be able to destabilize the Higgs potential. The up-type Higgs $H_u$ potential is not significantly affected by the large Yukawa coupling constants in the decoupling scenario, whereas there is no safe condition for the down-type Higgs $H_d$ potential which can be worsen by mixing with the flavon field $\phi$. The large Yukawa coupling constants $x$ introduces small values for the vev $v_3$ in the definition of $x$ and the energy scale is confirmed by order of $10\func{GeV}$ in our numerical analysis. On top of that, we also identified the appropriate sign of quartic coupling constants can make the Higgs potential stable. Therefore, the down type $H_d$ Higgs potential is stable by both the small vev $v_3$ and the appropriate quartic coupling constants in our BSM model. Lastly, we
discussed how we were able to explain both $(g-2)_{e,\mu}$ anomalies at $1\sigma$ constraint
and impact of the light non-SM scalar $H_1$. For the former, we first
simplified the prediction for both anomalies and used some numerical values at
the stage where no more analytic simplication is possible. For the latter,
we compared the cross section for the SM process $pp \rightarrow h$ and BSM process $pp \rightarrow H_1$ and included this comparison in Appendix~\ref{B}. 

We conclude that the proposed model of fermion mass hierarchies 
is able to successfully accommodate both the muon and electron anomalous magnetic moments within the $1\sigma$ experimentally allowed ranges, with the dominant contributions arising from one loop diagrams involving the 2HDM scalars and vector-like leptons.
The resulting model parameter space consistent with the $(g-2)_{e,\mu}$ anomalies requires masses of non-SM scalars and vector-like particles in the sub TeV and TeV ranges, thus making these particles accessible at the LHC and future colliders.
\chapter{The third BSM model - SM fermion mass hierarchies from one VL family with an extended 2HDM} \label{Chapter:The3rdBSMmodel}
In this chapter, we discuss our third BSM model, which is exactly same as the second BSM model except that one vector-like family is used instead of two for the purpose of diagonalizing the mass matrices without any assumptions. One vector-like family can provide two seesaw operators, so the first SM generation can not be massive in this model, however this is a good approximation taking into account the first SM generation is very light in both quark and lepton sectors. In this BSM model, we construct the SM $Z$ gauge coupling constants in the mass basis after enlarging the SM fermion sector by the fourth vector-like family.
\section{Introduction} \label{sec:I}

A great success of the energy frontier is the discovery of the Higgs particle by ATLAS and CMS collaborations at the Large Hadron Collider (LHC) on 4th July 2012~\cite{ATLAS:2012yve,CMS:2012qbp}. After that discovery, no new particle has been found so far by the experiments at 
LHC with $13\func{TeV}$ proton-proton centre of mass energy. This highlights the fact that  
not just the energy frontier but also the luminosity (intensity) frontier should be considered as of equal importance 
in the search for physics beyond the Standard Model (SM). 
For example, one may consider observables mediated by flavour-changing-neutral-currents (FCNCs), which are quite sensitive to new physics, since such FCNC observables are extremely suppressed in the Standard Model
(SM) due to the well-known Glashow-Illiopoulos-Maiani (GIM) mechanism. Another example of a highly suppressed process is provided by 
the branching ratio of $\mu \rightarrow e \gamma$ decay mediated by  
massive neutrinos at the one-loop level~\cite{Calibbi:2017uvl}:
\begin{equation}
\func{BR}\left( \mu \rightarrow e \gamma \right) \approx 10^{-55}.
\end{equation}
The experimentally known sensitivity for the branching ratio of $\mu \rightarrow e \gamma$ is
\begin{equation}
\func{BR}\left( \mu \rightarrow e \gamma \right)_{\func{EXP}} = 4.2 \times 10^{-13}.
\end{equation}
The large gap between the tiny rates of the flavour violating decays predicted by the SM
 and their experimental upper limits has motivated the construction of many flavour models with extended scalar, quark and leptonic spectrum aimed at enhancing those rates by several orders of magnitude up to an observable level within the reach of the sensitivity of the future experiments. A similar situation occurs for other rare FCNC decays such as, for instance $Z\to\mu\tau$ and $t\rightarrow cZ$, which are very suppressed in the SM, but in extensions of the SM, can acquire sizeable values, within the reach of the future experimental sensitivity.
Although various models with a heavy $Z^\prime$ boson have also 
received a lot of attention by the particle physics community as a new source of FCNCs, its properties, being not fully constrained, do not lead to definite predictions. For this reason we shall restrict ourselves to the SM $Z$ couplings in this paper.

In this paper we focus on the SM $Z$ FCNC interactions induced by tree-level gauge boson exchange in a model in which the fermion sector of the SM is enlarged with a fourth vector-like family. An interesting feature of this approach is all coupling constants of $Z$ interactions in this work are fixed by the known values of the SM $Z$ gauge boson interactions, together with mixing parameters.
Our main motivation for adding a fourth vector-like family is to explain quark and lepton mass hierarchies. We first forbid the SM Yukawa couplings with a global 
$U(1)'$ symmetry, then allow them to be generated effectively via mixing with the fourth vector-like family, a mechanism somewhat analogous to the seesaw mechanism for neutrino masses.
Consequently, the SM charged fermion masses are inversely proportional to the masses of the heavy vector-like leptons and directly proportional to the product of the couplings of Yukawa interactions that mix SM charged fermions with vector-like fermions. This implies that a small hierarchy in those couplings can yield a quadratically larger hierarchy in the
effective couplings. Combined with a moderate hierarchy in the vector-like masses, this allows us to naturally explain the SM charged fermion mass hierarchy and to predict the mass scale of  vector-like fermions.\\~\\
A similar model was discussed in
our previous works~\cite{Hernandez:2021tii}, although with two vector-like families, but the effect on the $Z$ and $W$ boson couplings was not studied.
In our previous work~\cite{Hernandez:2021tii}, whose purpose was to explain the muon and electron anomalous magnetic moments simultaneously, the main focus was on the 2HDM scalar sector, and the FCNCs arising from the $Z$ and $W$ boson couplings were not considered, 
since the full mass matrices were not accurately diagonalised, and hence such effects were beyond the approximations used there.
By contrast, the main goal of this work is to study the SM $Z$ and $W$ contributions to the FCNC observables at leading order to constrain the masses of vector-like fermions, and to explore other possible phenomenological signatures. The SM $W$ contributions to the CKM mixing matrix with the extended quark sector are also studied for the first time in this work. In order for these effects to be considered reliably, the mass matrices of each fermion sector are accurately diagonalized, both numerically and analytically, unlike the previous work where simple approximations were used which masked the 
effects we consider here. The results in this work are sufficiently accurate to enable the contribution of the $Z$ and $W$ boson couplings
to physics beyond the SM to be reliably considered for the first time.

In order to make the results completely transparent, we shall study the $Z$ and $W$ boson couplings in the presence of only one vector-like family 
so that mass matrices of this work can be straightforwardly diagonalized using both analytical and numerical methods. Since only one vector-like fermion family is used, the first generation of SM charged fermions do not acquire masses, which nonetheless is 
a very good approximation considering 
the SM fermions belonging to the first family 
are very light. Consequently, we restrict our attention to 
the second and third generations of SM fermions, 
as well as to several observables related to FCNC processes involving the second and third SM families.
In our approach, then, the SM is a low effective energy theory arising after 
integrating out a single heavy fourth vector-like family. 
In order to dynamically generate the hierarchical structure of SM fermion masses, the fermionic mass matrices given in \cite{Hernandez:2021tii} as well as the ones obtained in this work must be accurately and completely diagonalized, which, as mentioned above, has not been done previously.
The mass matrices for the charged lepton and up-quark sectors share the same structure, whereas the one for 
the down-type quark sector involves 
an additional non-zero element in a particular basis, although we later show that the results are basis independent. 
This reasoning does not apply to the neutrino sector, since this sector is treated independently. 
This different
feature of the down-type mass matrix, in the preferred basis, allows us to achieve
all mixings among the three generations of SM fermions even though the first one remains massless, and 
this leads to a prediction for the Cabbibo-Kobayashi-Maskawa (CKM) mixing matrix. In addition, due to the mixings between the SM quarks and the vector-like quarks, the CKM quark mixing matrix originating from the $W$ couplings
is not unitary, thus implying the need of relaxing the unitarity condition of the CKM mixing matrix, and we also study this feature.
\\~\\
This paper is organized as follows. In Section~\ref{sec:II} we introduce our model to explain the origin of the SM fermion's mass with a fourth vector-like family. In Section~\ref{sec:III} the mass matrices in both quark and lepton sectors are constructed and diagonalized using the mixing formalism. In Section~\ref{sec:IV} the  
$Z$ gauge boson interactions with fermions  
are determined from the mixing matrices used in the mass matrix diagonalization.  
Several FCNC observables for both lepton and quark sectors are analyzed in detail in Sections~\ref{sec:V} and \ref{sec:VI}, respectively. 
We state our conclusions in Section~\ref{sec:VII}. Several technical details are relegated to the Appendices. The perturbative analytical diagonalization of the mass matrices for the charged lepton, up type quark and down type quark sectors are discussed in detail in Appendices \ref{app:A}, \ref{app:B} and \ref{app:C}, respectively. The comparison between the numerical and approximate analytic diagonalization of the mass matrices for charged leptons and quarks is made in Appendices \ref{app:D} and \ref{app:E}, respectively.

\section{An extended model with a fourth vector-like family} \label{sec:II}

The origin of the pattern of SM fermion masses is interesting open question, not addressed by the SM. The mass parameters of the SM have been experimentally determined with good precision, 
and these experimentally observed mass parameters show a  
strong hierarchical structure of the SM fermion masses. The most extreme hierarchy is exhibited between the SM neutrino Yukawa coupling of about $10^{-12}$ and the top quark Yukawa coupling of about $1$. Regarding the tiny neutrino masses, many particle physicists regard their masses as most likely explained by the see-saw mechanism rather than by the Yukawa interactions, thus predicting 
the presence of the heavy right-handed neutrinos. The reason why the see-saw mechanism has received a large amount of attention by the particle physics community is that 
it provides a dynamical explanation of the tiny active neutrino masses. For a similar reason, it is interesting to speculate about the existence of a dynamical mechanism that produces the masses of all SM fermions via the exchange of heavy fermionic degrees of freedom thus implying that
 the SM is an effective low energy theory arising from some spontaneous breaking at higher energy scales  
 of a more complete underlying theory. 
  In order to specify a possible candidate of an underlying theory responsible for the generation of the SM fermion mass hierarchy,
  we shall consider a minimal extension 
  of the SM 
  consistent with the SM current experimental bounds. With this motivation in mind, we enlarge the SM fermion and scalar sectors by including a fourth vector-like family and an extra $SU(2)$ scalar doublet as well as a scalar singlet, respectively. Furthermore, we extend the SM gauge symmetry by adding a $U(1)^\prime$ global symmetry. The particle content of the proposed model is shown in Table \ref{tab:BSM_model}.
\begin{table}[H]
\resizebox{\textwidth}{!}{
\centering\renewcommand{\arraystretch}{1.3} 
\begin{tabular}{*{21}{c}}
\toprule
\toprule
Field & $Q_{iL}$ & $u_{iR}$ & $d_{iR}$ & $L_{iL}$ & $e_{iR}$ & $Q_{kL}$ & $%
u_{kR}$ & $d_{kR}$ & $L_{kL}$ & $e_{kR} $ & $\nu_{kR}$ & $\widetilde{Q}_{kR}$
& $\widetilde{u}_{kL}$ & $\widetilde{d}_{kL}$ & $\widetilde{L}_{kR}$ & $%
\widetilde{e}_{kL}$ & $\widetilde{\nu}_{kR}$ & $\phi$ & $H_u$ & $H_d$ \\ 
\midrule
$SU(3)_C$ & $\mathbf{3}$ & $\mathbf{3}$ & $\mathbf{3}$ & $\mathbf{1}$
& $\mathbf{1}$ & $\mathbf{3}$ & $\mathbf{3}$ & $\mathbf{3}$ & $\mathbf{1}$ & 
$\mathbf{1}$ & $\mathbf{1}$ & $\mathbf{3}$ & $\mathbf{3}$ & $\mathbf{3}$ & $%
\mathbf{1}$ & $\mathbf{1}$ & $\mathbf{1}$ & $\mathbf{1}$ & $\mathbf{1}$ & $%
\mathbf{1}$ \\ 
$SU(2)_L$ & $\mathbf{2}$ & $\mathbf{1}$ & $\mathbf{1}$ & $\mathbf{2}$
& $\mathbf{1}$ & $\mathbf{2}$ & $\mathbf{1}$ & $\mathbf{1}$ & $\mathbf{2}$ & 
$\mathbf{1}$ & $\mathbf{1}$ & $\mathbf{2}$ & $\mathbf{1}$ & $\mathbf{1}$ & $%
\mathbf{2}$ & $\mathbf{1}$ & $\mathbf{1}$ & $\mathbf{1}$ & $\mathbf{2}$ & $%
\mathbf{2}$ \\ 
$U(1)_Y$ & $\frac{1}{6}$ & $\frac{2}{3}$ & $-\frac{1}{3}$ & $-\frac{1%
}{2}$ & $1$ & $\frac{1}{6}$ & $\frac{2}{3}$ & $-\frac{1}{3}$ & $-\frac{1}{2}$
& $-1$ & $0$ & $\frac{1}{6}$ & $\frac{2}{3}$ & $-\frac{1}{3}$ & $-\frac{1}{2}
$ & $-1$ & $0$ & $0$ & $\frac{1}{2}$ & $-\frac{1}{2}$ \\ 
$U(1)^\prime$ & $0$ & $0$ & $0$ & $0$ & $0$ & $1$ & $-1$ & $-1$ & $1$
& $-1$ & $-1$ & $1$ & $-1$ & $-1$ & $1$ & $-1$ & $-1$ & $1$ & $-1$ & $-1$ \\ 
\bottomrule
\bottomrule
\end{tabular}}%
\caption{Particle assigments under the $SU(3)_C\times SU(2)_L\times U(1)_Y\times U(1)^\prime$ symmetry of the  
extended 2HDM theory 
with fourth vector-like family. 
The index $i=1,2,3$ denotes the 
 the $i$th SM fermion generation and $k=4$ stands for the 
 fourth vector-like family. }
\label{tab:BSM_model}
\end{table}
Our proposed theory is a minimal extended 2 Higgs Doublet Model (2HDM) where the SM fermion sector is enlarged by the inclusion of a fourth vector-like family and the scalar sector is augmented by an extra $SU(2)$ scalar doublet
and a singlet flavon and lastly the SM gauge symmetry is extended by the $U(1)^\prime$ global symmetry. As this model features the global $U(1)^\prime$ symmetry, there is no a  
neutral $Z^\prime$ gauge boson in the particle spectrum. 
Furthermore,  
the up-type quarks feature Yukawa interactions with the 
up-type SM Higgs $H_u$, whereas the extra scalar doublet $H_d$ couples with the SM down-type quarks and charged leptons. 
Our proposed model is especially motivated by the hierarchical structure of the SM and, in order to implement 
this hierarchy, 
we forbid the SM-type Yukawa interactions 
by appropiate $U(1)^\prime$ charge assignments of the scalar and fermionic fields. Then, for the above specified particle content, the following effective Yukawa interactions arise: 
\begin{equation}
\mathcal{L}_{\func{eff}}^{\func{Yukawa}} =y_{ik}^\psi (M_{\psi^\prime}^{-1})_{kl} {x_{lj}^{\psi^\prime}
\left\langle \phi \right\rangle} \overline{%
\psi}_{iL} \widetilde{H} \psi_{jR} + {x_{ik}^{\psi^\prime} \left\langle \phi
\right\rangle}(M_{\psi}^{-1} )_{kl}  y_{lj}^\psi  \overline{\psi}_{iL} \widetilde{H}
\psi_{jR} + \func{h.c.}  \label{eqn:the_effective_Yukawa_Lagrangian}
\end{equation}
where the indices $i,j=1,2,3$ and $k,l=4$ whereas $\psi,\psi^\prime = Q, u, d, L, e$ and $M$ means heavy vector-like mass. The masses of all SM fermions can be explained by this effective Lagrangian of Equation~\ref{eqn:the_effective_Yukawa_Lagrangian}, emphasizing their relative different masses are explained by the factor $\langle \phi \rangle/M \ll 1 \text{ ( apart from top quark )}$, except for the neutrinos which requires an independent approach to their mass. Feynman diagrams corresponding for the effective Lagrangian are shown in Figure \ref{fig:mass_insertion_diagrams}:
\begin{figure}[H]
\centering
\begin{subfigure}{0.48\textwidth}
	\includegraphics[width=1.0\textwidth]{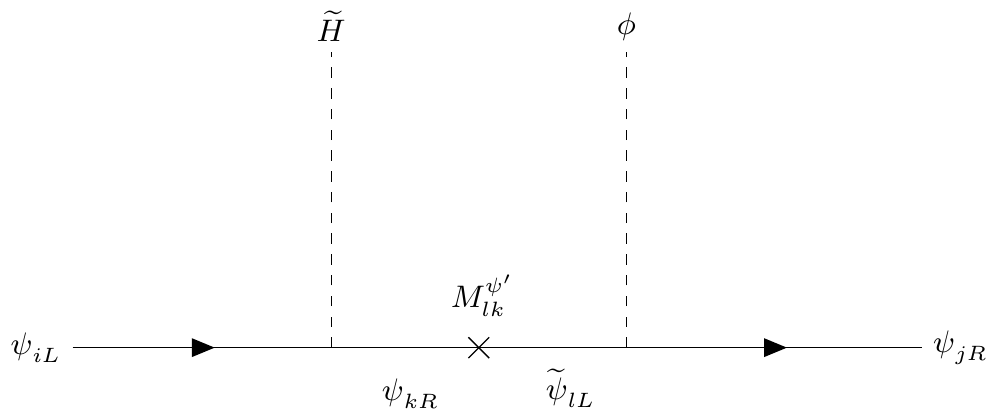}
\end{subfigure} \hspace{0.1cm} 
\begin{subfigure}{0.48\textwidth}
	\includegraphics[width=1.0\textwidth]{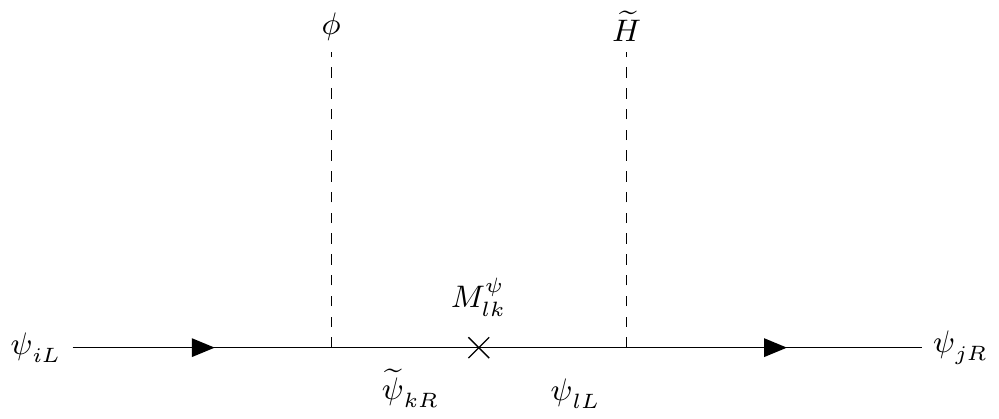}
\end{subfigure}
\caption{Feynman diagrams leading to the effective Yukawa
interactions, where $\protect\psi,\protect\psi^\prime = Q,u,d,L,e$ (neutrinos
will be treated separately), $i,j=1,2,3$, $k,l=4$, $M_{lk}$ is vector-like
mass and $\widetilde{H} = i\protect\sigma_2 H^*, H = H_{u,d}$}
\label{fig:mass_insertion_diagrams}
\end{figure}
The theory considered in this paper corresponds to the one given in  
one of our previous works~\cite{Hernandez:2021tii}, however one vector-like family is used instead of two so that the mass matrices for both quark and lepton sectors can be diagonalized much more economically than in our previous model of \cite{Hernandez:2021tii} at cost of having massless 
the first generation SM charged fermions (One of our main purposes is to diagonalize mass matrices for the quark and lepton sectors without any assumptions) and this is actually a good approximation taking into account that the first generation of SM charged fermions are very light.
\subsection{Effective Yukawa interactions for the SM fermions} \label{sec:II_1}
The renormalizable interactions of the quark sector in this model are given by:
\begin{equation}
\begin{split}
\mathcal{L}_{q}^{\func{Yukawa+Mass}} &= y_{ik}^{u} \overline{Q}_{iL} 
\widetilde{H}_u u_{kR} + x_{ki}^{u} \phi \overline{\widetilde{u}}_{kL}
u_{iR} + x_{ik}^Q \phi \overline{Q}_{iL} \widetilde{Q}_{kR} + y_{ki}^u 
\overline{Q}_{kL} \widetilde{H}_u u_{iR} \\
&+ y_{ik}^{d} \overline{Q}_{iL} \widetilde{H}_d d_{kR} + x_{ki}^{d} \phi 
\overline{\widetilde{d}}_{kL} d_{iR} + y_{ki}^d \overline{Q}_{kL} \widetilde{%
H}_d d_{iR} \\
&+ M_{kl}^{u} \overline{\widetilde{u}}_{lL} u_{kR} + M_{kl}^{d} \overline{%
\widetilde{d}}_{lL} d_{kR} + M_{kl}^Q \overline{Q}_{kL} \widetilde{Q}_{lR} + 
\func{h.c.}  \label{eqn:general_Quark_Yukawa__Mass_Lagrangian}
\end{split}%
\end{equation}
where $i,j=1,2,3$, $k,l=4$ and $\widetilde{H}=i\sigma _{2}H^{\ast }$. After the $U(1)^\prime$ symmetry is spontaneously broken by the vacuum expectation value (vev) of the 
 the singlet flavon $\phi$, and the heavy vector-like fermions are integrated out, the renormalizable Yukawa terms at higher energy scale give rise to 
 the effective Yukawa interactions which explain the current SM fermion mass hierarchy. The Feynman diagrams corresponding to the effective Yukawa interactions of the quark sector are shown in Figure~\ref{fig:diagrams_quark_mass_insertion}:
\begin{figure}[H]
\centering
\begin{subfigure}{0.48\textwidth}
	\includegraphics[width=1.0\textwidth]{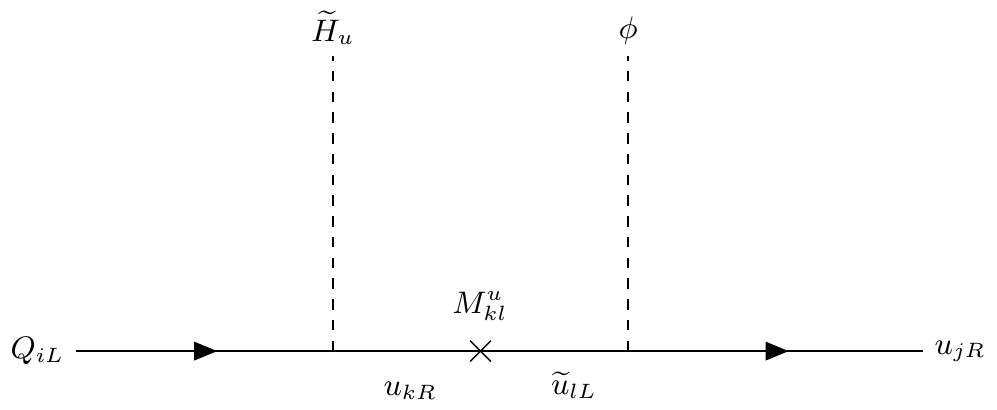}
\end{subfigure} \hspace{0.1cm} 
\begin{subfigure}{0.48\textwidth}
	\includegraphics[width=1.0\textwidth]{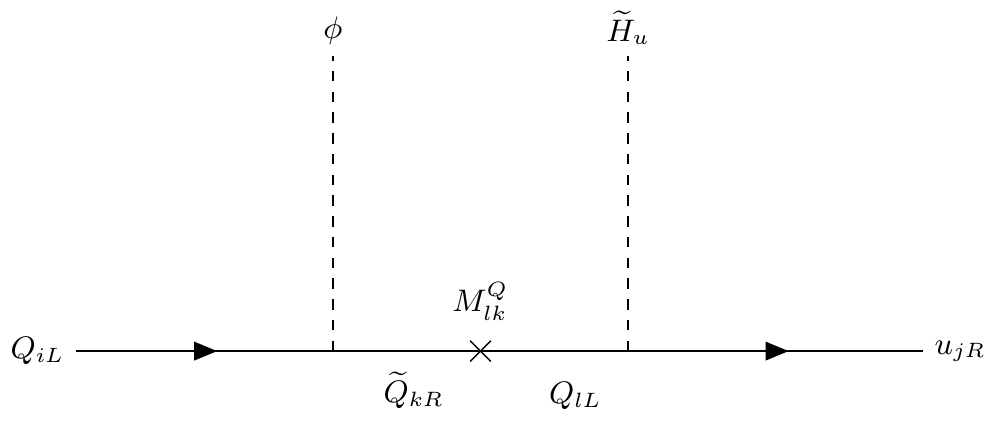}
\end{subfigure}
\begin{subfigure}{0.48\textwidth}
	\includegraphics[width=1.0\textwidth]{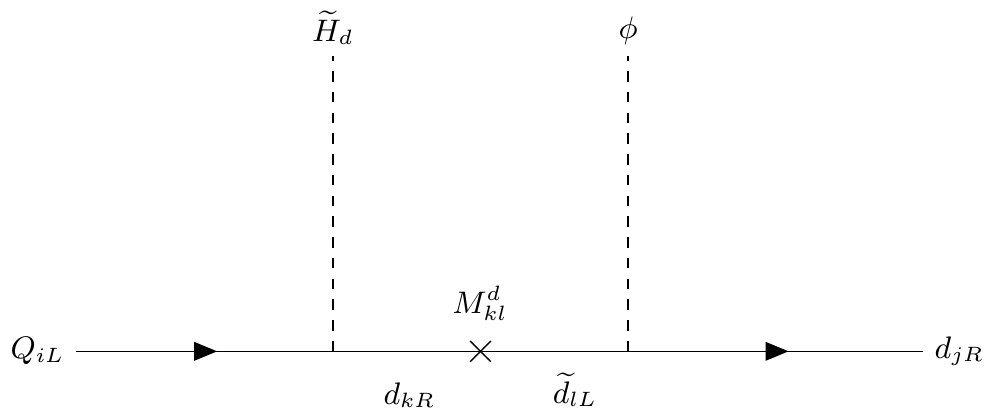}
\end{subfigure} \hspace{0.1cm} 
\begin{subfigure}{0.48\textwidth}
	\includegraphics[width=1.0\textwidth]{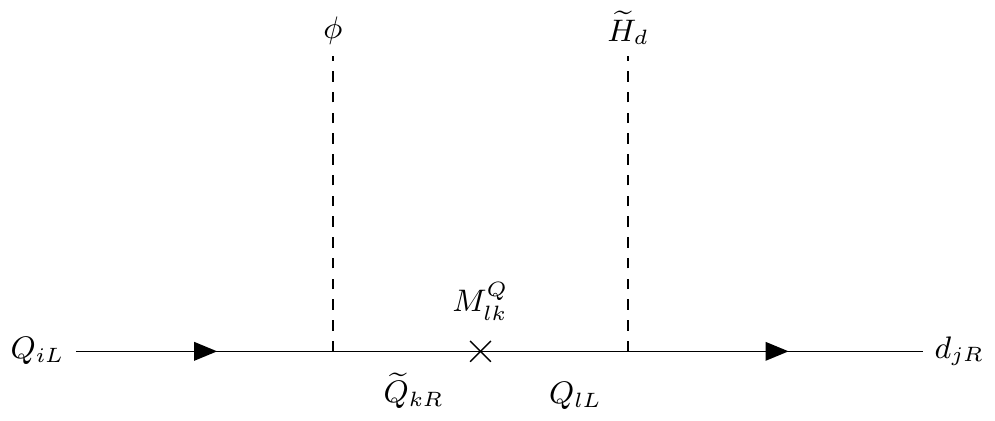}
\end{subfigure}
\caption{Feynman diagrams contributing to the up and down type quark's effective Yukawa
interactions in the mass insertion formalism. Here $i,j=1,2,3$ and $k,l=4$ and $M_{lk}$ is vector-like mass.}  
\label{fig:diagrams_quark_mass_insertion}
\end{figure}
The same approach can be applied to the SM charged lepton sector and the renormalizable charged lepton Yukawa interactions are given by:
\begin{equation}
\begin{split}
\mathcal{L}_{e}^{\func{Yukawa+Mass}} &= y_{ik}^{e} \overline{L}_{iL} 
\widetilde{H}_{d} e_{kR} + x_{ki}^{e} \phi \overline{\widetilde{e}}_{kL}
e_{iR} + x_{ik}^L \phi \overline{L}_{iL} \widetilde{L}_{kR} + y_{ki}^e 
\overline{L}_{kL} \widetilde{H}_{d} e_{iR} 
\\
&+ M_{kl}^{e} \overline{\widetilde{%
e}}_{lL} e_{kR} + M_{kl}^L \overline{L}_{kL} \widetilde{L}_{lR} + \func{h.c.},
\end{split}
\label{eqn:general_charged_lepton_Yukawa_Mass_Lagrangian}
\end{equation}
and its following effective Yukawa interactions read off in Figure~\ref{fig:diagrams_charged_leptons_mass_insertion}.
\begin{figure}[H]
\centering
\begin{subfigure}{0.48\textwidth}
	\includegraphics[width=1.0\textwidth]{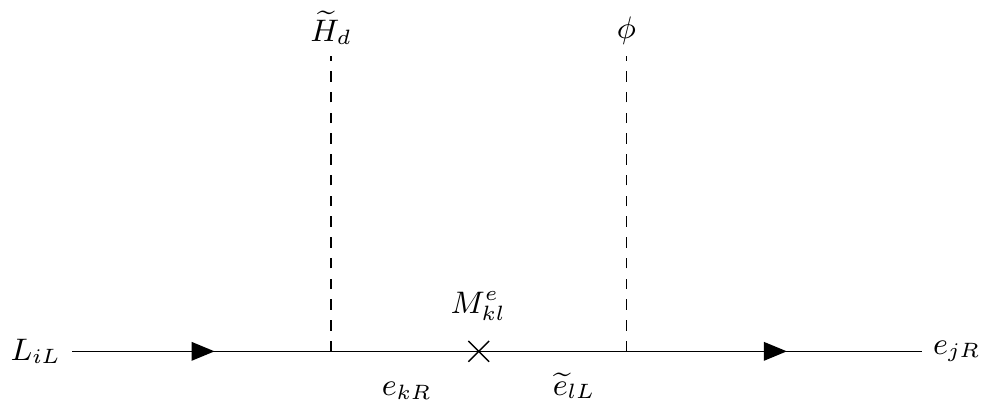}
\end{subfigure} \hspace{0.1cm} 
\begin{subfigure}{0.48\textwidth}
	\includegraphics[width=1.0\textwidth]{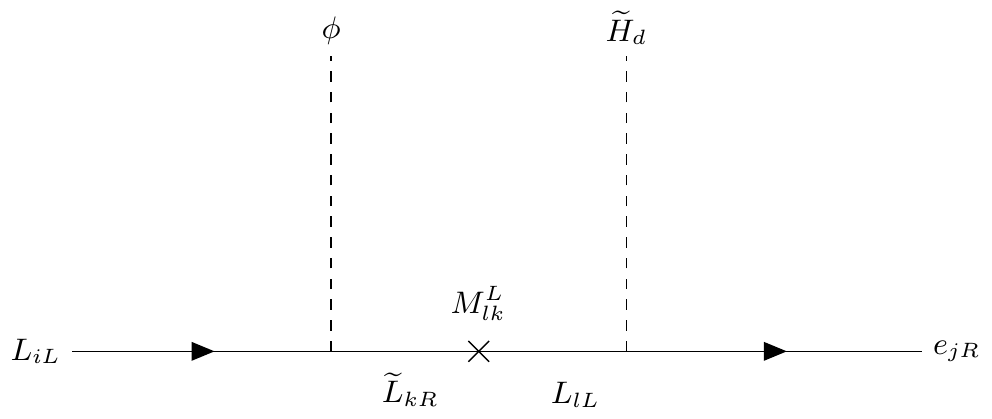}
\end{subfigure}
\caption{Feynman diagrams contributing to the charged lepton's effective Yukawa interactions in the mass insertion formalism. Here $i,j=1,2,3$ and $k,l=4$ and $M_{lk}$ is vector-like mass.}  
\label{fig:diagrams_charged_leptons_mass_insertion}
\end{figure}
It is possible to generate the masses of all SM charged fermions by the same method relying on 
effective Yukawa interactions. However, this is not the case for the SM light active neutrinos as they need to be independently treated since the simplest mechanism responsible for generating their tiny masses requires the inclusion of 
Majorana particles in the leptonic spectrum. In order to make the SM neutrinos massive, we made use of 
two important assumptions, one of which is that the SM neutrinos are Majorana particles and the other is they get masses via the type 1b seesaw mechanism~\cite{Hernandez:2021tii,Hernandez-Garcia:2019uof} mediated by the heavy vector-like neutrinos without considering the right-handed neutrinos $\nu_{iR}$. The renormalizable Yukawa interactions for the neutrino sector are given by:
\begin{equation}
\begin{split}
\mathcal{L}_{\nu}^{\func{Yukawa+Mass}} = y_{ik}^{\nu} \overline{L}_{iL} 
\widetilde{H}_u \nu_{kR} + x_{ik}^L \overline{L}_{iL} H_d \overline{%
\widetilde{\nu}}_{kR} + M_{kl}^{M} \overline{\widetilde{\nu}}_{lR} \nu_{kR}
+ \func{h.c.}
\end{split}
\label{eqn:general_neutrinos_Yukawa_Mass_Lagrangian}
\end{equation}
It is worth mentioning that 
the nature of the vector-like mass appearing in Equation~\ref{eqn:general_neutrinos_Yukawa_Mass_Lagrangian} is that the vector-like mass is different than the Majorana mass since the particles involved in vector-like mass terms 
are different, whereas the ones appearing in a Majorana mass terms does not. However, they share 
the common feature that both break the lepton number, which is confirmed 
 by checking each lepton number of $\nu_{kR}$ and $\widetilde{\nu}_{kR}$ in the two Yukawa interactions of Equation~\ref{eqn:general_neutrinos_Yukawa_Mass_Lagrangian}. We call this mechanism ``type 1b seesaw mechanism" and it can allow a different Yukawa interaction at each vertex as seen in Figure~\ref{fig:diagrams_neutrinos_mass_insertion}.
\begin{figure}[H]
\centering
\begin{subfigure}{0.48\textwidth}
	\includegraphics[width=1.0\textwidth]{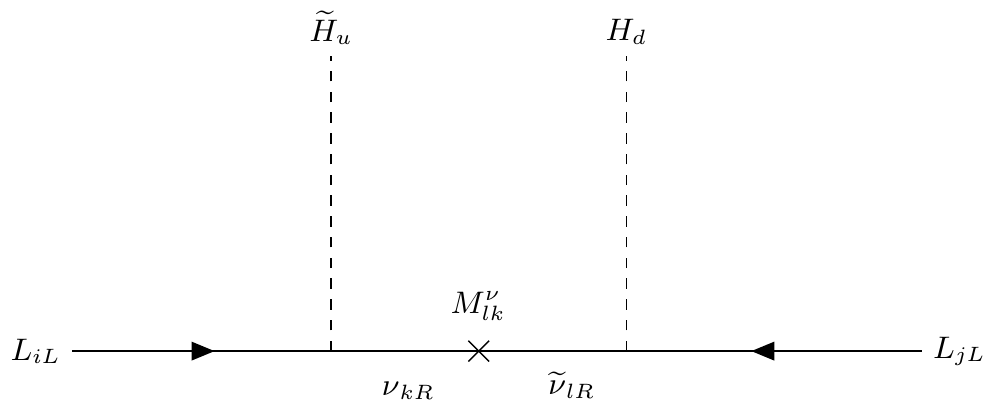}
\end{subfigure}
\caption{Type Ib seesaw diagram~\cite{Hernandez:2021tii,Hernandez-Garcia:2019uof} which leads to the effective Yukawa
interactions for the Majorana neutrinos in mass insertion formalism,
where $i,j=1,2,3$ and $k,l=4$ and $M_{lk}$ is vector-like mass.}
\label{fig:diagrams_neutrinos_mass_insertion}
\end{figure}
Allowing a different Yukawa interaction at each vertex of Figure~\ref{fig:diagrams_neutrinos_mass_insertion} means that one of the Yukawa interactions can have a very suppressed coupling constant, which can lower the expected order of magnitude of the right-handed Majorana neutrinos masses of the usual type I seesaw mechanism from $10^{14}\func{GeV}$ up to the $\func{TeV}$ scale. The most relevant 
features of the 
model considered in this paper are: 
\begin{enumerate}
\item 
It allows a dynamical and natural explanation of the origin of the observed SM fermion mass hierarchy
\item The model under consideration  
is economical in the sense that it includes a common mechanism for generating the masses of the SM charged fermions via 
 effective Yukawa interactions resulting after integrating out the heavy vector-like fermions.
 \item The expected right-handed neutrinos can have a much smaller mass compared to the ones mediating the usual type I seesaw mechanism, thus allowing to test our model at colliders as well as via charged lepton flavor violating processes.
\end{enumerate}
Now that we have discussed how the SM fermions get massive via the effective Yukawa interactions, the next task is to construct their mass matrices in the flavor basis and then to diagonalize those and this will be discussed in the next section.
\section{Effective Yukawa matrices using a mixing formalism}
\label{sec:III}
The effective Yukawa interactions discussed in section~\ref{sec:II} give rise to the following 
 mass matrix for fermions written in the flavor basis: 
\begin{equation}
M^{\psi }=\left( 
\begin{array}{c|ccccc}
& \psi _{1R} & \psi _{2R} & \psi _{3R} & \psi _{4R} & \widetilde{\psi }_{4R} \\[0.5ex] \hline
\overline{\psi }_{1L} & 0 & 0 & 0 & y_{14}^{\psi }\langle \widetilde{H}%
^{0}\rangle & x_{14}^{\psi }\langle \phi \rangle \\[1ex]
\overline{\psi }_{2L} & 0 & 0 & 0 & y_{24}^{\psi }\langle \widetilde{H}%
^{0}\rangle & x_{24}^{\psi }\langle \phi \rangle \\[1ex] 
\overline{\psi }_{3L} & 0 & 0 & 0 & y_{34}^{\psi }\langle \widetilde{H}%
^{0}\rangle & x_{34}^{\psi }\langle \phi \rangle \\[1ex]
\overline{\psi }_{4L} & y_{41}^{\psi }\langle \widetilde{H}%
^{0}\rangle & y_{42}^{\psi }\langle \widetilde{H}^{0}\rangle
& y_{43}^{\psi }\langle \widetilde{H}^{0}\rangle & 0 & 
M_{44}^{\psi } \\[1ex]
\overline{\widetilde{\psi }}_{4L} & x_{41}^{\psi ^{\prime }}\langle
\phi \rangle & x_{42}^{\psi ^{\prime }}\langle \phi \rangle
& x_{43}^{\psi ^{\prime }}\langle \phi \rangle & M_{44}^{\psi
^{\prime }} & 0 \\ 
\end{array}%
\right) ,  
\label{eqn:general_55_mass_matrix}
\end{equation}
where $\psi,\psi^{\prime} = Q,u,d,L,e$ and the zeros in the upper-left $3 \times 3$ block mean that the SM fermions 
acquire masses only via their mixing 
 with the fourth vector-like family. The other zeros appearing in the diagonal positions are forbidden by the $U(1)^{\prime}$ charge conservation. This mass matrix was obtained for the first time in 
\cite{King:2018fcg} and it can reveal the hierarchical structure of the SM since this mass matrix involves three different mass scales $\langle H^0 \rangle, \langle \phi \rangle$ and $M$. In order to dynamically reproduce the hierarchical structure of the SM fermion masses,  
  we need to maximally rotate this mass matrix and the resulting maximally rotated mass matrix should be a starting point for our analysis in this work. For the fully rotated mass matrix, the up-quark and charged lepton sectors share the same structure, whereas the down-type quark mass matrix has an  
  additional element since one of the quark doublet rotations was already used in the up quark sector, as it will be shown below.  
Regarding the diagonalization for each fermion sector, we will employ two methods for comparison; one of which is the numerical SVD diagonalization and the other is the analytical perturbative 
step-by-step diagonalization. We will make use of the numerical SVD diagonalization for the exact diagonalization as well as for our numerical scans in main body of this work, however it is important to look at the analytical approximated step-by-step diagonalization since   
it provides an analytical understanding  
on how the SM $Z$ gauge boson can induce the flavor violating interactions at tree-level and this analytic diagonalization will be covered in Appendix~\ref{app:A} to \ref{app:C}. Lastly, we have found that the analytic diagoanlization for each fermion sector is quite close to its numerical result with very small differences and this feature will be discussed in detail in Appendix~\ref{app:D} to \ref{app:E}.
\subsection{Diagonalizing the charged lepton mass matrix} \label{sec:III_1}
After all scalars of our proposed model acquire  
their vevs ($v_{d} = \langle H_d^0 \rangle$ and $v_{\phi} = \langle \phi \rangle$) from Equation~\ref{eqn:general_55_mass_matrix}, we otain the following 
fully rotated mass matrix for the charged lepton sector 
\begin{equation}
M^{e }=\left( 
\begin{array}{c|ccccc}
& e _{1R} & e _{2R} & e _{3R} & e _{4R} & \widetilde{L }_{4R} \\[0.5ex] \hline
\overline{L }_{1L} & 0 & 0 & 0 & 0 & 0 \\[1ex]
\overline{L }_{2L} & 0 & 0 & 0 & y_{24}^{e } v_{d} & 0 \\[1ex] 
\overline{L }_{3L} & 0 & 0 & 0 & y_{34}^{e } v_{d} & x_{34}^{L } v_{\phi} \\[1ex]
\overline{L }_{4L} & 0 & 0 & y_{43}^{e } v_{d} & 0 & M_{44}^{L } \\[1ex]
\overline{\widetilde{e }}_{4L} & 0 & x_{42}^{e} v_{\phi}
& x_{43}^{e} v_{\phi} & M_{44}^{e} & 0 \\ 
\end{array}%
\right), 
\label{eqn:cl_1}
\end{equation}
where we use this fully rotated basis as a starting point in order to easily explain the observed SM fermion mass hierarchy. 
This rotated basis is exactly consistent with the one given in~\cite{King:2018fcg} and we need to explain how the mass matrix of Equation~\ref{eqn:cl_1} is fully rotated. First of all, we rotate the left-handed leptonic fields $L_{1L}$ and $L_{3L}$ to turn off the entry $x_{14}^{L} v_{\phi}$ and then rotate $L_{2L}$ and $L_{3L}$ to trun off the next $x_{24}^{L} v_{\phi}$ entry. Next, we can rotate again the leptonic fields $L_{1L}$ and $L_{2L}$ to turn off the $y_{14}^{e} v_d$ entry. These rotations can be applied to the right-handed leptonic fields $e_{1,2,3R}$ in order to make the zeros appearing in the lower-left $2 \times 3$ block. This fully rotated mass matrix of Equation~\ref{eqn:cl_1} is our starting point to implement both the hierarchical structure of the SM fermion masses and to analyze the flavor violating interactions mediated by the SM $Z$ gauge boson. Before diagonalizing the mass matrix of Equation~\ref{eqn:cl_1}, it is convenient to rearrange the mass matrix by switching the Yukawa terms by mass parameters and then by swapping the fourth and fifth column in order to make the heavy vector-like masses locate in the diagonal positions as given in Equation~\ref{eqn:cl_2}. 
\begin{equation}
M^{e }
=
\left( 
\begin{array}{c|ccccc}
& e _{1R} & e _{2R} & e _{3R} & e _{4R} & \widetilde{L }_{4R} \\[0.5ex] \hline
\overline{L }_{1L} & 0 & 0 & 0 & 0 & 0 \\[1ex]
\overline{L }_{2L} & 0 & 0 & 0 & m_{24} & 0 \\[1ex] 
\overline{L }_{3L} & 0 & 0 & 0 & m_{34} & m_{35} \\[1ex]
\overline{L }_{4L} & 0 & 0 & m_{43} & 0 & M_{45}^{L } \\[1ex]
\overline{\widetilde{e }}_{4L} & 0 & m_{52} & m_{53} & M_{54}^{e} & 0 \\ 
\end{array}%
\right)
=
\left( 
\begin{array}{c|ccccc}
& e _{1R} & e _{2R} & e _{3R} & \widetilde{L }_{4R} & e _{4R} \\[0.5ex] \hline
\overline{L }_{1L} & 0 & 0 & 0 & 0 & 0 \\[1ex]
\overline{L }_{2L} & 0 & 0 & 0 & 0 & m_{24} \\[1ex] 
\overline{L }_{3L} & 0 & 0 & 0 & m_{35} & m_{34} \\[1ex]
\overline{L }_{4L} & 0 & 0 & m_{43} & M_{45}^{L } & 0 \\[1ex]
\overline{\widetilde{e }}_{4L} & 0 & m_{52} & m_{53} & 0 & M_{54}^{e} \\ 
\end{array}%
\right), 
\label{eqn:cl_2}
\end{equation}
We use two methods for diagonalizing 
the rotated mass matrix of Equation~\ref{eqn:cl_2}, 
one of which corresponds to the 
numerical diagonalization carried out by the singular value decomposition (SVD) and the other is an approximated analytical step-by-step diagonalization. We make use of the numerical SVD diagonalization for the exact diagonalization and perform numerical scans in main body of this work, however it is worth discussing the analytic step-by-step diagonalization as it gives an analytical understanding on how the SM $Z$ gauge boson can induce the flavor violating interactions at tree-level with the $SU(2)$ violating mixings, which will be defined in the analytic diagonalization covered in Appendix~\ref{app:A}. From the comparison between the analytic and numerical computations, we found that the former 
 works quite well and yields resuls close the ones obtained from the latter. The comparisons between the analytic and numerical computations for both lepton and quark sectors will be discussed in detail in   
 Appendices \ref{app:D} and \ref{app:E}, respectively. The charged lepton sector can be diagonalized by performing the SVD diagonalization as follows:
\begin{equation}
M^{e \prime} = \func{diag}\left( 0, m_{\mu}, m_{\tau}, M_{E_4}, M_{\widetilde{E}_4} \right) = V^{L} M^{e} (V^{e})^{\dagger},
\end{equation}
where $V^{L} (V^{e})$ is the mixing matrix for the left-handed (right-handed) leptonic fields, defined as follows:
\begin{equation}
\begin{split}
\begin{pmatrix}
e_{L} \\[0.5ex]
\mu_{L} \\[0.5ex]
\tau_{L} \\[0.5ex]
E_{4L} \\[0.5ex]
\widetilde{E}_{4L}
\end{pmatrix}
=
V^L  
\begin{pmatrix}
e_{1L} \\[0.5ex]
e_{2L} \\[0.5ex]
e_{3L} \\[0.5ex]
e_{4L} \\[0.5ex]
\widetilde{e}_{4L}
\end{pmatrix},
\qquad
\begin{pmatrix}
e_{R} \\[0.5ex]
\mu_{R} \\[0.5ex]
\tau_{R} \\[0.5ex]
\widetilde{E}_{4R} \\[0.5ex]
E_{4R}
\end{pmatrix}
=
V^e
\begin{pmatrix}
e_{1R} \\[0.5ex]
e_{2R} \\[0.5ex]
e_{3R} \\[0.5ex]
\widetilde{e}_{4R} \\[0.5ex]
e_{4R}
\end{pmatrix},
\label{eqn:cl_mixing}
\end{split}
\end{equation}
and the numerical mixing matrices $V^{L,e}$ can be expressed by an analytic expression consisting of a series of $V_{ij}^{L,e}$ which describes mixing between $i$th and $j$th fermion where $i,j=1,2,3,4,5$ and this will be discussed in Appendix~\ref{app:A}.
\subsection{Diagonalizing the up-type quark mass matrix} \label{sec:III_2}
The initial mass matrix for the up-type quark sector in the flavor basis is given by:
\begin{equation}
M^{u }
=
\left( 
\begin{array}{c|ccccc}
& u _{1R} & u _{2R} & u _{3R} & u _{4R} & \widetilde{Q }_{4R} \\[0.5ex] \hline
\overline{Q }_{1L} & 0 & 0 & 0 & 0 & 0 \\[1ex]
\overline{Q }_{2L} & 0 & 0 & 0 & y_{24}^{u } v_{u} & 0 \\[1ex] 
\overline{Q }_{3L} & 0 & 0 & 0 & y_{34}^{u } v_{u} & x_{34}^{Q } v_{\phi} \\[1ex]
\overline{Q }_{4L} & 0 & 0 & y_{43}^{u } v_{u} & 0 & M_{44}^{Q } \\[1ex]
\overline{\widetilde{u }}_{4L} & 0 & x_{42}^{u} v_{\phi}
& x_{43}^{u} v_{\phi} & M_{44}^{u} & 0 \\ 
\end{array}%
\right)
=
\left( 
\begin{array}{c|ccccc}
& u _{1R} & u _{2R} & u _{3R} & \widetilde{Q }_{4R} & u _{4R} \\[0.5ex] \hline
\overline{Q }_{1L} & 0 & 0 & 0 & 0 & 0 \\[1ex]
\overline{Q }_{2L} & 0 & 0 & 0 & 0 & m_{24}^{u} \\[1ex] 
\overline{Q }_{3L} & 0 & 0 & 0 & m_{35}^{u} & m_{34}^{u} \\[1ex]
\overline{Q }_{4L} & 0 & 0 & m_{43}^{u} & M_{44}^{Q } & 0 \\[1ex]
\overline{\widetilde{u }}_{4L} & 0 & m_{52}^{u}
& m_{53}^{u} & 0 & M_{44}^{u} \\ 
\end{array}%
\right)
, \label{eqn:uq_1}
\end{equation}
The mass matrix of Equation~\ref{eqn:uq_1} in the flavor basis is exactly consistent with the one corresponding to the charged lepton sector excepting for a few substitutions $y^{e} \rightarrow y^{u}$, $v_{d} \rightarrow v_{u}$, $x^{L} \rightarrow x^{Q}$ and $x^{e} \rightarrow x^{u}$. 
The analytic mixing matrix for the up-quark sector is exactly same as the one for the charged lepton sector, however unlike the charged lepton sector, 
diagonalizing the mass matrix for the up-quark sector requires more caution as some numerical off-diagonal elements of order unity. 
This feature is resulted from the 
some off-diagonal elements arising as a result of mixing between the heavy top quark mass and the other heavy exotic up-type quark masses are of order unitiry, not small enough to obtain precise results in the perturbative diagonalization when compared to the charged lepton sector. We discuss this feature in Appendix~\ref{app:E} by comparing a numerical mixing matrix obtained from the SVD with the one resulting from the analytic diagonalization. Then we can numerically diagonalize the up-type mass matrix by using the SVD diagonalization as follows:
\begin{equation}
\begin{split}
\begin{pmatrix}
u_{L} \\[0.5ex]
c_{L} \\[0.5ex]
t_{L} \\[0.5ex]
U_{4L} \\[0.5ex]
\widetilde{U}_{4L}
\end{pmatrix}
=
V_{L}^{u}  
\begin{pmatrix}
u_{1L} \\[0.5ex]
u_{2L} \\[0.5ex]
u_{3L} \\[0.5ex]
u_{4L} \\[0.5ex]
\widetilde{u}_{4L}
\end{pmatrix}
, \qquad
\begin{pmatrix}
u_{R} \\[0.5ex]
c_{R} \\[0.5ex]
t_{R} \\[0.5ex]
\widetilde{U}_{4R} \\[0.5ex]
U_{4R}
\end{pmatrix}
=
V_{R}^{u}
\begin{pmatrix}
u_{1R} \\[0.5ex]
u_{2R} \\[0.5ex]
u_{3R} \\[0.5ex]
\widetilde{u}_{4R} \\[0.5ex]
u_{4R}
\end{pmatrix}
\label{eqn:up_mixing}
\end{split}
\end{equation}
where 
the symbol $L$ means left-handed doublet and $e$ denotes right-handed singlet in the charged lepton sector, however  
it is worth mentioning that the above described notation used in the lepton sector becomes complicated in the quark sector since the mass matrices for the up- and down-type quark sectors have a different form, so we change the mixing notation by $V_{L(R)}^{u,d}$ instead of $V^{Q}$.
The analytic diagonalizations for the up-quark sector will be discussed in Appendix~\ref{app:B}.
\subsection{Diagonalizing the down-type quark mass matrix} \label{sec:III_3}
A nice  
feature of the model under consideration is that it can predict 
the CKM mixing matrix and this feature is mainly based on the mixings derived from the down-type quark mass matrix as we will see soon. A quite encouraging feature is that all the mixings among the three SM generations in the down-type quark sector can be accessible even though the first generation of the down-type quark sector remains massless and this feature is quite naturally attributed to this model with the vector-like family. We start from two mass matrices, one of which is for the up-quark sector whereas the another one is for the down-quark sector. 
\begin{equation}
\begin{split}
M^{u }&=\left( 
\begin{array}{c|ccccc}
& u _{1R} & u _{2R} & u _{3R} & u _{4R} & \widetilde{Q }_{4R} \\[0.5ex] \hline
\overline{Q }_{1L} & 0 & 0 & 0 & 0 & 0 \\[1ex]
\overline{Q }_{2L} & 0 & 0 & 0 & y_{24}^{u } v_{u} & 0 \\[1ex] 
\overline{Q }_{3L} & 0 & 0 & 0 & y_{34}^{u } v_{u} & x_{34}^{Q } v_{\phi} \\[1ex]
\overline{Q }_{4L} & 0 & 0 & y_{43}^{u } v_{u} & 0 & M_{44}^{Q } \\[1ex]
\overline{\widetilde{u }}_{4L} & 0 & x_{42}^{u} v_{\phi}
& x_{43}^{u} v_{\phi} & M_{44}^{u} & 0 \\ 
\end{array}%
\right), 
\\
M^{d }&=\left( 
\begin{array}{c|ccccc}
& d _{1R} & d _{2R} & d _{3R} & d _{4R} & \widetilde{Q }_{4R} \\[0.5ex] \hline
\overline{Q }_{1L} & 0 & 0 & 0 & y_{14}^{d } v_{d} & 0 \\[1ex]
\overline{Q }_{2L} & 0 & 0 & 0 & y_{24}^{d } v_{d} & 0 \\[1ex] 
\overline{Q }_{3L} & 0 & 0 & 0 & y_{34}^{d } v_{d} & x_{34}^{Q } v_{\phi} \\[1ex]
\overline{Q }_{4L} & 0 & 0 & y_{43}^{d } v_{d} & 0 & M_{44}^{Q } \\[1ex]
\overline{\widetilde{d }}_{4L} & 0 & x_{42}^{d} v_{\phi}
& x_{43}^{d} v_{\phi} & M_{44}^{d} & 0 \\ 
\end{array}%
\right), 
\label{eqn:diff_up_down}
\end{split}
\end{equation}
This difference between the mass matrices for the up-type and down-type quark sectors was noticed for the first time in~\cite{King:2018fcg}.  
The first property we need to focus on is the fifth column of both mass matrices is exactly same. The zeros appearing in the fifth column of both are the common elements shared by both up- and down-type quark mass matrices since the quark doublets as well as the fourth vector-like quark doublets contribute equally to both sectors. For the up-type quark sector, we were able to rotate further between $Q_{1L}$ and $Q_{2L}$ to vanish $y_{14}^{u} v_{u}$, however this rotation simply remixes $y_{14}^{d} v_{d}$ and $y_{24}^{d} v_{d}$, so both the down-type Yukawa terms survive. For the lower-left $2 \times 3$ block of the down-type quark mass matrix, the same zeros can appear since the down-type quarks $d_{1,2,3R}$ have a different mixing angle against that for the up-type quarks $u_{1,2,3R}$. The down-type mass matrix $M^{d}$ of Equation~\ref{eqn:diff_up_down} can be diagonalized by the numerical SVD diagonalization as follows:
\begin{equation}
\begin{split}
\begin{pmatrix}
d_{L} \\
s_{L} \\
b_{L} \\
D_{4L} \\
\widetilde{D}_{4L}
\end{pmatrix}
=
V_{L}^{d}  
\begin{pmatrix}
d_{1L} \\
d_{2L} \\
d_{3L} \\
d_{4L} \\
\widetilde{d}_{4L}
\end{pmatrix}
, \qquad
\begin{pmatrix}
d_{R} \\
s_{R} \\
b_{R} \\
D_{4R} \\
\widetilde{D}_{4R}
\end{pmatrix}
=
V_{R}^{d}
\begin{pmatrix}
d_{1R} \\
d_{2R} \\
d_{3R} \\
\widetilde{d}_{4R} \\
d_{4R}
\end{pmatrix}.
\label{eqn:down_mixing}
\end{split}
\end{equation}
The analytic diagonalization for the down-quark sector is discussed in Appendix~\ref{app:C}.
We will see that a numerical mixing matrix derived by the SVD is quite close to one by the analytic diagonalization in Appendix~\ref{app:E}. In Appendix~\ref{app:E} we confirm that even though the numerical mixing matrix $V_{L}^{d}$ can have all mixings among the three SM generations, the $Z$ coupling constants $D_{L}^{d \prime}$ in the mass basis will have zeros in the first column and row due to some internal cancellations. Therefore the whole structure of $D_{L}^{d \prime}$ is exactly same as the other $Z$ coupling constants $D_{L,R}^{u \prime}$ and $D_{R}^{d \prime}$ in the mass basis, so we verify that the SM $Z$ physics does not get affected by any specific basis we choose. This feature will be discussed again in the next section as well as in Appendix~\ref{app:E}.
\section{The SM $Z$ gauge boson interactions with the vector-like family}
\label{sec:IV}
One of our main motivations of this work is to study flavor violating 
processes mediated by the $Z$ gauge boson in order to constrain the mass range of the vector-like fermions. It is worth reminding that the neutral $Z^{\prime}$ gauge boson does not appear in the particle spectrum of this model due to the global $U(1)^{\prime}$ symmetry of the theory under consideration. It is worth mentioning that the tree-level flavor violating $Z$ decays are absent in the SM, indepently of the fermion mixings, as can be seen from Equation~\ref{eqn:SM_Z}
shown below:
\begin{equation}
\mathcal{L}_{\func{SM}}^{Z} = g Z_{\mu} J_{Z}^{\mu} = \frac{g}{c_w} Z_{\mu} \sum_{f=e,\mu,\tau} \overline{f} \gamma^{\mu} \left( T^3 - \sin^2\theta_w Q \right) f
\label{eqn:SM_Z}
\end{equation}
Factoring out the prefactor $g/c_{w}$, we can find matrices $D_{L,R}^{e}$, which determine the magnitude of the coupling constant for the $Z$ interactions to either the left-handed or right-handed SM fermions.
\begin{equation}
\begin{split}
D_{L}^{e}&= 
\left(
\begin{array}{c|ccc}
  & e_{1L} & e_{2L} & e_{3L} \\[0.5ex]
  \hline
\overline{e}_{1L} & \left( -\frac{1}{2}+\sin^2\theta_w \right) & 0 & 0 \\[1ex]
\overline{e}_{2L} & 0 & \left( -\frac{1}{2}+\sin^2\theta_w \right) & 0 \\[1ex]
\overline{e}_{3L} & 0 & 0 & \left( -\frac{1}{2}+\sin^2\theta_w \right) \\[1ex]
\end{array}
\right)
\\
D_{R}^{e}&= 
\left(
\begin{array}{c|ccc}
  & e_{1R} & e_{2R} & e_{3R} \\[0.5ex]
  \hline
\overline{e}_{1R} & \left(\sin^2\theta_w \right) & 0 & 0 \\[1ex]
\overline{e}_{2R} & 0 & \left(\sin^2\theta_w \right) & 0 \\[1ex]
\overline{e}_{3R} & 0 & 0 & \left(\sin^2\theta_w \right) \\[1ex]
\end{array}
\right)
\end{split}
\end{equation}
However, this SM $Z$ gauge boson can cause the renormalizable flavor violating interactions with the SM fermions by extending the SM fermion sector by the vector-like fermions as well as by considering the $SU(2)$ violating mixings defined in Appendix~\ref{app:A} together. This features will be discussed in detail in the following subsections.
\subsection{FCNC mediated by the SM $Z$ gauge boson in the charged lepton sector with the fourth vector-like charged leptons} \label{sec:IV_1}
We can construct an extended the SM $Z$ coupling constants in the charged lepton sector with the vector-like charged leptons in the flavor basis.
\begin{equation}
\begin{split}
D_{L}^{e} 
&= 
\scalemath{0.9}{
\left(
\begin{array}{c|ccccc}
  & e_{1L} & e_{2L} & e_{3L} & e_{4L} & \widetilde{e}_{4L} \\[1ex]
  \hline
\overline{e}_{1L} & \left( -\frac{1}{2}+\sin^2\theta_w \right) & 0 & 0 & 0 & 0 \\[1ex]
\overline{e}_{2L} & 0 & \left( -\frac{1}{2}+\sin^2\theta_w \right) & 0 & 0 & 0 \\[1ex]
\overline{e}_{3L} & 0 & 0 & \left( -\frac{1}{2}+\sin^2\theta_w \right) & 0 & 0 \\[1ex]
\overline{e}_{4L} & 0 & 0 & 0 & \left( -\frac{1}{2}+\sin^2\theta_w \right) & 0 \\[1ex]
\overline{\widetilde{e}}_{4L} & 0 & 0 & 0 & 0 & \left( \sin^2\theta_w \right) 
\end{array}
\right)}
\\
D_{R}^{e}
&= 
\left(
\begin{array}{c|ccccc}
  & e_{1R} & e_{2R} & e_{3R} & \widetilde{e}_{4R} & e_{4R} \\[1ex]
  \hline
\overline{e}_{1R} & \left( \sin^2\theta_w \right) & 0 & 0 & 0 & 0 \\[1ex]
\overline{e}_{2R} & 0 & \left( \sin^2\theta_w \right) & 0 & 0 & 0 \\[1ex]
\overline{e}_{3R} & 0 & 0 & \left( \sin^2\theta_w \right) & 0 & 0 \\[1ex]
\overline{\widetilde{e}}_{4R} & 0 & 0 & 0 & \left( -\frac{1}{2}+\sin^2\theta_w \right) & 0 \\[1ex]
\overline{e}_{4R} & 0 & 0 & 0 & 0 & \left( \sin^2\theta_w \right) 
\end{array}
\right)
\label{eqn:SMZ_DLe_DRe}
\end{split}
\end{equation}
where it is worth reminding that the order of the left-handed fermions is 12345, whereas that of the right-handed fermions is 12354 (This ordering is stressed in Appendix~\ref{app:A}). An important feature of this SM $Z$ coupling constants of Equation~\ref{eqn:SMZ_DLe_DRe} is the couplings constants are naturally determined, based on the nature of the vector-like charged leptons, without imposing neither any other symmetry nor other constraints. Therefore, the SM $Z$ coupling constants of Equation~\ref{eqn:SMZ_DLe_DRe} are not the identity matrix anymore unlike the case for 
the SM charged leptons. From these considerations, it follows that there can exist non-zero  
off-diagonal elements in the mass basis by operating the $SU(2)$ violating mixings. Reminding the mixing matrices for the left- or right-handed charged leptons of Equation~\ref{eqn:cl_umm_LH_RH}, the coupling constant in the mass basis ($D_{L,R}^{e \prime}$) can be written as follows:
\begin{equation}
\begin{split}
D_{L}^{e \prime} &= V^{L} D_{L}^{e} (V^{L})^{\dagger}
\\[1ex]
 &= V_{45}^{L} V_{23}^{L} V_{25}^{L} V_{35}^{L} V_{34}^{L} D_{L}^{e} (V_{34}^{L})^{\dagger} (V_{35}^{L})^{\dagger} (V_{25}^{L})^{\dagger} (V_{23}^{L})^{\dagger} (V_{45}^{L})^{\dagger}
\\[1ex]
D_{R}^{e \prime} &= V^{e} D_{R}^{e} (V^{e})^{\dagger}
\\[1ex]
 &= V_{54}^{e} V_{23}^{e} V_{25}^{e} V_{35}^{e} V_{24}^{e} V_{34}^{e} D_{R}^{e} (V_{34}^{e})^{\dagger} (V_{24}^{e})^{\dagger} (V_{35}^{e})^{\dagger} (V_{25}^{e})^{\dagger} (V_{23}^{e})^{\dagger} (V_{54}^{e})^{\dagger}
\label{eqn:SMZ_cl_mb_LR_1}
\end{split}
\end{equation}
It is possible to make the SM $Z$ coupling constants of Equation~\ref{eqn:SMZ_cl_mb_LR_1} simpler by using the $SU(2)$ conserving mixing as it just remixes an identity matrix.
\begin{equation}
\begin{split}
D_{L}^{e \prime} &= V_{45}^{L} V_{23}^{L} V_{25}^{L} V_{35}^{L} D_{L}^{e} (V_{35}^{L})^{\dagger} (V_{25}^{L})^{\dagger} (V_{23}^{L})^{\dagger} (V_{45}^{L})^{\dagger}
\\[1ex]
D_{R}^{e \prime} &= V_{54}^{e} V_{23}^{e} V_{25}^{e} V_{35}^{e} D_{R}^{e} (V_{35}^{e})^{\dagger} (V_{25}^{e})^{\dagger} (V_{23}^{e})^{\dagger} (V_{54}^{e})^{\dagger}
\label{eqn:SMZ_cl_mb_LR_2}
\end{split}
\end{equation}
However, the following mixing matrices after the $SU(2)$ violating mixings $V_{35}^{L,e}$ must be conserved since all of them contribute to the off-diagonal elements of the coupling constants in the mass basis. It is insightful to look at the coupling constants $D_{L,R}^{e \prime}$ in the mass basis (We substitute $\left( -1/2+\sin^2\theta_w \right)$ by $a$ and $\left( \sin^2\theta_w \right)$ by $b$ and assume the mixing angles $\theta_{35,25,23,45(54)}^{L,e}$ are small enough to approximate for simplicity).
\begin{equation}
\begin{split}
D_{L}^{e \prime} &\approx 
\begin{pmatrix}
a & 0 & 0 & 0 & 0 \\[1ex]
0 & a(1 + \theta_{23}^{L2}) + b \theta_{25}^{L2} & b \theta_{25}^{L} \theta_{35}^{L} & b \theta_{25}^{L} \theta_{45}^{L} & (a-b) \theta_{25}^{L} \\[1ex]
0 & b \theta_{25}^{L} \theta_{35}^{L} & a(1+\theta_{23}^{L2}) + b \theta_{35}^{L2} & b \theta_{35}^{L} \theta_{45}^{L} & (a-b) \theta_{35}^{L} \\[1ex]
0 & b \theta_{25}^{L} \theta_{45}^{L} & b \theta_{35}^{L} \theta_{45}^{L} & a + b \theta_{45}^{L2} & (a-b) \theta_{45}^{L} \\[1ex]
0 & (a-b) \theta_{25}^{L} & (a-b) \theta_{35}^{L} & (a-b) \theta_{45}^{L} & b + a(\theta_{25}^{L2} + \theta_{35}^{L2} + \theta_{45}^{L2})
\end{pmatrix}
\\[1ex]
D_{R}^{e \prime} &\approx 
\begin{pmatrix}
b & 0 & 0 & 0 & 0 \\[1ex]
0 & b(1 + \theta_{23}^{e2}) + a \theta_{25}^{e2} & a \theta_{25}^{e} \theta_{35}^{e} & (b-a) \theta_{25}^{e} & -a \theta_{25}^{e} \theta_{54}^{e} \\[1ex]
0 & a \theta_{25}^{e} \theta_{35}^{e} & b(1+\theta_{23}^{e2}) + a \theta_{35}^{e2} & (b-a) \theta_{35}^{e}  & -a \theta_{35}^{e} \theta_{54}^{e} \\[1ex]
0 & (b-a) \theta_{25}^{e} & (b-a) \theta_{35}^{e} & a + b (\theta_{25}^{e2}+\theta_{35}^{e2}+\theta_{54}^{e2}) & (a-b) \theta_{54}^{e} \\[1ex]
0 & -a \theta_{25}^{e} \theta_{54}^{e} & -a \theta_{35}^{e} \theta_{54}^{e} & (a-b) \theta_{54}^{e} & b + a \theta_{54}^{e2}
\end{pmatrix}
\label{eqn:simple_DLep_DRep}
\end{split}
\end{equation}
There are two important features we can read off from the SM $Z$ coupling constants of Equation~\ref{eqn:simple_DLep_DRep}; the first is the diagonal elements $(a,a,a,a,b)$ and $(b,b,b,a,b)$ get hardly affected by the small mixing angles and the second is magnitude of the off-diagonal elements are very weak as the mixing angles are defined by the ratio between the Yukawa and the vector-like masses. Therefore, we can imply the flavor violating mixing mediated by the SM $Z$ gauge boson in the mass basis. Using the SM $Z$ coupling constants in the mass basis, we can draw the Feynman diagram for $\tau \rightarrow \mu \mu \mu$ and $Z \rightarrow \mu \tau$ decay at tree-level and this will be discussed in the next section.
\subsection{FCNC mediated by the SM $Z$ gauge boson in the quark sector with the fourth vector-like quarks} \label{sec:IV_2}
The quark sector have two different mass matrices for the up- and down-type quark sector. We start from the up-type quark sector first. The SM $Z$ coupling constants for the up-type quark sector are given by:
\begin{equation}
\begin{split}
D_{L}^{u} &= 
\scalemath{0.9}{
\left(
\begin{array}{c|ccccc}
  & u_{1L} & u_{2L} & u_{3L} & u_{4L} & \widetilde{u}_{4L} \\[0.5ex]
  \hline
\overline{u}_{1L} & \left( \frac{1}{2}-\frac{2}{3}\sin^2\theta_w \right) & 0 & 0 & 0 & 0 \\[1ex]
\overline{u}_{2L} & 0 & \left( \frac{1}{2}-\frac{2}{3}\sin^2\theta_w \right) & 0 & 0 & 0 \\[1ex]
\overline{u}_{3L} & 0 & 0 & \left( \frac{1}{2}-\frac{2}{3}\sin^2\theta_w \right) & 0 & 0 \\[1ex]
\overline{u}_{4L} & 0 & 0 & 0 & \left( \frac{1}{2}-\frac{2}{3}\sin^2\theta_w \right) & 0 \\[1ex]
\overline{\widetilde{u}}_{4L} & 0 & 0 & 0 & 0 & \left( -\frac{2}{3}\sin^2\theta_w \right) \\[1ex]
\end{array}
\right)} \\[1ex]
D_{R}^{u} &= 
\scalemath{0.9}{
\left(
\begin{array}{c|ccccc}
  & u_{1R} & u_{2R} & u_{3R} & \widetilde{u}_{4R} & u_{4R} \\[0.5ex]
  \hline
\overline{u}_{1R} & \left( -\frac{2}{3}\sin^2\theta_w \right) & 0 & 0 & 0 & 0 \\[1ex]
\overline{u}_{2R} & 0 & \left( -\frac{2}{3}\sin^2\theta_w \right) & 0 & 0 & 0 \\[1ex]
\overline{u}_{3R} & 0 & 0 & \left( -\frac{2}{3}\sin^2\theta_w \right) & 0 & 0 \\[1ex]
\overline{\widetilde{u}}_{4R} & 0 & 0 & 0 & \left( \frac{1}{2}-\frac{2}{3}\sin^2\theta_w \right) & 0 \\[1ex]
\overline{u}_{4R} & 0 & 0 & 0 & 0 & \left( -\frac{2}{3}\sin^2\theta_w \right) \\[1ex]
\end{array}
\right)}
\label{eqn:SMZ_DLu_DRu}
\end{split}
\end{equation}
The up-type quark mass matrix is exactly same as the one for the charged lepton, so we can simply follow the mixing matrices given in Equation~\ref{eqn:ana_up_mixing}. Then the SM $Z$ gauge coupling constants in the mass basis are defined by:
\begin{equation}
\begin{split}
D_{L}^{u \prime} &= V_{L}^{u} D_{L}^{u} V_{L}^{u \dagger}
\\[1ex]
&= (V_{L}^{u})_{45} (V_{L}^{u})_{23} (V_{L}^{u})_{25} (V_{L}^{u})_{35} (V_{L}^{u})_{34} D_{L}^{u} (V_{L}^{u})_{34}^{\dagger} (V_{L}^{u})_{35}^{\dagger} (V_{L}^{u})_{25}^{\dagger} (V_{L}^{u})_{23}^{\dagger} (V_{L}^{u})_{45}^{\dagger}
\\[1ex]
&= (V_{L}^{u})_{45} (V_{L}^{u})_{23} (V_{L}^{u})_{25} (V_{L}^{u})_{35} D_{L}^{u} (V_{L}^{u})_{35}^{\dagger} (V_{L}^{u})_{25}^{\dagger} (V_{L}^{u})_{23}^{\dagger} (V_{L}^{u})_{45}^{\dagger}
\\[1ex]
D_{R}^{u \prime} &= V_{R}^{u} D_{R}^{u} V_{R}^{u \dagger}
\\[1ex]
&= (V_{R}^{u})_{54} (V_{R}^{u})_{23} (V_{R}^{u})_{25} (V_{R}^{u})_{35} (V_{R}^{u})_{24} (V_{R}^{u})_{34} D_{R}^{u} (V_{R}^{u})_{34}^{\dagger} (V_{R}^{u})_{24}^{\dagger} (V_{R}^{u})_{35}^{\dagger} (V_{R}^{u})_{25}^{\dagger} (V_{R}^{u})_{23}^{\dagger} (V_{R}^{u})_{54}^{\dagger}
\\
&= (V_{R}^{u})_{54} (V_{R}^{u})_{23} (V_{R}^{u})_{25} (V_{R}^{u})_{35} D_{R}^{u} (V_{R}^{u})_{35}^{\dagger} (V_{R}^{u})_{25}^{\dagger} (V_{R}^{u})_{23}^{\dagger} (V_{R}^{u})_{54}^{\dagger}
\label{eqn:SMZ_uq_mb_LR_1}
\end{split}
\end{equation}
The SM $Z$ gauge coupling constants for the up-quark sector in the mass basis can be seen by (We substitute $\left( 1/2-2/3\sin^2\theta_w \right)$ by $c$ and $\left( -2/3\sin^2\theta_w \right)$ by $d$ and assume the mixing angles $\theta_{35,25,23,45(54)L,R}^{u}$ are small enough to approximate for simplicity):
\begin{equation}
\begin{split}
D_{L}^{u \prime} &\approx 
\begin{pmatrix}
c & 0 & 0 & 0 & 0 \\[1ex]
0 & c & d \theta_{25L}^{u} \theta_{35L}^{u} & d \theta_{25L}^{u} \theta_{45L}^{u} & (c-d) \theta_{25L}^{u} \\[1ex]
0 & d \theta_{25L}^{u} \theta_{35L}^{u} & c & d \theta_{35L}^{u} \theta_{45L}^{u} & (c-d) \theta_{35L}^{u} \\[1ex]
0 & d \theta_{25L}^{u} \theta_{45L}^{u} & d \theta_{35L}^{u} \theta_{45L}^{u} & c & (c-d) \theta_{45L}^{u} \\[1ex]
0 & (c-d) \theta_{25L}^{u} & (c-d) \theta_{35L}^{u} & (c-d) \theta_{45L}^{u} & d
\end{pmatrix}
\\[1ex]
D_{R}^{u \prime} &\approx 
\begin{pmatrix}
d & 0 & 0 & 0 & 0 \\[1ex]
0 & d & c \theta_{25R}^{u} \theta_{35R}^{u} & (d-c) \theta_{25R}^{u} & -c \theta_{25R}^{u} \theta_{54R}^{u} \\[1ex]
0 & c \theta_{25R}^{u} \theta_{35R}^{u} & d & (d-c) \theta_{35R}^{u}  & -c \theta_{35R}^{u} \theta_{54R}^{u} \\[1ex]
0 & (d-c) \theta_{25R}^{u} & (d-c) \theta_{35R}^{u} & c & (c-d) \theta_{54R}^{u} \\[1ex]
0 & -c \theta_{25R}^{u} \theta_{54R}^{u} & -c \theta_{35R}^{u} \theta_{54R}^{u} & (c-d) \theta_{54R}^{u} & d
\end{pmatrix}
\label{eqn:simple_DLup_DRup}
\end{split}
\end{equation}
Next, we focus on the down-type quark mass matrix and the left-handed mixing matrices for that is different when compared to other left-handed mixing matrices for the up-type or charged lepton mass matrix in that it can reach to all mixings among the three SM generations. Keeping that in mind, we start from the SM $Z$ coupling constants for the down-type quarks.
\begin{equation}
\begin{split}
D_{L}^{d}
&= 
\scalemath{0.8}{
\left(
\begin{array}{c|ccccc}
  & d_{1L} & d_{2L} & d_{3L} & d_{4L} & \widetilde{d}_{4L} \\[0.5ex]
  \hline
\overline{d}_{1L} & \left( -\frac{1}{2}+\frac{1}{3}\sin^2\theta_w \right) & 0 & 0 & 0 & 0 \\[1ex]
\overline{d}_{2L} & 0 & \left( -\frac{1}{2}+\frac{1}{3}\sin^2\theta_w \right) & 0 & 0 & 0 \\[1ex]
\overline{d}_{3L} & 0 & 0 & \left( -\frac{1}{2}+\frac{1}{3}\sin^2\theta_w \right) & 0 & 0 \\[1ex]
\overline{d}_{4L} & 0 & 0 & 0 & \left( -\frac{1}{2}+\frac{1}{3}\sin^2\theta_w \right) & 0 \\[1ex]
\overline{\widetilde{d}}_{4L} & 0 & 0 & 0 & 0 & \left( \frac{1}{3}\sin^2\theta_w \right) 
\end{array}
\right)}
\\
D_{R}^{d}
&= 
\scalemath{0.9}{
\left(
\begin{array}{c|ccccc}
  & d_{1R} & d_{2R} & d_{3R} & \widetilde{d}_{4R} & d_{4R} \\[0.5ex]
  \hline
\overline{d}_{1R} & \left( \frac{1}{3}\sin^2\theta_w \right) & 0 & 0 & 0 & 0 \\[1ex]
\overline{d}_{2R} & 0 & \left( \frac{1}{3}\sin^2\theta_w \right) & 0 & 0 & 0 \\[1ex]
\overline{d}_{3R} & 0 & 0 & \left( \frac{1}{3}\sin^2\theta_w \right) & 0 & 0 \\[1ex]
\overline{\widetilde{d}}_{4R} & 0 & 0 & 0 & \left( -\frac{1}{2}+\frac{1}{3}\sin^2\theta_w \right) & 0 \\[1ex]
\overline{d}_{4R} & 0 & 0 & 0 & 0 & \left( \frac{1}{3}\sin^2\theta_w \right) 
\end{array}
\right)}
\label{eqn:SMZ_DLd_DRd}
\end{split}
\end{equation}
After simplifying the whole left-handed (right-handed) mixing matrices of Equation~\ref{eqn:down_mix2}, 
we obtain the following matrices of $Z$ couplings with quarks
\begin{equation}
\begin{split}
D_{L}^{d \prime} = V_{L}^{d} D_{L}^{d} (V_{L}^{d})^{\dagger} \\[1ex]
D_{R}^{d \prime} = V_{R}^{d} D_{R}^{d} (V_{R}^{d})^{\dagger} 
\label{eqn:DLdp_DRdp}
\end{split}
\end{equation}
The SM $Z$ coupling constants for the right-handed down-type quarks in the mass basis have the same form given in Equation~\ref{eqn:simple_DLep_DRep}, whereas those for the left-handed down-type quarks involves $12$ and $13$ mixings more, so it is worthwhile to look at its complete form (We again substitute $\left( -1/2+1/3 \sin^2\theta_w \right)$ by $e$ and $\left( 1/3 \sin^2\theta_w \right)$ by $f$ and assume all relevant mixing angles are small enough to approximate for simplicity)
\begin{equation}
\begin{split}
D_{L}^{d \prime} &\approx 
\scalemath{0.8}{
\begin{pmatrix}
e & 0 & 0 & 0 & 0 \\[1ex]
0 & e & e \theta_{12L}^{d} \theta_{13L}^{d} + f \theta_{25L}^{d} \theta_{35L}^{d} & -e \theta_{23L}^{d} \theta_{34L}^{d} + f \theta_{25L}^{d} \theta_{45L}^{d} & (e-f) \theta_{25L}^{d} \\[1ex]
0 & e \theta_{12L}^{d} \theta_{13L}^{d} + f \theta_{25L}^{d} \theta_{35L}^{d} & e & f \theta_{35L}^{d} \theta_{45L}^{d} & (e-f) \theta_{35L}^{d} \\[1ex]
0 & -e \theta_{23L}^{d} \theta_{34L}^{d} + f \theta_{25L}^{d} \theta_{45L}^{d} & f \theta_{35L}^{d} \theta_{45L}^{d} & e & (e-f) \theta_{45L}^{d} \\[1ex]
0 & (e-f) \theta_{25L}^{d} & (e-f) \theta_{35L}^{d} & (e-f) \theta_{45L}^{d} & f
\end{pmatrix}}
\\[1ex]
D_{R}^{d \prime} &\approx 
\begin{pmatrix}
f & 0 & 0 & 0 & 0 \\[1ex]
0 & f & e \theta_{25R}^{d} \theta_{35R}^{d} & (f-e) \theta_{25R}^{d} & -e \theta_{25R}^{d} \theta_{54R}^{d} \\[1ex]
0 & e \theta_{25R}^{d} \theta_{35R}^{d} & f & (f-e) \theta_{35R}^{d}  & -e \theta_{35R}^{d} \theta_{54R}^{d} \\[1ex]
0 & (f-e) \theta_{25R}^{d} & (f-e) \theta_{35R}^{d} & e & (e-f) \theta_{54R}^{d} \\[1ex]
0 & -e \theta_{25R}^{d} \theta_{54R}^{d} & -e \theta_{35R}^{d} \theta_{54R}^{d} & (e-f) \theta_{54R}^{d} & f
\end{pmatrix},
\label{eqn:simple_DLdp_DRdp}
\end{split}
\end{equation}
where it can confirm two relations hold for the zeros appearing in the first row and column of $D_{L}^{d \prime}$: $\theta_{13L}^{d} \simeq \theta_{12L}^{d} \theta_{23L}^{d}$ and $\theta_{15L}^{d} \simeq \theta_{12L}^{d} \theta_{25L}^{d}$. Following the analytic mixings given in Equation~\ref{eqn:down_mix2}, the left-handed down type quark sector can access to all mixings among the three SM generations and this feature is also reflected on a numerical mixing matrix $V_{L}^{d}$ of Equation~\ref{eqn:a_num_VCKM}. What this implies is the SM $Z$ physics does not get affected by any specific basis we choose and this will be verified again numerically in Appendix~\ref{app:E}. 
\chapter{Phenomenology in both quark and charged lepton sectors due to SM $Z$ guague boson FCNCs} \label{Chapter:The3rdpaper}
In this chapter, we discuss how to constrain our vector-like fermion fields  in the third BSM model covered in chapter~\ref{Chapter:The3rdBSMmodel} using the diverse FCNC observables. As the first SM generation remains massless, our main observables consist of the second and third generation of the SM. In the charged lepton sector, the FCNC observables such as $\tau \rightarrow \mu \gamma, \tau \rightarrow 3\mu$ and $Z \rightarrow \mu \tau$ are discussed and none of them can significantly constrain our predictions. In the quark sector, the main observables would be the rare $t \rightarrow c Z$ decay and the CKM mixing matrix and we discuss the most challenging part arises from fitting the CKM mixing matrix. Then, we conclude this paper, predicting each range of vector-like quarks and charged leptons.
\section{Phenomenology in the charged lepton sector of the SM} \label{sec:V}
Now that we have defined all required mixings and coupling constants in both quark and charged lepton sectors, it is time to discuss the relevant phenomenology. As mentioned in the introduction, one of our main goals is to study the FCNC observables to constrain the possible mass range of the vector-like fermions. 
Given that the second and third generations of SM charged leptons 
acquire masses through their mixings with the fourth vector-like charged leptons, we will restrict our analysis  
to the constraints imposed on the flavor violating decays involving  
the second and third charged lepton generations such as $\tau \rightarrow \mu \gamma, \tau \rightarrow \mu \mu \mu$ and lastly $Z \rightarrow \mu \tau$ decay.
\subsection{Analytic expression for $\tau \rightarrow \mu \gamma$ decay}
\begin{figure}[H]
\centering
\includegraphics[keepaspectratio,width=\textwidth]{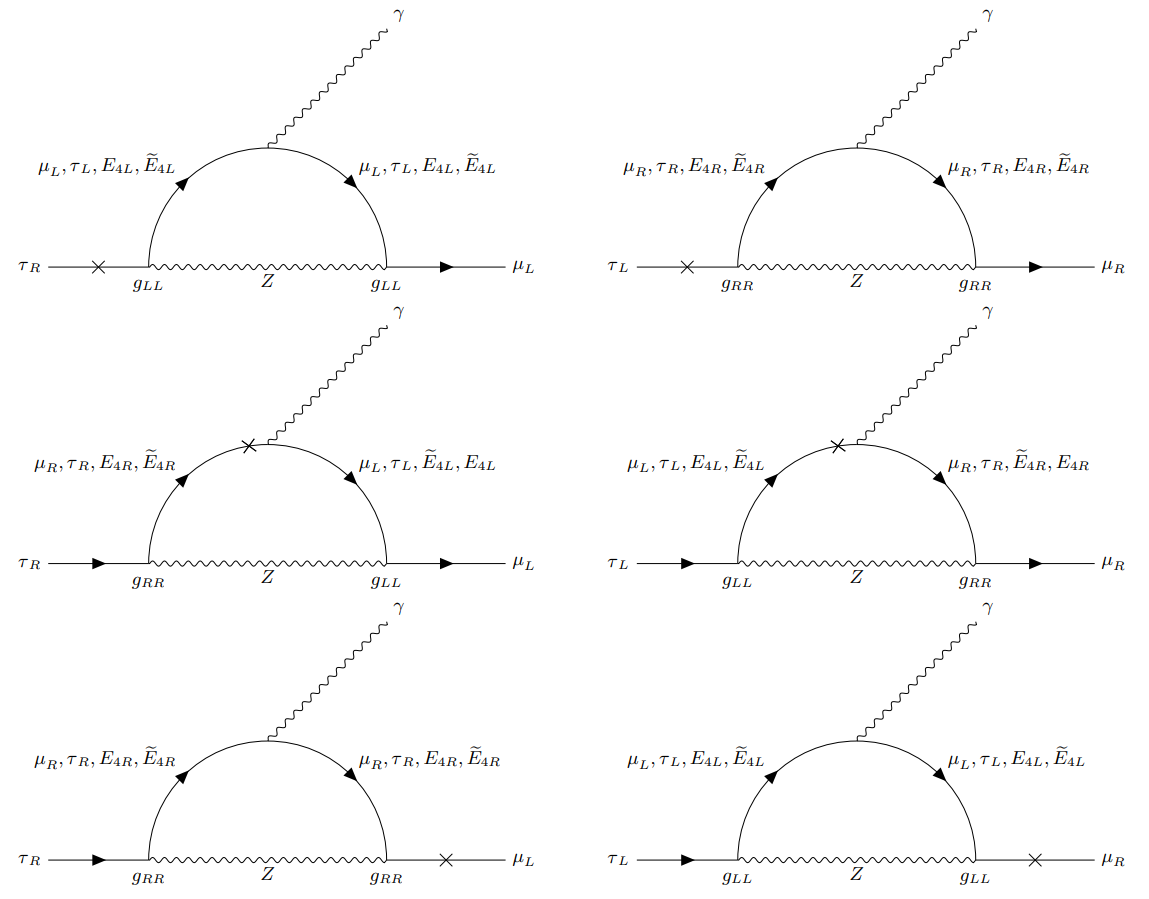}
\caption{Diagrams contributing to the charged lepton flavor violation (CLFV) $\tau \rightarrow \mu \gamma$ decay at one-loop level in the mass basis. The cross notation in each diagram means the helicity flip process.}
\label{fig:tmgdecay}
\end{figure}

The most important FCNC constraint corresponds to the charged lepton flavor violating (CLFV) $\mu \rightarrow e \gamma$ decay, however we can not make an appropriate prediction for the constraint as the electron does not acquire a mass in the model under consideration. This is due to the fact that the fermion sector of the model includes two heavy fermionic seesaw mediators. As previously mentioned, adding a fifth vector-like family to the fermion sector of the model will generate a nonvanishing electron. However in order to keep our model as economical as possible and to simplify our analysis corresponding to the FCNC constraints on vector-like masses, we restrict to the case of a fourth vector-like family in the fermionic spectrum. 
Therefore, in view of the aforementioned considerations, we first consider the CLFV decay $\tau \rightarrow \mu \gamma$ in order to determine how the model parameter space gets affected by the experimental constraint arising from this decay.
For the $\tau \rightarrow \mu \gamma$ decay, the  
leading order contribution appears in the one-loop diagrams since there is no possible contribution at tree-level. Then, all possible Feynman diagrams contributing to  
the $\tau \rightarrow \mu \gamma$ decay are given in Figure~\ref{fig:tmgdecay},
where $g_{LL(RR)}$ are the LH-LH (RH-RH) coupling constants defined in the mass basis for the charged lepton sector. These one-loop contributions mediated by the  
$Z^{\prime}$ boson were studied in one of our previous works~\cite{CarcamoHernandez:2019ydc} and their corresponding analytic prediction for the branching ratio of $\tau \rightarrow \mu \gamma$ decay is given in Equation~\ref{eqn:analytic_tmgdecay}~\cite{CarcamoHernandez:2019ydc,Raby:2017igl,Lindner:2016bgg,Lavoura:2003xp,Chiang:2011cv} shown below:
\begin{equation}
\begin{split}
&\func{BR} \left( \tau \rightarrow \mu \gamma \right) = \frac{\alpha_{\func{em}}}{1024\pi^4} \frac{m_{\tau}^5}{M_Z^4 \Gamma_{\tau}}
\\
&\times \Big( \lvert g_{\tau \mu}^L g_{\mu \mu}^L F(x_\mu) + g_{\tau \tau}^L g_{\tau \mu}^L F(x_\tau) + g_{\tau E_4}^L g_{E_4 \mu}^L F(x_{E_4}) + g_{\tau \widetilde{E}_4}^L g_{\widetilde{E}_4 \mu}^L F(x_{\widetilde{E}_4}) \\
&+ \frac{m_{\mu}}{m_{\tau}} g_{\tau \mu}^L g_{\mu \mu}^L F(x_\mu) + \frac{m_{\mu}}{m_{\tau}} g_{\tau \tau}^L g_{\tau \mu}^L F(x_\tau) + \frac{m_{\mu}}{m_{\tau}} g_{\tau E_4}^L g_{E_4 \mu}^L F(x_{E_4}) + \frac{m_{\mu}}{m_{\tau}} g_{\tau \widetilde{E}_4}^L g_{\widetilde{E}_4 \mu}^L F(x_{\widetilde{E}_4}) \\
&+ \frac{m_{\mu}}{m_{\tau}} g_{\tau \mu}^L g_{\mu \mu}^R G(x_{\mu}) + \frac{m_{\tau}}{m_{\tau}} g_{\tau \tau}^L g_{\tau \mu}^R G(x_{\mu}) + \frac{M_{E_4}}{m_{\tau}} g_{\tau E_4}^L g_{\widetilde{E}_4 \mu}^R G(x_{E_4}) + \frac{M_{\widetilde{E}_4}}{m_{\tau}} g_{\tau \widetilde{E}_4}^L g_{E_4 \mu}^R G(x_{\widetilde{E}_4}) \rvert^2 \\
&+ \lvert g_{\tau \mu}^R g_{\mu \mu}^R F(x_\mu) + g_{\tau \tau}^R g_{\tau \mu}^R F(x_\tau) + g_{\tau E_4}^R g_{E_4 \mu}^R F(x_{E_4}) + g_{\tau \widetilde{E}_4}^R g_{\widetilde{E}_4 \mu}^R F(x_{\widetilde{E}_4}) \\
&+ \frac{m_{\mu}}{m_{\tau}} g_{\tau \mu}^R g_{\mu \mu}^R F(x_\mu) + \frac{m_{\mu}}{m_{\tau}} g_{\tau \tau}^R g_{\tau \mu}^R F(x_\tau) + \frac{m_{\mu}}{m_{\tau}} g_{\tau E_4}^R g_{E_4 \mu}^R F(x_{E_4}) + \frac{m_{\mu}}{m_{\tau}} g_{\tau \widetilde{E}_4}^R g_{\widetilde{E}_4 \mu}^R F(x_{\widetilde{E}_4}) \\
&+ \frac{m_{\mu}}{m_{\tau}} g_{\tau \mu}^R g_{\mu \mu}^L G(x_{\mu}) + \frac{m_{\tau}}{m_{\tau}} g_{\tau \tau}^R g_{\tau \mu}^L G(x_{\mu}) + \frac{M_{\widetilde{E}_4}}{m_{\tau}} g_{\tau E_4}^R g_{\widetilde{E}_4 \mu}^L G(x_{\widetilde{E}_4}) + \frac{M_{E_4}}{m_{\tau}} g_{\tau \widetilde{E}_4}^R g_{E_4 \mu}^L G(x_{E_4}) \rvert^2 \Big),
\label{eqn:analytic_tmgdecay}
\end{split}
\end{equation}
where $\alpha_{\func{em}}$ is the fine structure constant, $\Gamma_{\tau}$ is the total decay width of the tau lepton ($\Gamma_{\tau} = 5 \times \Gamma(\tau_L^- \rightarrow \nu_\tau e_L^- \overline{\nu}_e) = 2.0 \times 10^{-12}$) and $F$ and $G$ are the loop functions defined by:
\begin{equation}
\begin{split}
F(x) &= \frac{5x^4 - 14x^3 + 39x^2 - 38x - 18x^2 \func{ln}x + 8}{12(1-x)^4} \\[1ex]
G(x) &= \frac{x^3 + 3x - 6x \func{ln}x - 4}{2(1-x)^3}, \qquad x = \frac{m_{\func{loop}}^2}{M_{Z}^2}
\end{split}
\end{equation}
where $m_{\func{loop}}$ is the propagating mass of the charged leptons in the loop. 
The most dominant contributions to the $\tau \rightarrow \mu \gamma$ branching ratio correspond to the terms proportional to $M_{E_4(\widetilde{E}_4)}/m_{\tau}$ because charged vector-like leptons are heavier than $200$ GeV, thus implying that the enhancement factor $M_{E_4(\widetilde{E}_4)}/m_{\tau}$ makes those contributions much larger than the ones not involving this factor. 
However, the contributions to the $\tau \rightarrow \mu \gamma$ decay rate involving the terms having the aforementioned proportionality factor do not keep increasing 
as the vector-like fermions get heavier 
since their flavor violating coupling constants get more suppressed at the same time by the small mixing angles, defined by the ratio between Yukawa and vector-like masses.
Therefore, these compensations provide some balanced relation between the vector-like mass and the coupling of the $Z$ gauge boson with a SM charged antilepton (lepton) and heavy charged vector-like (antilepton) lepton. 
The experimental bound for the branching ratio of $\tau \rightarrow \mu \gamma$ decay is given by:
\begin{equation}
\func{BR}\left( \tau \rightarrow \mu \gamma \right)_{\func{EXP}} = 4.4 \times 10^{-8}
\end{equation}
\subsection{Analytic expression for $\tau \rightarrow \mu \mu \mu$ decay}
\begin{figure}[H]
\centering
\includegraphics[keepaspectratio,width=\textwidth]{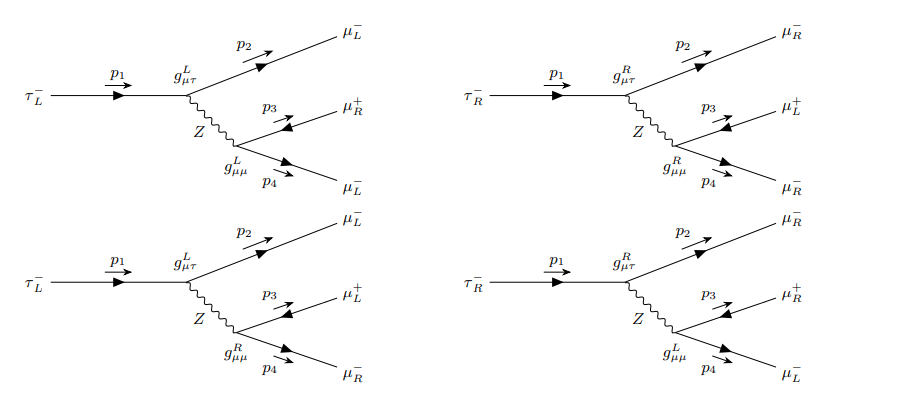}
\caption{Diagrams contributing to the charged lepton flavor violation (CLFV) $\tau \rightarrow 3\mu$ decay at tree-level. We refer to the top-left diagram as $\mathcal{M}_{LL}$ and the top-right as $\mathcal{M}_{RR}$ and similarly for the two below diagrams as $\mathcal{M}_{LR,RL}$, respectively.}
\label{fig:diagrams_tau3mu}
\end{figure}

The other interesting flavor violating decay mode is the $\tau \rightarrow \mu \mu \mu$ decay mediated by the SM $Z$ gauge boson. As the model under consideration 
 has 
 $Z$ mediated renormalizable flavor violating interactions, we can draw the Feynman diagrams for the $\tau \rightarrow \mu \mu \mu$ decay at tree-level as given in Figure~\ref{fig:diagrams_tau3mu}.
The contributions shown in Figure~\ref{fig:diagrams_tau3mu} are 
beyond Standard Model (BSM) 
effects, thus they need to be computed to set constraints on the model parameter space. 
In order to derive an analytic expression for the CLFV $\tau \rightarrow 3\mu$ decay rate mediated by the SM $Z$ gauge boson, 
 we start by writting down its definition as follows:
\begin{equation}
d\Gamma\left( \tau \rightarrow 3\mu \right) = \frac{1}{2m_{\tau}} \frac{d^3 p_2}{(2\pi)^3 2E_2} \frac{d^3 p_3}{(2\pi)^3 2E_3} \frac{d^3 p_4}{(2\pi)^3 2E_4} \lvert \mathcal{M} \rvert^2 \left( 2\pi \right)^4 \delta^{4} \left( p_1 - p_2 - p_3 - p_4 \right)
\end{equation}
Evaluating each polarized diagram in Figure~\ref{fig:diagrams_tau3mu}, it yields the following result: 
\begin{equation}
\begin{aligned}
\lvert \mathcal{M}_{LL} \rvert^2 &= \left( \frac{g_{\mu \tau}^L g_{\mu \mu}^L}{4M_Z^2} \right)^2 256\left( p_1 \cdot p_3 \right) \left( p_2 \cdot p_4 \right), \quad \lvert \mathcal{M}_{RR} \rvert^2 &= \left( \frac{g_{\mu \tau}^R g_{\mu \mu}^R}{4M_Z^2} \right)^2 256\left( p_2 \cdot p_4 \right) \left( p_1 \cdot p_3 \right) \\
\lvert \mathcal{M}_{LR} \rvert^2 &= \left( \frac{g_{\mu \tau}^L g_{\mu \mu}^R}{4M_Z^2} \right)^2 256\left( p_1 \cdot p_4 \right) \left( p_2 \cdot p_3 \right), \quad \lvert \mathcal{M}_{RL} \rvert^2 &= \left( \frac{g_{\mu \tau}^R g_{\mu \mu}^L}{4M_Z^2} \right)^2 256\left( p_1 \cdot p_4 \right) \left( p_2 \cdot p_3 \right)
\end{aligned}
\end{equation}
We are now ready to determine the squared amplitude averaged and summed over the initial and final spin states.
\begin{equation}
\begin{split}
\frac{1}{2} \sum_{\func{spin}} \lvert \mathcal{M} \rvert^2 &= \frac{1}{2} \left( \lvert \mathcal{M}_{LL} \rvert^2 + \lvert \mathcal{M}_{RR} \rvert^2 + \lvert \mathcal{M}_{LR} \rvert^2 + \lvert \mathcal{M}_{RL} \rvert^2 \right) \\
&= \frac{8}{M_Z^4} \left[ \left( g_{\mu \tau}^{2L} g_{\mu \mu}^{2L} + g_{\mu \tau}^{2R} g_{\mu \mu}^{2R} \right) \left( p_1 \cdot p_3 \right) \left( p_2 \cdot p_4 \right) + \left( g_{\mu \tau}^{2L} g_{\mu \mu}^{2R} + g_{\mu \tau}^{2R} g_{\mu \mu}^{2L} \right) \left( p_1 \cdot p_4 \right) \left( p_2 \cdot p_3 \right) \right]
\end{split}
\end{equation}
The momentum of the particles involved in the $\tau \rightarrow 3\mu$ are written in the rest frame as follows:
\begin{equation}
\begin{split}
p_1 &=  ( m_{\tau}, \vec{0} ) \\
p_2 &= \left( E_2, \vec{p}_2 \right) \\
p_3 &= \left( m_{\tau} - E_2 - E_4, -\vec{p}_2 -\vec{p}_4  \right) \\
p_4 &= \left( E_4, \vec{p}_4 \right) \\
\end{split}
\end{equation}
Then, we can carry out the inner products of momenta taking into account  
the momentum conservation ($p_1 = p_2 + p_3 + p_4$).
\begin{equation}
\begin{split}
p_1 \cdot p_3 &= m_{\tau} \left( m_{\tau} - E_2 - E_4 \right) \\
p_2 \cdot p_4 &= \frac{1}{2} \left( -m_{\tau}^2 - m_{\mu}^2 + 2m_{\tau} (E_2 + E_4) \right) \\
p_1 \cdot p_4 &= m_{\tau} E_4 \\
p_2 \cdot p_3 &= \frac{1}{2} \left( m_{\tau}^2 - m_{\mu}^2 - 2m_{\tau} E_4 \right)
\end{split}
\end{equation}
We can rewrite the squared amplitude in terms of the mass parameters after simplifying the summing over the diverse coupling constants to $g_{1,2}$, respectively ($g_1 = g_{\mu \tau}^{2L} g_{\mu \mu}^{2L} + g_{\mu \tau}^{2R} g_{\mu \mu}^{2R}, g_2 = g_{\mu \tau}^{2L} g_{\mu \mu}^{2R} + g_{\mu \tau}^{2R} g_{\mu \mu}^{2L}$).
\begin{equation}
\begin{split}
\frac{1}{2}\sum_{\func{spin}} \lvert \mathcal{M} \left( g_1, g_2, E_2, E_4 \right) \rvert^2 &= \frac{4}{M_Z^4} \Big[ g_1 \left( m_{\tau}^2 - m_{\tau} (E_2 + E_4) \right) \left( -m_{\tau}^2 - m_{\mu}^2 + 2m_{\tau} (E_2 + E_4) \right) \\
&+ g_2 \left( m_{\tau} E_4 \right) \left( m_{\tau}^2 - m_{\mu}^2 - 2m_{\tau} E_4 \right) \Big]
\end{split}
\end{equation}
Now it is time to evaluate the three body phase space integral by turning it into an effective two-body phase integral as follows (we simply drop out the prefactor $1/(2\pi)^5$ for simplicity in this derivation while keeping 
 the prefactor 
 in the computation of the aforementioned partial decay width).
\begin{equation}
\begin{split}
&\frac{d^3p_2}{2E_2} \frac{d^3p_3}{2E_3} \frac{d^3p_4}{2E_4} \delta^{4} \left( p_1 - p_2 - p_3 - p_4 \right) = \frac{d^3p_2}{2E_2} d^4p_3 \Theta \left( p_3^0 \right) \delta \left( p_3^2 \right) \frac{d^3p_4}{2E_4} \delta^{4} \left( p_1 - p_2 - p_3 - p_4 \right) \\
&= \frac{d^3p_2}{2E_2} \frac{d^3p_4}{2E_4} \Theta \left( p_1^0 - p_2^0 - p_4^0 \right) \delta \left( (p_1 - p_2 - p_4)^2 \right) \\
&=\frac{d^3p_2}{2E_2} \frac{d^3p_4}{2E_4} \Theta \left( p_1^0 - p_2^0 - p_4^0 \right) \delta \left( p_1^2 -2p_1 \cdot (p_2 + p_4) + (p_2^2 + 2p_2 \cdot p_4 + p_4^2) \right) \\
&= \frac{d^3p_2}{2E_2} \frac{d^3p_4}{2E_4} \Theta \left( p_1^0 - p_2^0 - p_4^0 \right) \delta \left( m_{\tau}^2 +2m_{\mu}^2 -2m_{\tau} (E_2 + E_4) + 2(E_2 E_4 - \lvert \vec{p}_2 \rvert \lvert \vec{p}_4 \rvert \cos\theta) \right) \\
&= \frac{d^3p_2}{2E_2} \frac{d^3p_4}{2E_4} \Theta \left( p_1^0 - p_2^0 - p_4^0 \right) \frac{1}{2\lvert \vec{p}_2 \rvert \lvert \vec{p}_4 \rvert} \delta \left( \frac{m_{\tau}^2 +2m_{\mu}^2 -2m_{\tau} (E_2 + E_4) + 2 E_2 E_4}{2\lvert \vec{p}_2 \rvert \lvert \vec{p}_4 \rvert} - \cos\theta \right)
\end{split}
\end{equation}
From the delta function, we can determine the integration range by assuming $E_2 \approx \lvert \vec{p}_2 \rvert, E_4 \approx \lvert \vec{p}_4 \rvert$. When $\cos\theta = 1$, the obtained result is
\begin{equation}
m_{\tau}^2 + 2m_{\mu}^2 - 2m_{\tau} (E_2 + E_4) = 0
\label{eqn:inside_deltafunction}
\end{equation}
From Equation~\ref{eqn:inside_deltafunction}, the integration range can be read off as follows:
\begin{equation}
\frac{m_{\mu}^2}{m_{\tau}} \leq E_2 \leq \frac{1}{2}m_{\tau}, \quad \frac{1}{2}m_{\tau} + \frac{m_{\mu}^2}{m_{\tau}} - E_2 \leq E_4 \leq \frac{1}{2}m_{\tau}
\end{equation}
It can be easily understood that once one mass parameter $E_2$ is set up by $\frac{1}{2}m_{\tau}$, the energy of the other mass parameters $E_4, E_3$ must be given by $\frac{m_{\mu}^2}{m_{\tau}}, \frac{1}{2}m_{\tau} - \frac{m_{\mu}^2}{m_{\tau}}$, respectively. Then, it remains 
to simplify the effective two body phase space integral as follows.
\begin{equation}
d^3p_2 d^3p_4 = 4\pi \lvert \vec{p}_2 \rvert^2 d\lvert \vec{p}_2 \rvert 2\pi \lvert \vec{p}_4 \rvert d\lvert \vec{p}_4 \rvert d\cos\theta, \quad \lvert \vec{p}_2 \rvert d\lvert \vec{p}_2 \rvert = E_2 dE_2, \quad \lvert \vec{p}_4 \rvert d\lvert \vec{p}_4 \rvert = E_4 dE_4
\end{equation}
Putting all pieces 
together, the decay width for the CLFV $\tau \rightarrow 3\mu$ decay at tree-level after carrying out the $\cos\theta$ integration is given by:
\begin{equation}
\Gamma\left( \tau \rightarrow 3 \mu \right) = \frac{1}{64 m_{\tau} \pi^3} \int_{m_{\mu}^2/m_{\tau}}^{\frac{1}{2}m_{\tau}} \int_{\frac{1}{2}m_{\tau}+\frac{m_{\mu}^2}{m_{\tau}}-E_2}^{\frac{1}{2}m_{\tau}} \left( \frac{1}{2}\sum_{\func{spin}} \lvert \mathcal{M} \left( g_1, g_2, E_2, E_4 \right) \rvert^2 \right) dE_4 dE_2
\end{equation}
The experimental bound for the $\tau \rightarrow 3\mu$ decay is given by:
\begin{equation}
\func{BR}\left( \tau \rightarrow 3\mu \right)_{\func{EXP}} = 2.1 \times 10^{-8}
\end{equation}
\subsection{Analytic expression for $Z \rightarrow \mu \tau$ decay}
\begin{figure}[H]
\centering
\includegraphics[keepaspectratio,width=\textwidth]{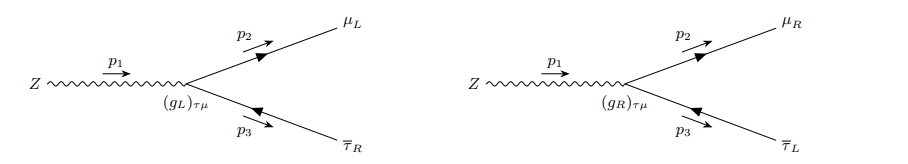}
\caption{Diagrams contributing to the charged lepton flavor violation (CLFV) $Z\tau\mu$ decay at tree-level}
\label{fig:diagrams_Zmutau}
\end{figure}

The last FCNC constraint we discuss is the $Z \rightarrow \mu \tau$ decay and diagrams contributing to the $Z \rightarrow \mu \tau$ decay are given in Figure~\ref{fig:diagrams_Zmutau}.
As in the CLFV $\tau \rightarrow 3\mu$ decay mediated by the SM $Z$ gauge boson, this CLFV $Z\rightarrow \mu\tau$ decay is also a new effect and it requires to derive its appropriate prediction from the ground. We can read off the invariant amplitude for each diagram given in Figure~\ref{fig:diagrams_Zmutau}. We refer to the left diagram as $\mathcal{M}_L$ and the right as $\mathcal{M}_R$. Then, the amplitudes are written as follows:
\begin{equation}
\begin{split}
i\mathcal{M}_L &= i (g_L)_{\tau\mu} \epsilon_\mu(p_1) \overline{u} (p_2) \gamma^\mu P_L v (p_3), \\[1ex]
i\mathcal{M}_R &= i (g_R)_{\tau\mu} \epsilon_\mu(p_1) \overline{u} (p_2) \gamma^\mu P_R v (p_3).
\label{eqn:amplitude_for_each_Zmutau}
\end{split}
\end{equation}
In order to have the squared amplitude averaged and summed, we square the amplitude given in Equation~\ref{eqn:amplitude_for_each_Zmutau} as follows:
\begin{equation}
\begin{split}
\frac{1}{3}\sum_{\func{spin}} \lvert \mathcal{M}_L \rvert^2 &= \frac{1}{3} (g_L)_{\tau\mu}^2 \left( -g_{\mu\nu} + \frac{p_{1\mu}p_{1\nu}}{M_Z^2} \right) \func{Tr} \left[ (\slashed{p}_2+m_{\mu}) \gamma^\mu P_L (\slashed{p}_3-m_{\tau}) \gamma^\nu P_L) \right] \\
&\simeq \frac{2}{3} (g_L)_{\tau\mu}^2 M_Z^2 \\
\frac{1}{3}\sum_{\func{spin}} \lvert \mathcal{M}_R \rvert^2 &= \frac{1}{3} (g_R)_{\tau\mu}^2 \left( -g_{\mu\nu} + \frac{p_{1\mu}p_{1\nu}}{M_Z^2} \right) \func{Tr} \left[ (\slashed{p}_2+m_{\mu}) \gamma^\mu P_R (\slashed{p}_3-m_{\tau}) \gamma^\nu P_R) \right] \\
&\simeq \frac{2}{3} (g_R)_{\tau\mu}^2 M_Z^2 \\
\frac{1}{3} \sum_{\func{spin}} \lvert \mathcal{M} \rvert^2 &= \frac{1}{3} \sum_{\func{spin}} \left( \lvert \mathcal{M}_L \rvert^2 + \lvert \mathcal{M}_R \rvert^2 \right) = \frac{2}{3}\left( (g_L)_{\tau\mu}^2 + (g_R)_{\tau\mu}^2 \right) M_Z^2 
\end{split}
\end{equation}
Then, the decay rate equation is given by:
\begin{equation}
\begin{split}
d\Gamma\left( Z\rightarrow\mu\tau \right) &= \frac{1}{2M_Z} \frac{d^3p_2}{(2\pi)^3 2E_2} \frac{d^3p_3}{(2\pi)^3 2E_3} \lvert \mathcal{M} \rvert^2 (2\pi)^4 \delta^{(4)} \left( p_1 - p_2 - p_3 \right) \\
\Gamma\left( Z\rightarrow\mu\tau \right) &= \frac{\lvert p^* \rvert}{32\pi^2 M_Z^2} \int \lvert \mathcal{M} \rvert^2 d\Omega \\
&= \frac{M_Z}{24\pi} \left( (g_L)_{\tau\mu}^2 + (g_R)_{\tau\mu}^2 \right)
\end{split}
\end{equation}
where $p^* \simeq M_Z/2$. Then, we are ready to write our prediction for the branching ratio of $Z\mu\tau$ decay at tree-level
\begin{equation}
\func{BR}\left(Z \rightarrow \mu\tau \right) = \frac{\Gamma\left( Z \rightarrow \mu\tau \right)}{\Gamma_{Z}} \simeq \frac{1}{2.5} \frac{M_Z}{24\pi} \left( (g_L)_{\tau\mu}^2 + (g_R)_{\tau\mu}^2 \right),
\end{equation}
where $\Gamma_{Z}$ is the total decay width of the SM $Z$ gauge boson ($\Gamma_{Z} \simeq 2.5\func{GeV}$). The experimental bound of the CLFV $Z \rightarrow \mu\tau$ decay is known as:
\begin{equation}
\func{BR}\left( Z \rightarrow \mu\tau \right)_{\func{EXP}} = 1.2 \times 10^{-5}
\end{equation}
\subsection{Numerical analysis for each prediction in the charged lepton sector}
We have discussed some 
relevant CLFV decay modes such as the $\tau \rightarrow \mu \gamma, \tau \rightarrow 3\mu$ and $Z \rightarrow \mu\tau$ from a theoretical point of view. 
By defining the renormalizable flavor violating interactions 
we showed that it is possible for the new physics to arise in a simple scenario thanks to the presence of vector-like charged leptons, which play a crucial role for these CLFV decay modes to happen. It is an encouraging feature that the mass range of the vector-like charged leptons can be constrained by the experimental bound of the CLFV decays and numerical scans for this feature will be discussed 
in detail in the following subsection.
\subsubsection{Free parameter setup}
For the numerical scan for the charged lepton sector, we first proceed to 
set up a possible mass range of the mass parameters of Equation~\ref{eqn:cl_2}.
\begin{center}
{\renewcommand{\arraystretch}{1.5} 
\begin{tabular}{cc}
\toprule
\toprule
\textbf{Mass parameter} & \textbf{Scanned Region($\func{GeV}$)} \\ 
\midrule
$y_{24}^{e} v_{d} = m_{24}$ & $\pm [1,10]$ \\
$y_{34}^{e} v_{d} = m_{34}$ & $\pm [1,10]$ \\
$y_{43}^{e} v_{d} = m_{43}$ & $\pm [1,10]$ \\
$x_{34}^{L} v_{\phi} = m_{35}$ & $\pm [50,200]$ \\
$x_{42}^{e} v_{\phi} = m_{52}$ & $\pm [50,200]$ \\
$x_{43}^{e} v_{\phi} = m_{53}$ & $\pm [50,200]$ \\
$M_{45}^{L}$ & $\pm [150,2000]$ \\
$M_{54}^{e}$ & $\pm [150,2000]$ \\
\bottomrule
\bottomrule 
\end{tabular}
\captionof{table}{Initial parameter setup for scanning the mass of the vector-like charged leptons} \label%
{tab:parameter_region_initial_scan}}
\end{center}
There are a few of features to be noticed before we start the numerical scan.
\begin{enumerate}
\item We assumed a vev for the SM up-type Higgs $H_u$ 
very close to $246\func{GeV}$, whereas the one 
of the SM down-type Higgs $H_d$ is assumed to be very small compared to the $v_u = \langle H_u \rangle$ and is ranged from $1$ to $10\func{GeV}$. The two vevs hold the relation $v_u^2 + v_d^2 = (246\func{GeV})^2$. We made that assumption since we are considering an scenario close to the decoupling limit where the neutral CP even part of $H_u$ is mostly identified with the $126$ GeV SM like Higgs boson. 
\item As the vev of the singlet flavon $\phi$ is a free parameter, we varied it in the 
range $[50,200]\func{GeV}$ whereas the mass parameters $m_{35,52,53}$ were varied in a range of values consistent 
with the observed hierarchical structure of the charged lepton masses. Furthermore,  
the vector-like masses are also other free parameters assumed to be larger or equal than 
$150\func{GeV}$ in order to successfully fulfill the 
experimental bounds on exotic charged lepton masses.
\item What we need to constrain in this numerical scan is the predicted muon and tau masses as well as the $23$ mixing angle. For the muon and tau masses, we required that the 
obtained values of the muon and tau masses to be in the range 
$[1 \pm 0.1] \times m_{\mu,\tau}$. Considering that the sizeable off-diagonal elements of the PMNS mixing matrix mainly arise from the neutrino sector, all mixing angles in the charged lepton sector are required to be  
as small as possible and 
thus we limit them to be lower than $0.2$.
\end{enumerate}
\subsubsection{Numerical scan result for the charged lepton sector}
\begin{figure}[H]
\centering
\begin{subfigure}{0.48\textwidth}
\includegraphics[keepaspectratio,width=0.95\textwidth]{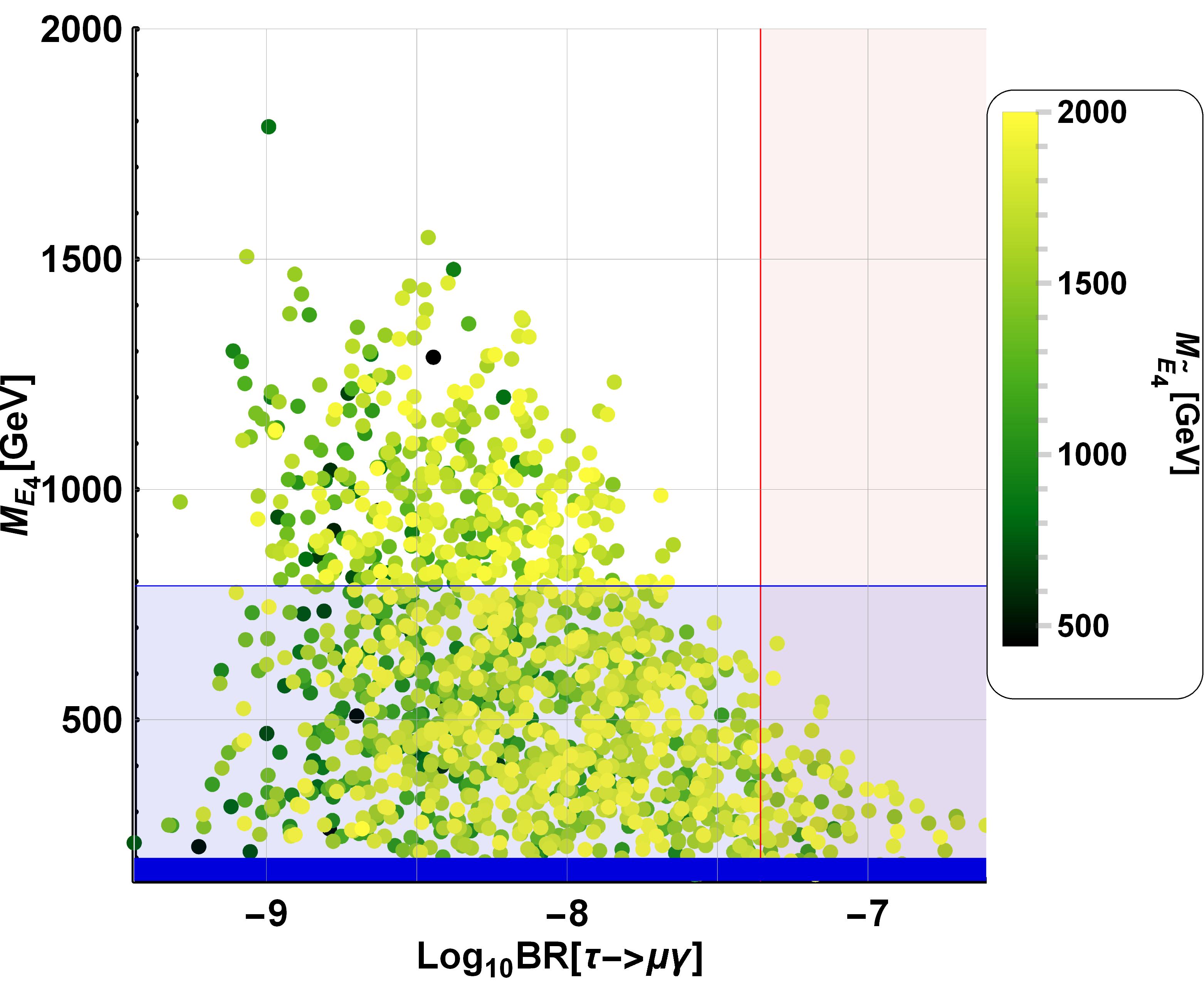}
\end{subfigure}
\begin{subfigure}{0.48\textwidth}
\includegraphics[keepaspectratio,width=0.95\textwidth]{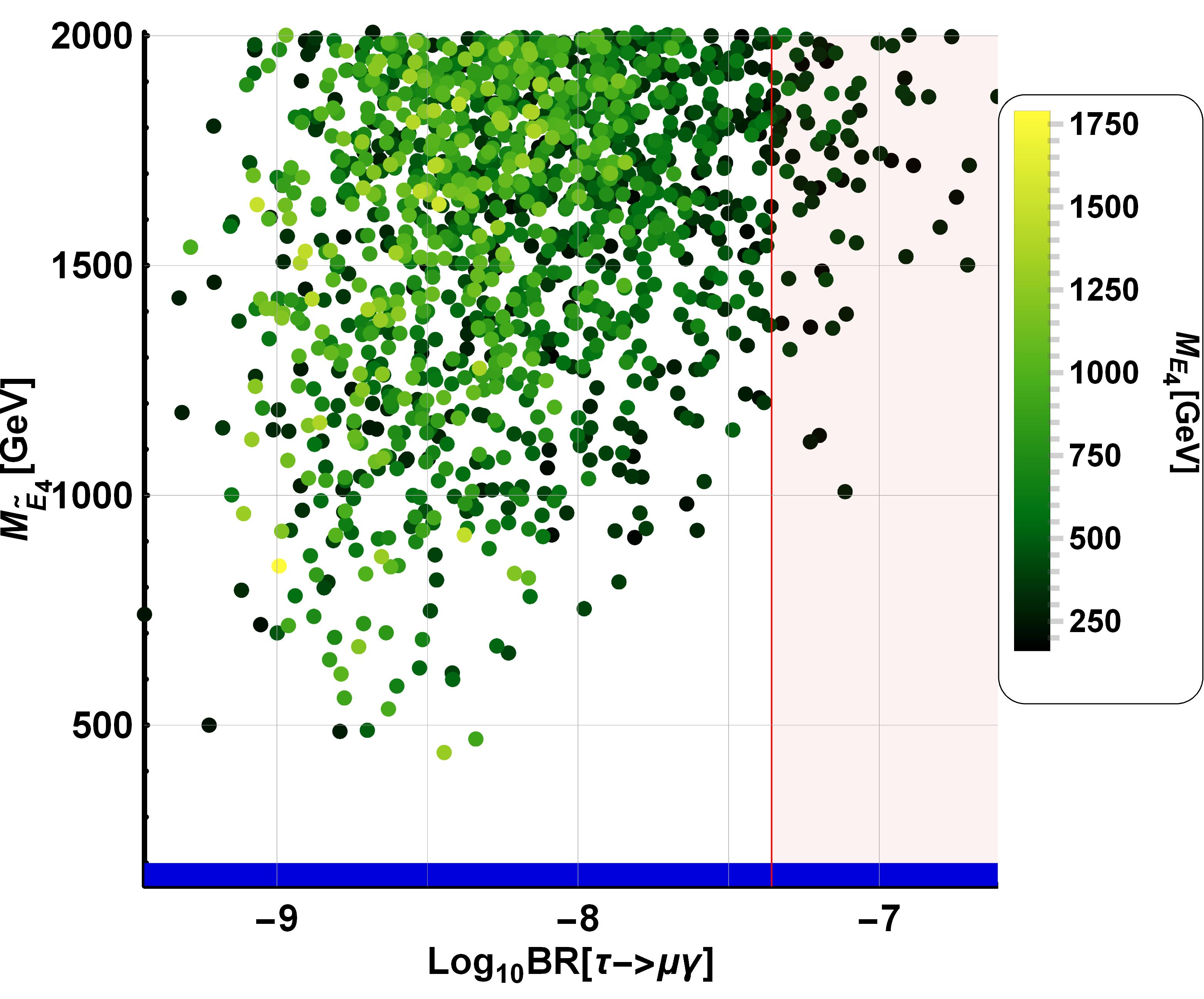}
\end{subfigure}
\begin{subfigure}{0.48\textwidth}
\includegraphics[keepaspectratio,width=0.95\textwidth]{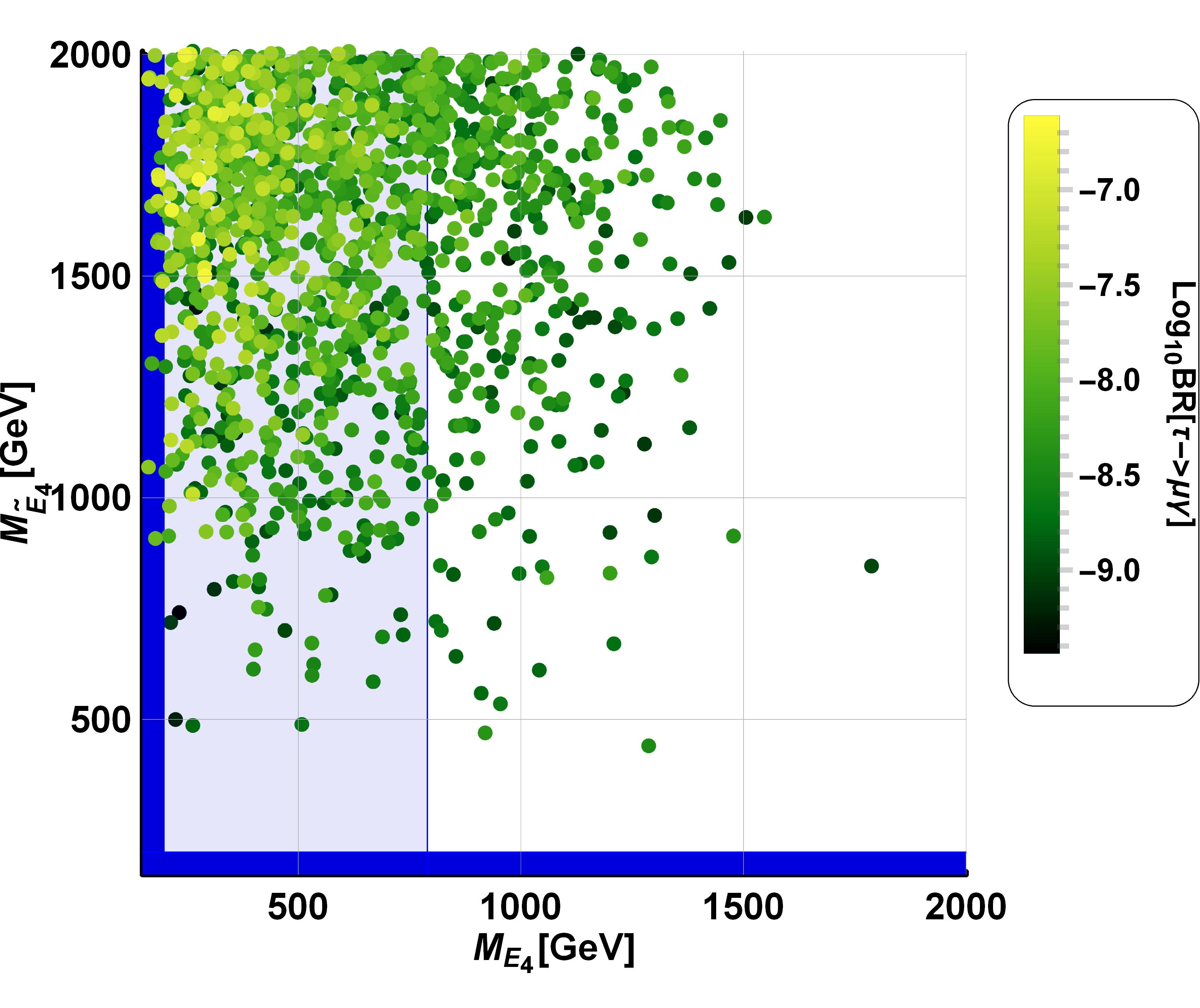}
\end{subfigure}
\begin{subfigure}{0.48\textwidth}
\includegraphics[keepaspectratio,width=0.95\textwidth]{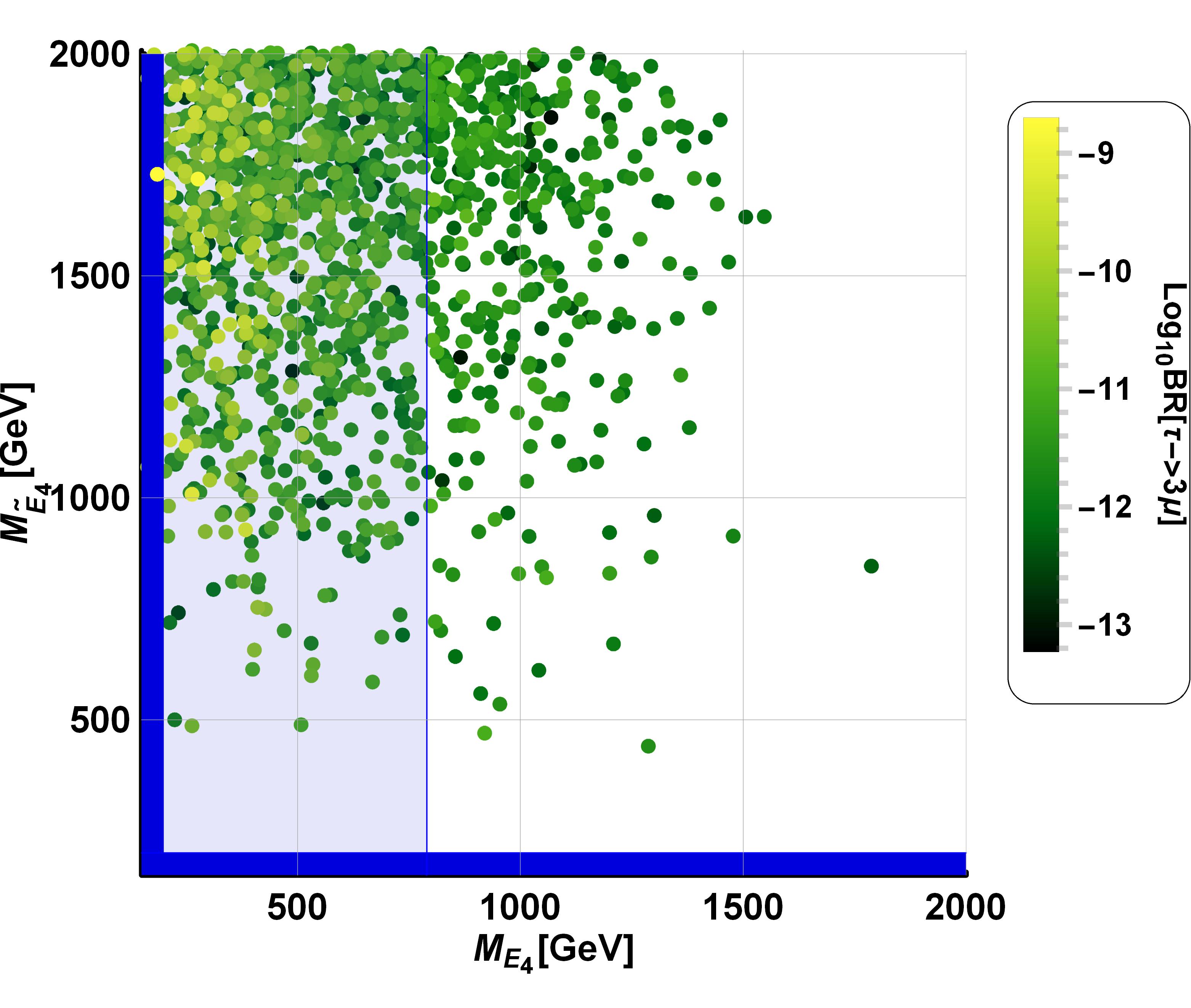}
\end{subfigure}
\begin{subfigure}{0.48\textwidth}
\includegraphics[keepaspectratio,width=0.95\textwidth]{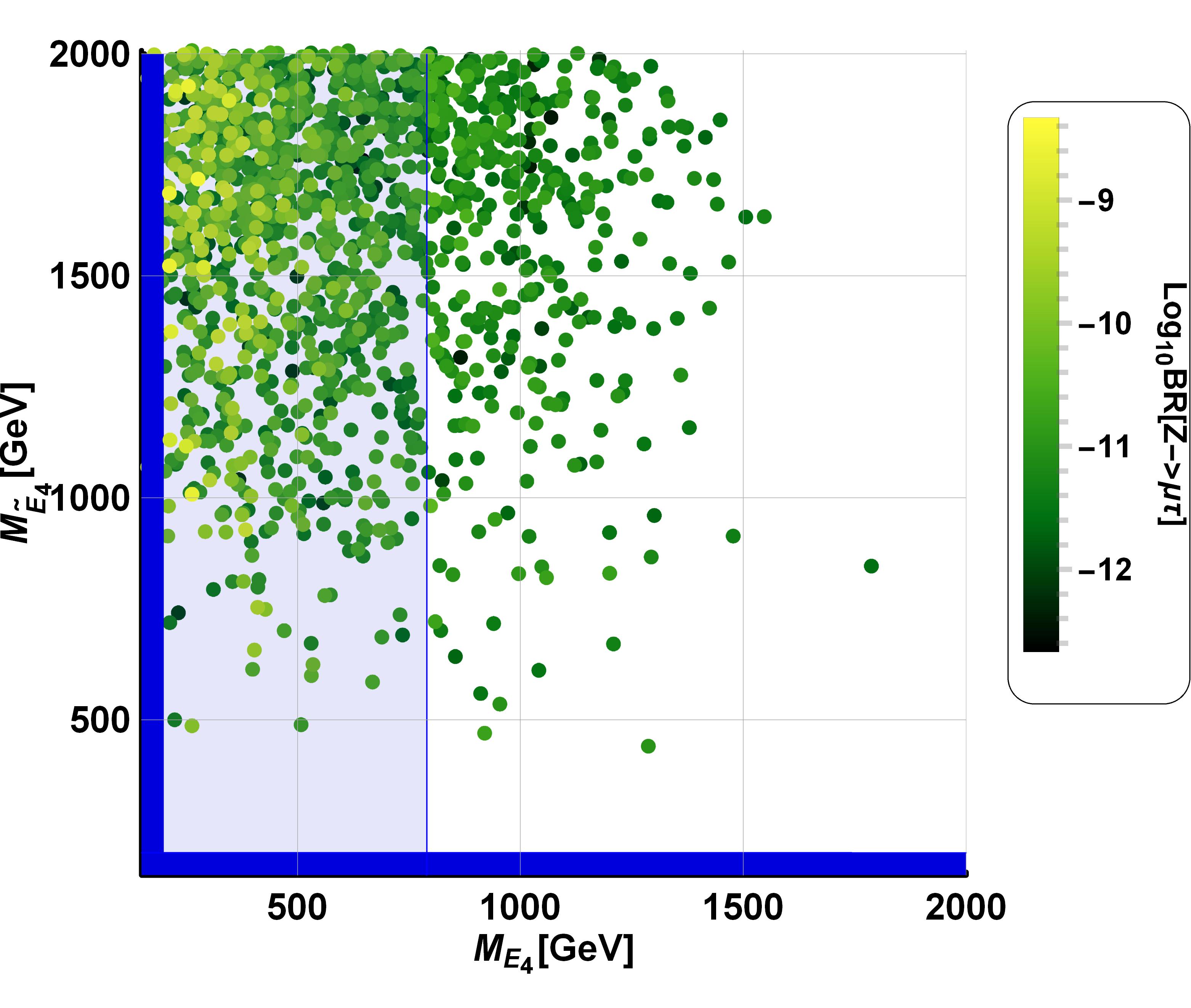}
\end{subfigure}
\caption{Scanned mass region of the vector-like charged leptons and contributions of the flavor violating interactions with the SM $Z$ gauge boson to the diverse CLFV decays $\tau \rightarrow \mu \gamma, \tau \rightarrow 3\mu$ and $Z \rightarrow \mu \tau$. The used constraints are the predicted muon and tau mass to be put between $[1 \pm 0.1] \times m_{\mu,\tau}$ and the $23$ mixing angle to be less than $0.2$. The darker blue region appearing in each diagram means either the singlet or doublet vector-like masses $M_{\widetilde{E}_4}, M_{E_4}$ are excluded up to $200\func{GeV}$ by reference~\cite{Xu:2018pnq}. The brighter blue region means the doublet vector-like mass $M_{E_4}$ is excluded up to $790\func{GeV}$ by the CMS~\cite{Bhattiprolu:2019vdu,CMS:2019hsm}. The brighter red region appearing in above two plots is the region excluded by the experimental bound for the $\func{BR}\left( \tau \rightarrow \mu \gamma \right)_{\func{EXP}} = 4.4 \times 10^{-8}$.}
\label{fig:cl_CLFV_decays}
\end{figure}

\begin{figure}[H]
\centering 
\begin{subfigure}{0.48\textwidth}
\includegraphics[keepaspectratio,width=0.95\textwidth]{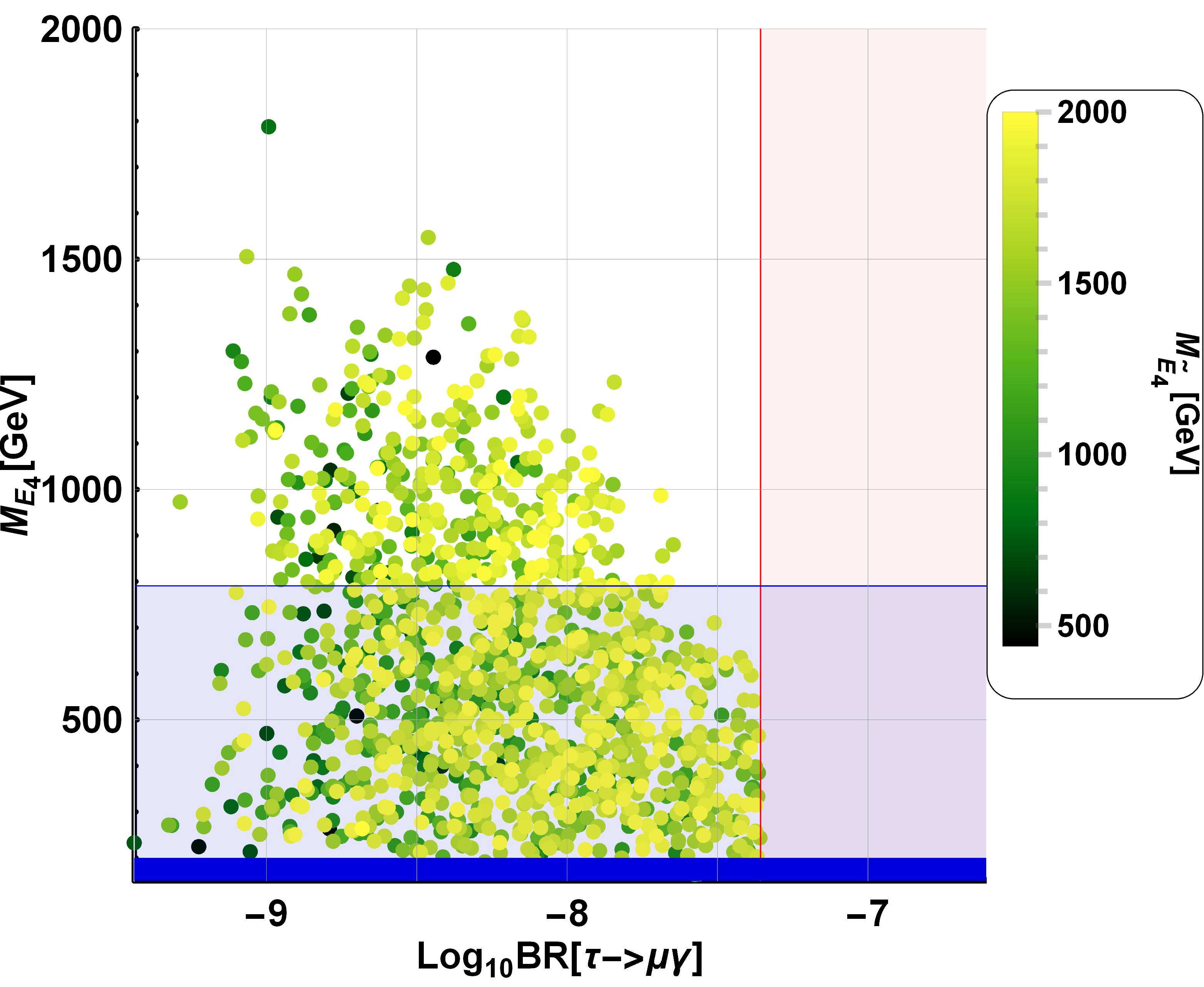}
\end{subfigure}
\begin{subfigure}{0.48\textwidth}
\includegraphics[keepaspectratio,width=0.95\textwidth]{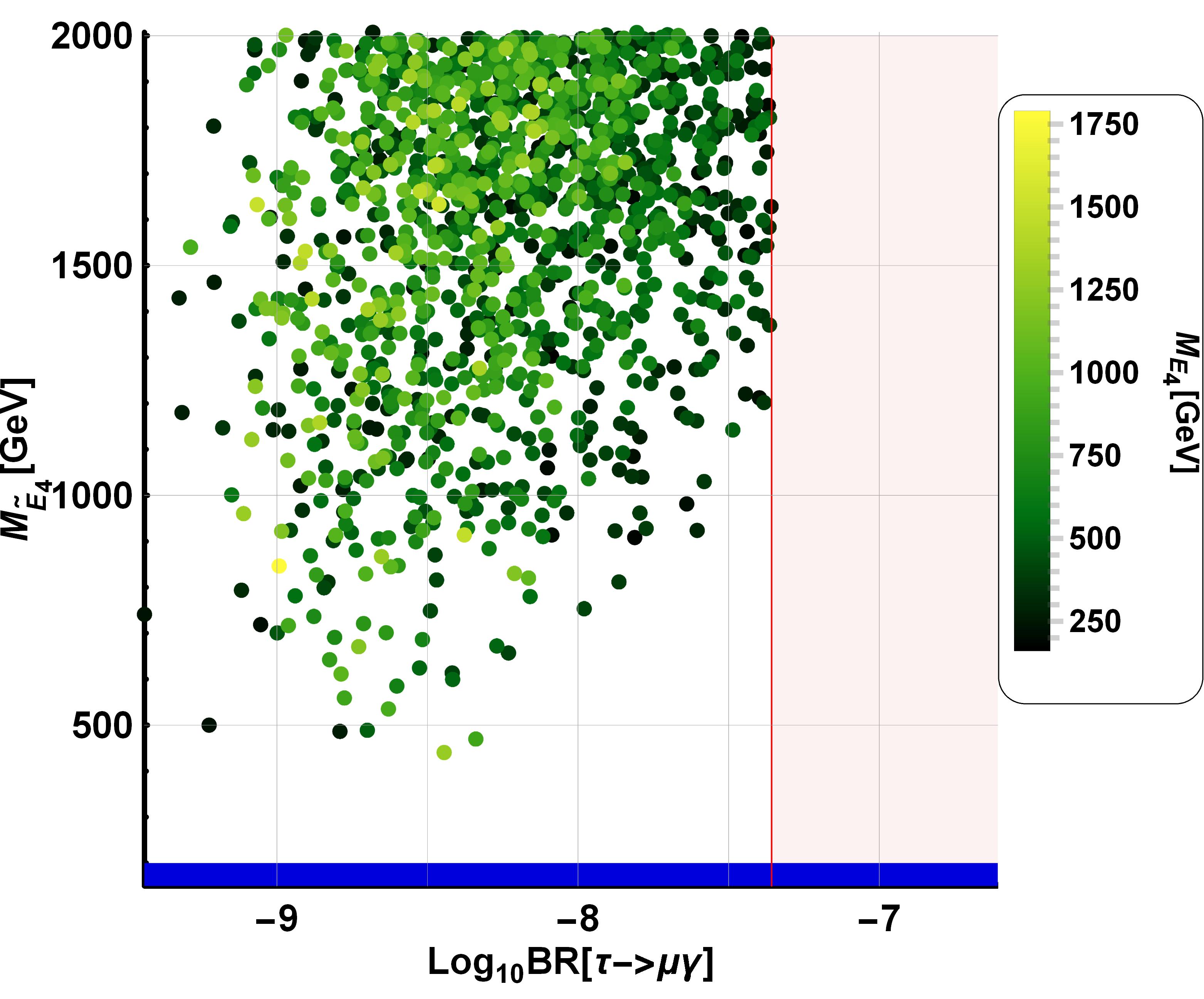}
\end{subfigure}
\begin{subfigure}{0.48\textwidth}
\includegraphics[keepaspectratio,width=0.95\textwidth]{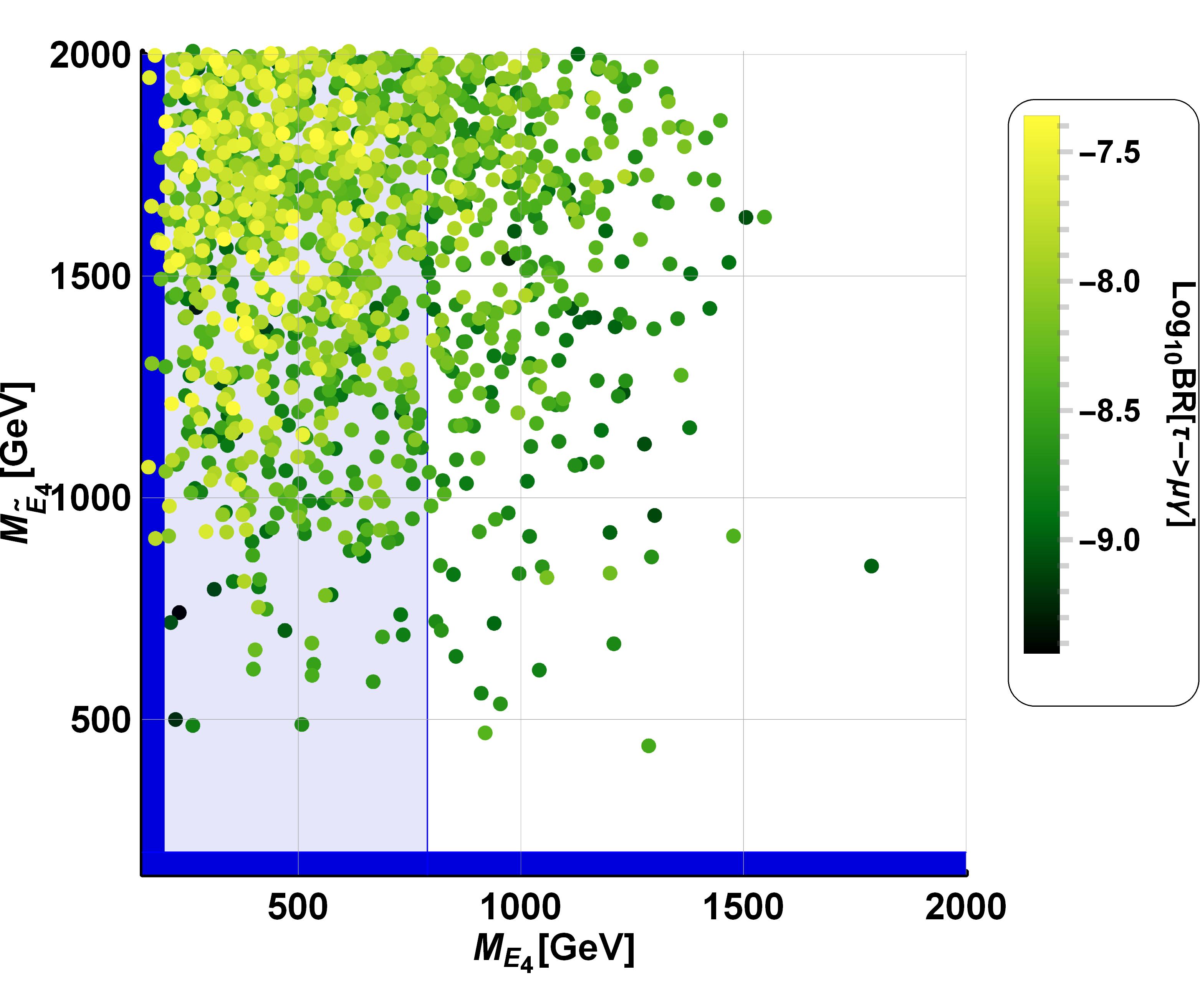}
\end{subfigure}
\begin{subfigure}{0.48\textwidth}
\includegraphics[keepaspectratio,width=0.95\textwidth]{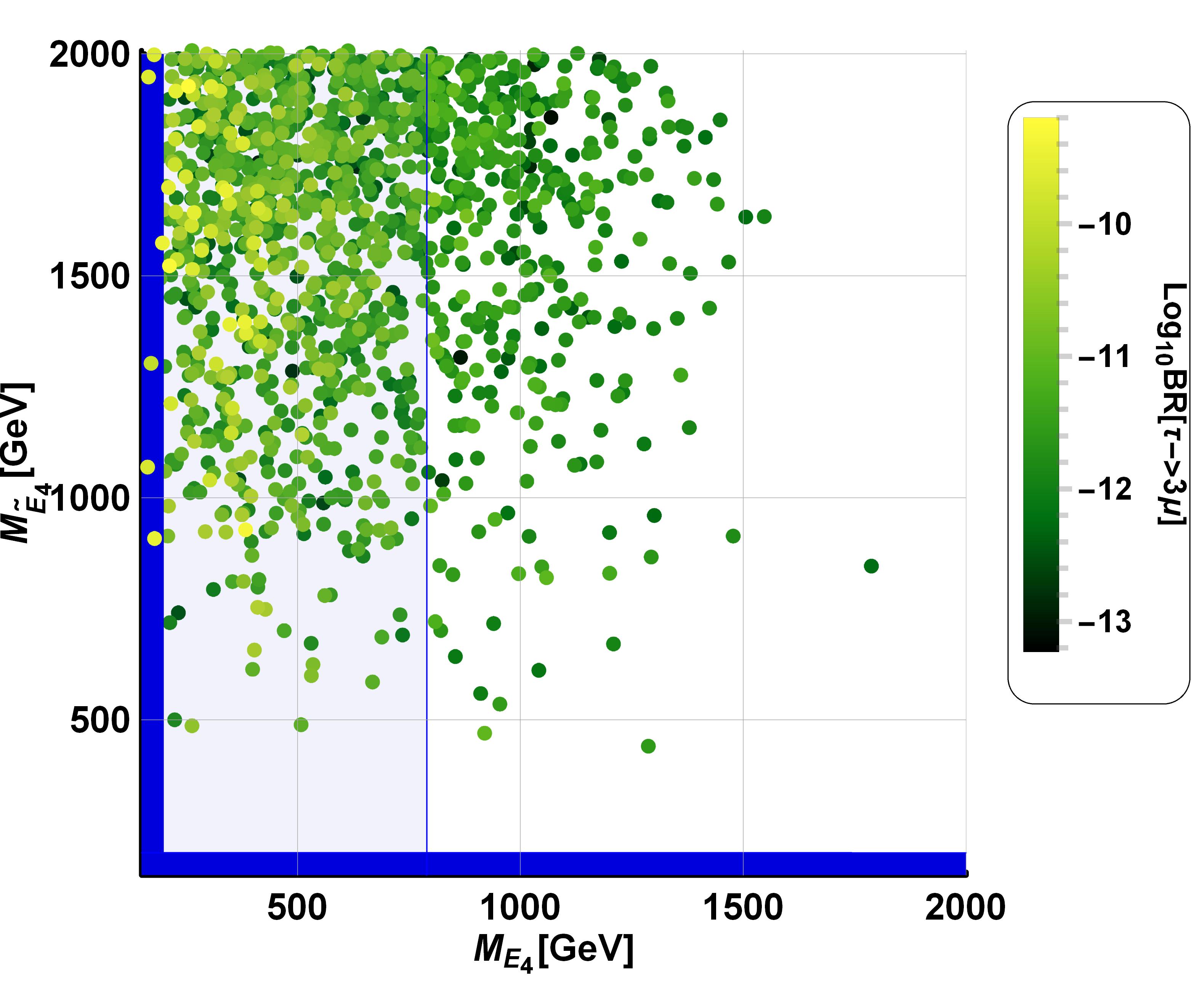}
\end{subfigure}
\begin{subfigure}{0.48\textwidth}
\includegraphics[keepaspectratio,width=0.95\textwidth]{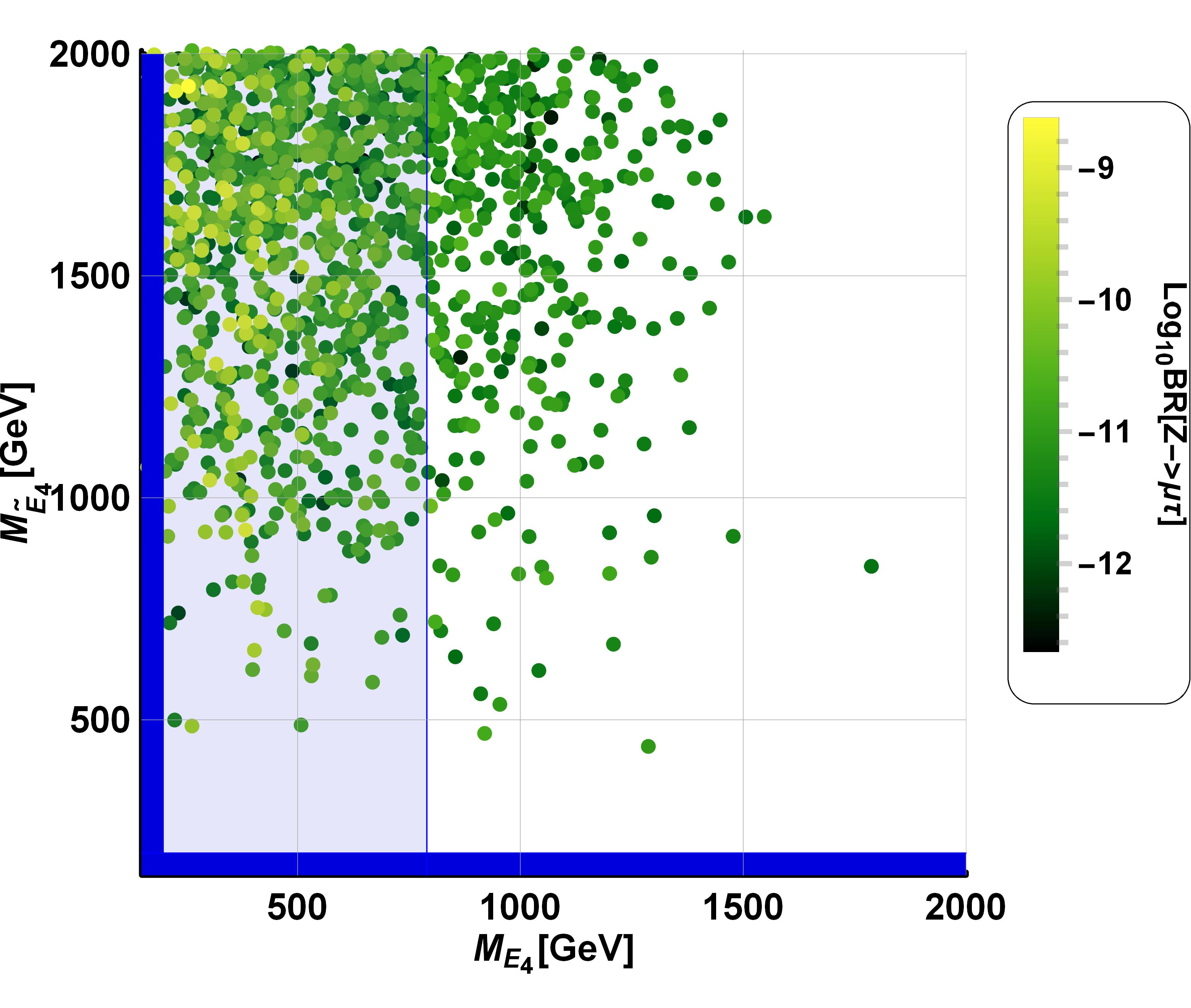}
\end{subfigure}
\caption{Reduced number of numerical predictions. The numerical predictions are constrained by the experimental value of the branching ratio of $\tau \rightarrow \mu \gamma$ decay. None of them are constrained by the branching ratio of $\tau \rightarrow 3\mu$ and $Z \rightarrow \mu \tau$ experimental bounds.}
\label{fig:cl_CLFV_second}
\end{figure}

The scanned mass range of the vector-like charged leptons are shown 
in Figure~\ref{fig:cl_CLFV_decays}.
The first we need to discuss is the experimental bounds for the vector-like charged leptons appearing in Figure~\ref{fig:cl_CLFV_decays}. The darker blue region is the excluded region for the vector-like charged leptons by $200\func{GeV}$~\cite{Xu:2018pnq}. The vector-like mass $M_{E_4}$ consists of the doublet vector-like charged leptons $E_{4L},\widetilde{E}_{4R}$, whereas the other vector-like mass $M_{\widetilde{E}_4}$ consists of the singlet vector-like charged leptons $\widetilde{E}_{4L}, E_{4R}$. Therefore, $M_{E_4}$ is the doublet vector-like mass, whereas $M_{\widetilde{E}_4}$ is the singlet vector-like mass, and the doublet vector-like mass is excluded by CMS up to $790\func{GeV}$~\cite{Bhattiprolu:2019vdu,CMS:2019hsm}, expressed by the brighter blue region of Figure~\ref{fig:cl_CLFV_decays}. The second is our predictions for the branching ratio of $\tau \rightarrow \mu \gamma$ in Figure~\ref{fig:cl_CLFV_decays}. The relevant experimental bound for each CLFV decay is given by:
\begin{equation}
\begin{split}
&\func{BR}\left( \tau \rightarrow \mu \gamma \right)_{\func{EXP}} = 4.4 \times 10^{-8} 
\\[1ex]
&-
\func{Log}_{10} \func{BR} \left( \tau \rightarrow \mu \gamma \right)_{\func{EXP}} \simeq -7.4 
\\[1ex] 
&\func{BR}\left( \tau \rightarrow 3\mu \right)_{\func{EXP}} = 2.1 \times 10^{-8} 
\\[1ex]
&- \func{Log}_{10} \func{BR} \left( \tau \rightarrow 3\mu \right)_{\func{EXP}} \simeq -7.7 
\\[1ex]
&\func{BR}\left( Z \rightarrow \mu \tau \right)_{\func{EXP}} = 1.2 \times 10^{-5} 
\\[1ex]
&- \func{Log}_{10} \func{BR} \left( Z \rightarrow \mu \tau \right)_{\func{EXP}} \simeq -4.9 
\end{split}
\end{equation}
Our predictions for the CLFV $\tau \rightarrow 3\mu$ and $Z \rightarrow \mu \tau$ decays are not excluded by the experimental bound, however those are not the case for the CLFV decay $\tau \rightarrow \mu \gamma$, 
which exceed its upper experimental bound in some parts of the parameter space. This is due to, in some parts of the parameter space, the dominant contributions to the $\tau \rightarrow \mu \gamma$ decay involving a charged exotic lepton as well as chirality flip in the internal line and proportional to $M/m_{\tau}$ because of the sizeable large value of the charged exotic lepton - SM charged lepton mass ratio. 
After removing all excluded points by the experimental bound of the $\tau \rightarrow \mu \gamma$ decay, 
we obtain Figure~\ref{fig:cl_CLFV_second}.
Looking at our numerical predictions for the branching ratio of the $\tau \rightarrow \mu \gamma$ decay shown in Figure~\ref{fig:cl_CLFV_second}, some 
of them are constrained by the experimental limit of this 
branching ratio, however most of them survive, which implies that  
our numerical predictions for the branching ratio of $\tau \rightarrow \mu \gamma$ are not significantly constrained by its experimental bound. Furthermore, 
none of the numerical predictions for the branching ratios of the $\tau \rightarrow 3\mu$ and $Z \rightarrow \mu \tau$ decays  
are constrained by its experimental bound. However, our numerical predictions for the different 
CLFV decays can significantly be constrained by the future LHC upgrades having higher center of mass energy and luminosity than the ones of the current LHC, which will allow to set tighest constraints on charged exotic vector-like masses, thus leading to stronger constraints on the model parameter space.  
Regarding the CLFV $Z \rightarrow \mu \tau$ decay, the FCC-ee experiment has planned to generate $10^{12}$ SM $Z$ gauge bosons, which will allow to probe our model since  
 the branching ratio of the $Z \rightarrow \mu \tau$ decay can reach values of the order of $10^{-10}$ in the allowed region of the parameter space. Thus, the $Z \rightarrow \mu \tau$ decay is within the reach of the FCC-ee experiment, whose  
$Z$ factory 
\cite{DeRomeri:2016gum,Calibbi:2021pyh} will be crucial to verify or rule out this model. Concluding this subsection, our numerical predictions are not significantly constrained by any of the CLFV $\tau \rightarrow \mu \gamma, \tau \rightarrow 3\mu$ and $Z \rightarrow \mu \tau$ decays and might be able to be seen from the $Z$ factory for the first time, predicting the doublet vector-like charged lepton mass which is ranged from $790$ to $1600\func{GeV}$ whereas the singlet vector-like charged lepton mass is ranged from $500$ to $2000\func{GeV}$ or above than that.
\section{Quark sector phenomenology}
\label{sec:VI}
\begin{figure}[H]
\centering
\includegraphics[keepaspectratio,width=\textwidth]{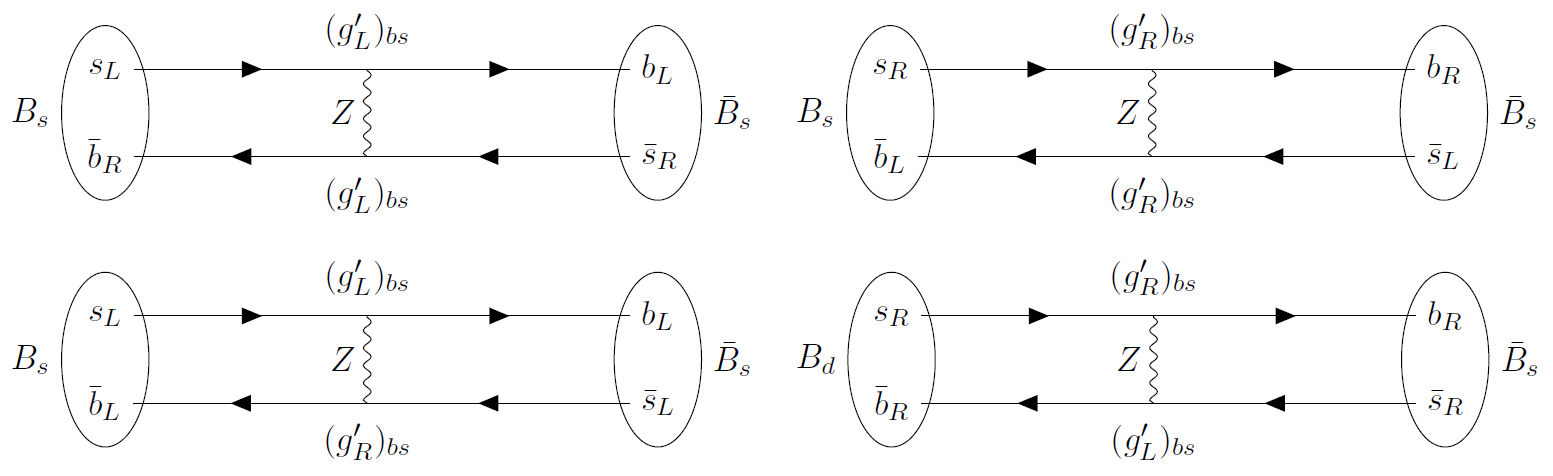}
\caption{Feynman diagrams contributing to the 
the $B_s-\bar{B}_s$ meson mixing involving the 
 tree-level exchange of the $Z$ gauge boson in the polarized basis}
\label{fig:BsBbars_mixing} 
\end{figure}

We have discussed the up- and down-type quark mass matrices pointed out that they have a different form, since the quark doublet rotation used in the up-type quark sector can not remove the down-type Yukawa term. This difference between up- and down-type quark mass matrices cause a distinct feature for each sector as follows:
\begin{itemize}
\item The up-type quark mass matrix can reach to the $23$ left (right)-handed mixing.
\item The down-type quark mass matrix can access to all left-handed mixings among the three SM generations, whereas the right-handed mixing can only have the $23$ mixing.
\end{itemize}
The interesting feature of the down-type quark mass matrix allows for flavor changing $Z$ interactions with down type quarks which yield 
neutral meson oscillations such as $K, B_d$ and $B_s$. Furthermore, an important feature to be mentioned is that the 
first generation of SM charged fermions do not acquire masses.  
Due to this property, our predictions for the neutral meson oscillations including the $d$ quark give a very suppressed energy difference corresponding to $10^{-40} \func{GeV}$, which is impossible to reach with the current experimental sensitivity. Then, the rest of the neutral meson oscillation $B_s$ is possible and an encouraging feature of the $B_s$ meson oscillation in our proposed model 
is the $B_s$ meson oscillation mediated by the SM $Z$ gauge boson can be calculated at tree-level as given in Figure~\ref{fig:BsBbars_mixing}.
From our numerical analysis we have found that the BSM contributions to the  
$B_s$ meson oscillation arising from 
the tree-level exchange of the  
$Z$ gauge boson yield the meson mass splitting of the order of 
$10^{-15}$ GeV or even less than that value, which is quite negligible compared to its corresponding   
experimental bound of $10^{-11} \func{GeV}$. The very suppressed  
new physics effect for the $B_s$ meson oscillation can be explained by considering   
the flavor violating coupling constants at each vertex of each diagram, whose value is about $10^{-6,-7}$ and this values are determined by 
the ratio between Yukawa and vector-like masses. For this reason, in the study of the phenomenological implications of our model in the flavor changing neutral interactions in the quark sector, we do not consider 
 the neutral meson oscillations as well as the $B_s \rightarrow \mu^+\mu^-$ decay. It is worth mentioning that the $B_s \rightarrow \mu^+\mu^-$ decay gives weaker effects than the neutral meson oscillations. Considering these facts, we conclude that the rare $t \rightarrow c Z$ decay and the CKM mixing matrix can constrain the quark sector of our  
 model, and thus  
 we discuss these two phenomenological aspects in the following subsections. 
\subsection{Analytic expression for the $t \rightarrow c Z$ decay}
\begin{figure}[H]
\centering
\includegraphics[keepaspectratio,width=\textwidth]{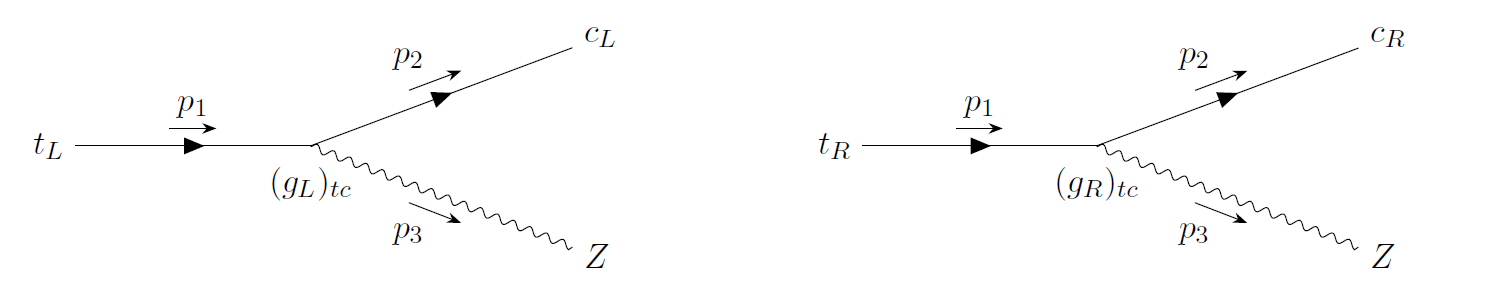}
\caption{tree-level Feynman diagrams contributing to the rare $t \rightarrow c Z$ decay}
\label{fig:tcZ}
\end{figure}

The $t \rightarrow c Z$ decay, which only appears at one-loop level in the SM,  
can take place at tree-level in our proposed model, thanks to the $Z$ mediated flavor changing neutral current interactions in the quark sector. In our proposed model, the $t \rightarrow cZ$ receives tree-level contributions which are depicted in 
Figure~\ref{fig:tcZ}.
Denoting the invariant amplitudes for the Feynman diagrams of the left and right panels of Figure~\ref{fig:tcZ} as $\mathcal{M}_L$ and $\mathcal{M}_R$, respectively, we find that they can be written as: 
\begin{equation}
\begin{split}
i\mathcal{M}_L &= i (g_L)_{tc} \epsilon_\mu^* (p_3) \overline{u} (p_2) \gamma^\mu P_L u (p_1) \\
i\mathcal{M}_R &= i (g_R)_{tc} \epsilon_\mu^* (p_3) \overline{u} (p_2) \gamma^\mu P_R u (p_1)
\label{eqn:amplitude_for_each_tcZ}
\end{split}
\end{equation}
In order to have the squared amplitude averaged and summed, we square the amplitudes given in Equation~\ref{eqn:amplitude_for_each_tcZ} and then sum over the different spin states, as follows:
\begin{equation}
\begin{gathered}
\frac{1}{2}\sum_{\func{spin}} \lvert \mathcal{M}_L \rvert^2 = (g_L)_{ct}^2 \left( (p_2 \cdot p_1) + \frac{2}{M_Z^2} (p_2 \cdot p_3)(p_1 \cdot p_3) \right) \\
\frac{1}{2}\sum_{\func{spin}} \lvert \mathcal{M}_R \rvert^2 = (g_R)_{ct}^2 \left( (p_2 \cdot p_1) + \frac{2}{M_Z^2} (p_2 \cdot p_3)(p_1 \cdot p_3) \right) \\
\frac{1}{2} \sum_{\func{spin}} \lvert \mathcal{M} \rvert^2 = \frac{1}{2} \sum_{\func{spin}} \left( \lvert \mathcal{M}_L \rvert^2 + \lvert \mathcal{M}_R \rvert^2 \right) = \frac{1}{2}\left( (g_L)_{ct}^2 + (g_R)_{ct}^2 \right) \left( (p_2 \cdot p_1) + \frac{2}{M_Z^2} (p_2 \cdot p_3)(p_1 \cdot p_3) \right) 
\end{gathered}
\end{equation}
Then, the decay rate equation is given by:
\begin{equation}
\begin{split}
d\Gamma\left( t \rightarrow c Z \right) &= \frac{1}{2m_t} \frac{d^3p_2}{(2\pi)^3 2E_2} \frac{d^3p_3}{(2\pi)^3 2E_3} \lvert \mathcal{M} \rvert^2 (2\pi)^4 \delta^{(4)} \left( p_1 - p_2 - p_3 \right) \\
\Gamma\left( t \rightarrow c Z \right) &= \frac{\lvert p^* \rvert}{32\pi^2 m_t^2} \int \lvert \mathcal{M} \rvert^2 d\Omega \\
&= \frac{1}{8\pi m_t^2} \frac{1}{2m_t}(m_t^2 - M_Z^2) ( (g_L)_{ct}^2 + (g_R)_{ct}^2 ) \\
&\times \left[ \frac{m_t^2 + m_c^2 - M_Z^2}{2} + \frac{2}{M_Z^2} (\frac{m_t^2 - m_c^2 - M_Z^2}{2}) (\frac{m_t^2 - m_c^2 + M_Z^2}{2}) \right]
\end{split}
\end{equation}
where $p^* \simeq \frac{1}{2m_t} \left( m_t^2 - M_Z^2 \right)$.
Then, we are ready to write down our prediction for the branching ratio of the tree-level $t \rightarrow c Z$ decay: 
\begin{equation}
\func{BR}\left(t \rightarrow c Z \right) = \frac{\Gamma\left( t \rightarrow c Z \right)}{\Gamma_{t}} < \func{BR} \left( t \rightarrow c Z \right)_{\func{EXP}} = 2.4 \times 10^{-4} \quad (95\% \func{CL})
\label{eqn:BRtcZ_exp}
\end{equation}
where $\Gamma_t = 1.32 \func{GeV}$.
\subsection{Analytic expression for the CKM mixing matrix}
In order to discuss the CKM mixing matrix, the first task we need to investigate is the $W$ current of the SM in order to see how the CKM mixing matrix can take place (we only consider the three SM generations at the moment).
\begin{equation}
\begin{split}
\mathcal{L}_{\func{SM}}^{W} &= g j_{\mu}^{W+} W^{\mu +} = \frac{g}{\sqrt{2}} \left( \overline{u}_{L}^{i} \gamma_{\mu} d_{L}^{i} \right) W^{\mu +} 
\\
&= \frac{g}{\sqrt{2}} \left( \overline{u}_{L}^{i} (V_{L}^{u\dagger} V_{L}^{u}) \gamma_{\mu} (V_{L}^{d\dagger} V_{L}^{d}) d_{L}^{i} \right) W^{\mu +} = \frac{g}{\sqrt{2}} \left( \overline{u}_{L}^{i \prime} \gamma_{\mu} (V_{\func{CKM}}) d_{L}^{i \prime} \right) W^{\mu +}  
\end{split}
\end{equation}
where $u_{L}^{i \prime}, d_{L}^{i \prime}$ are the up- and down-type quarks of the SM in the mass basis and $V_{\func{CKM}}$ is the CKM mixing matrix defined as $V_{L}^{u} V_{L}^{d \dagger}$. Now we extend the quark spectrum by considering the vector-like quarks, thus implying that 
the $W$ current takes the form:
\begin{equation}
\begin{split}
\mathcal{L}^{W} &= 
\frac{g}{\sqrt{2}}
\begin{pmatrix}
\overline{u}_{1L} & \overline{u}_{2L} & \overline{u}_{3L} & \overline{u}_{4L} & \overline{\widetilde{u}}_{4L}
\end{pmatrix}
\gamma_{\mu}
\begin{pmatrix}
1 & 0 & 0 & 0 & 0 \\
0 & 1 & 0 & 0 & 0 \\
0 & 0 & 1 & 0 & 0 \\
0 & 0 & 0 & 1 & 0 \\
0 & 0 & 0 & 0 & 0 
\end{pmatrix}
\begin{pmatrix}
d_{1L} \\
d_{2L} \\
d_{3L} \\
d_{4L} \\
\widetilde{d}_{4L}
\end{pmatrix}
W^{\mu +}
\\
&= 
\frac{g}{\sqrt{2}}
\begin{pmatrix}
\overline{u}_{1L} & \overline{u}_{2L} & \overline{u}_{3L} & \overline{u}_{4L} & \overline{\widetilde{u}}_{4L}
\end{pmatrix}
V_{L}^{u \dagger} V_{L}^{u}
\gamma_{\mu}
\begin{pmatrix}
1 & 0 & 0 & 0 & 0 \\
0 & 1 & 0 & 0 & 0 \\
0 & 0 & 1 & 0 & 0 \\
0 & 0 & 0 & 1 & 0 \\
0 & 0 & 0 & 0 & 0 
\end{pmatrix}
V_{L}^{d \dagger} V_{L}^{d}
\begin{pmatrix}
d_{1L} \\
d_{2L} \\
d_{3L} \\
d_{4L} \\
\widetilde{d}_{4L}
\end{pmatrix}
W^{\mu +}
\\
&= 
\frac{g}{\sqrt{2}}
\begin{pmatrix}
\overline{u}_{L} & \overline{c}_{L} & \overline{t}_{L} & \overline{U}_{4L} & \overline{\widetilde{U}}_{4L}
\end{pmatrix}
\gamma_{\mu}
V_{L}^{u}
\begin{pmatrix}
1 & 0 & 0 & 0 & 0 \\
0 & 1 & 0 & 0 & 0 \\
0 & 0 & 1 & 0 & 0 \\
0 & 0 & 0 & 1 & 0 \\
0 & 0 & 0 & 0 & 0 
\end{pmatrix}
V_{L}^{d \dagger}
\begin{pmatrix}
d_{L} \\
s_{L} \\
b_{L} \\
D_{4L} \\
\widetilde{D}_{4L}
\end{pmatrix}
W^{\mu +}
\end{split}
\label{eqn:CKM_analytic}
\end{equation}
where the CKM mixing prediction in our model is given by
\begin{equation}
V_{\func{CKM}} = 
V_{L}^{u}
\begin{pmatrix}
1 & 0 & 0 & 0 & 0 \\
0 & 1 & 0 & 0 & 0 \\
0 & 0 & 1 & 0 & 0 \\
0 & 0 & 0 & 1 & 0 \\
0 & 0 & 0 & 0 & 0 
\end{pmatrix}
V_{L}^{d \dagger}, \quad \text{for the upper-left $3 \times 3$ block}
\label{eqn:CKM_prediction}
\end{equation}
where $V_{L}^{u}$ is the mixing matrix for the up-type quarks defined in Equation~\ref{eqn:up_mixing} and $V_{L}^{d}$ is the one corresponding to  
the down-type quarks given in Equation~\ref{eqn:down_mixing}. The zero appearing in the middle matrix between $V_{L}^{u}$ and $V_{L}^{d \dagger}$ arises from the left-handed vector-like quark singlets $\widetilde{U}_{4L}$ and $\widetilde{D}_{4L}$ which do not interact with the $W$ currents, so our prediction for the CKM mixing matrix does not feature the unitarity requirement and this leads to the need of relaxing the unitarity constraint of the CKM quark mixing matrix. That deviation of unitarity of the CKM quark mixing matrix is due to the presence of heavy vector-like quarks and this aspect was studied in 
\cite{Branco:2021vhs} in the context of theory with different particle spectrum and symmetry than ours. 
Few sigma of SM deviations from the first row of the CKM mixing matrix without unitarity were analyzed in \cite{Branco:2021vhs}. 
The deviation from the Unitarity of the CKM will also be discussed in our numerical result and the experimental CKM mixing matrix without unitarity is given by~\cite{Branco:2021vhs,ParticleDataGroup:2020ssz}:
\begingroup
\setlength\arraycolsep{5pt}
\begin{equation}
\lvert K_{\func{CKM}} \rvert
=
\begin{pmatrix}
0.97370 \pm 0.00014 & 0.22450 \pm 0.00080 & 0.00382 \pm 0.00024 \\[1.5ex]
0.22100 \pm 0.00400 & 0.98700 \pm 0.01100 & 0.04100 \pm 0.00140 \\[1.5ex]
0.00800 \pm 0.00030 & 0.03880 \pm 0.00110 & 1.01300 \pm 0.03000 
\end{pmatrix}
\end{equation}
\endgroup
\subsection{Numerical analysis for each prediction in the quark sector}
When compared to the charged lepton sector simulation, the quark sector becomes much more complicated  
since we need 
to fit the masses of the $c,t,s$ and $b$ quarks simultaneously as well as the CKM mixing matrix without imposing the unitarity requirement. Therefore, we fit the masses of the four quarks first by using a fitting function $\chi_{\func{mass}}^{2}$ and then we start a second fitting procedure by using another fitting function $\chi_{\func{CKM}}^{2}$ and this will be discussed in detail  
in the following subsections.
\subsubsection{The fitting function $\chi^2$ and free parameter setup}
We set up our parameter region as follows:
\begin{center}
{\renewcommand{\arraystretch}{1.5} 
\begin{tabular}{cc}
\toprule
\toprule
\textbf{Mass parameter} & \textbf{Scanned Region($\func{GeV}$)} \\ 
\midrule
$y_{24}^{u} v_{u} = m_{24}^{u}$ & $\pm [10,50]$ \\[0.5ex]
$y_{34}^{u} v_{u} = m_{34}^{u}$ & $\pm [200,400]$ \\[0.5ex]
$y_{43}^{u} v_{u} = m_{43}^{u}$ & $\pm [200,400]$ \\[0.5ex]
$x_{34}^{Q} v_{\phi} = m_{35}^{Q}$ & $m_{35}^{Q}$ \\[0.5ex]
$x_{42}^{u} v_{\phi} = m_{52}^{u}$ & $\pm [500,700]$ \\[0.5ex]
$x_{43}^{u} v_{\phi} = m_{53}^{u}$ & $\pm [50,500]$ \\[0.5ex]
$M_{45}^{Q}$ & $M_{45}^{Q}$ \\[0.5ex]
$M_{54}^{u}$ & $\pm [1000,3000]$ \\[0.5ex]
\midrule
$y_{14}^{d} v_{d} = m_{14}^{d}$ & $\pm [1,10]$ \\[0.5ex]
$y_{24}^{d} v_{d} = m_{24}^{d}$ & $\pm [5,20]$ \\[0.5ex]
$y_{34}^{d} v_{d} = m_{34}^{d}$ & $\pm [10,30]$ \\[0.5ex]
$y_{43}^{d} v_{d} = m_{43}^{d}$ & $\pm [5,10]$ \\[0.5ex]
$x_{34}^{Q} v_{\phi} = m_{35}^{Q}$ & $\pm [10,100]$ \\[0.5ex]
$x_{42}^{d} v_{\phi} = m_{52}^{d}$ & $\pm [10,100]$ \\[0.5ex]
$x_{43}^{d} v_{\phi} = m_{53}^{d}$ & $\pm [10,100]$ \\[0.5ex]
$M_{45}^{Q}$ & $\pm [1000,3000]$ \\[0.5ex]
$M_{54}^{d}$ & $\pm [1000,3000]$ \\[0.5ex]
\bottomrule
\bottomrule 
\end{tabular}
\captionof{table}{Initial parameter setup for scanning mass of the vector-like quarks} 
\label{tab:parameter_region_initial_quarks}}
\end{center}
There are a few things to be noticed as in the charged lepton case.
\begin{enumerate}
\item The relation $v_{u}^{2} + v_{d}^{2} = (246\func{GeV})^2$ still holds and the mass parameters $m_{24,34,43}^{u}$ can not exceed the upper perturbative limit on 
the Yukawa constant $\sqrt{4\pi} \simeq 3.54$ multiplied by the vev 
$\approx 240\func{GeV}$,  
of the $H_{u}$ Higgs, thus yielding the bound of $850\func{GeV}$ for these mass parameters. These restrictions have been taken into account through the whole fitting process.
\item The down-type Higgs $H_{d}$ has a very small vev, which is about order of $10\func{GeV}$, based on our previous analysis 
~\cite{Hernandez:2021tii}, and the range of values of the mass parameters $m_{14,24,34,43}^{d}$ are considered under this assumption.
\item Since we do not know the correct scale of $v_{\phi}$, we considered  
$m_{35}^{Q}, m_{52,53}^{u}$ and $m_{52,53}^{d}$ as free parameters. For the same reason, the vector-like masses $M_{45}^{Q}$ and $M_{54}^{u,d}$ are considered free parameters as well.
\item The mass parameters $m_{35}^{Q}$ and $M_{45}^{Q}$ appear in a  
common term shared by both up- and down-type quark sector mass matrices and this feature has been discussed in the paragraph below  
Equation~\ref{eqn:diff_up_down}.
\end{enumerate}
The next thing to do is to set up the two fitting functions $\chi_{\func{mass}}^{2}$ and $\chi_{\func{CKM}}^{2}$ as follows:
\begin{equation}
\chi_{\func{mass}}^2 = \sum_{f=c,t,s,b} = \frac{(m_{f}^{\func{pred}}-m_{f}^{\func{EXP}})^2}{(\delta m_{f}^{\func{EXP}})^2}, \quad \chi_{\func{CKM}}^2 = \sum_{i,j=1,2,3} \frac{((V_{\func{CKM}}^{\func{pred}})_{ij} - (V_{\func{CKM}}^{\func{EXP}})_{ij})^2}{((\delta V_{\func{CKM}}^{\func{EXP}})_{ij})^2},
\end{equation}
where the superscript $\func{pred}$ means our prediction to its experimental value and the delta means error bar of the physical quantity at $1\sigma$. Our first goal is to fit the masses of the four quarks simultaneously. For the charged lepton case, we require that our obtained muon and tau masses to be in the range 
$[1 \pm 0.1] \times m_{\mu,\tau}$ and this requirement is also imposed for  
the $c,s$ and $b$ quarks excepting for the $t$ quark since the $t$ quark is too heavy. Besides that, we require that the obtained top quark mass 
to be in the range 
 $[1 \pm 0.01] \times m_{t}$ instead of $[1 \pm 0.1] \times m_{t}$. After the mass parameters have been converged to be put between the arranged range of each quark mass, 
 we use the other fitting function $\chi_{\func{CKM}}^{2}$ to fit our prediction for the CKM mixing matrix. Using one of the defined fitting functions, we need to vary the mass parameters of Table~\ref{tab:parameter_region_initial_quarks} by a factor of $\left[ 1 \pm \kappa \right]$ where $\kappa = 0.1$ in order to find better mass parameters. We rename the given parameters from the initial parameter setup by adding an subscript $r$ to the mass parameters. The varied mass parameters are given in Table~\ref{tab:parameter_region_initial_q_second}.
\begin{center}
{\renewcommand{\arraystretch}{1.5} 
\begin{tabular}{cc}
\toprule
\toprule
\textbf{Mass parameter} & \textbf{Scanned Region($\func{GeV}$)} \\ 
\midrule
$y_{24}^{u} v_{u} = m_{24}^{u}$ & $\left[ 1 \pm \kappa \right] \times m_{24r}^{u}$ \\[0.5ex]
$y_{34}^{u} v_{u} = m_{34}^{u}$ & $\left[ 1 \pm \kappa \right] \times m_{34r}^{u}$ \\[0.5ex]
$y_{43}^{u} v_{u} = m_{43}^{u}$ & $\left[ 1 \pm \kappa \right] \times m_{43r}^{u}$ \\[0.5ex]
$x_{34}^{Q} v_{\phi} = m_{35}^{Q}$ & $m_{35}^{Q}$ \\[0.5ex]
$x_{42}^{u} v_{\phi} = m_{52}^{u}$ & $\left[ 1 \pm \kappa \right] \times m_{42r}^{u}$ \\[0.5ex]
$x_{43}^{u} v_{\phi} = m_{53}^{u}$ & $\left[ 1 \pm \kappa \right] \times m_{43r}^{u}$ \\[0.5ex]
$M_{45}^{Q}$ & $M_{45}^{Q}$ \\[0.5ex]
$M_{54}^{u}$ & $\left[ 1 \pm \kappa \right] \times M_{54r}^{u}$ \\[0.5ex]
\midrule
$y_{14}^{d} v_{d} = m_{14}^{d}$ & $\left[ 1 \pm \kappa \right] \times m_{14r}^{d}$ \\[0.5ex]
$y_{24}^{d} v_{d} = m_{24}^{d}$ & $\left[ 1 \pm \kappa \right] \times m_{24r}^{d}$ \\[0.5ex]
$y_{34}^{d} v_{d} = m_{34}^{d}$ & $\left[ 1 \pm \kappa \right] \times m_{34r}^{d}$ \\[0.5ex]
$y_{43}^{d} v_{d} = m_{43}^{d}$ & $\left[ 1 \pm \kappa \right] \times m_{43r}^{d}$ \\[0.5ex]
$x_{34}^{Q} v_{\phi} = m_{35}^{Q}$ & $\left[ 1 \pm \kappa \right] \times m_{35r}^{Q}$ \\[0.5ex]
$x_{42}^{d} v_{\phi} = m_{52}^{d}$ & $\left[ 1 \pm \kappa \right] \times m_{42r}^{d}$ \\[0.5ex]
$x_{43}^{d} v_{\phi} = m_{53}^{d}$ & $\left[ 1 \pm \kappa \right] \times m_{43r}^{d}$ \\[0.5ex]
$M_{45}^{Q}$ & $\left[ 1 \pm \kappa \right] \times M_{45r}^{Q}$ \\[0.5ex]
$M_{54}^{d}$ & $\left[ 1 \pm \kappa \right] \times M_{54r}^{d}$ \\[0.5ex]
\midrule
$\kappa$ & $0.1$ \\
\bottomrule
\bottomrule 
\end{tabular}
\captionof{table}{Next parameter setup after the initial parameter setup to find better mass parameters} 
\label{tab:parameter_region_initial_q_second}}
\end{center}
We vary the parameter space given in Table~\ref{tab:parameter_region_initial_q_second} by first using the fitting function $\chi_{\func{mass}}^{2}$  
in order to find a suitable mass prediction for the four quarks $t,b,c$ and $b$. Once the obtained masses of these quarks are allocated in the ranges 
($\left[ 1 \pm 0.1 \right] \times m_{c,s,b}$ and $\left[ 1 \pm 0.01 \right] \times m_{t}$), we proceed
to fit the CKM quark mixing matrix once more by using the other fitting function $\chi_{\func{CKM}}^{2}$ and it is worth mentioning that  
fitting the CKM mixing matrix is much more challenging due to the very small experimental errors of the CKM matrix elements. We display a benchmark point most converged for the CKM mixing matrix at the next subsection to discuss the possible deviation from the SM result arising from the first row of the CKM mixing matrix.
\subsubsection{Numerical scan result for the quark sector}
We start with the most converged benchmark point ($\chi_{\func{CKM}}^{2} = 956.828$) after repeating the varying many times
\begin{equation}
\begin{split}
M^{u} &= 
\begin{pmatrix}
0 & 0 & 0 & 0 & 0 \\[0.5ex]
0 & 0 & 0 & 0 & 14.474 \\[0.5ex]
0 & 0 & 0 & 1206.340 & 277.563 \\[0.5ex]
0 & 0 & 273.503 & -1775.200 & 0 \\[0.5ex]
0 & 550.990 & 434.462 & 0 & -5624.050 
\end{pmatrix}
\\[2ex]
M^{d} &= 
\begin{pmatrix}
0 & 0 & 0 & 0 & -0.938 \\[0.5ex]
0 & 0 & 0 & 0 & -4.041 \\[0.5ex]
0 & 0 & 0 & 1206.340 & -27.427 \\[0.5ex]
0 & 0 & -5.636 & -1775.200 & 0 \\[0.5ex]
0 & 72.915 & -75.760 & 0 & 2623.620
\end{pmatrix}
\\[2ex]
M_{\func{diag}}^{u} &= 
\begin{pmatrix}
0 & 0 & 0 & 0 & 0 \\[0.5ex]
0 & 1.255 & 0 & 0 & 0 \\[0.5ex]
0 & 0 & 171.303 & 0 & 0 \\[0.5ex]
0 & 0 & 0 & 2155.890 & 0 \\[0.5ex]
0 & 0 & 0 & 0 & 5674.840 
\end{pmatrix}
\\[2ex]
M_{\func{diag}}^{d} &= 
\begin{pmatrix}
0 & 0 & 0 & 0 & 0 \\[0.5ex]
0 & 0.094 & 0 & 0 & 0 \\[0.5ex]
0 & 0 & 3.875 & 0 & 0 \\[0.5ex]
0 & 0 & 0 & 2146.190 & 0 \\[0.5ex]
0 & 0 & 0 & 0 & 2625.960 
\end{pmatrix},
\label{eqn:MuMd_diagMuMd}
\end{split}
\end{equation}
where the above two mass matrices of Equation~\ref{eqn:MuMd_diagMuMd} are the mass matrices for the up- and down-type quarks in the flavor basis, whereas the below two mass matrices are ones fully diagonalized, so revealing all propagating quark mass. From the mass matrices of Equation~\ref{eqn:MuMd_diagMuMd}, we have the mixing matrices $V_{L}^{u}$ and $V_{L}^{d}$ and arrive to our CKM prediction using the formula of Equation~\ref{eqn:CKM_prediction}.
\begin{equation}
V_{\func{CKM}}^{\func{pred}} 
=
\begin{pmatrix}
0.97409 & 0.22602 & 0.00799 & -6.38471 \times 10^{-6} & -0.00036 \\[0.5ex]
0.22615 & -0.97372 & -0.02697 & 3.80102 \times 10^{-5} & 0.00147 \\[0.5ex]
0.00166 & 0.02815 & -0.99880 & -0.00766 & 0.00874 \\[0.5ex]
0.00003 & 0.00019 & -0.00916 & 0.99919 & -0.01773 \\[0.5ex]
-0.00057 & 0.00112 & 0.03812 & 0.03539 & -0.00096 
\label{eqn:CKM_bestfit}
\end{pmatrix}
\end{equation} 
where a feature we should remember is the left-handed down-quark sector are able to reach to all mixings among the three SM generations, whereas the only left-handed $23$ mixing is allowed for the up-quark sector in this BSM model. The experimental CKM mixing matrix without unitarity is given in Equation~\ref{eqn:CKM_exp_wx_uni}.
\begingroup
\setlength\arraycolsep{5pt}
\begin{equation}
\lvert K_{\func{CKM}} \rvert
=
\begin{pmatrix}
0.97370 \pm 0.00014 & 0.22450 \pm 0.00080 & 0.00382 \pm 0.00024 \\[1.5ex]
0.22100 \pm 0.00400 & 0.98700 \pm 0.01100 & 0.04100 \pm 0.00140 \\[1.5ex]
0.00800 \pm 0.00030 & 0.03880 \pm 0.00110 & 1.01300 \pm 0.03000 
\end{pmatrix}
\label{eqn:CKM_exp_wx_uni}
\end{equation}
\endgroup
Restricting our attention to the upper-left $3\times 3$ block of Equation~\ref{eqn:CKM_bestfit}, it can be compared to its experimental bound given in Equation~\ref{eqn:CKM_exp_wx_uni}. In order to confirm that our prediction for the CKM mixing matrix is consistent with the experimental data, 
it requires for the upper-left $3 \times 3$ block of Equation~\ref{eqn:CKM_bestfit} to be inside the $3\sigma$ experimentally allowed range  
as follows:
\begin{equation}
\lvert (K_{\func{CKM}})_{ij} \rvert - 3\lvert (\delta K_{\func{CKM}})_{ij} \rvert < \lvert (V_{\func{CKM}}^{\func{pred}})_{ij} \rvert < \lvert (K_{\func{CKM}})_{ij} \rvert + 3\lvert (\delta K_{\func{CKM}} )_{ij}, \rvert, \quad \text{for $i,j=1,2,3$}
\end{equation}
and we confirm that the $13, 23, 31, 32$ elements in the CKM prediction of Equation~\ref{eqn:CKM_bestfit} cannot be fitted within the $3\sigma$ range with a small difference. 
From our numerical analysis we find that in our model the  
CKM quark mixing matrix mainly arises from the down type quark sector and has a subleading correction coming from the up type quark sector.
It is worth mentioning that the inclusion of an additional
vector-like family in our proposed model to provide masses for the 
first generation of SM charged fermions 
will lead to an improvement of our predictions related to the CKM quark mixing matrix.
However, that approach of having a fifth vector-like fermion family goes beyond the 
scope in this work and is deferred for a future publication.  
Furthermore, in this section we also discuss the possible deviation of the first row of the CKM mixing matrix without unitarity and this study is also covered in this reference~\cite{Branco:2021vhs} with an isosinglet vector-like quark in a model different than the one considered in this paper. According to 
\cite{Branco:2021vhs}, 
the deviation $\Delta$ of unitarity is defined as follows:
\begin{equation}
\Delta = 1 - \lvert V_{ud} \rvert^2 - \lvert V_{us} \rvert^2 - \lvert V_{ub} \rvert^2,
\end{equation}
and its experimental value is given by 
\cite{Belfatto:2019swo}.
\begin{equation}
\sqrt{\Delta} \sim 0.04
\label{eqn:dev_exp}
\end{equation}
Calculating the deviation of unitarity $\Delta$ from the best fitted CKM prediction of Equation~\ref{eqn:CKM_bestfit}, the result is
\begin{equation}
\begin{split}
\sqrt{\Delta} &\simeq 0.00035 
\end{split}
\end{equation}
Therefore, the deviation of unitarity derived from the model under consideration 
is too small to be observed compared to its experimental bound given in Equation~\ref{eqn:dev_exp}. Lastly, we discuss the rare $t \rightarrow c Z$ decay and collect all benchmark points satisfying $\chi_{\func{CKM}}^{2} < 980$ (notice that the most converged point reports $\chi^2_{\func{CKM}} = 956.828$).
\begin{figure}[H]
\centering
\begin{subfigure}{0.48\textwidth}
\includegraphics[keepaspectratio,width=0.9\textwidth]{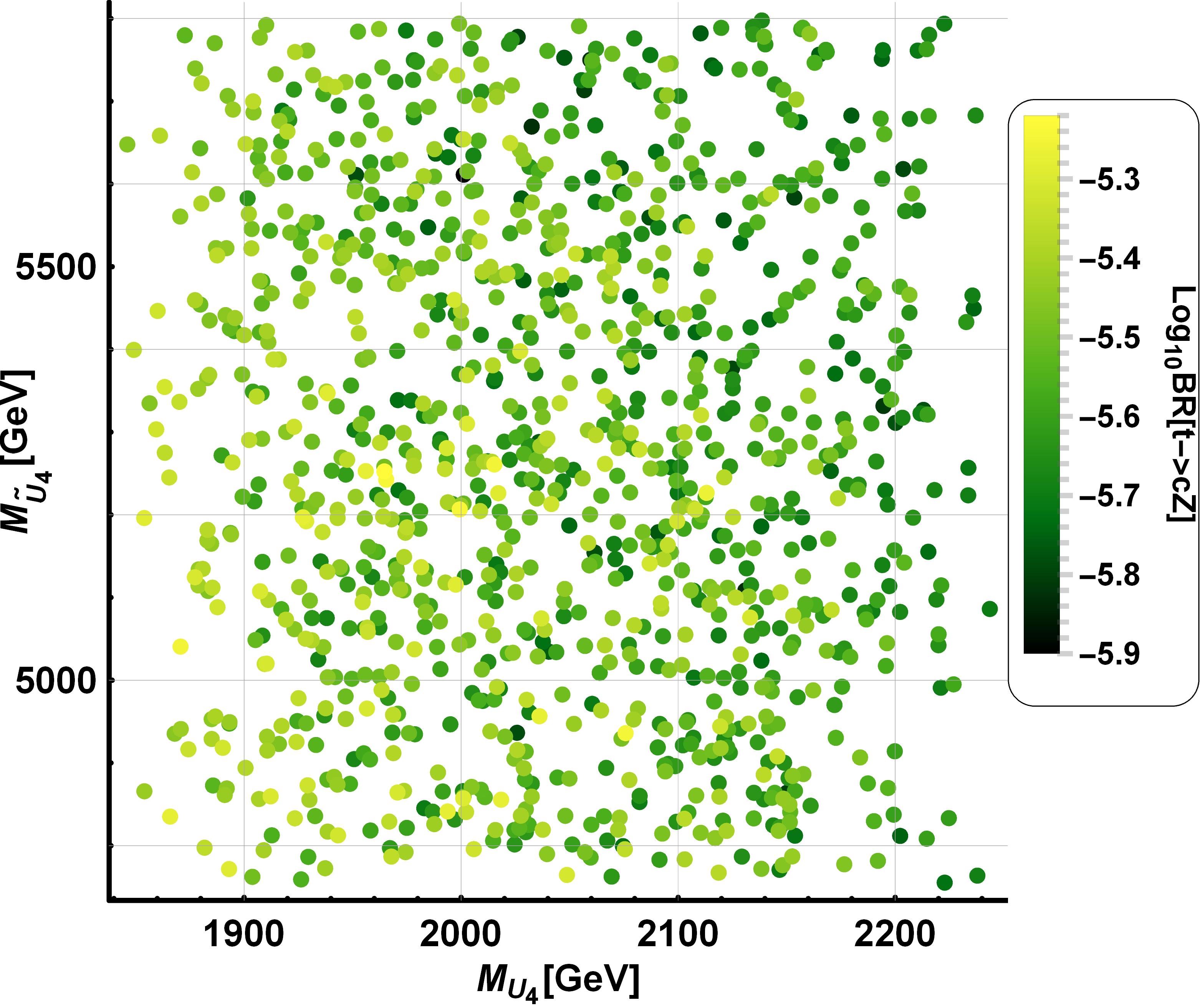}
\end{subfigure}
\hspace{0.1cm}
\begin{subfigure}{0.48\textwidth}
\includegraphics[keepaspectratio,width=0.9\textwidth]{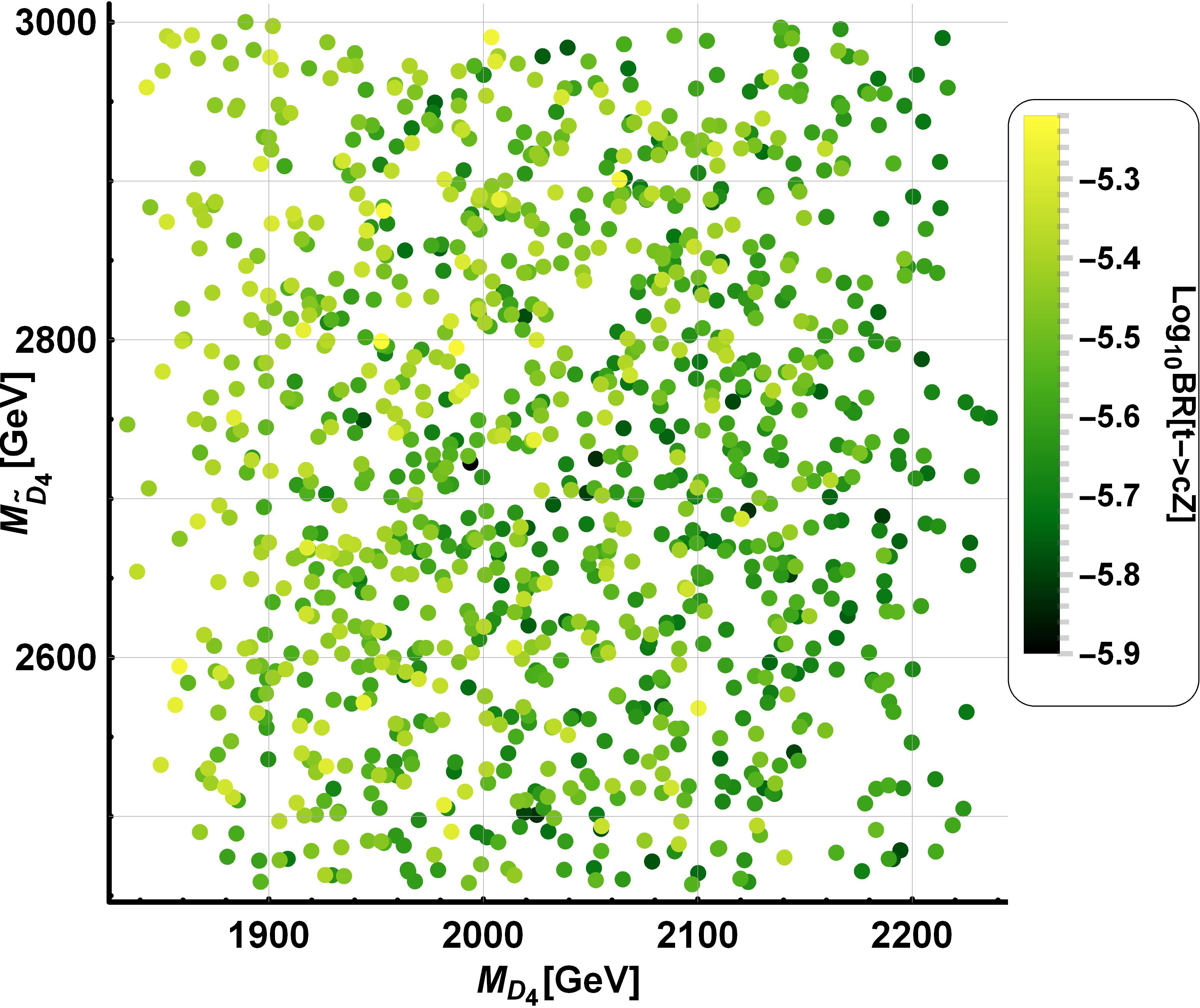}
\end{subfigure}
\caption{Scanned mass region of the vector-like quarks and contributions of the flavor violating interactions with the SM $Z$ gauge boson to the rare  $t \rightarrow c Z$ decay. The used constraints are the predicted $c,s,b$ and $t$ quark mass to be put between $[1 \pm 0.1] \times m_{c,s,b}$ and $[1 \pm 0.01] \times m_{t}$ and the CKM mixing matrix.}
\label{fig:tcZ_decay}
\end{figure}
Figure~\ref{fig:tcZ_decay} displays the allowed values of vector-like quark masses consistent with the constraints arising from the rare  $t \rightarrow c Z$ decay. Our obtained values for the vector-like quark masses are consistent with their lower experimental bound of $1000\func{GeV}$ arising from collider searches. In our numerical analysis the vector-like doublet up-type quark mass $M_{U_4}$ is ranged from $1850\func{GeV}$ up to about 
$2250\func{GeV}$ and the vector-like singlet up-type quark mass $M_{\widetilde{U}_4}$ is varied from $4750\func{GeV}$ to $5800\func{GeV}$. Regarding the exotic down type quark sector, we have varied  
the vector-like doublet down-type quark mass $M_{D_4}$ 
from $1850\func{GeV}$ up to $2250\func{GeV}$ and the vector-like singlet down-type quark mass $M_{\widetilde{D}_4}$ 
from $2450\func{GeV}$ up to $3000\func{GeV}$. As seen from Figure~\ref{fig:tcZ_decay} the order of magnitude of the obtained values for the branching ratio of the rare $t \rightarrow c Z$ decay range from $10^{-6}$ up to $10^{-5}$, which is consistent with 
its current experimental bound, whose logarithmic value is about $-3.6$ as indicated by 
Equation~\ref{eqn:BRtcZ_exp}.

\section{Conclusion} \label{sec:VII}
In this work we have considered a model where the SM fermion sector is extended by the inclusion of a 
fourth vector-like family and the scalar sector is augmented by the incorporation of an extra scalar doublet and a gauge singlet scalar. In addition, we have assumed a global $U(1)^{\prime}$ symmetry under which all particles are charged except the SM chiral quark and lepton fields. 
The model explains the hierarchical structure of the SM quark and lepton masses 
by assuming that the SM Yukawa interactions are forbidden by the $U(1)^{\prime}$ symmetry and arise effectively after it is spontaneously broken, due to 
induced mixing with the fourth vector-like family. This mixing also results in non-standard couplings of the $W$ and $Z$ gauge bosons which have been studied here for the first time.

This setup leads to sizeable branching fractions for the FCNC decays such as $\mu \rightarrow e \gamma$, $Z\to\mu\tau$ and $t\rightarrow cZ$, within the reach of the future experimental sensitivity. These FCNC decays are studied in detail in this work, in order to set constraints on the model parameter space. 
A great advantage of the approach taken in this work with respect to the ones considered in extensions of the SM having a $Z^{\prime}$ gauge boson 
is that it makes the study of the FCNC observables simpler than in the latter since in the former we can avoid assuming specific values for the unknown $U(1)^{\prime}$ coupling and  
$Z^{\prime}$ gauge boson mass. 
This makes the present phenomenology based on $W$ and $Z$ gauge boson couplings
more predictive than if the $U(1)^{\prime}$ were a spontaneously broken gauge symmetry, leading to a massive $Z'$.

Given that the hierarchical structure of the SM is implemented in our proposed model, the extended mass matrices for the charged lepton and quark sectors need to be completely and accurately diagonalised, as 
the starting point of our analytical and numerical analysis.
Since we only consider a fourth vector-like family, the model
cannot provide masses for the first generation of SM charged fermions, 
nevertheless this is a good approximation given that the first generation of the SM fermions are very light. For this reason, we mainly focus on the study of FCNC observables 
involving the second and third generations of SM fermions in both quark and lepton sectors. 

In the chosen convenient basis, the different shape of the down-type quark mass matrix allows all left-handed mixings between the three SM generations, whereas the up type quark sector can have only the $23$ left-handed mixing, while all quarks and charged lepton have the only $23$ right-handed mixing, and we have checked that the results are basis independent.
This feature implies that we can obtain
 a prediction for the CKM mixing matrix and this is one of main phenomenological aspects analyzed in this work.
 In order to diagonalize the fermionic mass matrices, in an analytic approximation,
 we have defined the $SU(2)$ conserving and $SU(2)$ violating mixings and we have shown that the $SU(2)$ violating mixing plays a crucial role for generating the $Z$ mediated flavor violating interactions. 
 Furthermore, the
 extension of the SM fermion sector by the inclusion of a vector-like family makes the matrices of $Z$ couplings with fermions 
 different than the identity matrix 
 due to the appearance of non-zero off-diagonal matrix elements of the $Z$ coupling matrices which will give rise to flavor violating $Z$ decays. The non-zero off-diagonal SM $Z$ gauge coupling constants are generally proportional to two of the small mixing angles, which are defined by the ratio between the SM fermion and vector-like masses, thus leading to small values.
\\~
Defining all the required $Z$ gauge coupling constants with fermions in the mass basis, as discussed above, we began by 
analyzing the FCNC processes of the charged lepton sector.
We have found that in the lepton sector, 
the following three 
FCNC decays are allowed: 
$\tau \rightarrow \mu \gamma, \tau \rightarrow 3 \mu$ and $Z \rightarrow \mu \tau$. Regarding 
the $\tau \rightarrow \mu \gamma$ decay, we discussed its leading contribution, which  
arises from the Feynman diagrams having a chirality flip in the internal fermionic lines and being proportional to 
$M/m_{\tau}$, where $M$ is the mass scale of the heavy charged vector-like leptons. However, the dominant terms cannot be as big as the vector-like masses get heavier since their coupling constants get suppressed at the same time, thus providing a 
balanced relation between the vector-like masses and their coupling constants. We have found that 
our predictions for the vector-like charged lepton masses are not severely constrained by the $\tau \rightarrow \mu \gamma$ decay since most of the obtained values for the $\tau \rightarrow \mu \gamma$ decay are consistent with its experimental upper bound. In the concerning to the  
$\tau \rightarrow 3\mu$ and $Z \rightarrow \mu \tau$ decays, we have derived an analytic expression for their corresponding rates at tree-level finding that  
none of  
our predictions is constrained by the experimental bounds of these decays. Considering the FCC-ee experiment which have planned to generate $10^{12}$ the $Z$ gauge bosons and our numerical prediction for the $Z \rightarrow \mu \tau$ branching ratio is of the  
order of $10^{-9}$ at most, thus implying that our model can be tested at the $Z$ factory via the $Z \rightarrow \mu \tau$ decay. 
 However, the CMS provided that the doublet vector-like mass can be constrained up to $790\func{GeV}$~\cite{Hernandez:2021tii,Xu:2018pnq} and our numerical predictions for the vector-like charged leptons are severely constrained by the CMS result. Therefore, we can expect that the vector-like charged lepton doublet mass is ranged from $790$ to nearly $1600 \func{GeV}$, whereas the vector-like charged lepton singlet mass is ranged from $500$ to $2000 \func{GeV}$ or above than that. 

Turning to the quark sector phenomenology, we have analyzed the rare $t \rightarrow c Z$ decay as well as the CKM mixing 
to set constraints on the quark sector parameters. It is worth mentioning that the
neutral meson $K, B_{d}, B_{s}$ oscillations do not set constraints on the quark sector parameters of our model since their new physics effects are quite negligible compared to the SM expectation.  
We have derived analytic expressions for the rare $t \rightarrow c Z$ decay as well as for the CKM mixing matrix. Due to the mixings between SM fermions and vector-like fermions, the CKM quark mixing matrix is not unitary, thus implying that the unitarity requirement has to be relaxed~\cite{Branco:2021vhs,ParticleDataGroup:2020ssz}. Using the most converged benchmark point, we showed how dominant the down-type quark mixing matrix plays a crucial role in the CKM mixing matrix and we discussed the deviation of unitarity arisen from the first row of the CKM mixing matrix, whose value is too small to be experimentally measured.
Finally, we investigate the branching ratio for the $t \rightarrow c Z$ decay and found that our numerical predictions are not excluded by its experimental bound, for 
 vector-like doublet up-type and down type quark masses $M_{U_4}$ and $M_{D_4}$ in the window \mbox{$1850$\text{GeV}$\leqslant M_{U_4},M_{D_4}\leqslant 2250\text{GeV}$} as well as 
 vector-like singlet up and down type quark masses $M_{\widetilde{U}_4}$ and $M_{\widetilde{D}_4}$ in the ranges \mbox{$4750$\text{GeV}$\leqslant M_{\widetilde{U}_4}\leqslant 5800\text{GeV}$}, 
and \mbox{$2450$\text{GeV}$\leqslant M_{\widetilde{D}_4}\leqslant 3000\text{GeV}$}, respectively.

In conclusion, we have analysed a range of FCNCs arising from non-standard $W$ and $Z$ gauge boson couplings in an extension of the SM with a fourth vector-like family, which can also address the hierarchy of quark and lepton masses, leading to several interesting rare decays which may be probed in future high luminosity experiments.
\chapter{Conclusions} \label{Chapter:Conclusions}
Many dedicated efforts to find an answer on what are the most fundamental particles and forces have shaped the awesome and beautiful SM. The CKM mixing matrix experimentally confirmed is one of the great successes of the SM and it reveals there exist a mixing between each generation of the SM. This mixing mechanism is applied not only to the SM fermion sector but also to the SM gauge particles. Even though, the SM is quite successful in both quark and lepton sector, however the SM has also some limitations such as masses of the SM neutrinos, a few of well-known anomalies, DM, DE, gravity, etc. and the limitations have been a strong motivation for the SM to be expanded. We start from this consideration: how can we expand the SM without violating the gauge symmetry and the current SM experimental bounds. A possible answer to the question is a minimal extension to the SM and to study the muon and electron $g-2$, which was a main target over my first and second works. The other choice could be to study the FCNC observables in a minimally extended SM, which are quite sensitive to new physics, which was discussed in my third project.
\\~\\
In chapter~\ref{Chapter:Introduction}, we discussed how successful the SM is with the mixing formalism and an approach to new physics in both theoretical and experimental aspects. In chapter~\ref{Chapter:TheSM}, a few of main features of the SM were discussed such as the Yukawa interactions, the spontaneous symmetry breaking, the broken gauge symmetry and CKM mixing matrix and a few of important limitations were also discussed.
\\~\\
In chapter~\ref{Chapter:The1stBSMmodel}, the common features appearing over my three works, which are the vector-like family, 2HDM and lastly $U(1)^{\prime}$ symmetry, were discussed as prerequisites. From the rest of chapter~\ref{Chapter:The1stBSMmodel} to chapter~\ref{Chapter:The1stpaper}, we covered our first work and our first BSM model, the Fermiophobic $Z^{\prime}$ model, in the first work was discussed in the rest of chapter~\ref{Chapter:The1stBSMmodel} as well as its mixing with the fourth vector-like family. In chapter~\ref{Chapter:The1stpaper}, we discuss main body of our first work. The $Z^{\prime}$ coupling constants in the mass basis were determined in terms of the mixing angles $\theta_{12L,R}, \theta_{14L,R}$ and $\theta_{24L,R}$ and then the analytic form of the CLFV $\mu \rightarrow e \gamma$, the muon and electron $g-2$ and the neutrino trident production process were discussed with the corresponding Feynman diagrams. Especially, we use the mass insertion approximation for the analytic form of the muon and electron $g-2$ and $\mu \rightarrow e \gamma$ decay, and it has two mass sources; the chirality flip $M_{4}^{C}$ and vector-like mass $M_{4}^{L}$, and both appear in the analytic form separately under the assumption $M_{4}^{L} \gg M_{4}^{C}$. Through the interplay among the muon and electron $g-2$ and $\mu \rightarrow e \gamma$ decay, we showed it is not possible to explain both anomalies analytically at the same time in the case of $M_{4}^{C} \gg m_{\mu}$, since the $\mu \rightarrow e \gamma$ gives rise to a very tight bound. It means we need to relax the condition $M_{4}^{C} \gg m_{\mu}$, and the case of either $M_{4}^{C}=0$ or $M_{4}^{C}=5m_{\mu}$ was discussed in our numerical scan. We investigated the parameter space for either the muon or electron $g-2$ versus $M_{Z}^{\prime}$ separately and then tried to explain both anomalies by considering some overlapped parameter regions and what we found was it is not still possible to explain the anomalies simultaneously no matter what the chirality flip mass between $0$ to $200\func{GeV}$ was considered. Next, we discussed the $Z^{\prime}$ mass bound, which is $48\func{GeV}$ at most following PDG, however this result is too old to trust. For this reason, we found a suitable experimental bound implemented by an effective four fermion vertex interaction given by LEP experiment and considered oblique corrections $S$ and $T$, however both can not be numerically determined due to lots of unknown coupling constant of $Z^{\prime}$ gauge boson, which implies our numerically predicted $Z^{\prime}$ mass $75\func{GeV}$ should not be excluded at the moment.
\\~\\
From chapter~\ref{Chapter:The2ndBSMmodel} to chapter~\ref{Chapter:The2ndpaper2}, my second work was discussed. In chapter~\ref{Chapter:The2ndBSMmodel}, we discussed a new BSM model, as we took the SM Lagrangian as an effective theory. An important difference between our first and second BSM model is the first BSM model allows general renormalizable Yukawa interactions, whereas the second BSM model does not since the SM-like Higgses are charged under the $U(1)^{\prime}$ global symmetry. Therefore, the second BSM model gives rise to the $5$ dimensional effective operator for the SM operators and the proportional factor $\langle \phi \rangle/M$ can explain the relative different mass of each fermion, as discussed in the mass insertion formalism. Another difference between my first and second BSM model is to start considering the hierarchical structure of the SM and this requires our mass matrices for quarks and leptons to be rotated maximally and this fully rotated mass matrices should be a starting point. We diagonalized mixing matrices using the mixing formalism, defining all the required mixing angles. In chapter~\ref{Chapter:The2ndpaper}, we investigated the non-SM $W$ contributions to the muon and electron $g-2$. In order to make our analysis simple, we assumed the vector-like neutrinos in the fifth vector-like family are too heavy to contribute, so they are not considered. Then, we constructed the type 1b seesaw mechanism and then defined the non-unitarity $\eta$, which plays a crucial role to the non-SM contributions to both anomalies as well as $\mu \rightarrow e \gamma$ decay. We derived the analytic form of both anomalies and $\mu \rightarrow e \gamma$ decay in terms of the non-unitarity $\eta$ and showed that the muon and electron $g-2$ prediction with the $W$ gauge boson are too small to its experimental bound, so we conclude the $W$ contributions can not explain both anomalies simultaneously. In chapter~\ref{Chapter:The2ndpaper2}, we discussed non-SM scalar contributions to the anomalies. For this task, the required sector is the charged lepton sector and we start from the diagonalization of the charged lepton sector. Using the assumption ($y_{34}^e = x_{43}^{e} = y_{15,25,35}^e = x_{51,52,53}^{e} = x_{25,35}^L = y_{52,53}^e = 0$) for the diagonalization of the charged leptons, we confirmed it is possible to derive one-loop diagrams for the anomalies by closing the scalar sector of mass insertion diagrams. Next, we constructed the 2HDM scalar potential, which is necessary to determine the physical scalars, and worked under the decoupling limit in order to make our analysis simpler. Then, we discussed the Higgs diphoton signal strength and the analytic form of the anomalies with the non-SM scalars, relevant for the numerical study for the non-SM scalar contributions to the anomalies. After building the fitting function, we fitted the relevant parameters and then proved both anomalies can be explained by the non-SM scalar contributions, predicting mass of the physical scalars and vector-like charged leptons. Finally, we discussed the vacuum stability of the 2HDM scalar potential. The up-type Higgs potential is stable due to the decoupling limit, however it needs to be determined for the stability of the down-type Higgs potential as it features mixing with the singlet flavon field. And then we proved that the down-type Higgs potential is stable with the small vev of $v_3$ ($\mathcal{O}(v_3) = 10\func{GeV}$) and with the suitable sign of quartic coupling constants in the Higgs potential.
\\~\\
In chapter~\ref{Chapter:The3rdBSMmodel}, we introduced our third BSM model, mainly motivated by the hierarchical structure of the SM, and discussed how this BSM model gives rise to the effective SM Yukawa interactions. For the purpose of correct diagonalization without any assumptions, we made use of only one vector-like family instead of two, and the first SM generation remains massless in this BSM model as a result. Using the mixing formalism, we showed that the mass matrices of quark and lepton sector in the flavor basis are diagonalized. The mass matrix for the charged lepton sector is exactly same as that for the up-type quark sector, however that for the down-type quark sector has an additional element due to the rotation already used in the diagonalization of the up-type quark sector and this leads to mixing with the first generation even though the first generation is massless. What this implement is we can build a prediction for the CKM mixing matrix and understand how dominant the down-type mixing angles are when compared to the up-type mixing angles. One of main motivations in this work is to study diverse FCNC observables to constrain vector-like fermions in both quark and charged lepton sector and we use the SM $Z$ gauge boson. With the SM fermions, we can not induce the renormalizable flavor violating interactions mediated by the $Z$ gauge boson since the SM $Z$ gauge coupling constants shape a identity matrix. If we extend the SM fermions by the fourth vector-like family, the identity matrix of $Z$ coupling constants is not the identity matrix any more and this matrix can give rise to the flavor violating interactions with the $SU(2)$ violating mixings. Then, we constructed the SM $Z$ coupling constants in the mass basis, having non-zero off-diagonal elements, in both quark and charged lepton sector. In chapter~\ref{Chapter:The3rdpaper}, we start investigating the charged lepton sector first and consider the FCNC observables such as $\tau \rightarrow \mu \gamma, \tau \rightarrow 3 \mu$ and $Z \rightarrow \mu \tau$. After finding out the analytic form of the FCNC observables, we carry out the numerical scan for the charged lepton sector and conclude our numerical predictions are not significantly constrained by the experimental bound for the FCNC observables, however the CMS experimental bound for the vector-like doublet charged lepton mass can significantly constrain our predictions if it turns out to be firmly established. In our numerical prediction, the vector-like doublet charged lepton mass is ranged from $790$ to $1600\func{GeV}$, whereas the vector-like singlet charged lepton mass is ranged from $500$ to $2000\func{GeV}$ or above than that. For the quark sector, we consider the rare $t \rightarrow c Z$ decay and the CKM mixing matrix and we find out that our numerical prediction for the quark sector is mainly constrained by the CKM mixing matrix, not by the experimental bound for the $t \rightarrow c Z$ decay, predicting mass of the vector-like doublet up-type quark mass is ranged from $1850$ to $2250\func{GeV}$ and the vector-like singlet up-type quark mass is ranged from $4750$ to $5800\func{GeV}$, whereas the vector-like doublet down-type quark mass is ranged from $1850$ to $2250\func{GeV}$ and the vector-like singlet down-type quark mass is ranged from $2450$ to $3000\func{GeV}$.
\\~\\
We discussed how successful the SM is in many senses, and however it has some important observables which can not be addressed by the SM at the same time, which have been the strong motivations to search for physics beyond the SM. All these works were worked, based on the principle of the minimal extension to the SM, and each work has its own surprising insight to new physics. As time goes on, the flavor physics and Higgs physics get more and more important, as they are likely to reveal new physics, and we will keep studying many well-motivated BSM models to search for new possibilities to physics beyond the SM.

\appendix
\chapter{Is it possible to explain the muon and electron $g-2$ in a $Z^\prime$ model?}
It is important to understand how the observables $\operatorname{BR} ( \mu \rightarrow e \gamma)$, muon $g-2$, electron $g-2$ and neutrino trident can be written in terms of the mixing angles. The coupling constants appearing in each observable consist of the mixing angles. The coupling constants are defined from Equation \eqref{eqn:mu_mu_zprime_coupling} to \eqref{eqn:mu_e_zprime_coupling} in Section \ref{sec:LFV_in_this_model}. 

\section{The branching ratio of $\mu \rightarrow e \gamma$}

The branching ratio of $\mu \rightarrow e \gamma$ is the following:
\begingroup
\begin{equation}
\operatorname{BR}(\mu\rightarrow e\gamma) = \frac{\alpha}{1024 \pi^4} \frac{m_{\mu}^5}{M_{Z^{\prime}}^4 \Gamma_\mu} (\left\vert \widetilde{\sigma }_{L}\right\vert ^{2}+\left\vert \widetilde{\sigma }_{R}\right\vert^{2})
\label{eqn:analytic_expression_for_muegamma_1}
\end{equation}
\endgroup
The $\widetilde{\sigma}_{L,R}$ are given by:
\begingroup
\begin{align}
\begin{split}
\widetilde{\sigma }_{L}& =\sum_{a=e,\mu,E}\left[(g_{L})_{ea}(g_{L})_{a\mu}F(x_{a})+\frac{m_{a}}{m_{\mu }}(g_{L})_{ea}(g_{R})_{a\mu}G(x_{a})\right] , \\
\widetilde{\sigma }_{R}& =\sum_{a=e,\mu,E}\left[ (g_{R})_{ea}(g_{R})_{a\mu}F(x_{a})+\frac{m_{a}}{m_{\mu }}(g_{R})_{ea}(g_{L})_{a\mu}G(x_{a})\right],\hspace{1.0cm}x_{a}=\frac{m_{a}^{2}}{M_{Z^{\prime }}^{2}}
\label{eqn:contributions_to_muegamma_(sigmas)_1}
\end{split}
\end{align}
\endgroup
Expanding the above $\widetilde{\sigma}_{L,R}$ in terms of electron, muon and fourth family:
\begingroup
\begin{align}
\begin{split}
\widetilde{\sigma}_L = \Big[ &\left( g_L \right)_{ee} \left( g_L \right)_{e\mu} F\left( x_e \right) + \frac{m_e}{m_\mu}\left( g_L \right)_{ee} \left( g_R \right)_{e\mu} G\left( x_e \right) \\
&\left( g_L \right)_{e\mu} \left( g_L \right)_{\mu\mu} F\left( x_\mu \right) + \frac{m_\mu}{m_\mu}\left( g_L \right)_{e\mu} \left( g_R \right)_{\mu\mu} G\left( x_\mu \right) \\
&\left( g_L \right)_{eE} \left( g_L \right)_{E\mu} F\left( x_{E} \right) + \frac{M_4^C}{m_\mu}\left( g_L \right)_{eE} \left( g_R \right)_{E\mu} G\left( x_{E} \right) \Big] \\
\widetilde{\sigma}_R = \Big[ &\left( g_R \right)_{ee} \left( g_R \right)_{e\mu} F\left( x_e \right) + \frac{m_e}{m_\mu}\left( g_R \right)_{ee} \left( g_L \right)_{e\mu} G\left( x_e \right) \\
&\left( g_R \right)_{e\mu} \left( g_R \right)_{\mu\mu} F\left( x_\mu \right) + \frac{m_\mu}{m_\mu}\left( g_R \right)_{e\mu} \left( g_L \right)_{\mu\mu} G\left( x_\mu \right) \\
&\left( g_R \right)_{eE} \left( g_R \right)_{E\mu} F\left( x_{E} \right) + \frac{M_4^C}{m_\mu}\left( g_R \right)_{eE} \left( g_L \right)_{E\mu} G\left( x_{E} \right) \Big]
\label{eqn:expansion_of_sigmas_with_each_family}
\end{split}
\end{align}
\endgroup
One important feature in Equation \eqref{eqn:expansion_of_sigmas_with_each_family} is the chirality-flipping mass was used instead of vector-like mass in the last line of Equation \eqref{eqn:expansion_of_sigmas_with_each_family}. It then is possible to turn the coupling constants in each $\widetilde{\sigma}$ into the mixing angles by using the Equations \eqref{eqn:mu_mu_zprime_coupling}-\eqref{eqn:mu_e_zprime_coupling}. It was assumed that $g^{\prime} q_{L4}$ in each coupling constant to be $1$.
\begingroup
\begin{align}
\begin{split}
\widetilde{\sigma}_L = \Big[ &\Big(\sin\theta_{12}^{L}\sin\theta_{24}^{L}+\cos\theta_{12}^{L}\cos\theta_{24}^{L}\sin\theta_{14}^{L}\Big)^2 \times \\
&\Big( \sin\theta_{12}^{L}\sin\theta_{24}^{L}+\cos\theta_{12}^{L}\cos\theta_{24}^{L}\sin\theta_{14}^{L}\Big)\Big(\cos\theta_{12}^{L}\sin\theta_{24}^{L}-\cos\theta_{24}^{L}\sin\theta_{12}^{L}\sin\theta_{14}^{L} \Big) F\left( x_1 \right) \\
& + \frac{m_1}{m_2} \Big(\sin\theta_{12}^{L}\sin\theta_{24}^{L}+\cos\theta_{12}^{L}\cos\theta_{24}^{L}\sin\theta_{14}^{L}\Big)^2 \times \\
& \Big( \sin\theta_{12}^{R}\sin\theta_{24}^{R}+\cos\theta_{12}^{R}\cos\theta_{24}^{R}\sin\theta_{14}^{R}\Big)\Big(\cos\theta_{12}^{R}\sin\theta_{24}^{R}-\cos\theta_{24}^{R}\sin\theta_{12}^{R}\sin\theta_{14}^{R} \Big) G\left( x_1 \right) \\
& + \Big( \sin\theta_{12}^{L}\sin\theta_{24}^{L}+\cos\theta_{12}^{L}\cos\theta_{24}^{L}\sin\theta_{14}^{L}\Big)\Big(\cos\theta_{12}^{L}\sin\theta_{24}^{L}-\cos\theta_{24}^{L}\sin\theta_{12}^{L}\sin\theta_{14}^{L} \Big) \times \\
& \Big(\cos\theta_{12}^{L}\sin\theta_{24}^{L}-\cos\theta_{24}^{L}\sin\theta_{12}^{L}\sin\theta_{14}^{L}\Big)^2 F\left( x_2 \right) \\
& + \frac{m_2}{m_2} \Big( \sin\theta_{12}^{L}\sin\theta_{24}^{L}+\cos\theta_{12}^{L}\cos\theta_{24}^{L}\sin\theta_{14}^{L}\Big)\Big(\cos\theta_{12}^{L}\sin\theta_{24}^{L}-\cos\theta_{24}^{L}\sin\theta_{12}^{L}\sin\theta_{14}^{L} \Big) \times \\
& \Big(\cos\theta_{12}^{R}\sin\theta_{24}^{R}-\cos\theta_{24}^{R}\sin\theta_{12}^{R}\sin\theta_{14}^{R}\Big)^2 G\left( x_2 \right) \\	
& + \cos\theta_{14}^{L}\cos\theta_{24}^{L} \Big(\sin\theta_{12}^{L}\sin\theta_{24}^{L}+\cos\theta_{12}^{L}\cos\theta_{24}^{L}\sin\theta_{14}^{L}\Big) \times \\
& \cos\theta_{14}^{L}\cos\theta_{24}^{L}\Big(\cos\theta_{12}^{L}\sin\theta_{24}^{L}-\cos\theta_{24}^{L}\sin\theta_{12}^{L}\sin\theta_{14}^{L}\Big) F\left( x_4 \right) \\
& + \frac{M_4^C}{m_2} \cos\theta_{14}^{L}\cos\theta_{24}^{L} \Big(\sin\theta_{12}^{L}\sin\theta_{24}^{L}+\cos\theta_{12}^{L}\cos\theta_{24}^{L}\sin\theta_{14}^{L}\Big) \times \\
& \cos\theta_{14}^{R}\cos\theta_{24}^{R}\Big(\cos\theta_{12}^{R}\sin\theta_{24}^{R}-\cos\theta_{24}^{R}\sin\theta_{12}^{R}\sin\theta_{14}^{R}\Big) G\left( x_4 \right) \Big] \\	
\widetilde{\sigma}_R = \Big[ &\Big(\sin\theta_{12}^{R}\sin\theta_{24}^{R}+\cos\theta_{12}^{R}\cos\theta_{24}^{R}\sin\theta_{14}^{R}\Big)^2 \times \\
&\Big( \sin\theta_{12}^{R}\sin\theta_{24}^{R}+\cos\theta_{12}^{R}\cos\theta_{24}^{R}\sin\theta_{14}^{R}\Big)\Big(\cos\theta_{12}^{R}\sin\theta_{24}^{R}-\cos\theta_{24}^{R}\sin\theta_{12}^{R}\sin\theta_{14}^{R} \Big) F\left( x_1 \right) \\
& + \frac{m_1}{m_2} \Big(\sin\theta_{12}^{R}\sin\theta_{24}^{R}+\cos\theta_{12}^{R}\cos\theta_{24}^{R}\sin\theta_{14}^{R}\Big)^2 \times \\
& \Big( \sin\theta_{12}^{L}\sin\theta_{24}^{L}+\cos\theta_{12}^{L}\cos\theta_{24}^{L}\sin\theta_{14}^{L}\Big)\Big(\cos\theta_{12}^{L}\sin\theta_{24}^{L}-\cos\theta_{24}^{L}\sin\theta_{12}^{L}\sin\theta_{14}^{L} \Big) G\left( x_1 \right) \\
& + \Big( \sin\theta_{12}^{R}\sin\theta_{24}^{R}+\cos\theta_{12}^{R}\cos\theta_{24}^{R}\sin\theta_{14}^{R}\Big)\Big(\cos\theta_{12}^{R}\sin\theta_{24}^{R}-\cos\theta_{24}^{R}\sin\theta_{12}^{R}\sin\theta_{14}^{R} \Big) \times \\
& \Big(\cos\theta_{12}^{R}\sin\theta_{24}^{R}-\cos\theta_{24}^{R}\sin\theta_{12}^{R}\sin\theta_{14}^{R}\Big)^2 F\left( x_2 \right) \\
& + \frac{m_2}{m_2} \Big( \sin\theta_{12}^{R}\sin\theta_{24}^{R}+\cos\theta_{12}^{R}\cos\theta_{24}^{R}\sin\theta_{14}^{R}\Big)\Big(\cos\theta_{12}^{R}\sin\theta_{24}^{R}-\cos\theta_{24}^{R}\sin\theta_{12}^{R}\sin\theta_{14}^{R} \Big) \times \\
& \Big(\cos\theta_{12}^{L}\sin\theta_{24}^{L}-\cos\theta_{24}^{L}\sin\theta_{12}^{L}\sin\theta_{14}^{L}\Big)^2 G\left( x_2 \right) \\	
& + \cos\theta_{14}^{R}\cos\theta_{24}^{R} \Big(\sin\theta_{12}^{R}\sin\theta_{24}^{R}+\cos\theta_{12}^{R}\cos\theta_{24}^{R}\sin\theta_{14}^{R}\Big) \times \\
& \cos\theta_{14}^{R}\cos\theta_{24}^{R}\Big(\cos\theta_{12}^{R}\sin\theta_{24}^{R}-\cos\theta_{24}^{R}\sin\theta_{12}^{R}\sin\theta_{14}^{R}\Big) F\left( x_4 \right) \\
& + \frac{M_4^C}{m_2} \cos\theta_{14}^{R}\cos\theta_{24}^{R} \Big(\sin\theta_{12}^{R}\sin\theta_{24}^{R}+\cos\theta_{12}^{R}\cos\theta_{24}^{R}\sin\theta_{14}^{R}\Big) \times \\
& \cos\theta_{14}^{L}\cos\theta_{24}^{L}\Big(\cos\theta_{12}^{L}\sin\theta_{24}^{L}-\cos\theta_{24}^{L}\sin\theta_{12}^{L}\sin\theta_{14}^{L}\Big) G\left( x_4 \right) \Big] \\	
\label{eqn:expansion_of_sigmas_with_mixing_angles}
\end{split}
\end{align}
\endgroup
\section{Anomalous muon $g-2$}

The anomalous muon $g-2$ is given by:
\begingroup
\begin{equation}
\begin{split}
\Delta a_{\mu}^{Z^{\prime}} &= -\frac{m_{\mu}^2}{8 \pi^2 M_{Z^{\prime}}^2} \sum_{a=e,\mu,E} \left[ \left( \lvert \left( g_L \right)_{\mu a} \rvert^2 + \lvert \left( g_R \right)_{\mu a} \rvert^2 \right) F(x_a) + \frac{m_a}{m_\mu} \operatorname{Re} \left[ \left( g_L \right)_{\mu a} \left( g_R^{*} \right)_{\mu a} \right] G(x_a) \right],
\\
x_{a}&=\frac{m_{a}^{2}}{M_{Z^{\prime }}^{2}}.
\label{eqn:analytic_expression_for_muong2}
\end{split}
\end{equation}
\endgroup
Expanding the above equation in terms of electron, muon and vector-like lepton couplings as per $\operatorname{BR}\left(\mu \rightarrow e \gamma \right)$:
\begingroup
\begin{align}
\begin{split}
\Delta a_{\mu}^{Z^{\prime}} = -\frac{m_{\mu}^2}{8 \pi^2 M_{Z^{\prime}}^2} \bigg[ &\left( \lvert \left( g_L \right)_{\mu e} \rvert^2 + \lvert \left( g_R \right)_{\mu e} \rvert^2 \right) F(x_e) + \frac{m_e}{m_\mu} \operatorname{Re} \left[ \left( g_L \right)_{\mu e} \left( g_R^{*} \right)_{\mu e} \right] G(x_e) \\
&+ \left( \lvert \left( g_L \right)_{\mu \mu} \rvert^2 + \lvert \left( g_R \right)_{\mu \mu} \rvert^2 \right) F(x_\mu) + \frac{m_\mu}{m_\mu} \operatorname{Re} \left[ \left( g_L \right)_{\mu \mu} \left( g_R^{*} \right)_{\mu \mu} \right] G(x_\mu) \\
&+ \left( \lvert \left( g_L \right)_{\mu E} \rvert^2 + \lvert \left( g_R \right)_{\mu E} \rvert^2 \right) F(x_{E}) + \frac{M_4^C}{m_\mu} \operatorname{Re} \left[ \left( g_L \right)_{\mu E} \left( g_R^{*} \right)_{\mu E} \right] G(x_{E}) \bigg]
\label{eqn:expanded_expression_for_muong2}
\end{split}
\end{align}
\endgroup
The chirality-flipping mass is used in the last line of equation \eqref{eqn:expanded_expression_for_muong2} similarly to Equation \eqref{eqn:expansion_of_sigmas_with_each_family}. It then is possible to represent $\Delta a_\mu$ in terms of mixing angles.
\begingroup
\begin{align}
\begin{split}
\Delta a_{\mu}^{Z^{\prime}} &= -\frac{m_{\mu}^2}{8 \pi^2 M_{Z^{\prime}}^2} \bigg[ \bigg( \lvert \Big(\sin\theta_{12}^{L}\sin\theta_{24}^{L}+\cos\theta_{12}^{L}\cos\theta_{24}^{L}\sin\theta_{14}^{L}\Big)\Big(\cos\theta_{12}^{L}\sin\theta_{24}^{L}-\cos\theta_{24}^{L}\sin\theta_{12}^{L}\sin\theta_{14}^{L}\Big) \rvert^2 \\
&+ \lvert \Big(\sin\theta_{12}^{R}\sin\theta_{24}^{R}+\cos\theta_{12}^{R}\cos\theta_{24}^{R}\sin\theta_{14}^{R}\Big)\Big(\cos\theta_{12}^{R}\sin\theta_{24}^{R}-\cos\theta_{24}^{R}\sin\theta_{12}^{R}\sin\theta_{14}^{R}\Big) \rvert^2 \bigg) F(x_1) \\
&+ \frac{m_1}{m_2} \Big(\sin\theta_{12}^{L}\sin\theta_{24}^{L}+\cos\theta_{12}^{L}\cos\theta_{24}^{L}\sin\theta_{14}^{L}\Big)\Big(\cos\theta_{12}^{L}\sin\theta_{24}^{L}-\cos\theta_{24}^{L}\sin\theta_{12}^{L}\sin\theta_{14}^{L}\Big) \\
&\times \Big(\sin\theta_{12}^{R}\sin\theta_{24}^{R}+\cos\theta_{12}^{R}\cos\theta_{24}^{R}\sin\theta_{14}^{R}\Big)\Big(\cos\theta_{12}^{R}\sin\theta_{24}^{R}-\cos\theta_{24}^{R}\sin\theta_{12}^{R}\sin\theta_{14}^{R}\Big) G(x_1) \\	
&+ \bigg( \lvert \Big(\cos\theta_{12}^{L}\sin\theta_{24}^{L}-\cos\theta_{24}^{L}\sin\theta_{12}^{L}\sin\theta_{14}^{L}\Big)^2 \rvert^2 + \lvert \Big(\cos\theta_{12}^{R}\sin\theta_{24}^{R}-\cos\theta_{24}^{R}\sin\theta_{12}^{R}\sin\theta_{14}^{R}\Big)^2 \rvert^2 \bigg) F(x_2) \\
&+ \frac{m_2}{m_2} \Big(\cos\theta_{12}^{L}\sin\theta_{24}^{L}-\cos\theta_{24}^{L}\sin\theta_{12}^{L}\sin\theta_{14}^{L}\Big)^2 \Big(\cos\theta_{12}^{R}\sin\theta_{24}^{R}-\cos\theta_{24}^{R}\sin\theta_{12}^{R}\sin\theta_{14}^{R}\Big)^2 G(x_2) \\
&+ \bigg( \lvert \cos\theta_{14}^{L}\cos\theta_{24}^{L}\Big(\cos\theta_{12}^{L}\sin\theta_{24}^{L}-\cos\theta_{24}^{L}\sin\theta_{12}^{L}\sin\theta_{14}^{L}\Big) \rvert^2 \\
&+ \lvert \cos\theta_{14}^{R}\cos\theta_{24}^{R}\Big(\cos\theta_{12}^{R}\sin\theta_{24}^{R}-\cos\theta_{24}^{R}\sin\theta_{12}^{R}\sin\theta_{14}^{R}\Big) \rvert^2 \bigg) F(x_4) \\
&+ \frac{M_4^C}{m_2} \cos\theta_{14}^{L}\cos\theta_{24}^{L}\Big(\cos\theta_{12}^{L}\sin\theta_{24}^{L}-\cos\theta_{24}^{L}\sin\theta_{12}^{L}\sin\theta_{14}^{L}\Big) \\
&\times \cos\theta_{14}^{R}\cos\theta_{24}^{R}\Big(\cos\theta_{12}^{R}\sin\theta_{24}^{R}-\cos\theta_{24}^{R}\sin\theta_{12}^{R}\sin\theta_{14}^{R}\Big) G(x_4)
\label{eqn:expanded_expression_for_muong2_mixing_angles}
\end{split}
\end{align}
\endgroup
\section{Anomalous electron $g-2$}

The anomalous electron $g-2$ is given by:
\begingroup
\begin{equation}
\begin{split}
\Delta a_{e}^{Z^{\prime}} &= -\frac{m_{e}^2}{8 \pi^2 M_{Z^{\prime}}^2} \sum_{a=e,\mu,E} \left[ \left( \lvert \left( g_L \right)_{ea} \rvert^2 + \lvert \left( g_R \right)_{ea} \rvert^2 \right) F(x_a) + \frac{m_a}{m_e} \operatorname{Re} \left[ \left( g_L \right)_{ea} \left( g_R^{*} \right)_{ea} \right] G(x_a) \right],
\\
x_{a}&=\frac{m_{a}^{2}}{M_{Z^{\prime }}^{2}}.
\label{eqn:analytic_expression_for_electrong2}
\end{split}
\end{equation}
\endgroup
Expanding the above equation in terms of electron, muon and vector-like lepton as previously, the form is
\begingroup
\begin{align}
\begin{split}
\Delta a_{e}^{Z^{\prime}} = -\frac{m_{e}^2}{8 \pi^2 M_{Z^{\prime}}^2} \bigg[ &\left( \lvert \left( g_L \right)_{ee} \rvert^2 + \lvert \left( g_R \right)_{ee} \rvert^2 \right) F(x_e) + \frac{m_e}{m_e} \operatorname{Re} \left[ \left( g_L \right)_{ee} \left( g_R^{*} \right)_{ee} \right] G(x_e) \\
&+ \left( \lvert \left( g_L \right)_{e\mu} \rvert^2 + \lvert \left( g_R \right)_{e\mu} \rvert^2 \right) F(x_\mu) + \frac{m_\mu}{m_e} \operatorname{Re} \left[ \left( g_L \right)_{e\mu} \left( g_R^{*} \right)_{e\mu} \right] G(x_\mu) \\
&+ \left( \lvert \left( g_L \right)_{eE} \rvert^2 + \lvert \left( g_R \right)_{eE} \rvert^2 \right) F(x_{E}) + \frac{M_4^C}{m_e} \operatorname{Re} \left[ \left( g_L \right)_{eE} \left( g_R^{*} \right)_{eE} \right] G(x_{E}) \bigg]
\label{eqn:expanded_expression_for_electrong2}
\end{split}
\end{align}
\endgroup
The chirality-flipping mass is used in the last line of Equation \eqref{eqn:expanded_expression_for_electrong2} similarly to the Equations \eqref{eqn:expansion_of_sigmas_with_each_family} or \eqref{eqn:expanded_expression_for_muong2}. It then is possible to represent anomalous electron $g-2$ in terms of mixing angles.
\begingroup
\begin{align}
\begin{split}
\Delta a_{e}^{Z^{\prime}} &= -\frac{m_{e}^2}{8 \pi^2 M_{Z^{\prime}}^2} \bigg[ \bigg( \lvert \Big(\sin\theta_{12}^{L}\sin\theta_{24}^{L}+\cos\theta_{12}^{L}\cos\theta_{24}^{L}\sin\theta_{14}^{L}\Big)^2 \rvert^2 
\\
&+ \lvert \Big(\sin\theta_{12}^{R}\sin\theta_{24}^{R}+\cos\theta_{12}^{R}\cos\theta_{24}^{R}\sin\theta_{14}^{R}\Big)^2 \rvert^2 \bigg) F(x_1) \\
&+ \frac{m_1}{m_1} \Big(\sin\theta_{12}^{L}\sin\theta_{24}^{L}+\cos\theta_{12}^{L}\cos\theta_{24}^{L}\sin\theta_{14}^{L}\Big)^2 \Big(\sin\theta_{12}^{R}\sin\theta_{24}^{R}+\cos\theta_{12}^{R}\cos\theta_{24}^{R}\sin\theta_{14}^{R}\Big)^2 G(x_1) \\
&+ \bigg( \lvert \Big(\sin\theta_{12}^{L}\sin\theta_{24}^{L}+\cos\theta_{12}^{L}\cos\theta_{24}^{L}\sin\theta_{14}^{L}\Big)\Big(\cos\theta_{12}^{L}\sin\theta_{24}^{L}-\cos\theta_{24}^{L}\sin\theta_{12}^{L}\sin\theta_{14}^{L}\Big) \rvert^2 \\
&+ \lvert \Big(\sin\theta_{12}^{R}\sin\theta_{24}^{R}+\cos\theta_{12}^{R}\cos\theta_{24}^{R}\sin\theta_{14}^{R}\Big)\Big(\cos\theta_{12}^{R}\sin\theta_{24}^{R}-\cos\theta_{24}^{R}\sin\theta_{12}^{R}\sin\theta_{14}^{R}\Big) \rvert^2 \bigg) F(x_2) \\
&+ \frac{m_2}{m_1} \Big(\sin\theta_{12}^{L}\sin\theta_{24}^{L}+\cos\theta_{12}^{L}\cos\theta_{24}^{L}\sin\theta_{14}^{L}\Big)\Big(\cos\theta_{12}^{L}\sin\theta_{24}^{L}-\cos\theta_{24}^{L}\sin\theta_{12}^{L}\sin\theta_{14}^{L}\Big) \\
&\times \Big(\sin\theta_{12}^{R}\sin\theta_{24}^{R}+\cos\theta_{12}^{R}\cos\theta_{24}^{R}\sin\theta_{14}^{R}\Big)\Big(\cos\theta_{12}^{R}\sin\theta_{24}^{R}-\cos\theta_{24}^{R}\sin\theta_{12}^{R}\sin\theta_{14}^{R}\Big) G(x_2) \\
&+ \bigg( \lvert \cos\theta_{14}^{L}\cos\theta_{24}^{L}\Big(\sin\theta_{12}^{L}\sin\theta_{24}^{L}+\cos\theta_{12}^{L}\cos\theta_{24}^{L}\sin\theta_{14}^{L}\Big) \rvert^2 \\
&+ \lvert \cos\theta_{14}^{R}\cos\theta_{24}^{R}\Big(\sin\theta_{12}^{R}\sin\theta_{24}^{R}+\cos\theta_{12}^{R}\cos\theta_{24}^{R}\sin\theta_{14}^{R}\Big) \rvert^2 \bigg) F(x_4) \\
&+ \frac{M_4^C}{m_1} \cos\theta_{14}^{L}\cos\theta_{24}^{L}\Big(\sin\theta_{12}^{L}\sin\theta_{24}^{L}+\cos\theta_{12}^{L}\cos\theta_{24}^{L}\sin\theta_{14}^{L}\Big) \\
&\times \cos\theta_{14}^{R}\cos\theta_{24}^{R}\Big(\sin\theta_{12}^{R}\sin\theta_{24}^{R}+\cos\theta_{12}^{R}\cos\theta_{24}^{R}\sin\theta_{14}^{R}\Big)G(x_4)
\label{eqn:expanded_expression_for_electrong2_mixing_angles}
\end{split}
\end{align}
\endgroup
\section{Neutrino trident}

The constraint from neutrino trident has a much simpler form compared to the other observables, as it only depends on coupling of the heavy $Z'$ to two muons.
\begingroup
\begin{equation}
\begin{split}
&\frac{\left( g_L \right)_{\mu\mu}^2 + \left( g_L \right)_{\mu\mu} \left( g_R \right)_{\mu\mu}}{M_{Z^{\prime}}^2} = \frac{\Big(\cos\theta_{12}^{L}\sin\theta_{24}^{L}-\cos\theta_{24}^{L}\sin\theta_{12}^{L}\sin\theta_{14}^{L}\Big)^4}{M_{Z^\prime}^2} \\ 
	&+ \frac{\Big(\cos\theta_{12}^{L}\sin\theta_{24}^{L}-\cos\theta_{24}^{L}\sin\theta_{12}^{L}\sin\theta_{14}^{L}\Big)^2 \Big(\cos\theta_{12}^{R}\sin\theta_{24}^{R}-\cos\theta_{24}^{R}\sin\theta_{12}^{R}\sin\theta_{14}^{R}\Big)^2}{M_{Z^{\prime}}^2}
\end{split}
\end{equation}
\endgroup
\chapter{Fermion mass hierarchies from vector-like families with an extended 2HDM and a possible explanation for the electron and muon anomalous magnetic moments}
We discuss the whole description of the diagonalization for the quark sector mass matrices in two bases and heavy scalar production at a proton-proton collider.
\section{Quark mass matrices in two bases}
\label{A}
As the lepton mass matrix is constructed in main body of this work, the
quark sector can be built in a similar way. Like the lepton sector, we make
use of two approaches to an effective lepton mass matrix, one of which is a
convenient basis and the other is a decoupling basis.
\subsection{A convenient basis for quarks}
Consider the $7 \times 7$ quark mass matrix rotated as in the lepton sector.
\begingroup
\begin{equation}
\begin{split}
M^{u}& =\left( 
\begin{array}{c|ccccccc}
& u_{1R} & u_{2R} & u_{3R} & u_{4R} & u_{5R} & \widetilde{Q}_{4R} & 
\widetilde{Q}_{5R} \\ \hline
\overline{Q}_{1L} & 0 & 0 & 0 & 0 & y_{15}^{u}v_{u} & 0 & x_{15}^{Q}v_{\phi }
\\ 
\overline{Q}_{2L} & 0 & 0 & 0 & y_{24}^{u}v_{u} & y_{25}^{u}v_{u} & 0 & 
x_{25}^{Q}v_{\phi } \\ 
\overline{Q}_{3L} & 0 & 0 & 0 & y_{34}^{u}v_{u} & y_{35}^{u}v_{u} & 
x_{34}^{Q}v_{\phi } & x_{35}^{Q}v_{\phi } \\ 
\overline{Q}_{4L} & 0 & 0 & y_{43}^{u}v_{u} & 0 & 0 & M_{44}^{Q} & M_{45}^{Q}
\\ 
\overline{Q}_{5L} & y_{51}^{u}v_{u} & y_{52}^{u}v_{u} & y_{53}^{u}v_{u} & 0
& 0 & 0 & M_{55}^{Q} \\ 
\overline{\widetilde{u}}_{4L} & 0 & x_{42}^{u}v_{\phi } & x_{43}^{u}v_{\phi }
& M_{44}^{u} & 0 & 0 & 0 \\ 
\overline{\widetilde{u}}_{5L} & x_{51}^{u}v_{\phi } & x_{52}^{u}v_{\phi } & 
x_{53}^{u}v_{\phi } & M_{54}^{u} & M_{55}^{u} & 0 & 0 \\ 
&  &  &  &  &  &  & 
\end{array}%
\right)  \\
M^{d}& =\left( 
\begin{array}{c|ccccccc}
& d_{1R} & d_{2R} & d_{3R} & d_{4R} & d_{5R} & \widetilde{Q}_{4R} & 
\widetilde{Q}_{5R} \\ \hline
\overline{Q}_{1L} & 0 & 0 & 0 & y_{14}^{d}v_{d} & y_{15}^{d}v_{d} & 0 & 
x_{15}^{Q}v_{\phi } \\ 
\overline{Q}_{2L} & 0 & 0 & 0 & y_{24}^{d}v_{d} & y_{25}^{d}v_{d} & 0 & 
x_{25}^{Q}v_{\phi } \\ 
\overline{Q}_{3L} & 0 & 0 & 0 & y_{34}^{d}v_{d} & y_{35}^{d}v_{d} & 
x_{34}^{Q}v_{\phi } & x_{35}^{Q}v_{\phi } \\ 
\overline{Q}_{4L} & 0 & 0 & y_{43}^{d}v_{d} & 0 & 0 & M_{44}^{Q} & M_{45}^{Q}
\\ 
\overline{Q}_{5L} & y_{51}^{d}v_{d} & y_{52}^{d}v_{d} & y_{53}^{d}v_{d} & 0
& 0 & 0 & M_{55}^{Q} \\ 
\overline{\widetilde{d}}_{4L} & 0 & x_{42}^{d}v_{\phi } & x_{43}^{d}v_{\phi }
& M_{44}^{d} & 0 & 0 & 0 \\ 
\overline{\widetilde{d}}_{5L} & x_{51}^{d}v_{\phi } & x_{52}^{d}v_{\phi } & 
x_{53}^{d}v_{\phi } & M_{54}^{d} & M_{55}^{d} & 0 & 0 \\ 
\end{array}%
\right) 
\end{split}%
\label{quark}
\end{equation}
\endgroup
Notice that the same rotations operated in the lepton sector is applied to
both up- and down-type quark sector except for $y_{14}^{d}$ since quark
doublet rotation is already used in the up-type quark sector. These two mass
matrices clearly tells that this model is an extended 2HDM in that the
up-type SM Higgs $H_u$ corresponds to up-type quark sector, while the
down-type SM Higgs $H_d$ corresponds for down-type quark sector. 
\subsection{A basis for decoupling heavy fourth and fifth vector-like family}
In this section, we treat the decoupling basis with quarks holding
an assumption $\left\langle \phi \right\rangle \approx M_{44}^Q$.
As in the charged lepton mass matrix, we can obtain the Yukawa matrix from
the $5\times 5$ upper blocks of Equation \ref{quark},
\begingroup
\begin{equation}
\widetilde{y}_{\alpha\beta}^{u} = 
\begin{pmatrix}
0 & 0 & 0 & 0 & y_{15}^{u} \\ 
0 & 0 & 0 & y_{24}^{u} & y_{25}^{u} \\ 
0 & 0 & 0 & y_{34}^{u} & y_{35}^{u} \\ 
0 & 0 & y_{43}^{u} & 0 & 0 \\ 
y_{51}^{u} & y_{52}^{u} & y_{53}^{u} & 0 & 0%
\end{pmatrix}%
, \hspace{0.1cm} \widetilde{y}_{\alpha\beta}^{d} = 
\begin{pmatrix}
0 & 0 & 0 & y_{14}^{d} & y_{15}^{d} \\ 
0 & 0 & 0 & y_{24}^{d} & y_{25}^{d} \\ 
0 & 0 & 0 & y_{34}^{d} & y_{35}^{d} \\ 
0 & 0 & y_{43}^{d} & 0 & 0 \\ 
y_{51}^{d} & y_{52}^{d} & y_{53}^{d} & 0 & 0%
\end{pmatrix}%
\end{equation}
\endgroup
where $\alpha$ and $\beta$ run from $1$ to $5$. The Yukawa matrices $%
\widetilde{y}_{\alpha \beta}^{u,d}$ can be diagonalized by the unitary
rotations $V$
\begingroup
\begin{equation}
\begin{split}
V_Q &= V_{45}^Q V_{35}^Q V_{25}^Q V_{15}^Q V_{34}^Q V_{24}^Q V_{14}^Q, 
\\
V_{u} &= V_{45}^{u} V_{35}^{u} V_{25}^{u} V_{15}^{u}
V_{34}^{u} V_{24}^{u} V_{14}^{u},
\\ 
V_{d} &= V_{45}^{d}
V_{35}^{d} V_{25}^{d} V_{15}^{d} V_{34}^{d} V_{24}^{d} V_{14}^{d}
\label{eqn:unitary_mixing_matrix}
\end{split}
\end{equation}
\endgroup
where each of the unitary matrices $V_{i4,5}$ are parameterized by a single
angle $\theta_{i4,5}$ featuring the mixing between the $i$th SM chiral quark
and the $4,5$th vector-like quark. In the rotated mass matrix, we need $%
(3,4), (1,5), (2,5), (3,5)$ mixing in the $Q$ sector and $(2,4), (3,4),
(1,5), (2,5), (3,5)$ mixing in the $u, d$ sectors to go to the decoupling
basis therefore the unitary mixing matrices $V$ are defined to be
\begingroup
\begin{equation}
\begin{split}
V_Q &= V_{35}^{Q} V_{25}^{Q} V_{15}^{Q} V_{34}^{Q} \\
&= 
\begin{pmatrix}
1 & 0 & 0 & 0 & 0 \\ 
0 & 1 & 0 & 0 & 0 \\ 
0 & 0 & c_{35}^Q & 0 & s_{35}^Q \\ 
0 & 0 & 0 & 1 & 0 \\ 
0 & 0 & -s_{35}^Q & 0 & c_{35}^Q%
\end{pmatrix}
\begin{pmatrix}
1 & 0 & 0 & 0 & 0 \\ 
0 & c_{25}^Q & 0 & 0 & s_{25}^Q \\ 
0 & 0 & 1 & 0 & 0 \\ 
0 & 0 & 0 & 1 & 0 \\ 
0 & -s_{25}^Q & 0 & 0 & c_{25}^Q%
\end{pmatrix}
\begin{pmatrix}
c_{15}^Q & 0 & 0 & 0 & s_{15}^Q \\ 
0 & 1 & 0 & 0 & 0 \\ 
0 & 0 & 1 & 0 & 0 \\ 
0 & 0 & 0 & 1 & 0 \\ 
-s_{15}^Q & 0 & 0 & 0 & c_{15}^Q%
\end{pmatrix}
\\
& \times
\begin{pmatrix}
1 & 0 & 0 & 0 & 0 \\ 
0 & 1 & 0 & 0 & 0 \\ 
0 & 0 & c_{34}^Q & s_{34}^Q & 0 \\ 
0 & 0 & -s_{34}^Q & c_{34}^Q & 0 \\ 
0 & 0 & 0 & 0 & 1%
\end{pmatrix}%
\approx 
\begin{pmatrix}
1 & 0 & 0 & 0 & s_{15}^Q \\ 
0 & 1 & 0 & 0 & s_{25}^Q \\ 
0 & 0 & 1 & s_{34}^Q & s_{35}^Q \\ 
0 & 0 & -s_{34}^Q & 1 & 0 \\ 
-s_{15}^Q & -s_{25}^Q & -s_{15}^Q & 0 & 1%
\end{pmatrix}%
, \\
&s_{34}^Q = \frac{x_{34}^Q \left\langle \phi \right\rangle}{\sqrt{\left(
x_{34}^Q \left\langle \phi \right\rangle \right)^2 + \left( M_{44}^Q
\right)^2}}, \hspace{0.5cm} s_{15}^Q = \frac{x_{15}^Q \left\langle \phi
\right\rangle}{\sqrt{\left( x_{15}^Q \left\langle \phi \right\rangle
\right)^2 + \left( M_{55}^Q \right)^2}}, \\
&s_{25}^Q = \frac{x_{25}^Q \left\langle \phi \right\rangle}{\sqrt{\left(
x_{25}^Q \left\langle \phi \right\rangle \right)^2 + \left( M_{55}^{\prime
Q} \right)^2}}, \hspace{0.5cm} s_{35}^Q = \frac{x_{35}^{\prime Q}
\left\langle \phi \right\rangle}{\sqrt{\left( x_{35}^{\prime Q} \left\langle
\phi \right\rangle \right)^2 + \left( M_{55}^{\prime\prime Q} \right)^2}}, \\
& x_{35}^{\prime Q} \left\langle \phi \right\rangle = c_{34}^{Q} x_{35}^{Q}
\left\langle \phi \right\rangle + s_{34}^{Q} M_{45}^{Q}, \hspace{0.5cm}
M_{45}^{\prime Q} = -s_{34}^{Q} x_{35}^{Q} \left\langle \phi \right\rangle +
c_{34}^{Q} M_{45}^{Q} \\
&\widetilde{M}_{44}^{Q} = \sqrt{\left( x_{34}^Q \left\langle \phi
\right\rangle \right)^2 + \left( M_{44}^Q \right)^2}, \\
&M_{55}^{\prime Q} = \sqrt{\left( x_{15}^Q \left\langle \phi \right\rangle
\right)^2 + \left( M_{55}^Q \right)^2}, \hspace{0.1cm} M_{55}^{\prime\prime
Q} = \sqrt{\left( x_{25}^Q \left\langle \phi \right\rangle \right)^2 +
\left( M_{55}^{\prime Q} \right)^2}, \hspace{0.1cm} 
\\
&\widetilde{M}_{55}^{Q} = 
\sqrt{\left( x_{35}^{\prime Q} \left\langle \phi \right\rangle \right)^2 +
\left( M_{55}^{\prime\prime Q} \right)^2}
\end{split}%
\end{equation}
\endgroup
\begingroup
\begin{equation}
\begin{split}
V_{u} &= V_{35}^{u} V_{25}^{u} V_{15}^{u} V_{34}^{u} V_{24}^{u} \\
&= 
\begin{pmatrix}
1 & 0 & 0 & 0 & 0 \\ 
0 & 1 & 0 & 0 & 0 \\ 
0 & 0 & c_{35}^{u} & 0 & s_{35}^{u} \\ 
0 & 0 & 0 & 1 & 0 \\ 
0 & 0 & -s_{35}^{u} & 0 & c_{35}^{u}%
\end{pmatrix}
\begin{pmatrix}
1 & 0 & 0 & 0 & 0 \\ 
0 & c_{25}^{u} & 0 & 0 & s_{25}^{u} \\ 
0 & 0 & 1 & 0 & 0 \\ 
0 & 0 & 0 & 1 & 0 \\ 
0 & -s_{25}^{u} & 0 & 0 & c_{25}^{u}%
\end{pmatrix}
\begin{pmatrix}
c_{15}^{u} & 0 & 0 & 0 & s_{15}^{u} \\ 
0 & 1 & 0 & 0 & 0 \\ 
0 & 0 & 1 & 0 & 0 \\ 
0 & 0 & 0 & 1 & 0 \\ 
-s_{15}^{u} & 0 & 0 & 0 & c_{15}^{u}%
\end{pmatrix}
\\
& \times 
\begin{pmatrix}
1 & 0 & 0 & 0 & 0 \\ 
0 & 1 & 0 & 0 & 0 \\ 
0 & 0 & c_{34}^{u} & s_{34}^{u} & 0 \\ 
0 & 0 & -s_{34}^{u} & c_{34}^{u} & 0 \\ 
0 & 0 & 0 & 0 & 1%
\end{pmatrix}
\begin{pmatrix}
1 & 0 & 0 & 0 & 0 \\ 
0 & c_{24}^{u} & 0 & s_{24}^{u} & 0 \\ 
0 & 0 & 1 & 0 & 0 \\ 
0 & -s_{24}^{u} & 0 & c_{24}^{u} & 0 \\ 
0 & 0 & 0 & 0 & 1%
\end{pmatrix}
\approx 
\begin{pmatrix}
1 & 0 & 0 & 0 & \theta_{15}^{u} \\ 
0 & 1 & 0 & \theta_{24}^{u} & \theta_{25}^{u} \\ 
0 & 0 & 1 & \theta_{34}^{u} & \theta_{35}^{u} \\ 
0 & -\theta_{24}^{u} & -\theta_{34}^{u} & 1 & 0 \\ 
-\theta_{15}^{u} & -\theta_{25}^{u} & -\theta_{35}^{u} & 0 & 1%
\end{pmatrix}%
, \\
&s_{24}^{u} \approx \frac{x_{42}^{u} \left\langle \phi \right\rangle}{%
M_{44}^{u}}, \hspace{0.5cm} s_{34}^{u} \approx \frac{x_{43}^{u} \left\langle
\phi \right\rangle}{M_{44}^{\prime u}}, s_{15}^{u} \approx \frac{x_{51}^{u}
\left\langle \phi \right\rangle}{M_{55}^{u}}, \hspace{0.5cm} s_{25}^{u}
\approx \frac{x_{52}^{\prime u} \left\langle \phi \right\rangle}{%
M_{55}^{\prime u}}, \hspace{0.5cm} s_{35}^{u} \approx \frac{x_{53}^{u}
\left\langle \phi \right\rangle}{M_{55}^{\prime\prime u}}, \\
& x_{52}^{\prime u} \left\langle \phi \right\rangle = c_{24}^{u} x_{52}^{u}
\left\langle \phi \right\rangle + s_{24}^{u} M_{54}^{u}, \hspace{0.5cm}
M_{54}^{\prime u} = -s_{24}^{u} x_{52}^{u} \left\langle \phi \right\rangle +
c_{24}^{u} M_{54}^{u}, \\
& x_{53}^{\prime u} \left\langle \phi \right\rangle = c_{34}^{u} x_{53}^{u}
\left\langle \phi \right\rangle + s_{34}^{u} M_{54}^{\prime u}, \hspace{0.5cm%
} M_{54}^{\prime\prime u} = -s_{34}^{u} x_{53}^{u} \left\langle \phi
\right\rangle + c_{34}^{u} M_{54}^{\prime u}, \\
&M_{44}^{\prime u} = \sqrt{\left( x_{42}^{u} \left\langle \phi \right\rangle
\right)^2 + \left( M_{44}^{u} \right)^2} \hspace{0.5cm}, \widetilde{M}%
_{44}^{u} = \sqrt{\left( x_{43}^{u} \left\langle \phi \right\rangle
\right)^2 + \left( M_{44}^{u} \right)^2}, \\
&M_{55}^{\prime u} = \sqrt{\left( x_{51}^{u} \left\langle \phi \right\rangle
\right)^2 + \left( M_{55}^{u} \right)^2}, \hspace{0.1cm} M_{55}^{\prime%
\prime u} = \sqrt{\left( x_{52}^{\prime u} \left\langle \phi \right\rangle
\right)^2 + \left( M_{55}^{\prime u} \right)^2}, \hspace{0.1cm} 
\\
&\widetilde{M}%
_{55}^{u} = \sqrt{\left( x_{53}^{\prime u} \left\langle \phi \right\rangle
\right)^2 + \left( M_{55}^{\prime\prime u} \right)^2}.
\end{split}%
\end{equation}
\endgroup
With the defined unitary mixing matrices in place, the $5 \times 5$ Yukawa
matrices in a mass basis (primed) are transformed by
\begingroup
\begin{equation}
\widetilde{y}_{\alpha\beta}^{\prime u} = V_Q \widetilde{y}_{\alpha\beta}^{u}
V_{u}^{\dagger}, \hspace{0.1cm} \widetilde{y}_{\alpha\beta}^{\prime d} = V_Q 
\widetilde{y}_{\alpha\beta}^{d} V_{d}^{\dagger},
\label{eqn:effective_Yukawa_constant}
\end{equation}
\endgroup
where tilde with prime means interaction basis whereas tilde alone corresponds to the mass
basis. The effective SM Yukawa couplings for the quarks then correspond to
the $3 \times 3$ upper block of $\widetilde{y}_{\alpha\beta}^{\prime u},%
\widetilde{y}_{\alpha\beta}^{\prime d}$, namely
\begingroup
\begin{equation}
y_{ij}^{u} \widetilde{H}_u \overline{Q}_{iL} u_{jR}, \hspace{0.1cm} y_{ij}^{d} 
\widetilde{H}_d \overline{Q}_{iL} d_{jR}, \hspace{0.1cm} \text{ with } y_{ij}^{u}
\equiv \widetilde{y}_{ij}^{\prime u}, \hspace{0.1cm} y_{ij}^{d} \equiv \widetilde{y}%
_{ij}^{\prime d}, \hspace{0.1cm} (i,j=1,2,3).
\end{equation}
\endgroup
The $3 \times 3$ SM Yukawa matrices for up- and down-type quark sector read:
\begingroup
\begin{equation}
\begin{split}
y_{ij}^u &= 
\begin{pmatrix}
s_{15}^Q y_{51}^u + y_{15}^u \theta_{15}^{u} & s_{15}^Q y_{52}^u + y_{15}^u
\theta_{25}^{u} & s_{15}^Q y_{53}^u + y_{15}^u \theta_{35}^{u} \\ 
s_{25}^Q y_{51}^u + y_{25}^u \theta_{15}^{u} & s_{25}^Q y_{52}^u + y_{24}^u
\theta_{24}^{u} + y_{25}^u \theta_{25}^{u} & s_{25}^Q y_{53}^u + y_{24}^u
\theta_{34}^{u} + y_{25}^u \theta_{35}^{u} \\ 
s_{35}^Q y_{51}^u + y_{35}^u \theta_{15}^{u} & s_{35}^Q y_{52}^u + y_{34}^u
\theta_{24}^{u} + y_{35}^u \theta_{25}^{u} & s_{34}^Q y_{43}^u + s_{35}^Q
y_{53}^u + y_{34}^u \theta_{34}^{u} + y_{35}^u \theta_{35}^{u}%
\end{pmatrix}
\\
y_{ij}^d &= 
\begin{pmatrix}
s_{15}^Q y_{51}^d + y_{15}^d \theta_{15}^{d} & s_{15}^Q y_{52}^d + y_{14}^d
\theta_{24}^{d} + y_{15}^d \theta_{25}^{d} & s_{15}^Q y_{53}^d + y_{14}^d
\theta_{34}^{d} + y_{15}^d \theta_{35}^{d} \\ 
s_{25}^Q y_{51}^d + y_{25}^d \theta_{15}^{d} & s_{25}^Q y_{52}^d + y_{24}^d
\theta_{24}^{d} + y_{25}^d \theta_{25}^{d} & s_{25}^Q y_{53}^d + y_{24}^d
\theta_{34}^{d} + y_{25}^d \theta_{35}^{d} \\ 
s_{35}^Q y_{51}^d + y_{35}^d \theta_{15}^{d} & s_{35}^Q y_{52}^d + y_{34}^d
\theta_{24}^{d} + y_{35}^d \theta_{25}^{d} & s_{34}^Q y_{43}^d + s_{35}^Q
y_{53}^d + y_{34}^d \theta_{34}^{d} + y_{35}^d \theta_{35}^{d}%
\end{pmatrix}%
\end{split}%
\end{equation}
\endgroup
\section{Heavy scalar production at a proton-proton collider}
\label{B}
We have confirmed that the mass of the non-SM CP even scalar $H_{1}$ is ranged from $%
200$ to $240\func{GeV}$ in Table \ref{tab:range_secondscan} and this light
mass of $H_{1}$ has not been observed at CERN or other experiments so far.
In order to see how big an impact of $H_{1}$ is when compared to that of SM
Higgs $h$, we studied a total cross section for the SM process $pp\rightarrow h$ and for BSM process $pp \rightarrow H_1$. The SM cross section for $pp\rightarrow h$ process is
\begingroup
\begin{equation}
\sigma_{\func{SM}} = \frac{\alpha_S^2 m_h^2}{64\pi v^2} \left( L\left( \frac{%
m_h^2}{m_t^2} \right) \right)^2 \frac{1}{S} \int_{\ln \sqrt{m_h^2/S}}^{-\ln 
\sqrt{m_h^2/S}} \func{PDF}(0,x_1(y),m_h) \func{PDF}(0,x_2(y),m_h) dy
\label{eqn:total_cross_section_pptoh}
\end{equation}
\endgroup
where $L$ is a loop integral
\begingroup
\begin{equation}
\begin{split}
L(a) &= \lvert \big[ 2a + (-4+a) \func{PolyLog}\left( 2,1/2\left( -\sqrt{-4+a%
} \sqrt{a} + a \right) \right) \\
&+ (-4+a) \func{PolyLog}\left( 2,1/2\left( \sqrt{-4+a} \sqrt{a} + a \right)
\right) \big]/a^2 \rvert,
\end{split}%
\end{equation}
\endgroup
$\alpha _{S}$ is the strong coupling constant, $v$ is the conventional SM Higgs vev $246.22\func{%
GeV}$, $m_{h}$ is the Higgs mass $125\func{GeV}$, $m_{t}$
is the top quark mass $173\func{GeV}$, $S$ is the squared LHC center of mass
energy $\left( 14\func{TeV}\right) ^{2}$, $\func{PDF}$ corresponds to the
parton distribution function where $0$ means 0th parton -
gluon, $x$ is the momentum fraction of the proton carried out by the gluon. 
Here the factorization scale has been taken to be equal to the SM like Higgs
boson mass $m_{h}$ 
and $x_{1,2}(y)$ are defined as follows:
\begingroup
\begin{equation}
x_1(y) = \frac{\sqrt{m_h^2/S}}{S} \exp(y), \quad x_2(y) = \frac{\sqrt{m_h^2/S%
}}{S} \exp(-y).
\end{equation}
\endgroup
With these defined functions and values, the total cross section for $pp
\rightarrow h$ is
\begingroup
\begin{equation}
\sigma_{\func{SM}} \simeq 18 \func{pb}.
\end{equation}
\endgroup
Next, the total cross section for $pp \rightarrow H_1$ process is
\begingroup
\begin{equation}
\begin{split}
\sigma \left( pp\rightarrow H_{1}\right) &=\frac{\alpha
_{S}^{2}m_{H_{1}}^{2}a_{hbb}^{2}}{64\pi v_2^2}\left( L\left( \frac{%
m_{H_{1}}^{2}}{m_{b}^{2}}\right) \right) ^{2}\frac{1}{S}\int_{\ln \sqrt{%
m_{H_{1}}^{2}/S}}^{-\ln \sqrt{m_{H_{1}}^{2}/S}}
\\
& \times \func{PDF}(0,x_{1}^{\prime
}(y),m_{H_{1}})\func{PDF}(0,x_{2}^{\prime }(y),m_{H_{1}})dy
\label{eqn:total_cross_section_pptoH1}
\end{split}
\end{equation}
\endgroup
where $m_{H_1}$ is mass of non-SM CP even scalar $H_1$, and $x_{1,2}^\prime$
are defined in a similar way:
\begingroup
\begin{equation}
x_1^\prime (y) = \frac{\sqrt{m_{H_1}^2/S}}{S} \exp(y), \quad x_2^\prime (y)
= \frac{\sqrt{m_{H_1}^2/S}}{S} \exp(-y)
\end{equation}
\endgroup
One main distinction between Equation \ref{eqn:total_cross_section_pptoh}
and Equation \ref{eqn:total_cross_section_pptoH1} is the non-SM scalar $H_1$
only interacts with down-type quark pair $b\bar{b}$ since it is a mixed
state between $h_d^0$ and $\phi$ while the SM Higgs $h$ can interact with
top-quark pair $t\bar{t}$. According to the mass range of $H_1$ reported in
Table \ref{tab:range_secondscan}, the total cross section for $pp
\rightarrow H_1$ is given in Figure \ref{fig:pptoH1at14TeV}.
\begingroup
\begin{figure}[H]
\centering
\includegraphics[keepaspectratio,scale=0.5]{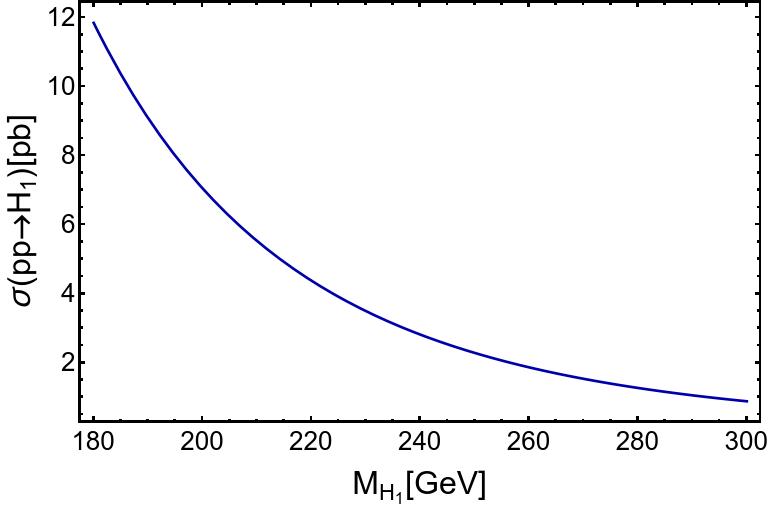}
\caption{The total cross section for $pp \rightarrow H_1$ at $14\func{TeV}$}
\label{fig:pptoH1at14TeV}
\end{figure}
\endgroup
The total cross section for $pp \rightarrow H_1$ runs
from nearly $8\func{pb}$ at $200 \func{GeV}$ to smaller values as mass of $%
H_1$ increases. The order of magnitude of this cross section for $pp \rightarrow H_1$ is
compatible to that of the SM process $pp \rightarrow h$, however the BSM
process is strongly suppressed since its single LHC production via gluon fusion mechanism is dominated by the triangular bottom quark loop as mentioned in Section \ref{sec:Numerical_analysis_of_scalars}. Therefore, our prediction with the
light non-SM scalar $H_1$ is possible to accommodate each anomaly constraint
at $1\sigma$.
\chapter{Constraining Vector-like fermion masses from $Z$ mediated FCNC observables in an extended 2HDM}
In this appendix, we discuss an analytic perturbative step-by-step diagonalization for each sector from Appendix~\ref{app:A} to \ref{app:C}. And then the numerical mixing matrix for each sector will be compared to its analytic result from Appendix~\ref{app:D} to \ref{app:E}, verifying the SM $Z$ physics is basis independent.
\section{Analytic approximated step-by-step diagonalization for the charged lepton sector} \label{app:A}
In order to diagonalize the mass matrix of Equation~\ref{eqn:cl_1} in an analytic way, we employ the method of 
mixing formalism and define intermediate mass basis.
The flavor basis is used when writing 
the initial mass matrix of Equation~\ref{eqn:cl_1}, whereas 
the true mass basis corresponds to 
 the fully diagonalized mass matrix, which reveals the masses of all propagating charged leptons. The intermediate mass basis is a basis where the heavy particles appearing in all terms generating the entries proportional to $v_{\phi}$ are integrated out remaining other terms unrotated. This separation makes the difference between $SU(2)$ conserving and $SU(2)$ violating mixings, which will be defined later, clear, which will become important when we consider the flavor violating interactions mediated by the SM $Z$ gauge boson. Before we carry out the digonalization step-by-step, it is convenient to rearrange the mass matrix of Equation~\ref{eqn:cl_1} by switching the Yukawa terms by mass parameters and by swapping the fourth and fifth column in order to make the heavy vector-like masses locate in the diagonal positions as given in Equation~\ref{eqn:cl_2s}
\begin{equation}
M^{e }
=
\left( 
\begin{array}{c|ccccc}
& e _{1R} & e _{2R} & e _{3R} & e _{4R} & \widetilde{L }_{4R} \\[0.5ex] \hline
\overline{L }_{1L} & 0 & 0 & 0 & 0 & 0 \\[1ex]
\overline{L }_{2L} & 0 & 0 & 0 & m_{24} & 0 \\[1ex] 
\overline{L }_{3L} & 0 & 0 & 0 & m_{34} & m_{35} \\[1ex]
\overline{L }_{4L} & 0 & 0 & m_{43} & 0 & M_{45}^{L } \\[1ex]
\overline{\widetilde{e }}_{4L} & 0 & m_{52} & m_{53} & M_{54}^{e} & 0 \\ 
\end{array}%
\right)
=
\left( 
\begin{array}{c|ccccc}
& e _{1R} & e _{2R} & e _{3R} & \widetilde{L }_{4R} & e _{4R} \\[0.5ex] \hline
\overline{L }_{1L} & 0 & 0 & 0 & 0 & 0 \\[1ex]
\overline{L }_{2L} & 0 & 0 & 0 & 0 & m_{24} \\[1ex] 
\overline{L }_{3L} & 0 & 0 & 0 & m_{35} & m_{34} \\[1ex]
\overline{L }_{4L} & 0 & 0 & m_{43} & M_{45}^{L } & 0 \\[1ex]
\overline{\widetilde{e }}_{4L} & 0 & m_{52} & m_{53} & 0 & M_{54}^{e} \\ 
\end{array}%
\right), 
\label{eqn:cl_2s}
\end{equation}
where the indices running from $1$ to $3$ correspond to the three SM families, the index $4$ labels the  
fourth vector-like particles 
and lastly the index $5$ denotes 
the tilde particles, which are a partner of the vector-like particles. Now we are ready to diagonalize the mass matrix of Equation~\ref{eqn:cl_2s} and the first step is to get the intermediate mass basis and to integrate out the particles generating the entries proportional to 
$v_{\phi}$ 
(equally, all mass terms involving index $5$). For this task, we first consider $34$ rotation in the left-handed fields to turn off the mass term $m_{35}$.
\begin{gather}
V_{34}^{L} M^{e }
=
\left( 
\begin{array}{c|ccccc}
& e _{1R} & e _{2R} & e _{3R} & \widetilde{L }_{4R} & e _{4R} \\[0.5ex] \hline
\overline{L }_{1L} & 0 & 0 & 0 & 0 & 0 \\[1ex]
\overline{L }_{2L} & 0 & 0 & 0 & 0 & m_{24} \\[1ex] 
\overline{L }_{3L}^{\prime} & 0 & 0 & -\frac{m_{35} m_{43}}{M_{45}^{L \prime}} & 0 & \frac{m_{34} M_{45}^{L}}{M_{45}^{L \prime}} \\[1ex]
\overline{L }_{4L}^{\prime} & 0 & 0 & \frac{m_{43} M_{45}^{L}}{M_{45}^{L \prime}} & M_{45}^{L \prime} & \frac{m_{34} m_{35}^{L}}{M_{45}^{L \prime}} \\[1ex]
\overline{\widetilde{e }}_{4L} & 0 & m_{52} & m_{53} & 0 & M_{54}^{e} \\ 
\end{array}%
\right), \\[2ex]
M_{45}^{L \prime} = \sqrt{M_{45}^{L 2} + m_{35}^2}, \quad s_{34}^{L} = \frac{m_{35}}{M_{45}^{L \prime}}, \quad c_{34}^{L} = \frac{M_{45}^{L}}{M_{45}^{L \prime}}, \quad V_{34}^{L} = \begin{pmatrix}
1 & 0 & 0 & 0 & 0 \\[1ex]
0 & 1 & 0 & 0 & 0 \\[1ex]
0 & 0 & c_{34}^{L} & -s_{34}^{L} & 0 \\[1ex]
0 & 0 & s_{34}^{L} & c_{34}^{L} & 0 \\[1ex]
0 & 0 & 0 & 0 & 1
\end{pmatrix},
\label{eqn:mm_cl_LH_34}
\end{gather}
where the primed fields correspond to the rotated fields. 
Throughout this whole work, the fields characterized by a capital letter are the ones belonging to a 
$SU(2)$ doublet under the SM gauge symmetry, whereas those ones denoted by a small letter are $SU(2)$ singlets. Then, the next rotation is $34$ rotation in the right-handed leptonic fields to turn off the $m_{53}$ entry. It is worth mentioning that 
the order used in the rotation of the left-handed fields is $12345$, whereas for the rotation of the right-handed fields the corresponding order is $12354$ since for consistency reasons we assigned the index $5$ for the tilde particle, 

\begin{gather}
V_{34}^{L} M^{e } (V_{34}^{e})^{\dagger}
=
\left( 
\begin{array}{c|ccccc}
& e _{1R} & e _{2R} & e _{3R}^{\prime} & \widetilde{L }_{4R} & e _{4R}^{\prime} \\[0.5ex] \hline
\overline{L }_{1L} & 0 & 0 & 0 & 0 & 0 \\[1ex]
\overline{L }_{2L} & 0 & 0 & -\frac{m_{24} m_{53}}{M_{54}^{e \prime}} & 0 & \frac{m_{24} M_{54}^{e}}{M_{54}^{e \prime}} \\[1ex] 
\overline{L }_{3L}^{\prime} & 0 & 0 & \frac{-m_{34} m_{53} M_{45}^{L} -m_{35} m_{43} M_{54}^{e}}{M_{45}^{L \prime} M_{54}^{e \prime}} & 0 & \frac{-m_{35} m_{43} m_{53} +m_{34} M_{45}^{L} M_{54}^{e}}{M_{45}^{L \prime} M_{54}^{e \prime}} \\[1ex]
\overline{L }_{4L}^{\prime} & 0 & 0 & \frac{-m_{34} m_{35} m_{53} +m_{43} M_{45}^{L} M_{54}^{e}}{M_{45}^{L \prime} M_{54}^{e \prime}} & M_{45}^{L \prime} & \frac{m_{43} m_{53} M_{45}^{L} +m_{34} m_{35} M_{54}^{e}}{M_{45}^{L \prime} M_{54}^{e \prime}} \\[1ex]
\overline{\widetilde{e }}_{4L} & 0 & m_{52} & 0 & 0 & M_{54}^{e \prime} \\ 
\end{array}%
\right), \\[2ex]
M_{54}^{e \prime} = \sqrt{M_{54}^{e 2} + m_{53}^2}, \quad s_{34}^{e} = \frac{m_{53}}{M_{54}^{e \prime}}, \quad c_{34}^{e} = \frac{M_{54}^{e}}{M_{54}^{e \prime}}, \quad V_{34}^{e} = \begin{pmatrix}
1 & 0 & 0 & 0 & 0 \\[1ex]
0 & 1 & 0 & 0 & 0 \\[1ex]
0 & 0 & c_{34}^{e} & 0 & -s_{34}^{e} \\[1ex]
0 & 0 & 0 & 1 & 0 \\[1ex]
0 & 0 & s_{34}^{e} & 0 & c_{34}^{e}
\end{pmatrix}.
\label{eqn:mm_cl_RH_34}
\end{gather}
The last step to arrive at the intermediate mass basis is the $24$ rotation in the right-handed fields.
\begin{gather}
V_{34}^{L} M^{e } (V_{34}^{e})^{\dagger} (V_{24}^{e})^{\dagger}
=
\\[2ex]
\scalemath{0.9}{
\left( 
\begin{array}{c|ccccc}
& e _{1R} & e _{2R}^{\prime} & e _{3R}^{\prime} & \widetilde{L }_{4R} & e _{4R}^{\prime \prime} \\[0.5ex] \hline
\overline{L }_{1L} & 0 & 0 & 0 & 0 & 0 \\[1ex]
\overline{L }_{2L} & 0 & -\frac{m_{24} m_{52} M_{54}^{e}}{M_{54}^{e \prime} M_{54}^{e \prime \prime}} & -\frac{m_{24} m_{53}}{M_{54}^{e \prime}} & 0 & \frac{m_{24} M_{54}^{e}}{M_{54}^{e \prime \prime}} \\[1ex] 
\overline{L }_{3L}^{\prime} & 0 & \frac{m_{52} (m_{35} m_{43} m_{53} -m_{34} M_{45}^{L} M_{54}^{e})}{M_{45}^{L \prime} M_{54}^{e \prime} M_{54}^{e \prime \prime}} & \frac{-m_{34} m_{53} M_{45}^{L} -m_{35} m_{43} M_{54}^{e}}{M_{45}^{L \prime} M_{54}^{e \prime}} & 0 & \frac{-m_{35} m_{43} m_{53} +m_{34} M_{45}^{L} M_{54}^{e}}{M_{45}^{L \prime} M_{54}^{e \prime \prime}} \\[1ex]
\overline{L }_{4L}^{\prime} & 0 & -\frac{m_{52} (m_{43} M_{45}^{L} m_{53} +m_{34} m_{35} M_{54}^{e})}{M_{45}^{L \prime} M_{54}^{e \prime} M_{54}^{e \prime \prime}} & \frac{-m_{34} m_{35} m_{53} +m_{43} M_{45}^{L} M_{54}^{e}}{M_{45}^{L \prime} M_{54}^{e \prime}} & M_{45}^{L \prime} & \frac{m_{43} m_{53} M_{45}^{L} +m_{34} m_{35} M_{54}^{e}}{M_{45}^{L \prime} M_{54}^{e \prime \prime}} \\[1ex]
\overline{\widetilde{e }}_{4L} & 0 & 0 & 0 & 0 & M_{54}^{e \prime \prime} \label{eqn:mm_intermediate} \\ 
\end{array}%
\right)}, \\[2ex]
M_{54}^{e \prime \prime} = \sqrt{M_{54}^{e \prime 2} + m_{52}^2}, \quad s_{24}^{e} = \frac{m_{52}}{M_{54}^{e \prime \prime}}, \quad c_{24}^{e} = \frac{M_{54}^{e \prime}}{M_{54}^{e \prime \prime}}, \quad V_{34}^{e} = \begin{pmatrix}
1 & 0 & 0 & 0 & 0 \\[1ex]
0 & c_{24}^{e} & 0 & 0 & -s_{24}^{e} \\[1ex]
0 & 0 & 1 & 0 & 0 \\[1ex]
0 & 0 & 0 & 1 & 0 \\[1ex]
0 & s_{24}^{e} & 0 & 0 & c_{24}^{e}
\end{pmatrix}.
\label{eqn:mm_cl_RH_24}
\end{gather}
The mass matrix given in Equation~\ref{eqn:mm_intermediate} is the intermediate mass basis and this diagonalization is exactly consistent with the one for the SM charged lepton sector in one of our works~\cite{Hernandez:2021tii}. When we diagonalized the charged lepton sector in \cite{Hernandez:2021tii}, we assumed all off-diagonal elements to be zero and this is actually a quite suitable assumption since the differences between the Yukawa induced mass terms and the vector-like masses are quite large. However, we consider all small mixings in order to get the fully diagonalized mass matrix in this work rather than setting 
them to zero, since we are interested in studying diverse FCNC constraints by scanning all possible and allowed mass ranges of the vector-like fermions in both SM quark and lepton sectors and the FCNC constraints are sensitive to the small mixings as it will be shown below.  
One more feature to be mentioned 
 in this diagonalization is that all the mixings have been made between 
  the same $SU(2)$ multiplets. In other words, the $SU(2)$ doublet left-handed fields are mixed with the another $SU(2)$ doublet left-handed fields, whereas the $SU(2)$ singlet right-handed fields are mixed with the another $SU(2)$ singlet right-handed fields, so we call this mixing ``$SU(2)$ conserving mixing". This $SU(2)$ conserving mixing can not cause the flavor violating interactions with the SM $Z$ gauge boson since they involve 
  an identity matrix resulting 
  from the SM $Z$ gauge interactions. Therefore, the next diagonalization process becomes especially important when we start exploring the FCNC constraints. Before we start the next diagonalization, it is convenient to reparameterize all elements of Equation~\ref{eqn:mm_intermediate} by a simpler one as given in Equation~\ref{eqn:mm_cl_rp}.
\begin{equation}
M^{e \prime} = V_{34}^{L} M^{e } (V_{34}^{e})^{\dagger} (V_{24}^{e})^{\dagger}
=
\left( 
\begin{array}{c|ccccc}
& e _{1R} & e _{2R}^{\prime} & e _{3R}^{\prime} & \widetilde{L }_{4R} & e _{4R}^{\prime \prime} \\[0.5ex] \hline
\overline{L }_{1L} & 0 & 0 & 0 & 0 & 0 \\[1ex]
\overline{L }_{2L} & 0 & m_{22}^{\prime} & m_{23}^{\prime} & 0 & m_{24}^{\prime} \\[1ex] 
\overline{L }_{3L}^{\prime} & 0 & m_{32}^{\prime} & m_{33}^{\prime} & 0 & m_{34}^{\prime} \\[1ex]
\overline{L }_{4L}^{\prime} & 0 & m_{42}^{\prime} & m_{43}^{\prime} & M_{45}^{L \prime} & m_{44}^{\prime} \\[1ex]
\overline{\widetilde{e }}_{4L} & 0 & 0 & 0 & 0 & M_{54}^{e \prime \prime} \\ 
\end{array}%
\right) 
\label{eqn:mm_cl_rp}
\end{equation}
We carry out first the $35$ rotation in the left-handed fields of mass matrix~\ref{eqn:mm_cl_rp} and this is a start of ``$SU(2)$ violating mixing". As already mentioned, since the difference between the Yukawa mass and vector-like mass is significantly sizeable, it is possible to simplify the mixing matrices in terms of the relevant small mixing angle $\theta_{35}^{L}$.
\begin{gather}
V_{35}^{L} M^{e \prime} 
=
\scalemath{0.8}{
\left( 
\begin{array}{c|ccccc}
& e _{1R} & e _{2R}^{\prime} & e _{3R}^{\prime} & \widetilde{L }_{4R} & e _{4R}^{\prime \prime} \\[0.5ex] \hline
\overline{L }_{1L} & 0 & 0 & 0 & 0 & 0 \\[1ex]
\overline{L }_{2L} & 0 & m_{22}^{\prime} & m_{23}^{\prime} & 0 & m_{24}^{\prime} \\[1ex] 
\overline{L }_{3L}^{\prime \prime} & 0 & m_{32}^{\prime} & m_{33}^{\prime} & 0 & m_{34}^{\prime} - M_{54}^{e \prime \prime} \theta_{35}^{L} \\[1ex]
\overline{L }_{4L}^{\prime} & 0 & m_{42}^{\prime} & m_{43}^{\prime} & M_{45}^{L \prime} & m_{44}^{\prime} \\[1ex]
\overline{\widetilde{e }}_{4L}^{\prime} & 0 & m_{32}^{\prime} \theta_{35}^{L} & m_{33}^{\prime} \theta_{35}^{L} & 0 & M_{54}^{e \prime \prime} + m_{34}^{\prime} \theta_{35}^{L} \\ 
\end{array}%
\right)
\approx
\left( 
\begin{array}{c|ccccc}
& e _{1R} & e _{2R}^{\prime} & e _{3R}^{\prime} & \widetilde{L }_{4R} & e _{4R}^{\prime \prime} \\[0.5ex] \hline
\overline{L }_{1L} & 0 & 0 & 0 & 0 & 0 \\[1ex]
\overline{L }_{2L} & 0 & m_{22}^{\prime} & m_{23}^{\prime} & 0 & m_{24}^{\prime} \\[1ex] 
\overline{L }_{3L}^{\prime \prime} & 0 & m_{32}^{\prime} & m_{33}^{\prime} & 0 & 0 \\[1ex]
\overline{L }_{4L}^{\prime} & 0 & m_{42}^{\prime} & m_{43}^{\prime} & M_{45}^{L \prime} & m_{44}^{\prime} \\[1ex]
\overline{\widetilde{e }}_{4L}^{\prime} & 0 & 0 & 0 & 0 & M_{54}^{e \prime \prime} \\ 
\end{array}%
\right)}
\\[2ex]
\theta_{35}^{L} = \frac{m_{34}^{\prime}}{M_{54}^{e \prime \prime}}, \quad
V_{35}^{L} = \begin{pmatrix}
1 & 0 & 0 & 0 & 0 \\[1ex]
0 & 1 & 0 & 0 & 0 \\[1ex]
0 & 0 & 1 & 0 & -\theta_{35}^{L} \\[1ex]
0 & 0 & 0 & 1 & 0 \\[1ex]
0 & 0 & \theta_{35}^{L} & 0 & 1 
\end{pmatrix}
\end{gather}
The next step is the $25$ rotation in the left-handed fields to turn off the mass term $m_{24}^{\prime}$.
\begin{gather}
V_{25}^{L} V_{35}^{L} M^{e \prime}
=
\scalemath{0.8}{
\left( 
\begin{array}{c|ccccc}
& e _{1R} & e _{2R}^{\prime} & e _{3R}^{\prime} & \widetilde{L }_{4R} & e _{4R}^{\prime \prime} \\[0.5ex] \hline
\overline{L }_{1L} & 0 & 0 & 0 & 0 & 0 \\[1ex]
\overline{L }_{2L}^{\prime} & 0 & m_{22}^{\prime} & m_{23}^{\prime} & 0 & m_{24}^{\prime} - M_{54}^{e \prime \prime} \theta_{25}^{L} \\[1ex] 
\overline{L }_{3L}^{\prime \prime} & 0 & m_{32}^{\prime} & m_{33}^{\prime} & 0 & 0 \\[1ex]
\overline{L }_{4L}^{\prime} & 0 & m_{42}^{\prime} & m_{43}^{\prime} & M_{45}^{L \prime} & m_{44}^{\prime} \\[1ex]
\overline{\widetilde{e }}_{4L}^{\prime \prime} & 0 & m_{22}^{\prime} \theta_{25}^{L} & m_{23}^{\prime} \theta_{25}^{L} & 0 & M_{54}^{e \prime \prime} + m_{24}^{\prime} \theta_{25}^{L} \\ 
\end{array}%
\right)
\approx
\left( 
\begin{array}{c|ccccc}
& e _{1R} & e _{2R}^{\prime} & e _{3R}^{\prime} & \widetilde{L }_{4R} & e _{4R}^{\prime \prime} \\[0.5ex] \hline
\overline{L }_{1L} & 0 & 0 & 0 & 0 & 0 \\[1ex]
\overline{L }_{2L}^{\prime} & 0 & m_{22}^{\prime} & m_{23}^{\prime} & 0 & 0 \\[1ex] 
\overline{L }_{3L}^{\prime \prime} & 0 & m_{32}^{\prime} & m_{33}^{\prime} & 0 & 0 \\[1ex]
\overline{L }_{4L}^{\prime} & 0 & m_{42}^{\prime} & m_{43}^{\prime} & M_{45}^{L \prime} & m_{44}^{\prime} \\[1ex]
\overline{\widetilde{e }}_{4L}^{\prime \prime} & 0 & 0 & 0 & 0 & M_{54}^{e \prime \prime} \\ 
\end{array}%
\right)}
\\[2ex]
\theta_{25}^{L} = \frac{m_{24}^{\prime}}{M_{54}^{e \prime \prime}}, \quad
V_{25}^{L} = \begin{pmatrix}
1 & 0 & 0 & 0 & 0 \\[1ex]
0 & 1 & 0 & 0 & -\theta_{25}^{L} \\[1ex]
0 & 0 & 1 & 0 & 0 \\[1ex]
0 & 0 & 0 & 1 & 0 \\[1ex]
0 & \theta_{25}^{L} & 0 & 0 & 1 
\end{pmatrix}
\end{gather}
The next is the $35$ rotation in the right-handed fields to turn off the mass term $m_{42}^{\prime}$.
\begin{gather}
V_{25}^{L} V_{35}^{L} M^{e \prime} (V_{35}^{e})^{\dagger} 
=
\scalemath{0.8}{
\left( 
\begin{array}{c|ccccc}
& e _{1R} & e _{2R}^{\prime} & e _{3R}^{\prime \prime} & \widetilde{L }_{4R}^{\prime} & e _{4R}^{\prime \prime} \\[0.5ex] \hline
\overline{L }_{1L} & 0 & 0 & 0 & 0 & 0 \\[1ex]
\overline{L }_{2L}^{\prime} & 0 & m_{22}^{\prime} & m_{23}^{\prime} & m_{23}^{\prime} \theta_{35}^{e} & 0 \\[1ex] 
\overline{L }_{3L}^{\prime \prime} & 0 & m_{32}^{\prime} & m_{33}^{\prime} & m_{33}^{\prime} \theta_{35}^{e} & 0 \\[1ex]
\overline{L }_{4L}^{\prime} & 0 & m_{42}^{\prime} & m_{43}^{\prime} - M_{45}^{L \prime} \theta_{35}^{e} & M_{45}^{L \prime} + m_{43}^{\prime} \theta_{35}^{e} & m_{44}^{\prime} \\[1ex]
\overline{\widetilde{e }}_{4L}^{\prime \prime} & 0 & 0 & 0 & 0 & M_{54}^{e \prime \prime} \\ 
\end{array}%
\right)
\approx
\left( 
\begin{array}{c|ccccc}
& e _{1R} & e _{2R}^{\prime} & e _{3R}^{\prime \prime} & \widetilde{L }_{4R}^{\prime} & e _{4R}^{\prime \prime} \\[0.5ex] \hline
\overline{L }_{1L} & 0 & 0 & 0 & 0 & 0 \\[1ex]
\overline{L }_{2L}^{\prime} & 0 & m_{22}^{\prime} & m_{23}^{\prime} & 0 & 0 \\[1ex] 
\overline{L }_{3L}^{\prime \prime} & 0 & m_{32}^{\prime} & m_{33}^{\prime} & 0 & 0 \\[1ex]
\overline{L }_{4L}^{\prime} & 0 & m_{42}^{\prime} & 0 & M_{45}^{L \prime} & m_{44}^{\prime} \\[1ex]
\overline{\widetilde{e }}_{4L}^{\prime \prime} & 0 & 0 & 0 & 0 & M_{54}^{e \prime \prime} \\ 
\end{array}%
\right)}
\\[2ex]
\theta_{35}^{e} = \frac{m_{43}^{\prime}}{M_{45}^{L \prime}}, \quad
V_{35}^{e} = \begin{pmatrix}
1 & 0 & 0 & 0 & 0 \\[1ex]
0 & 1 & 0 & 0 & 0 \\[1ex]
0 & 0 & 1 & -\theta_{35}^{e} & 0 \\[1ex]
0 & 0 & \theta_{35}^{e} & 1 & 0 \\[1ex]
0 & 0 & 0 & 0 & 1 
\end{pmatrix}
\end{gather}
After performing 
the right-handed $25$ rotation, we have a mass matrix, whose form is block diagonal.
\begin{gather}
V_{25}^{L} V_{35}^{L} M^{e \prime} (V_{35}^{e})^{\dagger} (V_{25}^{e})^{\dagger} =
\\[2ex]
\left( 
\begin{array}{c|ccccc}
& e _{1R} & e _{2R}^{\prime \prime} & e _{3R}^{\prime \prime} & \widetilde{L }_{4R}^{\prime \prime} & e _{4R}^{\prime \prime} \\[0.5ex] \hline
\overline{L }_{1L} & 0 & 0 & 0 & 0 & 0 \\[1ex]
\overline{L }_{2L}^{\prime} & 0 & m_{22}^{\prime} & m_{23}^{\prime} & m_{22}^{\prime} \theta_{25}^{e} & 0 \\[1ex] 
\overline{L }_{3L}^{\prime \prime} & 0 & m_{32}^{\prime} & m_{33}^{\prime} & m_{32}^{\prime} \theta_{25}^{e} & 0 \\[1ex]
\overline{L }_{4L}^{\prime} & 0 & m_{42}^{\prime} - M_{45}^{L \prime} \theta_{25}^{e} & 0 & M_{45}^{L \prime} + m_{42}^{\prime} \theta_{25}^{e} & m_{44}^{\prime} \\[1ex]
\overline{\widetilde{e }}_{4L}^{\prime \prime} & 0 & 0 & 0 & 0 & M_{54}^{e \prime \prime} \\ 
\end{array}%
\right)
\approx
\left( 
\begin{array}{c|ccccc}
& e _{1R} & e _{2R}^{\prime \prime} & e _{3R}^{\prime \prime} & \widetilde{L }_{4R}^{\prime \prime} & e _{4R}^{\prime \prime} \\[0.5ex] \hline
\overline{L }_{1L} & 0 & 0 & 0 & 0 & 0 \\[1ex]
\overline{L }_{2L}^{\prime} & 0 & m_{22}^{\prime} & m_{23}^{\prime} & 0 & 0 \\[1ex] 
\overline{L }_{3L}^{\prime \prime} & 0 & m_{32}^{\prime} & m_{33}^{\prime} & 0 & 0 \\[1ex]
\overline{L }_{4L}^{\prime} & 0 & 0 & 0 & M_{45}^{L \prime} & m_{44}^{\prime} \\[1ex]
\overline{\widetilde{e }}_{4L}^{\prime \prime} & 0 & 0 & 0 & 0 & M_{54}^{e \prime \prime} \\ 
\end{array}%
\right)
\\[2ex]
\theta_{25}^{e} = \frac{m_{42}^{\prime}}{M_{45}^{L \prime}}, \quad
V_{25}^{e} = \begin{pmatrix}
1 & 0 & 0 & 0 & 0 \\[1ex]
0 & 1 & 0 & -\theta_{25}^{e} & 0 \\[1ex]
0 & 0 & 1 & 0 & 0 \\[1ex]
0 & \theta_{25}^{e} & 0 & 1 & 0 \\[1ex]
0 & 0 & 0 & 0 & 1 
\end{pmatrix}
\end{gather}
We arrive at the fully diagonalized mass matrix by diagonalizing the upper-left $3 \times 3$ block as well as the lower-right $2 \times 2$ block as shown below in Equation~\ref{eqn:mass_pro_cl}.
\begin{equation}
\begin{split}
&V_{45}^{L} V_{23}^{L} V_{35}^{L} V_{25}^{L} M^{e \prime} (V_{35}^{e})^{\dagger} (V_{25}^{e})^{\dagger} (V_{23}^{e})^{\dagger} (V_{54}^{e})^{\dagger} = \func{diag}\left( 0, m_{\mu}, m_{\tau}, M_{E_4}, M_{\widetilde{E}_4} \right) \\
&V_{45}^{L} V_{23}^{L} V_{35}^{L} V_{25}^{L} V_{34}^{L} M^{e} (V_{34}^{e})^{\dagger} (V_{24}^{e})^{\dagger} (V_{35}^{e})^{\dagger} (V_{25}^{e})^{\dagger} (V_{23}^{e})^{\dagger} (V_{54}^{e})^{\dagger} = \func{diag}\left( 0, m_{\mu}, m_{\tau}, M_{E_4}, M_{\widetilde{E}_4} \right) 
\label{eqn:mass_pro_cl}
\end{split}
\end{equation}
As mentioned in the introduction, the SM charged lepton belonging to the first family, namely, the electron does not acquire a mass 
with one vector-like family as seen in Equation~\ref{eqn:mass_pro_cl}. This is due to the fact that the model under consideration has two leptonic seesaw mediators, which provide tree-level masses to the muon and tau leptons. It is worth mentioning that the number of seesaw mediators has to be larger or equal than the number of SM fermion families in order to provide masses to the SM fermions. A non vanishing electron mass can be generated by introducing one extra vector-like family as done in the reference~\cite{Hernandez:2021tii}. Then, we can easily confirm how the SM charged leptons in the flavor basis are connected with those ones in the mass basis via the following unitary mixing matrices.
\begin{equation}
\begin{split}
\begin{pmatrix}
e_{L} \\[0.5ex]
\mu_{L} \\[0.5ex]
\tau_{L} \\[0.5ex]
E_{4L} \\[0.5ex]
\widetilde{E}_{4L}
\end{pmatrix}
&=
\begin{pmatrix}
e_{1L} \\[0.5ex]
e_{2L}^{\prime} \\[0.5ex]
e_{3L}^{\prime \prime} \\[0.5ex]
e_{4L}^{\prime} \\[0.5ex]
\widetilde{e}_{4L}^{\prime \prime}
\end{pmatrix}
=
V^L  
\begin{pmatrix}
e_{1L} \\[0.5ex]
e_{2L} \\[0.5ex]
e_{3L} \\[0.5ex]
e_{4L} \\[0.5ex]
\widetilde{e}_{4L}
\end{pmatrix}
=
V_{45}^L V_{23}^L V_{25}^L V_{35}^L V_{34}^L  
\begin{pmatrix}
e_{1L} \\[0.5ex]
e_{2L} \\[0.5ex]
e_{3L} \\[0.5ex]
e_{4L} \\[0.5ex]
\widetilde{e}_{4L}
\end{pmatrix}
,
\\
\begin{pmatrix}
e_{R} \\[0.5ex]
\mu_{R} \\[0.5ex]
\tau_{R} \\[0.5ex]
\widetilde{E}_{4R} \\[0.5ex]
E_{4R}
\end{pmatrix}
&=
\begin{pmatrix}
e_{1R} \\[0.5ex]
e_{2R}^{\prime \prime} \\[0.5ex]
e_{3R}^{\prime \prime} \\[0.5ex]
\widetilde{e}_{4R}^{\prime \prime} \\[0.5ex]
e_{4R}^{\prime \prime}
\end{pmatrix}
=
V^e
\begin{pmatrix}
e_{1R} \\[0.5ex]
e_{2R} \\[0.5ex]
e_{3R} \\[0.5ex]
\widetilde{e}_{4R} \\[0.5ex]
e_{4R}
\end{pmatrix}
=
V_{54}^e V_{23}^e V_{25}^e V_{35}^e V_{24}^e V_{34}^e
\begin{pmatrix}
e_{1R} \\[0.5ex]
e_{2R} \\[0.5ex]
e_{3R} \\[0.5ex]
\widetilde{e}_{4R} \\[0.5ex]
e_{4R}
\end{pmatrix}.
\label{eqn:cl_umm_LH_RH}
\end{split}
\end{equation}
The left-handed $34$ mixing $V_{34}^{L}$ and right-handed $24,34$ mixings $V_{24,34}^{e}$ are the $SU(2)$ conserving mixings, whereas the left-handed $25,35$ mixings $V_{25,35}^{L}$ and right-handed $25,35$ mixings $V_{25,35}^{e}$ are the $SU(2)$ violating mixings. We will see that these $SU(2)$ violating mixings play a crucial role in generating the renormalizable flavor violating mixings mediated by the SM $Z$ gauge boson in section~\ref{sec:IV}. It is worth mentioning that 
this step-by-step diagonalization is 
a quite good approximation to the corresponding 
numerical diagonalization carried out by the singular value decomposition (SVD) method since the former yields similar results to the ones obtained from the latter, with very small differences due to the fact that 
all off-diagonal elements resulting from the step-by-step diagonalization are quite negligible and thus they can be approximated to zero, as discussed in detail  
 in Appendix~\ref{app:D}.
\section{Analytic approximated step-by-step diagonalization for the up-quark sector} \label{app:B}
The initial mass matrix for the up-type quark sector in the flavor basis is given by:
\begin{equation}
M^{u }=\left( 
\begin{array}{c|ccccc}
& u _{1R} & u _{2R} & u _{3R} & u _{4R} & \widetilde{Q }_{4R} \\[0.5ex] \hline
\overline{Q }_{1L} & 0 & 0 & 0 & 0 & 0 \\[1ex]
\overline{Q }_{2L} & 0 & 0 & 0 & y_{24}^{u } v_{u} & 0 \\[1ex] 
\overline{Q }_{3L} & 0 & 0 & 0 & y_{34}^{u } v_{u} & x_{34}^{Q } v_{\phi} \\[1ex]
\overline{Q }_{4L} & 0 & 0 & y_{43}^{u } v_{u} & 0 & M_{44}^{Q } \\[1ex]
\overline{\widetilde{u }}_{4L} & 0 & x_{42}^{u} v_{\phi}
& x_{43}^{u} v_{\phi} & M_{44}^{u} & 0 \\ 
\end{array}%
\right)
=
\left( 
\begin{array}{c|ccccc}
& u _{1R} & u _{2R} & u _{3R} & \widetilde{Q }_{4R} & u _{4R} \\[0.5ex] \hline
\overline{Q }_{1L} & 0 & 0 & 0 & 0 & 0 \\[1ex]
\overline{Q }_{2L} & 0 & 0 & 0 & 0 & m_{24}^{u} \\[1ex] 
\overline{Q }_{3L} & 0 & 0 & 0 & m_{35}^{u} & m_{34}^{u} \\[1ex]
\overline{Q }_{4L} & 0 & 0 & m_{43}^{u} & M_{44}^{Q } & 0 \\[1ex]
\overline{\widetilde{u }}_{4L} & 0 & m_{52}^{u}
& m_{53}^{u} & 0 & M_{44}^{u} \\ 
\end{array}%
\right), \label{eqn:uq_1s}
\end{equation}
The mass matrix of Equation~\ref{eqn:uq_1s} in the flavor basis is exactly consistent with the one corresponding to the charged lepton sector excepting for a few substitutions $y^{e} \rightarrow y^{u}$, $v_{d} \rightarrow v_{u}$, $x^{L} \rightarrow x^{Q}$ and $x^{e} \rightarrow x^{u}$. However, these substitutions do not change the whole structure of the mass matrix, so we do not need to derive all the required mixings from the initial mass matrix, instead the given mixings in the charged lepton sector can be reused as follows (For the charged lepton sector, it is enough to notice the symbol $L$ means left-handed doublet and $e$ means right-handed singlet. However, it becomes complicated in the quark sector since the mass matrices in the up- and down-type quark have a different form, so we change the mixing notation by $V_{L(R)}^{u,d}$ instead of 
$V^{Q}$.):
\begin{equation}
\begin{split}
\begin{pmatrix}
u_{L} \\[0.5ex]
c_{L} \\[0.5ex]
t_{L} \\[0.5ex]
U_{4L} \\[0.5ex]
\widetilde{U}_{4L}
\end{pmatrix}
=
\begin{pmatrix}
u_{1L} \\[0.5ex]
u_{2L}^{\prime} \\[0.5ex]
u_{3L}^{\prime \prime} \\[0.5ex]
u_{4L}^{\prime} \\[0.5ex]
\widetilde{u}_{4L}^{\prime \prime}
\end{pmatrix}
&=
V_{L}^{u}  
\begin{pmatrix}
u_{1L} \\[0.5ex]
u_{2L} \\[0.5ex]
u_{3L} \\[0.5ex]
u_{4L} \\[0.5ex]
\widetilde{u}_{4L}
\end{pmatrix}
=
(V_{L}^{u})_{45} (V_{L}^{u})_{23} (V_{L}^{u})_{25} (V_{L}^{u})_{35} (V_{L}^{u})_{34}  
\begin{pmatrix}
u_{1L} \\[0.5ex]
u_{2L} \\[0.5ex]
u_{3L} \\[0.5ex]
u_{4L} \\[0.5ex]
\widetilde{u}_{4L}
\end{pmatrix}
,
\\
\begin{pmatrix}
u_{R} \\[0.5ex]
c_{R} \\[0.5ex]
t_{R} \\[0.5ex]
\widetilde{U}_{4R} \\[0.5ex]
U_{4R}
\end{pmatrix}
=
\begin{pmatrix}
u_{1R} \\[0.5ex]
u_{2R}^{\prime \prime} \\[0.5ex]
u_{3R}^{\prime \prime} \\[0.5ex]
\widetilde{u}_{4R}^{\prime \prime} \\[0.5ex]
u_{4R}^{\prime \prime}
\end{pmatrix}
&=
V_{R}^{u}
\begin{pmatrix}
u_{1R} \\[0.5ex]
u_{2R} \\[0.5ex]
u_{3R} \\[0.5ex]
\widetilde{u}_{4R} \\[0.5ex]
u_{4R}
\end{pmatrix}
=
(V_{R}^{u})_{54} (V_{R}^{u})_{23} (V_{R}^{u})_{25} (V_{R}^{u})_{35} (V_{R}^{u})_{24} (V_{R}^{u})_{34}
\begin{pmatrix}
u_{1R} \\[0.5ex]
u_{2R} \\[0.5ex]
u_{3R} \\[0.5ex]
\widetilde{u}_{4R} \\[0.5ex]
u_{4R}
\end{pmatrix}.
\label{eqn:ana_up_mixing}
\end{split}
\end{equation}
As mentioned in subsection~\ref{sec:III_2}, this approximated step-by-step diagonalization for the up-quark sector requires more caution since some of the off-diagonal elements 
being of order unity and appearing as a result of mixings can be sizeable due to the heavy top quark mass and the heavy exotic up type quark masses thus requiring  
the use of the numerical SVD technique for the correct diagonalization and the SVD diagonalization will be used in our numerical scans in the main body of this work. 
The comparison between a numerical mixing matrix derived from the SVD method and the one obtained from the analytic perturbative diagonalization will be discussed in Appendix~\ref{app:E}.
\section{Analytic approximated step-by-step diagonalization for the down-quark sector} \label{app:C}
We start from the initial down-type mass matrix given in Equation~\ref{eqn:diff_up_down} in the flavor basis.
\begin{equation}
M^{d }=\left( 
\begin{array}{c|ccccc}
& d _{1R} & d _{2R} & d _{3R} & d _{4R} & \widetilde{Q }_{4R} \\[0.5ex] \hline
\overline{Q }_{1L} & 0 & 0 & 0 & y_{14}^{d } v_{d} & 0 \\[1ex]
\overline{Q }_{2L} & 0 & 0 & 0 & y_{24}^{d } v_{d} & 0 \\[1ex] 
\overline{Q }_{3L} & 0 & 0 & 0 & y_{34}^{d } v_{d} & x_{34}^{Q } v_{\phi} \\[1ex]
\overline{Q }_{4L} & 0 & 0 & y_{43}^{d } v_{d} & 0 & M_{44}^{Q } \\[1ex]
\overline{\widetilde{d }}_{4L} & 0 & x_{42}^{d} v_{\phi}
& x_{43}^{d} v_{\phi} & M_{44}^{d} & 0 \\ 
\end{array}%
\right)
\end{equation}
As in the charged lepton case, it is convenient to rearrange the Yukawa mass terms by mass parameters and to swap the fourth and fifth column.
\begin{equation}
M^{d }
=
\left( 
\begin{array}{c|ccccc}
& d _{1R} & d _{2R} & d _{3R} & d _{4R} & \widetilde{Q }_{4R} \\[0.5ex] \hline
\overline{Q }_{1L} & 0 & 0 & 0 & m_{14}^{d} & 0 \\[1ex]
\overline{Q }_{2L} & 0 & 0 & 0 & m_{24}^{d} & 0 \\[1ex] 
\overline{Q }_{3L} & 0 & 0 & 0 & m_{34}^{d} & m_{35}^{d} \\[1ex]
\overline{Q }_{4L} & 0 & 0 & m_{43}^{d} & 0 & M_{44}^{Q } \\[1ex]
\overline{\widetilde{d }}_{4L} & 0 & m_{52}^{d}
& m_{53}^{d} & M_{44}^{d} & 0 \\ 
\end{array}%
\right)
=
\left( 
\begin{array}{c|ccccc}
& d _{1R} & d _{2R} & d _{3R} & \widetilde{Q }_{4R} & d _{4R} \\[0.5ex] \hline
\overline{Q }_{1L} & 0 & 0 & 0 & 0 & m_{14}^{d} \\[1ex]
\overline{Q }_{2L} & 0 & 0 & 0 & 0 & m_{24}^{d} \\[1ex] 
\overline{Q }_{3L} & 0 & 0 & 0 & m_{35}^{d} & m_{34}^{d} \\[1ex]
\overline{Q }_{4L} & 0 & 0 & m_{43}^{d} & M_{44}^{Q } & 0 \\[1ex]
\overline{\widetilde{d }}_{4L} & 0 & m_{52}^{d}
& m_{53}^{d} & 0 & M_{44}^{d} \\ 
\end{array}%
\right)
\end{equation}
In order to proceed from the flavor basis to the intermediate mass basis, the first thing to do is to carry out the $SU(2)$ conserving mixings $\theta_{34L}^{d}$ and $\theta_{24,34R}^{d}$ and we display the intermediate mass matrix for the down-type quarks without middle steps since the process is exactly same as the charged lepton case (After calculating all mixings required, we simplified the calculated mass parameters by $m^{\prime}$).
\begin{equation}
M^{d \prime}
=
V_{34L}^{d} M^{d} (V_{34R}^{d})^{\dagger} (V_{24R}^{d})^{\dagger} 
= 
\left(
\begin{array}{c|ccccc}
 & d_{1R} & d_{2R}^{\prime} & d_{3R}^{\prime} & \widetilde{d}_{4R} & d_{4R}^{\prime \prime} \\
 \hline
\overline{d}_{1L} & 0 & m_{12}^{d \prime} & m_{13}^{d \prime} & 0 & m_{14}^{d \prime} \\
\overline{d}_{2L} & 0 & m_{22}^{d \prime} & m_{23}^{d \prime} & 0 & m_{24}^{d \prime} \\
\overline{d}_{3L}^{d \prime} & 0 & m_{32}^{d \prime} & m_{33}^{d \prime} & 0 & m_{34}^{d \prime} \\
\overline{d}_{4L}^{\prime} & 0 & m_{42}^{d \prime} & m_{43}^{d \prime} & M_{45}^{Q \prime}
   & m_{44}^{d \prime} \\
\overline{\widetilde{d}}_{4L} & 0 & 0 & 0 & 0 & M_{54}^{d \prime \prime} \\
\end{array}
\right)
\end{equation}
We should carry out the $SU(2)$ violating mixings to turn off the mass parameters $m_{14,24,34,42,43}^{d \prime}$ and the mixing angles are very suppressed by the ratio between Yukawa and vector-like masses. Then the block diagonal form of this mass matrix appears as follows:
\begin{equation}
M^{d \prime \prime} = V_{15L}^{d} V_{25L}^{d} V_{35L}^{d} M^{d \prime} (V_{35R}^{d})^{\dagger} (V_{25R}^{d})^{\dagger}
= 
\left(
\begin{array}{c|ccccc}
 & d_{1R} & d_{2R}^{\prime \prime} & d_{3R}^{\prime \prime} & \widetilde{d}_{4R}^{\prime \prime} & d_{4R}^{\prime \prime} \\
 \hline
\overline{d}_{1L}^{\prime} & 0 & m_{12}^{d \prime} & m_{13}^{d \prime} & 0 & 0 \\
\overline{d}_{2L}^{\prime} & 0 & m_{22}^{d \prime} & m_{23}^{d \prime} & 0 & 0 \\
\overline{d}_{3L}^{\prime \prime} & 0 & m_{32}^{d \prime} & m_{33}^{d \prime} & 0 & 0 \\
\overline{d}_{4L}^{\prime} & 0 & 0 & 0 & M_{45}^{Q \prime}
   & m_{44}^{d \prime} \\
\overline{\widetilde{d}}_{4L}^{\prime \prime \prime} & 0 & 0 & 0 & 0 & M_{54}^{d \prime \prime} \\
\end{array}
\right)
\label{eqn:md_block_diagonal}
\end{equation}
An important feature of the mass matrix of Equation~\ref{eqn:md_block_diagonal} is the mass parameters of the first row is proportional to those of the second row by a factor (In other words, $m_{12}^{d \prime} / m_{22}^{d \prime} = m_{13}^{d \prime} / m_{23}^{d \prime}$. We follow the convention to diagonalize the upper-left $3 \times 3$ block~\cite{King:2002nf} rather than simply rotating the upper-left block. As the mass matrix of Equation~\ref{eqn:md_block_diagonal} consists of only real numbers, we can exclude the complex numbers in the convention and the convention is given by:
\begin{equation}
V_{12L}^{d} V_{13L}^{d} V_{23L}^{d} M^{d \prime \prime} (V_{23R}^{d})^{\dagger} (V_{13R}^{d})^{\dagger} (V_{12R}^{d})^{\dagger} = \func{diag} \left( 0 , m_{s}, m_{b}, M_{D_4}, M_{\widetilde{D}_4} \right)
\end{equation}
and then we arrive to the fully diagonalized mass matrix, which reveals all propagating mass for the down-type quarks. Then, the connection from the flavor to mass basis for the down-type quarks can be seen via the unitary mixing matrices as follows (notice that the right-handed down-type quark mixing matrices $(V_{R}^{d})_{12,13}$ remain as an identity matrix as the relevant mass matrix has the form of 
$\begin{pmatrix}
0 & m_a \\
0 & m_b
\end{pmatrix}$ 
and this form generally induces only left-handed mixing matrices).
\begin{equation}
\begin{split}
\begin{pmatrix}
d_{L} \\
s_{L} \\
b_{L} \\
D_{4L} \\
\widetilde{D}_{4L}
\end{pmatrix}
=
\begin{pmatrix}
d_{1L}^{\prime \prime \prime} \\
d_{2L}^{\prime \prime \prime} \\
d_{3L}^{\prime \prime \prime \prime} \\
d_{4L}^{\prime \prime} \\
\widetilde{d}_{4L}^{\prime \prime \prime \prime}
\end{pmatrix}
&=
V_{L}^{d}
\begin{pmatrix}
d_{1L} \\
d_{2L} \\
d_{3L} \\
d_{4L} \\
\widetilde{d}_{4L}
\end{pmatrix}
=
(V_{L}^{d})_{45} (V_{L}^{d})_{12} (V_{L}^{d})_{13} (V_{L}^{d})_{23} (V_{L}^{d})_{15} (V_{L}^{d})_{25} (V_{L}^{d})_{35} (V_{L}^{d})_{34}  
\begin{pmatrix}
d_{1L} \\
d_{2L} \\
d_{3L} \\
d_{4L} \\
\widetilde{d}_{4L}
\end{pmatrix}
\\
\begin{pmatrix}
d_{R} \\
s_{R} \\
b_{R} \\
D_{4R} \\
\widetilde{D}_{4R}
\end{pmatrix}
=
\begin{pmatrix}
d_{1R} \\
d_{2R}^{\prime \prime \prime} \\
d_{3R}^{\prime \prime \prime} \\
\widetilde{d}_{4R}^{\prime \prime \prime} \\
d_{4R}^{\prime \prime \prime}
\end{pmatrix}
&=
V_{R}^{d}
\begin{pmatrix}
d_{1R} \\
d_{2R} \\
d_{3R} \\
\widetilde{d}_{4R} \\
d_{4R}
\end{pmatrix}
=
(V_{R}^{d})_{54} (V_{R}^{d})_{23} (V_{R}^{d})_{25} (V_{R}^{d})_{35} (V_{R}^{d})_{24} (V_{R}^{d})_{34}
\begin{pmatrix}
d_{1R} \\
d_{2R} \\
d_{3R} \\
\widetilde{d}_{4R} \\
d_{4R}
\end{pmatrix}
\label{eqn:down_mix2}
\end{split}
\end{equation}
\section{Numerical comparison for the charged lepton sector} \label{app:D}
We have previously stressed that the analytical charged lepton mixing matrix 
is quite close to the numerical one 
and we will compare them 
in this Appendix. For this comparison, we start from the 
charged lepton mass matrix in the flavor basis, evaluated in one of the benchmark points used in our numerical scans:
\begin{equation}
M^{e}
=
\begin{pmatrix}
0 & 0 & 0 & 0 & 0 \\
0 & 0 & 0 & 0 & -2.151 \\
0 & 0 & 0 & 161.657 & 3.955 \\
0 & 0 & 4.600 & 536.050 & 0 \\
0 & 51.135 & 97.915 & 0 & 696.178
\end{pmatrix}
\end{equation}
Firstly, we evaluate the mixing matrices $V^{L,e}$ using the analytic mixings of Equation~\ref{eqn:cl_umm_LH_RH}. The analytic mixing matrices $V_{\func{ana}}^{L,e}$ are given by: 
\begin{equation}
\begin{split}
V_{\func{ana}}^{L} 
&=
\left(
\begin{array}{ccccc}
 1. & 0. & 0. & 0. & 0. \\
 0. & 0.985598 & 0.161888 & -0.0488209 & 0.00211728 \\
 0. & 0.169076 & -0.943613 & 0.284567 & 0.00548067 \\
 0. & 0.00002015 & 0.288689 & 0.957399 & -0.00668649 \\
 0. & -0.00301343 & 0.00675946 & 0.00494558 & 0.99996 \\
\end{array}
\right)
\\
V_{\func{ana}}^{e}
&=
\left(
\begin{array}{ccccc}
 1. & 0. & 0. & 0. & 0. \\
 0. & 0.986254 & -0.157386 & 0.00134068 & -0.050305 \\
 0. & 0.14846 & 0.97769 & -0.00738828 & -0.148413 \\
 0. & -0.000610678 & 0.00669679 & 0.999958 & -0.00627468 \\
 0. & 0.0725407 & 0.138946 & 0.00531106 & 0.987625 \\
\end{array}
\right)
\end{split}
\end{equation}
Notice that the mixing matrices $V_{L,R}^{u}$ and $V_{R}^{d}$ have exactly the same structure than the charged lepton mixing matrix since
 all off-diagonal elements in the first row and column are zero, however the mixing matrix $V_{L}^{d}$ is different as it can have mixings with the down-type first generation as seen in Equation~\ref{eqn:down_mixing})

The mixing matrices $V_{\func{num}}^{L,e}$ derived by the numerical SVD are given by:
\begin{equation}
\begin{split}
V_{\func{num}}^{L} 
&=
\left(
\begin{array}{ccccc}
 1. & 0. & 0. & 0. & 0. \\
 0. & 0.985598 & 0.161888 & -0.0488206 & 0.00211728 \\
 0. & 0.169076 & -0.94362 & 0.284543 & 0.00548077 \\
 0. & -0.0000241021 & -0.288666 & -0.957407 & 0.00668612 \\
 0. & -0.00301342 & 0.00675933 & 0.00494538 & 0.99996 \\
\end{array}
\right)
\\
V_{\func{num}}^{e}
&=
\left(
\begin{array}{ccccc}
 1. & 0. & 0. & 0. & 0. \\
 0. & 0.986254 & -0.157386 & 0.00134068 & -0.0503047 \\
 0. & 0.148461 & 0.977693 & -0.00738841 & -0.148398 \\
 0. & 0.000610624 & -0.00669703 & -0.999958 & 0.00627442 \\
 0. & 0.0725382 & 0.138932 & 0.00531088 & 0.987627 \\
\end{array}
\right)
\end{split}
\end{equation}
The difference between the mixing matrices can be easily seen by subtracting one from another after taking absolute value.
\begin{equation}
\begin{split}
\lvert V_{\func{ana}}^{L} \rvert - \lvert V_{\func{num}}^{L} \rvert
&=
\scalemath{0.9}{
\left(
\begin{array}{ccccc}
 0 & 0 & 0 & 0 & 0 \\
 0 & 0 & -7.58945\times 10^{-8} & 2.51664\times 10^{-7} & 5.97371\times 10^{-10} \\
 0 & 3.46904\times 10^{-10} & -7.19169\times 10^{-6} & 2.38502\times 10^{-5} & -1.00783\times 10^{-7} \\
 0 & -3.95207\times 10^{-6} & 2.35474\times 10^{-5} & -7.10253\times 10^{-6} & 3.68376\times 10^{-7} \\
 0 & 9.97344\times 10^{-9} & 1.29353\times 10^{-7} & 2.03736\times 10^{-7} & -1.91205\times 10^{-9} \\
\end{array}
\right)}
\\
\lvert V_{\func{ana}}^{e} \rvert - \lvert V_{\func{num}}^{e} \rvert
&=
\scalemath{0.9}{
\left(
\begin{array}{ccccc}
 0 & 0 & 0 & 0 & 0 \\
 0 & -2.39211\times 10^{-8} & 4.58026\times 10^{-8} & 9.72974\times 10^{-10} & 3.25661\times 10^{-7} \\
 0 & -1.06479\times 10^{-6} & -2.03783\times 10^{-6} & -1.29588\times 10^{-7} & 1.44968\times 10^{-5} \\
 0 & 5.47244\times 10^{-8} & -2.43141\times 10^{-7} & 0 & 2.58046\times 10^{-7} \\
 0 & 2.50398\times 10^{-6} & 1.42997\times 10^{-5} & 1.84594\times 10^{-7} & -2.19659\times 10^{-6} \\
\end{array}
\right)}
\end{split}
\end{equation}
Therefore we have confirmed that the analytic mixing matrix for the charged lepton sector is quite close to one obtained from
the numerical SVD diagonalization. Using the numerical mixing matrices derived by the SVD diagonalization, we confirm the following $D_{L,R}^{e \prime}$ matrices of $Z$ couplings with leptons 
of Equation~\ref{eqn:simple_DLep_DRep} as follows (we included here the pre-factor $g/c_w$):
\begin{equation}
\begin{split}
D_{L}^{e \prime} 
&=
\scalemath{0.9}{
\left(
\begin{array}{ccccc}
 -2.01645\times 10^{-1} & 0. & 0. & 0. & 0. \\
 0. & -2.01643\times 10^{-1} & 4.22223\times 10^{-6} & 5.1508\times 10^{-6} & 7.70341\times 10^{-4} \\
 0. & 4.22223\times 10^{-6} & -2.01634\times 10^{-1} & 1.33333\times 10^{-5} & 1.9941\times 10^{-3} \\
 0. & 5.1508\times 10^{-6} & 1.33333\times 10^{-5} & -2.01629\times 10^{-1} & 2.43264\times 10^{-3} \\
 0. & 7.70341\times 10^{-4} & 1.9941\times 10^{-3} & 2.43264\times 10^{-3} & 1.62175\times 10^{-1} \\
\end{array}
\right)}
\\
D_{R}^{e \prime}
&=
\scalemath{0.9}{
\left(
\begin{array}{ccccc}
 1.62204\times 10^{-1} & 0. & 0. & 0. & 0. \\
 0. & 1.62203\times 10^{-1} & 3.60409\times 10^{-6} & 4.87783\times 10^{-4} & -2.59066\times 10^{-6} \\
 0. & 3.60409\times 10^{-6} & 1.62184\times 10^{-1} & -2.68815\times 10^{-3} & 1.4277\times 10^{-5} \\
 0. & 4.87783\times 10^{-4} & -2.68815\times 10^{-3} & -2.01614\times 10^{-1} & 1.93227\times 10^{-3} \\
 0. & -2.59066\times 10^{-6} & 1.4277\times 10^{-5} & 1.93227\times 10^{-3} & 1.62194\times 10^{-1} \\
\end{array}
\right)}
\end{split}
\end{equation}
\section{Numerical comparison for the quark sector} \label{app:E}
As we did in Appendix~\ref{app:D}, we carry out the same approach with the most converged numerical point ($\chi_{\func{CKM}}^2 = 956.828$) for the up- and down-type quark sectors. 
\begin{equation}
\begin{split}
M^{u} &= 
\begin{pmatrix}
0 & 0 & 0 & 0 & 0 \\[0.5ex]
0 & 0 & 0 & 0 & 14.474 \\[0.5ex]
0 & 0 & 0 & 1206.340 & 277.563 \\[0.5ex]
0 & 0 & 273.503 & -1775.200 & 0 \\[0.5ex]
0 & 550.990 & 434.462 & 0 & -5624.050 
\end{pmatrix}
\\
M^{d} 
&= 
\begin{pmatrix}
0 & 0 & 0 & 0 & -0.938 \\[0.5ex]
0 & 0 & 0 & 0 & -4.041 \\[0.5ex]
0 & 0 & 0 & 1206.340 & -27.427 \\[0.5ex]
0 & 0 & -5.636 & -1775.200 & 0 \\[0.5ex]
0 & 72.915 & -75.760 & 0 & 2623.620
\end{pmatrix}
\end{split}
\end{equation}
For the comparison, we find the analytic mixing matrices $(V_{L,R}^{u,d})_{\func{ana}}$ and the numerical mixing matrices $(V_{L,R}^{u,d})_{\func{num}}$ and then subtract one from another after taking absolute value. The numerical differences are given by:
\begin{equation}
\begin{split}
\lvert (V_{L}^{u})_{\func{ana}} \rvert - \lvert (V_{L}^{u})_{\func{num}} \rvert 
&= 
\scalemath{0.9}{
\left(
\begin{array}{ccccc}
 0 & 0 & 0 & 0 & 0 \\
 0 & -9.63899\times 10^{-10} & -8.4547\times 10^{-6} & 1.12194\times 10^{-5} & 5.10645\times 10^{-7} \\
 0 & 8.60152\times 10^{-7} & -4.562\times 10^{-3} & 6.12434\times 10^{-3} & -3.20031\times 10^{-4} \\
 0 & -4.32016\times 10^{-5} & 6.11329\times 10^{-3} & -4.57314\times 10^{-3} & 3.70205\times 10^{-4} \\
 0 & -6.95382\times 10^{-8} & 3.44869\times 10^{-5} & 4.09197\times 10^{-5} & -2.39443\times 10^{-6} \\
\end{array}
\right)}
\\
\lvert (V_{R}^{u})_{\func{ana}} \rvert - \lvert (V_{R}^{u})_{\func{num}} \rvert 
&=
\scalemath{0.9}{
\left(
\begin{array}{ccccc}
 0 & 0 & 0 & 0 & 0 \\
 0 & -5.36541\times 10^{-8} & 4.20284\times 10^{-8} & 3.73725\times 10^{-8} & 4.77149\times 10^{-7} \\
 0 & -1.56456\times 10^{-4} & -4.71406\times 10^{-5} & -7.74521\times 10^{-4} & 1.34741\times 10^{-3} \\
 0 & 1.27331\times 10^{-4} & -7.96888\times 10^{-4} & 6.8947\times 10^{-5} & 5.76999\times 10^{-5} \\
 0 & 1.86671\times 10^{-4} & 1.32945\times 10^{-3} & 2.03335\times 10^{-5} & -1.39892\times 10^{-4} \\
\end{array}
\right)}
\\
\lvert (V_{L}^{d})_{\func{ana}} \rvert - \lvert (V_{L}^{d})_{\func{num}} \rvert 
&=
\scalemath{0.8}{
\left(
\begin{array}{ccccc}
 0 & 0 & 0 & 0 & 0 \\
 7.62437\times 10^{-9} & 3.28418\times 10^{-8} & -7.80564\times 10^{-7} & -5.52416\times 10^{-7} & 9.20155\times 10^{-9} \\
 -2.16144\times 10^{-7} & -9.31034\times 10^{-7} & -2.34437\times 10^{-6} & 3.16788\times 10^{-6} & -5.46568\times 10^{-8} \\
 -2.49349\times 10^{-8} & -1.07406\times 10^{-7} & 3.14516\times 10^{-6} & -2.37145\times 10^{-6} & 9.52623\times 10^{-8} \\
 0 & 3.34474\times 10^{-10} & 5.01567\times 10^{-8} & 3.51026\times 10^{-8} & -1.24719\times 10^{-9} \\
\end{array}
\right)}
\\
\lvert (V_{R}^{d})_{\func{ana}} \rvert - \lvert (V_{R}^{d})_{\func{num}} \rvert 
&=
\scalemath{0.9}{
\left(
\begin{array}{ccccc}
 0 & 0 & 0 & 0 & 0 \\
 0 & 6.97453\times 10^{-6} & -4.09035\times 10^{-5} & -7.75726\times 10^{-8} & 1.54578\times 10^{-6} \\
 0 & -4.13664\times 10^{-5} & 6.52624\times 10^{-6} & -1.24584\times 10^{-7} & 1.44401\times 10^{-5} \\
 0 & 6.55855\times 10^{-8} & -3.67395\times 10^{-7} & -3.73704\times 10^{-10} & 6.10094\times 10^{-8} \\
 0 & 2.63961\times 10^{-6} & 1.5191\times 10^{-5} & 4.55174\times 10^{-8} & -5.58627\times 10^{-7} \\
\end{array}
\right)}
\end{split}
\end{equation} 
Here we can see that the differences for the charged lepton or down-type quark sectors are at most of the order of maximally order of 
$10^{-5}$, whereas the maximal difference for the up-quark sector goes up to the order of $10^{-3}$ due to the sizeable off-diagonal $\mathcal{O}(1)$ element but it is still a good approximation. Now we confirm that the numerical matrices of $Z$ couplings with quarks 
$D_{L,R}^{u,d \prime}$ in the mass basis are given by: 
\begin{equation}
\begin{split}
D_{L}^{u \prime} 
&= 
\scalemath{0.9}{
\left(
\begin{array}{ccccc}
 2.55713\times 10^{-1} & 0 & 0 & 0 & 0 \\
 0 & 2.55711\times 10^{-1} & -3.14988\times 10^{-5} & 3.18699\times 10^{-5} & -7.55591\times 10^{-4} \\
 0 & -3.14988\times 10^{-5} & 2.55083\times 10^{-1} & 6.37525\times 10^{-4} & -1.51148\times 10^{-2} \\
 0 & 3.18699\times 10^{-5} & 6.37525\times 10^{-4} & 2.55068\times 10^{-1} & 1.52929\times 10^{-2} \\
 0 & -7.55591\times 10^{-4} & -1.51148\times 10^{-2} & 1.52929\times 10^{-2} & -1.06859\times 10^{-1} \\
\end{array}
\right)}
\\
D_{R}^{u \prime}
&=
\scalemath{0.9}{
\left(
\begin{array}{ccccc}
 -1.08136\times 10^{-1} & 0 & 0 & 0 & 0 \\
 0 & -1.07941\times 10^{-1} & -7.52723\times 10^{-4} & 8.3786\times 10^{-3} & 1.3608\times 10^{-4} \\
 0 & -7.52723\times 10^{-4} & -1.05225\times 10^{-1} & -3.24002\times 10^{-2} & -5.26224\times 10^{-4} \\
 0 & 8.3786\times 10^{-3} & -3.24002\times 10^{-2} & 2.52512\times 10^{-1} & 5.85742\times 10^{-3} \\
 0 & 1.3608\times 10^{-4} & -5.26224\times 10^{-4} & 5.85742\times 10^{-3} & -1.08041\times 10^{-1} \\
\end{array}
\right)}
\\
D_{L}^{d \prime}
&=
\scalemath{0.9}{
\left(
\begin{array}{ccccc}
 -3.09781\times 10^{-1} & 0 & 0 & 0 & 0 \\
 0 & -3.0978\times 10^{-1} & 4.35109\times 10^{-6} & 8.49277\times 10^{-6} & 4.62349\times 10^{-4} \\
 0 & 4.35109\times 10^{-6} & -3.09749\times 10^{-1} & 6.287\times 10^{-5} & 3.42266\times 10^{-3} \\
 0 & 8.49277\times 10^{-6} & 6.287\times 10^{-5} & -3.09658\times 10^{-1} & 6.68059\times 10^{-3} \\
 0 & 4.62349\times 10^{-4} & 3.42266\times 10^{-3} & 6.68059\times 10^{-3} & 5.39124\times 10^{-2} \\
\end{array}
\right)}
\\
D_{R}^{d \prime}
&=
\scalemath{0.9}{
\left(
\begin{array}{ccccc}
 5.4068\times 10^{-2} & 0 & 0 & 0 & 0 \\
 0 & 5.40678\times 10^{-2} & 4.0453\times 10^{-7} & 2.15284\times 10^{-4} & -3.09181\times 10^{-6} \\
 0 & 4.0453\times 10^{-7} & 5.40667\times 10^{-2} & -6.8355\times 10^{-4} & 9.81685\times 10^{-6} \\
 0 & 2.15284\times 10^{-4} & -6.8355\times 10^{-4} & -3.09705\times 10^{-1} & 5.22435\times 10^{-3} \\
 0 & -3.09181\times 10^{-6} & 9.81685\times 10^{-6} & 5.22435\times 10^{-3} & 5.39929\times 10^{-2} \\
\end{array}
\right)}
\end{split}
\label{eqn:numerical_DLdp}
\end{equation}
where the pre-factor $g/c_w$ was included in those matrices. The most interesting case of Equation~\ref{eqn:numerical_DLdp} is $D_{L}^{d \prime}$ since we know that the left-handed down-type quark sector can access to all mixings among the three SM generations. The used numerical mixing matrix $V_{L}^{d}$ is given by:
\begin{equation}
V_{L}^{d}
=
\scalemath{0.9}{
\left(
\begin{array}{ccccc}
 9.74095\times 10^{-1} & -2.26141\times 10^{-1} & 0 & 0 & 0 \\
 2.26\times 10^{-1} & 9.73488\times 10^{-1} & -2.81626\times 10^{-2} & -2.12226\times 10^{-2} & 1.27099\times 10^{-3} \\
 7.97153\times 10^{-3} & 3.43371\times 10^{-2} & 7.98097\times 10^{-1} & 6.01423\times 10^{-1} & 9.40883\times 10^{-3} \\
 -6.67815\times 10^{-6} & -2.87659\times 10^{-5} & 6.01585\times 10^{-1} & -7.98598\times 10^{-1} & 1.83648\times 10^{-2} \\
 -3.62201\times 10^{-4} & -1.56017\times 10^{-3} & -1.85253\times 10^{-2} & 9.03631\times 10^{-3} & 9.99786\times 10^{-1} \\
\end{array}
\right)},
\label{eqn:a_num_VCKM}
\end{equation}
and we can confirm all elements of the first row and column of $D_{L}^{d \prime}$ cancel each other, so identifying the given result in Equation~\ref{eqn:numerical_DLdp}. What we found in this Appendix verifies the fact that the SM $Z$ physics does not get affected by any specific choice of basis.
\backmatter


\end{document}